\documentclass[a4paper,fleqn,longmktitle]{cas-sc}
\usepackage[numbers]{natbib}
\usepackage{import}
\usepackage{enumitem}
\usepackage{graphicx}
\usepackage{epstopdf}
\usepackage{amsmath,amssymb}
\usepackage{braket}
\usepackage{hyperref}
\usepackage{empheq}
\usepackage{bm}
\usepackage{float}
\usepackage{soul} 
\usepackage{subcaption}

\numberwithin{equation}{section}
\usepackage{chngcntr}
\counterwithin{figure}{section}

\usepackage{titlesec}
\titleformat{\section}{\bfseries\large}{\thesection.}{0.5em}{}
\titleformat{\subsection}{}{\thesubsection}{1em}{\itshape}
\titleformat{\subsubsection}{}{\thesubsubsection}{1em}{\itshape}

\titlespacing*{\subsection}
{0pt}{5.5ex plus 1ex minus .1ex}{2.3ex plus .1ex}

\def\tsc#1{\csdef{#1}{\textsc{\lowercase{#1}}\xspace}}
\tsc{WGM}
\tsc{QE}
\tsc{EP}
\tsc{PMS}
\tsc{BEC}
\tsc{DE}

\DeclareMathOperator*{\SumInt}{%
\mathchoice%
  {\ooalign{$\displaystyle\sum$\cr\hidewidth$\displaystyle\int$\hidewidth\cr}}
  {\ooalign{\raisebox{.14\height}{\scalebox{.7}{$\textstyle\sum$}}\cr\hidewidth$\textstyle\int$\hidewidth\cr}}
  {\ooalign{\raisebox{.2\height}{\scalebox{.6}{$\scriptstyle\sum$}}\cr$\scriptstyle\int$\cr}}
  {\ooalign{\raisebox{.2\height}{\scalebox{.6}{$\scriptstyle\sum$}}\cr$\scriptstyle\int$\cr}}
}

\begin{document}

\let\WriteBookmarks\relax
\def\floatpagepagefraction{1}
\def\textpagefraction{.001}
\shorttitle{Superconductivity in high-$T_c$ and related strongly correlated systems \ldots}
\shortauthors{J. Spa{\l}ek et~al.}

\title[mode = title]{Superconductivity in high-$T_c$ and related strongly correlated systems from variational perspective: Beyond mean field theory}                      

\author[1]{J. Spa{\l}ek}[type=author,
                        auid=000,bioid=1,
                        prefix=,
                        role=,
                        orcid=0000-0003-3867-8493]
\cormark[1]
\ead{jozef.spalek@uj.edu.pl}

\author[1]{M. Fidrysiak}[type=author,
                        auid=000,bioid=1,
                        prefix=,
                        role=,
                        orcid=0000-0003-2097-5828]

\author[2]{M. Zegrodnik}[type=author,
                        auid=000,bioid=1,
                        prefix=,
                        role=,
                        orcid=0000-0001-5801-5174]

\author[2]{A. Biborski}[type=author,
                        auid=000,bioid=1,
                        prefix=,
                        role=,
                        orcid=0000-0002-8372-5562]

\address[1]{Institute of Theoretical Physics, Jagiellonian University, ul. {\L}ojasiewicza 11, 30-348 Krak{\'o}w, Poland}

\address[2]{Academic Centre for Materials and Nanotechnology, AGH University of Science and Technology, \\ Al.~Mickiewicza~30,~30-059~Krak\'{o}w,~Poland}


\begin{abstract}
The principal purpose of this topical review is to single out some of the universal features of high-temperature (high-$T_c$) and related strongly-correlated systems, which can be compared with experiment in a quantitative manner. The description starts with the concept of exchange-interaction-mediated (real-space) pairing, combined with strong correlations among narrow band electrons, for which the reference state is that of the Mott-Hubbard insulator. The physical discussion of concrete properties relies on variational approach, starting from generalized renormalized mean-field theory (RMFT) in the form of statistically-consistent Gutzwiller approximation (SGA), and its subsequent generalization, i.e., a systematic Diagrammatic Expansion of the Variational (Gutzwiller-type) Wave Function (DE-GWF). The solution leads to the two energy scales, one involving unconventional quasiparticles close to the Fermi energy, and the other reflecting the fully correlated state, involving electrons deeper below the Fermi surface. Those two regimes are separated by a kink in the dispersion relation, which is observed in photoemission. As a result, one obtains both the doping dependent properties in the correlated state, and effective renormalized quasiparticles. The reviewed ground-state characteristics for high-$T_c$ systems encompass high-$T_c$ superconductivity, nematicity, charge- (and pair-) density-wave effects, as well as non-BCS kinetic energy gain in the paired state. The calculated dynamic properties are: the universal Fermi velocity, the Fermi wave-vector, the effective mass enhancement, the pseudogap; and the $d$-wave gap magnitude. We discuss that, within the variational approach, the minimal realistic model is represented by the so-called $t$-$J$-$U$ Hamiltonian. Inadequacy of the $t$-$J$ and Hubbard models, particularly of their RMFT versions, is discussed explicitly. For heavy fermion systems, modeled by the \textit{Anderson lattice model} and with the DE-GWF approach, we discuss the phase diagram, encompassing superconducting, Kondo insulating, ferro- and anti-ferromagnetic states. The superconducting state is then a two-$d$-wave gap system. If the orbital degeneracy of $f$-electrons is included, the coexistent ferromagnetic- (spin-triplet) superconducting phases appear and match those observed for $\mathrm{UGe_2}$ in a semiquantitative manner. Finally, in the second part, we generalize our approach to the collective spin and charge fluctuations in high-$T_c$ systems, starting from variational approach, defining the saddle-point state, and combined with $1/N$ expansion. The present scheme differs essentially from that starting from the saddle-point Hartree-Fock approximation, and incorporating the fluctuations in the random phase approximation (RPA). The spectrum of collective spin and charge excitations is determined for the Hubbard and $t$-$J$-$U$ models, and subsequently compared quantitatively with recent experiments. The Appendices provide formal details to make this review self-contained.
\end{abstract}


\maketitle

\tableofcontents

\section{Introduction and fundamental features of strongly correlated many-particle systems} \label{sec:introduction}

\subsection{Scope of the review}

The principal results discussed in the present report are:

\begin{enumerate}
\item We overview the proposed systematic extension of the mean-field theory, starting form a \emph{renormalized mean-field theory} (RMFT) in its \emph{statistically consistent version} (SGA). The obtained results for static properties in the superconducting state are compared directly and in quantitative manner with experiment.
\item The theoretical method is based on \emph{diagrammatic expansion of the Gutzwiller} (or related) \emph{variational wave function} for superconducting phase which leads to convergence in the third-to-fifth order of expansion and is also in good agreement with the results of variational Monte-Carlo.
\item Within the method the physical state, whenever possible, is shown to be described by two-stage description: (\emph{i}) The self-consistently determined wave uncorrelated wave function which describes the quasiparticle properties close to the Fermi energy and pseudogap, and (\emph{ii}) the fully correlated wave function, describing the macroscopic equilibrium properties, such as the $d$-wave-type superconducting gap in the nodal direction and other properties.
\item The model selected is either of the single-band nature, the so-called $t$-$J$-$U$-$(V)$ model, or the three-band $d$-$p_{x/y}$ model. The relation between the two is discussed in the physical terms. In the former case, the Hubbard- and $t$-$J$-mode limits are discussed as particular limits and their difference with the results obtained from the general $t$-$J$-$U$ model case compared.
\item The collective dynamic excitations (\emph{paramagnons} and \emph{plasmons}) are treated within the variational wave function + $1/N$ expansion in the first nontrivial order (large-$N$ approximation). The results agree in a fully quantitative manner with those obtained experimentally from either resonant inelastic $x$-ray (RIXS) or magnetic neutron scattering. In other words, the agreement is achieved by starting from the saddle-point (SGA) variational solution for the correlated state. The comparison with experiment, both here and in point 2., are carried out for a single set of microscopic parameters and this circumstance allows for a discussion of universality withing the cuprate family.
\item Other strongly correlated systems: heavy-fermions, spin-triplet superconductors (on the example of $\mathrm{UGe_2}$), and twisted bilayer-graphene, are also briefly discussed, within the method described in point 2 in its simplest-SGA-version.
\end{enumerate}

\subsection{Relation to other approaches: A brief overview}

The interest in unconventional mechanisms of pairing was stimulated decisively by discoveries of high-temperature (high-$T_c$) superconductivity in $\mathrm{La_{1-\delta}Ba_{\delta}CuO_4}$ (LBCO, 214) \cite{BednorzZPhysB1986}, $\mathrm{\mathrm{La_{2-\mathit{x}}Sr_\mathit{x}CuO_4}}$ \cite{KishioChemLett1987,CavaPhysRevLett1987} and in $\mathrm{YBa_2Cu_3O_{7-\delta}}$ (YBCO, 123) \cite{WuPhysRevLett1987} systems. In this respect, the ideas of Anderson \cite{AndersonScience1987,AndersonProcIntSchollEnricoFermi1988} about the role of strong correlations and their modeling via Hubbard \cite{HubbardProcRoySoc1963} or $t$-$J$ \cite{ChaoJPCM1977,SpalekPhysRevB1988_2,ZhangPhysRevB1988} models, as well as the proposal of the $d$-\emph{wave} form of superconducting gap \cite{BickersIntJModPhys1987,MonthouxPhysRevLett1991,ShenPhysRevLett1993,InuiPhysRevB1988} with an essentially two-dimensional character, forming in a square $\mathrm{CuO_2}$ lattice, provided the basis for an intensive analysis of those systems, both in normal and superconducting states. Both the Gutzwiller variational (GA) \cite{BaskaranSolStateCommun1987} and slave-boson (SBA) \cite{RuckensteinPhysRevB1987} approaches have been applied shortly after the discoveries and turned out to be equivalent if GA is reformulated to the statistically-consistent (SGA) form \cite{JedrakPhsRevB2011}. Parenthetically, the SGA version of what we call hereinafter \emph{mean-field theory}, contains only self-consistent averages of physical fields, in distinction to SBA, which in turn contains auxiliary Bose fields, allowing for their spurious Bose-Einstein condensation, at least at the saddle-point level.

The first reviews of the pairing based on GA \cite{AndersonJPhysCondensMatter2004,EdeggerAdvPhys2007,OgataRepProgPhys2008,RanderiaBook2011} and SBA \cite{LeeRevModPhys2006,LeeRepProgPhys2007} concepts and their achievements involve two results. First, it has become evident that within GA (and SGA) one can define two gaps: first corresponding to the uncorrelated wave function reference state, and the second called correlated gap in the correlated state. Second, within SBA, an illustrative phase diagram was constructed involving both the so-called superconducting dome and the non-Fermi (non-Landau) fermionic liquid in the normal state. Another class of models (e.g., spin-fermion models) are based on spin-fluctuation picture, with an early review based on RPA-type of approach and its extensions to the fluctuations \cite{ChubukovChapter2003,AbanovAdvPhys2003,PlakidaPhysicaC2016,ScalapinoRevModPhys2012} or on the so-called self-consistent renormalization theory \cite{MoriyaAdvPhys2000}. Related to these approaches is that based on the equation of motion method for the Green functions and immediately connected with it decoupling scheme \cite{PlakidaBook2010,FengIntJModPhysB2015,KoikegamiJPCM2021}. In this method, the equations for the single- and two-particle Green functions are self-consistently decoupled and, in principle, should describe both static and collective properties. This approach has been overviewed separately and qualitative feature of results detailed.

It should be underlined that a crucial step has been undertaken in the last decade with the advancements of resonant inelastic $x$-ray technique (RIXS) [for detailed references, see Sec.~\ref{sec:fluctuations}] which, in conjunction with magnetic neutron scattering, provided a quantitative characterization of quantum fluctuations of spin and charge character (paramagnons and plasmons, respectively). The existence of well defined paramagnons can be immediately related to the fundamental role of exchange interaction and strong correlations, as discussed later, whereas specific properties of plasmons require invoking of long-range (three-dimensional) Coulomb interaction, in addition to strong short-range correlations (cf. Sec.~\ref{sec:fluctuations}). Incorporation of collective effects into a unified picture and with the same or close values of microscopic parameters as those taken when comparing theory to experiment, forms in our view, a prerequisite of a consistent macroscopic theory, particularly when reliable electronic structure, obtained earlier, is reproduced properly at the same time.

In brief, here we return to the original variational approach, starting from SGA as a properly defined renormalized mean-field theory and overview the systematic expansion of the variational wave function beyond the mean-field level, using a specially designed diagrammatic expansion (DE-GWF method, cf. Sec.~\ref{sec:vwf_solution}). Furthermore and foremost, we compare selected theoretical DE-GWF results for the cuprates with experiment in a fully quantitative manner (cf. Sec.~\ref{sec:selected_equilibrium_properties}). Those results are obtained for a  single set of microscopic parameters to test their consistency and degree of universality. The latter task requires also discussing how to select a proper microscopic Hamiltonian and check its applicability in describing the data within the DE-GWF analysis. Therefore, we have decided to employ the general $t$-$J$-$U$ model, which formally encompasses both the $t$-$J$ and Hubbard models as limiting situations. It is also essential to stress two additional aspects of the review. First, we have selectively overviewed application of the variational scheme to selected other correlated systems (cf. Sec.~\ref{sec:related_correlated_systems}). Second, we subsequently review the extension of the DE-GWF method and apply it to extensively studied collective quantum excitations (fluctuations): paramagnons and plasmons within DE-GWF+$1/N$ method, cf. Sec.~\ref{sec:fluctuations}. In brief, Secs.~\ref{sec:vwf_solution}-\ref{sec:fluctuations}, together with Appendices~\ref{appendix:derivation_of_the_tj_model}-\ref{appendix:de-gwf}, represent the core of this report.

At the end of this introductory section we would like to emphasize that our review does not address by any means all other methods, such as determinant quantum Monte-Carlo (DQMC) or functional renormalization group (fRG), and others. Instead, as the reviewed approach applies to practically infinite lattices, we have concentrated on its applicability in describing concrete experimental results in a quantitative manner. In other words, this review cannot be regarded as an exhaustive overview of the whole subject from a methodological perspective, which would be an overly ambitious task, given the tremendous theoretical effort put into the field  over the past 30 years, containing well over $10\,000$ papers. Due to the abundance of reasonable (yet approximate to lesser or larger degree) theoretical results, we believe that the task, formulated above and based on turning to concrete experimental results, is worth undertaking, if not indispensable at this point of field advanced developments.

\subsection{Generalities: What is a correlated system? Universal characteristics}

Conceptually, the main overall features of the correlated systems can be characterized briefly by the following characteristics.

\begin{enumerate}
\item The ground state energy of a periodic condensed system of fermions can be described by starting from the system atomic configuration and subsequently adding other dynamic interactions which appear in the emerging condensed state. Namely, its energy per atomic state can be expressed in the form

  \begin{align}
    \frac{E_G}{N} = \epsilon_a + \langle T \rangle + \langle V \rangle + \langle V_{1, 2} \rangle \equiv E_1 + E_2,
  \end{align}

  \noindent
  where $\epsilon_a$ is the single-particle in an atomic (Wannier) state, $\langle T\rangle$ and $\langle V\rangle$ are the average kinetic and potential energies, whereas $\langle V_{12}\rangle$ is the expectation value of the two-particle interaction. Thus, the single-particle part $E_1$ contains the first three terms, and $E_2 \equiv \langle V_2\rangle$. In the periodic system we will be assuming that $\epsilon_a \approx 0$; it acquires a constant (reference) value and will be often disregarded unless stated explicitly. In this manner, the remaining terms characterize solely the contributions in the condensed state. Also, note that usually $E_1 < 0$. Now, one can define three physically distinct situations:

  \begin{enumerate}[label=\arabic*${}^{\circ}$]
  \item $|E_1| \gg E_2$: Fermi-liquid (metallic) regime,
  \item $|E_1| \sim E_2$: localization-delocalization (Anderson-Mott-Hubbard) regime,
   \label{item:correlation_limit}
  \item $|E_1| \ll E_2$: strong-correlation (Mott) regime.
  \end{enumerate}

Here we focus almost exclusively on the regimes $2^{\circ}$ and $3^{\circ}$, which favors atomic (Wannier) representation of the states and interactions, whereas in the situation $1^{\circ}$ the starting point is that described by a gas of fermions (or Landau Fermi liquid) and associated with them momentum representation of the states and the Fermi-Dirac statistics both in their canonical form.

  \begin{figure}
    \centering
    \includegraphics[width=\textwidth]{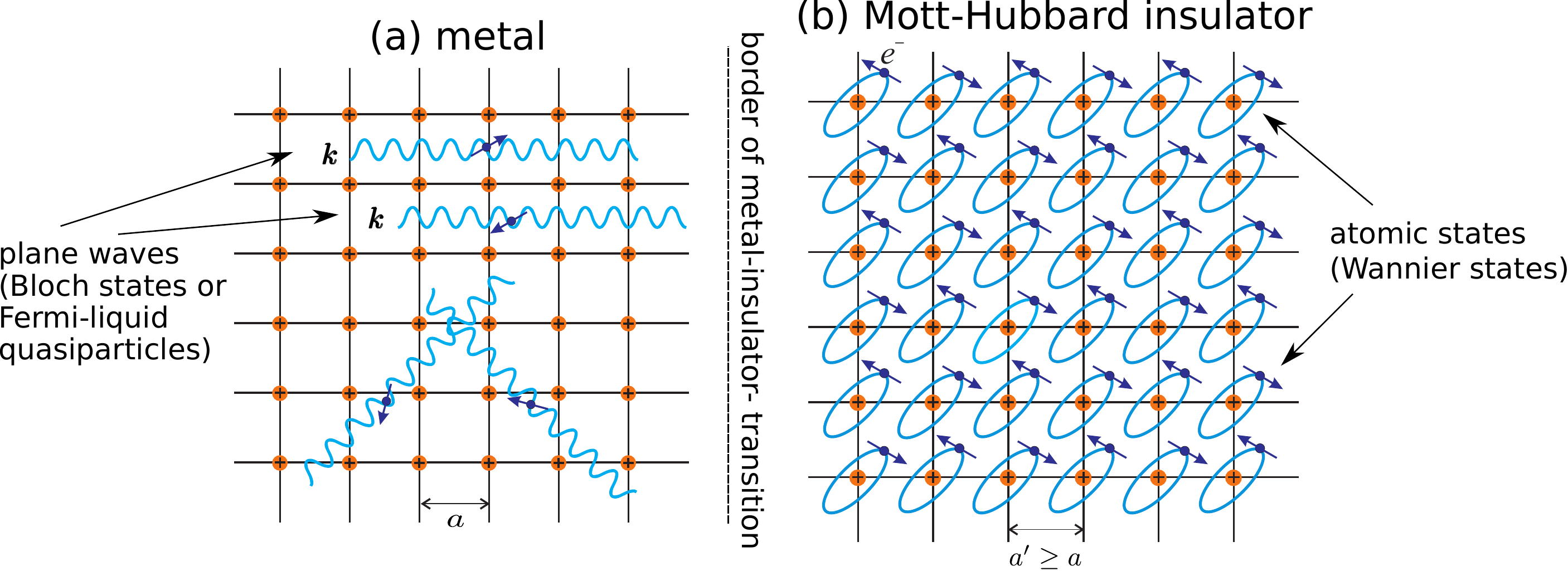}.
    \caption{Schematic representation of the metallic (a) and quasi-atomic (Mott-insulating) states of a planar lattice with one valence electron per atom (b). The Mott-Hubbard (metal-insulator) boundary is marked in the middle. The localized (insulating) state (b) is that of antiferromagnetic insulator (AFI). The transition is usually discontinuous.}
    \label{fig:metal-isulator-transition}
  \end{figure}

\item The above division into the three regimes is illustrated in Fig.~\ref{fig:metal-isulator-transition}, where their complementary nature in the quantum-mechanical sense is represented on an example of a solid with metallic (delocalized) state of electrons (a) or correlated (Mott or Mott-Hubbard) state (b) with one valence electron per parent atom. Additionally, we have marked the dividing line (Mott-Hubbard boundary) between the two states. Important remark should be provided already here. First, the momentum representation is described by Bloch functions $\left\{\Psi_{\mathbf{p}\sigma}(\mathbf{r})\right\}$ of particle with (quasi)momentum $\mathbf{p} = \hbar \mathbf{k}$ and the spin quantum number $\sigma = \pm 1 \equiv \uparrow, \downarrow$, whereas the position representation is expressed by the corresponding set of Wannier states $\left\{w_{i\sigma}(\mathbf{r})\right\}$ with atomic position $i \equiv \mathbf{R}_i$ as quantum number, in the single-band (single-orbital) situation. These two representations are usually regarded as equivalent in the sense that they are related by the lattice Fourier transformation. However, in the situation depicted in Fig.~\ref{fig:metal-isulator-transition}, when we have a sharp boundary (usually first-order line) between the states shown in (a) and (b), this representation equivalence is broken. The macroscopic state (a) near the transition is represented, strictly speaking, by a modified Landau-Fermi liquid (the so-called \emph{almost localized Fermi liquid}), whereas the Mott-insulating state is well accounted for by that of the localized-spin (Heisenberg) antiferromagnet \cite{SpalekPhysRevLett1987,AndersonPhysRev1959,AndersonSolidStatePhysics1963}. The transition is usually of discontinuous (first-order) character.

  \begin{figure}
    \centering
    \includegraphics[width=0.8\textwidth]{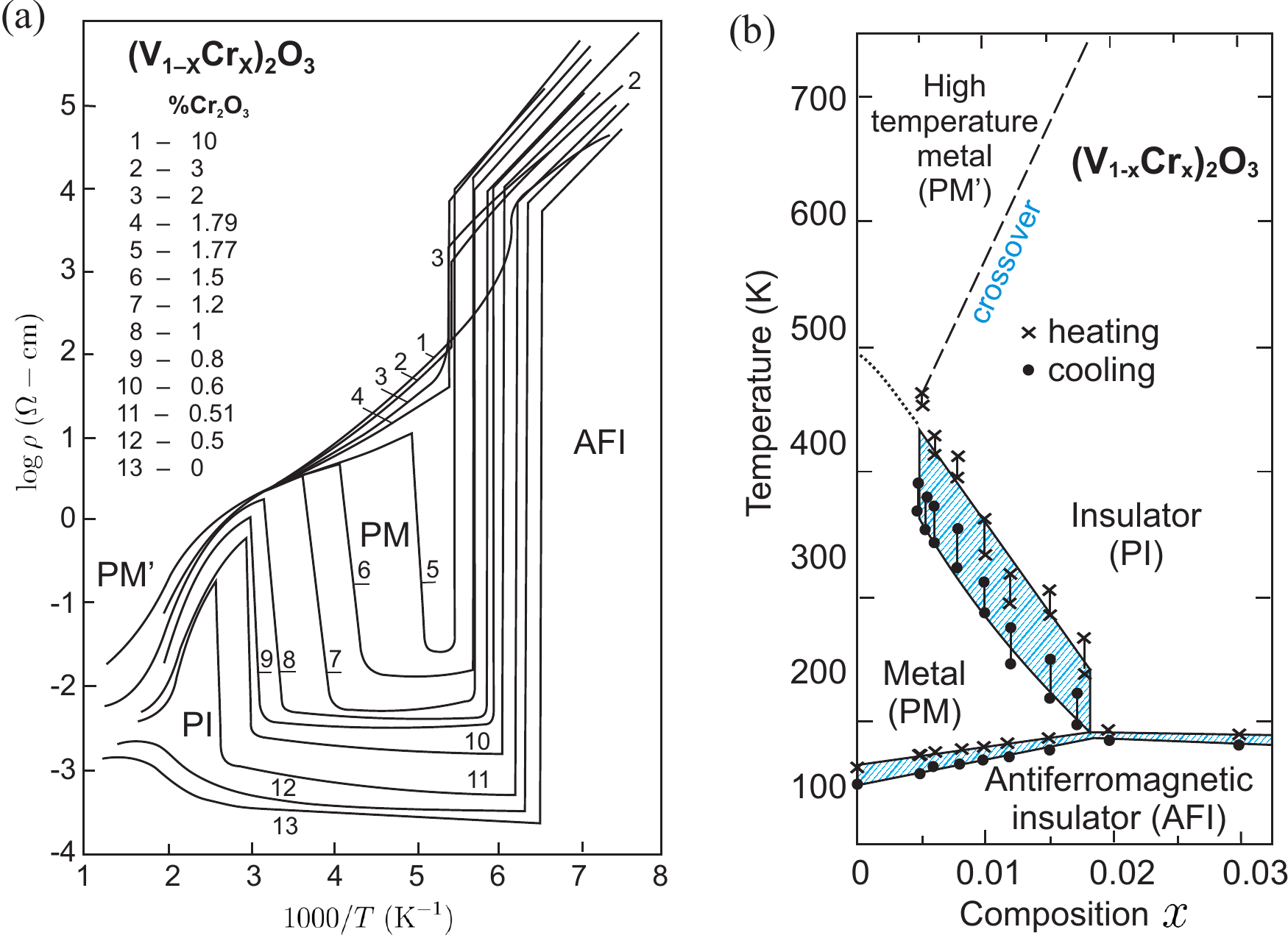}.
    \caption{(a) Temperature dependence of the electrical resistivity (in logarithmic scale) vs. $1/T$ for lightly Cr-doped $\mathrm{V_2}\mathrm{O_3}$. A very sharp transition from antiferromagnetic insulating (AFI) to paramagnetic metallic (PM) phase is followed by a reverse PM $\rightarrow$ paramagnetic insulator (PI) at higher temperature, which in turn is followed by PI $\rightarrow$ PM' crossover transition to a reentrant metallic (PM') phase at even higher temperatures; (b) phase diagram for the same system on $T$-$x$ plane; the hatched region depicts the hysteretic behavior accompanying the discontinuous transitions (taken from Ref.~\cite{KuwamotoPhysRevB1980}, with small modifications). Both AFI $\rightarrow$ PM and PM $\rightarrow$ PI transitions represent examples of the Mott-Hubbard transition (see main text).}
    \label{fig:1.3}
  \end{figure}

  From the above qualitative picture, one can infer that with approaching metal$\rightarrow$insulator (delocalization$\rightarrow$localization) boundary with formation of the localized-spin state, the kinetic energy of the renormalized-by-interaction particle progressive motion throughout the system is drastically reduced and in the insulating state it reduces to zero. Effectively, one can say that the Landau quasiparticle effective mass $m^{*} \rightarrow \infty$. This feature shows that strong enough inter-particle interactions (called in this context \emph{strong correlations}) limit the stability of the Landau-Fermi quasiparticle picture, as exemplified explicitly by the appearance of the Mott-Hubbard phase transition. Also, a proper quantitative description of the Mott insulator requires a model incorporating effective exchange interactions (kinetic exchange in the one-band case \cite{AndersonPhysRev1959,AndersonSolidStatePhysics1963} or superexchange in the multiple-orbital situation \cite{ZaanenJSolStateChem1990}). In the subsequent section we provide a quantitative analysis of these statements. The starting point of these considerations is the parametrized microscopic Hamiltonian analyzed briefly below (cf. also Appendix~\ref{appendix:hubbard_model}).

    \begin{figure}
    \centering
    \includegraphics[width=0.8\textwidth]{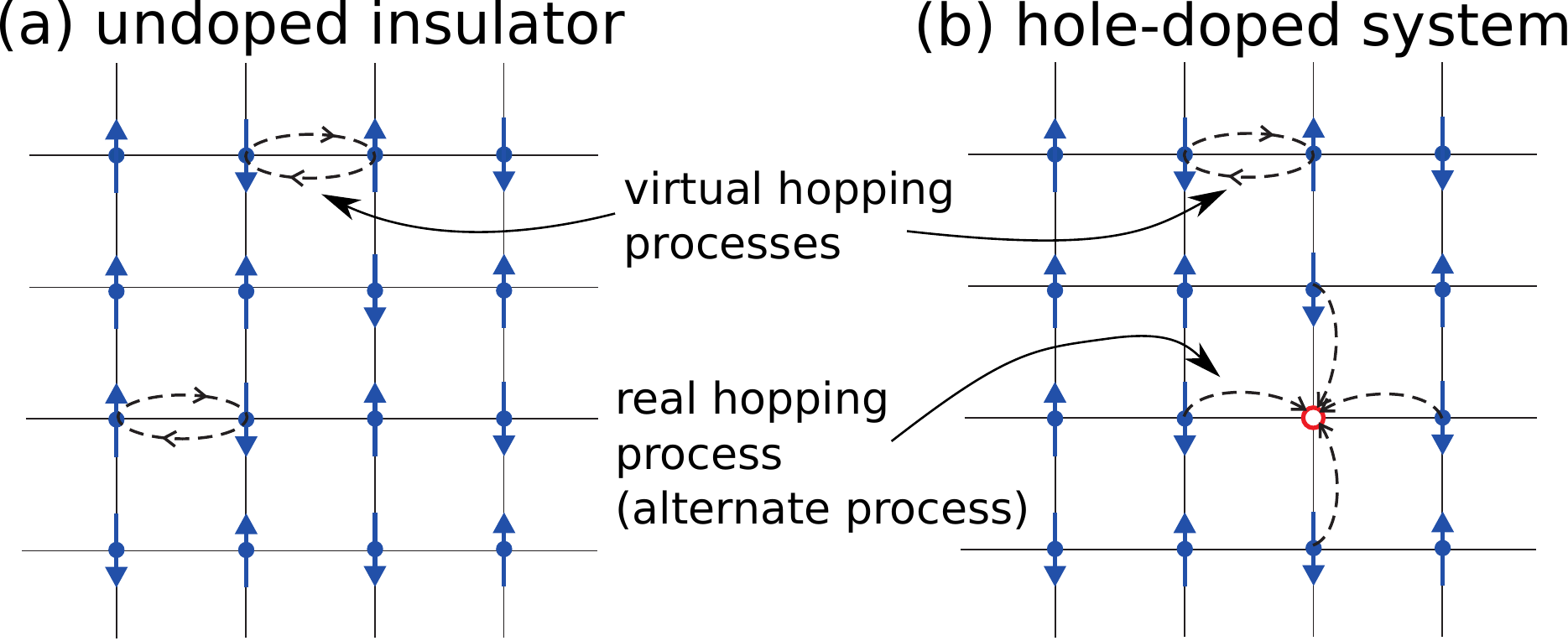}.
    \caption{Schematic representation of the particle dynamics in terms of hopping processes (dashed arrows) in the Mott-insulating state (a) and the strongly-correlated metal phase (b). Virtual hopping involves two consecutive direct hopping processes and occurs in the cases (a) and (b). The direct hopping results in real motion of holes and occurs in the strongly-correlated metal phase (b). In the strong-correlation regime, the direct hopping processes via doubly occupied configurations $|\uparrow\downarrow\rangle$ are precluded. In the last case we speak about extreme strong correlations.}
    \label{fig:hopping-processes}
  \end{figure}

\item Strictly speaking, \emph{the Mott-Hubbard transition} takes place when we have one electron per relevant correlated-valence orbital (the filling $n=1$), i.e., a half-filled band configuration when looking at it from the metallic side. The Mott insulating state, appearing in such a situation, is thus completely different from that of Bloch-Wilson band insulator, where the number of valence electrons per relevant orbital is $n = 2$ (even number in many-orbital situation) so the band is full and separated from other bands. This difference is exhibited explicitly by the circumstance that the Mott insulator has unpaired spins and thus is a magnetic (usually antiferromagnetic) insulator, whereas the Bloch-Wilson insulator is \emph{diamagnetic} (with zero net spin moment). A fundamental question is what happens if mobile holes are introduced into the Mott insulator, either by extrinsic doping or by self-doping. The situation is presented schematically in Fig.~\ref{fig:hopping-processes} on example of square lattice, where we mark virtual second-order hopping processes in the case of Mott insulator (a) and the hole-doped system (b). In the doped case (b), in addition to the virtual hopping, also the real hopping is admissible. It was shown first by Anderson \cite{AndersonPhysRev1959} that the virtual processes, depicted in (a), lead to the antiferromagnetic \emph{kinetic exchange} and, in consequence, to the \emph{antiferromagnetic ordering} in majority of Mott insulators. Those considerations have been subsequently generalized to the case of the doped insulator \cite{ChaoJPCM1977,SpalekPhysicaB1977} and to emerging $t$-$J$ or $t$-$J$-$U$ models of high-temperature superconductivity. The model plays a prominent role in this review.
\item One of the very specific features of the correlated systems, apart from Mott-Hubbard localization, is the \emph{spin-direction dependence of the heavy-quasiparticle mass}, first proposed theoretically \cite{SpaekPhysRevLett1990,KorbelPhysRevB1995} and subsequently observed experimentally \cite{SheikinPhysRevB2003,McCollamPhysRevLett2005}. This phenomenon has, among others \cite{SpalekPhysicaB2006}, two fundamental implications. First of them is connected with a substantial effect on superconductivity in large magnetic field and, in particular, on the second critical magnetic field and the appearance of the Fulde-Ferrel-Larkin-Ovchinnikov (FFLO) phase \cite{MaskaPhysRevB2010,KaczmarczykJPCM2010,KaczmarczykPhysRevB2011,KaczmarczykPhysRevB2009}. Namely, the FFLO phase is favored in a wider field ($H_a$) and temperature ($T$) range, in turn leading to the higher critical field $H_{c2}$ value and deviation from the conventional behavior in the low-$T$ range. The second implication is of more fundamental character and is associated with the circumstance that strong deviation from initially ($H_a =0$) spin-independent masses upon increasing $H_a$ transforms the system of quantum-mechanically \emph{indistinguishable} particles into their \emph{distinguishable} components. This issue is currently under investigation \cite{SpalekStudHistScien2020,SpalekInPreparation}. None of the two mentioned implications is dwelt upon any further in this report.
\end{enumerate}

\subsection{Reference point: From Landau-Fermi liquid to the Mott-insulator boundary}

In accordance with the division into the regimes $1^{\circ}$--$3^{\circ}$, in the foregoing subsection we overview briefly the results for the regimes $2^{\circ}$ and $3^{\circ}$ which form a basis for a subsequent theoretical discussion in the remaining part of this review. 

The oldest and clearest example of experimentally obtained from Fermi-liquid Mott-insulating transition is provided by an example of pure and doped vanadium sesquioxide ($\mathrm{V_2O_3}$). In Fig.~\ref{fig:1.3} we present the results for Cr-doped $\mathrm{V_2O_3}$ \cite{KuwamotoPhysRevB1980} (complementing, the original Bell-group results \cite{McWhanPhysRevB1973} with the reentrant metallic (PM) phase). Figure~\ref{fig:1.3}(a) shows the temperature dependence of resistivity, encompassing the antiferromagnetic-insulator (AFI), paramagnetic metal (PM), paramagnetic (quasi) insulator (PI), and reentrant paramagnetic metal (PM') states. The AFI $\rightarrow$ PM and PM $\rightarrow$ PI transitions in Fig.~\ref{fig:1.3}(a) are clearly discontinuous. On the basis of those and other measurements \cite{CarterPhysRevB1993}, the phase diagram shown in Fig.~\ref{fig:1.3}(b) has been established, from which one draws the conclusion that the AFI $\rightarrow$ PM and PM $\rightarrow$ PI transitions are discontinuous (note well developed hysteresis), whereas the high-temperature transition is of crossover character.

  \begin{figure}
    \centering
 \includegraphics[width=0.8\textwidth]{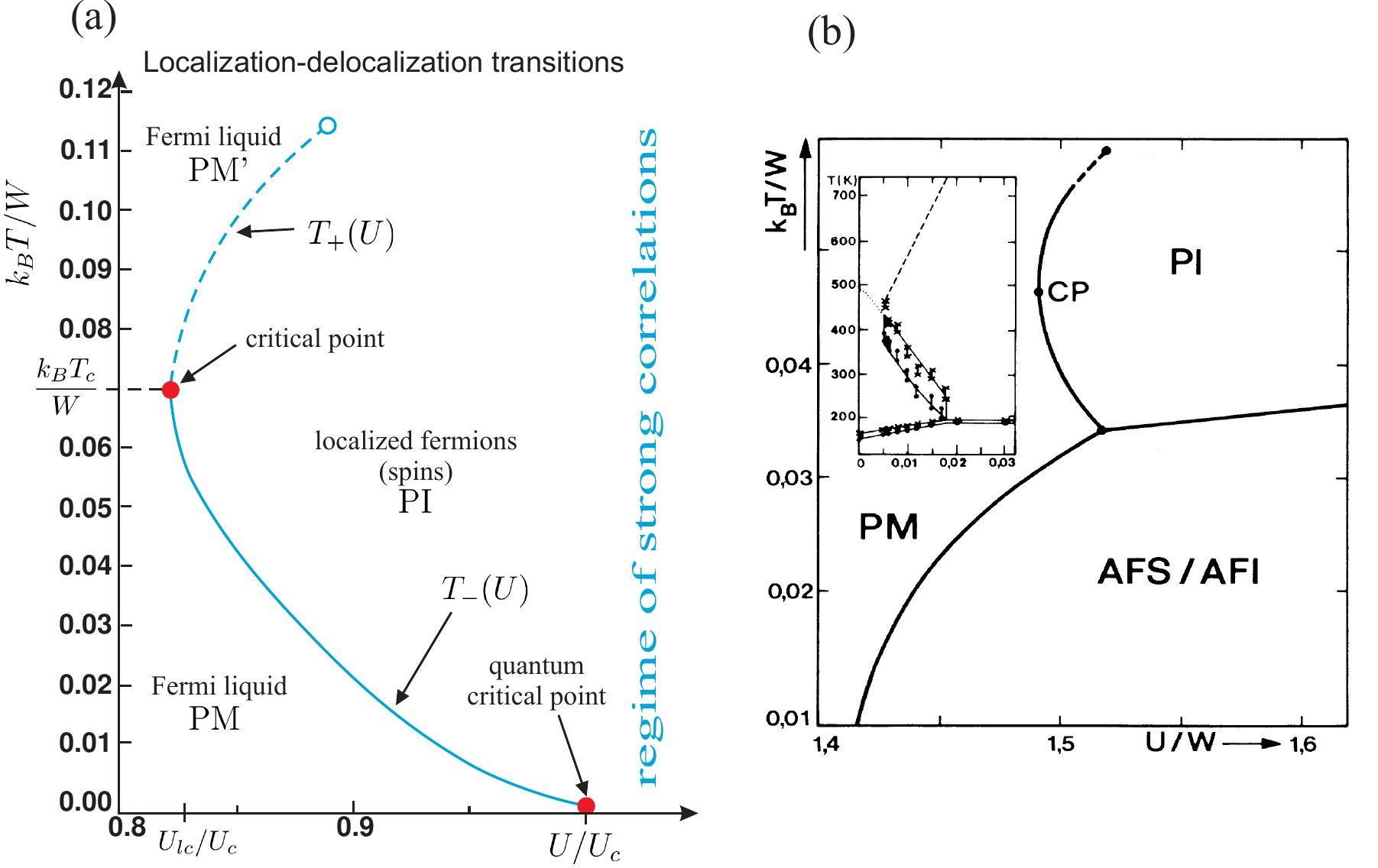}.
    \caption{Phase diagram (a) on temperature $T$ \emph{vs.} relative interaction $U/U_c$ plane, containing the first-order metal-insulator transition lines (continuous part). Note the presence of two critical point: Classical at $T=T_c$ and quantum at $T=0$. (b) Phase diagram with inclusion of antiferromagnetic Slater (AFS) and Mott (AFI) insulating phases. Note that $W_c=U_c/2$. Inset: Experimental \cite{KuwamotoPhysRevB1980} phase diagram on the $T$-$x$ plane for $\mathrm{(V_{1-\mathit{x}}Cr_\mathit{x})_{2}O_{3}}$ (cf. Fig~\ref{fig:1.3}(b)). The quantum critical (lower) point appearance is wiped out by the presence of the magnetic order. The dashed curve in (a) marks a supercritical behavior.}
    \label{fig:1.4}
  \end{figure}

The low-$T$ transition was identified as the Mott (Mott-Hubbard) transition, albeit complicated by the presence of antiferromagnetic (AFI) phase, which takes the form of nondegenerate (wide-gap) semiconductor (with resistivity, $\rho$, scaling as $\ln \rho \sim 1/T$). However, such a transition is followed by a second ``anti-Mott'' transition from a bad-metal (PM) phase to the (quasi) paramagnetic insulator (PI), which is yet continued by a crossover transition back to high-temperature metallic (PM') state. The question then is whether this seemingly involved behavior can be interpreted, at least qualitatively, within a simple picture of correlated electrons. It turns out that the phase diagram, depicted in Fig.~\ref{fig:1.3}(b), can be rationalized within a relatively simple one-band picture of correlated electrons. From that picture the rich behavior observed in Fig.~\ref{fig:1.3}(a) should follow, at least in a qualitative manner. 

An elementary reasoning, rationalizing the behavior depicted in Fig.~\ref{fig:1.3}, is as follows. We start from the Hubbard model in the half-filled-band situation ($n=1$). We accept the first Gutzwiller-type interpretation of the transition PM $\rightarrow$ PI \cite{BrinkmanPhysRevB1970}, which plays a role of a mean-field approach at $T=0$. Within this approach, the expression for the ground state energy $E_G$, effective mass enhancement $m^{*}/m_{0}$ in the correlated state, and the static magnetic susceptibility are \cite{BrinkmanPhysRevB1970}:

\begin{empheq}
[left = \empheqlbrace]{align}
    \frac{E_{G}}{N}&=\left(1-\frac{U}{U_c} \right)^2 \bar{\epsilon}, \label{eq:1.2}\\
    \frac{m^{*}}{m_0}&=\frac{1}{1-\left(\frac{U}{U_c} \right)^2} \equiv 1+\frac{1}{3}F_1^s, \label{eq:1.3}\\
    \frac{\chi}{\chi_0}&= \frac{1}{1-\left(\frac{U}{U_c} \right)^2}\; 
    \frac{1}{1-\rho_0(\mu)U\frac{1+U/2U_c}{1+(U/U_c)^2}} \equiv \frac{1}{1+F_0^a}. \label{eq:1.4}
\end{empheq}

\noindent
In these expressions, $\bar{\epsilon}$ is the expectation value of bare band energy, $\rho_0(\mu)$ is the density of states associated with the bare band (taken at the Fermi energy $\mu$), and $U_c\equiv 8|\bar{\epsilon}|$ is the critical Hubbard interaction for a continuous PM $\rightarrow$ PI transition as $U \rightarrow \ U_c - 0^{+}$. Note that the Wilson ratio $\chi/\gamma$ remains finite as $U \rightarrow \ U_c - 0^{+}$, where $\gamma \equiv \gamma_0 \frac{m^{*}}{m_0}$ is the linear specific-heat coefficient in the metallic  phase. This means that the transition is not magnetically driven if the renormalized Stoner  criterion is not met before the PM $\rightarrow$ PI boundary is reached. This transition is thus the effect of interelectronic correlations and is customarily called the Brinkman-Rice transition \cite{BrinkmanPhysRevB1970}. The expressions after the second equality sign of Eqs.~\eqref{eq:1.3} and \eqref{eq:1.4} follow from the Landau-Fermi-liquid theory. Thus $F_1^s\rightarrow\infty$ represents the instability point of the corresponding Fermi liquid.

This elementary theory has been subsequently generalized to the nonzero temperature ($T>0$) regime \cite{SpalekPhysRevLett1987,SpalekPhysRevB1986,SpalekPhysRevB1989}. The basic concept here is that of renormalized-by-correlations quasiparticle energy $E_{\mathbf{k} \sigma} = q_{\sigma} \epsilon_{\mathbf{k}\sigma}$, where $\epsilon_{{\bf k}\sigma}$ denotes 
the bare quasiparticle energy and $q_{\sigma}$ is the so-called band-narrowing factor. The latter is related to (in general spin-dependent) effective mass enhancement in the correlated state which at $T = 0$ is $q_{\sigma} \equiv m_{0}/m_\sigma^{*} = 1-(U/U_c)^{2}$ \cite{SpaekPhysRevLett1990}. In effect, the whole approach may be generalized to nonzero temperature, where now the Landau-type expression for the free energy leads to the phase diagram depicted in Fig.~\ref{fig:1.4} in the paramagnetic state (a) and that including the antiferromagnetic ordering (b) \cite{SpalekPhysRevLett1987}. In Fig.~ \ref{fig:1.4}(a), the regime of strong correlations ($U>U_c$) is also marked. 

  \begin{figure}
    \centering
    \includegraphics[width=0.8\textwidth]{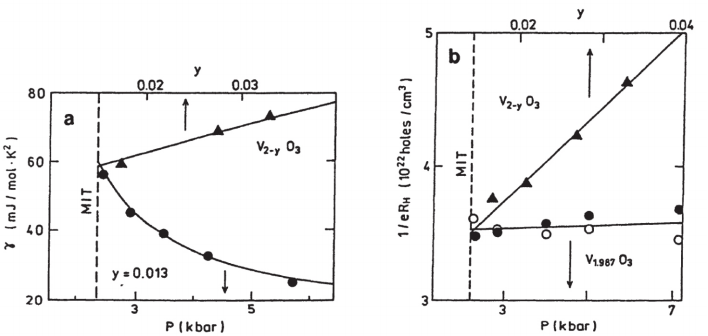}.
    \caption{(a) The linear coefficient $\gamma$ of the heat capacity versus nonstoichiometry parameter $y$ and pressure, and (b) the carrier density versus $y$ and pressure, both for V$_{2-y}$O$_{3}$ near the metal-insulator transition. Both types of behavior (either vs. $y$ or pressure $p$) converge at the MIT (Mott-Hubbard) boundary. Parameter $y$ reflects the degree non-stoichiometry deviation from the nominal compound $\mathrm{V_2O_3}$. The first-order metal-insulator transition (MIT) is marked by the vertical dashed line. After~\cite{CarterPhysRevB1993}.}
    \label{fig:1.5}
  \end{figure}

To illustrate explicitly that the regime of strong correlations emerges from a Fermi liquid to strong-correlation system, in Fig.~\ref{fig:1.5}(a)-(b) we have plotted the results for related vanadium-deficient system V$_{2-y}$O$_3$ with a creation of holes in $3d^2$ states, apart from a minute atomic disorder. The measured quantities are the linear specific heat coefficient $\gamma$ (a) and the effective carrier concentration $n_c=\frac{1}{eR_H}$ (b), determined from the classical Hall coefficient $R_H$, both as a function of either pressure $p$ (lower abcissa) or nonstoichiometry parameter $y$ (upper $x$-axis). Both dependencies terminate at the Mott-Hubbard metal-insulator transition (MIT). The elementary interpretation of those results is as follows. The lower curves correspond to constant $y$ (i.e., presumably at constant $n_c$). This means that, as the MIT is approached, the increase of $\gamma$ can be attributed to the increase of $y/U_c$. On the other hand, the corresponding lower curve in panel (b) is dead flat, meaning that the carrier concentration is constant. The opposite behavior is observed as a function of $\gamma$, where the carrier concentration is $n_c \sim y$.

An elementary interpretation of those observations is as follows. By taking the Drude expression for static electrical resistivity $\rho = m^*/(n_c e^2 \tau)$, where $\tau$ is the relaxation time, one can infer that to reach the localization point ($\rho \rightarrow \infty$) one of the following conditions has to be fulfilled: \emph{(i)} $m^* \rightarrow \infty$ or \emph{(ii)} $n_c \tau \rightarrow \infty$. The case \emph{(i)} occurs at the Mott transition, whereas the second is associated with Anderson localization transition. Here we are interested only in the former situation near the Mott transition, i.e., disregard the effect of atomic disorder.

In summary, this overall picture has been confirmed later both experimentally and by DMFT calculations \cite{GeorgesRevModPhys1996}. In essence, it composes a canonical explanation of physics of the metal-insulator transition induced solely by the electronic correlations, and not strongly influenced by either the atomic disorder or by deviation from half-filling. None of them takes place in the case of high-$T_c$ cuprates, for which $U$ is large and may be filling-dependent, i.e.,  $U > U_c \equiv U_c(n)$. Also, it does not include intersite correlations, which are vital in the strong-correlation regime, as will be detailed in Sec.~\ref{sec:theoretical_models}. But first, we summarize the basic properties of high-$T_c$ cuprates.

\subsection{Basic experimental properties of the high-$T_c$ cuprates}

   \begin{figure}
    \centering
    \includegraphics[width=\textwidth]{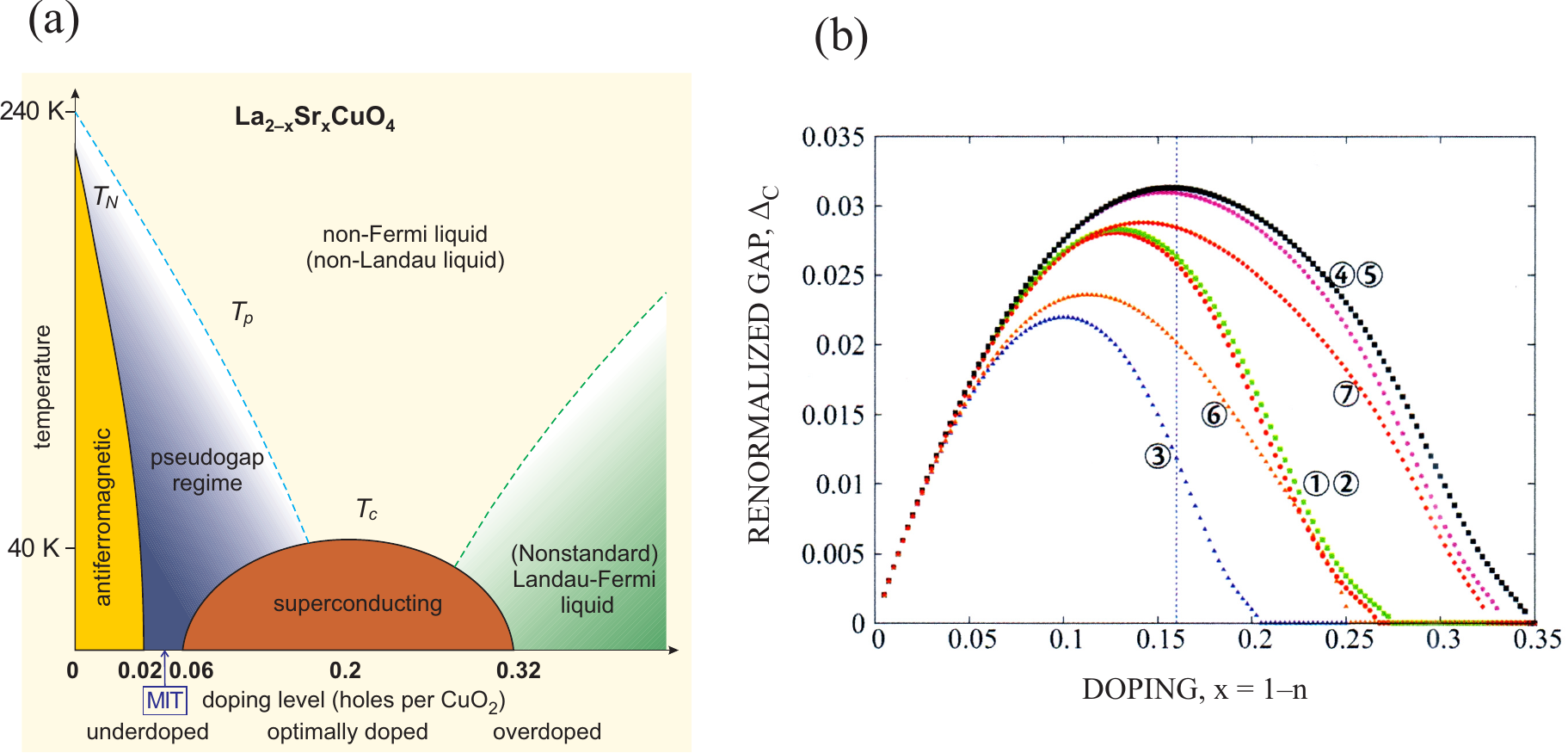}.
    \caption{(a) Schematic phase diagram on the temperature-doping plane for $\mathrm{La_{2-\mathit{x}}Sr_{\mathit{x}}CuO_4}$ ($x$ denoted hole concentration). (b) Renormalized superconducting gap vs. doping $x$, obtained within mean-field approximation SGA \cite{JedrakPhsRevB2011} for the $t$-$J$ model (for details, see the main text). The respective curves in (b) correspond to various choices of the model parameters. $T_N$, $T_c$, and $T_p$ characterize N\'eel temperature for antiferromagnetic ordering, superconducting critical temperature, and that for the onset of the pseudogap, respectively. The phase diagram (a) should contain also other phases, such as charge density wave; those will be discussed later (cf. Sec.~\ref{subsection:superconducting_properties}). Note that the gap $\Delta_c$ is dimensionless; it describes the magnitude of real space correlation $\langle{\hat{a}^\dagger_{i\uparrow}} \hat{a}^\dagger_{j\downarrow}\rangle$, where $\langle i, j\rangle$ are nearest neighbors.}
    \label{fig:1.6}
  \end{figure}

\begin{figure}
    \centering
    \includegraphics[width=0.9\textwidth]{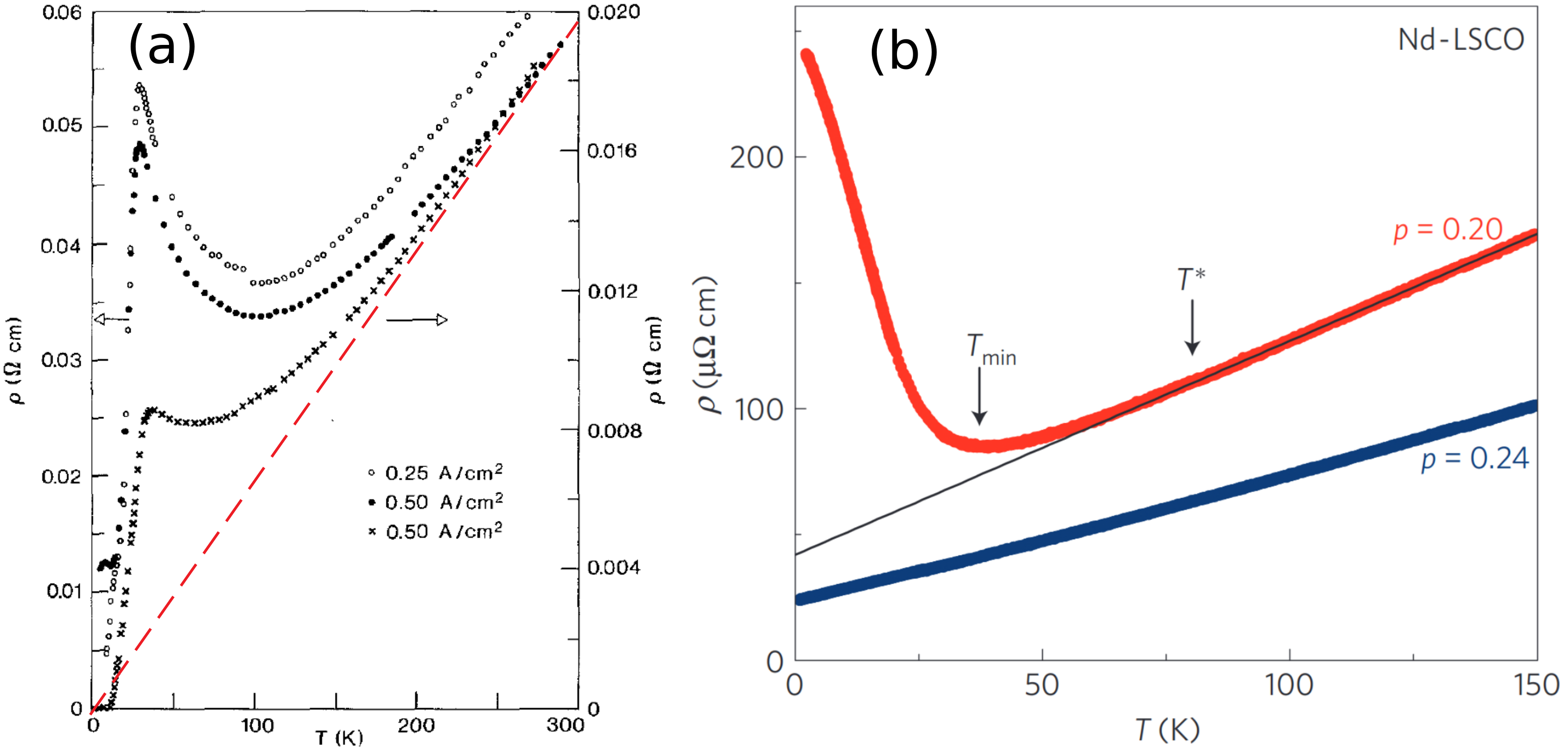}.
    \caption{Temperature dependence of the resistivity: original data of Bednorz and M\"{u}ller \cite{BednorzZPhysB1986} (a) for LaBaCuO and those for electron-doped Nd-LSCO as compared to LSCO \cite{DaouNatPhys2008}. The dashed/dotted straight lines illustrate the extrapolate linear-$T$ behavior, 
    which may signal the fundamental nature of the carrier transport lifetime $\tau \sim k_BT$ \cite{ZaanenBook2015}. The extrapolated residual resistivity $\rho_0 \equiv \rho (T\rightarrow 0)$ may be associated with either atomic disorder increasing with the doping $\delta$ or with the carrier localization with the decreasing doping.}
    \label{fig:1.7}
  \end{figure}

  The appearance of high-$T_c$ superconductivity in lanthanum \cite{BednorzZPhysB1986} and yttrium \cite{ChuPhysRevLett1987} cuprates opened up a new area of research of strongly correlated systems. Namely, starting from the parent antiferromagnetic Mott insulator state for, e.g., $\mathrm{La_2CuO_4}$ or $\mathrm{YBa_2Cu_3O_6}$ and doping it (creating holes in nominally $\mathrm{Cu_2O_2^{2-}}$ states of Cu$-$O plane, one observes the insulator$\rightarrow$metal transition accompanied by the appearance of superconducting state. The overall phase diagram of $\mathrm{La_{2-\delta}Sr_{\delta}CuO_4}$ is illustrated in Fig.~\ref{fig:1.6}(a), with exemplary model results shown in Fig.~\ref{fig:1.6}(b). In panel (a) we see that the superconductivity (SC) appears at a (lower) critical concentration of holes $\delta_{c_1}\sim 0.055$ (for $\mathrm{La_{2-\delta}Sr_{\delta}CuO_4}$), shortly after the doping-induced metallization occurs, and subsequently disappears at the (upper) critical concentration $\delta_{c_2}\sim 0.25$-$0.3$. The maximal superconducting transition temperature is reached at the optimal concentration $\delta_\mathrm{opt}\sim 0.18$-$0.2$. A more detailed phase diagram, encompassing the charge-density-wave (CDW) and related states, is and discussed later. The characteristic features to explain at this level are: \emph{(i)} suppression of the SC state near the Mott-insulator limit at $\delta_{c1}=0.05$, \emph{(ii)} the dome-like shape of superconducting transition temperature $T_c(\delta)$, followed up by the disappearance of SC at $\delta_{c_2}$, and \emph{(iii)} the presence of the so-called \emph{pseudogap} for $T \lesssim T_p$ \cite{HashimotoNatPhys2014} and its disappearance in the unconventional (non-Fermi-liquid) metallic phase, i.e., for $\delta \rightarrow \delta_{c_2}$. Any explanation of the properties should take into account the correlated character of the underlying electronic states. An exotic nature of the normal state of those systems is reflected directly by the approximately linear-$T$ behavior of static electric resistivity (cf. Fig.~\ref{fig:1.7}(a)), particularly in the optimal doping regime, as shown in Fig.~\ref{fig:1.7}(b). However, the linear in $T$ behavior is questioned in the pseudogap regime \cite{BarisicPNAS2013}. In that situation, a $T^2$-term admixture occurs in the pseudogap temperature range before gradual crossover to $\sim T$ dependence and higher $T$. These basic characteristics will be supplemented with additional detailed features when comparing more quantitatively theory with experiment.  
 
At the end of this introductory overview of the physical properties of the high-$T_c$ cuprates, we would like to mention two unusual 
properties, by which those materials differ from all the previously known superconductors (other will be elaborated later when comparing the theory with experimental results later on).

\begin{figure}
    \centering
    \includegraphics[width=0.7\textwidth]{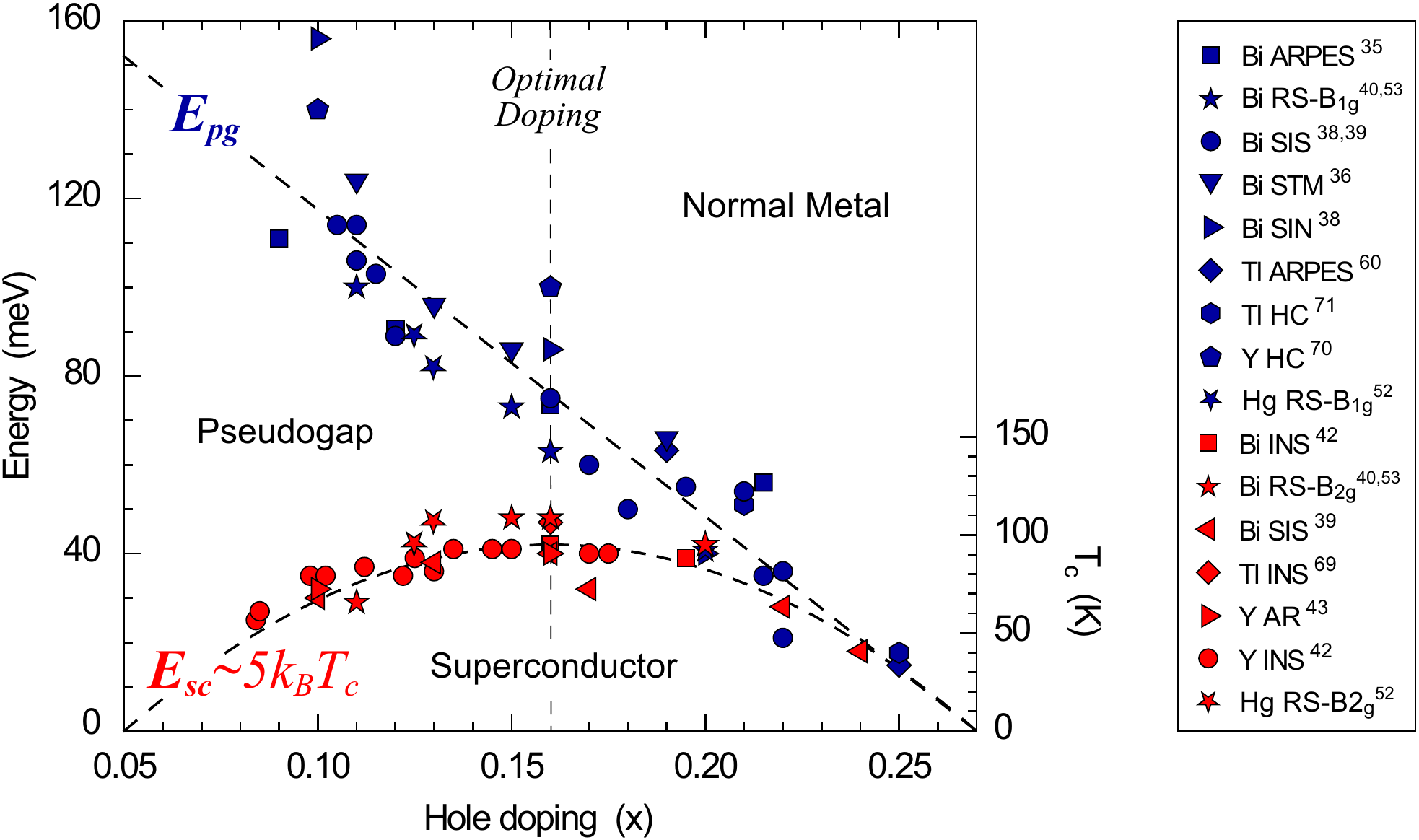}
    \caption{Superconducting gap (lower part) and the pseudogap ($E_\mathrm{pg}$) for various high-$T_c$ SC cuprates (for details, see~\cite{HufnerRepProgPhys2008}). The region of merging of the two curves in not sharply defined. The optimal-doping vertical line divides the whole phase diagram into the underdoped (left) and oberdoped (right) regimes. After~\cite{HufnerRepProgPhys2008}.}
    \label{fig:1.9}
\end{figure}

The first of the them is associated with the appearance of pseudogap roughly at temperature $T_p$ which should be distinguished clearly from the true superconducting transition $T_c$, both marked schematically in Fig.~\ref{fig:1.6}(a). A more accurate data of those two temperatures for different high-$T_c$ systems are displayed in Fig.~\ref{fig:1.9} (after Ref.~\cite{HufnerRepProgPhys2008}), although the regime, where $T_c$ and $T_p$ merge into each other, is not agreed upon universally as yet. The basic question still is whether the two are related (see later).

\begin{figure}
    \centering
    \includegraphics[width=0.5\textwidth]{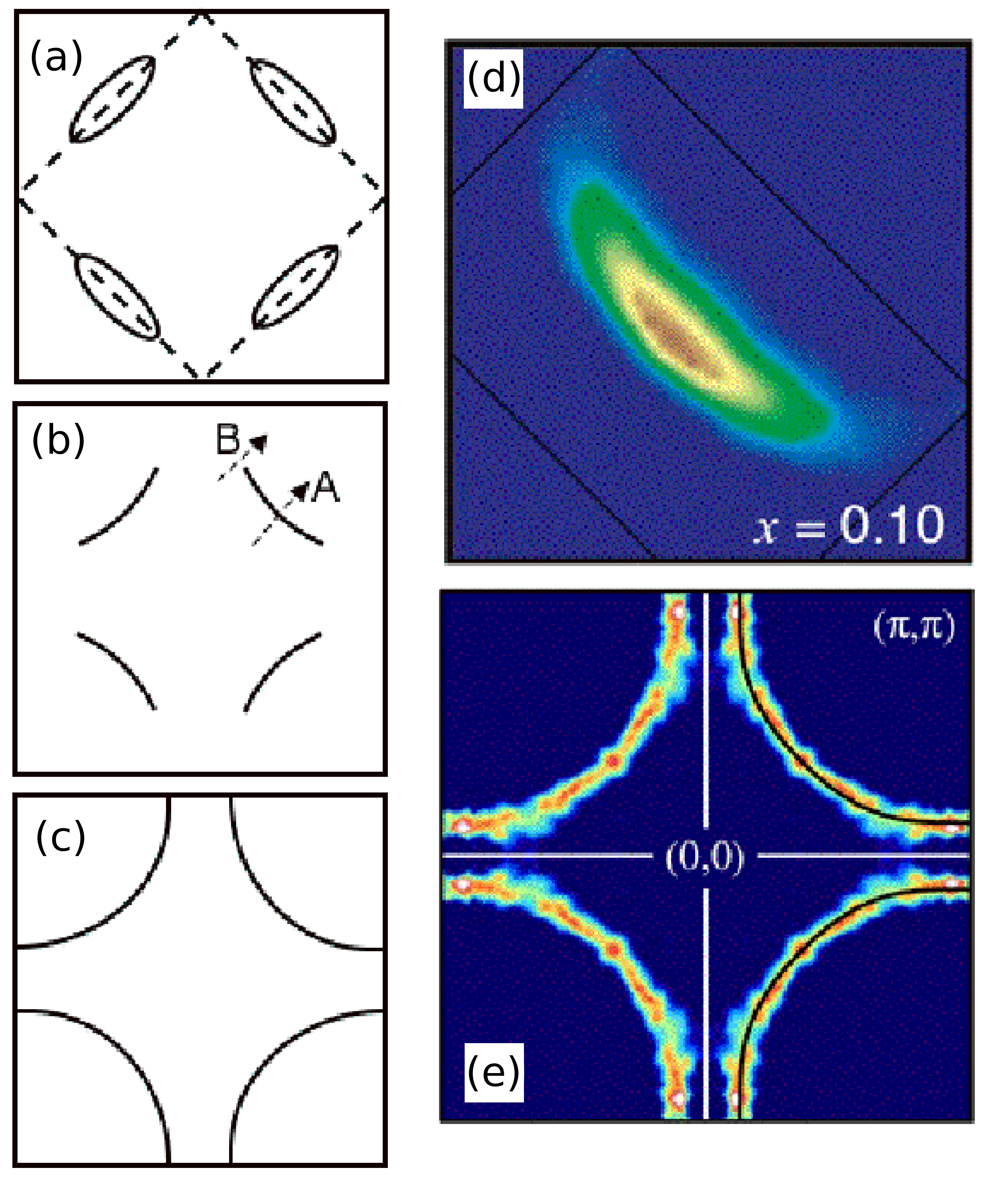}
    \caption{The evolution of the two-dimensional Fermi surface with increasing doping. At $\delta \rightarrow 0$ the Fermi pockets are formed (a). In the underdoped case, only Fermi arcs appear (b). Finally, in the overdoped range, the full Fermi surface is restored (c). Panels (d) and (e) show experimental results of Refs.~\cite{PlatePhysRevLett2005,ShenScience2005} in the two last cases, respectively. Reproduced from Ref.~\cite{LeeRepProgPhys2007}.}
    \label{fig:1.8}
\end{figure}

The second feature is associated with the unconventional evolution of the Fermi-surface topology with the doping of the parent compound. In the 
undoped case, such as La$_2$CuO$_4$ or YBa$_2$Cu$_3$O$_6$, the system is, to a good accuracy, a Mott insulator for which no Fermi surface exists. At large doping $\delta \gtrsim 1/4$, the system becomes a Fermi liquid, with a full Fermi surface of a quasi-two-dimensional ($d=2$) metal \cite{LeeRevModPhys2006,LeeRepProgPhys2007}. The basic question is how it evolves between those two limits. The representative topologies in the overdoped and underdoped regimes are shown in Fig.~\ref{fig:1.8}. Note that in the underdoped case (low doping), the arcs appear around the nodal direction only and with a finite carrier lifetime, whereas in the former situation (for large nonzero doping) the whole open Fermi surface is well defined. Other related properties will be discussed later.

\subsection{From localized spins to statistical spin liquid (non-Fermi liquid)}

One of the fundamental questions is how the correlated system evolves from the Mott insulating state to the correlated liquid upon doping. As we have seen on example of $\mathrm{V_2O_3}$, with the decreasing value of interaction $U$, the system transforms sharply from the Mott (or Mott-Hubbard) state the (almost localized) Fermi liquid. In contrast to that, from the experimental phase diagram for high-$T_c$ cuprates (cf. Fig.~\ref{fig:1.6}(a)) there are no signatures of any clear phase transition from antiferromagnetic insulator to a correlated metal, though the transition is relatively sharp (in La$_{2-\delta}$Sr$_{\delta}$CuO$_4$ for $\delta \approx 0.05$ and in YBa$_2$Cu$_3$O$_{6+\delta}$ for $\delta \approx 0.035$), even though both strong correlations and a sizable atomic disorder are present. To understand this crossover-type behavior, the following elementary argument can be put forward. 

The number of distinct microscopic configurations of the system $N_e = n N$ particles distributed among $N$ atomic sites with no double occupancies reads $2^{N_e} \binom{N}{N_e}$, so the entropy $S_m$ of such hopping spins (per site) is 

\begin{align}
  S_m = n \ln 2 -n \ln n - (1-n) \ln (1-n).
  \label{eq:entropy_from_configurations}
\end{align}

\noindent
In the Mott insulating limit ($n=1$), this expression reduces correctly to $S_m = N k_B \ln 2$. Equation~\eqref{eq:entropy_from_configurations} differs remarkably from that obtained using the Fermi-Dirac statistics for coherent quasiparticle states, and it represents the entropy of the so-called \emph{statistical spin liquid}. In the situation with $\delta = 1-n$, free energy of such a liquid can be estimated as $F_{m/N} = 
- |\bar{\epsilon}| (1-n)/(1-\frac{n}{2}) - TS_m/N$ and compared with that of the localized states $F_{e}/N = E_l - k_B T \ln 2$, 
where $\bar{\epsilon}$ is the average bare band energy with bandwidth $W$ and 
$E_l$ denotes the exchange energy of localized spins. By equating those two free energy expressions, we can estimate the crossover hole concentration $\delta$ 
for the transition from localized to itinerant state. The boundary value $\delta=\delta_c$ can be then determined from the condition $|z||t|(1-n)/(1-n/2)=E_l \approx -(J/4) z$, where $J$ is the exchange integral for interspin antiferromagnetic interaction and $z$ is the nearest-neighbor number. Taking $|t| = 0.3\,\mathrm{eV}$, $z=4$, and $J = 0.1\,\mathrm{eV}$ we obtain $\delta = J/(8|t|) \approx 0.04$, a surprisingly good value if we regard a transition from the paramagnetic insulating state to the statistical-spin (non-Fermi) liquid. A systematic theoretical approach to the insulator to metal transition as a function of doping $\delta$ has not been formulated as yet. Examples of such analysis for the spin-liquid state are provided in Refs.~\cite{SpalekPhysRevB1988_3,SpalekPhysicaB1990}. One additional exception is the work on the role of disorder in the $t$-$J$ model and its semiquantitative agreement with observed fluctuations in electron tunneling data in high-$T_c$ cuprates \cite{MaskaPhysRevLett2007}.

\section{Theoretical models and their relation to local pairing: From one- to three-band model} \label{sec:theoretical_models}

\begin{figure}
    \centering
    \includegraphics[width=\textwidth]{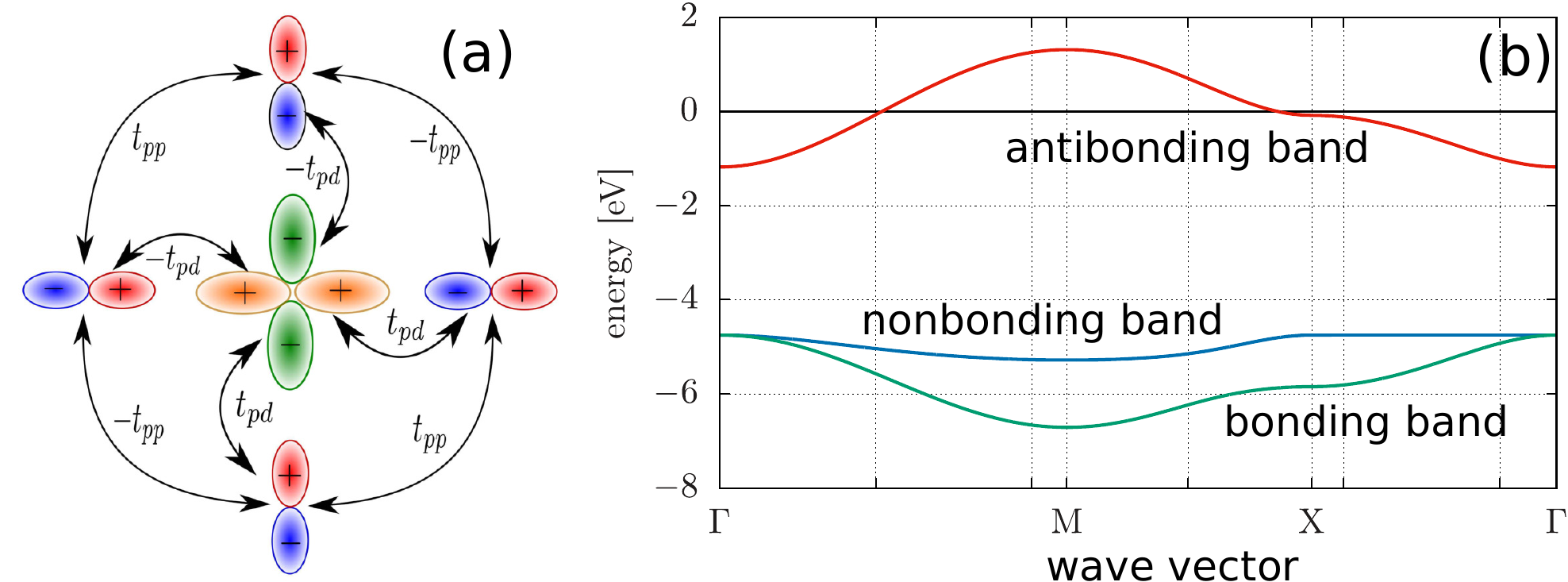}.
    \caption{(a) Definition of the hopping parameters between the $p_x$, $p_y$, and $d_{x^2-y^2}$ orbitals, with the sign convention for the antibonding orbital structure. This structural unit forms a basis of three-band model of the $\mathrm{CuO_2}$ plane in the cuprates. (b) Bare band structure of the three-band ($d$-$p$) model with microscopic parameters: $t_{pd} \approx 1.13\,\mathrm{eV}$, $t_{pp} \approx 0.49\,\mathrm{eV}$, and $\epsilon_{pd} \sim 3.57\,\mathrm{eV}$. The Fermi energy is taken as the reference value and corresponds to the filling $n=5$ per $\mathrm{Cu}^{2+} \mathrm{O}_2^{2-}$ complex, corresponding to half-filled antibonding band. This band is split off by about $\epsilon_{pd}$ from the remaining filled bands and reflects a single-hybridized (bare) band in the situation with interelectronic interactions ignored.}
    \label{fig:dp-model}
  \end{figure}

The principal structural unit for modeling high-$T_c$ SC is the copper-oxygen square lattice, composed of $3d_{x^2-y^2}$ states for the unpaired valence electron of $\mathrm{Cu}^{2+}$ ions and the hybridized antibonding $p_{x}$/$p_{y}$ states due to oxygen ions, as illustrated in Fig.~\ref{fig:dp-model}(a). The corresponding hopping parameters and the sign convention for the antibonding states is also specified there: $t_{pd}$ express the $p$-$d$ intersite hybridization, whereas $t_{pp}$ the hopping between neighboring $p_x$ and $p_y$ states. The remaining $p$ states are disregarded in such a three-orbital (three-band) model. In the same manner, the $2p_z$ and other split-off orbitals ($3d_{z^2}$, etc.) are neglected in the \emph{standard} version of the model. Effectively, in the parent (Mott insulating) situation, we start with $\mathrm{Cu^{2+}O_2^{2-}}$ complex forming electronic basis containing $5$ electrons per formula unit: singly-occupied $3d_{x^2-y^2}$ state representing $\mathrm{Cu}^{2+}$ ion and $4$ electrons filling antibonding $p_{x}$/$p_{y}$ states and representing individual electronic configurations of each of the two oxygen $\mathrm{O}^{2-}$ ions, respectively. The resultant orbital structure of the $\mathrm{CuO_2}$ units, composing the cooper-oxygen plane, is detailed in Fig.~\ref{fig:dp-model}(a). Below we characterize the basic correlated models, starting from the single narrow-band case, combining into one those hybridized $d$-$p$ states  that form the antibonding band. The three-band model is discussed later, cf. Fig.~\ref{fig:dp-model}(b), where we display the bare band structure of the two-dimensional three-band (or $d$-$p$) model. We will then argue that the antibonding (uppermost) band indeed plays the most significant role in the effective one-band model description. As the system is substantially doped by holes, the antibonding band becomes less than half-filled, whereas the other (lower) bands are assumed to remain filled and thus inert on the low-energy scale and in the range of filling considered here, i.e., with doping $\delta \lesssim \frac{1}{2}$.

\subsection{One-band model of high-$T_c$ superconductivity and $t$-$J$ model}

Before turning to multi-band situation, we discuss a strictly one-band model with the short-range interactions included (both intraatomic and between the nearest neighbors). The most general single-band model of spin-$\frac{1}{2}$ fermions with \emph{all pair-site} interaction terms in real space (cf. Appendix~\ref{appendix:hubbard_model}) takes the following form

  \begin{align}
    \label{eq:general_model}
    \hat{\mathcal{H}} = &\sideset{}{'}\sum \limits_{ij\sigma}  t_{ij} \hat{a}^\dagger_{i\sigma} \hat{a}_{j\sigma} + U \sum\limits_i \hat{n}_{i\uparrow} \hat{n}_{i\downarrow} + \sideset{}{'} \sum \limits_{ij} \tilde{J}_{ij} \hat{\mathbf{S}}_i \cdot \hat{\mathbf{S}}_j + \frac{1}{2} \sideset{}{'}\sum \limits_{ij} \left(V_{ij} - \frac{1}{2} \tilde{J}_{ij} \right) \hat{n}_i \hat{n}_j  \nonumber \\ & + \frac{1}{2} \sideset{}{'} \sum \limits_{ij\sigma} \tilde{V}_{ij} (\hat{n}_{i\bar{\sigma}} + \hat{n}_{j\bar{\sigma}}) \left( \hat{a}^\dagger_{i\sigma} \hat{a}_{j\sigma} + \hat{a}^\dagger_{j\sigma} \hat{a}_{i\sigma} \right) + \sideset{}{'} \sum_{ij} J^\prime_{ij} \hat{a}^\dagger_{i\uparrow} \hat{a}^\dagger_{i\downarrow} \hat{a}_{j\downarrow} \hat{a}_{j\uparrow},
  \end{align}

\noindent
where the primed summation indicates $i \neq j$. The microscopic parameters express, respectively, the hopping integral $t_{ij}$, the Hubbard 
intraatomic interaction $U$, the kinetic effective exchange integral $\tilde{J}_{ij}$ (to be defined below), the interatomic Coulomb interaction $V_{ij}$, the correlated hopping $\tilde{V}_{ij}$, and the pair-hopping term $J_{ij}^\prime$. Here, the main role is attached to the first three terms, with the kinetic superexchange (the third term) of antiferromagnetic character. Additionally, if the Wannier functions used to define $J_{ij}^\prime$ are real, then $J_{ij}^\prime = J_{ij}^H$. Note that $\tilde{J}_{ij} \equiv - J^H_{ij} + J_{ij}^{\mathrm{kex}}$, where $J^H_{ij}$ is the direct exchange integral (here negligible) and $J_{ij}^{\mathrm{kex}}$ is that expressing the Anderson kinetic exchange integral. In what follows, it is assumed that $\tilde{J}_{ij} = J_{ij}^\mathrm{kex} \equiv J_{ij}$.
  
The general model, represented by the Hamiltonian~\eqref{eq:general_model}, contains a number of microscopic parameters: The first two hopping integrals are usually taken as $t \approx 0.35\,\mathrm{eV}$ and $t^\prime \approx 0.25 |t|$; the other parameters are: $U \sim 7 \div 10\,\mathrm{eV}$, ${J}_{ij} \sim 0.1$-$0.15\,\mathrm{eV}$, and $V_{\langle ij \rangle} - \frac{1}{2} \tilde{J}_{\langle{ij}\rangle} \sim \left(\frac{1}{3} \div \frac{1}{4}\right) U \sim 2 \div 3\,\mathrm{eV}$, where $\langle ij\rangle$ labels pairs of nearest neighbors. This means that the bare bandwidth may be estimated as $W \sim 2 z |t| = 8 |t| \sim 3\,\mathrm{eV} \sim U/3$. Hence, the value of $U$ is substantially larger than $W$, but within the same order of magnitude. Relatively less is known about the correlated hopping magnitude $\tilde{V}_{ij}$. Its importance is in presumption \cite{HirschPhysRevB2000} that it may lead to pairing in the hole-doping case. Indeed, the role of this term may elucidated when considering at the same time the superconducting phase diagram on both hole- and electron-doping sides (see, e.g., \cite{ZegrodnikPhysRevB2017,WysokinskiJPCM2017,KoikegamiJPCM2021}). Here we discuss almost exclusively the hole-doped regime, so the effect of this term is disregarded. We usually regard such a system as strongly correlated and, therefore, transform this Hamiltonian according to the strong-correlation limit assumption, $U \gg W$; this should be justified \textit{a posteriori}. In that limit of strong electronic correlations, we first provide the effective Hamiltonian representing the first four canonically transformed terms, as obtained in the leading nontrivial (second) order in $t_{ij}/U$ (cf. Appendix.~\ref{appendix:derivation_of_the_tj_model}), which is

\begin{equation}
  \label{eq:effective_hamiltonian}
  \tilde{H} = P_1 \left\{ \sideset{}{'} \sum \limits_{ij\sigma} t_{ij} \hat{b}^\dagger_{i\sigma} \hat{b}_{i\sigma} +  \frac{1}{2} \sideset{}{'}\sum \limits_{ij} {J}_{ij} \left(\hat{\mathbf{S}}_i \cdot \hat{\mathbf{S}}_j - \frac{1}{4} \hat{\nu}_i \hat{\nu}_j\right) + \frac{1}{2} \sideset{}{'} \sum_{ij} V_{ij} \hat{\nu}_i \hat{\nu}_j \right\} P_1 + \left[\textrm{three-site terms}\right],
\end{equation}

\noindent
where the local operators have the projected form, excluding onsite double occupancies, i.e., 

\begin{align}
  \label{eq:projection_heff}
  \left\{
  \begin{array}{ll}
  \hat{b}^\dagger_{i\sigma} &\equiv \hat{a}^\dagger_{i\sigma} (1 - \hat{n}_{i\bar{\sigma}}), \\
  \hat{b}_{i\sigma} &\equiv \left(\hat{b}_{i\sigma}^{\dagger}\right)^\dagger, \\
  \hat{\nu}_{i\sigma} &\equiv \hat{b}^\dagger_{i\sigma} \hat{b}_{i\sigma} = \hat{n}_{i\sigma}(1- \hat{n}_{i\bar{\sigma}}), \\
    \hat{\nu}_i &\equiv \sum_{\sigma} \hat{\nu}_{i\sigma}.
  \end{array}
                  \right.
\end{align}

\noindent
Equation~\eqref{eq:effective_hamiltonian} represents the so-called $t$-$J$-$U$ Hamiltonian projected onto the subspace of singly-occupied and empty Wannier states (of effectively $d_{x^2-y^2}$ character for the high-$T_c$ superconducting cuprates). This Hamiltonian is the simplest one used to study for the hole- and electron-doped superconductors. In the Mott-insulator limit, all carriers are localized and the states on their parent $\mathrm{Cu}^{2+}$ ions are singly occupied, i.e., $\hat{n}_i  \equiv 1$ condition is effectively imposed \emph{in the operator form}. Therefore, the only nontrivial term which remains is the Heisenberg exchange Hamiltonian with the exchange integral ${J}_{ij} = 4 t^2_{ij}/(U - V_{ij})$. However, in the situation with nonzero number of holes introduced into in the Mott insulator, i.e., for $\delta \equiv 1 - n > 0$, all the first three terms in Eq.~\eqref{eq:effective_hamiltonian} become relevant. In that situation, it is better to start from the general $t$-$J$-$U$ model and recover the $t$-$J$ model in the $U \rightarrow \infty$ limit (for explanation see below). For $U \rightarrow \infty$ and $V_{ij} = 0$ it takes the canonical form of the $t$-$J$ model, but also encompasses the Hubbard-model limit for $J \equiv 0$ (cf. Appendix~\ref{appendix:sga_and_slave_bosons} for details). 

A few methodological remarks are in place at this point, namely the $t$-$J$ model contains specific ingredients, among them the projected fermion operators \eqref{eq:projection_heff} have non-canonical (non-fermion) anticommutation relations, i.e., 

\begin{align}
  \left\{ \hat{b}_{i\sigma},\hat{b}^{\dagger}_{i\sigma'}\right\} = \delta_{ij} \left[(1-\hat{n}_{i\bar{\sigma}})\delta_{\sigma\sigma'} + S^{\bar{\sigma}}_i(1 - \delta_{\sigma \sigma'}) \right].
  \label{eq:projected_fermion_commutation_relations}
\end{align}

\noindent
This feature complicates the applicability of standard methods such as the perturbation expansion, Green-function analysis, and related techniques. In the next section we propose a way to avoid an explicit projection of the fermion operators, by starting from the $t$-$J$-$U$-$(V)$ model. The strongly correlated with the projection of the corresponding states is then easier to analyze by taking $U \rightarrow \infty$ limit explicitly (see also below).

\subsection{$t$-$J$-$U$-$V$ model: Interpolation between the Hubbard and $t$-$J$ models and real-space pairing operators}

As is apparent from Eq.~\eqref{eq:projected_fermion_commutation_relations}, the projected fermion operators have complicated anticommutation relations (cf. also Appendix~\ref{appendix:derivation_of_the_tj_model}).  In the consideration of the $t$-$J$ model~\eqref{eq:effective_hamiltonian} and its extensions, very often the projection operators $\hat{P}_1$ are disregarded. Nonetheless, even with such a drastic approximation, some interesting qualitative properties of high-$T_c$ superconductors can be reproduced. Therefore, to preserve the principal properties (enforced by the projection operator $\hat{P}_1$) and make the model tractable, we have proposed \cite{SpalekPhysRevB2017} a general $t$-$J$-$U$-$V$ model in which the most relevant features of the $t$-$J$ model and strong correlations are preserved, while the model contains only the ordinary fermionic operators and thus makes the whole algebra easier to tackle. Physically, we propose the model which, on one hand, contains the essential features of correlated $3d$ states due to $\mathrm{Cu}^{2+}$ ions and, on the other, takes implicitly into account the \emph{superexchange interactions} via anionic $2p$ shells of $\mathrm{O}^{2-}$ ions (cf. Appendix~\ref{appendix:sga_and_slave_bosons}). The starting Hamiltonian, encompassing all the above features, has the following single-band form \cite{SpalekPhysRevB2017}

  \begin{align}
    \label{eq:tjuv_model}
    \hat{\mathcal{H}} = \sideset{}{'}\sum_{i\sigma} t_{ij} \hat{a}_{i\sigma}^\dagger \hat{a}_{i\sigma} + U \sum_i \hat{n}_{i\uparrow} \hat{n}_{i\downarrow} +  \sideset{}{'}\sum_{ij} J_{ij} \hat{\mathbf{S}}_i \cdot \hat{\mathbf{S}}_j + \frac{1}{2}\sideset{}{'}\sum_{ij} \left(V_{ij} - \frac{1}{2} J_{ij}\right) \hat{n}_i \hat{n}_j.
  \end{align}

\begin{figure}
    \centering
    \includegraphics[width=0.7\textwidth]{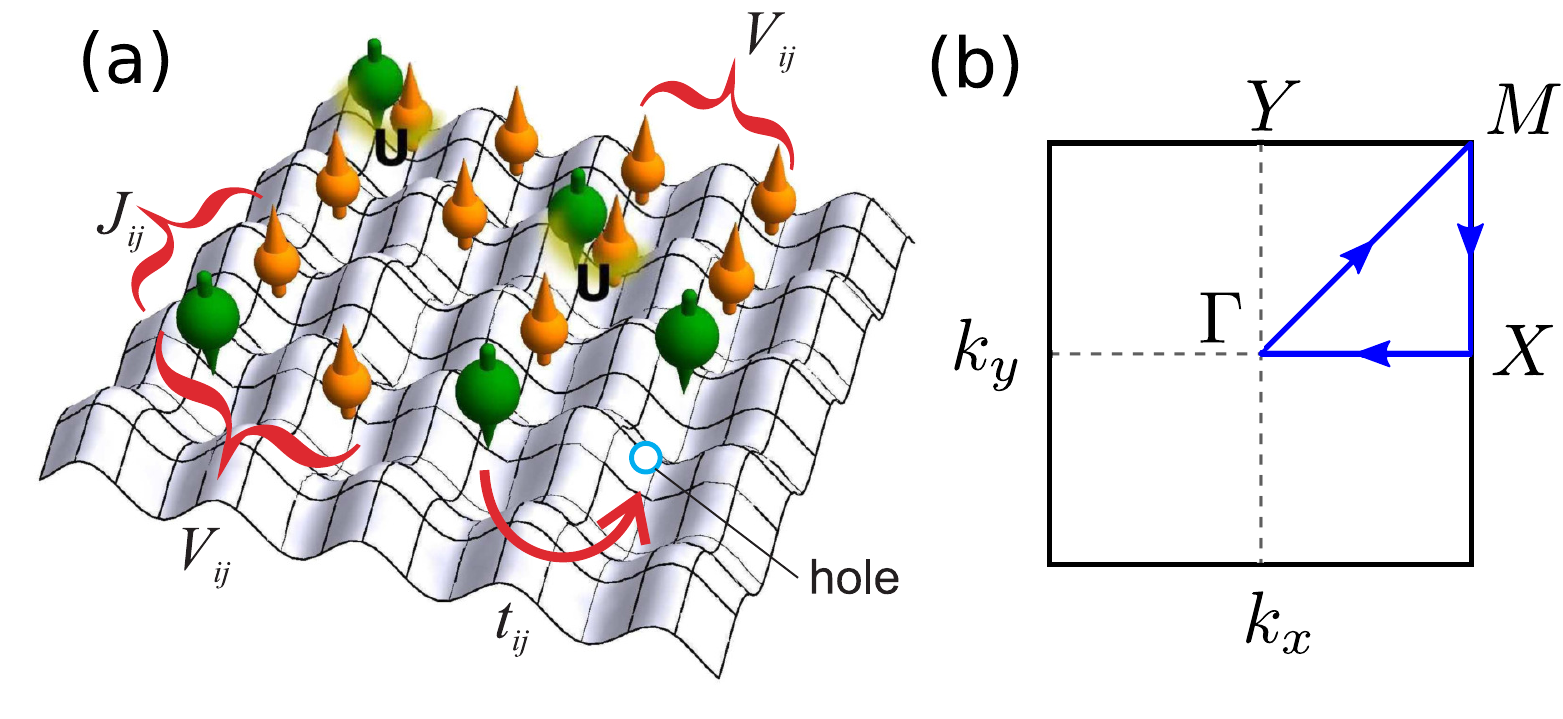}
    \caption{(a) Schematic representation of single plane of Cu ions with active electrons occupying the $3d_{x^2-y^2}$ orbitals and ``dressed'' with $2p_{\sigma}$ orbitals (not shown), as proposed by Anderson \cite{AndersonScience1987}. The onsite Coulomb repulsion, $U$, is attached to the doubly occupied sites. The other parameters are defined in the main text. (b) The Brillouin zone with the high-symmetry points marked. The line $\Gamma$-$M$ is called the nodal direction for the case of pure $d$-wave superconductivity, whereas the $\Gamma$-$X$ and $\Gamma$-$Y$ directions are the antinodal ones.}
    \label{fig:single_band_dynamics}
  \end{figure}
  
\noindent
The model~\eqref{eq:tjuv_model} is schematically illustrated in Fig.~\ref{fig:single_band_dynamics}. It has a few specific features. First, from the formal point of view, it can be regarded as interpolating between the Hubbard-model ($V_{ij} \equiv J_{ij} \equiv 0$) and the $t$-$J$-model ($V_{ij} \equiv 0$, $U \rightarrow \infty$) limits. In the latter case, the double occupancies are eliminated automatically if only one selects a reliable method of solving this $t$-$J$-$U$-$V$ model. Second, the kinetic exchange interaction, discussed in the single-band model (cf. Appendix~\ref{appendix:derivation_of_the_tj_model}), results from the first two terms of Eq.~\eqref{eq:tjuv_model}. Now, to make the model physically feasible, we have to assume that the exchange part results also from the remaining virtual hopping processes via other bands, e.g., by the superexchange via $2p_\sigma$ states due to $\mathrm{O}^{2-}$ ions. However, this means that in the effective single-band model \eqref{eq:tjuv_model} other bands appear still in an implicit manner. This last assumption is necessary, since if the realistic parameters for the $3d_{x^2-y^2}$ states are to be taken, the kinetic exchange part is \cite{SpalekPSS1981} $J_{\langle ij\rangle} = 4 t^2/(U - V_{\langle ij\rangle})$ for $|t| \approx 0.3\,\mathrm{eV}$, $U = 8\,\mathrm{eV}$, $V_{\langle ij\rangle} \approx 2\,\mathrm{eV}$. In effect, $J_{\langle ij\rangle} \approx 0.05\,\mathrm{eV} \approx 300\,\mathrm{K}$ takes far too small value as compared to that determined experimentally, $J_{\langle ij\rangle} \approx 1100$-$1300\,\mathrm{K}$. This elementary estimate shows that the $t$-$J$-$U$-$V$ model is not, strictly speaking, a formally generalized version of either the Hubbard or $t$-$J$ model. Namely, if we want to keep the microscopic parameters $t$, $U$, and $V$ as those for realistic $d_{x^2-y^2}$ orbitals near the atomic limit, then minimally we have to include in it the superexchange interaction via the nearest neighboring anions. In the following sections we discuss the solution of the model~\eqref{eq:tjuv_model} and its limiting regimes, before turning to the extended three-band ($d$-$p$) model. Theoretical estimate of the $J_{ij}$, based on the three band model, is provided in Appendix~\ref{appendix:sga_and_slave_bosons}.
  
Next, we can relate directly the Hamiltonian \eqref{eq:tjuv_model} to the local pairing by introducing the local singlet operators (defined in a similar manner in the projected space and introduced originally in Ref.~\cite{SpalekPhysRevB1988})

  \begin{align}
    \label{eq:local_singlet_operators}
    \left\{
    \begin{tabular}{c}
    $\hat{B}_{ij}^\dagger \equiv \frac{1}{\sqrt{2}} \left( \hat{a}_{i\uparrow}^\dagger \hat{a}_{j\downarrow}^\dagger -  \hat{a}_{i\downarrow}^\dagger \hat{a}_{j\uparrow}^\dagger \right)$, \\
    $\hat{B}_{ij} \equiv \frac{1}{\sqrt{2}} \left( \hat{a}_{i\downarrow} \hat{a}_{j\uparrow} -  \hat{a}_{i\uparrow} \hat{a}_{j\downarrow} \right),$
    \end{tabular}
    \right.
  \end{align}

  \noindent
  and rewrite the $t$-$J$-$U$-$V$ Hamiltonian in its general form as

  \begin{align}
    \label{eq:hamiltonian_in_terms_of_singlet_operators}
    \hat{\mathcal{H}} = \sideset{}{'}\sum_{ij\sigma} t_{ij} \hat{a}^\dagger_{i\sigma} \hat{a}_{j\sigma} + \frac{1}{2} U \sum_i \hat{B}^\dagger_{ii} \hat{B}_{ii} - \sideset{}{'}\sum_{ij} J_{ij} \hat{B}^\dagger_{ij} \hat{B}_{ij} + \frac{1}{2} \sideset{}{'} \sum_{ij} V_{ij} \hat{n}_i \hat{n}_j.
  \end{align}

  \noindent
In the above, $\langle\hat{B}_{\langle ij\rangle}^\dagger\rangle$ represent the of nearest-neighbor spin-singlet pairing amplitude and $\langle\hat{B}^\dagger_{ii}\rangle$ is the intraatomic one. The pairing amplitudes are typically assumed to fulfill the condition $\langle \hat{a}_{i\uparrow}^\dagger \hat{a}^\dagger_{j\downarrow}\rangle = - \langle \hat{a}^\dagger_{i\downarrow} \hat{a}^\dagger_{j\uparrow}\rangle$ and thus the corresponding pairing order-parameter is described by either $\langle \hat{a}^\dagger_{i\uparrow} \hat{a}^\dagger_{j\downarrow}\rangle$ or, equivalently, by $\langle\hat{a}^\dagger_{i\downarrow} \hat{a}^\dagger_{j\uparrow}\rangle$; one can then select just one of them to characterize the real-space spin-singlet pairing. Nonetheless, such an assumption may not be fulfilled if we have an inhomogeneous pairing, coexisting with pair-density wave \cite{ZegrodnikPhysRevB2018}. This situation is also analyzed later for both $V_{ij} \equiv 0$ and $V_{ij} \neq 0$. Note that the third term in Eq.~\eqref{eq:hamiltonian_in_terms_of_singlet_operators} diminishes the system energy if the intersite pairing amplitude $\langle \hat{B}_{ij}^\dagger\rangle \neq 0$; in this situation the real-space pairing takes place. Also, the intraatomic pairing amplitudes appear naturally in the so-called negative-$U$ models \cite{MicnasRevModPhys1990}. However, this is not the case for high-$T_c$ superconducting cuprates as in that situation $U > 0$ and represents the largest energy scale for the system.
  
\subsection{High-$T_c$ cuprates: Three-band model}

In Fig.~\ref{fig:dp-model} we present three-orbital model of the copper-oxygen plane. Following the general line of reasoning outlined in Appendix~\ref{appendix:hubbard_model}, we discuss more formally the main features of the model. The so-called $d$-$p$ model, depicted in Fig.~\ref{fig:dp-model}, may be expressed by the Hamiltonian

\begin{align}
  \label{eq:d-p-hamiltonian}
  \hat{\mathcal{H}} = \sum \limits_{\langle il, jl^\prime\rangle} t_{ij}^{ll^\prime} \hat{a}^\dagger_{il\sigma} \hat{a}_{jl^\prime\sigma} + \sum \limits_{il} (\epsilon_l - \mu) \hat{n}_{il} +  \sum \limits_{il} U_l \hat{n}_{il\uparrow} \hat{n}_{il\downarrow} + \frac{1}{2} \sum_{ill^\prime} U_{ll^\prime} \hat{n}_l \hat{n}_{l^\prime},
\end{align}

\noindent
where $\hat{a}^\dagger_{il\sigma}$ ($\hat{a}_{il\sigma}$) creates (annihilates) the electron with spin $\sigma$ at the $i$-th atomic site and the orbital $l = d_{x^2 - y^2}, p_x, p_y$, the summation $\langle il, il^\prime\rangle$ comprises the interorbital nearest-neighbor hoppings. The second term defines the position of atomic orbitals with their energy positions $\epsilon_{p_x} = \epsilon_{p_y} \equiv \epsilon_p$ and $\epsilon_d - \epsilon_p \equiv \epsilon_{dp}$, with respect to chemical potential $\mu$. The intraatomic interaction parameters are $U_d$ and $U_{p_x} = U_{p_y} \equiv U_p$. All the remaining interaction terms are neglected.

In effect, the explicit form of the three-band Hamiltonian considered in the later part of this review is 

\begin{align}
    \hat{\mathcal{H}} = \sum_{ij\alpha \sigma} t^{pp}_{ij}  \hat{p}_{i\alpha\sigma}^{\dagger}\hat{p}_{j\alpha\sigma} + \epsilon_{dp} \sum_{i\sigma} \hat{d}_{i\sigma}^{\dagger}\hat{d}_{i\sigma} + \sideset{}{'}\sum_{ij\alpha\sigma} t_{ij}^{pd}(\hat{d}^{\dagger}_{i\sigma}\hat{p}_{i\alpha\sigma} + \textrm{H.c.}) + 
  U_d \sum_i \hat{n}_{di\uparrow}\hat{n}_{di\downarrow} + U_p \sum_i \hat{n}_{pi\alpha\uparrow}\hat{n}_{pi\alpha\downarrow}.
  \label{eq:explicit_for_of_the_dp_model}
\end{align}

\noindent
The zero-order hopping processes are taking place only between $p$ electrons, with $t_{ij}^{pd} \approx 1.1 \div 1.3 \, \textrm{eV}$ expressing the $p$-$d$ hybridization process inducing the $d$-$d$ hopping at higher orders (for sign convention for interesting us antibonding states, cf. Fig.~\ref{fig:dp-model}) the charge-transfer energy is $\epsilon_{dp} = 3.6\,\mathrm{eV}$, $U_p \approx 4$-$6\, \textrm{eV}$ and $U_d \approx 8$-$10\, \mathrm{eV}$. The neglected $U_{pd}$ interaction is of magnitude $\sim 2\,\mathrm{eV}$. The remaining symbols have standard meaning. Parenthetically, the magnitude of $p$-$d$ interaction is estimated as $U_{pd} \approx 2\,\mathrm{eV}$ and hence disregarded. 

A methodological remark is in place here. The quantity $\equiv t_{ij}^{pd}$ represents the $p$-$d$ interatomic hybridization. Strictly speaking, such a model of coherent states requires us to start from a hybridized basis of the single particle states entering the parameter $t_{pd}$, in which the original (atomic $|3d_{x^2-y^2}\rangle$ and $|p_x/p_y\rangle$ states are already mixed and orthogonalized, i.e., $|3d_{x^2-y^2}\rangle \rightarrow |w_{x^2-y^2}\rangle$ and $|p_x/p_y\rangle \rightarrow |w_{p_x}/w_{p_y}\rangle$, where $w_\alpha$ form a hybridized-basis of Wannier states with the spatial symmetry consistent with that of the original non-orthogonal, non-hybridized atomic orbitals. Only under this condition, we can use the original atomic phrasing about occupancies and symmetry of the resultant Wannier orbitals. These mutually orthogonal and normalized orbitals, in turn, permit to define field operators of these coherent states in the form

\begin{align}
  \label{eq:field_operators_in_wannier_representation}
  \hat{\varPsi}_{l\sigma}(\mathbf{r}) = \sum_i w_{i\sigma}^{(l)}(\mathbf{r}) \hat{a}_{il\sigma},
\end{align}

\noindent
with $l = d, p_x, p_y$. Only under this proviso the creation and annihilation operators have the usual fermionic anticommutation relations (cf., e.g., \cite{BookFeynman1972} in the situation of nonorthogonal basis), i.e., 

\begin{align}
  \label{eq:anticommutation_relations}
  \left\{\hat{a}_{il\sigma}^\dagger, \hat{a}_{jl^\prime\sigma^\prime}\right\} = \delta_{ij} \delta_{ll^\prime} \delta_{\sigma\sigma^\prime}, \hspace{2em} \left\{\hat{a}_{il\sigma}, \hat{a}_{jl^\prime\sigma^\prime}\right\} = 0.
\end{align}

A methodological remark is in place here. Namely, Hamiltonian~\eqref{eq:explicit_for_of_the_dp_model} represents an extended version of the multiorbital Hubbard model in its pristine form, as it does not contain explicit kinetic superexchange interaction. This interaction is discussed in Appendix~\ref{appendix:derivation_of_the_tj_model}. However, in the multiorbital situation, the $d$-$p$ exchange also appears and is known under the name of antiferromagnetic Kondo (kinetic) exchange, particularly in the context of heavy-electron systems \cite{SpalekPhysRevB1988,SpalekJPhysParis1989,WysokinskiPhysRevB2016,KadzielawaMajorActaPhysPolon2014} (see also below). In the context of high-temperature superconductivity, the last interaction is usually projected out by assuming that the $3d$ spins of $\mathrm{Cu}^{2+}$ ions and $2p$ electrons are strongly bound and form the so-called Zhang-Rice singlet states \cite{ZhangPhysRevB1988}. There is no complete clarity as to that issue \cite{AdolphsPhysRevLett2016}. In the later part of the paper we show that a successful analysis of selected properties of cuprate superconductivity in the three-band model does not require addressing explicitly that question, as $t^{pd}$ is included explicitly.

\subsection{Heavy fermion systems: Anderson lattice model with local spin-singlet pairing}
\label{subsection:anderson_lattice_model}

A particularly interesting quantum materials are represented by the heavy fermion systems. In this case, the originally atomic $4f$ or $5f$ states (due to Ce or U ions, respectively) are hybridized with strongly itinerant ones (of electron-gas type) to produce hybridized states of quasiparticles with extremely large effective masses, $10^2$-$10^3 m_e$, where $m_e$ is the free-electron mass. As said in the foregoing section, those electrons are close to the border of their localization, the state in which their effective mass $m^{*}/m_e$ can thus become almost divergent when the hybridized conduction electron and $f$ states have dominant $f$-electron contribution . The Hamiltonian describing a model situation, presumably $4f^1$ configuration due to $\mathrm{Ce}^{3+}$ ions, is mixed with the conduction (\textit{c}) states of $5d$-$6s$ character. Its explicit form in the real-space representation is \cite{CzychollPhysRep1986}

\begin{align}
  \label{eq:anderson_model}
  \hat{\mathcal{H}} = \sum_{ij\sigma} t_{ij} \hat{c}_{i\sigma}^\dagger \hat{c}_{j\sigma} + \epsilon_f \sum_{i\sigma} \hat{f}_{i\sigma}^\dagger \hat{f}_{i\sigma} + U \sum_i \hat{n}_{i\uparrow}^f \hat{n}_{i\downarrow}^f + \sum_{ij\sigma} V_{ij} \left(\hat{f}_{i\sigma}^\dagger \hat{c}_{j\sigma} + \hat{c}_{i\sigma}^\dagger \hat{f}_{j\sigma}\right).
\end{align}

\begin{figure}
    \centering
    \includegraphics[width=0.4\textwidth]{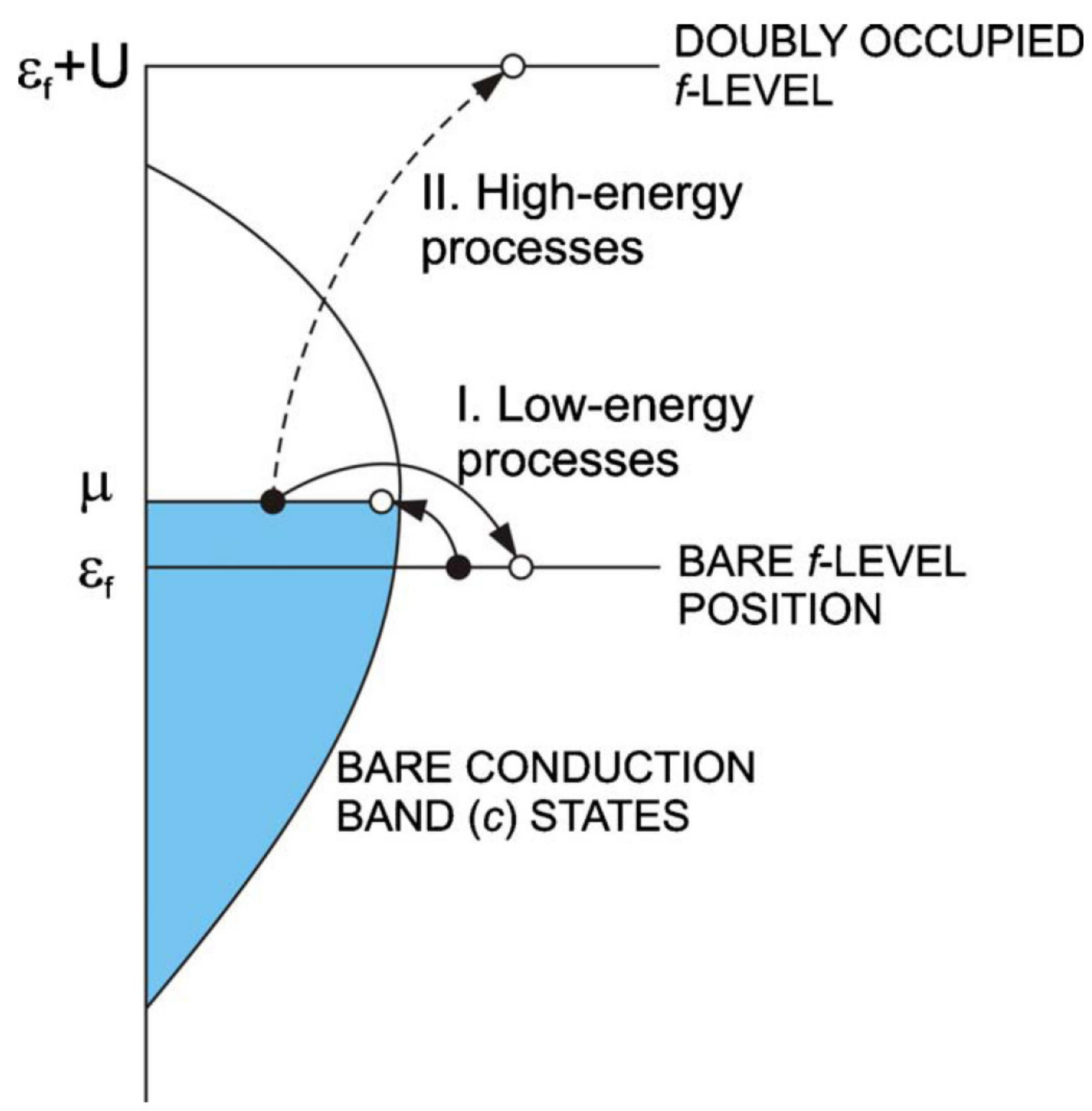}
    \caption{Schematic representation of hybridization (\textit{c-f} mixing) processes
    as \textit{f-c} hoppings and their division into low- and high- energy processes within the periodic Anderson model. 
    The processes I lead to formation of the heavy-quasiparticle states, 
    the other (II) lead to both Kondo-type and $f$-$f$ exchange couplings, which in turn are 
    expressed as real-space hybrid pairing in the second order in $V/(U + 
    \epsilon_f)$ and the $f$-$f$ pairing in the fourth order (see main text). Within such scheme the heavy quasiparticle mass, the Kondo and superexchange interactions, as well as the spin-singlet real-space pairing can be all accounted for on equal footing, at least in model situation with starting $4f^1$ configuration for $\mathrm{Ce}$ compounds.}
    \label{efU}
\end{figure}

\noindent
This Hamiltonian represents the \emph{Anderson-lattice or periodic Anderson model} (labeled as ALM or PAM, respectively). In the case of cerium compounds, the spin index of $f$-electrons represents the pseudospin describing the lowest lying doublet with total angular momentum $J=5/2$ and the Land\'{e} factor $g = 6/7$. The simplest version of the model comprises an additional assumption that the hybridization matrix elements are of intraatomic character $V_{ij} \equiv V \delta_{ij}$. This situation is schematically illustrated in Fig.~\ref{efU}. In essence, we have originally atomic and strongly correlated $f$-electrons ($U$ is by far the largest parameter in the system) mixed with conduction electrons. Typical parameter values are $U \simeq 5$-$6\,\mathrm{eV}$, the atomic level position relative to the Fermi energy of $c$-electrons is $\epsilon_f \simeq -2\,\mathrm{eV}$, $|V| \simeq 0.1$-$0.5\,\mathrm{eV}$, and $t_{ij}$ is a variable parameter and assumes values decisively larger than $V$.

This Hamiltonian does not contain explicitly either spin or pairing operators. To visualize those, we proceed in a similar manner as in Appendix~\ref{appendix:derivation_of_the_tj_model}. Namely, we decompose the hybridization as follows

\begin{align}
  \label{eq:hybridization_decomposition}
  \hat{f}^\dagger_{i\sigma} \hat{c}_{i\sigma} = (1 - \hat{n}_{i\bar{\sigma}}^f + \hat{n}_{i\bar{\sigma}}^f) \hat{f}^\dagger_{i\sigma} \hat{c}_{i\sigma} = (1 - \hat{n}_{i\bar{\sigma}}^f) \hat{f}^\dagger_{i\sigma} \hat{c}_{i\sigma} + \hat{n}_{i\bar{\sigma}}^f \hat{f}^\dagger_{i\sigma} \hat{c}_{i\sigma}, \text{etc.}
\end{align}

\noindent
In this decomposition we separate the processes which do not involve double $f$-level occupancies (the first term) from those which do. As the second term involves the high-energy scale $\sim U$, we include them only as virtual hopping processes. In effect, we transform out those processes canonically from \eqref{eq:anderson_model} and replace them with the corresponding low-energy terms containing those virtual-hopping processes. Leaving the details to Appendix~\ref{appendix:kondo-exchange}, we obtain the following effective Hamiltonian to the fourth order in $V_{ij}/(U + \epsilon_f)$ (and with the chemical-potential term included):

\begin{align}
  \nonumber
\tilde{\mathcal{H}} -\mu \hat{N}
&\simeq 
\sum_{ij \sigma} \left( t_{ij} - \mu \delta_{ij}  \right) \hat{c}_{i \sigma}^\dagger  \hat{c}_{j \sigma}^{}
+ \epsilon^{f} \sum_{i,\sigma} \hat{\nu}_{i\sigma}^{f}      
+ \sum_{i,m,\sigma} \left( V_{im} \left( 1 - \hat{n}_{i\bar{\sigma}}^{f} \right) \hat{f}_{i\sigma}^{\dagger} \hat{c}_{m\sigma}^{} + \mathrm{H.c.}\right)   
  \\
  \nonumber
&\qquad 
+ \sum_{i,m} J_{im}^{K} \left( \hat{\mathbf{S}}_i^f \cdot \hat{\mathbf{S}}_m^c - \frac{ \hat{n}_m^c  \hat{\nu}_i^f }{4} \right)	
+ \sum_{i \neq j,\sigma} J_{ij}^{H}    \left( \mathbf{ \hat{S}}_i^f \cdot \mathbf{ \hat{S}}_j^f - \frac{ \hat{\nu}_i^f  \hat{\nu}_j^f}{4} \right) 
\\
\label{eq:anderson_hamiltonian_fourth_order_canonical_transformation}
&\qquad
            +2 i \sum_{\langle mi \rangle \langle mj \rangle} J_{ij}^{H} \left(1 +\frac{n^{f}}{n^{c}} \right)  \mathbf{ \hat{s}}_m^c \cdot \left( \mathbf{ \hat{S}}_j^f \times \mathbf{ \hat{S}}_i^f \right),
\end{align}

\noindent
where $\epsilon^f \equiv \epsilon_f - \mu$, $\hat{\nu}^f_{\sigma} \equiv \hat{n}^f_{i\sigma} (1 - \hat{n}^f_{i\bar{\sigma}})$, and $\hat{\nu}_i^f = \sum_\sigma \hat{\nu}^f_{i\sigma}$. In Eq.~\eqref{eq:anderson_hamiltonian_fourth_order_canonical_transformation}, $\hat{\mathbf{S}}_i^f$ and $\hat{\mathbf{S}}_m^c$ are the local spin operators in the fermion representation for $f$ and $c$ electrons, respectively, whereas $J_{im}^{K}$ and $J_{ij}^{H}$ are exchange integrals; their explicit form in the fourth order terms is discussed in Appendix~\ref{appendix:kondo-exchange}. This Hamiltonian represents the so-called \emph{Anderson-Kondo lattice} (AKL).

Before turning to local-pairing representation, a few important remarks are in order. First, the effective Hamiltonian results from a modified Schrieffer-Wolff transformation, in which only the part involving high-energy processes (cf. Fig.~\ref{efU}) is transformed out and replaced by the corresponding low-energy (virtual-hopping) process. In other words, the residual part of hybridization (the third term) is nonzero only if $f$-electrons are assumed itinerant from the start, i.e., when the number of $f$-electrons is not conserved independently from that of $c$-electrons, i.e., $\hat{\nu}^f_i \neq 1$. Second, the Kondo ($\sim J_{im}^{K}$) and $f$-$f$ ($\sim J_{ij}^{H}$) exchange terms contain full exchange operators. The last term is unusual as it is the Dzyaloshinskii-Moriya-type interaction of purely electronic origin and is mediated by the conduction electrons. The form of this term is approximate as we have that the charge fluctuations have a negligible effect on it, i.e., we have assumed that $n^f_i \rightarrow \langle \hat{\nu}_i^f\rangle$ and $n^c_i \rightarrow \langle \hat{n_i^c}\rangle$. The whole term is disregarded in the following discussion, as we discuss only either pure singlet-paired and/or collinear magnetically ordered phases. Also, in \eqref{eq:anderson_hamiltonian_fourth_order_canonical_transformation} we have disregarded  a  small renormalization $\sim V^2/(U + \epsilon_f)$ of the hopping term $\sim t_{mn}$ \cite{HowczakJPCM2012}.

Next, as before, we express the exchange part via the corresponding local hybrid pairing operators which are

\begin{align}
  \label{eq:local_pairing_operators_anderson}
  \left\{
  \begin{tabular}{c}
    $\hat{b}^\dagger_{im} \equiv \frac{1}{\sqrt{2}} \left(\hat{\tilde{f}}^\dagger_{i\uparrow} \hat{c}^\dagger_{m\downarrow} - \hat{\tilde{f}}^\dagger_{i\downarrow} \hat{c}^\dagger_{m\uparrow}\right) \equiv (\hat{b}_{im})^\dagger,$ \\
        $\hat{B}^\dagger_{ij} \equiv \frac{1}{\sqrt{2}} \left(\hat{\tilde{f}}^\dagger_{i\uparrow} \hat{\tilde{f}}^\dagger_{j\downarrow} - \hat{\tilde{f}}^\dagger_{i\downarrow} \hat{\tilde{f}}^\dagger_{j\uparrow}\right) \equiv (\hat{B}_{ij})^\dagger,$
  \end{tabular}
  \right.
\end{align}

\noindent
where $\hat{\tilde{f}}_{i\sigma}^\dagger \equiv \hat{f}^\dagger_{i\sigma}(1 - \hat{n}_{i\sigma}^f)$. Those operators correspond to the hybrid $f$-$c$ and $f$-$f$ pairing, respectively. In effect, the effective Hamiltonian \eqref{eq:anderson_hamiltonian_fourth_order_canonical_transformation} acquires a more compact form

\begin{align}
  \label{eq:anderson_hamiltonian_by_pairing_operators}
  \tilde{\mathcal{H}} = \sideset{}{'}\sum_{mn\sigma} t_{mn} \hat{c}^\dagger_{m\sigma} \hat{c}_{m\sigma} + \epsilon_f \sum_{i\sigma} \hat{\nu}_{i\sigma} + \sum_{im\sigma} V_{im} \left(\hat{\tilde{f}}_{i\sigma} \hat{c}_{m\sigma} + \text{H.c.}\right) - \sum_{\substack{\langle im\rangle \\ \langle in\rangle}} J_{im}^{(K)} \hat{b}^\dagger_{im} \hat{b}_{in} - \sideset{}{'}\sum_{\substack{\langle ik\rangle \\ \langle jk\rangle}} J_{im}^{H} \hat{B}^\dagger_{ik} \hat{B}_{jk}.
\end{align}

\noindent
In this Hamiltonian we have included both two- and three-site pairing terms for the sake of completeness. It contains both the Kondo ($f$-$c$) and $f$-$f$ pairings which are of antiferromagnetic type to the fourth order. From Appendix~\ref{appendix:kondo-exchange} we see that $J_{\langle im\rangle}^{K}$ is by two orders of magnitude larger than $J_{\langle ij\rangle}^{H}$. In general, both pairing channels should be important, as will be discussed later. It should be noted that, in the case of a single electron pair, the local-pair binding appears in the system even if the last term in Eq.~\eqref{eq:anderson_hamiltonian_by_pairing_operators} is absent \cite{ByczukPhysRevB1992}. This is a precursor of superconducting pairing by the Kondo-type interactions.

The question remains as to what happens in the localized-moment limit ($n^f \rightarrow 1$), when the whole hybridization term \eqref{eq:hybridization_decomposition} can be transformed out. This situation was analyzed by Schrieffer and Wolff (1966) \cite{SchriefferPhysRev1966}  up to the second order in $V$. In this limit our transformed Hamiltonian reflects the so-called \emph{Kondo-lattice limit} and it takes the effective form

\begin{align}
  \label{eq:kondo_lattice_limit}
  \tilde{\mathcal{H}} = \sum_{nm\sigma} t_{mn} \hat{c}^\dagger_{m\sigma} \hat{c}_{n\sigma} + \sum_{\langle im\rangle} J_{im}^{(K)} \hat{\mathbf{S}}_i^f \cdot \hat{\mathbf{s}}_m^c + \sum_{\langle ij\rangle} J_{ij}^{H} \hat{\mathbf{S}}_i^f \cdot \hat{\mathbf{S}}_j^f.
\end{align}

\noindent
Strictly speaking, this model describes the system of localized spins $\{ \hat{\mathbf{S}}_i^f\}$ coupled to uncorrelated carriers via Kondo-type exchange coupling. To phrase it differently, in the full model represented by Hamiltonian \eqref{eq:anderson_model} only the total number of particles is conserved, i.e.,

\begin{align}
  \label{eq:particle_conservation_condition}
  \sum_{i\sigma} \hat{n}_{i\sigma}^f + \sum_{m\sigma} \hat{n}^c_{m\sigma} = \text{const.} = N_e,
\end{align}

\noindent
whereas in the Kondo-lattice limit \eqref{eq:kondo_lattice_limit} each of two numbers is conserved separately, as can be checked readily by commuting the Hamiltonian~\eqref{eq:kondo_lattice_limit} with the corresponding particle-number operators for $c$ and $f$ electrons. Therefore, the spin operator in \eqref{eq:kondo_lattice_limit} is essentially the atomic spin (in the case of spin-$\frac{1}{2}$: $\hat{\mathbf{S}}_i = \hat{\boldsymbol{\sigma}}/2$, where $\hat{\boldsymbol{\sigma}} = (\hat{\sigma}_x, \hat{\sigma}_y, \hat{\sigma}_z)$ are the Pauli matrices). To see that explicitly, we start from the fermion representation of $f$-electrons $\hat{\mathbf{S}}_i = \left(\hat{f}_{i\uparrow}^\dagger \hat{f}_{i\downarrow}, \hat{f}_{i\downarrow}^\dagger \hat{f}_{i\uparrow}, \frac{1}{2} \left( \hat{n}_{i\uparrow}^f - \hat{n}_{i\downarrow}^f\right)\right)$ and calculate $\left\langle \left(\hat{\mathbf{S}}_i^f\right)^2\right\rangle$ for $n^f \rightarrow 1$ and obtain

\begin{align}
  \label{eq:spin_squared_in_localized_limit}
  \left\langle \left(\hat{\mathbf{S}}_i^f\right)^2\right\rangle = \frac{3}{4} (1 - 2 \langle \hat{n}_{i\uparrow}^f\hat{n}_{i\downarrow}^f \rangle) \rightarrow \frac{1}{2} \left(\frac{1}{2} + 1\right)  \text{\hspace{1em}for\hspace{1em}}  \langle \hat{n}_{i\uparrow}^f \hat{n}_{i\downarrow}^f\rangle \rightarrow 0.
\end{align}

\noindent
Hence, if $U \rightarrow \infty$, then $\langle \hat{n}_i^f \rightarrow 1\rangle$ and $\left\langle \left(\hat{\mathbf{S}}_i^f\right)^2\right\rangle = S(S + 1) = \frac{3}{4}$, where the spin $S = \frac{1}{2}$. This illustrates again the leading role of strong correlations in achieving the localization of strongly correlated $f$-electrons.

\subsubsection*{Remark: Choice of the starting Hamiltonian: From Kondo-lattice to Anderson-Kondo lattice}

\begin{figure}
    \centering
    \includegraphics[width=0.6\textwidth]{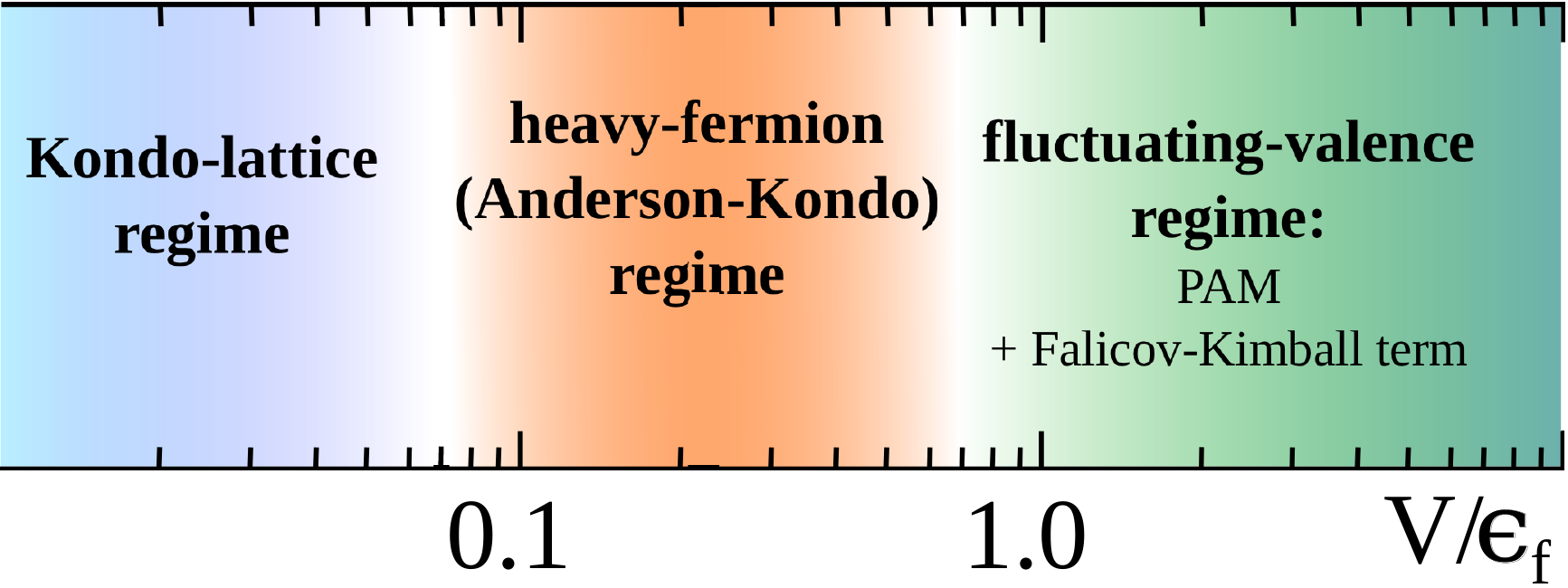}
    \caption{Illustrative representation of various physical regimes as a
    function of hybridization to \textit{f}-level position ratio that lead to different effective model Hamiltonians in the large-$U$ limit, starting from Anderson-lattice model. The borders between the regions are not sharply defined. This classification is also relevant for various versions of the renormalized mean-field analysis. After~\cite{HowczakJPCM2012}.}
    \label{Fig10}
\end{figure}

In Fig.~\ref{Fig10} we specify schematically three physically distinct regimes as a function $V/\epsilon_f$ for the periodic Anderson model in the limit $U \rightarrow \infty$. For small $V/\epsilon_f \sim 0.1$ (the Kondo-lattice regime with the bare $f$-level position deeply below the Fermi surface) the effective Hamiltonian \eqref{eq:kondo_lattice_limit} is a good starting point. In the opposite, mixed valence, regime, $V/\epsilon_f > 1$, the full Anderson-lattice Hamiltonian must be used explicitly, whereas in the case $V/\epsilon_f \sim 1$, the Anderson-Kondo-lattice regime, the start with the effective Hamiltonian \eqref{eq:anderson_hamiltonian_fourth_order_canonical_transformation} is appropriate. Obviously, the three forms of the Hamiltonian~\eqref{eq:anderson_model} should be practically equivalent in proper limits. Usage of the effective Hamiltonians, instead of the general Anderson-lattice Hamiltonian, may lead to qualitatively new results already in the lowest (mean-field-like) level, as will be discussed later. The reason for this is the circumstance that construction of the effective Hamiltonian via canonical transformation leads to physically most important dynamic processes (e.g., virtual hopping) which are partly included already on the mean-field level for its effective form.

\subsection{Multiorbital models with Hund's rule interaction}
\label{subsection:models_with_hunds_rule}

In the case of multiorbital systems, discussed above, we usually have the conduction (uncorrelated) electrons hybridized with the orbitally-degenerate and correlated  ($f$ or $d$) states. In that situation, the conduction band and hybridization parts take the standard single-particle form (see, e.g., the first two terms of \eqref{eq:anderson_model}, where the hybridization takes place with each of the degenerate states $\sim (\hat{c}^\dagger_{i\sigma} \hat{f}^{(l)}_{i\sigma} + \text{H.c.})$, with $l$ being the orbital index for degenerate $f$-electrons. On the other hand, in the simplest nontrivial case, the orbitally degenerate part has the form provided in Appendix~\ref{appendix:kondo-exchange}). In effect, the starting total Hamiltonian is

\begin{align}
  \label{eq:hamiltonian_anderson_multiorbital}
  \hat{\mathcal{H}} - &\mu \hat{N}_e = \sum_{i j l \sigma} t_{ij} \hat{c}^{(l)\dagger}_{i\sigma} \hat{c}^{(l)}_{j\sigma} + V \sum_{i l \sigma} \left( \hat{f}^{(l)\dagger}_{i\sigma} \hat{c}^{(l)}_{i\sigma} + \mathrm{H.c.}\right)  + \epsilon_f \sum_{i l} \hat{n}^{f(l)}_i + U \sum_{il} \hat{n}^{f(l)}_{i\uparrow} \hat{n}^{f(l)}_{i\downarrow} + U' \sum_{i} \hat{n}^{f(1)}_i \hat{n}^{f(2)}_i - \nonumber \\ & - 2 J^H \sum_i \left( \mathbf{\hat{S}}_i^{f(1)} \cdot \mathbf{\hat{S}}_i^{f(2)} + \frac{1}{4} \hat{n}_i^{f(1)} \hat{n}_i^{f(2)} \right) - \mu \hat{N}_e.
\end{align}

\noindent
In the most elementary case, $l=1, 2$. Correlations are determined by the intraorbital intraatomic $f$-$f$ repulsion $U$ of the Hubbard type, interorbital repulsion $U^\prime$, and by the ferromagnetic Hund's-rule intraatomic interorbital coupling $J^H$. Hereafter, we restrict ourselves to the situation, relevant also to $d$ electrons, when $U^\prime = U - 2J$.

To identify the dominant pairing channel, we rewrite the interaction in $\mathcal{\hat{H}}$ in terms of the spin-triplet- and the spin-singlet-pairing operators $\hat{A}_{i m}^\dagger$ and $\hat{B}_{i }^\dagger$, respectively. The spin-triplet operators are defined in the interorbital form 

\begin{align}
\hat{A}_{i m}^\dagger
&\equiv
\begin{cases}
\hat{f}_{i\uparrow}^{(1)\dagger}		\hat{f}_{i \uparrow}^{(2)\dagger}		& \quad m=1,	\\
\frac{1}{\sqrt{2}}	
\left(
\hat{f}_{i\uparrow}^{(1)\dagger}		\hat{f}_{i \downarrow}^{(2)\dagger} 
+
\hat{f}_{i\downarrow}^{(1)\dagger}		\hat{f}_{i \uparrow}^{(2)\dagger}	
\right)	&	\quad m=0, \\
\hat{f}_{i\downarrow}^{(1)\dagger}		\hat{f}_{i \downarrow}^{(2)\dagger} 	&	\quad m=-1, \\
\end{cases}
\label{eq:triplet_pairing_operator}
\end{align}

\noindent

and the spin-singlet operators by intraatomic operators

\begin{align}
\hat{B}_{i}^{\dagger} &\equiv \frac{1}{\sqrt{2}}	
\left(
\hat{f}_{i\uparrow}^{(1)\dagger}		\hat{f}_{i \downarrow}^{(2)\dagger} 
-
\hat{f}_{i\downarrow}^{(1)\dagger}		\hat{f}_{i \uparrow}^{(2)\dagger}	
\right),
\label{eq:singlet_pairing_operator}
\end{align}

\begin{figure}
	\centering
	\includegraphics[width=0.6\textwidth]{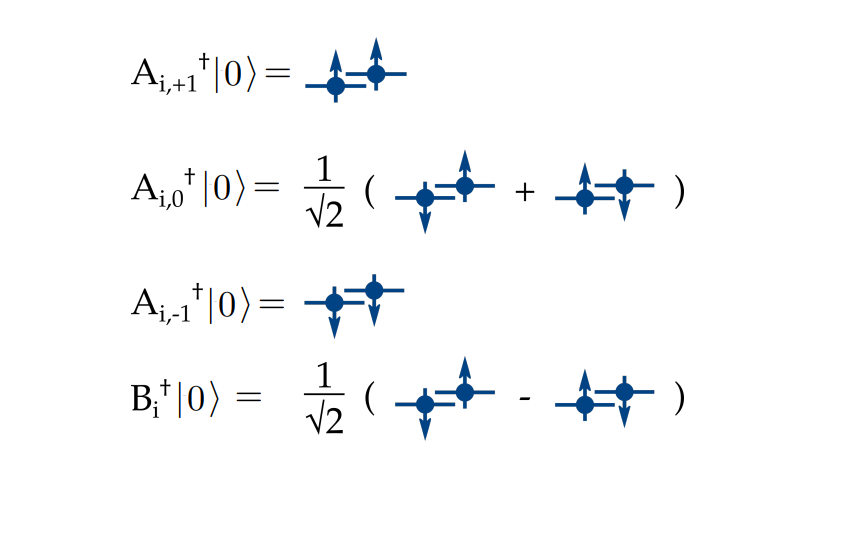}
	\caption{Schematic picture of spin-triplet intraatomic real-space pairing related to definition \eqref{eq:triplet_pairing_operator}.}
	\label{fig:pairing}
\end{figure}

\noindent
where the three triplet components correspond to $z$-axis spin projection $m=1, 0, -1$ for pair of orbitals, see Fig.~\ref{fig:pairing} for illustration. The above pairing operators can be expressed in terms of the spin-spin and density-density interactions as follows:

\begin{align}
\displaystyle\sum \limits_{	 m = -1 }^{1}
\hat{A}^{\dagger}_{	i m	}	\hat{A}^{}_{	i m 	}
\equiv
\mathbf{\hat{S}}_{i}^{f(1)} \cdot \mathbf{\hat{S}}_{i}^{f(2)} 
+ \frac{3}{4} 	\hat{n}^{f(1)}_{	i 	}	\hat{n}^{f(2)}_{	i	} ,           \label{eq:triplet_pairing_relation}
\\
\hat{B}_{i}^\dagger \hat{B}_{i }^{} 
\equiv
-\left(
\mathbf{\hat{S}}_{i }^{f(1)} \cdot \mathbf{\hat{S}}_{i }^{f(2)}
- \frac{1}{4} 	\hat{n}^{f(1)}_{	i 	}	\hat{n}^{f(2)}_{	i 	} 
\right).           \label{eq:singlet_pairing_relation}
\end{align}

\noindent
With the help of this representation, one can rewrite the $f$-electron part of the Hamiltonian~\eqref{eq:hamiltonian_anderson_multiorbital}, using above spin-triplet- and spin-singlet-pairing operators as \eqref{eq:triplet_pairing_operator}-\eqref{eq:singlet_pairing_operator} 

\begin{align}
\hat{\mathcal{H}}^{f}		=
\epsilon^{f}	\displaystyle\sum \limits_{	i	l	\sigma	} \hat{n}^{f(l)}_{	i		\sigma	}	
+	U	\displaystyle\sum \limits_{	i l	} 
\hat{n}^{f(l)}_{	i 	\uparrow	}	\hat{n}^{f(l)}_{	i 	\downarrow	} 
+	\left( U' + J \right)	\displaystyle\sum \limits_{	i }^{}
\hat{B}^{\dagger}_{	i 	}	\hat{B}^{}_{	i 	} 			
+\left( U' - J \right)	\displaystyle\sum \limits_{	i m }^{}
\hat{A}^{\dagger}_{	i m	}	\hat{A}^{}_{	i m 	}.              	\label{eq:Hf3}
\end{align}

\noindent
Since $U$ and $U^\prime$ are positive and not too large, the spin-triplet pairing part is at first the most important new channel of pairing for not too large $U$. Strictly speaking, one can think of a pure two-band model with Hamiltonian (\ref{eq:Hf3}) with separate band states sharing Fermi surface and additionally supplied with the hopping between \textit{f} (or \textit{d}) orbitals. 

In summary, in this section we  have introduced the concept of real-space pairing for the model Hamiltonians of correlated fermions reflecting various physical situations. Note that, in each case, the exchange interaction (kinetic, Kondo, superexchange, or Hund's rule) has been transformed to the form of the local (intrasite or intersite) character. Now, the next fundamental question is whether this type of pairing, in conjunction with the electronic correlations, can lead to stable paired states. This is the subject of the next sections. Note also that, formally, the spin and local-pairing representations are completely equivalent. It is the interelectron-correlation aspect which differentiates between the resultant magnetic and superconducting states, as we discuss next. 

\section{Variational wavefunction systematic solution: Saddle-point (mean-field) approximation and beyond} \label{sec:vwf_solution}
\label{sec:de_gwf}

\subsection{Single narrow-band case: DE-GWF expansion}
\label{sec:de_gwf_expansion}

We start with the Hubbard Hamiltonian rewritten in a slightly different form

\begin{align}
\hat{\mathcal{H}} = \sideset{}{'}\sum_{ij\sigma} t_{ij} \hat{a}^\dagger_{i\sigma} \hat{a}_{j\sigma} + U \sum_{i} \hat{n}_{i\uparrow}\hat{n}_{i\downarrow} \equiv \mathcal{H}_0 + U \sum_i \hat{d}_i.
\end{align}

\noindent
As we have noted earlier, the local correlation function $d_i^2 \equiv\langle\hat{d}_i\rangle \equiv \langle\hat{n}_{i\uparrow} \hat{n}_{i\downarrow}\rangle$ plays a principal role in the whole approach, particularly for $U$ much larger than $W \equiv 2z |\sum_{j(i)}t_{ij}|$. In such a situation, it is favorable energetically to reduce the weight of local double occupancies in the many-particle wave function. Gutzwiller \cite{GutzwillerPhysRevLett1963,GutzwillerPhysRev1965} proposed, right at the birth of the Hubbard model, the wave function in the form

\begin{align}
  \label{eq:gutzwiller_wave_function}
  |\Psi_G\rangle \equiv \hat{P} |\Psi_0\rangle \equiv \prod_i \hat{P}_i |\Psi_0\rangle = \prod_i \left[1 - (1-g_i)\hat{d}_i \right] |\Psi_0\rangle,
\end{align}

\noindent
where $0 \leq g_i \leq 1$ are variational parameters which allow for an interpolation between the electron gas (or metallic; $g_i \equiv 1$) and atomic ($g_i \equiv 0$) limits. $|\Psi_0\rangle$ thus represents an uncorrelated (single-particle) state, to be defined later in a self-consistent manner.

The most general local form of the correlation operator $\hat{P}$ is given by \cite{BuenemannEurophysLett2012} 

\begin{equation}
 \hat{P}=\prod_i\hat{P}_i=\prod_i\sum_{\Gamma}\lambda_{i,\Gamma} |\Gamma\rangle_{ii}\langle\Gamma|,
  \label{eq:Gutz_operator}
\end{equation}

\noindent
with the variational parameters $\lambda_{i,\Gamma}\in\{\lambda_{i\emptyset},\lambda_{i\uparrow},\lambda_{i\downarrow},\lambda_{i\uparrow\downarrow}\}$ corresponding to states from the local (site) many-particle basis $\ket{\emptyset}_i, \ket{\uparrow}_i, \ket{\downarrow}_i, \ket{\uparrow\downarrow}_i$ ($i$ is the lattice-site index). Due to such choice of basis, the $\lambda_{i,\Gamma}$ parameters weight the probability amplitude of a given local state appearance. In the limiting situation of infinite onsite Coulomb repulsion, $U\rightarrow \infty$, all the double occupancies, $\{\ket{\uparrow\downarrow}_i\}$, should be absent in the system ($\lambda_{\uparrow\downarrow}\equiv 0$). In the following we consider first the tranlationally invariant paremagnetic case which means that $\lambda_{i,\Gamma}\equiv\lambda_{\Gamma}$ and $\lambda_{\uparrow}=\lambda_{\downarrow}\equiv\lambda_{1}$.

The expectation value of the energy in the Gutzwiller state is
\begin{equation}
 \langle\mathcal{\hat{H}} \rangle_G=\frac{\langle\Psi_G|\mathcal{\hat{H}}|\Psi_G \rangle}{\langle\Psi_G|\Psi_G \rangle}=\frac{\langle\Psi_0|\hat{P}\mathcal{\hat{H}}\hat{P}|\Psi_0 \rangle}{\langle\Psi_0|\hat{P}^2|\Psi_0 \rangle},
 \label{eq:H_expectation}
\end{equation}
where the norm of the wave function is explicitly introduced in the denominator as the correlation operator is not unitary. It has been proposed by B\"unemann \emph{et al.} \cite{BuenemannEurophysLett2012} that this expectation value can be effectively evaluated by using the diagrammatic method in which one imposes first the following ansatz for the $\hat{P}_i$ operator
\begin{equation}
 \hat{P}_i^2\equiv1+x\hat{d}^{\textrm{HF}}_i,
 \label{eq:P2_diag}
\end{equation}

\noindent
where $x$ is yet another variational parameter and 
$\hat{d}^{\textrm{HF}}_i\equiv\hat{n}_{i\uparrow}^{\textrm{HF}}\hat{n}_{i\downarrow}^{\textrm{HF}}$, $\hat{n}_{i\sigma}^{\textrm{HF}}\equiv\hat{n}_{i\sigma}-n_{0}$
with $n_{0}\equiv\langle\Psi_0|\hat{n}_{i\sigma}|\Psi_0\rangle$. To determine the relation between $\lambda_{\Gamma}$ and $x$, we write down the explicit form of the $\hat{P}^2_i$ operator (cf. Eq.~\eqref{eq:Gutz_operator})

\begin{equation}
\hat{P}^2_i=\lambda_{\emptyset}^2+(\lambda_1^2-\lambda_{\emptyset}^2)(\hat{n}_{i\uparrow}+\hat{n}_{i\downarrow})+(\lambda_{\emptyset}^2-2\lambda_1^2+\lambda^2_{\uparrow\downarrow})\hat{n}_{i\uparrow}\hat{n}_{i\downarrow},
 \label{eq:P2_mod}
\end{equation}

\noindent
where we made use of orthonormality of the local basis ($\langle\Gamma|\Gamma\rangle=\delta_{\Gamma\Gamma'}$) and the explicit representation of the projection operators 

\begin{equation}
  \left\{
 \begin{split}
  &\ket{\emptyset}_{ii}\bra{\emptyset}=(1-\hat{n}_{i\uparrow})(1-\hat{n}_{i\downarrow}),\\
  &\ket{\uparrow}_{ii}\bra{\uparrow}=\hat{n}_{i\uparrow}(1-\hat{n}_{i\downarrow}),\\
  &\ket{\downarrow}_{ii}\bra{\downarrow}=\hat{n}_{i\downarrow}(1-\hat{n}_{i\uparrow}),\\
  &\ket{\uparrow\downarrow}_{ii}\bra{\uparrow\downarrow}=\hat{n}_{i\uparrow}\hat{n}_{i\downarrow}.
  \label{eq:local_proj}
\end{split}
\right.
\end{equation}

\noindent
At the same time, from (\ref{eq:P2_diag}), we have
\begin{equation}
 \hat{P}^2_i=1+x\hat{n}_0^2-xn_0(\hat{n}_{i\uparrow}+\hat{n}_{i\downarrow})+x\hat{n}_{i\uparrow}\hat{n}_{i\downarrow}.
 \label{eq:P2_mod2}
\end{equation}

\noindent
By comparing Eqs.~\eqref{eq:P2_mod} and \eqref{eq:P2_mod2} one can express the variational perameters $\lambda_{\Gamma}$ in terms of the parameter $x$ as

\begin{equation}
  \left\{
\begin{split}
 &\lambda^2_d=1+x(1-n_0)^2,\\
 &\lambda^2_{1}=1-xn_0(1-n_0),\\
 &\lambda^2_{\emptyset}=1+xn_0^2,
\end{split}
\right.
 \label{eq:lambda_x}
\end{equation}

\noindent
which means that we are left with a single variational parameter, with respect the which the energy of the system has to be minimized. Such an analysis is illustrated first on example of the Hubbard model.

Having proposed the correlator in the form~\eqref{eq:P2_diag}, we can write all the relevant expectation values which appear during the evaluation of \eqref{eq:H_expectation} in the form of a systematic expansion. The result is as follows \cite{KaczmarczykPhysStatSol2015}

\begin{equation}
  \left\{
\begin{split}
 &\langle\Psi_G|\Psi_G\rangle=\sum_{k=0}^{\infty}\frac{x^k}{k!}\sideset{}{'}\sum_{l_1 \ldots l_k}\langle \hat{d}^{\textrm{HF}}_{l_1 \ldots l_k} \rangle_0,\\
 &\langle\Psi_G|\hat{c}^{\dagger}_{i\sigma}\hat{c}_{j\sigma}|\Psi_G\rangle=\sum_{k=0}^{\infty}\frac{x^k}{k!}\sideset{}{'}\sum_{l_1 \ldots l_k}\langle \tilde{c}^{\dagger}_{i\sigma}\tilde{c}_{j\sigma}\hat{d}^{\textrm{HF}}_{l_1 \ldots l_k} \rangle_0,\\
 &\langle\Psi_G|\hat{n}_{i\sigma}\hat{n}_{j\sigma'}|\Psi_G\rangle=\sum_{k=0}^{\infty}\frac{x^k}{k!}\sideset{}{'}\sum_{l_1 \ldots l_k}\langle \tilde{n}^{\dagger}_{i\sigma}\tilde{n}_{j\sigma}\hat{d}^{\textrm{HF}}_{l_1 \ldots l_k} \rangle_0,\\
 &\langle\Psi_G|\hat{s}^{\dagger}_{i\sigma}\hat{s}_{j\bar{\sigma}}|\Psi_G\rangle=\left(\lambda_{1}\right)^4\sum_{k=0}^{\infty}\frac{x^k}{k!}\sideset{}{'}\sum_{l_1 \ldots l_k}\langle \hat{s}^{\dagger}_{i\sigma}\hat{s}_{j\bar{\sigma}}\hat{d}^{\textrm{HF}}_{l_1 \ldots l_k} \rangle_0,\\
  &\langle\Psi_G|\hat{d}_i|\Psi_G\rangle=\left(\lambda_d\right)^2\sum_{k=0}^{\infty}\frac{x^k}{k!}\sideset{}{'}\sum_{l_1 \ldots l_k}\langle \hat{d}_i\hat{d}^{\textrm{HF}}_{l_1 \ldots l_k} \rangle_0,
\end{split}
\right.
\label{eq:expectation_val_terms}
\end{equation}

\noindent
where $\hat{s}_{i\sigma}=\hat{c}^{\dagger}_{i\sigma}\hat{c}_{i\bar{\sigma}}$, 
$\langle  \ldots  \rangle_0=\langle\Psi_0| \ldots |\Psi_0\rangle$, $\hat{d}^{\textrm{HF}}_{l_1 \ldots l_k}\equiv\hat{d}^{\textrm{HF}}_{l_1} \ldots \;\hat{d}^{\textrm{HF}}_{l_k}$ 
with $\hat{d}_{\emptyset}^{\textrm{HF}}\equiv 1$, and the primed sums impose the summation restrictions $l_p\neq l_{p'}$, $l_p\neq i,j$. Also, the following notation has been used $\tilde{c}^{(\dagger)}_{i\sigma}=\hat{P}_i\hat{c}^{(\dagger)}_{i\sigma}\hat{P}_i$, $\tilde{n}_{i\sigma}=\hat{P}_i\hat{n}_{i\sigma}\hat{P}_i$. 
Note that in Eqs.~\eqref{eq:expectation_val_terms}, the expectation value in the correlated state (left-hand side) is expressed in terms of the expectation values in the non-correlated state (right-hand side).

Since $\braket{ \ldots }_0$ comprises averages in single-particle state, the latter can be decomposed by the use of the Wick's theorem
and expressed in the form of diagrams with internal vertices positioned on lattice sites $l_1 \ldots l_k$ and external vertices on site $i$ and $j$. Lines connecting those verices correspond to the expectation values

\begin{equation}
 P_{ij}\equiv\langle \hat{c}^{\dagger}_{i\sigma} \hat{c}_{j\sigma}\rangle_0, \quad S_{ij}\equiv\langle \hat{c}^{\dagger}_{i\uparrow} \hat{c}^{\dagger}_{j\downarrow}\rangle_0,
 \label{eq:lines}
\end{equation}

\noindent
and the summation over $l_1 \ldots l_k$ corresponds to attaching the 
internal vertices to different lattice sites (for details see \cite{KaczmarczykNewJPhys2014}).

As has been shown in \cite{BuenemannEurophysLett2012}, due to the condition \eqref{eq:P2_diag} all the diagrams with lines that leave and enter the same internal vertex vanish. This reduces significantly the number of terms which have to be included in the calculations. In order to achieve the same for the external verices we must express $\hat{d}_i$, $\tilde{n}_{i\sigma}$, and $\tilde{c}^{(\dagger)}_{i,\sigma}$ in 
Eqs.~\eqref{eq:expectation_val_terms} with the use of $\hat{n}^{\textrm{HF}}_{i,\uparrow}$ and $\hat{d}^{\textrm{HF}}_{i}$ operators

\begin{equation}
\left\{
 \begin{split}
  \hat{d}_i&=(1-xd_0)\hat{d}^{\textrm{HF}}_i+n_0(\hat{n}^{\textrm{HF}}_{i,\uparrow}+\hat{n}^{\textrm{HF}}_{i,\downarrow})+n^2_0\hat{P}_i^2\\
   \tilde{n}_{i\sigma}&=\hat{n}^{\textrm{HF}}_{i\sigma}+\beta\hat{n}^{\textrm{HF}}_{i\bar{\sigma}}+\gamma\hat{d}^{\textrm{HF}}_{i}+n_0\hat{P}_i^2,\\
  \tilde{c}^{(\dagger)}_{i,\sigma}&=q\hat{c}^{(\dagger)}_{i\sigma}+\alpha\hat{c}^{(\dagger)}_{i\sigma}\hat{n}^{\textrm{HF}}_{i\bar{\sigma}},\\
  \end{split}
  \right.
\end{equation}

\noindent
where
\begin{equation}
\left\{
\begin{split}
 q&=\lambda_1(\lambda_dn_0+\lambda_{\emptyset}(1-n_0)),\\
 \alpha&=\lambda_1(\lambda_d-\lambda_{\emptyset}),\\
 \beta&=\frac{n_0(\lambda^2_d-1)}{(1-n_0)},\\
 \gamma&=\frac{(\lambda_d^2-1)(1-2n_0)}{(1-n_0)^2}.
\end{split}
\right.
\end{equation}

\noindent
As a result, we obtain the following formulas for the expectation values in the correlated state 
\begin{equation}
\left\{
\begin{split}
&\langle\hat{c}^{\dagger}_{i\sigma}\hat{c}_{j\sigma}\rangle_G=q^2T_{ij}^{(1)(1)}+q\alpha T_{ij}^{(1)(3)}+\alpha^2T_{ij}^{(3)(3)}, \\
& \langle\hat{n}_i\hat{n}_j \rangle_G=8(1+\beta)\big(\gamma I_{ij}^{(24)}+n_0I_i^{(2)}\big)+4\gamma^2I_{ij}^{(44)} + 8\gamma n_0 I_i^{(4)}+2(1+\beta)^2\big(I^{(22)}_{ij\;\uparrow\uparrow}+I^{(22)}_{ij\;\uparrow\downarrow}\big) + 4n_0^2, \\
& \langle\hat{s}_{i\sigma}\hat{s}_{j\bar{\sigma}}\rangle_G=\lambda_1^4S_{ij}^{(22)}, \\
& \langle\hat{d}_{i}\rangle_G=\lambda_d^2\bigg( \frac{1-\lambda_d^2n_0^2}{1-n_0^2}I_i^{(4)}+2n_0I_i^{(2)}+n_0^2 \bigg),
\end{split}
\right.
\end{equation}

\noindent
where the symbols $T^{(1)(1)}_{ij}$, $T^{(1)(3)}_{ij}$, $T^{(3)(3)}_{ij}$, 
$I^{(24)}_{ij}$, $I^{(2)}_i$, $I^{(44)}_{ij}$, $I^{(4)}_i$, $I^{(22)}_{ij\;\uparrow\uparrow}$, 
$I^{(22)}_{ij\;\uparrow\downarrow}$ correspond to the so-called diagrammatic sums. They all have the form analogical to the one presented below for general operator $\hat{o}$, i.e., 

\begin{equation}
 \mathcal{S}\equiv\sum_{k=1}^{\infty}\frac{x^k}{k!}\sum_{l_1...l_k}\langle\hat{o} \;\hat{d}^{\textrm{HF}}_{l_1...l_k} \rangle_0^c, 
 \label{eq:diag_sums_general}
\end{equation}

\noindent
where $\langle...\rangle^c_0$ denotes that only connected diagrams are included. This is due to the fact 
that  in the expectation values $\langle...\rangle_G=\langle\Psi_G|...|\Psi_G\rangle/\langle\hat{P}^2\rangle_0$
the disconnected terms are systematically canceled out by the denominator. In effect, only the connected diagrams remain in the expression. For the sums
$T^{(1)(1)}_{ij}$, $T^{(1)(3)}_{ij}$, $T^{(3)(3)}_{ij}$, 
$I^{(24)}_{ij}$, $I^{(2)}_i$, $I^{(44)}_{ij}$, $I^{(4)}_i$, $I^{(22)}_{ij\uparrow\uparrow}$, 
$I^{(22)}_{ij\uparrow\downarrow}$, the symbol $\hat{o}$ corresponds to 
$\hat{c}^{\dagger}_{i\sigma}\hat{c}_{j\sigma}$, 
$\hat{c}^{\dagger}_{i\sigma}\hat{n}^{\textrm{HF}}_{j\bar{\sigma}}\hat{c}_{j\sigma}$,
$\hat{n}^{\textrm{HF}}_{i\bar{\sigma}}\hat{c}^{\dagger}_{i\sigma}\hat{n}^{\textrm{HF}}_{j\bar{\sigma}}\hat{c}_{j\sigma}$,
$\hat{n}^{\textrm{HF}}_{i\sigma}\hat{d}^{\textrm{HF}}_{j}$,
$\hat{n}^{\textrm{HF}}_{i\sigma}$,
$\hat{d}^{\textrm{HF}}_{i}\hat{d}^{\textrm{HF}}_{j}$,
$\hat{d}^{\textrm{HF}}_{i}$,
$\hat{n}^{\textrm{HF}}_{i\uparrow}\hat{n}^{\textrm{HF}}_{i\downarrow}$,
$\hat{n}^{\textrm{HF}}_{i\uparrow}\hat{n}^{\textrm{HF}}_{i\uparrow}$, respectively.

As a result of these formal manipulations, the expectation value of the Hubbard Hamiltonian in the nonmagnetic and nonsuperconducting state can be written in the following form

\begin{align}
    \braket{\hat{\mathcal{H}}}_G =  2\sum_{ij} t_{ij} 
    \left[q^2T_{ij}^{(1)(1)} + q\alpha T^{(1)(3)} + \alpha^2
    T_{ij}^{(3)(3)} + NU\lambda^2_d[(1-xd^2)I_i^{(4)}] + 2n_0I_i^{(2)} + d^2_0\right].
\end{align}

\noindent
We also define the grand-canonical potential $\mathcal{F}$ at zero
temperature as

\begin{align}
    \mathcal{F} \equiv \braket{\hat{\mathcal{H}}}_G - 2\mu_G n_GN, 
\end{align}

\noindent
where

\begin{align}
    n_G \equiv \braket{\hat{n}_{i\sigma}}_G = \lambda^2_d 
    [d^2_0 + I^{(4)}_2 (1-xd_0^2) + 2n_0 I_1^{(2)}] 
    + \lambda^2_1 [n_0(1-n_0)  I^{(2)}_i (1-2n_0) - I^{(4)}_i
    (1+xn_0(1-n_0))],
\end{align}

\noindent
with $d_0^2 \equiv \braket{\hat{n}_{i\uparrow}\hat{n}_{i\downarrow}} = n_0^2$. Note that, in general, the particle-number expectation values, $n_G$ and $n_0$, may differ (e.g., they do in the superconducting phase), so sometimes an additional constraint to the expression for $\mathcal{F}$ must be introduced $\sim \mu^{\prime} N(n_G-n_0)$, where $\sim \mu^{\prime}$ is a new parameter to be optimized. Also, one sees that both the hopping and Hubbard terms are renormalized by intersite correlations in the higher expansion order. 

The whole further procedure is built up on the principle that the wave function $\ket{\Psi_0}$ is the ground-state wave function of an uncorrelated state and therefore, an effective diagrammatic expansion, based on the Wick-theorem in the direct space, can be constructed. Such an expansion may involve a huge number of connected diagrams at higher ($k\geq 3$) orders. Due to the lattice translational invariance, we have $T_{ij}^{(1)(3)} = T_{i-j}^{(1)(3)}$, etc., and
$I^{(2)(4)}_i = I^{(2)(4)} $. In practice, the computations are carried out to fixed order $k$, where $k$ denoted the number of internal vertices in respective graphs. Alternatively, one can calculate the diagrammatic sums up to the order $k$ and stop if the $(k+1)$th order introduces only minor numerical corrections regarded as negligible, which is a natural convergence criterion. Also, the lines $S_{ij}$ and $T_{ij}$ are accounted for up to certain cutoff distance in real space, which need to specified in each case. In effect, the infinite summation in~\eqref{eq:expectation_val_terms} turns into a finite (executable) summation of diagrams.

The interesting us quantities are the average particle number $\langle \hat{n}_{i\sigma}\rangle$, $d^2$, the pairing amplitude $\langle \hat{a}^\dagger_{i\sigma} \hat{a}^\dagger_{j\bar{\sigma}}\rangle_G$, and hopping amplitude $\langle \hat{a}^\dagger_{i\sigma} \hat{a}_{j\bar{\sigma}}\rangle_G$. For achieving that, we have to detemrine first their counterparts in the uncorrelated state $|\Psi_0\rangle$.

\subsection{The effective single-particle Hamiltonian approach to the diagrammatic expansion and $|\Psi_0\rangle$ determination}
\label{sec:de_gwf_effective_hamiltonian}

As mentioned earlier, the uncorrelated (``unprojected'') wave function $\ket{\Psi_0}$ should be determined separately not only to close the whole procedure, but also to make it realistic and rapidly convergent. To determine this wave function we use the variational principle

\begin{align}
    \frac{\delta}{\delta\bra{\Psi_0}} \left\{ \mathcal{F} - 
    \lambda \left(\braket{\Psi_0|\Psi_0} - 1\right)\right\} = 0 \label{vareq}
\end{align}

\noindent
where $\lambda$ is a Lagrange multiplier responsible for the wave function normalization. An identical procedure is used when deriving variationally the Schr\"odinger wave equation for correlated electrons \cite{SpalekPhysRevB2000}, in which the parameter $\lambda$ plays the role of the energy eigenvalue for the ground state. Here, the role of action is played by the effective grand-canonical potential playing the role of the Landau functional in the theory of phase transitions at temperature $T = 0$ it is just the ground-state-energy functional. Next, we rewrite the equation
\eqref{vareq} in the form

\begin{align}
    \frac{\delta \mathcal{F}}{\delta P_{ij}} \cdot 
    \frac{\delta P_{ij}}{\delta \bra{\Psi_0}} + 
     \frac{\delta \mathcal{F}}{\delta S_{ij}} \cdot 
    \frac{\delta S_{ij}}{\delta \bra{\Psi_0}} = \lambda \ket{\Psi_0}.
\end{align}

\noindent
Note that we have assumed here a spin nonpolarized but paired state. Now, with the definitions $P_{ij} \equiv \bra{\Psi_0}\hat{a}^{\dagger}_{i\sigma}\hat{a}_{j\sigma}\ket{\Psi_0}$ and $S_{ij} \equiv \bra{\Psi_0}\hat{a}^{\dagger}_{i\uparrow}\hat{a}^{\dagger}_{j\downarrow}\ket{\Psi_0}$ as the relevant variational variables (physically important quantities), we can explicitly evaluate derivatives over $\bra{\Psi_0}$ as

\begin{align}
    \left[\sideset{}{'} \sum_{ij\sigma} \frac{\delta\mathcal{F}}{\delta P_{ij}} \hat{a}^{\dagger}_{i\sigma}
    \hat{a}_{j\sigma} + \sideset{}{'}\sum_{ij} \left(
    \frac{\delta\mathcal{F}}{\delta S_{ij}}
    \hat{a}^{\dagger}_{i\uparrow}
    \hat{a}_{j\downarrow} + \mathrm{H.c.}\right)\right] \ket{\Psi_0} = \lambda \ket{\Psi_0},
\end{align}

\noindent
or briefly

\begin{align}
  \mathcal{\hat{H}}^\mathrm{eff}\ket{\Psi_0} = E^\mathrm{eff} \ket{\Psi_0},
\end{align}

\noindent
where we have introduced the effective single-particle Hamiltonian

\begin{align}
      \mathcal{\hat{H}}^{\mathrm{eff}} \equiv \sideset{}{'}\sum_{ij\sigma} t^\mathrm{eff}_{ij}
     \hat{a}^{\dagger}_{i\sigma}\hat{a}_{j\sigma} +
  \sideset{}{'}\sum_{ij} (\Delta^\mathrm{eff}_{ij}\hat{a}^{\dagger}_{i\uparrow}\hat{a}_{j\downarrow} + \Delta_{ij}^{\mathrm{eff}*}\hat{a}^{\dagger}_{i\downarrow}\hat{a}_{j\uparrow}) - \mu_0 n_0 N.
  \label{eq:effective_single_particle_hamiltonian}
\end{align}

\noindent
In Eq.~\eqref{eq:effective_single_particle_hamiltonian}, the the effective hopping and pairing-gap amplitudes coefficients are explicitly defined as 

\begin{align}
    t^\mathrm{eff}_{ij} \equiv \frac{\delta \mathcal{F}}{\delta P_{ij}}
\end{align}

\noindent
and

\begin{align}
    \Delta^\mathrm{eff}_{ij} = \frac{\delta \mathcal{F}}{\delta S_{ij}}.
\end{align}

\begin{figure}
  \centering
  \includegraphics[width=0.5\textwidth]{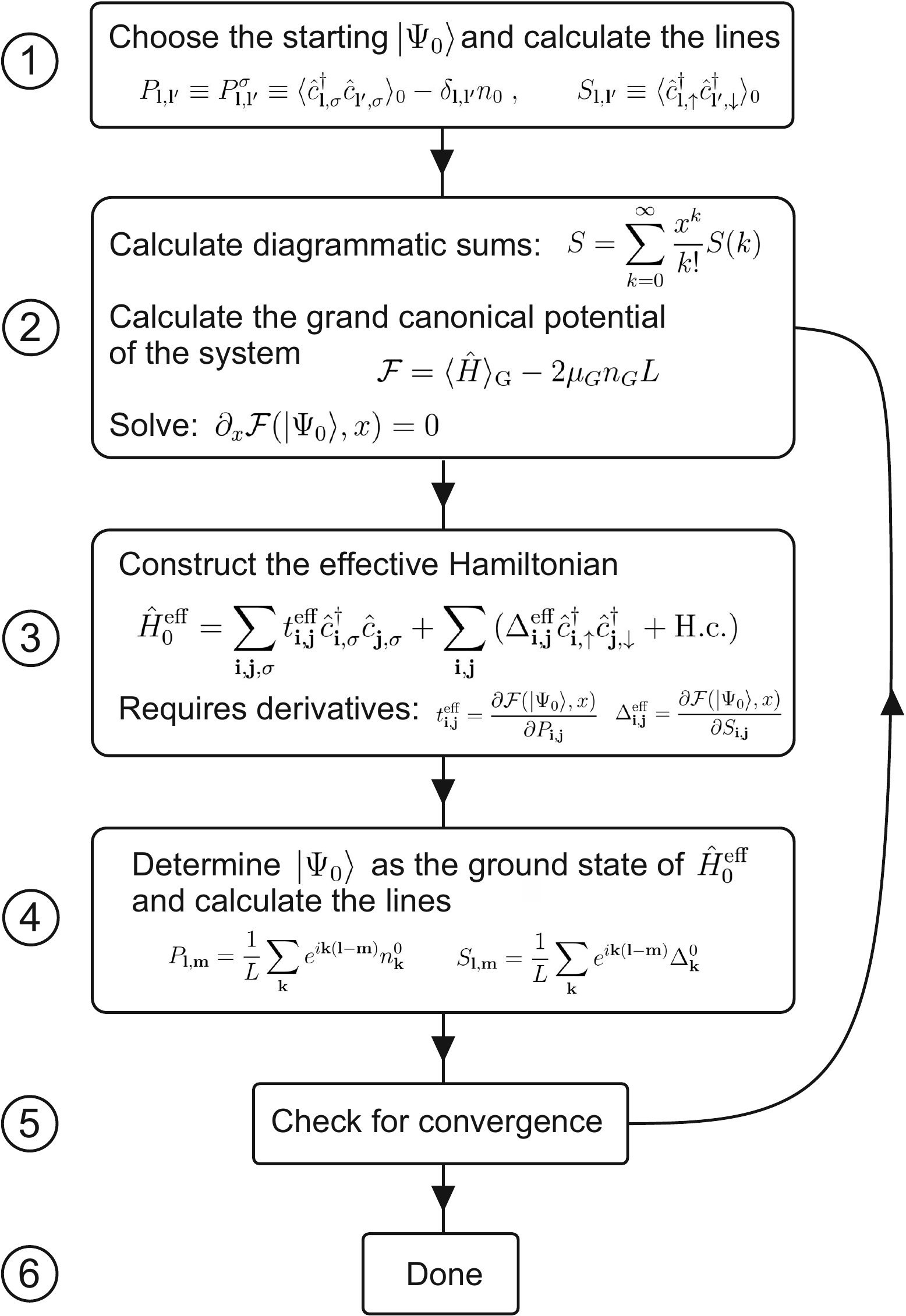}
  \caption{Flowchart illustrating the procedure of calculating the effective uncorrelated characteristics which, in turn, determine those for the correlated state. The effective hopping ($t^\mathrm{eff}$) and gap ($\Delta^\mathrm{eff}$) parameters are input quantities in determining the corresponding quantities in the correlated state, obtained via DE-GWF expansion. After~\cite{KaczmarczykPhilMag2014}.}
    \label{fig:flowchart_de_gwf}
\end{figure}

\noindent
Both of them are complex functions of $P_{ij}$ and $S_{ij}$, fulfilling $t^\mathrm{eff}_{ij} = t^\mathrm{eff}_{ji}$ and $\Delta^\mathrm{eff}_{ij} = \Delta^\mathrm{eff}_{ji}$.  The effective Hamiltonian $\mathcal{\hat{H}}^\mathrm{eff}$ has a single-particle from, with the correlations implicitly contained in the functional $\mathcal{F}$, and it defines quasiparticles in the correlated state. The quantities describing the correlated state are represented by averages taken with the wave function $|\Psi_G\rangle$. In effect, we determine $\ket{\Psi_0}$ self-consistently by solving equations for $S_{ij}$ and $P_{ij}$, and other quantities in the correlated state. The procedure is outlined in the flowchart, shown in Fig.~\ref{fig:flowchart_de_gwf}. In the next section, we present the explicit solution of the BCS-type Hamiltonian~\eqref{eq:effective_single_particle_hamiltonian} as an intermediate step in the procedure.

\subsection{BCS approach to effective single-particle Hamiltonian}

Next step is to diagonalize Hamiltonian~\eqref{eq:effective_single_particle_hamiltonian} in the paired state. For this purpose we introduce the space Fourier transforms of $t^{\mathrm{eff}}_{ij}$ and $\Delta^{\mathrm{eff}}_{ij}$ as

\begin{equation}
\left\{
    \begin{split}
        \epsilon_\mathbf{k} \equiv \frac{1}{N} \sum_{ij} t^{\mathrm{eff}}_{ij} e^{i\mathbf{k}(\mathbf{R}_i-\mathbf{R}_j)} \\ 
        \Delta_\mathbf{k} \equiv \frac{1}{N} \sum_{ij} \Delta^{\mathrm{eff}}_{ij}
        e^{i\mathbf{k}(\mathbf{R}_i-\mathbf{R}_j)} 
    \end{split}
    \right.
\end{equation}

\noindent
Note that since both $t^\mathrm{eff}_{ij}$ and $\Delta^\mathrm{eff}_{ij}$ are nonzero only for $i \neq j$, thus excluding a simple isotropic pairing of BCS type (since in the strongly correlated limit $\Delta_{ii} \equiv 0$). In $\mathbf{k}$ space, the Hamiltonian can be brought to the usual Boguliubov-Nambu form

\begin{align}
    \mathcal{\hat{H}}^{\mathrm{eff}} - \mu N = \sum_\mathbf{k}
  \begin{array}{r}
    \begin{pmatrix}
    \hat{a}^{\dagger}_{\mathbf{k},\uparrow},  \hat{a}_{-\mathbf{k},\downarrow}
    \end{pmatrix} \\ {}
  \end{array}
    \begin{pmatrix}
\epsilon_\mathbf{k} & \Delta_\mathbf{k} \\
\Delta^{*}_{\mathbf{k}} & -\epsilon_\mathbf{k} 
\end{pmatrix}
\begin{pmatrix}
    \hat{a}_{\mathbf{k},\uparrow} \\ \hat{a}^{\dagger}_{-\mathbf{k},\downarrow}
    \end{pmatrix}+ \sum_\mathbf{k}\epsilon_\mathbf{k},
\end{align}

\noindent
where we have made the substitution $\epsilon_\mathbf{k} \equiv \epsilon_\mathbf{k}-\mu$. The diagonalization transformation

\begin{align}
    \begin{pmatrix}
\hat{a}_{\mathbf{k}\uparrow} \\
\hat{a}^{\dagger}_{-\mathbf{k}\downarrow}\end{pmatrix} = 
\begin{pmatrix}
u_\mathbf{k} & -v_\mathbf{k} \\
v_{\mathbf{k}} & u_\mathbf{k} 
\end{pmatrix}
 \begin{pmatrix}
\hat{\alpha}_\mathbf{k} \\
\hat{\beta}^{\dagger}_\mathbf{k}, 
\end{pmatrix}
\end{align}

\noindent
with the eigenvalues $\lambda_{\mathbf{k}}^{1,2} \equiv \pm E_\mathbf{k}$ and excitation energy $E_\mathbf{k} = \left(\epsilon^2_\mathbf{k}+\Delta^2_\mathbf{k}\right)^{\frac{1}{2}}$. Also, the transformation coefficients acquire the form

\begin{align}
\begin{pmatrix}
u_\mathbf{k} \\
v_\mathbf{k}, 
\end{pmatrix} = \frac{1}{\sqrt{2}} \left(1\pm \frac{\epsilon_\mathbf{k}}{E_\mathbf{k}}\right)^{\frac{1}{2}}.
\end{align}

\noindent
Note that $u_\mathbf{k}$ approaches the maximal value of unity for $\epsilon_\mathbf{k}\rightarrow E_\mathbf{k}$ ($\epsilon_\mathbf{k}\rightarrow\mu$ physically), whereas $v_\mathbf{k}$ when $-\epsilon_\mathbf{k} \rightarrow E_\mathbf{k}$. Hence, they represent mixture coefficients (Bogoliubov coherence factors) of electron- and hole states, respectively.

The same sums which are used here for the case of $t$-$J$-$U$-$V$ model (with nonzero double occupancies) are 
also used for the case of the $t$-$J$ model (with zero double occupancies) and their form is given in 
Ref.~\cite{KaczmarczykNewJPhys2014}. Within the above approach all the diagrammatic sums, and hence the system ground state energy, can be expressed through the expectation values (\ref{eq:lines}) and the variational parameter $x$ .

In those asymptotic limits, excited quasiparticle (broken-pair) excitations are electrons and holes, respectively. This follows from the fact that the diagonalized Hamiltonian can be represented by either of the two forms

\begin{align}
    \mathcal{\hat{H}}^\mathrm{eff} - \mu N = \sum_l E_\mathbf{k} (\alpha^{\dagger}_\mathbf{k}\alpha_\mathbf{k} - \beta_\mathbf{k}\beta^{\dagger}_\mathbf{k}) + \mathrm{const.} = 
    \sum_\mathbf{k} E_\mathbf{k} (\alpha^{\dagger}_\mathbf{k}\alpha_\mathbf{k} + \beta^{\dagger}_\mathbf{k}\beta_\mathbf{k}) + \mathrm{const.}
\end{align}

\noindent
So, we can represent operator $\beta_\mathbf{k}$ as a creation of a hole with negative energy (below the Fermi surface) or creation of
excited state creation with positive energy.

To close the solution, we need to specify equation for $\Delta_\mathbf{k}$ and $\mu$. For that purpose, we assume without loss of generality that $\Delta_{ij} \equiv V_{ij}\braket{\hat{a}_{i\uparrow}\hat{a}_{j\downarrow}}_0$, where $V_{ij}$ is the pairing potential ($\Delta_{ij}^\mathrm{eff}$ here, as in \eqref{eq:effective_single_particle_hamiltonian}). In that situation $\Delta_\mathbf{k} = \frac{1}{N}
\sum_\mathbf{k} V_{\mathbf{k}\mathbf{k}'}\Delta^0_{\mathbf{k}'}$, where

\begin{align}
    \Delta^0_\mathbf{k} = \frac{1}{N} \sum_{ij} e^{i\mathbf{k}(\mathbf{R}_i-\mathbf{R}_j)} \braket{\hat{a}_{i\uparrow}\hat{a}_{j\downarrow}}_0.
\end{align}

\noindent
Then, the self-consistent equation for $\Delta^0_\mathbf{k}$ takes the form

\begin{align}
     \Delta^0_\mathbf{k} \equiv \sum_{\mathbf{k}'} V_{\mathbf{k}\mathbf{k}'} \frac{\Delta^0_{\mathbf{k}'}}{2E_\mathbf{k}} \tanh\left(\frac{E_{\mathbf{k}'}}{2k_BT}\right).
\end{align}

\noindent
In the above formulation, $\Delta^0_\mathbf{k}$ is dimensionless. In the  $t$-$J$ model with three-site terms included, we have that $V_{\mathbf{k}\mathbf{k}'} = -V \gamma_\mathbf{k}\gamma_{\mathbf{k}'}$, i.e., the pairing potential is separable into product of $\gamma_\mathbf{k}$ and $\gamma_{\mathbf{k}^\prime}$, $V = \frac{4t^2}{U}$, and $E_\mathbf{k} = \left[(\epsilon_\mathbf{k} - \mu)^2 + |\Delta_\mathbf{k}|^2\right]^{\frac{1}{2}}$.

What differentiates between high-$T_c$- and conventional (low-$T$) superconductors is that now we have to adjust the chemical potential $\mu = \mu (T)$ in the normal phase. This means we have to determine it from the condition for the conservation of particles, which takes the usual form

\begin{align}
    \frac{1}{N} \sum_\mathbf{k} n^0_\mathbf{k} = \frac{1}{N} \frac{1}{\exp(\beta E_\mathbf{k})+1} = n_e, 
\end{align}

\noindent
where $n_e\equiv N_e/N$ is the band filling. The quasiparticle distribution function is defined by the space Fourier transform of $P_{ij}$

\begin{align}
    n^0_\mathbf{k} = \frac{1}{N} \sum_{ij} \exp\left(i\mathbf{k}(\mathbf{R}_i-\mathbf{R}_j)\right) P_{ij}.
\end{align}

\noindent
The outlined analysis of physical properties is based on the above two paradigms, by which those systems differ from that for the low-temperature superconductors. Those are (\emph{i}) explicit $\mathbf{k}$-dependence of the pairing gap, and (\emph{ii})the necessity to adjust chemical potential to each phase considered.  The remaining factor is the reduced dimensionality ($d=2$), although this factor requires a more detailed analysis because, e.g., of the long-range nature of the Coulomb interaction in the third dimension caused by reduced and emerging plasmon screening. 

\subsubsection{Two pairing gaps}

At this point, one can introduce two distinct pairing gaps. The physical (``true'') amplitude of real-space pairing is defined as an expectation value taken in the correlated state, i.e.,

\begin{align}
    \Delta_{G\,{ij}} \equiv \frac{\braket{\Psi_G|\hat{a}^{\dagger}_{i\uparrow}\hat{a}^{\dagger}_{j\downarrow}  |\Psi_G}}{\langle\Psi_G|\Psi_G\rangle}
    \equiv \frac{\braket{\Psi_0|\hat{P}\hat{a}^{\dagger}_{i\uparrow}\hat{a}^{\dagger}_{j\downarrow}\hat{P}|\Psi_0}}{\langle\Psi_0|\hat{P}_G^2|\Psi_0\rangle}.
\end{align}

\noindent
In other words, leaving for now the effect of normalization factor in the denominator, the pairing-amplitude operator may be defined as

\begin{align}
    \hat{\Delta}_{ij} = \hat{P}\hat{a}^{\dagger}_{i\uparrow}\hat{a}^{\dagger}_{j\downarrow}\hat{P} = 
     (\hat{P}_i\hat{a}^{\dagger}_{i\uparrow}\hat{P}_i) (\hat{P}_j\hat{a}^{\dagger}_{j\downarrow}\hat{P}_j) \hat{P}^\prime,
     \equiv \hat{b}^{\dagger}_{i\uparrow}\hat{b}^{\dagger}_{j\downarrow}\hat{P}^{\prime}
\end{align}

\noindent
where projected $\hat{b}^{\dagger}_{i\sigma}$ and $\hat{b}_{i\sigma}$ are introduced (cf. Appendix~\ref{appendix:derivation_of_the_tj_model}), and $\hat{P}' \equiv \prod_{l\neq i, j} \hat{P}_l^2$. Note that, formally, the pairing operators have explicit spin-singlet form, i.e., 

\begin{align}
    \hat{\Delta}_{ij} \rightarrow \hat{B}^{\dagger}_{ij} \equiv \frac{1}{\sqrt{2}} \hat{P}
    \left(\hat{a}^{\dagger}_{i\uparrow}\hat{a}^{\dagger}_{j\downarrow}-
    \hat{a}^{\dagger}_{i\downarrow}\hat{a}^{\dagger}_{j\uparrow}\right)\hat{P} = 
    \frac{1}{\sqrt{2}} \left(\hat{b}^{\dagger}_{i\uparrow}\hat{b}^{\dagger}_{j\downarrow}-
    \hat{b}^{\dagger}_{i\downarrow}\hat{b}^{\dagger}_{j\uparrow}\right)\hat{P}^{\prime}.
\end{align}

\noindent
In these formulas $\hat{P}^{\prime}$ denotes projections in which neither $\hat{P}_i$ nor $\hat{P}_j$ appear. For $i\neq j$ and 
for the case when the interchange of indices $i \leftrightarrow j$ is a symmetry operation, we have that $\hat{\Delta}_{ij} = \sqrt{2}\hat{b}^{\dagger}_{i}\hat{b}_{j}$. The same type of reasoning concerning the projection is applied to the hopping part $\sim \hat{a}^{\dagger}_{i\sigma}\hat{a}_{j\sigma}$. 

Apart from the correlated gap $\braket{\hat{\Delta}_{ij}}_G$, we have defined also the gap $\Delta^0$ for uncorrelated state according to 

\begin{align}
    \Delta^0_{ij} = \braket{\Psi_0|\hat{a}^{\dagger}_{i\uparrow}\hat{a}^{\dagger}_{j\downarrow}|\Psi_0}.
\end{align}

\noindent
Whereas $\Delta_G \equiv \langle \Psi_G|\hat{a}_{i\uparrow} \hat{a}_{j\downarrow} |\Psi_G\rangle$ reflects the true $d$-wave SC behavior, its uncorrelated counter-part $\Delta^0$ follows qualitatively the doping dependence of the pseudogap, as  will be discussed at the end of this Report.

\subsection{Gutzwiller, renormalized mean field and statistically consistent Gutzwiller approximations: A brief overview}

At the end of this section, we briefly refer to the earlier SGA approach \cite{JedrakArXiV2010,JedrakPhsRevB2011} to the correlated fermions and to the high-$T_c$ superconductivity in particular. The standard Gutzwiller-type approximation relies on reducing the projected hopping part to the form

\begin{align}
    \hat{P}\left\{ \sum_{ij\sigma} t_{ij} \hat{a}^{\dagger}_{i\sigma}\hat{a}_{j\sigma}\right\}\hat{P} \rightarrow \sum_{ij\sigma}q_{\sigma}\hat{a}^{\dagger}_{i\sigma}\hat{a}_{j\sigma},
\end{align}

\noindent
representing the first term $\sim T^{(1)(1)}_{ij}$ (the renormalization coefficient $q_{\sigma}$ is called \emph{the band narrowing factor}, fulfilling the condition $0 < q_\sigma \leq 1$). Similar procedure is carried out for the terms $\sim U$ and $n_G$ in the formulas~\eqref{eq:expectation_val_terms}. Therefore, they represent the zeroth-order contributions in the present diagrammatic expansion, DE-GWF. In practical analysis, $(\hat{P}_i\hat{a}^{\dagger}_{i\sigma}\hat{P}_i)
(\hat{P}_j\hat{a}^{\dagger}_{j\sigma}\hat{P}_j)$ and $(\hat{P}_i\hat{a}^{\dagger}_{i\uparrow}\hat{P}_i)
(\hat{P}_j\hat{a}^{\dagger}_{j\downarrow}\hat{P}_j)$ may be regarded as proportional to $\hat{a}^{\dagger}_{i\sigma}\hat{a}_{j\sigma}$ and $\hat{a}^{\dagger}_{i\uparrow}\hat{a}_{j\downarrow}^\dagger$, respectively. In effect, the single-particle hopping is renormalized by the factor $q_{\sigma} \equiv q_{\sigma}\{ d^2, P_{ij}, S_{ij}, ... \} $ and the Hubbard Hamiltonian takes the approximate form in the paramagnetic phase

\begin{align}
    \mathcal{\hat{H}} \simeq \sum_{ij\sigma} q_{\sigma} 
    t_{ij} \hat{a}^{\dagger}_{i\sigma}\hat{a}_{j\sigma} + 
    Ud^2 N, \label{paramham}
\end{align}

\noindent
where, as before, $d^2 \equiv \braket{\hat{n}_{i\uparrow}\hat{n}_{i\downarrow}}$ becomes an additional (variational) parameter which is determined by minimizing the ground state energy $E_0 = \braket{\mathcal{\hat{H}}}_0$ or, at $T>0$, the Gibbs (or free)
energy in the form $\mathcal{F} = E_G - TS_0$. In this expression, the configuration entropy can be thus taken as in the Landau-Fermi-liquid theory, in usual fermionic form

\begin{align}
    S_0 = -k_B \left\{\sum_{k\sigma} f_{\textbf{k}\sigma}
    \ln f_{\textbf{k}\sigma} + (1-f_{\textbf{k}\sigma})
    \ln(1-f_{\textbf{k}\sigma}) \right\},
\end{align}

\noindent
where $f_{\textbf{k}\sigma} \equiv f(E_{\textbf{k}\sigma})$
is the quasipaticle energy $E_{\textbf{k}\sigma} = E_{\textbf{k}} 
= q_{\sigma}\epsilon_{\textbf{k}}$, and $\epsilon_{\textbf{k}}$ is the bare band energy. In general, the band narrowing factor in the homogeneous spin-polarized situation reads \cite{GutzwillerPhysRevLett1963,GutzwillerPhysRev1965,BrinkmanPhysRevB1970}

\begin{align}
    q_{\sigma} \equiv q_{\sigma}(d,n_{\uparrow}, n_{\downarrow}) 
    = \frac{\left\{\left[(n_{\sigma}-d^2)( 1-n_{\sigma}-n_{\bar{\sigma}} +
    d^2) 
    \right]^{\frac{1}{2}} + d(n_{\bar{\sigma}} -d^2)^{\frac{1}{2}}
    \right\}^2}
    {n_{\sigma}(1-n_{\sigma})}.
\end{align}

The physics behind the form \eqref{paramham} of the effective quasiparticle Hamiltonian replacing the original Hubbard Hamiltonian is as follows \cite{SpalekPhysRevLett1987,SpalekPhysRevB1989}. With the increasing interaction we reach the point $U \sim W$, when the hopping and interaction
terms are of comparable amplitude. Under these circumstances, there is no small energy scale in the system. The proposal was to
define the correlation factor $\braket{\hat{n}_{i\uparrow} \hat{n}_{i\downarrow}}$ as the new basic parameter and evaluate the renormalized band energy by the factor $q$. The resultant correlated state with renormalized Fermi-liquid characteristics is obtained for the physical state after minimizing $E_G$ or $\mathcal{F}$ with respect to $d^2$. At nonzero temperature, such an approach leads to the results depicted in Fig.~\ref{fig:1.4}.

\subsubsection{Renormalized mean field theory (RMFT)}

The renormalized mean field theory is essentially based on the simplest (not statistically consistent) Gutzwiller approximation, this time applied for the
$t$-$J$ model, obtained from the Hubbard model in the strong-correlation limit (for detailed derivation of $t$-$J$-$(V)$ model see Appendix~\ref{appendix:derivation_of_the_tj_model}). This full $t$-$J$ Hamiltonian, containing also the three-site-term contributions has the form

\begin{align}
\hat{P}_{1}\widetilde{\mathcal{\hat{H}}}\hat{P}_{1}= \sum_{ij\sigma}\!^{'} t_{ij}\, \hat{b}_{i\sigma}^{\dagger} \hat{b}_{j\sigma}+\frac{1}{2} \sum_{ij}\!^{'}
V_{ij} \hat{\nu}_{i}\, \hat{\nu}_{j}+\sum_{ij}\!^{'}\, \tilde{J}_{ij}\: \hat{\mathbf{S}}_{i} \cdot \hat{\mathbf{S}}_{j}+\; \mbox{(3-site terms)},
\label{eq:full_tj_hamiltonian_wuth_three_site_terms}
\end{align}

\noindent
where $\hat{P}_i$ is the part of the total projector $\hat{P}$ with the part containing doubly occupied configurations neglected. We can rewrite this Hamiltonian by introducing the projected creation and annihilation operators, $\hat{b}_{i\sigma} ^{\dagger} \equiv \hat{a}_{i\sigma}^{\dagger}(1-\hat{n}_{i\bar{\sigma}})$ and $\hat{b}_{i\sigma} = (\hat{b}_{i\sigma}^{\dagger})^{\dagger}$, respectively. In that representation, the $t$-$J$ Hamiltonian acquires a more compact form

 \begin{align}
\mathcal{\hat{H}}_{t\text{-}J} \equiv \hat{P}_{1}\widetilde{\hat{\mathcal{H}}}\hat{P}_{1}=\hat{P}_{1} \left\{ \sum_{ij}t_{ij}\hat{b}_{i\sigma}^{\dagger}\hat{b}_{j\sigma} - \sum_{ijk}\frac{2t_{ij}t_{kj}}{U} \hat{B}_{ij}^{\dagger}\hat{B}_{kj} \right\} \hat{P}_{1}.
\label{eq:tj_model_hamiltonian_compact_form}
\end{align}

\noindent
 Note that the spin operators $\{ \hat{\mathbf{S}}_i\}$ have identical form in terms of either operators $\{ \hat{a}_{i\sigma}, \hat{a}_{i\sigma}^{\dagger}\}$ or $\{ \hat{b}_{i\sigma}, \hat{b}_{i\sigma}^{\dagger}\}$.

Now, RMFT relies on the assumption that the operator $\hat{P}_1
^{\prime} = \mathbb{1}$, i.e., the correlations beyond two-sites 
 are disregarded. However, we still have to deal
with projected operators, $\hat{b}_{i}$ and $\hat{b}_{i}^{\dagger}$, 
which obey non-fermionic anticommutation relations.  In the present 
situation, the projected hopping and the exchange-interaction
terms are of comparable amplitude. Again, we renormalized both 
the parameters, $t_{ij}$ and $\delta_{ij}$, in the following 
manner: $t_{ij}$ is reduced by the Gutzwiller factor 
$q_{\sigma}$ in the limit $d^2\rightarrow 0$, i.e., 

\begin{align}
    q_{\sigma} = \frac{1-n}{1-n_{\sigma}} 
    \rightarrow g =\frac{1-n}{1-\frac{n}{2}} = \frac{2\delta}{1+\delta}, 
\end{align}

\noindent
with $\delta \equiv 1 - n$ is the hole doping. On the other hand, the full exchange operator is

\begin{align}
    -\left(\hat{\mathbf{S}}_i \cdot \hat{\mathbf{S}}_j - \frac{1}{4}\hat{n}_i\hat{n}_j\right) = +1,
\end{align}

\noindent
in the spin singlet state and for $n=1$. For $n<1$ this value  is diminished by factor $(1-\delta)^2$ due to the holes. But that is not the whole story. In the mean-field approximation. Hence, to represent the full exchange value, we have to do the following renormalization

\begin{align}
    J \rightarrow 4(1-\delta)^2 J \simeq 4J(1-2\delta).
\end{align}

\noindent
The numerical factor $4$ before $J$ can also be justified in the following way. The full exchange part in the real-space 
operators has the form (cf. Appendix~\ref{appendix:derivation_of_the_tj_model})

\begin{align}
    -J \sum_{\braket{ij}} \hat{b}^{\dagger}_{ij}
    \hat{b}_{ij} = -4J\sum_{\braket{ij}} \hat{b}^{\dagger}_{i\uparrow}\hat{b}^{\dagger}_{j\downarrow}
    \hat{b}_{i\uparrow}\hat{b}_{j\downarrow},
\end{align}

\noindent
in which each pair of neighboring sites $\braket{ij}$ is taken 
only once. Summarizing, the renormalization factors are introduced 
to disregard the projected operators $\hat{b}_{i\sigma}$ and 
$\hat{b}_{i\sigma}^{\dagger}$ and replace them by the original
operators (unprojected) $\hat{a}_{i\sigma}$ and $\hat{a}_{i\sigma}^{\dagger}$ in the hopping term. Furthermore, 
having those factors in $J$ we can decouple the pairing part in the 
mean-field (BCS) way, i.e., replace $\hat{b}^{\dagger}\hat{b}^{\dagger}\hat{b}\hat{b}$ with 
$\hat{b}^{\dagger}\hat{b}^{\dagger}\braket{\hat{b}\hat{b}}
+ \hat{b}\hat{b}\braket{\hat{b}^{\dagger}\hat{b}^{\dagger}} - \langle\hat{b}^\dagger \hat{b}^\dagger\rangle \langle\hat{b} \hat{b}\rangle$.

We summarize next the most advanced form of RMFT, namely the statistically consistent Gutzwiller approximation (SGA) before
discussing the related theory beyond RMFT.

\subsection{Supplement: Statistically consistent Gutzwiller approximation (SGA) as corrected GA}

Now, we are ready to summarize the most advanced formulation of the approach based on Gutzwiller (or Jastrow \cite{ZegrodnikPhysRevB2019,BiborskiPhysRevB2020}) variational wave 
function, namely the statistically consistent Gutzwiller approximation (SGA). This approach forms a basis to the systematic diagrammatic expansion of the Gutzwiller (or related) wave function (DE-GWF method).

\subsubsection{The case of the Hubbard model in normal phase}

The first question we have to address is why we need a modification of the original Gutzwiller approximation (GA) already when considering the Hubbard model. This question has been tackled in detail in an early report \cite{JedrakArXiV2010}. We take an example of uniformly 
magnetized state. According to GA, the ground state energy is 

\begin{align}
    \frac{E_G}{N} \equiv \frac{\braket{\psi_G|\mathcal{H}|\psi_G}}{\braket{\psi_G|\psi_G}} = 
    \sum_{\sigma} q_{\sigma}(d,n_{\sigma},n_{\bar{\sigma}}) \bar{\epsilon}_{\sigma} + Ud^2, \label{groundga}
\end{align}

\noindent
where

\begin{align}
    \bar{\epsilon}_{\sigma} = \frac{1}{N} \braket{\psi_0|\sum_{ij}t_{ij} \hat{a}_{i\sigma}^{\dagger}\hat{a}_{j\sigma}|\psi_0} =
    \frac{1}{N}\sum_\textbf{k}\epsilon_{\textbf{k}}
\end{align}

\noindent
is the part of the average bare band energy filled with electron of spin $\sigma$. It is more convenient to consider 
the variables $n=\sum_{\sigma}n_{\sigma}\equiv \frac{1}{N}\sum_i\braket{\hat{n}_{i\sigma}}$ and $m \equiv \sum_{\sigma} n_{\sigma}$, 
respectively. In that case, we can rewrite \eqref{groundga} in a slightly more general Hamiltonian form under GA approximation

\begin{align}
    \mathcal{H}_\mathrm{GA} (d,n,m) = \sum_{\textbf{k},\sigma} \{ q_{\sigma}\epsilon_{\textbf{k}} - \sigma \alpha \} \hat{n}_{\textbf{k} \sigma}
    + NUd^2, 
\end{align}

\noindent
where the Zeeman term $g\mu_BH_a \equiv \alpha$ has been added. Obviously, $\mathcal{H}_\mathrm{GA} \ket{\psi_0} = E_G\ket{\psi_0}$. 

In a standard formulation, we construct the grand potential in the form 

\begin{align}
\mathcal{F}^{(\mathrm{GA})} = k_BT \sum_{\textbf{k}\sigma} \ln[1+e^{-\beta E_{\textbf{k}\sigma^{(GA)}}}] + Ud^2, 
\end{align}

\noindent
with 
\begin{align}
    E_{\textbf{k}\sigma} \equiv q_{\sigma} \epsilon_{\textbf{k}} - \sigma \alpha - \mu.
\end{align}

\noindent
To close the whole procedure, in GA we have to minimize the functional with respect to $d$, which leads to

\begin{align}
 \frac{\partial \mathcal{F}^{(GA)}}{\partial d} = 2UNd + \sum_{\textbf{k}\sigma} \frac{\partial q_{\sigma}}{\partial d}
 f(E^{GA}_{\textbf{k}\sigma})\epsilon_{\textbf{k}}, 
\end{align}

\noindent
which is supplemented by the self-consistent equation for m and $\mu$ obtained from their definition, i.e.,

\begin{equation}
\left\{
    \begin{split}
        m = \frac{1}{N} \sum_{\textbf{k}\sigma} \sigma f\left(E^{(\mathrm{GA})}_{\textbf{k}\sigma}\right), \\ 
        n = \frac{1}{N} \sum_{\textbf{k}\sigma} f\left(E^{(\mathrm{GA})}_{\textbf{k}\sigma}\right),
    \end{split}
    \right.
\end{equation}

\noindent
where $f(E)$ is the Fermi-Dirac function. So, GA provides an essentially renormalized single-particle picture with additional
variational minimization with respect to $d$. 

Here appears a principal problem. Namely, we could equally well determine polarization variationally. A straightforward 
calculation shows that

\begin{align}
    \frac{\partial \mathcal{F}^{(GA)}}{\partial m} =  \sum_{\textbf{k}\sigma} \frac{\partial q_{\sigma}}{\partial m} \epsilon_{\textbf{k}}
\end{align}

\noindent
is nonzero, so therefore the formulation violates what we have called Bogoliubov theorem which states that the results obtained 
variationally should coincide with those obtained from direct statistical-mechanical description (self-consistent equations). 
Only under this proviso the quasiparticles are defined in a consistent manner. Originally it has been formulated within the Hartree-Fock self-consistent approach, where self-consistent fields appears as a signature of mutual interactions. Here we have in the simplest ordered situation the appearance of nontrivial form of band narrowing factor $q_{\sigma}$. 

To make the variational $\alpha$ (GA) approach consistent in the above sense we intrude additional constraints. Those constraints will be taken care of automatically in the lowest order of DE-GWF expansion, which should reduce then to SGA. Explicitely, we define the effective SGA Hamiltonian in the form

\begin{align}
    \mathcal{H}_{\lambda} = \mathcal{H}_{GA} - \lambda_m(\hat{M} - M) - \lambda_n(\hat{N} - N),  
\end{align}

\noindent
where 

\begin{equation}
\left\{
    \begin{split}
        &\hat{M} = \sum_{i\sigma}\sigma \hat{n}_{i\sigma} = \sum_{\textbf{k}\sigma} \sigma \hat{n}_{\textbf{k}\sigma}, \\ 
        &\hat{N} = \sum_{i\sigma}\hat{n}_{i\sigma} = \sum_{\textbf{k}\sigma} \hat{n}_{\textbf{k}\sigma},
    \end{split}
    \right.
\end{equation}

\noindent
$M=\braket{\hat{M}}$, $N_e = \braket{\hat{N}_e}$, and the Lagrange multipliers $\lambda_m$ and $\lambda_n$ play the role of homogeneous mean fields, which are dual to the total spin polarization and the particle number, respectively. Note that the multipliers are associated with the physical quantities appearing 
as extra variables in a self-consistent treatment. 

The definition of Hamiltonian $\mathcal{H}_{\lambda}$ is used to construct the generalized grand-potential functional 
$\mathcal{F}^{(\mathrm{SGA})}$ which is of the form

\begin{align}
  \mathcal{F}^{(SGA)}  \equiv -k_BT \ln Z_{\lambda}
\end{align}

\noindent
with

\begin{align}
    Z_{\lambda}  \equiv \mathrm{Tr} \left\{ \exp(-\beta(\mathcal{\hat{H}})_{\lambda} -\mu\hat{N} )\right\}.
\end{align}

\noindent
Explicitly, one obtains

\begin{align}
    \mathcal{F}^{(\mathrm{SGA})} = -k_BT \sum_{\textbf{k}\sigma} \ln\left(1+\exp\left(-E^{(SGA)}_{\textbf{k}\sigma}\right)\right) + 
    N\left(\lambda_nn + \lambda_mm + Ud^2\right), 
\end{align}

\noindent
with 

\begin{align}
    E^{(\mathrm{SGA})}_{\textbf{k}} = q_{\sigma}\left(\epsilon_{\textbf{k}}-\sigma(h+dm)-(\mu -\lambda_n)\right).
\end{align}

\noindent
We see that $\lambda_m$ indeed plays the role of effective magnetic field, whereas $\lambda_n$ introduces chemical potential 
shift. The stationary values of $m$, $\mu$, $\lambda_m$, $\lambda_n$ are obtained from the necessary conditions

\begin{align}
    \frac{\partial \mathcal{F}^{(GA)}}{\partial m} = \frac{\partial \mathcal{F}^{(GA)}}{\partial n} =\frac{\partial \mathcal{F}^{(GA)}}{\partial \lambda_m} =\frac{\partial \mathcal{F}^{(GA)}}{\partial \lambda_n} =\frac{\partial \mathcal{F}^{(GA)}}{\partial d}
    = 0.
\end{align}

\noindent
The equilibrium properties are provided by the solution for which $\mathcal{F}^{(\mathrm{SGA})}$ attains its minimal value. The 
multiplier $\lambda_n$ adjusts additionally the chemical potential $\mu$. Explicitly, the above necessary minimum condition may be written in  
a closed forms as follows

\begin{equation}
\left\{
    \begin{split}
        &\lambda_n = - \frac{1}{N}\sum_{\textbf{k}\sigma} 
        \frac{\partial q_{\sigma}}{\partial n}f(E_{\textbf{k}\sigma}), \\ 
        &\lambda_m = - \frac{1}{N}\sum_{\textbf{k}\sigma}\epsilon_{\textbf{k}\sigma}
        \frac{\partial q_{\sigma}}{\partial m}f(E_{\textbf{k}\sigma})\epsilon_{\textbf{k}\sigma},\\ 
        &n = \frac{1}{N}\sum_{\textbf{k}\sigma}f(E_{\textbf{k}\sigma}), \\
        &m = \frac{1}{N}\sum_{\textbf{k}\sigma} \sigma f(E_{\textbf{k}\sigma}).
    \end{split}
    \right.
\end{equation}

\noindent
This means that the SGA procedure provides the correct form of self-consistent equations for both $n$ and $m$. 

To summarize, the SGA method can be regarded as a sophisticated version of the Gutzwiller approximation, and it is the only mean-field method which provides a stable solution in the situation with nonzero order parameters. Additionally, it can be shown that it is equivalent to the saddle-point solution within the slave-boson approach \cite{JedrakArXiV2010}. However, SGA has an advantage over the latter method that it contains physically interpretable extra field, i.e., it does not contain the ghost Bose fields which, at the saddle-point level may undergo Bose-Einstein condensation. Our goal in the remaining part of this review is to start from the SGA approximation
and systematically incorporate higher-order corrections which turn up to play a crucial role in bringing theoretical results to agreement with experiment. But first, we discuss the leading-order SGA solution for the $t$-$J$ model. These two subsections will form a point of reference in discussing DE-GWF in the $t$-$J$-$U$ model, which represents a merger of the $t$-$J$ and Hubbard models. 

\subsubsection{$t$-$J$ model in SGA}

Discussion of the SGA method in the case of $t$-$J$ model has one principal advantage. Namely, it allows for analysis of high-$T_c$ 
superconducting phase in two dimensions and, at the same time, visualize the principal difference with \textbf{R}enormalized \textbf{M}ean-\textbf{F}ield \textbf{T}heory (RMFT). Its added value is connected with that it is the zero-order part of the DE-GWF systematic approach beyond mean-field approximation. 

We start with general form of the $t$-$J$ model

\begin{align}
    \mathcal{\hat{H}}_{t\text{-}J} = \hat{P}_1\left(
    \sum_{ij\sigma} t_{ij} \hat{a}_{i\sigma}^{\dagger}\hat{a}_{j\sigma} + J \sum_{\braket{ij}}\left(
    \hat{\mathbf{S}}_i\cdot \hat{\mathbf{S}}_j - \frac{c_1}{4} \hat{n}_i\hat{n}_j + c_3\mathcal{\hat{H}}_3\right
  )\right) \hat{P}_1.
  \label{eq:general_form_of_the_tj_model}
\end{align}

\noindent
The first term is the kinetic energy as the hopping term with the hopping integrals between nearest neighbors, $t \equiv 
t_{\braket{ij}} = - |t|$ and next nearest neighbors (n.n.) $t^{\prime} = 0.25|t|$, the second is the n.n. kinetic exchange with 
exchange integral $J=4t^2/U$ (each n.n. pair $\braket{ij}$ is then taken only once in the sum), and the last term is its three-site part, i.e.,

\begin{align}
    \mathcal{\hat{H}}_3 = - \sum_{ijk\sigma} \frac{t_{ij}t_{jk}}{U} \left (\hat{b}_{i\sigma}^{\dagger}\hat{\nu}_{j\bar{\sigma}}\hat{b}_{k\sigma} -
    \hat{b}_{i\sigma}^{\dagger}\hat{S}^{\bar{\sigma}}_j\hat{b}_{k\bar{\sigma}} 
    \right),
\end{align}

\noindent
where, among the sites in $i$, $j$, $k$, $\braket{ij}$ and $\braket{kj}$ denote pairs of n.n. with $k\neq i$. The extra coefficients, $c_1$ or 
$c_2$, are equal to either $0$ or $1$, depending on the approximation adopted (see below). In accordance with our general SGA philosophy, we obtain the effective 
single-particle Hamiltonian with renormalized microscopic parameters, that is supplemented with statistical-consistency constraint. In effect, we arrive at

\begin{align}
    \hat{\mathcal{H}}^\mathrm{ren}_{tj} = \sum_{ij\sigma} g^t_{ij} t_{ij} \hat{a}^\dagger_{i\sigma} \hat{a}_{j\sigma} + \sum_{\langle ij\rangle}g^J_{ij} J_{ij} \hat{\mathbf{S}}_i \cdot \hat{\mathbf{S}}_j + \hat{\mathcal{H}}_\mathrm{constr},
\end{align}

\noindent
where $\mathcal{\hat{H}}_\mathrm{constr}$ contains the statistical consistency constraints (see below). Note that now we have no projection operator $\hat{P}$, but instead we have renormalized parameters, $g_{ij}^t$ and $g_{ij}^s$,  which are evaluated via expectation values of the corresponding projection operators next to them \cite{JedrakPhDthesis}. not that we have shown  only the first two terms of Eq.~\eqref{eq:general_form_of_the_tj_model}.

To proceed with SGA, we introduce the effective Hamiltonian in the explicit form which defines the renormalized mean-field theory in this case, i.e., 

\begin{align}
\hat{\mathcal{H}}^\mathrm{eff} & = \sum_{\langle ij\rangle \sigma} \left(t_{ij} g^t_{ij} \hat{a}^\dagger_{i\sigma} \hat{a}_{j\sigma} + \mathrm{H.c.} \right) - \sum_{\langle ij\rangle \sigma} \frac{3}{4} J_{ij} g_{ij}^J \left( P^0_{ji} \hat{a}^\dagger_{i\sigma} \hat{a}_{j\sigma} + \mathrm{H.c.} - |P^0_{ij}|^2\right) \nonumber \\ &- \sum_{\langle ij\rangle \sigma} \frac{3}{4} J_{ij} g_{ij}^J \left( \Delta^0_{ij} \hat{a}_{j\sigma}^\dagger \hat{a}_{i\bar{\sigma}}^\dagger + \mathrm{H.c.} - |\Delta^0_{ij}|^2 \right) + \hat{\mathcal{H}}_\mathrm{constr}
\end{align}

\noindent
with $P^0_{ij} \equiv \langle \hat{a}_{i\sigma} \hat{a}_{j\sigma}\rangle_0$ and $\Delta^0_{ij} \equiv \langle \hat{a}_{i\sigma} \hat{a}_{j\sigma}\rangle_0$.

The further analysis is too lengthy to be quoted here, see \cite{JedrakPhDthesis}. Here we only show the principal results. First, the exmplary diagram involving the magnitude of correlated gap $\Delta_G \equiv \braket{\psi_G|\hat{\Delta}_{ij}|\psi_G}$ vs the doping 
$\delta = 1-n$ has been displayed earlier as Fig.~\ref{fig:1.6}(b). The meaning of the curves 1-3 (4-6) corresponds to the three model situations:
\emph{(i)} $c_1 = c_2 = 0$, \emph{(ii)} $c_1 =1,$ $c_2 =0$, and \emph{(iii)} $c_1=c_2=1$. Physically, those selections differ by taking either simplified or full exchange part, each for $t^{\prime} =0$ and $t^{\prime} = 0.25|t|$. One can see that, depending on the parameters values, the upper critical concentrations for disappearance of $d$-wave superconductivity is between $\delta \equiv 0.2$ and $0.35$, in good qualitative agreement with the experimental data for La$_{2-\delta}$Sr$_{\delta}$CuO$_4$ and related cuprates. The antiferromagnetic solution is not considered here.

The maximal value of $\Delta_{ij}^0$ varies between $0.02$ and $0.03$, which for $|t| = \frac{1}{3} \mathrm{eV}$ is equivalent to
$\Delta^{0}_{\langle ij\rangle} \simeq 70 \mathrm{K}$ and $105 \mathrm{K}$ in physical units. However, the dynamical quantities do not agree with the results of SGA for the $t$-$J$ model \cite{JedrakPhsRevB2011}. 

To summarize this subsection, we see that while the overall properties such as the phase diagram in Fig.~\ref{fig:1.6}(b) are reproduced correctly, the microscopic properties such as the gap magnitude and the Fermi velocity are not. These problems have motivated us to consider higher-order correlations as a possible cure for those inefficiencies of the SGA approach, even though it has already offered an essential improvement over RMFT (for review see, e.g., \cite{EdeggerAdvPhys2007}). 

The next section is devoted to a detailed study of the results obtained within DE-GWF expansion. We first discuss the DE-GWF solution for the $t$-$J$-$U$ model and, subsequently, we turn out attention to selected equilibrium properties of the cuprates, as well as compare the results with those of the Hubbard- and $t$-$J$ models,  regarded as limiting cases of the $t$-$J$-$U$-($V$) model. This circumstance will allow us to discuss also some other physical properties of interest, description of which has also been insufficient when analyzed within SGA.

\section{Selected equilibrium properties of the cuprates within DE-GWF: Results and comparison with experiment}\label{sec:selected_equilibrium_properties}

\begin{figure}
    \centering
    \includegraphics[width=0.5\textwidth]{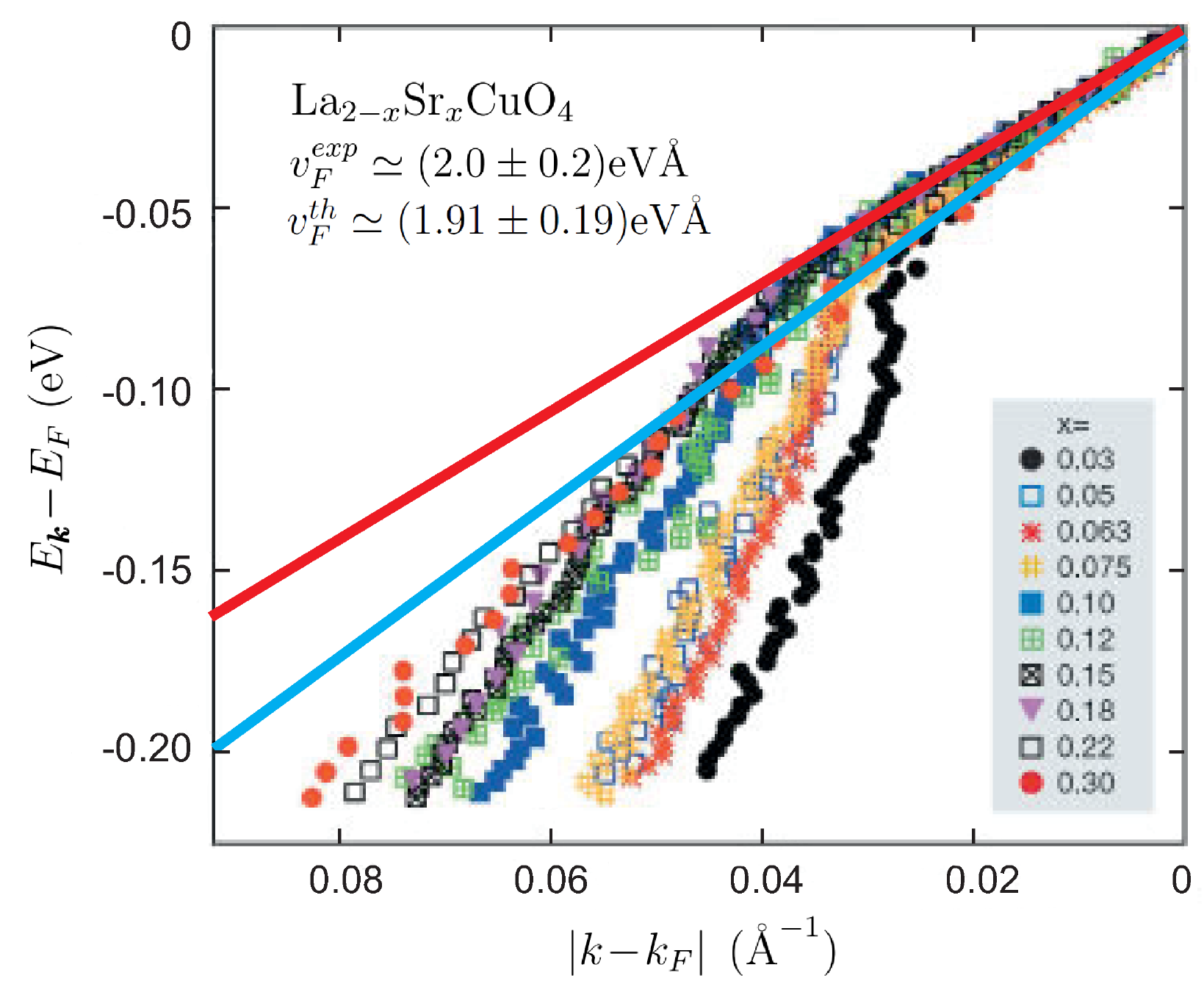}
    \caption{Universal (almost doping-independent) Fermi velocity $v_F$ measured in the nodal direction as determined from the linear part of the relative particle energy $E_\mathbf{k} - F_F \approx v_F |k - k_f|$ (data from~\cite{ZhouNature2003}). The straight lines determine range of $v_F(\delta)$ variation. The extracted experimental and calculated theoretical values are specified. The stoichiometry parameter, $x$, in the legend characterizes the hole concentration $\delta$. Note also the kink in the dispersion relation, appearing at $\sim 0.04 \, \text{\AA}^{-1}$ ($\sim -100\,\mathrm{meV}$) which will be dwelt on later. After Ref.~\cite{SpalekPhysRevB2017}.}
    \label{fig:universal_fermi_velocity}
\end{figure}

\begin{figure}[pos=hbtp]
  \centering
    \includegraphics[width=0.85\textwidth]{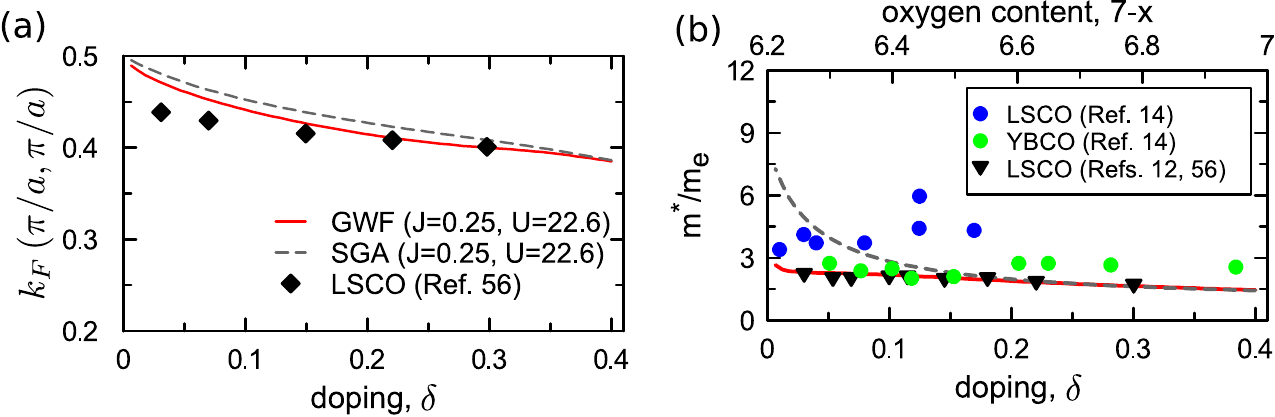}
    \caption{ Fermi momentum $k_F$ (a) and the effective relative mass $m^{*}/m_0$ (b) vs. $\delta$. Solid and dashed lines represent DE-GWF and SGA solutions, respectively, both compared with experimental data (after Ref.~\cite{SpalekPhysRevB2017}). Both quantities are determined for the nodal direction.}
    \label{fig:fermi_momentum_and_effective_mass}
    \includegraphics[width=0.8\textwidth]{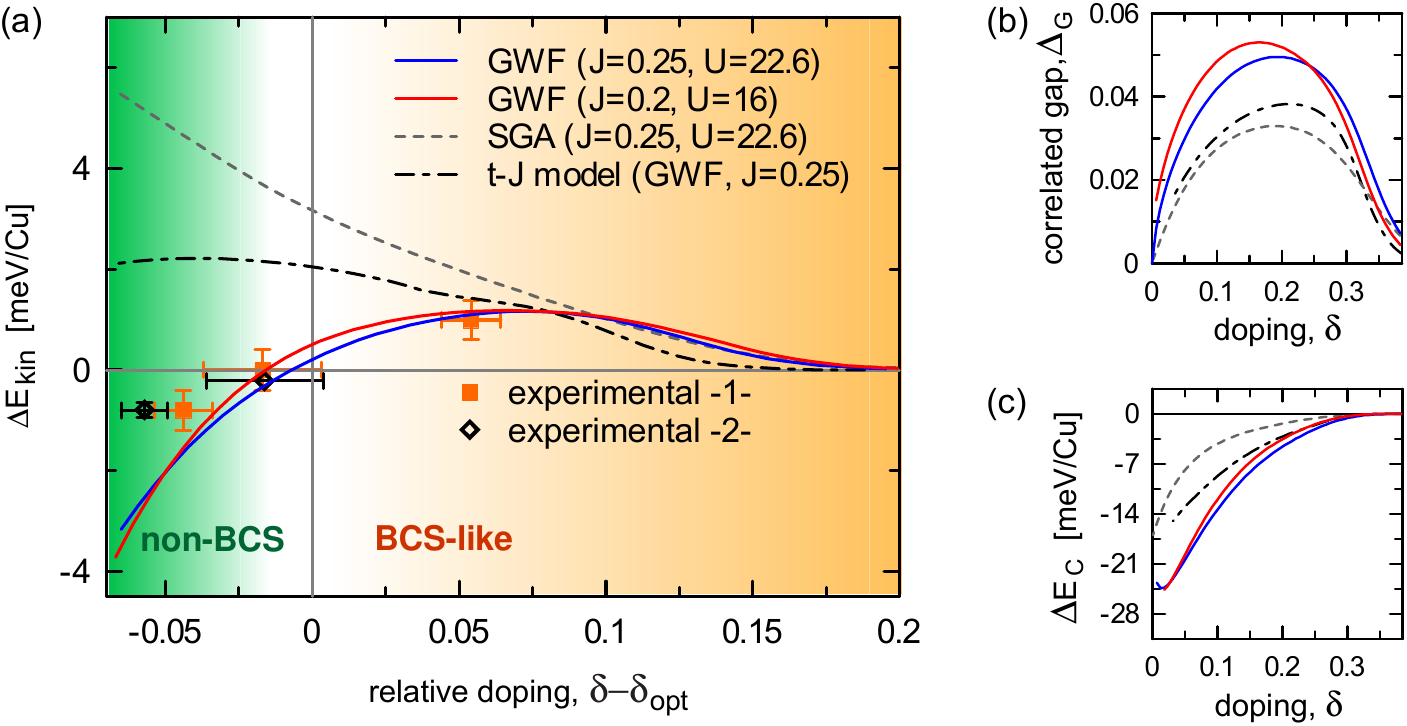}
    \caption{Selected superconducting properties: (a) Kinetic energy gain $\Delta E_\mathrm{kin}$ vs. relative doping $\delta - \delta_\mathrm{opt}$ ($\delta_\mathrm{opt}$ is the optimal doping). The microscopic parameters are $J = 0.2 |t|$, $U = 22.6 |t|$ (blue solid lines) and $J = 0.2 |t|$, $U = 16 |t|$ (red solid lines); the experimental points are taken from Ref.~\cite{DeutscherPhysRevB2005}. For comparison, the results obtained with SGA method (gray dashed line) and those for the $t$-$J$ model ($J = 0.25 |t|$) in DE-GWF approximation (dash-dotted line) are also included. Note that only the $t$-$J$-$U$ model solution describes the data in a quantitative manner. (b) correlated-gap magnitude $\Delta_G$ and (c) the condensation energy $\Delta E_c \equiv E_G^\mathrm{SC} - E_G^\mathrm{PM}$, both vs. $\delta$, are also drawn for the respective values of microscopic parameters and models. The results analyzed in Figs.~\ref{fig:universal_fermi_velocity}-\ref{fig:fermi_momentum_and_effective_mass} are for the same or close values of the microscopic parameters, proving the consistency of our theoretical results. After Ref.~\cite{SpalekPhysRevB2017}.}
    \label{fig:kinetic_energy_gain}
    \includegraphics[width=0.95\textwidth]{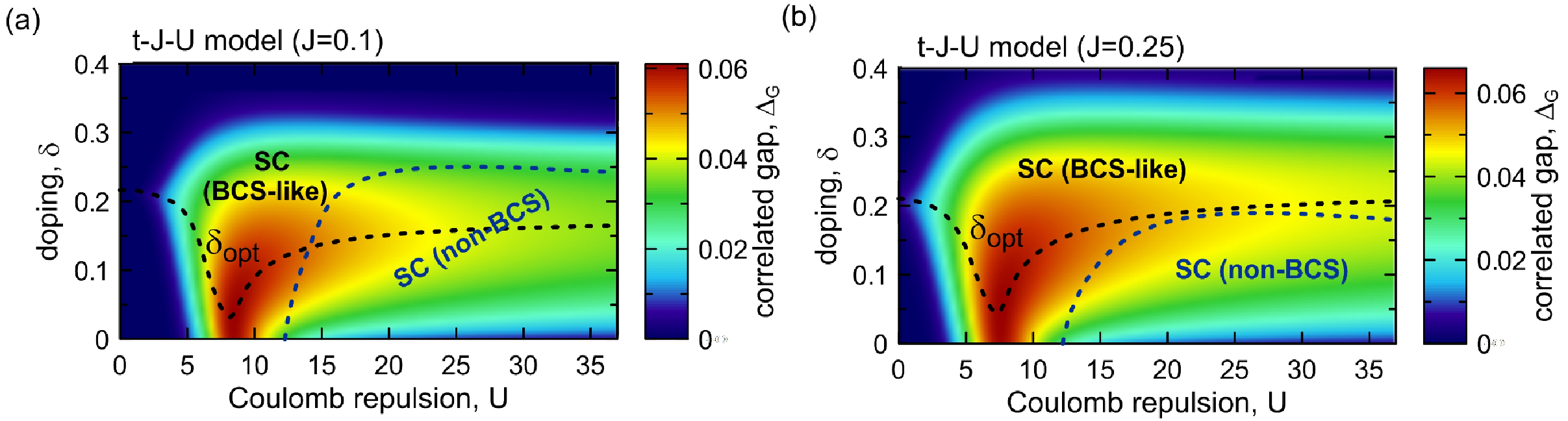}
    \caption{Phase diagram for superconductivity appearance with (a) and (b) covering both BCS-like and non-BCS superconducting states for $J = 0.1 |t|$ and $0.25 |t|$, respectively. For the realistic value of $J$ (b), the non-BCS state appears only in the underdoped regime. Note that the superconducting phase is accentuated for $\delta \lessapprox 0.3$ and for values of $U$ not too large ($U \gtrapprox W$, $W$ is the bare bandwidth, i.e., when the exchange interaction $\sim J$ is the strongest). The last property provides an \emph{a posteriori} justification of the $t$-$J$-$U$ model. After Ref.~\cite{SpalekPhysRevB2017}.}
    \label{fig:bcs_non-bcs_phase_diagram}
  \end{figure}

After introducing our DE-GWF method in the preceding section as a systematic generalization of the sophisticated version of RMFT (SGA), as well as discussing briefly the insufficiency of SGA description of experimental results, we concentrate next on detailed discussion of the results. In particular we overview DE-GWF as a tool for quantitative description of selected experimental data. In most important instances, we also address the results that have not be discussed as yet within SGA. 

For the first part of the discussion, we select the $t$-$J$-$U$-$(V)$ model, which encompasses both the Hubbard and $t$-$J$ models as particular limiting cases (Secs.~\ref{subsection:properties_of_the_tjuv_model}-\ref{subsection:supplementary_results_single_band}), whereas those of the three-band ($d$-$p$) model are analyzed in Sec.~\ref{subsec:results_for_the_3_band_model}. 

\subsection{Properties of cuprates: $t$-$J$-$U$-$(V)$ single-band model within DE-GWF}
\label{subsection:properties_of_the_tjuv_model}

In the following subsections we discuss results for the single-band $t$-$J$-$U$-$(V)$ model of correlated  electrons. From formal point of view, this model covers both the Hubbard-model ($J = 0$) and the $t$-$J$-model ($U \rightarrow \infty$) limits. Its physical meaning is elaborated in Appendix~\ref{appendix:sga_and_slave_bosons}. The main message coming from this discussion is that a realistic value of $U \approx 8$-$10\,\mathrm{eV}$ may be retained to quantitatively match experiment, even in the single-band cases. In particular, $U/|t| \sim 18$-$22$ need not to be lowered unrealistically to $U \approx 2$-$3\,\mathrm{eV}$, with the simultaneous value of $|t| \approx 0.35\,\mathrm{eV}$, which would mean that $U \sim W$ (in which case, the $t$-$J$ model validity would be questionable for high-$T_c$ cuprates). In other words, we regard the superexchange process via $2p$ states due to the oxygen as an additional and important contribution, in addition to that of the kinetic exchange coming from $d$-$d$ virtual hopping between the hybridized $d$-$p$ orbitals. 

We start with the one-band $t$-$J$-$U$ which we rewrite in the following form (cf. Appendix~\ref{appendix:sga_and_slave_bosons})

\begin{align}
    \mathcal{\hat{H}} = \sum_{ij\sigma} t_{ij} \hat{a}^{\dagger}_{i\sigma}     \hat{a}_{j\sigma} + \frac{1}{2} \sum_{ij} J_{ij} \hat{\mathbf{S}}_i \cdot \hat{\mathbf{S}}_j + U \sum_i \hat{n}_{i\uparrow} \hat{n}_{i\downarrow} + \frac{1}{2} \sum_{ij} V_{ij} \hat{n}_i \hat{n}_j, \label{ham4}
\end{align}

\noindent
and analyze the results as a function of parameters $J$, $U$, $V$, all in units of $|t|$, unless stated otherwise.

\subsection{Selected physical results: Fermi velocity, wave vector, and effective mass}
\label{subsection:discusssion_of_physical_results}

First, we would like to note that there are two type of states introduced in DE-GWF approach. Namely, we have \emph{(i)} the effective 
single particle Hamiltonian $\mathcal{\hat{H}}_\mathrm{eff}$ and the corresponding uncorrelated wave function $\ket{\Psi_0}_i$, and \emph{(ii)} correlated wave function $\ket{\Psi_0}_G$ and the corresponding ground state energy $E_G = \braket{\Psi_G|\mathcal{\hat{H}}|\Psi_G}/\braket{\Psi_G|\Psi_G}$. We shall show that both the states represented by the two wave functions have a physical significance.

\begin{figure}
    \centering
    \includegraphics[width=0.65\textwidth]{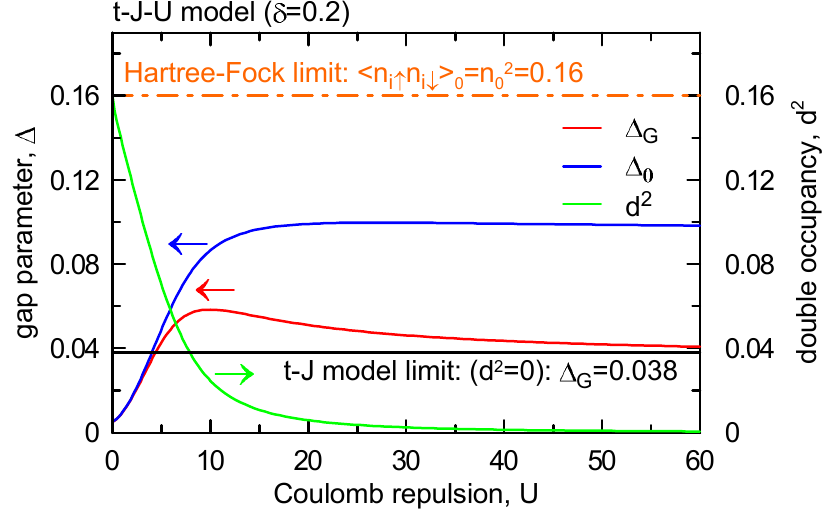}
    \caption{The $U$ dependence of the correlated gap ($\Delta_G$, red line), effective uncorrelated ($\Delta_0$, blue line), and the double occupancy probability ($d^2$, green line) near the optimal doping ($\delta = 0.2$) and for $J = 0.25 |t|$. The vanishing value of $d^2 < 10^{-3}$ for $U > 10$ means that we have a strongly correlated system for appreciable $\delta \sim 0.2$, even though the $U$ value is finite, i.e., $U \sim 2$-$3\,\mathrm{eV}$.}
    \label{fig:correlated_and_bare_gaps_vs_U}
\end{figure}

To illuminate the role of the state $\ket{\Psi_G}$, we plot in Fig.~\ref{fig:universal_fermi_velocity} experimentally determined nodal-direction dispersion relations in the from $|E_\mathbf{k}-E_F|=v_F|k-k_F|$. There are two approximate straight-line parts; the values in the Figure  reflect an almost universal character, roughly independent of doping, close to the Fermi level $E_F$. The theoretical value is that obtained from diagonalization of $\mathcal{H}_\mathrm{eff}$. The agreement between experiment and theory is quantitative. Note that the theoretical dispersion resulting $\mathcal{H}_\mathrm{eff}$ can be regarded as representing quasiparticles very close to the Fermi surface. Interpretation of the approximate straight-line behavior below the kink is provided later, where the correlations are taken into account explicitly within $\mathbf{k}$-space formulation of DE-GWF approach.

To complete the single-particle picture in the nodal direction, we have plotted in Fig.~\ref{fig:fermi_momentum_and_effective_mass}(a)-(b) the doping dependence of the Fermi wave vector $k_F$ and the relative effective mass $m^{*}/m_e$ as a function of doping, calculated within DE-GWF and compared those results with available experimental data. For reference, the corresponding SGA results are also displayed. The data trend is well reproduced except from the data concerning $m^{*}/m_e$ for LSCO. The latter deviation may signal an onset of CDW ordering (see below). It is rewarding to see that the renormalized Fermi-Liquid behavior, described by $\mathcal{\hat{H}}_\mathrm{eff}$, is appropriate for quantitative description of those properties also in an underdopped regime. An attractive suggestion is that Fermi-arc-type behavior may be due to the factors not included in this picture. One of them may be the neglected lifetime effects due to spin/charge excitations (see also the discussion on the effect of quantum fluctuations in the later part).

\subsection{Superconducting properties: Single-band case}
\label{subsection:superconducting_properties}

\begin{figure}[pos=hbtp]
    \centering
    \includegraphics[width=0.77\textwidth]{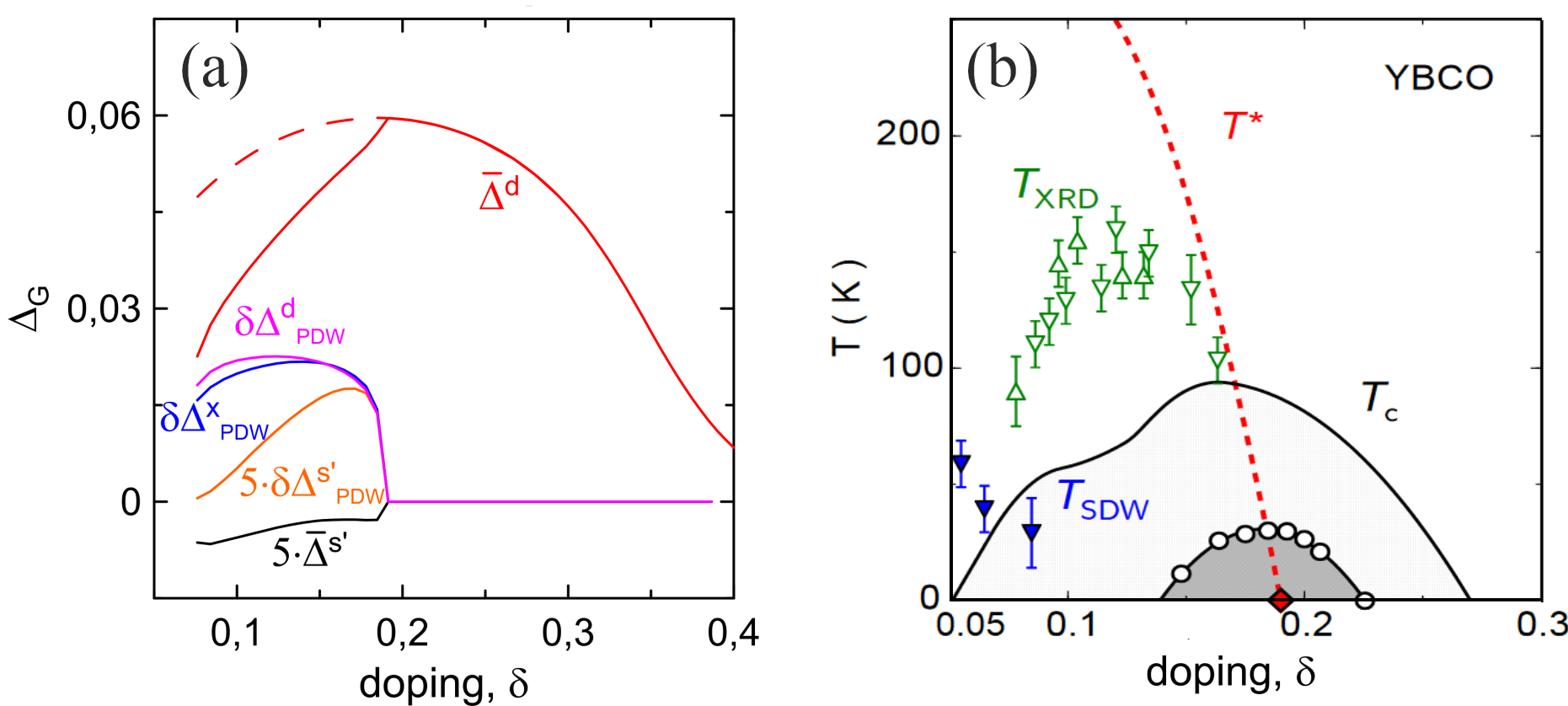}
    \caption{The phase diagram comprising various charge-density-wave states: (a) theory and (b) experiment \cite{BadouxNature2016}. For detailed discussion of various order-parameter components see \cite{ZegrodnikPhysRevB2018}. Note that the onset  of pair-density wave (PDW) induces also a small $s$-wave type of ordering in the system with the primary $d$-wave SC ordering. Pure $d$-wave superconducting phase appears only at and above the optimal doping, as observed. After \cite{ZegrodnikPhysRevB2018}.}
    \label{fig:phase_diagram_with_pdw/cdw_states}
    \includegraphics[width=0.7\textwidth]{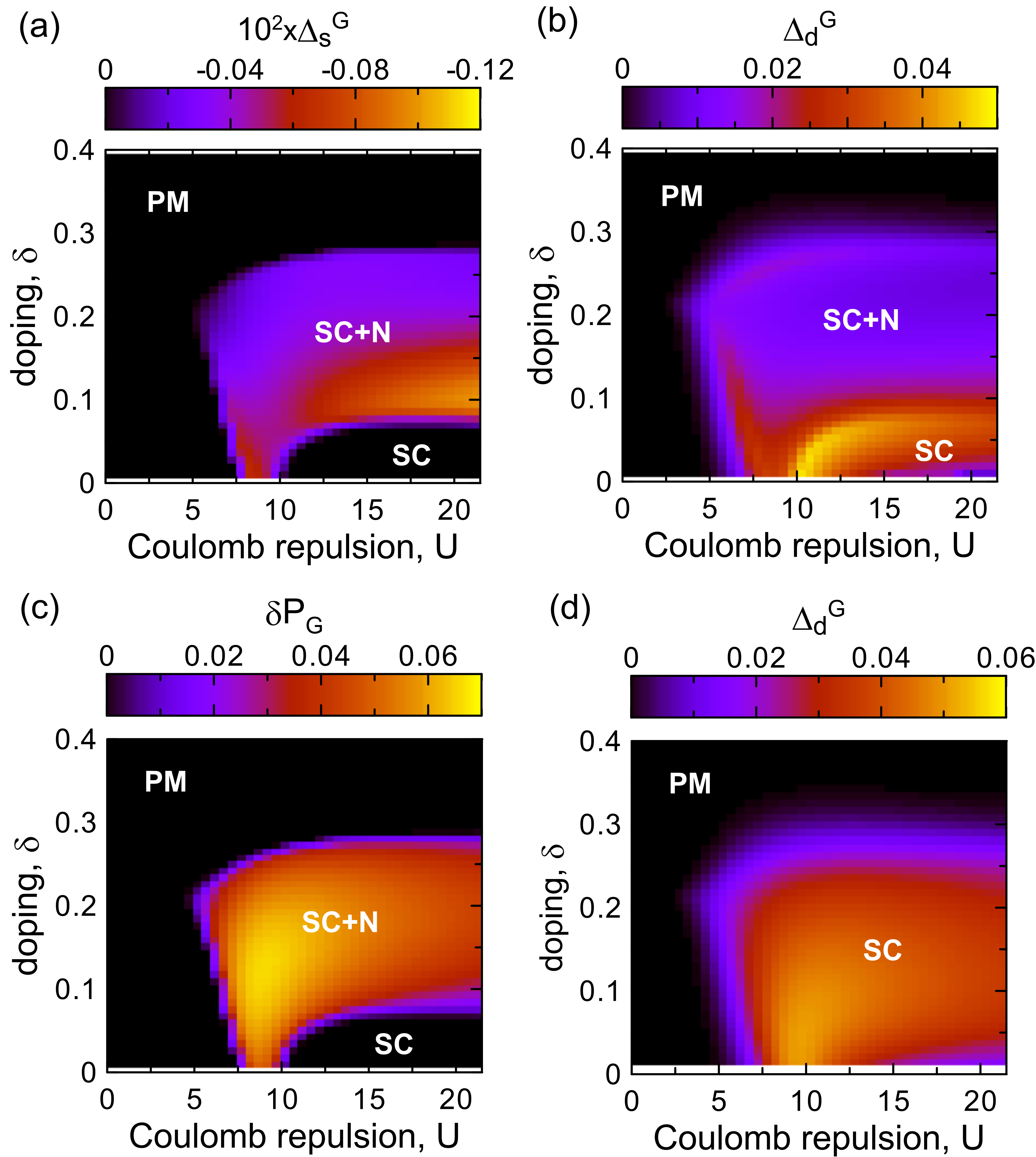}
    \caption{(a)+(b): Phase diagram including superconducting (SC), nematic (N), and the coexistent SC+N phases, all as a function of $U$: $s$-wave (a) and $d$-wave components of the correlated gap, and the nematicity parameter $\delta P_G$ in the coexistent SC+N phase. For comparison, the same diagram for a pure $d$-wave phase is shown in (d). Note that, in panel (a), non-vanishing $s$-wave  component is present in the coexisting SC+N phase, with amplitude comparable to that of the $d$-wave (b). No additional coexisting CDW or PDW order has been included in this analysis. After~\cite{ZegrodnikNewJPhys2018}.}
    \label{fig:phase_diagram_with_sc+nematic_phases}
  \end{figure}

In Fig.~\ref{fig:kinetic_energy_gain} we have assembled the principal physical properties in the superconducting state. In panel (a) we plot the kinetic energy gain in the superconducting state. This energy is defined as the difference between hopping energy in the SC state and that in the normal (PM) state, i.e., 

\begin{align}
  \Delta E_\mathrm{kin} \equiv E^\mathrm{SC} - E^\mathrm{PM},
\end{align}

\noindent
The circumstance that $\Delta E_\mathrm{kin} < 0$ is regarded as a sign of non-BCS nature of the state. The DE-GWF results (continuous lines) describe the experimental data \cite{DeutscherPhysRevB2005} quantitatively. On the contrary, neither the SGA approximation for the $t$-$J$-$U$ model (dashed line) nor DG-GWF results for the $t$-$J$ model match the experiment. Panels (b) and (c) of Fig.~\ref{fig:kinetic_energy_gain} describe theoretical results for the same set of parameters and models. They concern the doping dependence of the correlated gap, $\Delta_G$, and the condensation energy in the ground state, $E^\mathrm{SC}_G-E^\mathrm{PM}_G$, where $E^\mathrm{\mathrm{SC}, \mathrm{PM}}$ characterizes the ground-state energy in the SC and normal states, respectively. Unfortunately, the gap magnitude $\Delta_G = \braket{\hat{a}^{\dagger}_{i\uparrow}\hat{a}^\dagger_{j\downarrow}}$ calculation does not determine the effective coupling constant, as this is an essentially non-BCS approach, which may be regarded as analogous to the Eliashberg-type extension of BCS theory.

Next, we plot other SC characteristics for the case of pure $d$-wave SC phase. Namely, in Fig.~\ref{fig:bcs_non-bcs_phase_diagram} we display exemplary phase diagrams on $\delta$-$U$ plane for the two values of $J$. We see that the SC state appears above a threshold value ($U\equiv U/|t| \sim 5$) and persists to very large values of $U$, particularly for larger $J$ value. This is because in our $t$-$J$-$U$ model, we have decoupled $J$ from $U$. Note that the largest correlated gap magnitudes are achieved in the regime $U \sim W$, i.e., near the onset of strong correlations, where the Hubbard-split subband structure sets in. This may provide a hint, which systems should possess the highest value of $T_c$. For the sake of completeness, in Fig.~\ref{fig:correlated_and_bare_gaps_vs_U} we show both the correlated and uncorrelated gaps, $\Delta_G$ and $\Delta_0$, respectively, as well as the calculated double-occupancy probability $d^2 \equiv \braket{n_{i\uparrow}n_{J\downarrow}}_G$. The higher-order correlations (beyond SGA) reduce the SC gap by a factor of two. Also, the Hartree-Fock and $t$-$J$-model limits are marked to single out the remarkable evolution of all involved quantities with the increasing $U$. Note that by the ``non-BCS regime'' we mean the region with $\Delta E_\mathrm{kin} < 0$, since in the BCS theory $\Delta E_\mathrm{kin} \geq 0$.

\begin{figure}
    \centering
    \includegraphics[width=0.6\textwidth]{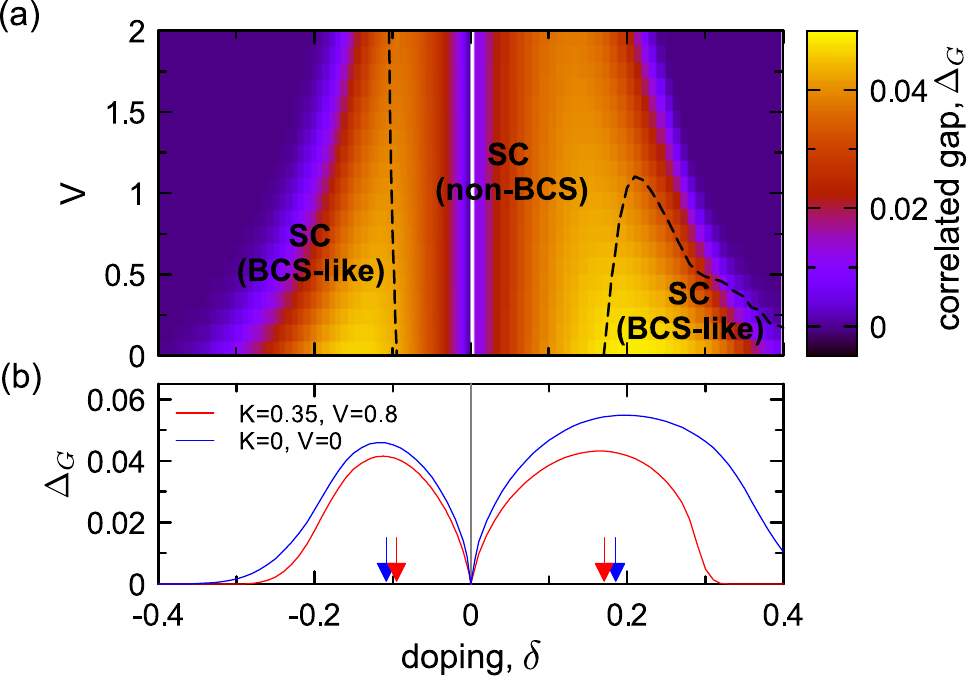}
    \caption{Phase diagram for extended $t$-$J$-$U$-$V$ model for both hole- and electron-doping sides: (a) Correlated gap as a function of intersite Coulomb repulsion $V$ and doping $\delta$, for $J=0.25$, $U=21$, and correlated-hopping parameter $K=0.35$. The borders between the BCS-like and non-BCS regimes are marked by the dashed lines. (b) Correlated gap $\Delta_G$ as a function of doping for two selected sets of model parameters ($K=0.35$, $V=0.8$ and $K=0$, $V=0$). The vertical arrows show the dopings for which the crossover between the BCS-like and non-BCS states appear for the corresponding model parameters. After~\cite{ZegrodnikPhysRevB2017}.}
    \label{fig:extended_tjuv_model}
\end{figure}

\subsubsection*{Complex features of phase diagram: Charge-density-wave and nematic phases}

The presented so far discussion of phase diagram and other superconducting properties of the high-$T_c$ cuprates is focused in the single-band model and ignores the 
possibility of charge-density wave and/or nematic ordering appearance. These types of ordering, as well as the multi-band effects are the subject 
of the next two subsections. We start with the first of them.

The importance of the charge fluctuations has been noted shortly after the discovery of the high-$T_c$ superconductivity with its characteristic dome-like behavior of doping, cf. Fig.~\ref{fig:1.6}(b). Namely, several authors \cite{ArpaiaScience2019,CastellaniZPhysB1996} have proposed that the pseudogap may be a sign of a quantum critical behavior of charge fluctuations with a hidden quantum critical point located right in the middle of the dome. This work was complementary in a sense to the works on inhomogeneous charge-spin ordering in the form of stripes and are still under debate. We have analyzed the possibility of appearance of a robust charge ordering in the form of either charge (CDW)- or pair (PDW)-density wave \cite{ZegrodnikPhysRevB2018}. In Fig.~\ref{fig:phase_diagram_with_pdw/cdw_states} we present a representative phase diagram in the form of the correlated gap magnitude $\Delta_G$ dependence on the doping. We observe that the PDW state appears in the underdoped-to-optimal regime compares, at least qualitatively well with the overall phase diagram displayed side-by-side with the theoretical results. Note, however, that a small $s$-wave component ($\Delta^S$) of the superconducting gap appears. This last feature may present a problem if the clear nodal behavior persists also in the mixed PDW+SC state. We have no simple answer at this point to this basic question at this time. The results agree quantitatively with experiment \cite{BadouxNature2016}.

The second feature is the onset of nemacity, i.e., spontaneous appearance of the anisotropic-hopping amplitudes. Such a solution solution is characterized by the hopping probability $P_{ij} \equiv \braket{ \hat{a}_{i\sigma}^{\dagger}\hat{a}_{j\sigma}}_G$ that acquires distinct values when calculated along $x$ and $y$ directions. The typical phase diagram, analyzed as a function of the Coulomb interaction $U$ and doping $\delta$, is displayed in Fig.~\ref{fig:phase_diagram_with_sc+nematic_phases}. Note, that the mixed superconducting + nematic (SC+N) phase appears inside the ordinary dome-like behavior, cf. panel (d). Also, the weakening of the $d$-wave SC phase with $C_4$ symmetry breaking is apparent, which demonstrates the competitive nature of SC and N phase interplay \cite{ZegrodnikNewJPhys2018}. The panels (a) and (b) illustrate this competition when compared with panels (c) and (d). Importantly, we exhibit only the stable solutions: For example the coexistent SC+N phase is stable with respect to the SC phase if the ground-state-energy difference $\Delta E = E_{\mathrm{SC}+\mathrm{N}} - E_{\mathrm{SC}} < 0$. The results presented in Fig.~\ref{fig:phase_diagram_with_sc+nematic_phases} were obtained for the Hubbard model. The discussion of the extended Hubbard model is provided in Ref.~\cite{ZegrodnikNewJPhys2018}, where also full quantitative analysis as a function of doping has been carried out. In conclusion, these features should be tested further as, for example, the study of SC+PDW+N coexistence. Neither theoretical nor experimental studies of this sort have been carried out so far. 

\subsection{Supplementary results for the single-band case}
\label{subsection:supplementary_results_single_band}

\begin{figure}
    \centering
    \includegraphics[width=0.5\textwidth]{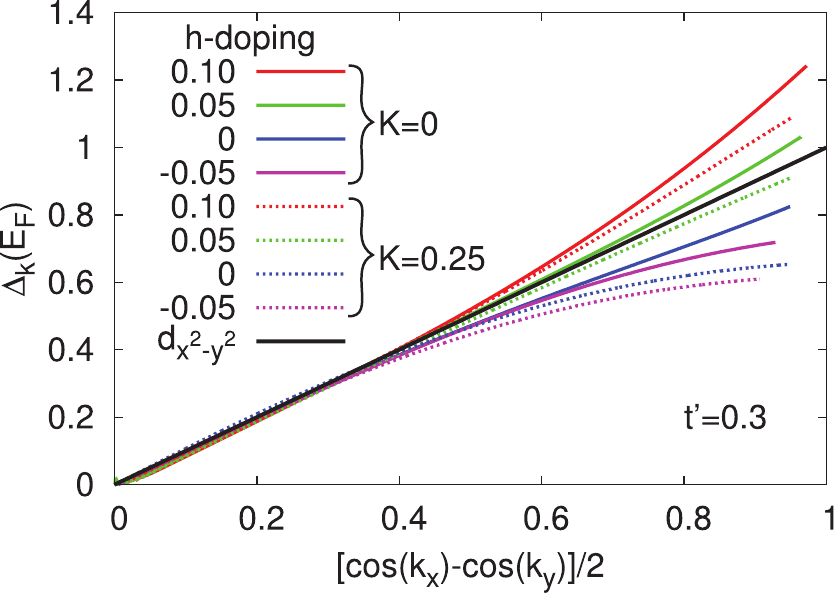}
    \caption{Deviation of the gap shape $\Delta_\mathbf{k}$ from the pure $d_{x^2 - y^2}$ symmetry (straight black line) for selected values of e- and h-dopings and for $K = 0$ and $K = 0.25$. Sole $t^\prime$-term introduces an e-h asymmetric behavior of the gap shape in the vicinity of the antinodal direction. This leads to a qualitative comparison with the experimental results for e-doped cuprates, where in the underdoped regime there is a strong suppression of the gap in the antinodal direction. For all curves $t^\prime=0.3$. After Ref.~\cite{WysokinskiJPCM2017}}
    \label{fig:non_d_wave_gap}
\end{figure}

In Fig.~\ref{fig:correlated_and_bare_gaps_vs_U} we compare the evolution with increasing $U$ of various quantities: the double occupancy probability $d^2$ (green line) and correlated ($\Delta_G$) and uncorrelated gap magnitudes. In the $U \rightarrow \infty$ ($U \gg |t|$) limit, the gaps are determined by the nonzero value of $J = 0.25 |t|$. The important feature is that, in the limit, where relevant physical quantities (cf. preceding section) are well described by taking $U \sim 20 |t|$, the value of $d^2 \sum 10^{-3}$, i.e., it is nonzero. The limiting range of $t$-$J$ model applicability is thus not reached as yet (see, e.g., the $\Delta_G$ dependence). This observation, and the analysis of other properties, leads us to the conclusion that the high-$T_c$ cuprates may be regarded as strongly correlated ($U/W \sim 3$), but not extremely correlated ($U/W \gg 1$) systems. This circumstance speaks in favor of $t$-$J$-$U$ model \emph{\'{a} posteriori}.

\begin{figure}
    \centering
    \includegraphics[width=\textwidth]{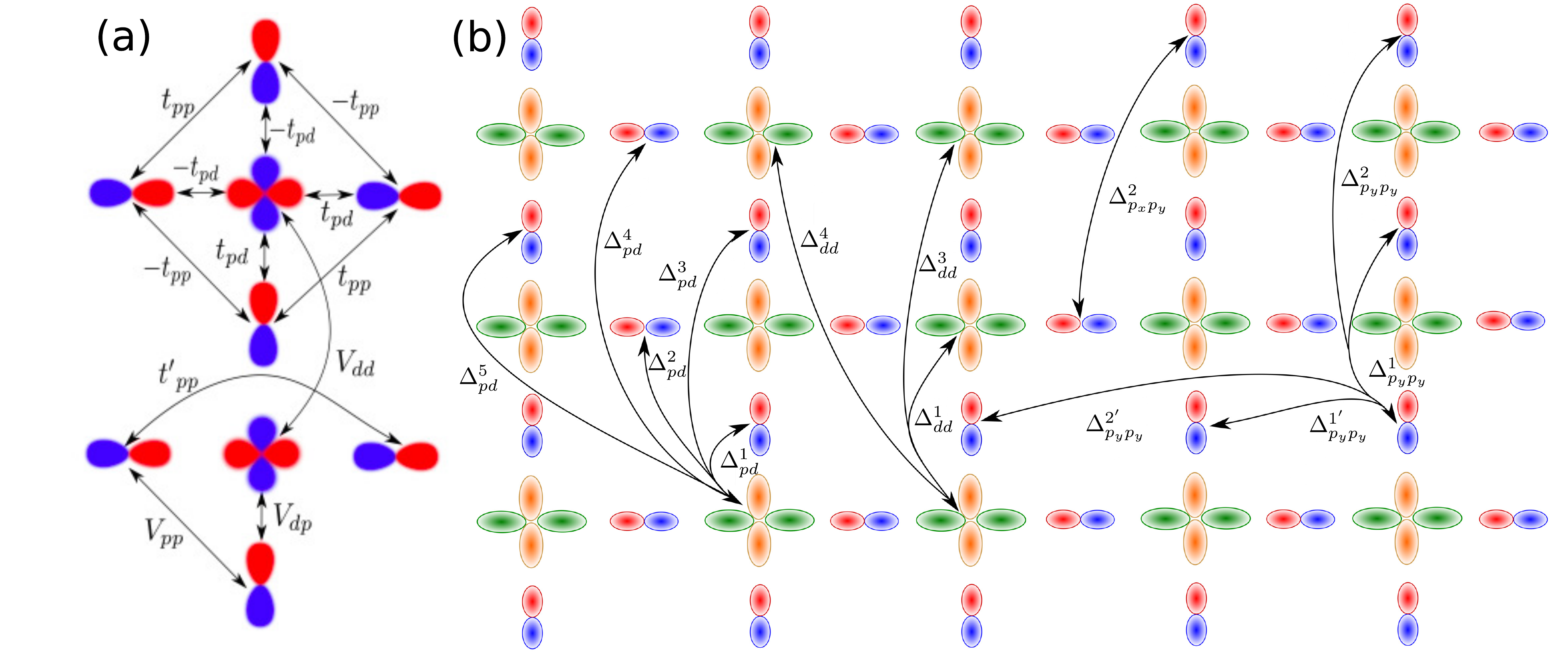}
    \caption{(a) Hopping parameters between the three types of orbitals ($d_{x^2 - y^2}$, $p_x$, $p_y$) and the sign convention corresponding to the antibonding orbital structure. In the bottom part of (a), the relevant intersite interactions ($V_{pp}$, $V_{pd}$) are also marked. After~\cite{ZegrodnikJPCM2021}. (b) Definition of the gap-component amplitudes within the DE-GWF scheme. The superscripts correspond to the consecutive neighbor contributions The subscripts characterize $d$-$d$, $p$-$d$, and $p$-$p$ pairing components. The multiplicity of introduced component gaps still reproduces to a good accuracy a single (antibonding) band behavior (see main text) and Fig.~\ref{fig:sc_gap_components_dp_model}. After~\cite{ZegrodnikPhysRevB2019}.}
    \label{fig:detailed_map_of_pairing_components}
  \end{figure}

Finally, one may try to compare the phase diagrams on the hole- and electron-doped sides. This feature has been also analyzed within our DE-GWF method \cite{WysokinskiJPCM2017,ZegrodnikPhysRevB2017} by including both the intersite Coulomb interaction $\sim V$ and the correlated hopping terms $\sim K$ (cf. Appendix~\ref{appendix:hubbard_model}) in the Hubbard model, which breaks the electron-hole symmetry. In Fig.~\ref{fig:extended_tjuv_model} we draw the phase diagram on the plane in $V$-$\delta$ (a) and the correlated gap $\Delta_G$, both versus doping $\delta$ (negative values of $\delta$ mean that we have the electron-doping case).

We see that essentially the same type of overall behavior takes place on either electron- or hole-doping side and, paradoxically, the two sides become similar (cf.~\ref{fig:extended_tjuv_model}(b)) when the correlated hopping parameter $K \equiv \tilde{V}_{ij}$ between n.n., $(ij) \equiv \langle ij\rangle$, for $K \neq 0$. Parenthetically, the circumstance that incorporation of the n.n. interactions, $V$ and $K$, does not change the principal features of the phase diagram, discussed earlier within the $t$-$J$-$U$ model, speaks in favor of a degree of universality of our models of correlated electrons in conjunction with approaches such as DE-GWF which include both intra- and inter-site correlations.

Finally, we would like to raise one important point of the results within the DE-GWF method. Namely, the superconducting state is not of pure $d$-wave character \cite{KaczmarczykPhysRevB2013,WysokinskiJPCM2017}, either in the Hubbard of $t$-$J$-$U$-$V$ models. In Fig.~\ref{fig:non_d_wave_gap} we plot the correlated gap $\Delta_\mathbf{k}$ vs. $(\cos k_x - \cos k_y)/2$, which should follow a straight line for a pure $d$-wave gap. Clearly, we see that the $\mathbf{k}$-dependence of the gap deviates significantly from linear behavior as we move away from the nodal direction, though in the antinodal direction the appearance of the pseudogap may obscure the picture. This deviation is caused by the circumstance that the further intersite correlations produce additional Fourier components to the simple $d$-wave-type solution. This behavior agrees quantitatively with experiment for electron doped superconductors \cite{ArmitageRevmodphys2010}.

\subsection{Results for three-band ($d$-$p$) model: A brief overview}
\label{subsec:results_for_the_3_band_model}

The standard version of the $d$-$p$ model has been introduced in Sec.~\ref{sec:theoretical_models}. Here we discuss its extended form illustrated in Fig.~\ref{fig:detailed_map_of_pairing_components}(a), in which explicit $d$-$d$ interaction ($V_{dd}$) between the nearest neighboring $\mathrm{Cu}^{2+}$ $d_{x^2-y^2}$ orbitals, as well as the next-nearest-neighbor hopping between oxygen $2p_{x,y}$ orbitals, have been included. In this section we restrict ourselves to summarizing only the principal results and refer to the papers \cite{ZegrodnikPhysRevB2019,ZegrodnikJPCM2021}, where a detailed analysis is done. Typical values of microscopic parameters are: $t_{dp} = 1.0\,\mathrm{eV}$, $\epsilon_{dp} = 3.2\,\mathrm{eV}$, $U_{d} = 11\,\mathrm{eV}$, and $U_{p} = 4\,\mathrm{eV}$.

  \begin{figure}
    \centering
    \includegraphics[width=0.6\textwidth]{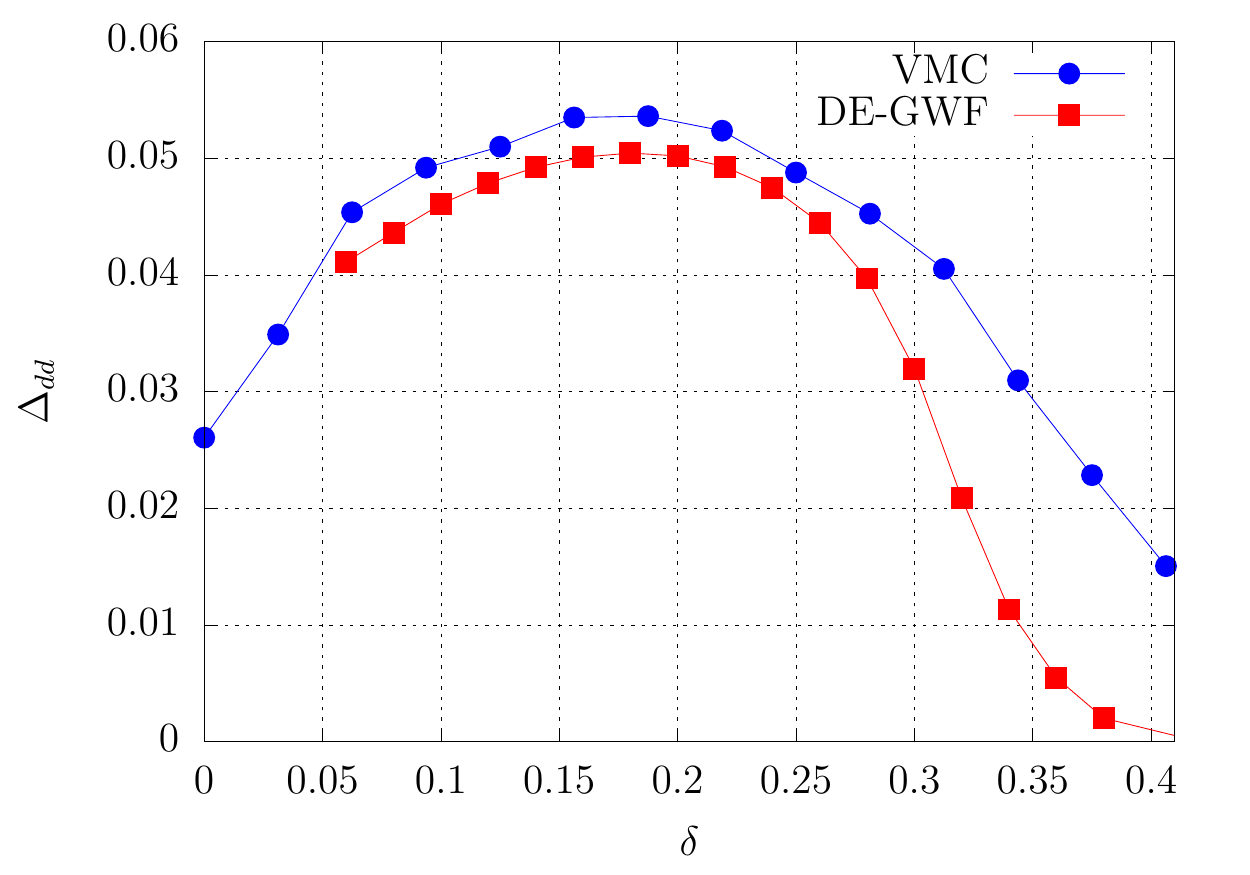}
    \caption{The exemplary comparison of $\Delta_{dd}(\delta)$ obtained within DE-GWF (lower curve) and VMC (upper curve) approaches. The parameters are: $t_{pd} = 1.1\,\mathrm{eV}$, $t_{pp} = 0.49\,\mathrm{eV}$, $\epsilon_{pd} = 3.57\,\mathrm{eV}$, $U_{d} = 10.3\,\mathrm{eV}$, and $U_{p} = 4.1\,\mathrm{eV}$.  After~\cite{BiborskiPhysRevB2020}.}
    \label{fig:sc_degwf_vs_vmc}
\end{figure}

\begin{figure}
    \centering
    \includegraphics[width=0.5\textwidth]{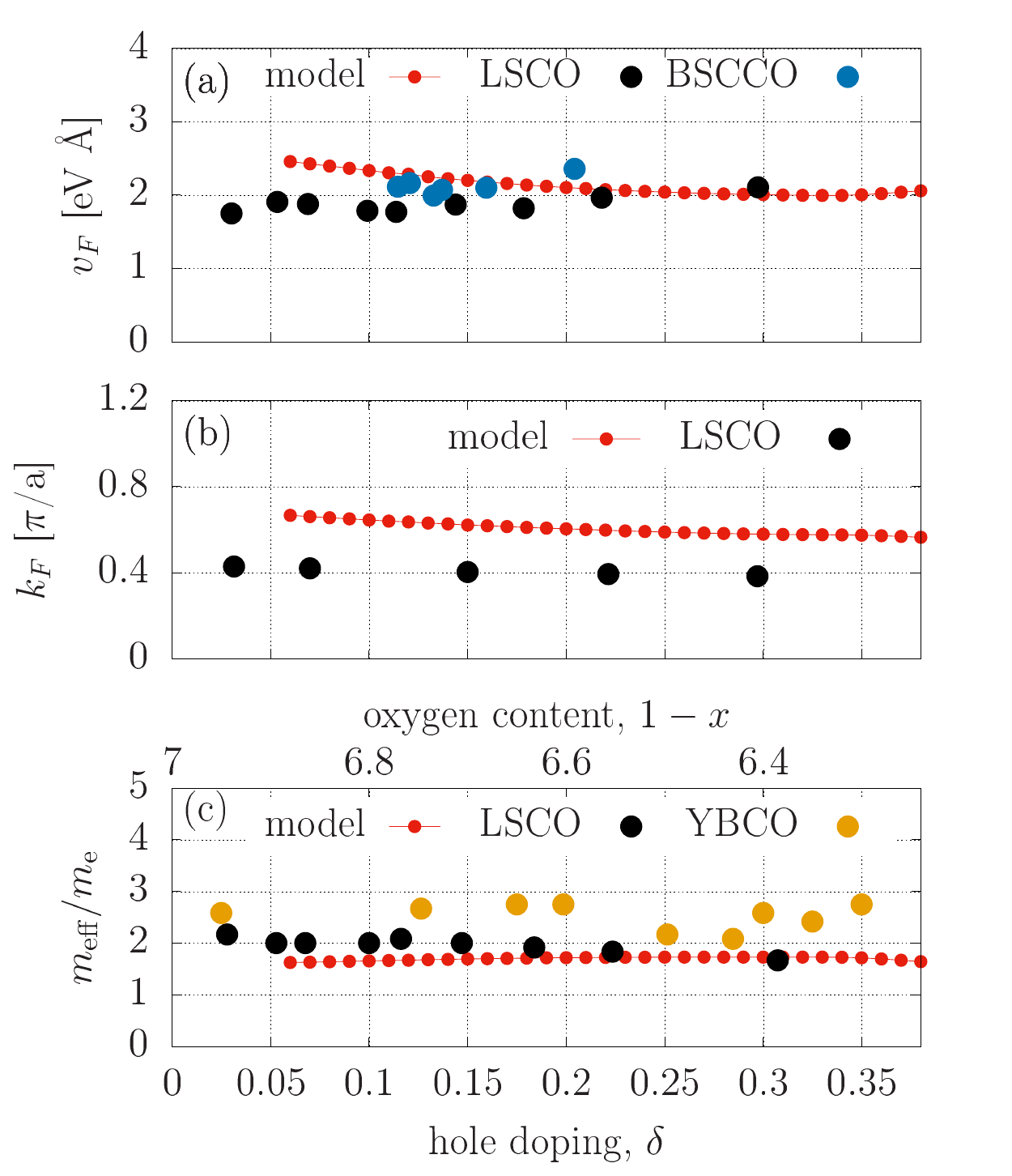}
    \caption{The basic characteristics calculated for the three-band model. From top to bottom: Fermi velocity $v_F$, Fermi wave-vector $k_F$, and effective mass enhancement $m_\mathrm{eff}/m_e$; all as a function of hole doping $\delta$. The parameters are: $t_{pd} = 1\,\mathrm{eV}$, $t_{pp} = 0.4\,\mathrm{eV}$, $\epsilon_{pd} = 3.2\,\mathrm{eV}$, $U_{d} = 11\,\mathrm{eV}$, and $U_{p} = 4.1\,\mathrm{eV}$.  For comparison with the corresponding single-band results, see Figs.~\ref{fig:universal_fermi_velocity}-\ref{fig:fermi_momentum_and_effective_mass}. After~\cite{ZegrodnikJPCM2021}.}
    \label{fig:principal_quantities_for_the_3_band_model}
\end{figure}

In Fig.~\ref{fig:sc_degwf_vs_vmc} we show the dominant $d$-wave SC gap amplitude vs. $\delta$ and compare this result with that coming from extensive \textbf{V}ariational-\textbf{M}onte-\textbf{C}arlo (VMC) analysis \cite{BiborskiPhysRevB2020}, which included also a Jastrow-type projector. The results agree semiquantitatively among those two techniques. Note that the nonzero gap at $\delta = 0$ and above $\delta \sim 0.4$ results from the limited ($4 \times 4$) $\mathrm{CuO_2}$) cluster size. Nonetheless, Fig.~\ref{fig:sc_degwf_vs_vmc} illustrates a rather universal character of the dome-like behavior and has a similar form in both one- and three-band models. The solution depicted in Fig.~\ref{fig:principal_quantities_for_the_3_band_model} has been used to recalculate the basic features ($v_F$, $k_F$, $m^{*}/m_0$) discussed already for one-band model: They are shown in Fig.~\ref{fig:fermi_momentum_and_effective_mass}. We see that the results are of the same quality as analyzed previously within one-band model. This speaks again for a degree of similarity between those two models, albeit for proper values of parameters that have to be selected for each case separately.

\begin{figure}  
    \centering
    \includegraphics[width=0.7\textwidth]{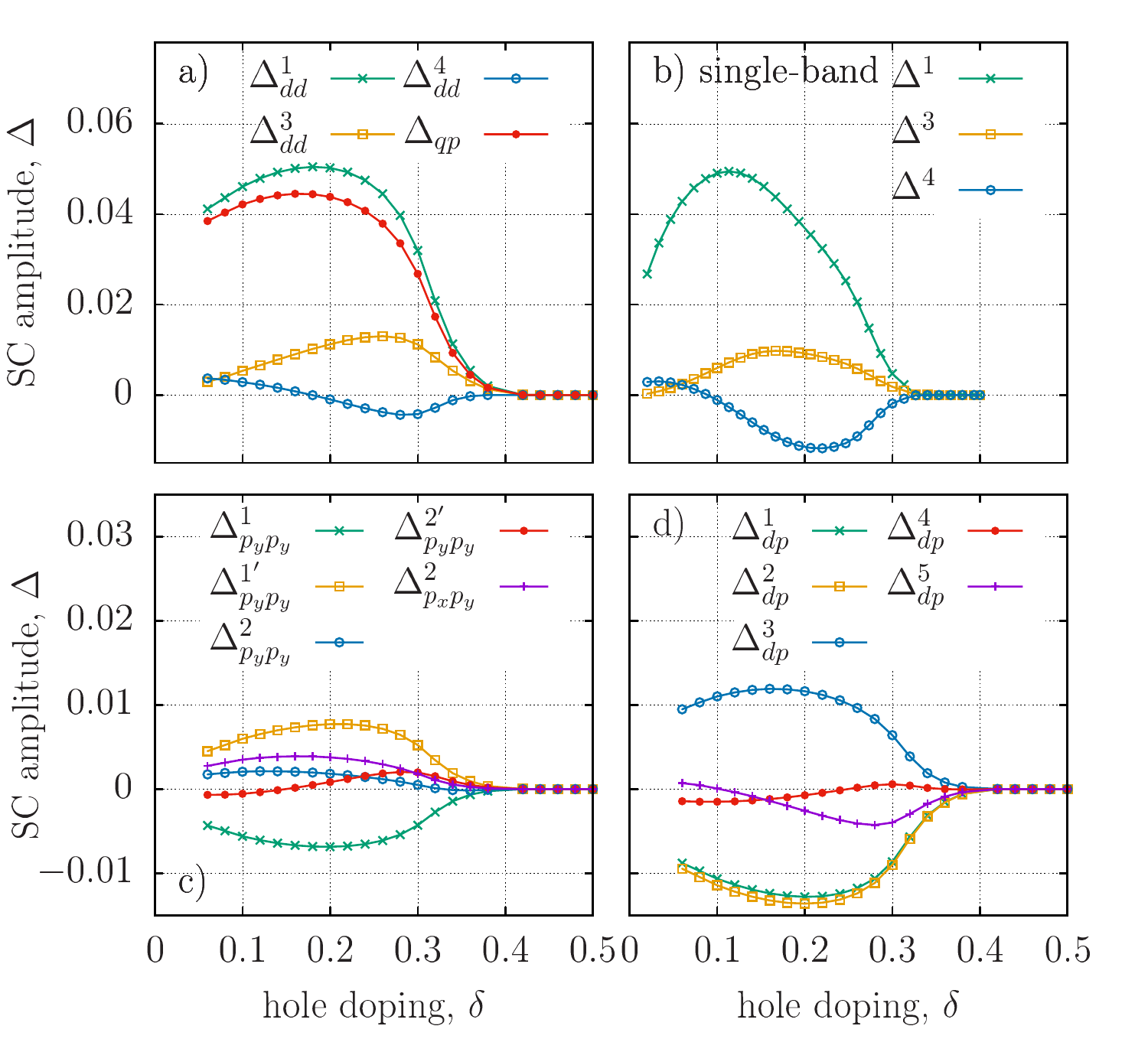}
    \caption{Doping dependence of the respective pairing amplitudes, depicted in Fig.~\ref{fig:detailed_map_of_pairing_components}. $\Delta_\mathrm{qp}$ is the gap characterizing solely the antibonding band. For comparison, the results for the one-band model (b) are also included. The parameters are: $t_{pd} = 1.13\,\mathrm{eV}$, $t_{pp} = 0.49\,\mathrm{eV}$, $\epsilon_{pd} = 3.57\,\mathrm{eV}$, $U_{d} = 10\,\mathrm{eV}$, and $U_{p} = 4.1\,\mathrm{eV}$. The single-band (Hubbard-model) results in (b) are shown for parameter values: $|t| = 0.36\,\mathrm{eV}$, $t^\prime = 0.25 |t|$, and $U = 6 |t|$. Note that in order to obtain a fair agreement with single-band Hubbard model, the $U$ value must be taken anomalously small ($6|t|$). Such nonphysical feature does not appear when we start from the $t$-$J$-$U$ model. After~\cite{ZegrodnikPhysRevB2019}.}
    \label{fig:sc_gap_components_dp_model}
  \end{figure}

The above conclusions motivated us to calculate a detailed map of the pairing components, defined in Fig.~\ref{fig:detailed_map_of_pairing_components}(b), as our method allows for such a comprehensive analysis. Explicitly, we have included $d$-$d$, $p$-$d$, and $p$-$p$ components between consecutive neighbors $n=1, 2, 3$ as our model is fully microscopic and there is no necessity to make simplifying \emph{a priori} assumptions, such as Zhang-Rice spin-singlet formation. In fact, the itinernacy of all involved electrons comes out naturally. In effect, the particular gap components vs. $\delta$ are show in the panel in Fig.~\ref{fig:sc_gap_components_dp_model}(a), (b), and (d). In panels (a) and (b) we display the $d$-$d$ components for one- and three-band cases, and we can see clearly that those two models provide mutually consistent results, at least as far as the $d$-$d$ paring components are concerned. Additionally, we have added in (a) the gap $\Delta_{qp}$ arising from antibonding $d$-$p$ band, as it may represent the naive picture of one-band model with three-band hybridized picture. And it indeed does, when we compare the red curve in panel (a) with the green one in (b). Furthermore, the $p$-$p$ and $p$-$d$ components are essentially smaller from the corresponding dominant $d$-$d$ gap. The nematic phase in the three-band model has been also studied recently and results are published separately \cite{ZegrodnikEurPhysJB2020}.

\begin{figure}
    \centering
    \includegraphics[width=0.75\textwidth]{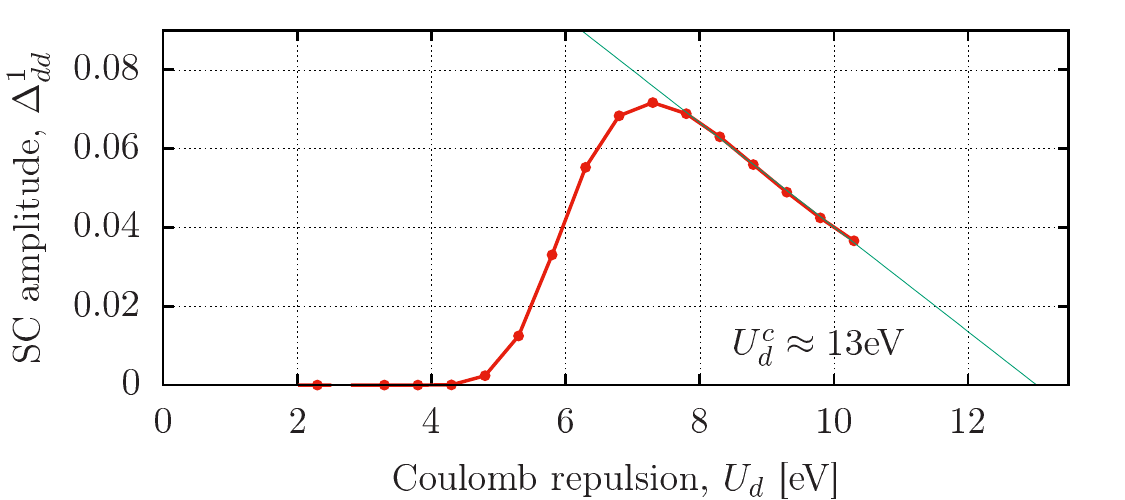}
    \caption{The nearest-neighbor pairing amplitude $\Delta_{dd}^1$ for the case of half filling ($\delta = 0$), as a function of $U_d$ for fixed $V_p = 4.1\,\mathrm{eV}$. The extrapolated limiting value for the Mott boundary is $U_d^c \approx 13\,\mathrm{eV}$. This critical $U_d^c$ value $> W$ illustrates once more the fact that SC is related to Mottness, i.e., it is the strongest on the strong-correlation side, but not too far away from the Mott-Hubbard boundary. After~\cite{ZegrodnikPhysRevB2019}.}
    \label{fig:sc_lower_value_of_u}
\end{figure}

Two methodological remarks are in place at this point. First of them concerns the lowest value of $U \equiv U_{dd}$, above which the SC amplitude $\Delta^{\prime}_{dd}$ reaches zero. This has been shown in Fig.~\ref{fig:sc_lower_value_of_u} taking $U_p = 4.1\,\mathrm{eV}$. As for $U_{dd} > 10\,\mathrm{eV}$ we cannot reach easily 
reach convergence of the results, we have made a linear extrapolation of $\Delta^{\prime}_{dd}$ (blue line) and obtained the critical value $U^c_{dd} = 13\,\mathrm{eV}$. 
This is the reason why the results, displayed in Fig.~\ref{fig:sc_gap_components_dp_model}(d) for smaller $U \simeq 11 \mathrm{eV}$, can be taken only as an estimate of the behavior strong correlation limit. Nonetheless, such limitations are not as strict also then for a single band model, where we can analyze the results in much wider doping range, though the analysis of the $\delta = 0$ limit poses technical problems.

The second remark concerns the electron counting in the three-orbital $d$-$p$ model. For the Mott (charge transfer) insulating state, the total number of electrons $n_d+2n_p \equiv n_d + n_{p_x}+ n_{p_y} = 5$. The corresponding total number of holes is $\tilde{n}_d = 2-n_d$, $\tilde{n}_p = 2-n_p$. Also, we have 
$\delta = 5 -n_\mathrm{tot}$ holes in the doped situation. The suggestion coming from experiment \cite{TabisArXiV2021,BarisicPNAS2013} that the number of holes changes from $\delta$ to $1+\delta$ as 
the doping exceeds the optimal value $\delta_\mathrm{opt}$ requires a separate analysis as this may involve a Mott-type transition with doping. Analysis of this point is one of the prospective problems to be investigated further. 

At the end, we would like to  remark that the effects of interlayer Josephson and related Coulomb couplings have also been studied in the $t$-$J$-$U$ model, within the DE-GWF approach \cite{ZegrodnikPhysRevB2017_2}. In particular, the nodal-direction electronic-structure splitting for the double-layered cuprates has been determined and compared quantitatively to experiment \cite{KordyukPhysRevB2004}. 

\subsubsection{Outlook: Models vs realistic model}

To summarize, this section presents the systematic way to test the microscopic models of high-$T_c$ superconductivity by comparing the results beyond RMFT with experiment. This is necessary in view of the fact, that the values microscopic parameters may vary widely from one-band to the three-band cases. In this respect, the effective value of $U\sim 4 \div 6\,\mathrm{eV}$ in the one-band Hubbard model is unrealistic, as the system then is not strongly correlated, since $W \sim 3\,\mathrm{eV}$ for $|t| \simeq 0.35\,\mathrm{eV}$). To improve this formal deficiency we have proposed and used the $t$-$J$-$U$ model as the minimal realistic description. In that situation, the one-band value of kinetic exchange constant $4t^2/U$ is much smaller than the experimental value $J\simeq 0.12\,\mathrm{eV}$ when realistic $U$-value in range $8$-$11\,\mathrm{eV}$ is taken, but then its main component is due to superexchange taking place via $2p$-bands (cf. also Appendix~\ref{appendix:sga_and_slave_bosons}). In that situation, one can take realistic value of $U \sim 10\,\mathrm{eV}$ at the expense of introducing $J$ as an independent microscopic parameter. In other words, in $t$-$J$-$U$ model $t \equiv t_{dd}$ is caused by single-particle $p$-$d$ hybridization ($t_{dd} \sim t_{pd}^2/\epsilon_{pd}$) and the value of $J$ is mainly provided by superexchange ($J \sim t_{pd}$), rather than by kinetic exchange $J \sim t_{dd}^2/U$. The $t$-$J$-$U$ model provides us also with one extra advantage, namely, we can analyze the $t$-$J$- and the Hubbard-model limits as particular cases of that more general formulation. Clearly, a further work to single out the minimal quantitative model of high-$T_c$ superconductivity is required. The present review provides already a clear step towards this goal by presenting a quantitative comparison with selected experimental data.

\subsection{Reciprocal-space DE-GWF approach: Single-particle spectral properties}
\label{subsection:k-DE-GWF}

The preceding sections contain real-space treatment and allow to include correlations in real space, extending to few ($3$-$5$) lattice constants. Here we discuss a variant of the variational approach for parameterized models (on example of the Hubbard model) that is based on the same diagrammatic expansion of the variational (Gutzwiller-type) wave function, this time formulated in its reciprocal-space version. Within the present approach, we can define, among others, quasiparticle characteristics. Furthermore, we obtain the related quantities, such as the two characteristic velocities near the Fermi energy and $\mathbf{k}$-dependent magnetic susceptibility enhancement. In view of the above, this analysis offers a substantial progress in the sense that the $\mathbf{k}$-dependent quantities can be obtained for extended (practically infinite systems) in a systematic manner. Finally, the presented below approach can be applied to a wide class of models of correlated electrons.

\subsubsection{Moments of the electronic spectral function as characteristic of correlated quasiparticles}

The complete information regarding zero-temperature one-particle dynamics of the correlated system is contained in the imaginary time-ordered Green's function, defined as 

\begin{align}
  \label{eq:lehmann_representation}
  G(\mathbf{k}, \omega) &\equiv \frac{1}{i} \int \limits_{-\infty}^{\infty} dt \cdot \mathrm{exp}(i \omega t) \cdot \langle \mathcal{T} \hat{a}_{\mathbf{k}\sigma}(t) \hat{a}^\dagger_{\mathbf{k}\sigma} \rangle = \int \limits_{-\infty}^{\infty} dE \frac{\mathcal{A}(\mathbf{k}, E)}{\omega - E + i\epsilon\cdot\mathrm{sgn}(E)},
\end{align}

\noindent
where $\hat{a}_{\mathbf{k}\sigma} ^\dagger \equiv \sum_i \exp(i\mathbf{k} \mathbf{r}_i) \hat{a}_{i\sigma}^\dagger$ is the Fourier-transformed creation operator,

\begin{align}
\hat{a}_{\mathbf{k}\sigma}(t) \equiv \mathrm{e}^{i (\hat{\mathcal{H}} - \mu \hat{N})t} \hat{a}_{\mathbf{k}\sigma} \mathrm{e}^{-i (\hat{\mathcal{H}} - \mu \hat{N})t},
\end{align}

\noindent
and $\hat{N}$ is the total particle number operator. Moreover, $\mathcal{T}$ denotes the time-ordering operator that (for operator bilinears) is defined as follows follows: $\mathcal{T} [\hat{o}_1(t_1) \hat{o}_2(t_2)] = \hat{o}_1(t_1) \hat{o}_2(t_2)$ for $t_1 > t_2$ and $\mathcal{T} [\hat{o}_1(t_1) \hat{o}_2(t_2)] = - \hat{o}_2(t_2) \hat{o}_1(t_1)$ for $t_1 < t_2$. The operators $\hat{o}_i$ may be either fermionic creation- or annihilation operators. The last term on the right of Eq.~\eqref{eq:lehmann_representation} defines the so-called Lehmann representation. The quantity in the numerator, $\mathcal{A}(\mathbf{k}, E)$, is referred to as spectral density function that carries the same dynamical information as the Green's function. This can be demonstrated by noting that once $\mathcal{A}(\mathbf{k}, E)$ is known, $G(\mathbf{k}, E)$ can be calculated explicitly by evaluating the integral on the right-hand-side of Eq.~\eqref{eq:lehmann_representation}. On the other hand, the Green's function determines the spectral function with the use of identity 

\begin{align}
\frac{1}{x \pm i \epsilon} = \mathcal{P} \frac{1}{x} \mp i \pi \delta(x),
\end{align}

\noindent
where $\mathcal{P}$ indicates principle value integral, so that $\mathcal{A}(\mathbf{k}, E) = - \pi^{-1} \mathrm{sgn}(E) \cdot \mathrm{Im} G(\mathbf{k}, E)$. As will be shown below,  $\mathcal{A}(\mathbf{k}, \omega)$ is somewhat more convenient to work with in the present context. Moreover, the spectral function is almost directly probed by angle-resolved photoemission experiments, providing valuable insight into electronic excitations \cite{HufnerRepProgPhys2008}. Note that, in general situation, the spectral function in Eq.~\eqref{eq:lehmann_representation} may carry explicitly the spin index. Here we limit ourselves to the magnetically disordered state, where the two spin directions are equivalent.

By time differentiation and taking the Fourier-transform of Eq.~\eqref{eq:lehmann_representation}, one can derive a series of identities

\begin{align}
  \label{eq:commutator_identities}
 \frac{1}{i} \lim_{t \rightarrow 0^{-}} \frac{d^n}{dt^n} \langle \mathcal{T} \hat{a}_{\mathbf{k}\sigma} (t) \hat{a}_{\mathbf{k}\sigma}^\dagger \rangle = & \int \limits_{-\infty}^\infty  \frac{d\omega}{2\pi} (-i\omega)^n e^{-i\omega t} \int \limits_{-\infty}^\infty dE \frac{\mathcal{A}(\mathbf{k}, E)}{\omega - E + i\epsilon\cdot\mathrm{sgn}(E)}
\end{align}

\noindent
or, equivalently, 

\begin{align}
   \label{eq:commutator_identities_2}
  i^{n+1} \left\langle \hat{a}^\dagger_{\mathbf{k}\sigma} \underbrace{\Big[ \hat{\mathcal{H}} - \mu \hat{N} \big[\hat{\mathcal{H}} - \mu \hat{N} \ldots [\hat{\mathcal{H}} - \mu \hat{N}}_{\text{$\hat{\mathcal{H}} - \mu \hat{N}$ appears $n$ times}}, \hat{a}_{\mathbf{k}\sigma}]\big]\Big]\right\rangle = i \cdot (-i)^n \int_{-\infty}^0 dE \mathcal{A}(\mathbf{k}, E) E^n.
\end{align}

\begin{figure}
    \centering
    \includegraphics[width=0.6\textwidth]{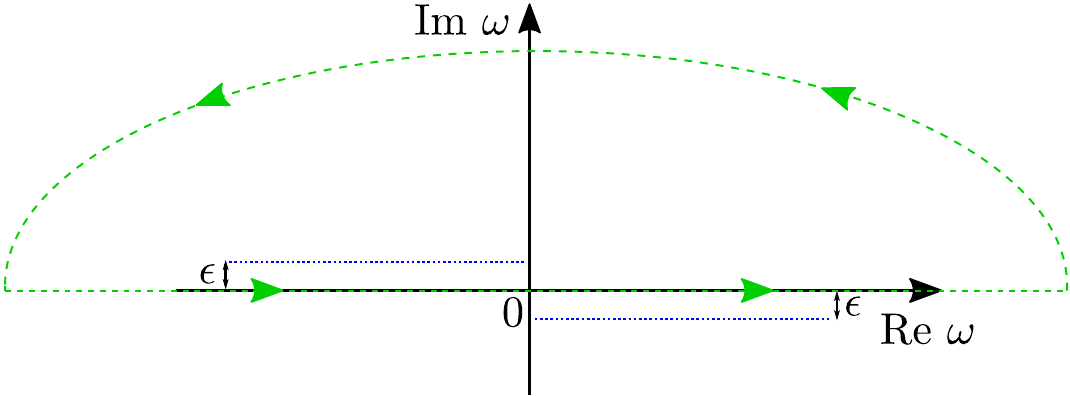}.
    \caption{Integration contour in the complex frequency plane, $(\mathrm{Re} \omega, \mathrm{Im} \omega)$ used to evaluate the right-hand-side of integral~\eqref{eq:commutator_identities} (green dashed line). Green arrows indicate contour orientation. The poles of the Green's function are located on the blue dotted lines, shifted by $\epsilon$ off the real axis. For $t < 0$, the contour must be closed in the upper half-plane to make sure that the exponential factor $\exp(-i\omega t)$ does not diverge for $|\omega| \rightarrow \infty$.}
    \label{fig:complex_contour}
  \end{figure}

\noindent
The right-hand side of Eq.~\eqref{eq:commutator_identities_2} has been obtained by closing the integration contour over $\omega$ in the upper complex half-plane, as schematically illustrated in Fig.~\ref{fig:complex_contour}. This contour has been selected in order to suppress $e^{-i\omega t}$ as $\mathrm{Im}\, \omega \rightarrow + \infty$ (recall that $t < 0$). Equation~\eqref{eq:commutator_identities_2} relates equal-time expectation values to the consecutive moments of the electron spectral function, defined as

\begin{align}
  \label{eq:spectral_moment_definition}
  \mathcal{M}_n(\mathbf{k}) \equiv \int \limits_{-\infty}^0 d\omega\, \omega^n \mathcal{A}(\mathbf{k}, \omega).
\end{align}

\noindent
Note that the moments $\mathcal{M}_n(\mathbf{k})$, to some extent, characterize the single-particle dynamics and thus also excited states of the system. In particular, the two lowest-order identities ($n = 0, 1$) take the form

\begin{align}
  \label{eq:n_k}
  \mathcal{M}_0(\mathbf{k}) \equiv & \int_{-\infty}^0 dE \mathcal{A}(\mathbf{k}, E) = \langle \hat{n}_\mathbf{k}\rangle \equiv \langle \hat{a}_{\mathbf{k}\sigma}^\dagger \hat{a}_{\mathbf{k}\sigma}\rangle, \\
  \mathcal{M}_1(\mathbf{k}) \equiv & \int_{-\infty}^0 dE \cdot E \mathcal{A}(\mathbf{k}, E) = -\langle \hat{a}_{\mathbf{k}\sigma}^\dagger \cdot [\hat{\mathcal{H}} - \mu \hat{N}, \hat{a}_{\mathbf{k}\sigma}]\rangle.   \label{eq:m1_k}
\end{align}

\noindent
An alternative way to derive Eqs.~\eqref{eq:n_k} and \eqref{eq:m1_k} is to integrate the explicit expression for the zero-temperature spectral function

\begin{align}
  \label{eq:spectral_function_definition}
  \mathcal{A}(\mathbf{k}, E) \equiv \sum_m \delta(E - E_m + E_0) \langle 0 | \hat{a}_{\mathbf{k}\sigma} |m \rangle \langle m| \hat{a}^\dagger_{\mathbf{k}\sigma}|0\rangle + \sum_m \delta(E + E_m - E_0) \langle 0 | \hat{a}^\dagger_{\mathbf{k}\sigma} |m \rangle \langle m| \hat{a}_{\mathbf{k}\sigma}|0\rangle,
\end{align}

\noindent
where $|0\rangle$ is true ground state of the system, $\{|m\rangle\}$ for $m \neq 0$ denotes the set of all excited states, and $\epsilon$ is an infinitesimal positive number. As follows from Eq.~\eqref{eq:spectral_function_definition}, the spectral weight fulfills the sum rule $\int_{-\infty}^\infty d\omega \mathcal{A}(\mathbf{k}, \omega) = 1$, which is a direct consequence of fermionic anticommutation relations $\left\{\hat{a}_{\mathbf{k}\sigma}, \hat{a}_{\mathbf{k}^\prime\sigma^\prime}^\dagger\right\} = \delta_{\mathbf{k}\mathbf{k}^\prime}\delta_{\sigma\sigma^\prime}$ and the completeness condition $\sum_m |m\rangle\langle m| = 1$. 

For the case of Landau Fermi liquid, even in presence of strong correlations, the spectral function can be decomposed \cite{LuttingerPhysRev1960} as follows

\begin{align}
  \label{eq:spectral_function_decomposition}
  \mathcal{A}(\mathbf{k}, E) = Z_\mathbf{k} \delta(E - \epsilon_\mathbf{k}^\mathrm{corr}) + \mathcal{A}^\mathrm{inc}(\mathbf{k}, E),
\end{align}

\noindent
where $Z_\mathbf{k}$ is the quasiparticle weight measuring coherence in the fermionic subsystem, $\epsilon_\mathbf{k}^\mathrm{corr}$ is correlated (dressed in the interactions) energy of quasiparticles. In particular, $Z_\mathbf{k} = 1$ and $\mathcal{A}^\mathrm{inc} \equiv 0$ for non-interacting lattice fermions. The first term on the right-hand side of Eq.~\eqref{eq:spectral_function_decomposition} accounts for the most of the relevant low-energy physics of normal metals. The (smooth) incoherent part,  $\mathcal{A}^\mathrm{inc}(\mathbf{k}, E)$, emerges naturally due to sum rules (e.g., $\int_{-\infty}^\infty d\omega \mathcal{A}(\mathbf{k}, \omega) = 1$) as the loss of the coherent spectral weight ($Z_\mathbf{k} < 1$) must be somehow compensated.

By substituting Eq.~\eqref{eq:spectral_function_decomposition} into the spectral moment definition \eqref{eq:spectral_function_definition}, one obtains

\begin{align}
  \label{eq:n_k_integrated}
  \mathcal{M}_0(\mathbf{k}) = & Z_\mathbf{k} \theta(-\epsilon_\mathbf{k}^\mathrm{corr}) + \int_{-\infty}^0 dE \mathcal{A}^\mathrm{inc}(\mathbf{k}, E), \\
  \mathcal{M}_1(\mathbf{k}) =& Z_\mathbf{k} \epsilon_\mathbf{k}^\mathrm{corr} \theta(-\epsilon_\mathbf{k}^\mathrm{corr}) +  \int_{-\infty}^0 dE \cdot E \mathcal{A}^\mathrm{inc}(\mathbf{k}, E).   \label{eq:m1_k_integrated}
\end{align}

\begin{figure}
    \centering
    \includegraphics[width=0.6\textwidth]{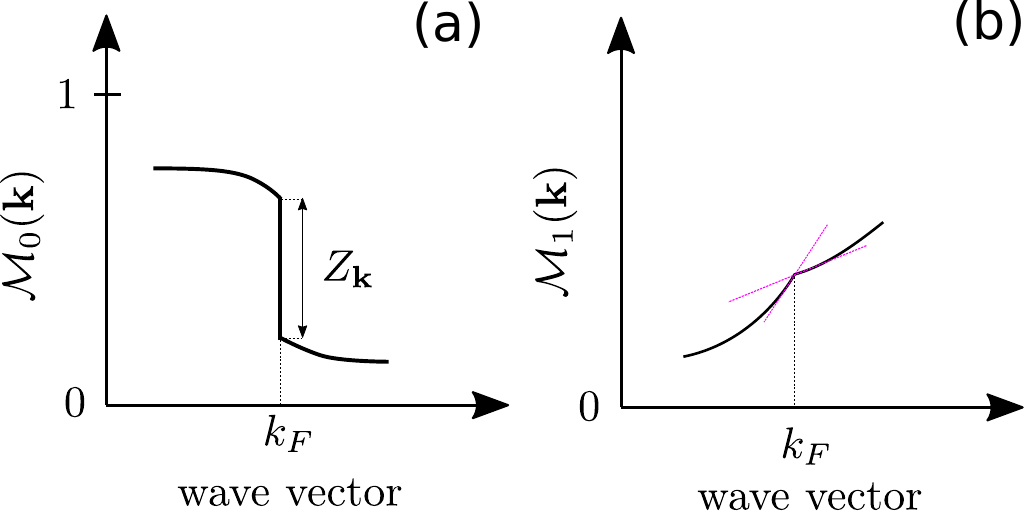}.
    \caption{Schematic representation of the two leading spectral function moments in the vicinity of Fermi wave vector $k_F$, $\mathcal{M}_0(\mathbf{k}) = n_\mathbf{k}$ (a) and $\mathcal{M}_1(\mathbf{k})$ (b). The former has a jump discontinuity of magnitude $Z_\mathbf{k}$ at the Fermi surface, whereas the latter is continuous, but exhibits discontinuity of the first derivative equal to $Z_\mathbf{k} v_F$, where $v_F$ is the value of Fermi velocity. Purple lines in (b) represent the tangent lines on both sides of the cusp.}
    \label{fig:moments_schematic}
  \end{figure}

\noindent
From Eq.~\eqref{eq:n_k_integrated}, it follows that the statistical distribution function $\langle \hat{n}_\mathbf{k}\rangle = \mathcal{M}_0(\mathbf{k})$ has a jump discontinuity of magnitude $Z_\mathbf{k}$ at the Fermi surface (defined by the condition $\epsilon_\mathbf{k}^\mathrm{corr} \equiv 0$), as elaborated by Luttinger \cite{LuttingerPhysRev1960}. More subtle information is encoded in Eq.~\eqref{eq:m1_k_integrated} \cite{RanderiaPhysRevB2004}. By linearizing the quasiparticle dispersion near the Fermi surface, $\epsilon_\mathbf{k}^\mathrm{corr} = \mathbf{v}_F \cdot (\mathbf{k} - \mathbf{k}_F) + O[(\mathbf{k} - \mathbf{k}_F)^2]$, where $\mathbf{v}_F(\mathbf{k}_F) \equiv \nabla \epsilon^\mathrm{corr}_\mathbf{k} |_{\mathbf{k} = \mathbf{k}_F}$ is the Fermi velocity (in general dependent on the choice of position on the Fermi surface, $\mathbf{k}_F$). By substituting the latter formula into Eq.~\eqref{eq:m1_k_integrated}, one finds that the \emph{first derivative} of $\mathcal{M}_1(\mathbf{k})$ has a discontinuity of the magnitude $Z_\mathbf{k} |\mathbf{v}_F| = Z_\mathbf{k} v_F$ at the Fermi surface, resulting in a cusp feature. By combining Eqs.~\eqref{eq:n_k_integrated} and \eqref{eq:m1_k_integrated}, one can thus evaluate both $Z_\mathbf{k}$ and $v_F$, which are important characteristics of correlated quasiparticles. The expected behavior of $n_\mathbf{k}$ and $\mathcal{M}_1(\mathbf{k})$ near the Fermi surface is illustrated in Fig.~\ref{fig:moments_schematic}.

\subsubsection{Evaluation of the diagrammatic sums without real-space cutoff}

\begin{figure}
    \centering
    \includegraphics[width=\textwidth]{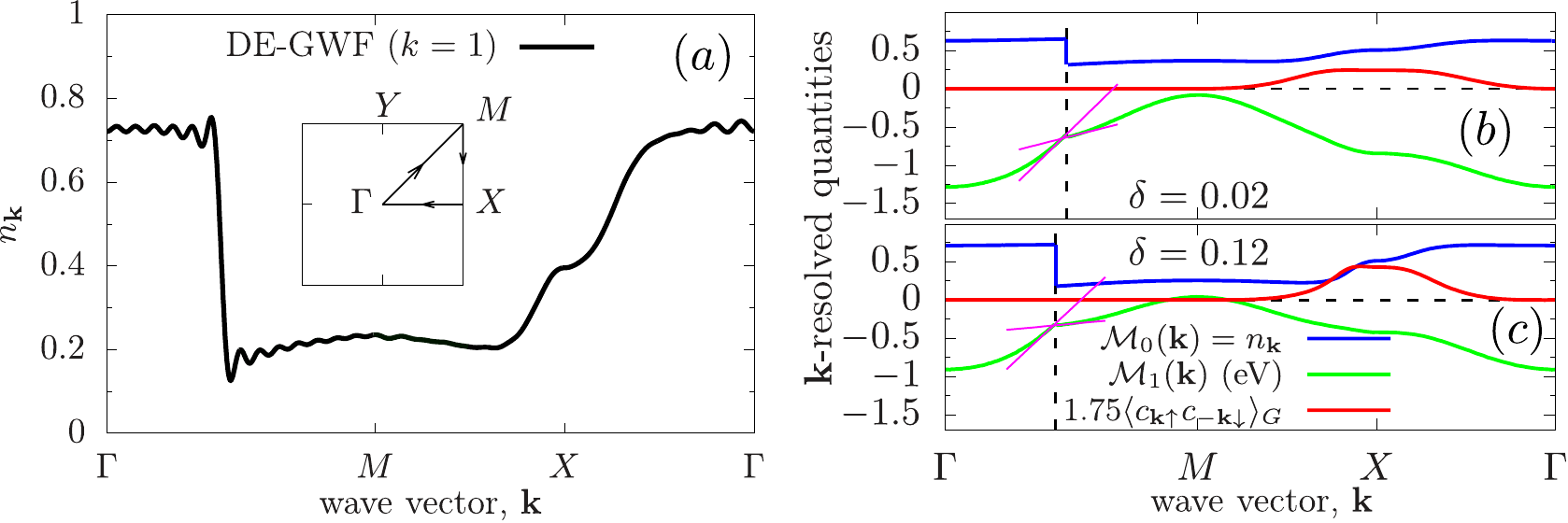}.
    \caption{(a) Statistical distribution function, $n_\mathbf{k}$, evaluated within the real-space DE-GWF method along the representative contour in the Brillouin zone (cf. inset). The function is strictly continuous and exhibits distinct Gibbs phenomenon (oscillations) close to the anticipated Fermi discontinuity along $\Gamma$-$M$ line. The calculations were carried out for the $t$-$J$-$U$ model with the following parameters: $t^\prime/|t| = -0.25$, $U/|t| = 20$, $J/|t| = 1/3$. (b)-(c) Statistical distribution function (blue lines) and first moments of the electron spectral function ($\mathcal{M}_1$) for hole doping levels $\delta = 0.02$ and $\delta = 0.12$, respectively, evaluated using the $\mathbf{k}$-DE-GWF method. The red lines indicate $d$-wave superconducting amplitude. Note the lack of Gibbs oscillations here and a clear discontinuity of the spectral-function-moment derivative. The calculations have been carried out for the Hubbard with the parameters: $t^\prime/|t| = -0.25$, $U/|t| = 12$. Adapted from Ref.~\cite{FidrysiakJPhysCondensMatter2018}.}
    \label{fig:spectral_moments}
  \end{figure}

As the first attempt to evaluate right-hand-sides of Eqs.~\eqref{eq:n_k} and \eqref{eq:m1_k} is to  approximate exact ground state $|0\rangle$ with a normalized variational wave function $| \Psi_G \rangle$ and evaluate expectation values using real-space diagrammatic expansion, introduced in previous sections. This is, however, not feasible due to an unavoidable finite cutoff on the range of internal lines in each of the diagrams. As a representative example, let us inspect the Fourier-transformed diagrammatic sum

\begin{align}
  \label{eq:fourier_transformed_diagrammatic_sum_example}
  \langle{\hat{n}_{\mathbf{k}\sigma}}\rangle_G = & \sum_{\boldsymbol{\delta}} e^{i \mathbf{k} \boldsymbol{\delta}}\langle \hat{c}_{i+\delta, \sigma}^\dagger \hat{c}_{i\sigma} \rangle_G = \langle \hat{n}_i \rangle_G + \sum_{\boldsymbol{\delta} \neq \mathbf{0}} e^{i \mathbf{k} \boldsymbol{\delta}} \langle \hat{c}_{i+\boldsymbol{\delta}, \sigma}^\dagger \hat{c}_{i\sigma} \rangle_G  \nonumber\\ & = \langle \hat{n}_i \rangle_G + \sum_{\boldsymbol{\delta} \neq \mathbf{0}} e^{i \mathbf{k} \boldsymbol{\delta}} \left(q^2 T^{11}_{i+\boldsymbol{\delta}, i} + 2 q \alpha T^{13}_{i+\boldsymbol{\delta}, i} + \alpha^2 T^{33}_{i+\boldsymbol{\delta}, i}  \right),
\end{align}

\noindent
where $q = \lambda_\emptyset \lambda_1 + n_0 (\lambda_d \lambda_1 - \lambda_\emptyset \lambda_1)$ and $\alpha = \lambda_d \lambda_1 - \lambda_\emptyset \lambda_1$ are coefficients, expressed in terms of the variational correlator parameters $\lambda_\emptyset$, $\lambda_1$, and $\lambda_d$ (cf. Eq.~\eqref{eq:lambda_x} for explicit definition), and $n_0$ is density per spin (once again, we assume spin-isotropic situation and drop the index $\sigma$). The symbols, $T^{11}_{ij}$, $T^{13}_{ij}$, and $T^{33}_{ij}$, denote diagrammatic sums introduced in Sec.~\ref{sec:de_gwf_expansion}. For any finite cutoff $|\boldsymbol{\delta}| < R_\mathrm{max}$ in Eq.~\eqref{eq:fourier_transformed_diagrammatic_sum_example}, $n_{\mathbf{k}} \equiv \langle\hat{n}_{\mathbf{k}\sigma}\rangle$ turns out to be a smooth function of wave vector, $\mathbf{k}$. In particular, there is no discontinuity at the Fermi surface and $Z_\mathbf{k}$ cannot be calculated. Exemplary DE-GWF results for $n_\mathbf{k}$, illustrating this behavior for the Hubbard model, are depicted in Fig.~\ref{fig:spectral_moments}(a). The true jump may emerge only in the limit $R_\mathrm{max} \rightarrow \infty$. Note characteristic oscillations caused by the cutoff in the correlations range, seen in Fig.~\ref{fig:spectral_moments}(a). Such behavior close to discontinuities is known as the Gibbs phenomenon. 

  The finite-cutoff-related problems, outlined above, can be eliminated by reformulating the method in such a way that that real-space summation is not necessary at all, e.g., by switching to the wave vector representation at the diagrammatic-expansion level. Such a modification is called $\mathbf{k}$-space \textbf{D}iagrammatic \textbf{E}xpansion of the \textbf{G}utzwiller \textbf{W}ave \textbf{F}unction, or $\mathbf{k}$-DE-GWF. Starting from Eq.~\eqref{eq:fourier_transformed_diagrammatic_sum_example}, by adding and subtracting local ($\boldsymbol{\delta} = \mathbf{0}$) terms in the sum, we can write

  \begin{align}
    \label{eq:1}
    \langle{\hat{n}_{\mathbf{k}\sigma}}\rangle_G =& \langle \hat{n}_i \rangle_G + \sum_{\boldsymbol{\delta}} e^{i \mathbf{k} \boldsymbol{\delta}} \left(q^2 T^{11}_{i+\boldsymbol{\delta}, i} + 2 q \alpha T^{13}_{i+\boldsymbol{\delta}, i} + \alpha^2 T^{33}_{i+\boldsymbol{\delta}, i}  \right) - \left( q^2 T^{11}_{i, i} + 2 q \alpha T^{13}_{i, i} + \alpha^2 T^{33}_{i, i} \right) = \nonumber \\ &  \langle \hat{n}_i \rangle_G + q^2 T^{11}(\mathbf{k}) + 2 \alpha q T^{13}(\mathbf{k}) + \alpha^2 T^{33}(\mathbf{k}) - \int \frac{d^d \mathbf{q}}{(2 \pi)^d} \left( q^2 T^{11}(\mathbf{q}) + 2 \alpha q T^{13}(\mathbf{q}) + \alpha^2 T^{33}(\mathbf{q}) \right),
  \end{align}

  \noindent
  where $T^{\alpha\beta}(\mathbf{k}) \equiv \sum_{\boldsymbol{\delta}} e^{i\mathbf{k}\boldsymbol{\delta}} \cdot T^{\alpha\beta}_{i+\boldsymbol{\delta}, i}$ are the diagrammatic sums transformed to the $\mathbf{k}$-space. Similar analysis can be carried out for higher moments of the spectral function which, however, requires specifying a concrete microscopic Hamiltonian ($\hat{\mathcal{H}}$ enters the expressions for $\mathcal{M}_n(\mathbf{k})$ with $n \geq 1$ as follows from Eq.~\eqref{eq:m1_k_integrated}). For simplicity, in the remaining part of this section we restrict to the Hubbard model, i.e., with solely on-site Coulomb repulsion included. After some algebra, one arrives at

\begin{align}
  \mathcal{M}_1(\mathbf{k}) \approx&  -\langle \hat{a}_\mathbf{k \sigma}^\dagger [\mathcal{H} - \mu \hat{N}, \hat{a}_\mathbf{k \sigma}] \rangle_G = (\epsilon_\mathbf{k} - \mu) \cdot \langle \hat{n}_{\mathbf{k}\sigma}\rangle_G + U \cdot \sum\limits_{\boldsymbol{\delta}} \mathrm{e}^{i \mathbf{k} \boldsymbol{\delta}} \langle \hat{a}^\dagger_{i+\boldsymbol{\delta}, \sigma} \hat{n}_{i \bar{\sigma}} \hat{a}_{i\sigma}\rangle_G = \\ &  (\epsilon_\mathbf{k} - \mu) \cdot n_G + U d^2_G + [(\epsilon_\mathbf{k} - \mu) \cdot q^2 + U q q'] \times \left(T^{11}(\mathbf{k}) - \int \frac{d^d\mathbf{q}}{(2\pi)^d}T^{11}(\mathbf{q})\right) + \nonumber \\ &   \left[2 (\epsilon_\mathbf{k} - \mu) \cdot q\alpha + U (q \alpha' + q' \alpha)\right] \times  \left(T^{13}(\mathbf{k}) - \int \frac{d^d\mathbf{q}}{(2\pi)^d}T^{13}(\mathbf{q})\right) +  \nonumber\\ &+  [(\epsilon_\mathbf{k} - \mu) \cdot \alpha^2 + U \alpha \alpha'] \times  \left(T^{33}(\mathbf{k}) - \int \frac{d^d\mathbf{q}}{(2\pi)^d}T^{33}(\mathbf{q})\right).
\end{align}

\noindent
The remaining task is to calculate $\mathbf{k}$-space diagrammatic sums, which is not straightforward due to constraints in summation over internal vertices present in Eqs.~\eqref{eq:expectation_val_terms}. By referring to combinatoric arguments, it can be shown that all the summation restrictions can be lifted if shifted paramagnetic lines are used $\tilde{P}_{ij} \equiv \langle  \hat{a}^\dagger_{i\sigma} \hat{a}_{j\sigma} \rangle_0 - \delta_{ij} \langle  \hat{a}^\dagger_{i\sigma} \hat{a}_{j\sigma} \rangle_0$ in place of the original ones $P_{ij} \equiv \langle  \hat{a}^\dagger_{i\sigma} \hat{a}_{j\sigma} \rangle_0$ (spin indices are dropped). Also, the shifted vertices $\hat{d}_i^\mathrm{HF}$ should be replaced with original ones, $\hat{d}_i$. In $\mathbf{k}$-space, the shifted lines take the form

\begin{align}
  &\tilde{P}_{\mathbf{k}\sigma} \equiv  \sum_i e^{i\mathbf{k}\boldsymbol{\delta}} \langle  \hat{a}^\dagger_{i+\boldsymbol{\delta}, \sigma} \hat{a}_{j\sigma} \rangle_0   - \langle{\hat{n}_{i\sigma}}\rangle_0, \\
  &\tilde{S}_{\mathbf{k}\sigma} = S_{\mathbf{k}\sigma} \equiv \sum_i e^{i\mathbf{k}\boldsymbol{\delta}} \langle  \hat{a}^\dagger_{i+\boldsymbol{\delta}, \sigma} \hat{a}_{j\sigma}^\dagger \rangle_0.
\end{align}

\noindent
As an example, we write down explicitly the expression for the statistical distribution function, $n_\mathbf{k}$, to the first expansion order in the paramagnetic state (i.e., with all anomalous expectation values $\tilde{S}_\mathbf{k}$ set to zero)

  \begin{figure}
    \centering
    \includegraphics[width=\textwidth]{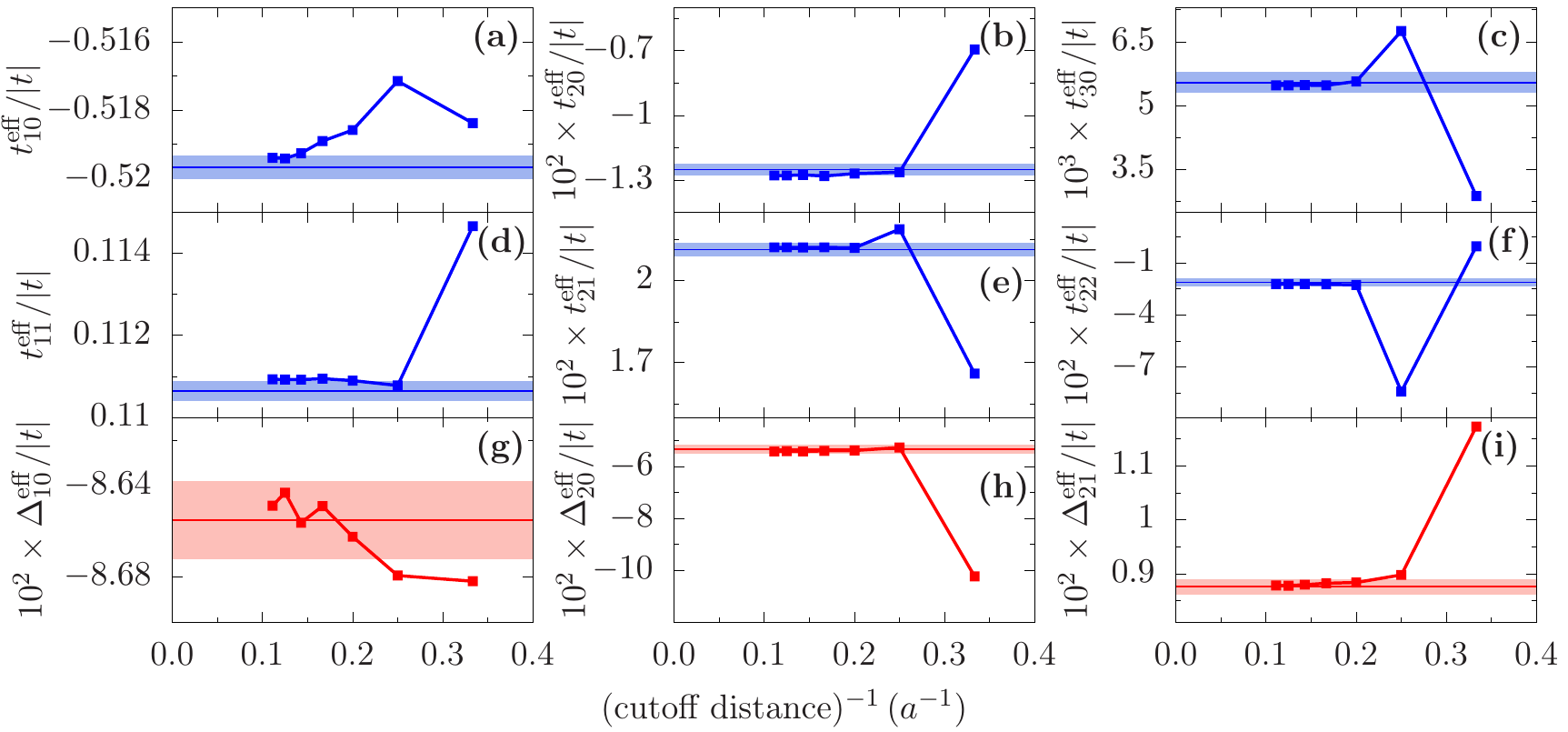}.
    \caption{(a)-(i) Scaling of selected effective Hamiltonian parameters DE-GWF as a function of the inverse of cutoff for the real-space diagram size (points). Blue color represents paramagnetic hopping and anomalous (superconducting) terms are marked in red. The solid lines are the corresponding values obtained in a fully self-consistent manner by $\mathbf{k}$-DE-GWF method (in this case, there is no cutoff and correlations up to infinite range are incorporated). The shaded region marks uncertainties due to Monte-Carlo integration. The calculations have been performed to the order $k=2$ for the $t$-$J$-$U$ model wit the parameters: $t^\prime/|t|$, $U/|t| = 20$, $J = |t|/3$, and doping $\delta \approx 0.20$.} 
    \label{fig:de_gwf_scaling}
  \end{figure}

\begin{align}
  \label{eq:explicit_expression_nk_1st_order}
  & n_\mathbf{k}  \approx  n_G + q^2\tilde{P}_\mathbf{k} + 2 q x \alpha \int \frac{d^2\mathbf{q}_0}{(2\pi)^2} \frac{d^2\mathbf{q}_1}{(2\pi)^2} \tilde{P}_\mathbf{k} \tilde{P}_{\mathbf{k} + \mathbf{q}_0 + \mathbf{q}_1}\tilde{P}_{-\mathbf{q}_0} \tilde{P}_{-\mathbf{q}_1}  - \alpha^2 \int \frac{d^2\mathbf{q}_0}{(2\pi)^2} \frac{d^2\mathbf{q}_0}{(2\pi)^2} \tilde{P}_{\mathbf{k} + \mathbf{q}_0 + \mathbf{q}_1}\tilde{P}_{-\mathbf{q}_0} \tilde{P}_{-\mathbf{q}_1} + \nonumber \\ & - 2 \alpha^2 x \int \frac{d^2\mathbf{q}_0}{(2\pi)^2} \frac{d^2\mathbf{q}_1}{(2\pi)^2} \frac{d^2\mathbf{q}_2}{(2\pi)^2} \tilde{P}_{-\mathbf{k}_0 - \mathbf{k}_1 - \mathbf{k}_2} \tilde{P}_{\mathbf{k}_2} \tilde{P}_{-\mathbf{k}_0} \tilde{P}_{-\mathbf{k}_1} \tilde{P}_{\mathbf{k} + \mathbf{k}_0 + \mathbf{k}_1} - \nonumber \\ & - q^2 \int \frac{d^2\mathbf{k}}{(2\pi)^2} \tilde{P}_\mathbf{k} - 2 q \alpha x \int \frac{d^2\mathbf{k}}{(2\pi)^2} \frac{d^2\mathbf{q}_0}{(2\pi)^2} \frac{d^2\mathbf{q}_1}{(2\pi)^2} \tilde{P}_{\mathbf{k} + \mathbf{q}_0 + \mathbf{q}_1}\tilde{P}_{-\mathbf{q}_0} \tilde{P}_{-\mathbf{q}_1} + \alpha^2 \int \frac{d^2\mathbf{k}}{(2\pi)^2} \frac{d^2\mathbf{q}_0}{(2\pi)^2} \frac{d^2\mathbf{q}_0}{(2\pi)^2} \tilde{P}_{\mathbf{k} + \mathbf{q}_0 + \mathbf{q}_1}\tilde{P}_{-\mathbf{q}_0} \tilde{P}_{-\mathbf{q}_1} + \nonumber \\ & + 2 \alpha^2 x \int \frac{d^2\mathbf{k}}{(2\pi)^2} \frac{d^2\mathbf{q}_0}{(2\pi)^2} \frac{d^2\mathbf{q}_1}{(2\pi)^2} \frac{d^2\mathbf{q}_2}{(2\pi)^2} \tilde{P}_{-\mathbf{k}_0 - \mathbf{k}_1 - \mathbf{k}_2} \tilde{P}_{\mathbf{k}_2} \tilde{P}_{-\mathbf{k}_0} \tilde{P}_{-\mathbf{k}_1} \tilde{P}_{\mathbf{k} + \mathbf{k}_0 + \mathbf{k}_1}.
\end{align}

\noindent
The above expression can be constructed by inspecting the diagrammatic expansion for $T^{11}$, $T^{13}$, and $T^{33}$, presented in Appendix~\ref{appendix:de-gwf}. Inclusion of the superconducting lines of $d$-wave symmetry proceeds along the same lines, but leads to cumbersome expressions. In Eq.~\eqref{eq:explicit_expression_nk_1st_order}, $\mathbf{k}$ denotes external wave vector, whereas integration is performed over internal quasi-momenta, $\mathbf{q}_i$. The essential observation, based on Eq.~\eqref{eq:explicit_expression_nk_1st_order}, is that $n_\mathbf{k}$ has now a \emph{true discontinuity} at the Fermi surface, from which quasiparticle weight $Z_\mathbf{k}$ can be extracted. In the particular case of paramagnetic state, this can be readily seen, since then $\tilde{P}_\mathbf{k} = P_\mathbf{k} - \langle \hat{n}_i\rangle_0 = \theta(-\epsilon^\mathrm{eff}_\mathbf{k}) - \langle \hat{n}_i\rangle_0$, where $\theta(x)$ is Heaviside step function and $\epsilon^\mathrm{eff}_\mathbf{k}$ denotes energy spectrum of the effective Hamiltonian, introduced in Sec.~\ref{sec:de_gwf_effective_hamiltonian}. To the lowest order of expansion, i.e., discarding all integrals on the right-hand side of Eq.~\eqref{eq:explicit_expression_nk_1st_order}, one obtains $n_\mathbf{k} \approx (1-q^2) \langle\hat{n}_{i\sigma} \rangle_0 + q^2 \theta(-\epsilon^\mathrm{eff}_\mathbf{k})$ (note that $n_G = \langle\hat{n}_{i\sigma} \rangle_0$ to this order). The correlated Fermi surface thus coincides exactly with that of effective quasiparticles, governed by the effective Hamiltonian $\hat{\mathcal{H}}_\mathrm{eff}$, yet the quasiparticle weight $Z_\mathbf{q} = q^2$ becomes less than one. The higher order terms in expansion \eqref{eq:explicit_expression_nk_1st_order} do not alter the analytic structure of the propagator (the Fermi surface remains unchanged at arbitrary finite order), but $Z_\mathbf{k}$ is non-trivially renormalized in a wave-vector-dependent manner. We will return to this point below. Physically, we have included correlations up to infinite range.

In Fig.~\ref{fig:spectral_moments}(b) and (c), the results of exemplary diagrammatic calculation for the Hubbard model across the first Brillouin zone are displayed for hole-doping levels $\delta = 0.02$ and $\delta = 0.12$, respectively. For the sake of concreteness, the model parameters have been selected as $t^\prime/|t| = 0.25$, $U/|t| = 12$. The calculated statistical distribution function $n_\mathbf{k}$ (blue solid line) now has a true discontinuity, seen along $\Gamma$-$M$ direction. Its magnitude defines the value of $Z_\mathbf{k}$. Note lack of Gibbs oscillations. For small hole doping $\delta = 0.02$ (panel (a)), the quasiparticle weight is substantially suppressed as compared to that for the highly-doped system at $\delta = 0.12$. This indicates loss of single-particle coherence as half-filling is approached, which is compensated by the appearance of featureless continuum background. This behavior is due to increasing role of strong-correlations close to $\delta = 0$. Naively one would expect to encounter such a discontinuity twice while moving along a closed contour, such as the $\Gamma$-$X$-$M$-$\Gamma$ employed in Fig.~\ref{fig:spectral_moments}(b)-(c). Only one discontinuity is, nonetheless, present in this case due to $d$-wave superconducting order that gaps the Fermi surface everywhere except for nodal ($\Gamma$-$M$) line. The superconducting amplitude related to the gap is displayed in Fig.~\ref{fig:spectral_moments}(b)-(c) as a red line. Finally, the green lines represent the first moment of the electronic spectral function $\mathcal{M}_1(\mathbf{k})$, calculated within the $\mathbf{k}$-DE-GWF approach. This function exhibits discontinuity of the first derivative at the Fermi surface, which is indicated by different slopes on each side (purple lines are respective tangents). This is consistent with the general theory outlined above and allows to extract Fermi velocity of fully correlated quasiparticles.

  \begin{figure}
    \centering
    \includegraphics[width=0.9\textwidth]{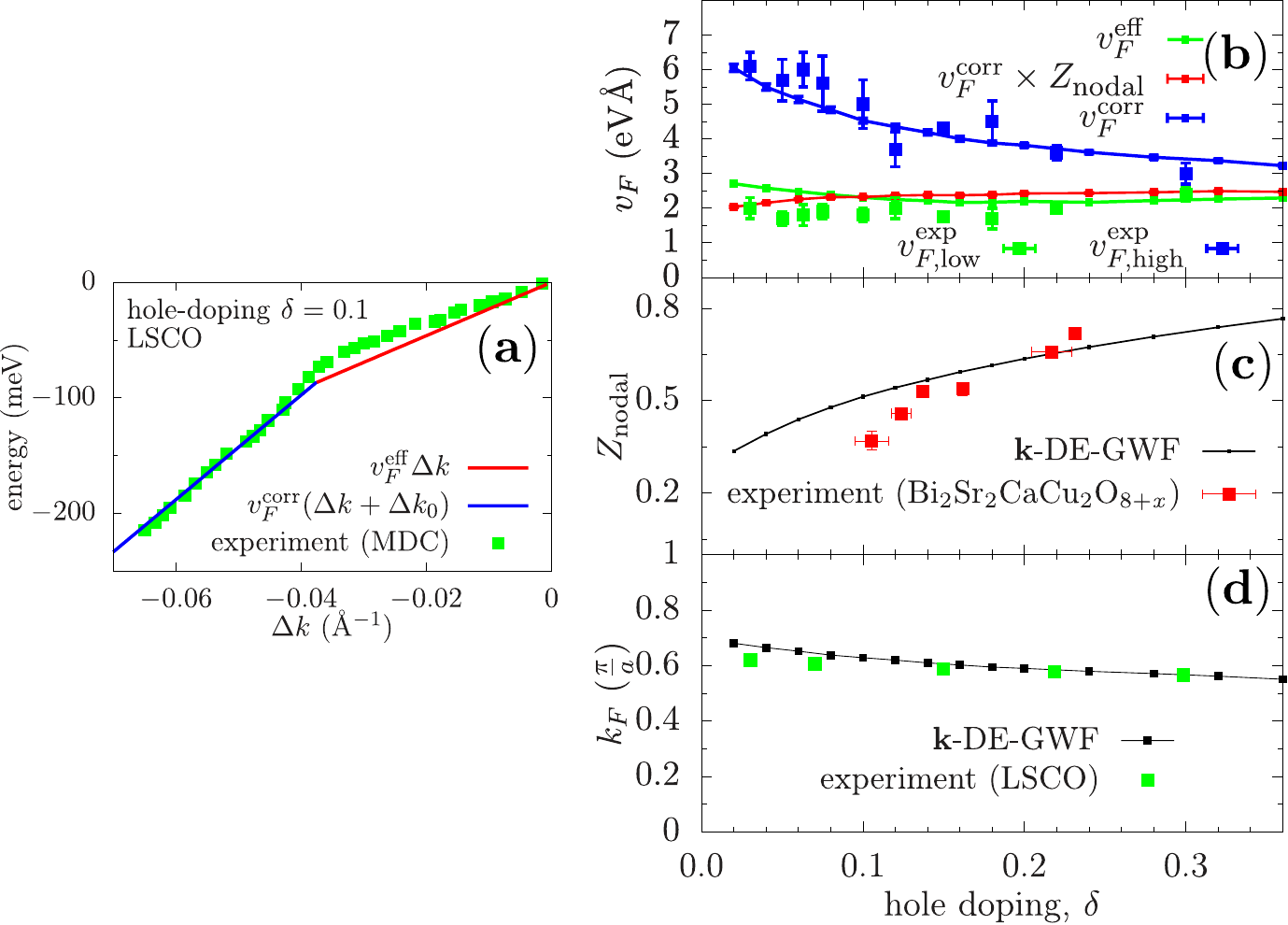}
    \caption{(a) Experimental energy dispersion along the nodal direction for $\mathrm{La_{1.9}Sr_{0.1}CuO_4}$ extracted from Ref.~\cite{ZhouNature2003}. The slopes of solid lines are obtained theoretically from the effective Hamiltonian (red) and first moment of the electron spectral function (blue) for $\delta = 0.1$. (b) Doping-dependence of quasiparticle velocities above and below the kink (green and blue squares, respectively), extracted from Refs.~\cite{ZhouNature2003,MatsuyamaPhysRevB2017}. Corresponding green and blue lines represent calculated effective- and correlated velocities calculated using $\mathbf{k}$-DE-GWF method. The red line is the correlated velocity multiplied by the calculated quasiparticle weight $Z_\mathbf{k}$. (c) Calculated $Z_\textbf{k}$ as a function of doping (black points and line), compared with experimental data for $\mathrm{Bi_2Sr_2CaCu_2O_{8+\mathit{x}}}$ (red points, extracted from Refa.~\cite{JohnsonPhysRevLett2001,RanderiaPhysRevB2004}). (d) Calculated Fermi wave vector along the nodal direction compared with data \cite{HashimotoPhysRevB2008} for $\mathrm{La_{2-\delta}Sr_{\delta}CuO_4}$. Note that $k_F(\delta)$, as in Figs.~\ref{fig:fermi_momentum_and_effective_mass}(a) and \ref{fig:principal_quantities_for_the_3_band_model}(b), is well defined for arbitrary $\delta > 0$, i.e., also deep in the underdoped metallic regime. Figures adapted from Ref.~\cite{FidrysiakJPhysCondensMatter2018}.} 
    \label{fig:fermi_velocities}
  \end{figure}

Once the procedure of evaluating diagrammatic sums to infinite spatial range has been established, it is natural to modify the self-consistent DE-GWF procedure accordingly. In the $\mathbf{k}$-DE-GWF method, all the correlation functions (lines, either paramagnetic or superconducting) are evaluated by Fourier-transforming the respective $\mathbf{k}$-space diagrammatic sums and integrating over internal momenta $\mathbf{q}_i$. The dimension of such $\mathbf{k}$-space integrals grows rapidly with the expansion order, $k$, which makes the calculations difficult at orders $k > 3$. Nonetheless, it should be emphasized that the $\mathbf{k}$-space lines are smooth except for jump discontinuities. The latter observation renders Monte-Carlo techniques suitable for evaluation  of the expressions. We do not elaborate here on this technical aspect; for details the reader is referred to Ref.~\cite{FidrysiakJPhysCondensMatter2018}. It is now instructive to contrast these $\mathbf{k}$-DE-GWF self-consistent results with those obtained from the standard DE-GWF calculation for various correlation cutoff range. It is naturally expected that the methods should coincide for large cutoff values. Such a comparison for selected model parameters (cf. caption) is presented in Fig.~\ref{fig:de_gwf_scaling}. Note that differences are sound for the smallest considered cutoffs. As the correlation distance is increased, the real-space calculations becomes equivalent to the $\mathbf{k}$-space one within the error bars, as it should be. For this demonstration, we have restricted to the second diagrammatic order only ($k=2$), since otherwise real-space computations with large cutoff ($\sim 10$ lattice sites) would be not possible due to numerical limitations. A non-trivial observation, based on Fig.~\ref{fig:de_gwf_scaling}, is that the nearest-neighbor effective hopping integrals saturate slowly with increasing correlations range. 

\subsubsection{Multiple velocity scales and dispersion kinks in correlated electron systems}

In contrast to normal metals, quasiparticle energies in strongly correlated electron systems exhibit ubiquitous kink-like features in their dispersion. These can be directly identified in angle-resolved photoemission spectroscopy, either in the the momentum- of energy-distribution curves (abbreviated as MDCs and EDCs, respectively). Exemplary experimental dispersion \cite{ZhouNature2003}, as extracted from the MDCs for $\mathrm{La_{2-\delta}Sr_{\delta}CuO_4}$ at $\delta = 0.1$, is displayed in Fig.~\ref{fig:fermi_velocities}(a) (green points). On the vertical axis zero represents the Fermi level, and on the horizontal one Fermi wave vector. For well defined binding energy $\approx 70\,\mathrm{meV}$, the curve abruptly changes its slope. This feature is commonly referred to as ``kink'' in the photoemission literature, and it has been also observer in other classes of correlated materials \cite{HuPhysRevLett2019}. The two distinct velocity scales, defined by the dispersion slopes below- and above the kink, have very different doping-dependence with the former larger in magnitude than the latter. This can be seen in Fig.~\ref{fig:fermi_velocities}(b). The green and blue points refer to the quasiparticle velocity above- and below the kink, respectively. The true Fermi velocity (green points) appear to be practically insensitive to the chemical doping. This behavior is inconsistent with a naive analysis based on tight-binding Hamiltonian, where doping-dependence of Fermi momentum (and, in turn, Fermi velocity) is enforced by the Luttinger theorem. On the other hand, as was pointed out in previous subsections, this behavior is faithfully reproduced by DE-GWF analysis of the $t$-$J$-$U$ model (the value of Fermi velocity matches that obtained from the effective Hamiltonian). A qualitatively different trend is visible for higher-energy quasiparticles corresponding to binding energy exceeding $70\,\mathrm{meV}$. By inspecting the doping-dependence of the blue squares in Fig.~\ref{fig:fermi_velocities}, we can observe that \textit{(i)} the high-energy velocity monotonically increases as one moves towards half-filled case (i.e., Mott-insulating state, $\delta = 0$), and \textit{(ii)} these two scales converge on the overdoped normal-metal side of the phase diagram. Such a behavior suggests that strong electronic correlations and Mottness play a decisive role in this dual behavior.

\subsubsection{Selected $\mathbf{k}$-DE-GWF results for the Hubbard model and comparison with experiment}

  By employing the tools developed above one can now address essential correlated quasiparticle parameters using the $\mathbf{k}$-DE-GWF method. We will discuss Fermi velocity and spectral weight across the cuprate phase diagram, employing the Hubbard model with $t^\prime/|t| = 0.25$, $U/|t| = 12$ and allowing for the $d$-wave superconductivity. All thermodynamic quantities are evaluated by Monte-Carlo integration in the $\mathbf{k}$-space. To make the discussion quantitative and make comparison with experiment possible, we also set the absolute energy scale by taking the nearest-neighbor hopping integral $t = -0.35\,\mathrm{eV}$ and the lattice spacing {$a = 3.70\,$\AA}  (the latter, along with Planck's constant $\hbar$, is necessary to express the calculated velocity in physical units).

  In Fig.~\ref{fig:fermi_velocities}(a) the slopes of the red and blue lines correspond to calculated effective and ``correlated'' (i.e., that obtained from the moment of the spectral function) velocity scales, $v_F^\mathrm{eff}$ and $v_F^\mathrm{corr}$, respectively. Their values match remarkably well the quasiparticle velocities below- and above the kink. In Fig.~\ref{fig:fermi_velocities}(b) theoretical values of the velocities represented by solid green and blue lines. Note excellent agreement with experimental data (solid squares) in entire available doping range. The effective velocity (green line) depends very weakly on doping, which is consistent with DE-GWF result discussed in subsection~\ref{subsection:properties_of_the_tjuv_model}. The blue line, on the other hand, follows closely the high-energy quasiparticle dispersion. Panel (d) shows the quasiparticle $Z_\mathbf{k}$ and corresponding experimental data for $\mathrm{Bi_2Sr_2CaCu_2O_{8+\mathit{x}}}$ (this compound has been selected here due to availability of estimates for the absolute value of $Z_\mathbf{k}$). Panel (d) provides an analogous comparison for the Fermi wave vector in the nodal direction.

\section{Related strongly correlated systems} \label{sec:related_correlated_systems}
  \begin{figure}
    \centering
    \includegraphics[width=0.85\textwidth]{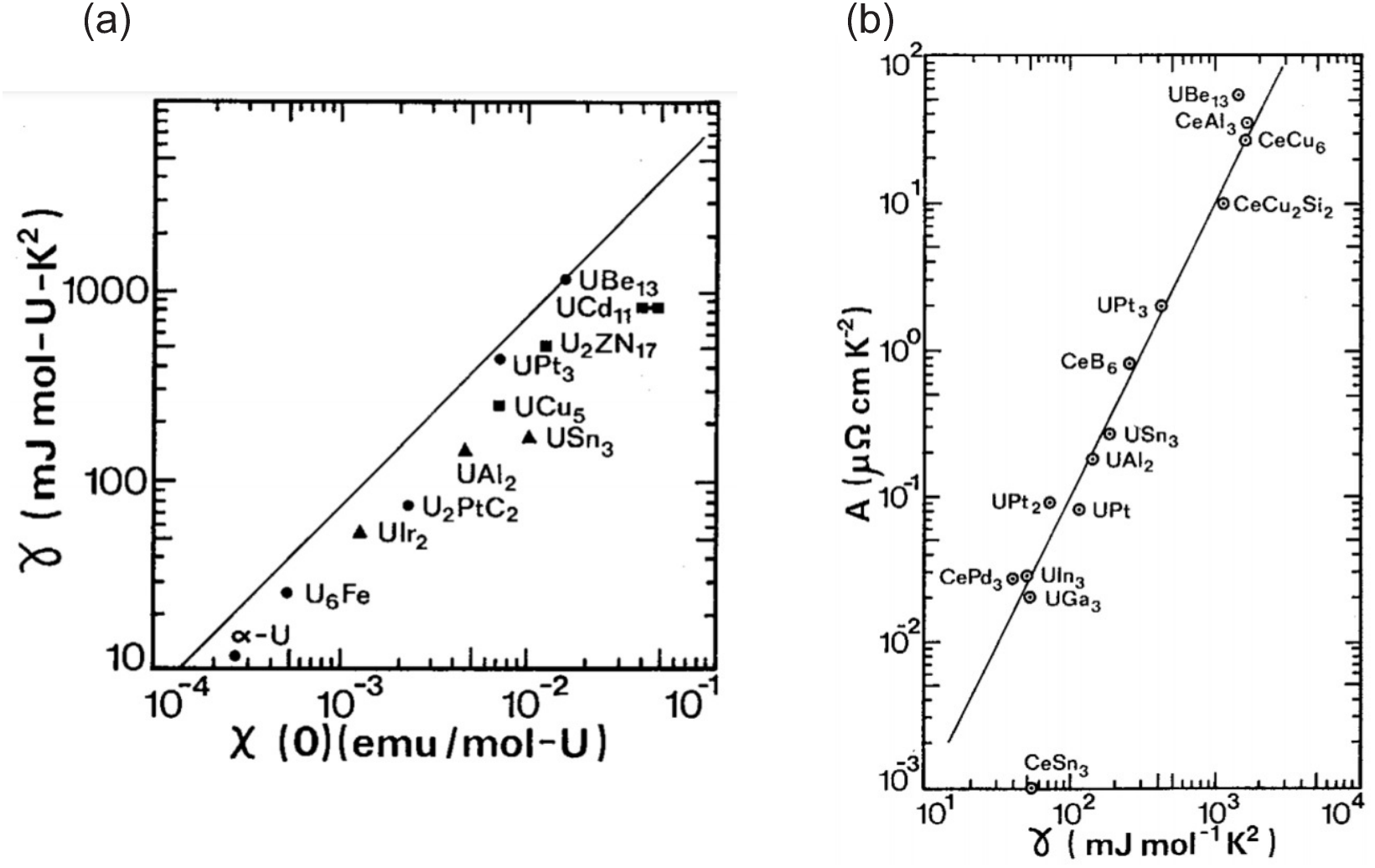}
    \caption{(a) Universal scaling of the coefficient $\gamma$ vs. static magnetic susceptibility, $\chi(0)$, for heavy fermion systems. The straight line represents the trend for ideal fermion gas. (b) Universal scaling of the $A$ coefficient in resistivity of the Fermi liquid ($\rho = \rho_0 + A T^2$) vs. the coefficient $\gamma$. The straight line represents the so-called Kadowaki-Woods scaling law $A \sim \gamma^2$. After~\cite{LeeCommCondensMattPhys}. For additional discussion, see also \cite{SpalekEurJPhys2000}. The proportionality $A \sim \gamma^2 \sim \rho(\epsilon_F)^2$ means that $T^2$ term in resistivity is indeed due to the electron-electron interaction (i.e., the Baber-Landau-Pomeranchuk scattering). The plots reflect a rather standard metallic behavior (the straight lines), but with largely enhanced resistivity coefficient values of $\gamma$, $\chi(0)$, $A$.}
    \label{fig:gamma_and_A}
  \end{figure}

    \begin{figure}
    \centering
    \includegraphics[width=0.95\textwidth]{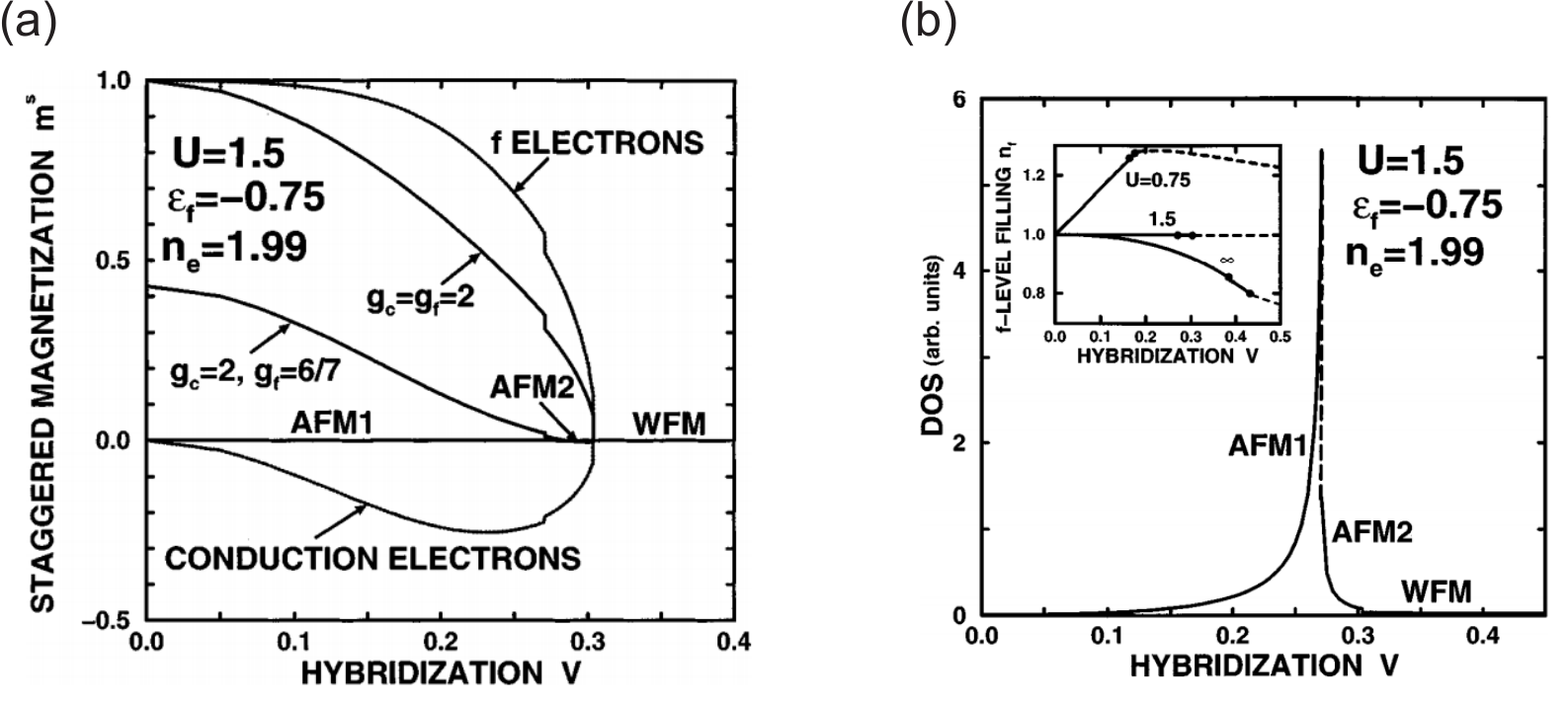}
    \caption{(a) Sublattice magnetic moments for $f$ and conduction electrons and the total magnetization for the Land\'{e} factors $g_f = 2$ and
      $g_f = 6/7$, respectively. (b) The density of quasiparticle states at Fermi energy (relative to the value in the bare band, and reduced by a factor $10^{-2}$). The inset displays the $f$-level occupancy for various $U$. All quantities are a function of bare hybridization $V$. AFM1 and AFM2 label two distinct antiferromagnetic states (see \cite{DoradzinskiPhysRevB1997}), whereas WFM denotes a weak ferromagnetic phase. Note that these results for intraatomic hybridization $V \sim 0.3$ we have the system with almost compensated magnetic moments and very high density of states at the Fermi level. After~\cite{DoradzinskiPhysRevB1997}}
    \label{fig:heavy_fermions_dos_and_moments}
  \end{figure}

\subsection{Heavy fermion systems: Basic properties}
\label{subsec:basic_properties_heavy_fermions}

\subsubsection{Basic properties at a glance}
\label{subsubsec:basic_properties_at_glance}

In the present context, by heavy-fermion (or heavy electron) system we understand a compound with very heavy effective masses $m^{*} = 10^2$-$10^3m_e$ (with $m_e$ being the electron mass), but still exhibiting metallic (Landau Fermi-liquid) properties. Those system contain correlated either $4f$ or $5f$ electrons, hybridized with conduction ($6d$/$6s$ or other quasi-free electron states). The canonical properties of such systems, containing cerium and uranium, are displayed in Fig.~\ref{fig:gamma_and_A}. In panel (a) we have plotted (on the logarithmic scale) the dependence of linear specific heat coefficient, $\gamma$, vs. zero-temperature Pauli susceptibility. Panel (b) shows the coefficient $A$ in the $A T^2$ contribution to the static electric conductivity at low temperature vs. $\gamma$. From this simple illustration, a few important conclusions may be inferred. First, both the values of $\gamma$ and $\chi$ in panel (a) as by $2$-$3$ orders of magnitude larger than those for the ordinary or ferromagnetic transition metals. Even for $\mathrm{V_2O_{3-\delta}}$ close to the Mott localization (cf. Fig.~\ref{fig:gamma_and_A}(a)), the value of $\gamma$ (and $\chi$) reaches up to one order of magnitude larger values. Yet, the system remains a normal metal as can be verified by showing from the linear dependence $\gamma \sim \chi(0)$ that $(\gamma/\gamma_0)/(\chi/\chi_0) \sim 2$, i.e., it is close to the free-electron value ($\gamma_0$ and $\chi_0$ are the free electron values, both proportional to the density of states at the Fermi energy, $\rho(\epsilon_F)$). It looks as though the heavy-fermion system is that of metal with heavy quasiparticles indeed; the similar situation takes place for liquid ${}^3\mathrm{He}$ in the normal state \cite{DobbsBook2001}.

Second, turning to Fig.~\ref{fig:gamma_and_A}(b) one sees that $A \sim \gamma^2$ (the so-called Kadowaki-Woods scaling \cite{KadowakiSolStatCommun1986}), i.e., the $T^2$ term in the resistivity is due to electron-electron scattering and is anomalously large. Because of that, it is called the Landau-Pomeranchuk-Baber term as it is directly proportional to $\rho_0(\epsilon_F)$, exemplifying the electron-electron scattering process.

The question is then what is the source of those extraordinary properties. Relatively early, it has been established that the correct modeling relies relies on starting from the periodic Anderson (or Anderson-lattice) model introduced in Sec.~\ref{subsection:anderson_lattice_model}. As a representative example of the results for the Anderson lattice model, obtained within the mean-field-type approach, we show in Fig.~\ref{fig:heavy_fermions_dos_and_moments} the evolution from the magnetic metallic (AFM1, AFM2, WFM) states to the normal metallic phase (Fig.~\ref{fig:heavy_fermions_dos_and_moments}(a)) and associated with it enhancement of the density of states at the resultant Fermi energy (Fig.~\ref{fig:heavy_fermions_dos_and_moments}(b)). The results in Fig.~\ref{fig:heavy_fermions_dos_and_moments} are displayed as a function of the hybridization amplitude $|V|$. The transition is associated with the essential reduction of the magnetic-moment amplitude of $f$-electrons, partly compensated by the conduction ($c$) electrons. In  effect, the heavy-fermion state emerges at the instability of the high-moment state, mimicking the localization-delocalization transition of $f$-electron states. At such an instability, a notable quantum-critical behavior, and the associated with it onset of other (superconducting) state, do appear. Here we focus on the stable unconventional superconducting state.

\subsubsection{Superconducting properties}

    \begin{figure}
    \centering
    \includegraphics[width=0.9\textwidth]{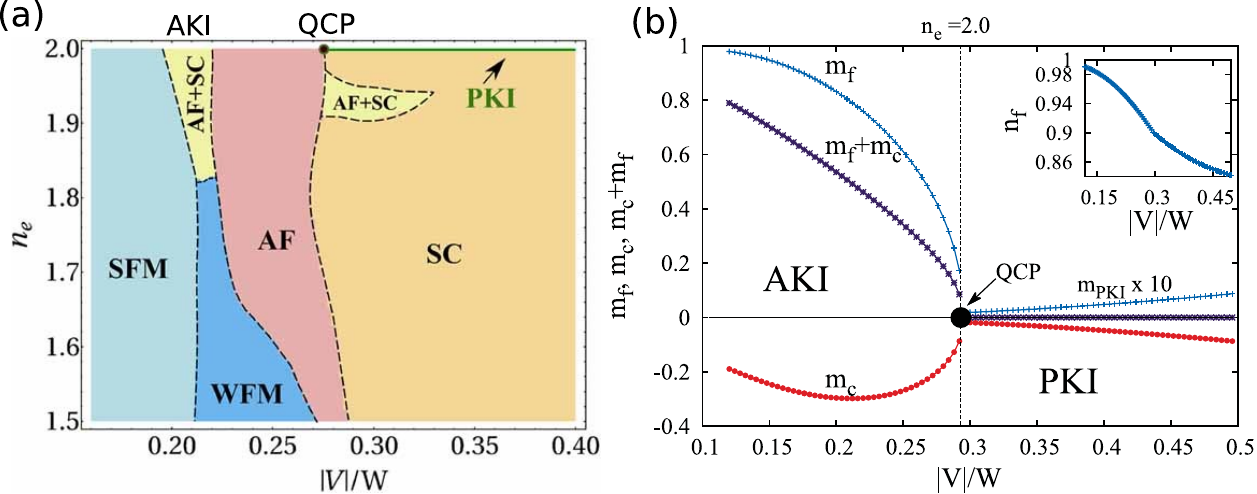}
    \caption{(a) Phase diagram encompassing magnetic and SC states on hybridization ($|V|$) - number of
electrons ($n_e$) plane in the case with interatomic hybridization (of extended $s$-wave form) and anisotropic SC gap of d-wave type. The
model parameters are $\epsilon_f = -0.85$ and $U = 2$. The solid point corresponds to the QCP. Apart from magnetic and superconducting states, the system exhibits the antiferromagnetic (AKI) and paramagnetic (PKI) Kondo insulating states. (b) f-electron ($m_f \equiv n^{(f)}_{\uparrow} - n^{(f)}_{\downarrow}$), conduction ($m_c$), and total ($m_f + m_c$) magnetic moments in the Kondo insulating state, all vs. intra-atomic hybridization magnitude. The inset exhibits the $f$-level occupancy as a function of $|V|$. On panel (a) transition from AF (AKI) to paramagnetic (PKI) Kondo-insulating state for $n_e = 2$, induced by the change of hybridization
strength and located at the QCP, is shown. Note a complete compensation of the spin moments in PKI state, which represents a parent state for the superconductivity appearing for $|V| > 0.4$ and $n_e < 2$. For the sake of clarity the magnitude of the magnetic moments was multiplied by a factor of 10 in the PKI and the SC cases. After \cite{HowczakPSS2013}.}
\label{fig:heavy_fermions_phase_diargam_with_sc}
    \centering
    \includegraphics[width=0.9\textwidth]{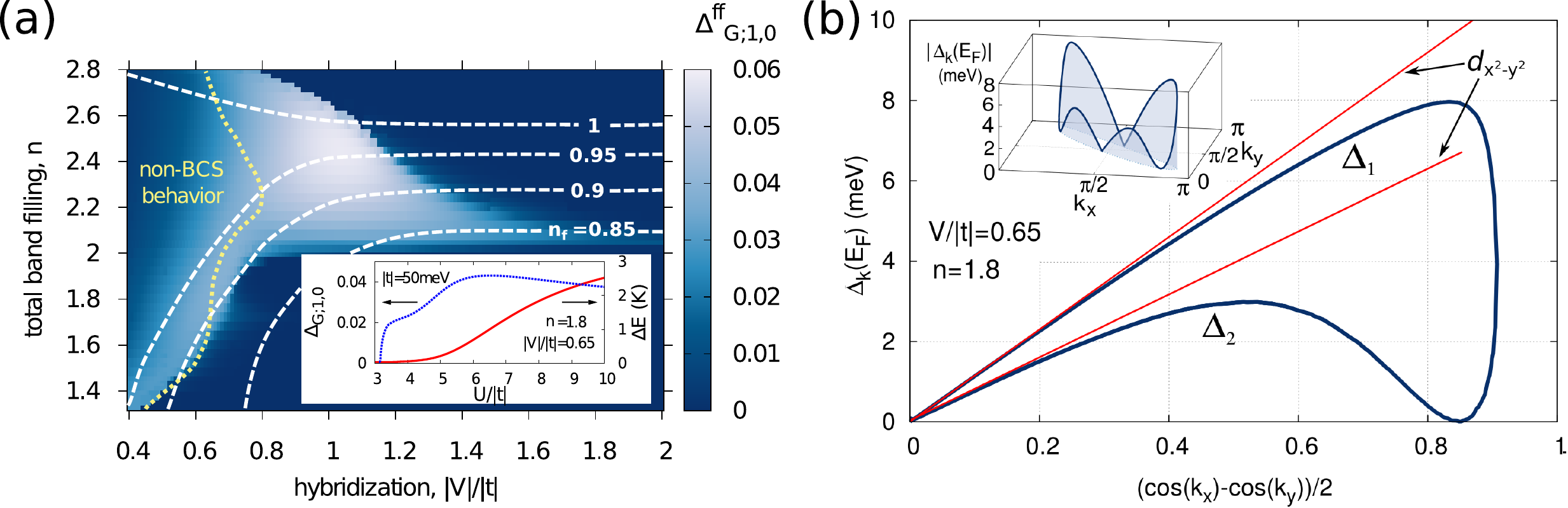}
    \caption{(a) Phase diagram on total band filling -- hybridization magnitude plane, with the dominant, $d_{x^2-y^2}$-wave superconducting order parameter for $f$ electrons, $\Delta_{G;1,0}^{ff}$ (see main text). The dashed lines mark the selected $f$-orbital isovalents. The dotted line singles out the {\it non-BCS} region defined by the gain in the kinetic energy, $\Delta E_{kin}>0$ in SC state. For the selected point ($n=1.8$, $|V|/|t|=0.65$) we show in the inset that SC appears with the increasing $f$-$f$ interaction. (b) Angle dependence of the SC gaps at the Fermi surface for selected parameters $n=1.8$ and $|V|/|t|=0.65$. The larger gap, $\Delta_1$ follows $d_{x^2-y^2}$ dependence. The inset shows the quarter of the Brillouin zone with explicitly drawn absolute values of the SC gaps. Not that in Figs.~\ref{fig:heavy_fermions_dos_and_moments}-\ref{fig:heavy_fermion_non_bcs} the $f$-level occupancy is always close to unity. This reflects the situation in Ce heavy-electron compounds for which the $f$-electron configuration is close to the value $n_c = 1$, i.e., the valency of cerium is $\mathrm{Ce}^{3+\delta}$ with $\delta \ll 1$. After~\cite{WysokinskiPhysRevB2016}.}
    \label{fig:heavy_fermion_non_bcs}
  \end{figure}

Extensive literature has accumulated on the analysis of the superconducting state in heavy-fermion systems \cite{RiseboroughBook2008,PfleidererRevModPhys2009}, starting from the pioneering paper in 1979 \cite{SteglichPhysRevLett1979}. Its most remarkable qualitative feature qualitative feature is pairing of the heavy quasiparticles near the threshold of $f$-electron localization, where often (anti)ferromagnetism sets in. Here we only discuss briefly the results for the superconducting phase and its coexistence with magnetism, as well as mention the limit of Kondo insulating phase with a characteristic quantum critical point.

We start with an overview of the results in the SGA approximation. The most general form of the effective Hamiltonian in the strong-correlation limit is quite involved, though its analysis involves the same degree of complexity as the single narrow-band case, presented in Sec.~\ref{sec:vwf_solution}. The explicit form of the Hamiltonian, involving only the most important terms, is given by Eq.~\eqref{eq:anderson_hamiltonian_fourth_order_canonical_transformation}. Next, by application the SGA method, we obtain both the quasiparticle energies and the $d$-wave solution for the $f$-$f$ pairing, which is the dominant component. In Fig.~\ref{fig:heavy_fermions_phase_diargam_with_sc}(a) we exhibit an exemplary overall phase diagram on the total electron number per site ($n_f + n_c = n_e$) vs. hybridization magnitude plane. The situation for $n_e = 2$ corresponds to the Kondo insulators: antiferromagnetic (AKI) and nonmagnetic (PKI), which meet at the quantum critical point (QCP). The partial to total-magnetic-moment compensation of $f$-electrons ($m_f$) and $c$-electrons ($m_c$) is illustrated in Fig.~\ref{fig:heavy_fermions_phase_diargam_with_sc}(b). The inset in that figure is the $f$-level occupancy; note that it is close to unity in the range of interest, which defines the heavy-electron regime. The strong (SFM) and weak (WFM) ferromagnetic phases, as well as that with antiferromagnetic order (AFM), label the corresponding orientation of the $f$-electron magnetic moments. The most relevant to us here are purely superconducting (SC) and coexisting (AFM+SC) phases, the latter appearing near the QCP. For a more detailed discussion of the phase diagram and related physical properties, see Ref.~\cite{HowczakPSS2013,HowczakJPCM2012}. Note that these mean-field-type results reflect qualitatively the properties of quasi-two-dimensional system $\mathrm{Ce_{1-\mathit{x}}Rh_\mathit{x}In_f}$ as a function of pressure \cite{KnebelJPhysSocJapan2011,KnebelPhysStatSolidi2010}. The $f$-electron occupancy, $n_f$, corresponds to the modeled fraction valence of cerium via the formula $\mathrm{Ce}^{+4 - n_f}$.

The decisive question at this point is how the phase diagram, depicted in Fig.~\ref{fig:heavy_fermions_phase_diargam_with_sc}(a), changes upon inclusion of nonlocal correlations within DE-GWF scheme. This problem has not been addressed as yet by including all the principal phases appearing there. Here we present only the DE-GWF phase diagram \cite{WysokinskiPhysRevB2016} on the plane filling-hybridization ($n$-$|V|$), including both the hybrid $f$-$c$ and $f$-$f$ pairing amplitudes in the correlated state, $\Delta_2$ and $\Delta_1$, respectively. These are the results obtained within DE-GWF. We see that the phase diagram \ref{fig:heavy_fermion_non_bcs}(a) contains both non-BCS and BCS-type solution, as for the case of the cuprates (cf. Sec.~\ref{sec:selected_equilibrium_properties}). The SC instability is particularly well visible in the regime near the heavy-fermion ($n_f \approx 1$) to mixed-valence crossover transition. Again, the unconventional superconducting state appears close to the localized-moment regime and has a definite $d$-wave character close to the nodal direction, similarly as in the cuprates (cf. Fig.~\ref{fig:heavy_fermion_non_bcs}(b)). Those two features, common for the (single-band) cuprates and (multiple-band) heavy-fermion systems speak in favor of the universality of the unconventional superconductivity in strongly correlated systems. Detailed studies, involving the also the systems close to the Mott-Hubbard localization of $f$-electrons, are required to draw definite conclusions.

\subsection{Spin-triplet pairing and ferromagnetism-superconductivity coexistence: The case of $\mathrm{UGe_2}$}
\label{subsec:fm_sc_coexistence_uge2}

In this subsection, we would like show to what degree the exchange-mediated pairing can be applied to the description of spin-triplet pairing, particularly when it is enhanced by the inter-electron correlations. This discussion is based on a concrete example of $\mathrm{UGe_2}$, for which the spin fluctuations \emph{may not} play the principal role. This statement may not be valid for other related systems ($\mathrm{UCoGe}$, $\mathrm{UIr}$, $\mathrm{URhGe}$), as explained briefly below.

It is necessary to make a methodological here. Namely, the model situation for the spin-triplet superconductivity in liquid ${}^3\mathrm{He}$, where this type of (single-band) pairing takes the form of $p$-wave (orbitally antisymmetric state with moment $l=1$) character \cite{Vollhardt1990}. In the case of solid the situation is more controversial (e.g. for $\mathrm{UPt_3}$ or $\mathrm{Sr_2RuO_4}$ \cite{MachidaJSupoercond1999}) due to the fact that those systems are multiorbital and hence, the multiband interactions/correlation effects are important, even for the triplet pairing. In effect, the orbitally symmetric (interorbital) pairing is an attractive possibility \cite{MachidaPhysicaB2000}. Here will \emph{not} dwell upon those types of orderings and concentrate instead on the specific example of $\mathrm{UGe_2}$, where spin-triplet superconductivity has not been questioned. A brief overview of the situation in related systems is provided at the section end.

\subsubsection{The concept of local spin-triplet pairing}

In subsection~\ref{subsection:models_with_hunds_rule}, the local exchange-type of spin-triplet pairing has been introduced on the basis of analogy to the spin-single pairing that was the source of the high-$T_c$ superconductivity theory in the cuprates, based on $t$-$J$ and related models. However, in the multi-orbital systems, Hund's rule ferromagnetic exchange may become stronger than any intersite exchange contribution. Its consequences, particularly near the transition to the localized-moment regime, may be thus discussed separately from any other pairing contribution.

The Hund's-rule-induced pure interorbital pairing leads, in the BCS approximation, to the following Bogoliubov-Nambu-de Gennes form of the effective Hamiltonian in the two-band case \cite{SpalekPhysRevB2001}

\begin{align}
  \hat{\mathcal{H}}_\mathrm{BCS} = \sum_\mathbf{k} \mathbf{f}_\mathbf{k}^\dagger \hat{\mathcal{H}}_\mathbf{k} \mathbf{f}_\mathbf{k}  + \sum_\mathbf{k} E_\mathbf{k},
  \label{eq:bcs_hamiltonian_hund} 
\end{align}

\noindent
where

\begin{align}
  \hat{\mathcal{H}}_\mathbf{k} =
  \begin{pmatrix}
    E_\mathbf{k}  & 0 & \Delta_1 & \Delta_0 \\
    0 & E_\mathbf{k}  & \Delta_0 & \Delta_{-1} \\
     \Delta^{*}_1 & \Delta^{*}_0 & -E_\mathbf{k} & 0 \\
     \Delta^{*}_0 & \Delta^{*}_{-1} & 0 & -E_\mathbf{k} \\ 
  \end{pmatrix}
  \label{eq:bcs_hamiltonian_matrix_hund}
\end{align}

\noindent
and $\mathbf{f}^\dagger_\mathbf{k} \equiv \left( \hat{f}^{(1)\dagger}_{\mathbf{k}\uparrow}, \hat{f}^{(1)\dagger}_{\mathbf{k}\downarrow}, \hat{f}^{(2)}_{-\mathbf{k}\uparrow}, \hat{f}^{(2)}_{-\mathbf{k}\downarrow} \right)$. In Eq.~\eqref{eq:bcs_hamiltonian_matrix_hund} the gap parameters are $\Delta_m \equiv -2J \sum_\mathbf{k} \langle\hat{f}_{\mathbf{k}\sigma}^{(1)\dagger}\hat{f}_{\mathbf{k}\sigma^\prime}^{(2)\dagger}\rangle$, with $m = (\sigma+\sigma^\prime)/2$. This Hamiltonian possesses three types of solutions for $\Delta_m$: (\emph{i}) $A_1$ phase with $\Delta_{+1} \neq 0$ and $\Delta_{0} = \Delta_{-1} = 0$; (\emph{ii}) $A$ phase with $\Delta_{+1} = \Delta_{-1} \neq 0$ and $\Delta_0 = 0$; and (\emph{iii}) $B$ phase with all the amplitudes equal. This labeling is in accordance with that introduced in the context of superfluid ${}^3\mathrm{He}$. However, it must be emphasized that here the pairing is of $s$-wave, even-parity, orbital-singlet, and $\mathbf{k}$-independent.

      \begin{figure}
    \centering
    \includegraphics[width=0.4\textwidth]{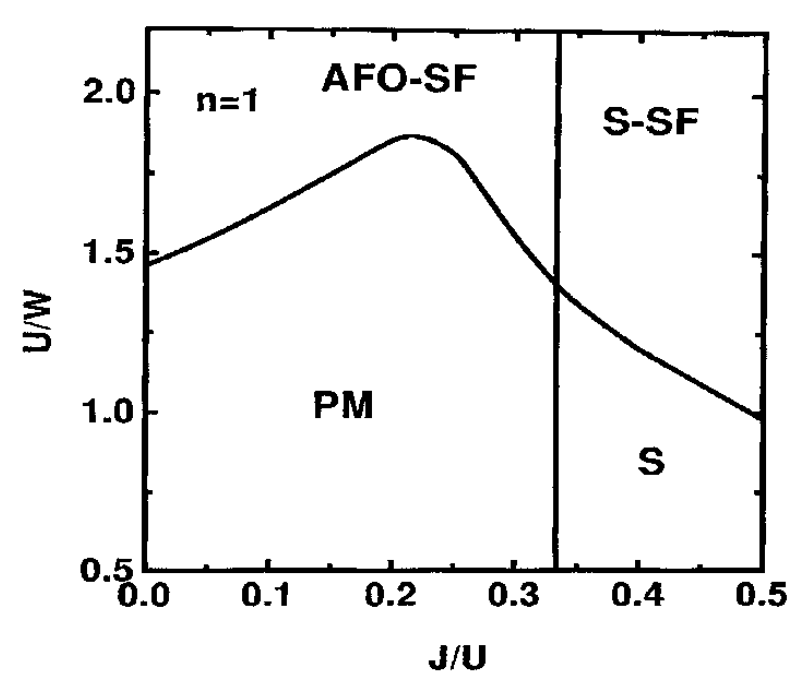}
    \caption{The phase diagram for the two-band model on the plane $U/W$-$J/U$, for $n = 1$.The superconducting (S) and ferromagnetic superconducting (S-SF)
phases for $J>U/3$ are also included. AFO labels antiferromagnetic orbital ordering, SF -- saturated ferromagnetic and paramagnetic metallic states, whereas $S$ labels even spin-triplet superconducting state. After \cite{KlejnbergPhysRevB2000}.}
    \label{fig:kleinberg_phase_diagram}
  \end{figure}

A detailed phase diagram in the SGA approximation can be provided only by incorporating also magnetic solutions and including the direct Coulomb interactions (the Hubbard and local interorbital terms). This has been discussed separately \cite{ZegrodnikNewJPhys2013,ZegrodnikNewJPhys2014}. In general, for the doubly degenerate band, ferromagnetism (FM) and spin triplet superconductivity do not coexist and saturated ferromagnetism (SF) phase transforms into coexistent antiferromagnetic-spin-triplet phase, which eventually evolves into Mott-Hubbard insulating antiferromagnetic state at half-filling ($n = 2$). Here we display in Fig.~\ref{fig:kleinberg_phase_diagram} only the superconducting (SC), coexisting SC-saturated ferromagnetic (FM), and antiferromagnetic orbital (AO)-SF, and normal paramagnetic (PM) phase, all on the plane $U/W$ vs. $J/U$ and for quarter filling ($n=1$) case \cite{KlejnbergPhysRevB2000}. All the borderlines are of discontinuous (first-order) type. on this example one sees clearly the abundance of magnetically-, orbitally-ordered, and superconducting phases, as the corresponding ground-state energies are all of comparable magnitude.

\subsubsection{The $\mathrm{UGe_2}$ case}
In subsection~\ref{subsec:basic_properties_heavy_fermions}  we have addressed the case of spin-singlet pairing in two-orbital ($c$-$f$) Anderson lattice, bearing in mind the situation in $\mathrm{Ce}$-based heavy-fermion compounds, such as $\mathrm{CeCoIn_5}$ or $\mathrm{CeRhIn_5}$, in which cerium ions are in $\mathrm{Ce}^{+3+\delta}$ ($4f^{1-\delta}$) electronic configuration. Here we analyze briefly the situation in uranium compounds, such as $\mathrm{UGe_2}$ or $\mathrm{URhGe}$, in which the uranium ions are approximately in between $5f^2$ and $5f^3$ configuration. To describe such systems, one needs to start minimally from a doubly-degenerate $5f$ states, each hybridized separately with conduction-band electrons. The appropriate starting Hamiltonian has been formulated in Sec.~\ref{subsec:fm_sc_coexistence_uge2} (cf. also Eq.~\eqref{eq:hamiltonian_anderson_multiorbital}). However, since we address only the spin-triplet pairing observer in $\mathrm{UGe_2}$, the local-spin singlet pairing part may be safely ignored.

      \begin{figure}
    \centering
    \includegraphics[width=0.6\textwidth]{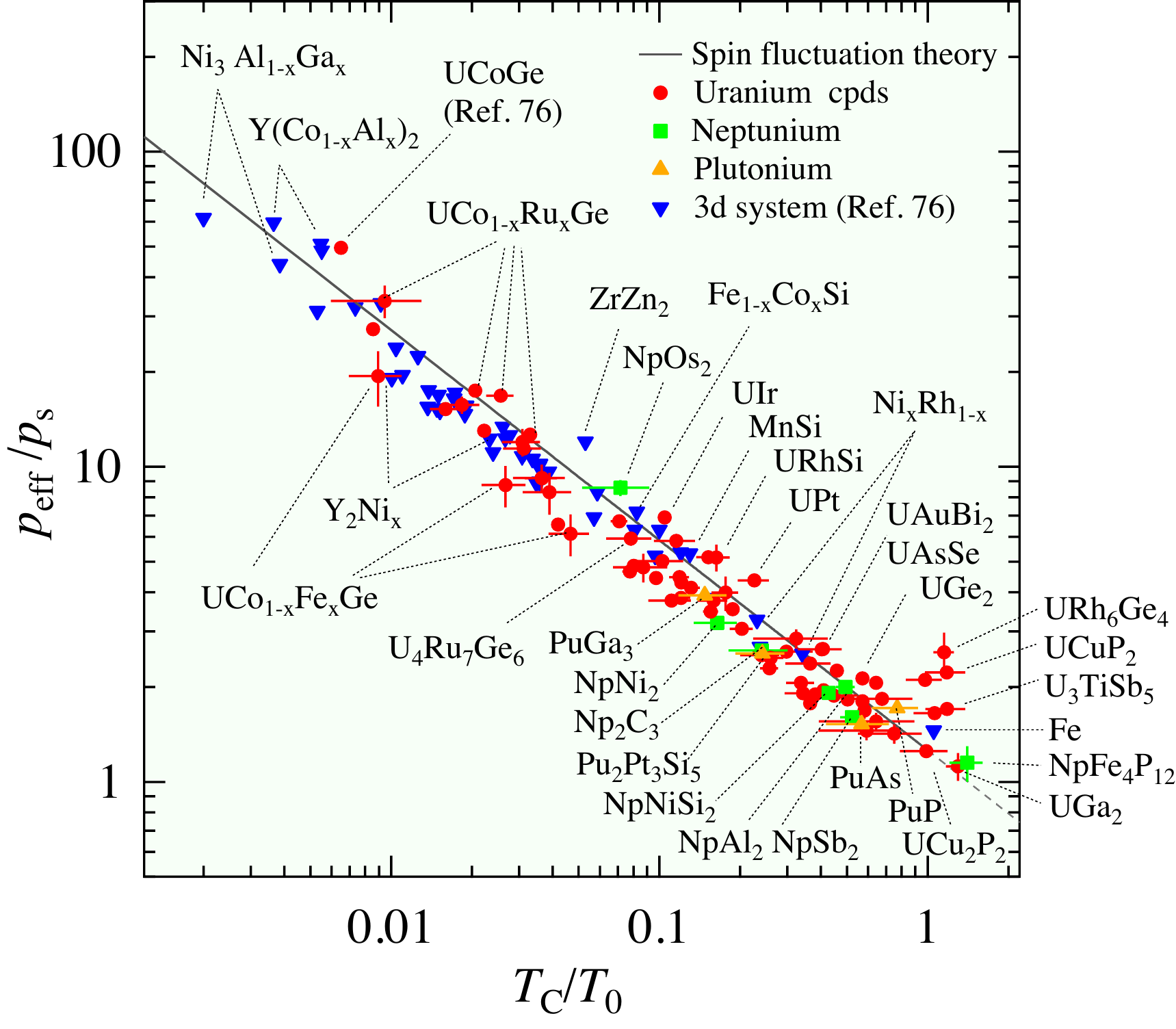}
    \caption{Generalized Rhodes-Wohlfarth plot. Here $p_\mathrm{eff}$ and $p_s$ denote the magnitude of magnetic moments (effective spins) in paramagnetic ($T > T_c$) and ferromagnetic ($T \rightarrow 0$) states. Data points for uranium, neptunium and plutonium compounds are plotted as circles, squares and triangles, respectively. The abscissa is the Curie temperature ($T_c$) divided by a temperature characterizing the spin fluctuations. The data has been taken from Refs.~\cite{TakahashiJPSJ1986,TakahashiJPCM2001,TakahashiBook2013}. The solid line represents the spin-fluctuation theory calculation. After \cite{TateiwaPhysRevB2017}.}
    \label{fig:generalized_rhodes_wohlfarth_plot}
  \end{figure}

The reason for selecting the model with local pairing is illustratively justified in Fig.~\ref{fig:generalized_rhodes_wohlfarth_plot}, where we have plotted the ratio of paramagnetic to zero-temeprature paramagnetic moment against the Curie to spin-fluctuation ratio (the so-called Rhodes-Wohlfarth plot \cite{TateiwaPhysRevB2017}). The moment ratio speaks about the relative strength of fluctuations against the effective magnetic field, and its value close to unity (the case of $\mathrm{UGe_2}$) means that the effect of fluctuations of the resultant ferromagnetic moment is relatively small. This suggests that the saddle-point approximation (SGA here) provides a reasonable first-order description, at least in qualitative terms, of the overall properties over a relatively wide temperature range. We assume that this working principle is valid also when the spin-triplet superconductivity is included.

Starting from the Hamiltonian~\eqref{eq:hamiltonian_anderson_multiorbital} and utilizing the local-pairing representation~\eqref{eq:Hf3}, we obtain the effective Hamiltonian in the mean-field approximation (SGA) \cite{KadzielawaMajorPhysRevB2018}

\begin{align}
	\hat{\mathcal{H}}_\mathrm{eff}  = \sum_{\mathbf{k}, \sigma}^{} \Psi_{\mathbf{k} \sigma}^\dagger \left( \begin{array}{cccc}
		\epsilon_{ \mathbf{k} } & 0 &  q_\sigma V&  0\\
		0 & - \epsilon_{ \mathbf{k} } & 0 & - q_\sigma V\\
		q_\sigma V & 0 & \epsilon_{ \sigma }^{f} & \Delta_{ \sigma \sigma }^{ff} \\
		0 & - q_\sigma V & \Delta_{ \sigma \sigma }^{ff} & -\epsilon_{ \sigma }^{f} \\
	\end{array}
	\right) \Psi_{\mathbf{k} \sigma}
  + E_0,
  \label{eq:uge2_heff}
\end{align}

\noindent
where $\Psi_{\mathbf{k}\sigma }^\dagger \equiv \left( \hat{c}^{(1)\dagger}_{ \mathbf{k} \sigma }, \hat{c}^{(2)}_{ -\mathbf{k} \sigma }, \hat{f}^{(1)\dagger}_{ \mathbf{k} \sigma }, \hat{f}^{(2)}_{ -\mathbf{k} \sigma } \right)$, $\epsilon_\mathbf{k}$ denotes bare $c$-electron energy, $\epsilon_\sigma^f$ is an effective $f$-level position, $\Delta^{ff}_{\sigma\sigma} \equiv \mathcal{V}_\sigma \langle\hat{f}^{(1)}{i\sigma}\hat{f}^{(2)}{i\sigma} \rangle$ is the effective $f$-$f$ equal-spin SC gap parameter, $\mathcal{V}_\mathrm{\sigma} \equiv  - U'  g_{1\sigma} + (J - U')  g_{2\sigma}$ denotes effective pairing coupling, and $E_0$ is a constant. The renormalization factors $q_\sigma$, $g_{1\sigma}$, and $g_{2\sigma}$ account for the correlation effects (for details see \cite{KadzielawaMajorPhysRevB2018}). Amazingly, the $4\times 4$ matrix in Eq.~\eqref{eq:uge2_heff} may be diagonalized analytically and, therefore, the quasiparticle energies take the following analytic form

	\begin{align}
	E_{ \mathbf{k} \sigma }^{(\lambda)} =& \pm \Bigg(
	q_\sigma^2 V^2 
	+ \frac{1}{2} \left[ \left( \Delta^{ff}_{ \sigma \sigma } \right)^2 + \left( \epsilon^f_\sigma \right) ^2 + \epsilon_\mathbf{ k }^2 \right] \nonumber \\&
	\pm \frac{1}{2} \sqrt{
		\left[
			\left( \Delta^{ff}_{ \sigma \sigma } \right)^2 + \left( \epsilon^f_\sigma \right)^2 - \epsilon_\mathbf{ k }^2 
		\right]^2 
		+ 4 q_\sigma^2 V^2   	
		\left[
		\left( \Delta^{ff}_{ \sigma \sigma } \right)^2 
		+ 
		\left(
		\epsilon_\mathbf{ k }
		+ \epsilon^f_\sigma
		\right)^2  
		\right]} 
	\Bigg)^{\frac{1}{2}}.
	\end{align}

        \noindent
        The gap $\Delta_{\sigma\sigma}^{ff}$ can be expressed to a good accuracy as

        \begin{align}
\Delta_\mathbf{k}^2 = \frac{\epsilon_\mathbf{k}^2}{(\epsilon_\mathbf{k} + \epsilon_\sigma^f)^2} \times (\Delta_{\sigma\sigma}^{ff})^2 + o\left[(\Delta_{\sigma\sigma}^{ff})^2\right].
\end{align}

      \begin{figure}
    \centering
    \includegraphics[width=0.95\textwidth]{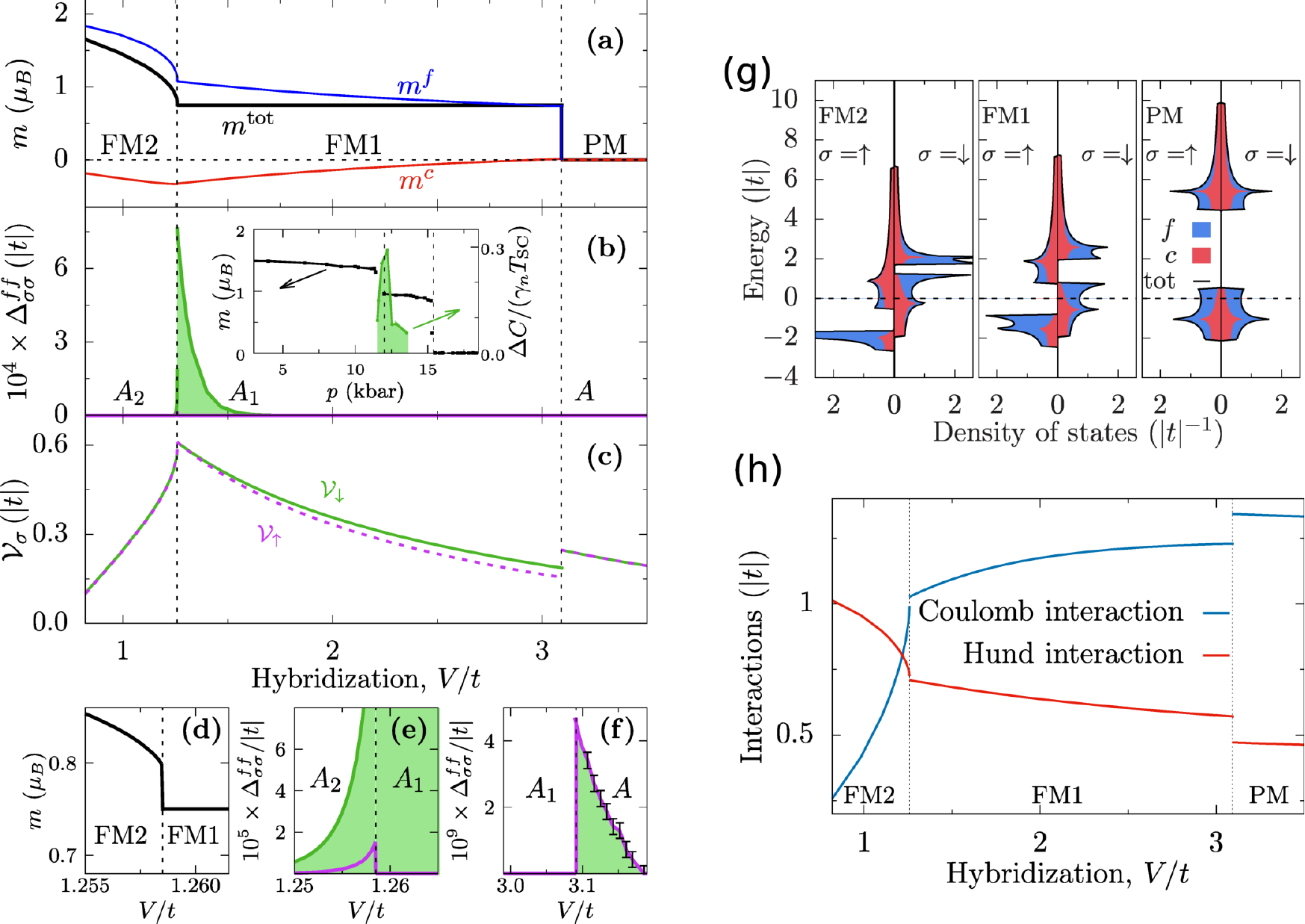}
    \caption{(a)-(f) Calculated zero-temperature phase diagram of $\mathrm{UGe_2}$ for  Hund's coupling $J / |t|= 1.1$ versus $f$-$c$ hybridization $V$. The remaining parameters read: $t' /|t| = 0.25$, $U / |t|= 3.5$, $\epsilon^f / |t|= -4$, and $n^\mathrm{tot} = 3.25$. (a) Total magnetic moment $m^\mathrm{tot}$ per formula unit (black solid line), and the corresponding $f$ and $c$ electron magnetization $m^f$ and $m^c$ (blue and red lines, respectively). $m^c$ represents a residual Kondo compensating cloud. (b) Triplet $f$-$f$ SC gap component $\Delta^{ff}_{\uparrow\uparrow}$ (purple shading) and $\Delta^{ff}_{\downarrow\downarrow}$ (green shading). Three distinct SC phases  $A_2$, $A_1$, and $A$ are marked. The $A$-phase gaps ($\sim 10^{-9}|t|$) are not visible in panel (b). Inset shows experimental  magnetization for $\mathrm{UGe_2}$ \cite{PfleidererPhysRevLett2002} and the specific-heat jump at the SC transition temperature $T_\mathrm{SC}$ (normalized by $T_\mathrm{SC}$ and the linear specific-heat coefficient $\gamma_n$) \cite{TateiwaPhysRevB2004}. (c) Effective coupling constant $\mathcal{V}_\sigma$ for spin-up (purple) and spin-down (green) triplet pairing. Note that value of coupling is the largest near the $A_2 \rightarrow A_1$ transition. (d) Total magnetic moment near the FM2$\rightarrow$FM1 metamagnetic transition. (e)-(f) SC gap components near the FM2$\rightarrow$FM1 and FM1$\rightarrow$PM transition points, respectively. (g) Spin- and orbital-rtesolved density of states in (moving from left to the right) FM2, FM1, and PM state. (h) Comparison of the intraorbital Coulomb (Hubbard term) and Hund's rule exchange contribution to the total ground-state energy. After~\cite{KadzielawaMajorPhysRevB2018}.}
    \label{fig:uge2_phase_diagram}
  \end{figure}

The optimal equilibrium values of all interesting us quantities are determined variationally. Also, as in $\mathrm{UGe_2}$ the spin-triplet superconductivity appears in the vicinity of the transition between the two distinct ferromagnetic phases (high-moment, FM2, and low-moment, FM1, phases), each partially compensated by the negative Kondo polarization of conduction electrons, the optimization process is quite involved. Additionally, there are three $A$-type SC phases possible: $A_2$ with $\Delta^{ff}_{\uparrow\uparrow} \neq \Delta^{ff}_{\downarrow\downarrow}$, $A_1$ with only $\Delta^{ff}_{\downarrow\downarrow} \neq 0$, and the symmetric $A$ phase with $\Delta^{ff}_{\uparrow\uparrow} = \Delta^{ff}_{\downarrow\downarrow} = \Delta$. The phase diagram, encompassing also overall representation of other properties, is presented in Fig.~\ref{fig:uge2_phase_diagram}. The inset illustrates the corresponding experimental properties: magnetic moment $m$ per U atom and the specific-heat jump at the SC transition close to the FM1-FM2 border. 

A few important features features should be pointed out. First, it is worth emphasizing that the principal experimentally observed properties are qualitatively reproduced by theoretical calculation \cite{KadzielawaMajorPhysRevB2018}, cf. Fig.~\ref{fig:uge2_phase_diagram}(a)-(b). Second, the effective pairing potential (Fig.~\ref{fig:uge2_phase_diagram}(c)) is only weakly-dependent on spin direction. The insets (d)-(f) illustrate the details of the phase diagram close to the phase-transition points. Third, the system in the FM1 phase is  \emph{half-metallic} and the pairing at the Fermi-level takes place only between the spins with magnetic moments opposite to the total magnetization (cf. panel (g)), again in accordance with experiment \cite{AokiJPSJ2019}. Finally, the system at the $\mathrm{FM2}\rightarrow\mathrm{FM1}$ transition transforms from the Hund-metal regime to the correlated metal, as illustrated explicitly in Fig.~\ref{fig:uge2_phase_diagram}(h), where the corresponding energies are shown. All the figures presented in this subsection show the results as a function of intraatomic hybridization magnitude, $|V|$, to emulate the real system evolution as a function of pressure. Also, the present formulation should be supplemented with discussion of quantum spin/charge fluctuations, as well as of the correlations beyond SGA (within the DE-GWF procedure). Those extensions require much more involved analysis, both analytic and numerical, to make the comparison quantitative.

      \begin{figure}
    \centering
    \includegraphics[width=0.55\textwidth]{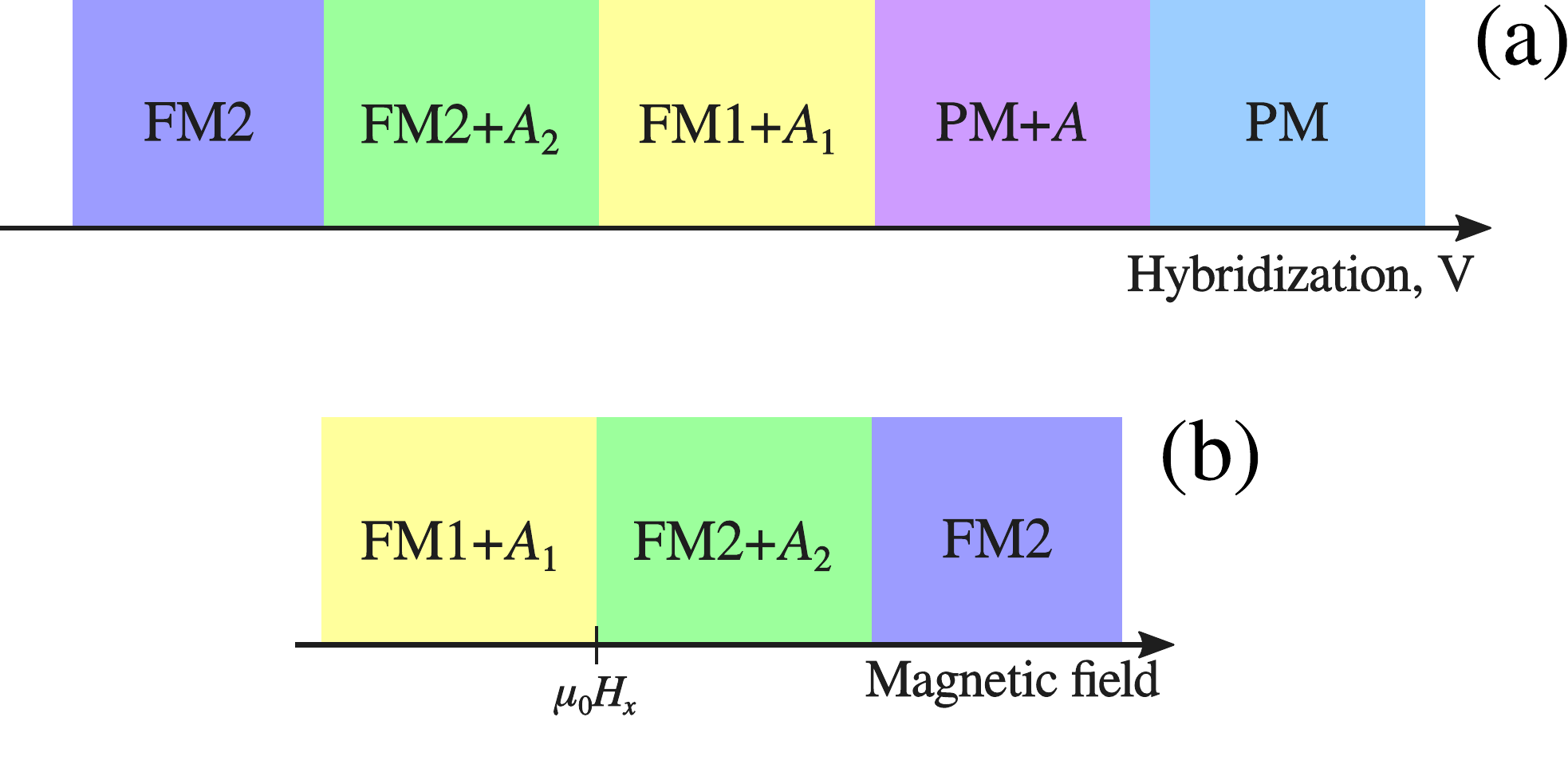}
    \caption{(a) Sequence of ferromagnetic (FM2, FM1) and nonmagnetic (PM) phases, coexisting with  superconducting (${A_2}$, ${A_1}$, $A$) states, as a function of increasing hybridization magnitude (emulating pressure change\cite{AbramJMagnMagnMat2016}) for zero external magnetic field. (b) The same as in (a), but for fixed hybridization and varying pressure, with the most prominent FM1+$A_1$ phase taken as the starting point. The boundaries mark transition points in both the magnetic and  superconducting sectors. After~\cite{FidrysiakPhysRevB2019}.}
    \label{fig:uge2_sequence_of_phases}
  \end{figure}

At the end, in Fig.~\ref{fig:uge2_sequence_of_phases} we have provided a sketch of the phase sequence as a function of hybridization (a) and applied Zeemann magnetic field (b) \cite{FidrysiakPhysRevB2019}. We see that those phase transitions provide an opportunity to study various superconducting phases also in the systems with a weaker or absent magnetic ordering ($\mathrm{URhGe}$, $\mathrm{UCoGe}$, $\mathrm{UIr}$). In those cases, the role of quantum spin fluctuations in SC state may be enhanced (cf. Fig.~\ref{fig:generalized_rhodes_wohlfarth_plot}) and easier to examine. This is particularly important, since within the present approach the magnetism and superconductivity have the same microscopic origin.

\subsection{Twisted bilayer graphene}

      \begin{figure}
    \centering
    \includegraphics[width=0.4\textwidth]{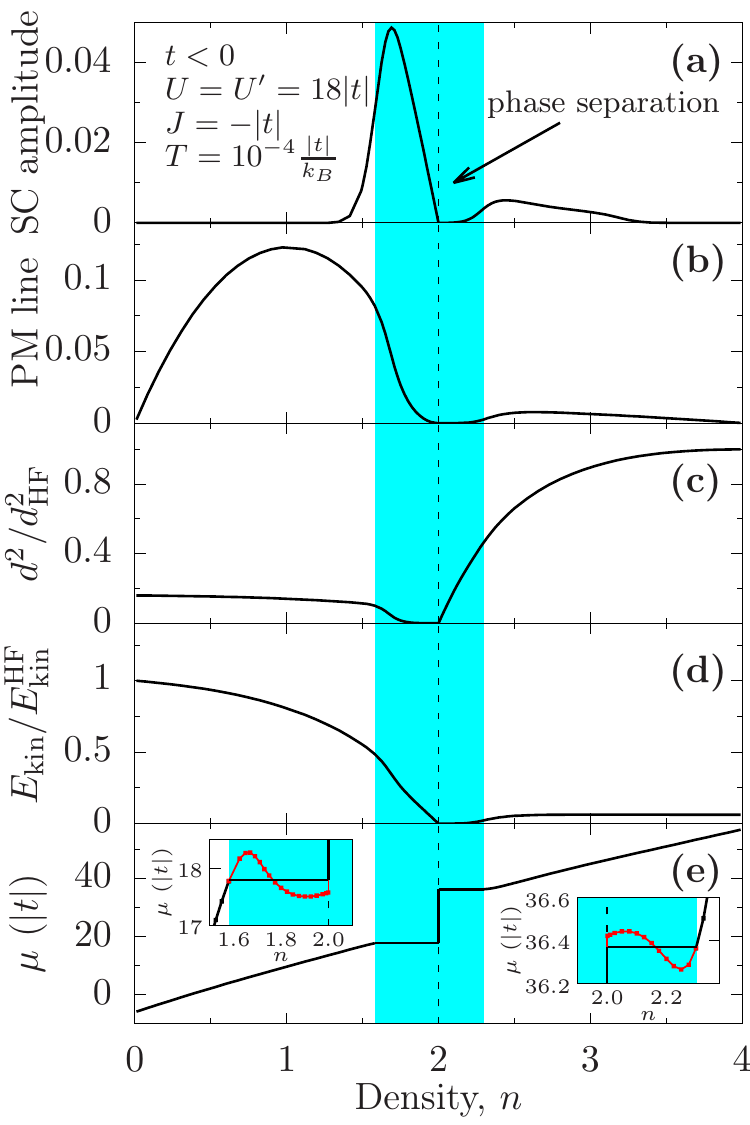}
    \caption{Phase diagram for two-orbital toy model of twisted bilayer graphene. (a) Doping dependence of the dimensionless superconducting gap amplitude component. The shaded area marks the phase-separation region, where the superconducting state appears separated spatially from the Mott insulating phase emerging near the half-filling. (b) Hopping probability $\langle c^{(l)\dagger}_{i\sigma} c^{(l)}_{j\sigma} \rangle_G$ which represents the electron itineracy. (c) Probability $d^2$ of the double site occupancy, normalized to its Hartree-Fock value $d^2_\mathrm{HF}$. (d) The ratio of kinetic energies calculated within the SGA and Hartree-Fock approximations. (e) Chemical potential $\mu$ as a function of electron density. Note that $\mu$ is constant in the phase-separation region and its value is determined by Maxwell construction. The close-ups of the phase-separation regions below and above the half-filling are displayed in the insets. The squares are computational data points and the red solid lines mark unstable spatially homogeneous solutions. After~\cite{FidrysiakPhysRevB2018}.}
    \label{fig:tbg_two_orbital}
  \end{figure}

The twisted by-magic-angle bilayer graphene forms the so-called moir\'{e} lattice and exemplifies a flat band structure, which may be regarded as one of the strongly-correlated system with the Mott-insulator phases and electron-concentration-dependent superconductivity, emulating together the behavior of two-dimensional superconducting cuprates \cite{CaoNature2018_MI,CaoNature2018_SC,AndreiNatMat2020}. This case represents a highly prospective field of research \cite{AndreiNatMat2020}. Here we mention only selected early SGA results of modeling the above states \cite{FidrysiakPhysRevB2018} within a two-band model with variable electron concentration (in real system this variation is realized by adjusting the gate voltage). If Fig.~\ref{fig:tbg_two_orbital} we reproduce a portion of the phase diagram covering SC pairing amplitude of $d_x + id_y$ character (a), as well as electron hopping (b) and double-occupancy probabilities (c), both vanishing for the filling $n=2$ and singling the onset of Mott insulator. The shaded area near $n=2$ marks the phase separation, as indicated by an anomalous nonmonotonic  behavior of the chemical potential $\mu$ in that region (cf. insets in panel (e)). the kinetic energy, depicted in (d), is reduced with respect to its Hartree-Fock (HF) value, demonstrating the effect of electronic correlations. A further, more realistic calculations, involving also the possibility of ferromagnetism/antiferromagentism appearance (intertwinned with the Mott insulating states) in the proper filling range are required, since the spin configurations may be strongly frustrated and the standard type of phase transition may not be observed.

\subsection{Outlook}

Summarizing this section, one can say that the variational approach may be useful to a broad class of narrow (multiple-)band systems, in which the local correlations play significant, if not essential role. The associated exchange interactions cooperate or compete with them, depending on the system, and provide a unified microscopic picture of magnetism, superconductivity, and Mott-Hubbard metal-insulator transition. As already said, the SGA approach provides a proper form of mean-field theory for those states and corrects the statistical inconsistencies of either original Gutzwiller of renormalized mean-field theory. The SGA method, in turn, can be generalized to the DE-GWF form in order to account for nonlocal correlations in those systems. The final generalization step requires taking into account quantum spin and charge fluctuations, starting from the  correlated saddle-point state (as obtained either from SGA or DE-GWF).

In the remaining part of our article, we overview this last step within the SGA+$1/\mathcal{N}_f$ combined approach. The full DE-GWF+$1/\mathcal{N}_f$ theory of strongly correlated systems must still wait for its successful formulation, perhaps in the near future.

\section{Beyond real-space pairing: Role of quantum spin \& charge fluctuations on example of one-band model and comparison to experiment} \label{sec:fluctuations}

As outlined in Secs.~\ref{sec:introduction}-\ref{sec:related_correlated_systems}, a number of principal features observed in high-temperature superconductors and related strongly-correlated electron systems can be systematically described within local-electronic-correlation and local-pairing scenarios, accounted for within variational scheme. The relevant microscopic models capturing this physics are based on Hubbard, $t$-$J$, $t$-$J$-$U$-$(V)$ single-band Hamiltonians, with possible multi-orbital extensions. Nonetheless, variational calculations alone might not provide a complete picture as they cover predominantly thermodynamic and single-particle properties, whereas new spectroscopic experiments evidence that collective spin \cite{LeTaconNatPhys2011,DeanPhysRevLett2013,DeanNatMater2013,DeanPhysRevB2013,LeTaconPhysRevB2013,ZhouNatCommun2013,IshiiNatCommun2014,LeeNatPhys2014,DeanPhysRevB2014,GuariseNatCommun2014,JiaNatCommun2014,WakimotoPhysRevB2015,MinolaPhysRevLett2015,PengPhysRevB2015,EllisPhysRevB2015,HuangSciRep2016,GretarssonPhysRevLett2016,MinolaPhysRevLett2017,IvashkoPhysRevB2017,MeyersPhysRevB2017,ChaixPhysRevB2018,Robarts_arXiV_2019,FumagalliPhysRevB2019,PengPhysRevB2018,RobartsArXiV2020} and charge \cite{IshiiPhysRevB2017,HeptingNature2018,IshiiJPhysSocJapan2019,LinNPJQuantMater2020,SinghArXiV2020,NagPhysRevLett2020} excitations are ubiquitous and robust throughout the phase diagram of high-$T_c$ cuprates and other correlated materials, including iron-based compounds \cite{ZhouNatCommun2013,PelliciariNatcommun2021} and iridates \cite{GretarssonPhysRevLett2016,LuPhysRevLett2017,ClancyPhysRevB2019}. We overview those aspects of dynamics below. There are also purely theoretical arguments supporting a non-trivial role the collective fluctuations may play in correlated electronic systems. A comparison of the Hubbard model phase diagrams, obtained using Monte-Carlo and dynamical mean field theory (DMFT) \cite{SchaferPhysRevB2015}, shows that metal to insulator transition is substantially affected by the circumstance whether the fluctuation effects are incorporated or not (cf. also \cite{KimPhysRevLett2020}). Indeed, the essential difference between those computational schemes is that DMFT handles poorly long-range collective excitations (DMFT extensions to address this aspect have been proposed though \cite{AyralPhysRevB2016}). Also, neglecting long-wavelength fluctuation effects may lead to violation of Mermin-Wagner theorem at finite temperature, as well as to overestimating the stability of the long-range-ordered phases and associated with them critical temperatures. The above arguments point towards the necessity of incorporating both the collective-mode excitations and local correlations into a single theoretical framework and on the same footing. The goal of this section is to demonstrate how such a program may be realized by starting from the variational-wave-function approach. Whereas a number of analytical and numerical techniques have been developed specifically to describe quantum dynamics, we treat the problem selectively and focus predominately on those constituting a direct extension of those introduced in Secs.~\ref{sec:introduction}-\ref{sec:related_correlated_systems}. To provide the context, we first briefly overview the relevant experimental techniques applied to determine collective excitations in correlated materials, present the available data, as well as discuss how these compare to theoretical predictions within the scheme.\footnote{Subsections~\ref{subsec:probes_of_collective_excitations}-\ref{sec:large_n_expansion} are, in part, based on Ref.~\cite{FidrysiakPhDThesis2016}.}

\subsection{Selected experimental probes of collective excitations}
\label{subsec:probes_of_collective_excitations}

Two experimental techniques that stand out in the context of collective spin and charge excitations in solids are inelastic neutron scattering (INS) and resonant inelastic $x$-ray scattering (RIXS). They are based on different principles and are capable of mapping directly collective excitations in complementary wave-vector and energy ranges. 

\subsubsection{Inelastic neutron scattering}

Neutrons, being neutral and spin-$\frac{1}{2}$ particles, may be used to effectively study magnetic structure and dynamical properties of solid-state materials. Specifically, \emph{inelastic} neutron scattering (INS) involves energy exchange with the magnetic subsystems and thus probes magnon and paramagnon dynamics. The unpolarized INS cross-section is given by the formula \cite{BookSquires}

\begin{align}
\frac{d^2 \sigma}{d \Omega d\omega} = \left(\frac{\gamma r_0}{2}\right)^2  g^2 |f(\mathbf{k})|^2 \mathrm{e}^{-2 W} \frac{k_f}{k_i} \sum \limits_{\alpha \beta} (\delta_{\alpha \beta} - \hat{k}_\alpha \hat{k}_\beta) S^{\alpha 
\beta}(\omega, \mathbf{k}), \label{pni_eq:neutron_cross_section}
\end{align}

\noindent
where $\gamma = 1.913$, $r_0 = 2.818 \cdot 10^{-15}\,\mathrm{m}$, $f(\mathbf{k})$ is the magnetic structure factor of magnetic ion, $\mathrm{e}^{-2 W}$ denotes the Debye-Waller factor, and $g$ is Land\'{e} $g$-factor. In equation~(\ref{pni_eq:neutron_cross_section}) $\hbar \mathbf{k}_i$ and $\hbar \mathbf{k}_f$ are initial and final neutron momenta, $\hbar \mathbf{k} = \hbar \mathbf{k}_f - \hbar \mathbf{k}_i$ is the momentum transfer, and $\hat{\mathbf{k}}$ denotes the unit vector parallel to ${\mathbf{k}}$. The dynamical spin structure factor

\begin{align}
 S^{\alpha 
  \beta}(\mathbf{k}, \omega) \equiv \frac{1}{2\pi \hbar} \int dt \mathrm{e}^{- i \omega t}  \left< \delta \hat{S}^\alpha_{\mathbf{k}}(0) \delta \hat{S}^\beta_{-\mathbf{k}}(t)\right>,
 \label{pni_eq:dyn_str_factor}
\end{align}

\noindent
entering equation~(\ref{pni_eq:neutron_cross_section}), incorporates information about time-dependent spin correlations. The latter may be related to imaginary part of the dynamical magnetic susceptibility

\begin{align}
\chi^{\alpha 
\beta}(\mathbf{k}, \omega) = \frac{i}{\hbar} \int \limits_0^\infty dt \left<[\hat{S}^\alpha_{\mathbf{k}}(t),  \hat{S}^\beta_{-\mathbf{k}}(0)]\right> \mathrm{e}^{i \omega t} \label{pni_eq:real_time_susc}
\end{align}

\noindent
with the use of fluctuation-dissipation theorem 

\begin{align}
 S^{\alpha 
\alpha}(\mathbf{k}, \omega) = \frac{1}{\pi} \frac{1}{1 - \exp(-\beta \omega)} \chi^{\alpha 
\alpha''}(\mathbf{k}, \omega).\label{pni_eq:fluctuation_dissipation_theorem}
\end{align}

\noindent
In Eq.~\eqref{pni_eq:dyn_str_factor}, a modified spin operator $\delta \hat{S}^\alpha_\mathbf{k}(t) \equiv \hat{S}^\alpha_\mathbf{k}(t) - \left<\hat{S}^\alpha_\mathbf{k}\right>$ has been introduced to single-out inelastic contribution. The INS experiments are thus capable of probing directly the dissipative part of the dynamical magnetic susceptibility $\chi^{\alpha \alpha\prime\prime}(\mathbf{k}, \omega)$, multiplied by an additional factor term $1/(1 - \exp(-\beta \omega))$. The latter arises due to the so-called principle of detailed balance that relates the absorption and emission processes. 

\subsubsection{Resonant inelastic $x$-ray scattering}

Resonant inelastic $x$-ray scattering (RIXS) is a complementary experimental technique relying on electromagnetic radiation rather than neutrons to excite the system. The setup may be described as follows: The incoming $x$-ray excites an electron from one of the core levels to the valence shell. The outgoing photon is then produced by recombination of the valence electron with the unoccupied core-level state. Once the recombination has taken place, there is no hole in the core shell and hence the change in the photon energy is connected directly with excitations in the valence-electron sector.

The essential obstacle in applying this technique to map the magnetic dynamics is that photons do not cause direct single-spin-flip transitions that are associated with elementary magnetic excitations. Nonetheless, such processes might occur \emph{indirectly} due to the spin-orbit coupling in the intermediate state, yet selection rules for might prevent those transitions \cite{GrootPhysRevB1998}. In the context of cooper oxide superconductor,  it has been shown that  single-spin-flip processes become feasible if the spin moment is not orthogonal to the plane determined by $d_{x^2 - y^2}$ orbital \cite{AmentPhysRevLett2009}, as is the case for antiferromagnetically ordered cuprates, such as $\mathrm{La_{2}CuO_4}$. In effect, RIXS has then been used extensively to map magnetic and charge excitations for multiple copper-oxide, iron-pnicitide, and iridate compounds. One of the advantage of RIXS that has proven instrumental for the studies of correlated materials is its ability to probe high-energy excitations and access the Brillouin-zone region far away from the magnetic ordering wave vector. The INS scattering rate decreases with energy, making it less effective in this important regime. Since RIXS is based on a relatively complex process, derivation of closed formulas for scattering rate is more involved than those for INS \cite{HaverkortPhysRevLett2010}. A comparison between theoretically calculated dynamical susceptibilities with measured spectral intensities requires a requires thus considerable care \cite{TsutsuiPhysRevB2016,JiaPhysRevX2016}. 

\subsection{Collective excitations in strongly-correlated electron systems as seen in experiment}

\begin{figure}
    \centering
    \includegraphics[width=\textwidth]{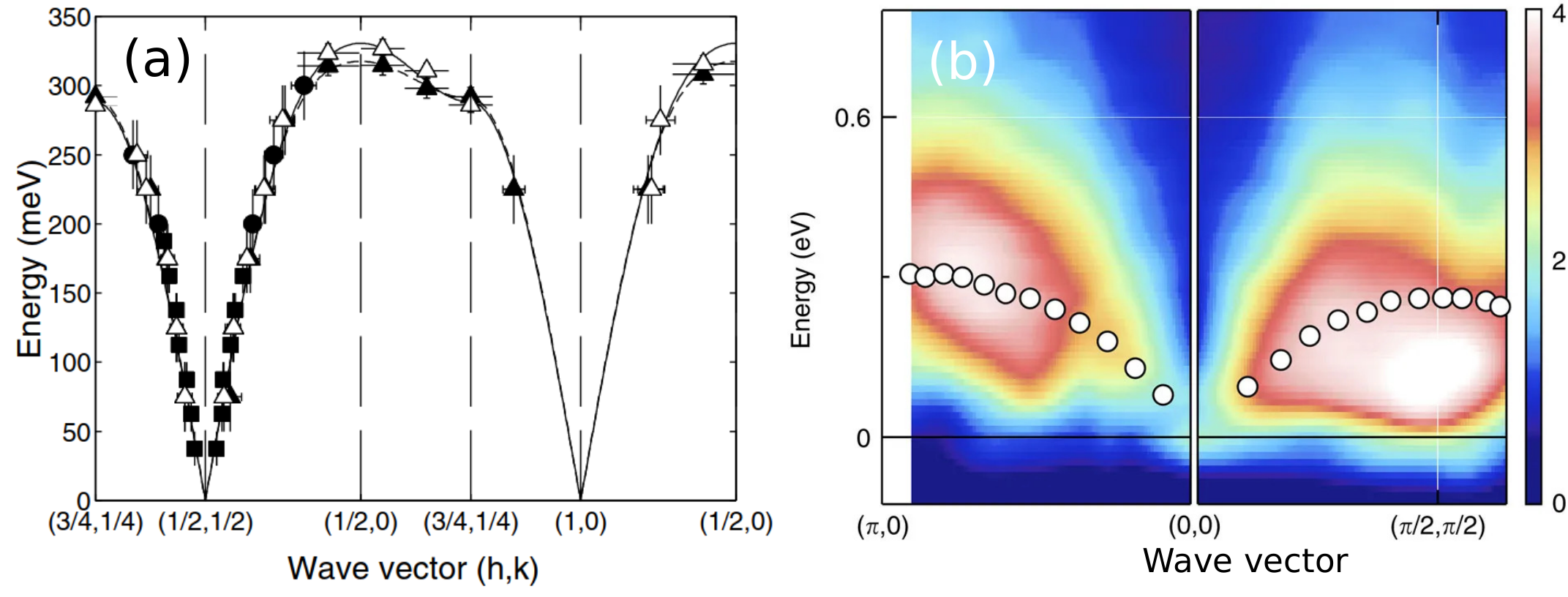}
    \caption{(a) High-resolution spin-wave energy across the Brillouin zone, obtained for insulating $\mathrm{La_2CuO_4}$ using inelastic neutron scattering technique. Open and solid symbols correspond to temperatures $T = 10\,\mathrm{K}$ and $T = 295\,\mathrm{K}$, respectively. Adapted with permission from Ref.~\cite{ColdeaPhysRevLett2001}. (b) RIXS intensity with subtracted elastic peak for underdoped $\mathrm{Bi_{1.6}Pb_{0.4}Sr_2Ca_{0.95}Y_{0.05}Cu_2O_{8+\delta}}$. The circles represent spin-wave dispersion for insulating $\mathrm{Bi_{1.6}Pb_{0.4}Sr_2YCu_2O_{8+\delta}}$. The spectra in (b) are taken for $20\,\mathrm{K}$. Adapted from Ref.~\cite{GuariseNatCommun2014}.}
    \label{fig:ins_rixs}
  \end{figure}

The INS approach is utilized extensively to map the magnetic excitation spectra in magnetic insulators, such as parent compounds of high-temperature superconductors. Exemplary low-temperature spectra of Ref.~\cite{ColdeaPhysRevLett2001} for antiferromagnetic $\mathrm{La_2CuO_4}$ are displayed in Fig.~\ref{fig:ins_rixs}(a). Near the antiferromagnetic zone center, $(h, k) = (\frac{1}{2}, \frac{1}{2})$, linear dispersion is clearly visible. From theory perspective, the character of those magnetic excitations is now well understood. This may be attributed to substantial simplifications resulting from freezing out the low-energy charge dynamics in the Mott insulating state, which allows for an effective spin-only description.

Much less is known about the character of the collective spin- and charge fluctuations in weakly- and moderately-doped systems. This problem is of particular interest due to its interrelation with unconventional states of matter emerging upon doping, such as pseudogap, high-temperature superconductivity, or spin/charge density wave phases. Already early neutron scattering experiments  revealed the presence of incommensurate spin excitations in underdoped and moderately doped the cuprates \cite{CheongPhysRevLett1991,MasonPhysRevLett1992,ThurstonPhysRevB1992}. Detailed analysis shows that those modes follow an uncommon ``\emph{hourglass}'' dispersion composed of both downwards and upwards dispersing branches \cite{AraiPhysRevLett1999,MatsudaPhysRevLett2008,StockPhysRevB2005,VignolleNatPhys2007}. The hourglass spectum is ubiquitous among the cuprates, but  has been also noted in other compounds \cite{BoothroydNature2011}. The theory of low-energy magnetic dynamics is still in the making, with proposals including static \cite{VojtaPhysRevLett2004,SeiboldPhysRevLett2005} and fluctuating stripes \cite{VojtaPhysRevLett2006}, as well as quantum criticality \cite{KharkovPhysRevB2019}.

Another aspect of doped copper oxide compounds concerns the high-energy part of the magnetic spectra and has come to light due to advancements in RIXS technique and its ability to efficiently map this energy sector. The doping evolution of the magnetic excitations is highly anisotropic and depends on Brillouin zone contour, but its overall features are remarkably universal among variety of cuprate families. Representative paramagnon energies along two Brillouin zone directions, extracted from RIXS data \cite{GuariseNatCommun2014}, are displayed in Fig.~\ref{fig:ins_rixs}(b). The points refer to the parent compound $\mathrm{Bi_{1.6}Pb_{0.4}Sr_2YCu_2O_{8+\delta}}$, whereas color map corresponds to the underdoped $\mathrm{Bi_{1.6}Pb_{0.4}Sr_2Ca_{0.95}Y_{0.05}Cu_2O_{8+\delta}}$. The maximum of intensity for the $\Gamma$-$X$ direction in the doped case follows closely that for the parent system, yet notable damping may be observed. Such a lack of softening is difficult to explain from the perspective of weak-coupling theory analysis that yields rapid Stoner overdamping of paramagnons in the metallic phase. The situation is quite different for the $\Gamma$-$M$ line, where substantial paramagnon softening is indeed observed in doped sample. Analogous anisotropic damping of spin modes is not unique to this compound and has been observed also in other copper oxides \cite{DeanNatMater2013,IshiiNatCommun2014,LeeNatPhys2014,PengNatPhys2017, WakimotoPhysRevB2015,IvashkoPhysRevB2017,ChaixPhysRevB2018,MinolaPhysRevLett2017,Robarts_arXiV_2019}, as well as in iron pnictides \cite{ZhouNatCommun2013} and iridates \cite{GretarssonPhysRevLett2016}. The ubiquitous presence of paramagnons as propagating modes has motivated revival of \cite{LeTaconNatPhys2011,JiaNatCommun2014,PengPhysRevB2018} of early ideas that they may contribute in a decisive manner to the pairing mechanism (cf., e.g., \cite{ScalapinoRevModPhys2012,ChubukovChapter2003}). This fluctuation-driven scenario is physically distinct from local pairing induced by local exchange interaction. Putting aside the paramagnons, RIXS has also provided access to dispersive low-energy charge modes in layered copper oxides \cite{IshiiPhysRevB2017,HeptingNature2018,IshiiPhysRevB2017,Lin_arXiV_2019,NagPhysRevLett2020}. The latter have been interpreted in terms of quantum critical, particle-hole-excitation-driven, and plasmon scenarios (cf. \cite{GrecoCommPhys2019}), out of which the latter stands now out as the most viable candidate due to their substantially three-dimensional character \cite{HeptingNature2018,NagPhysRevLett2020}.

Another class of materials, where collective mode fluctuations may play a notable role on par with local electronic correlations are uranium-based heavy-fermion compounds $\mathrm{UGe_{2}}$ \cite{SaxenaNature2000,TateiwaJPhysCondensMatter2001,PfleidererPhysRevLett2002,HuxleyPhysRevB2001}, UGhGe \cite{AokiNature2001}, UCoGe \cite{HuyPhysRevLett2007}, and UIr \cite{KobayashiPhysicaB2006}, where bulk superconductivity coexists with ferromagnetic ordering (cf. also Sec.~\ref{sec:related_correlated_systems}). This is a rare occurrence, as magnetism tends to suppress superconductivity in normal circumstances. Moreover, the situation here is particularly non-trivial as those uranium-based materials are inherently multi-orbital, allowing for complex order parameters, in close analogy to $A$, $A_1$, and $A_2$ states of superfluid ${}^3\mathrm{He}$. The origin of SC in these systems is not yet fully understood, with various pairing mechanisms proposed, including those based on critical spin fluctuations \cite{TadaJPhysConfSeries2013,WuNatCommun2017,HaradaPhysRevB2007,AokiJPSJ2014}. It should be, however, noted that the superconducting domes in uranium-based materials typically appear close to \emph{first order} magnetic or metamagnetic transitions (as indicated by a jump of the ordered moment) and are influenced by sizable uniaxial magnetic anisotropy, hence no critical magnetic critical point exists within the superconducting dome. This does not exclude the contribution of spin excitations to formation of Cooper pairs. Indeed, resistivity measurements suggest strongly enhanced spin fluctuations close to those discontinuous transitions in uranium materials, particularly in those exhibiting weak ferromagnetism, such as $\mathrm{UCoGe}$ \cite{HattoriPhysRevLett108,WuNatCommun2017,BastienPhysRevB2016} with ordered moment $\sim 0.05\,\mu_B$. This is consistent with soft magnetism and weak-first-order behavior, making spin fluctuations a viable pairing mechanism. On the other end, there exists a group of large-moment systems, represented by $\mathrm{UGe_2}$ (with moments $\sim 1\,\mu_B$ and $1.5\,\mu_B$ in FM1 and FM2 phases, respectively). The magnetization jump $\sim 0.5\,\mu_B$ near the metamagnetic transition, where superconducting dome occurs, is substantial and attributing of pairing a due to spin fluctuations is not as well-founded as it is for the case of, e.g., $\mathrm{UCoGe}$. The sequence of phases observed in $\mathrm{UGe_2}$, both as a function of pressure and applied magnetic field, has been  reproduced qualitatively within the Anderson-lattice-type model within statistically-consistent Gutzwiller method \cite{KadzielawaMajorPhysRevB2018,FidrysiakPhysRevB2019} that is based on local correlations, but those studies do not account for long-wavelength spin-fluctuation effects. Within the latter scheme, pairing is governed by the direct Hund's exchange between $f$-orbitals, combined with electronic correlations, cf. Sec.~\ref{subsec:fm_sc_coexistence_uge2} for details. Interpolating between these two regimes of small and large magnetic moments is challenging from theory perspective. To get a consistent picture of superconductivity and magnetism in the class of uranium superconductors, covering the whole spectrum of observed magnetic properties, a theoretical approach capable of including both local correlations and long-wavelength fluctuations near weak- first-order phase transitions seems thus necessary.

\subsection{Supplement: Local-moment magnetism from the spin-wave theory perspective}
\label{subsection:SWT}

One of the basic schemes to study magnetic excitations in local-moment insulating magnetic materials is spin-wave theory. This relatively simple approach allows to capture essential features of magnetic spectra for many magnetic materials and, with appropriate renormalization, constitutes a reliable tool for a quantitative analysis. In this subsection we summarize briefly selected spin-wave-theory predictions for the antiferromagnetic Heisenberg Hamiltonian

\begin{align}
  \label{eq:HeisenbergHamiltonian}
  \hat{\mathcal{H}} = J \sum\limits_{\langle i, j\rangle} \hat{\mathbf{S}}_i \cdot \hat{\mathbf{S}}_j,
\end{align}

\noindent
where $\langle i, j \rangle$ indicates summation over nearest-neighbor sites of the square lattice (each pair of indices is taken once), and $\hat{\mathbf{S}}_i$ denotes spin operator acting on site $i$. The model~\eqref{eq:HeisenbergHamiltonian} is regarded as a basis for more realistic models for the description of high-$T_c$ copper-oxide superconductors.

\subsubsection{Holstein-Primakoff spin-wave theory and $1/S$ expansion}

One of the variants of spin-wave theory is based on the Holstein-Primakoff \cite{HolsteinPhysRev1940} transformation

\begin{align}
  \label{eq:holstein_primakoff_representation}
  \left\{
  \begin{tabular}{l}
    $\hat{S}_i^{+} = \sqrt{2S - \hat{n}_i} \cdot \hat{b}_i$, \\
    $\hat{S}_i^{-} = \hat{b}_i^\dagger \cdot \sqrt{2S - \hat{n}_i}$, \\
    $\hat{S}_i^z = S - \hat{n}_i$,
  \end{tabular}
  \right.
\end{align}

\noindent
where $\hat{S}_i^\alpha$ are spin operators acting on site $i$, $\hat{S}_i^{\pm} \equiv \hat{S}_i^x \pm \hat{S}_i^y$, and $S$ denotes spin magnitude. Note that $\hbar$ is not included in the definition of spin operators. The bosonic creation and annihilation operators $\hat{b}_i$ and $\hat{b}_i^\dagger$ fulfill $\left[b_i, b_j^\dagger\right] = \delta_{ij}$ and $\left[b_i, b_j\right] = 0$. The boson number operator $\hat{n}_i \equiv \hat{b}_i^\dagger \hat{b}_i$ is also defined. The above commutation relations for $\hat{b}_i$ imply the algebra of spin operators $[\hat{S}_i^\alpha, \hat{S}_i^\beta] = i \epsilon_{\alpha \beta \gamma} \hat{S}_i^\gamma$. Here we focus on $S=1/2$ systems that emerge as effective models for the strong-coupling regime of the half-filled Hubbard model. The transformation \eqref{eq:holstein_primakoff_representation} is nonlinear due to the presence of particle number operators ($\hat{n}_i$) within the square root terms, which are thus expand in the power series in $\hat{n}_i/(2S)$ as

\begin{align}
  \label{eq:largeS_expansion}
  \hat{S}_i^{+} = \sqrt{2S} \hat{a}_i - \frac{1}{2\sqrt{2S}} \hat{n}_i \hat{a}_i + \text{[higher  order terms]}.
\end{align}

\noindent
The series~\eqref{eq:largeS_expansion} is a formal expansion in $1/(2S)$ so that the spin magnitude plays a role of the control parameter. The scheme should is thus applicable in the limit $\langle \hat{n}_i\rangle/(2S) \ll 1$.

Another feature of the mapping \eqref{eq:holstein_primakoff_representation} is that the solution should be sought in the restricted Hilbert space defined by the operator condition $0 \leq \hat{n}_i \leq 2S$. Those kind of inequality constraints are difficult to handle, and are ignored here. This is physically admissible if the spin fluctuation amplitude is small as compared to the spin, i.e., for $\langle \hat{n}_i\rangle/(2S) \ll 1$. More refined transformations, containing local  \emph{equality constraints} that may be imposed by means of gauge fields, have been also developed. We briefly comment on them here, since they are better suited for systems with large quantum fluctuations than the Holstein-Primakoff representation, thus providing insights into the limitations of the latter technique. Specifically, one can take

\begin{align}
  \label{eq:schwinger_boson_representation}
  \left\{
  \begin{tabular}{l}
    $\hat{S}_i^{+} = \hat{b}_{i\uparrow}^\dagger \hat{b}_{i\downarrow}$, \\
    $\hat{S}_i^{-} = \hat{b}_{i\downarrow}^\dagger \hat{b}_{i\uparrow}$, \\
    $\hat{S}_i^z = \frac{1}{2} (\hat{a}^\dagger_{i\uparrow} \hat{a}_{i\uparrow} - \hat{a}^\dagger_{i\downarrow} \hat{a}_{i\downarrow})$,
  \end{tabular}
  \right.
\end{align}

\noindent
with the equality constraint $\sum_\sigma \hat{b}_{i\sigma}^\dagger \hat{b}_{i\sigma} \equiv 2S$. In case $\hat{b}_{i\downarrow}$ ($\hat{b}_{i\downarrow}^\dagger$) fulfill bosonic commutations relations, this is the so-called Schwinger-boson representation \cite{AuerbachBook1994}. The square roots are now absent on the right-hand side of Eq.~\eqref{eq:schwinger_boson_representation}, which comes at a price of introducing two species of creation and annihilation operators ($\sigma = \pm 1$) instead of one. Remarkably, contrary to the Holstein-Primakoff formulation, in Eq.~\eqref{eq:schwinger_boson_representation} there no natural expansion parameter appears, and the problem needs to be treated in a non-perturbative manner. A common approach is based on artificially extending the number of $\hat{a}_{i\sigma}$ operators from one to $N$ and threading $1/N$ as a formal small parameter \cite{ReadPhysRevLett1989,AuerbachPhysRevB1991,AffleckPhysRevB1988}. The spin-symmetry group is enlarged from $SU(2)$ to $SU(N)$ accordingly. In order to reproduce correct spin commutation relations, $\hat{b}_{i\sigma}$ may be considered either as bosonic or fermionic (for $S=\frac{1}{2}$) annihilation operators. This statistical ambiguity leads to two distinct techniques, called Schwinger boson and slave fermion methods, respectively. The latter formulation is particularly suitable for studying the physics of spinons in the quantum spin liquids (see, e.g., \cite{HermelePhysRevB2008}). 

The parent compounds of high-temperature cuprate superconductors are antiferromagnets with ordered staggered magnetic moments $0.62 \pm 0.04 \mu_B$ \cite{ManousakisRevModPhys1991}, i.e., $\sim 60\%$ of the saturation value. The reduction due to quantum zero-point fluctuations is substantial, but the system remains still away from the quantum criticality. This is in contrast to other systems, such as a dimerized antiferromagnet $\mathrm{TlCuCl_3}$, that may be driven through a quantum critical point under applied pressure \cite{RueggPhysRevLett2008,MerchantNatPhys2014}. In the present case, the use of Holstein-Primakoff approach is thus admissible.

\subsubsection{Spin dynamics in a local-moment antiferromagnet}

We now briefly summarize basic linear spin-wave-theory predictions antiferromagnetic Heisenberg model with nearest-neighbor coupling $J_{ij} = J = 4 t^2 /U$ that comes out of the canonical transformation of the corresponding one-orbital Hubbard model at half-filling (cf. Appendix~\ref{appendix:derivation_of_the_tj_model}). The latter will serve us as a reference point for more sophisticated analysis below. Due to antiferromagnetic ordering, two sublattices emerge, say $A$ and $B$, defined as $A \equiv \{n_1 \mathbf{a}^\prime_1 + n_2 \mathbf{a}^\prime_2 : n_1, n_2 \in \mathbb{Z}\}$ and $B \equiv \{n_1 \mathbf{a}^\prime_1 + n_2 \mathbf{a}^\prime_2 + \boldsymbol{\delta} : n_1, n_2 \in \mathbb{Z}\}$, where $\mathbf{a}^\prime_1 \equiv (1, -1)$ and $\mathbf{a}^\prime_2 \equiv (1, 1)$ define the magnetic supercell (cf. Fig.~\ref{fig:metal-isulator-transition}(b)), where we set lattice spacing to one ($a = 1$). Enlargement of the unit cell leads to a reduction of the original to magnetic Brillouin zone (MBZ), as illustrated in Fig.~\ref{fig:hubbard_fs}. 

For each sublattice, Holstein-Primakoff transformation \eqref{eq:holstein_primakoff_representation} is carried out separately, but with a caveat that the ordered moment points in opposite direction in sublattices $A$ and $B$. This is achieved by rotating the spin by angle $\theta = \pi$ around $x$ axis in the $B$ sublattice only or, equivalently, by setting $\hat{S}^z_j = -S + \hat{b}_i^\dagger \hat{b}_i$, $\hat{S}_j^{+} = \sqrt{2S} \hat{b}^\dagger_j$, and $\hat{S}_j^{-} = \sqrt{2S} \hat{b}_j$ for $j \in B$. By including only the leading $O(S^2)$ and $O(S)$ terms, the Heisenberg Hamiltonian~\eqref{eq:HeisenbergHamiltonian} takes then the form

\begin{align}
  \label{eq:heisenberg_holstein_primakoff}
  \hat{\mathcal{H}} = - \frac{S^2 z J N}{2} + S z J \sum_\mathbf{k} \left[ \hat{a}_\mathbf{k}^\dagger \hat{a}_\mathbf{k} + \hat{b}_\mathbf{k}^\dagger \hat{b}_\mathbf{k} + \gamma_\mathbf{k} \left( \hat{a}_\mathbf{k}  \hat{b}_{-\mathbf{k}} + \hat{b}_\mathbf{-k}^\dagger \hat{a}_{\mathbf{k}}^\dagger \right) \right],
\end{align}

\noindent
where the wave vector summation is performed over MBZ, $z = 4$ is the number on nearest neighbors in square lattice (for a hypercubic lattice $z = 2 d$, where $d$ is spatial dimension), and $\gamma_\mathbf{k} \equiv \frac{2}{z} \sum_i \cos(k_i)$. This is a bosonic BCS-type Hamiltonian that is readily diagonalized by a Bogoliubov transformation

\begin{align}
  \label{eq:Bogoliubov_transformation}
  \left\{
  \begin{tabular}{l}
    $\hat{a}_\mathbf{k} \equiv \cosh(\theta_\mathbf{k}) \hat{\alpha}_\mathbf{k} - \sinh(\theta_\mathbf{k}) \hat{\beta}_{-\mathbf{k}}^\dagger$, \\
    $\hat{b}^\dagger_{-\mathbf{k}} \equiv -\sinh(\theta_\mathbf{k}) \hat{\alpha}_\mathbf{k} + \cosh(\theta_\mathbf{k}) \hat{\beta}_{-\mathbf{k}}^\dagger$,
  \end{tabular}
  \right.
\end{align}

\noindent
so that

\begin{align}
  \label{eq:heisenberg_holstein_primakoff_diagonalized}
  \hat{\mathcal{H}} = - \frac{S^2 z J N}{2} + \sum_\mathbf{k} \mathcal{E}_\mathbf{k} \left(\hat{\alpha}_\mathbf{k}^\dagger \hat{\alpha}_\mathbf{k} + \hat{\beta}_\mathbf{k}^\dagger \hat{\beta}_\mathbf{k} \right) - E_c,
\end{align}

\noindent
where $\mathcal{E}_\mathbf{k} = S z J \sqrt{1 - \gamma_\mathbf{k}^2}$ is the spin-wave dispersion, and $E_c = \frac{1}{2} JSz \left(1 - \sqrt{1 - \gamma_\mathbf{k}^2}\right) > 0$ is condensation energy. The first term on the right-hand-side, $-z S^2 J N / 2$,  is the classical energy contribution, evaluated as if $\hat{S}_i$ was a three-component classical vector of length $S$. The latter scales as $S^2$ and hence it constitutes dominant term in the large-$S$ series. In this sense, the $1/S$ expansion may be viewed as semi-classical approximation. The anomalous terms, $\hat{a}_\mathbf{k} \hat{b}_\mathbf{-k}$ and $\hat{b}_\mathbf{-k}^\dagger \hat{a}^\dagger_\mathbf{k}$, arise due to rotation of the spin in sublattice $B$. The Bogoliubov transformation coefficients are

\begin{align}
  \label{eq:Bogoliubov_coeffs}
  \left\{
  \begin{tabular}{l}
    $\cosh(\theta_\mathbf{k}) = \sqrt{\frac{1}{2} \left( \left(1 - \gamma_\mathbf{k}^2\right)^{-\frac{1}{2}} + 1\right)}$, \\
    $\sinh(\theta_\mathbf{k}) = \mathrm{sgn} \gamma_\mathbf{k} \cdot \sqrt{\frac{1}{2} \left( \left(1 - \gamma_\mathbf{k}^2\right)^{-\frac{1}{2}} - 1\right)}$.
  \end{tabular}
  \right.
\end{align}

\noindent
Close to the magnetic wave vector, $\mathbf{q}_\mathrm{AF} \equiv (\pi, \pi)$, the energy $\mathcal{E}_\mathbf{k}$ can be expanded as $\mathcal{E}_\mathbf{k} \approx c |\delta\mathbf{k}|$ with $\delta\mathbf{k} \equiv \mathbf{k} - \mathbf{q}_\mathrm{AF}$, i.e., the dispersion is linear and determined by spin-wave velocity $c \approx 2 S J \sqrt{d}$.

The imaginary part of the transverse magnetic susceptibility, given by Eq.~\eqref{pni_eq:real_time_susc}, can be readily evaluated. In the zero-temperature limit, one arrives at

\begin{align}
  \label{eq:transverse_susceptibility_swt}
  \chi^{+-}(\mathbf{k}, \omega) = S \pi \frac{1 - \gamma_\mathbf{k}}{\sqrt{1 - \gamma_\mathbf{k}^2}} \left[ \delta(\omega - \mathcal{E}_\mathbf{k}) - \delta(\omega + \mathcal{E}_\mathbf{k}) \right].
\end{align}

\noindent
Close to the magnetic ordering wave vector ($\mathbf{k} \approx \mathbf{q}_\mathrm{AF}$), Eq.~\eqref{eq:transverse_susceptibility_swt_low_energy} may be approximated as

\begin{align}
  \label{eq:transverse_susceptibility_swt_low_energy}
  \chi^{+-}(\mathbf{k}, \omega) \approx 4 d S^2 \pi \frac{J}{\mathcal{E}_\mathbf{k}} \left[ \delta(\omega - \mathcal{E}_\mathbf{k}) - \delta(\omega + \mathcal{E}_\mathbf{k}) \right],
\end{align}

\noindent
where $\mathcal{E}_\mathbf{k} \approx c |\delta \mathbf{k}|$ and $d = 2$ is spatial dimension. Equations~\eqref{eq:transverse_susceptibility_swt} and \eqref{eq:transverse_susceptibility_swt_low_energy} describe coherent and dispersive excitations, weighted by a wave-vector-dependent prefactor. The intensity becomes substantial at low energies close to $\mathbf{q}_\mathrm{AF}$ and diverges as $\mathcal{E}_{\mathbf{k}}^{-1}$. Such a scaling makes INS a suitable technique for studying low-energy excitations. On the other hand, the coherence factor $\frac{1 - \gamma_\mathbf{k}}{\sqrt{1 - \gamma_\mathbf{k}^2}}$ in Eq.~\eqref{eq:transverse_susceptibility_swt} is equal to zero for $\mathbf{k} = \mathbf{0}$, which indicates suppression of the INS intensity close to the $\Gamma$ point. Such a variation of intensities between $\mathbf{k}$-space regions, connected by $\mathbf{q}_\mathrm{AF}$, results from unfolding of the MBZ to the paramagnetic one.

Similar analysis can be carried out for the zero-temperature longitudinal spin susceptibility in the vicinity of $\mathbf{k} = \mathbf{q}_\mathrm{AF}$. One arrives at

\begin{align}
  \label{eq:longitudinal_susceptibility_swt}
  \chi^{zz}(\mathbf{k}, \omega) \propto 4 \pi S^2 d^2 J^2 \frac{1}{N} \sum_\mathbf{k} \frac{1}{\mathcal{E}_\mathbf{k} \mathcal{E}_{\mathbf{k}+\mathbf{p}}} \left[ \delta(\omega - \mathcal{E}_\mathbf{k} - \mathcal{E}_{\mathbf{k}+\mathbf{p}}) - \delta(\omega + \mathcal{E}_\mathbf{k} + \mathcal{E}_{\mathbf{k}+\mathbf{p}}) \right].
\end{align}

\noindent
Physically, the structure of Eq.~\eqref{eq:longitudinal_susceptibility_swt} is determined by two-magnon scattering processes \cite{HeilmannPhysRevB1981} that are also observed in neutron scattering experiments \cite{HubermanPhysRevB2005}. For a comprehensive discussion of those aspects and magnon decay processes the reader is referred to the topical review \cite{ZhitomirskyRevModPhys2013}. Finally, we remark  that $\chi^{+-}(\mathbf{k}, \omega)$ and $\chi^{zz}(\mathbf{k}, \omega)$ can be classified in the formal $1/S$ expansion as $O(S)$ and $O(1)$ terms, respectively. This follows from the observation that $\mathcal{E}_\mathbf{k}$ is $O(S)$. The longitudinal magnetic repose appears thus as a subleading effect in the spin-wave-theory analysis.

\subsubsection{Spin-waves in high-$T_c$ copper oxides}

  \begin{figure}
    \centering
    \includegraphics[width=0.6\textwidth]{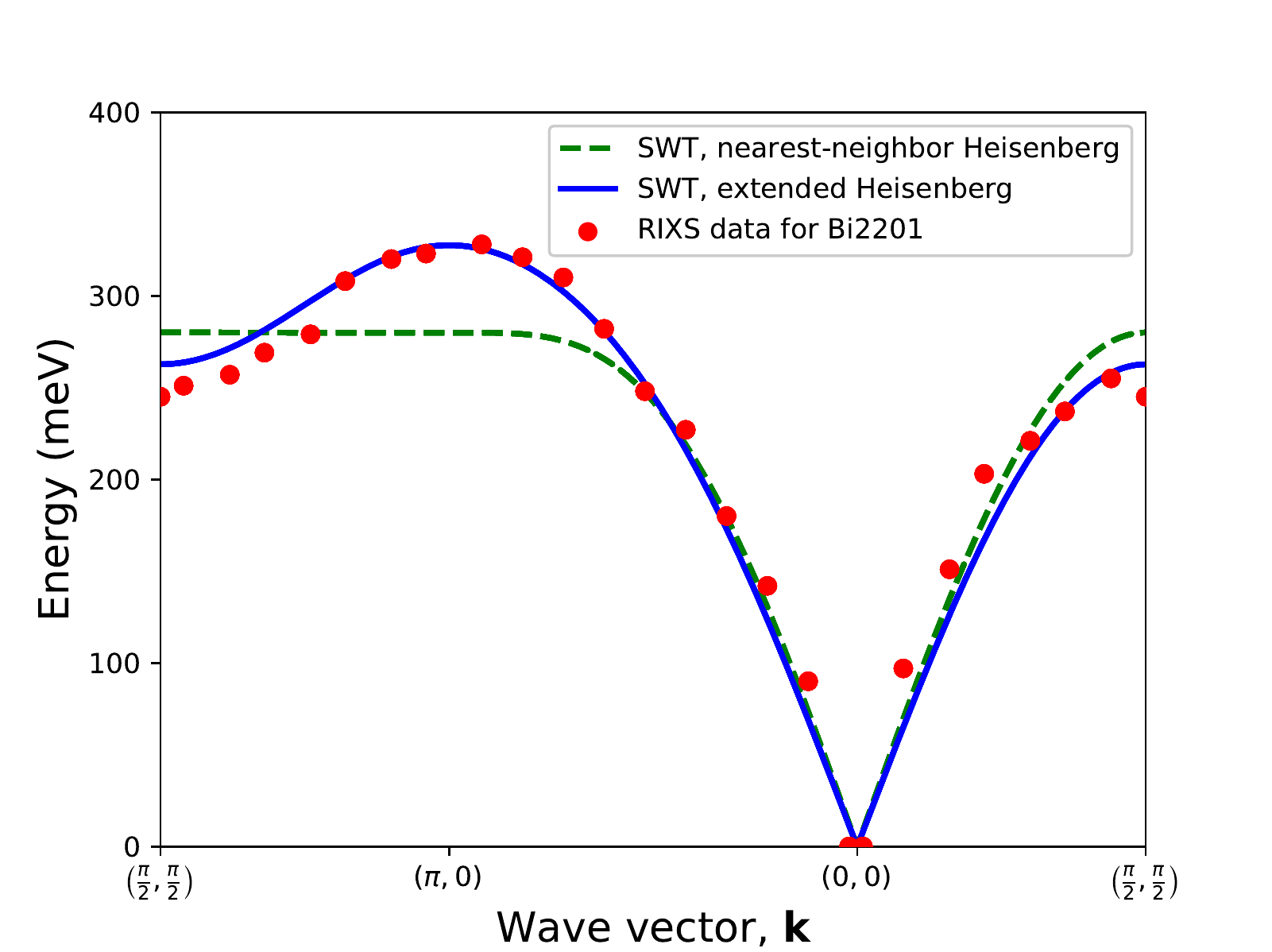}
    \caption{Experimental \cite{PengNatPhys2017} RIXS spin-wave energies for antiferromagnetic cuprate compound $\mathrm{Bi_2Sr_{0.9}La_{1.1}CuO_{6+\delta}}$ (red circles). The dashed green line is the linear spin-wave-theory (SWT) result for the plain Heisenberg model with nearest-neighbor exchange coupling  $J = 140\,\mathrm{meV}$, whereas the blue solid line represents similar calculation for extended model with ring exchange included \eqref{eq:ring_exchnage_hamiltonian} (model parameters are: $J = 140\,\mathrm{meV}$, $J^\prime = J^{\prime\prime} = 4.6\,\mathrm{meV}$, and $J_c = 91.6\,\mathrm{meV}$). In the latter case, the dispersion was scaled by standard correction $Z_c = 1.18$ \cite{SinghPhysRevB1989}.}
    \label{fig:swt_exp}
  \end{figure}

We are now in position to discuss how the predictions of  the Heisenberg model~\eqref{eq:HeisenbergHamiltonian} and relate it to experiments for high-$T_c$-cuprate parent compounds. In Fig.~\ref{fig:swt_exp}, the spin-wave dispersion for antiferromagnetically ordered $\mathrm{Bi_2Sr_{0.9}La_{1.1}CuO_{6+\delta}}$ (Bi2201), obtained using the RIXS technique \cite{PengNatPhys2017}, is displayed (red circles) and compared with that obtained from spin-wave theory calculation. The excitation dispersion is linear close to the $\Gamma$ point ($\mathbf{k} = (0, 0)$), in agreement with the Holstein-Primakoff spin-wave theory result of Eq.~\eqref{eq:transverse_susceptibility_swt_low_energy}. The dispersion, $\mathcal{E}_\mathbf{k} = SzJ\sqrt{1 - \gamma_\mathbf{k}^2}$, for $J = 140\,\mathrm{meV}$ is plotted in Fig.~\ref{fig:swt_exp} by green dashed line and indeed matches the data in the low-energy regime. At larger energies, the spin-wave-theory calculation does not follow the data. In particular, $\mathcal{E}_\mathbf{k}$ is constant along the magnetic Brillouin zone boundary ($(0.5\pi, 0.5\pi)$-$(\pi, 0)$ line), whereas the experimental points exhibits substantial dispersion along this line. This deficiency is not the fault of Holstein-Primakoff method itself, but it points towards the conclusion that the nearest-neighbor Heisenberg model is insufficient to describe spin waves in antiferromagnetic cuprates. Numerous extensions of the latter have been proposed over the years, including the model with extended exchange interactions

  \begin{align}
    \label{eq:ring_exchnage_hamiltonian}
\hat{\mathcal{H}}_c =& J \sum_{\langle i, j\rangle} \hat{\mathbf{S}}_i \cdot \hat{\mathbf{S}}_j + J^{\prime} \sum_{\langle i, j\rangle^\prime} \hat{\mathbf{S}}_i \cdot \hat{\mathbf{S}}_j + J^{\prime\prime} \sum_{\langle i, j\rangle^{\prime\prime}} \hat{\mathbf{S}}_i \cdot \hat{\mathbf{S}}_j \nonumber\\&+ J_c \sum_{\langle i, j, k, l\rangle} \left[ (\hat{\mathbf{S}}_i \cdot \hat{\mathbf{S}}_j) (\hat{\mathbf{S}}_k \cdot \hat{\mathbf{S}}_l) + (\hat{\mathbf{S}}_i \cdot \hat{\mathbf{S}}_l) (\hat{\mathbf{S}}_k \cdot \hat{\mathbf{S}}_j) - (\hat{\mathbf{S}}_i \cdot \hat{\mathbf{S}}_k) (\hat{\mathbf{S}}_j \cdot \hat{\mathbf{S}}_l) \right],
  \end{align}

  \noindent
  where $J$, $J^\prime$, and $J^{\prime\prime}$ refer to nearest-, next-nearest-, and next-next-nearest neighbor Heisenberg exchange, and the last term is the so-called cyclic (or ring) exchange, operating on four-site plaquettes of square lattice \cite{TakahashiJPhysC1977,RogerPhysRevB1989,MacDonaldPhysRevB1988,HeadingsPhysRevB2010,DelannoyPhysRevB2008}. Primes in summation indices indicate farther neighbors.

A general remark is in place here. Namely, extended models of this type are derived from Hubbard Hamiltonians by employing canonical perturbation expansion beyond the leading order \cite{ChaoPhysRevB1978}.  Application of linear spin-wave theory to the extended Hamiltonian yields modified dispersion, marked in Fig.~\ref{fig:swt_exp} by blue solid line, which matches experiment notably better than the Heisenberg-model result. The parameters used have been obtained in Ref.~\cite{PengNatPhys2017} by a least-squares fit to experimental data, yielding $J = 140\,\mathrm{meV}$, $J^\prime = J^{\prime\prime} = 4.6\,\mathrm{meV}$, and $J_c = 91.6\,\mathrm{meV}$ (additionally, the multiplicative correction $Z_c = 1.18$ has been applied to the spin-wave energies). Note that the ring exchange $J_c$ is the same order as the nearest-neighbor antiferromagentic coupling and thus cannot be neglected, even though formally $J$ and $J_c$ are $O(t^2/U)$ and $O(t^4/U^3)$, respectively. Detailed analysis shows \cite{DelannoyPhysRevB2008}, however, that $J_c = 80 t^4 / U ^3$ involves a large numeric prefactor that may compensate for the additional $t^2/U^2$ factor. The necessity to include those higher-order corrections into the description to get qualitative agreement with experimental data suggests that the localized model fails to incorporate some of the relevant aspects of antiferromagentic cuprates, even in the insulating state. This conjecture is further supported by observation that by restoring the electronic degrees of freedom, the spin wave dispersion can be accurately reproduced already within the one-band Hubbard model with only nearest-neighbor hopping included. This was explicitly demonstrated in Ref.~\cite{PeresPhysRevB2002} within random-phase approximation, starting from the spin-density-wave state. However, random-phase approximation does not provide a satisfactory equilibrium description of local electronic correlations and associated with them metal-insulator transition, yet it still may serve as a starting point for more accurate analysis. Below, we discuss a field-theoretical $1/\mathcal{N}_f$ technique that allows to systematically study corrections to the random-phase-approximation-type starting point. 

\subsection{A brief overview of collective excitations in itinerant-electron systems: Insights from $1/\mathcal{N}_f$ expansion with Hubbard-Stratonovich transform}
\label{sec:large_n_expansion}

The $1/\mathcal{N}_f$ expansion, with $\mathcal{N}_f$ being the number of fermionic species (flavors), is a useful tool to study collective excitations in fermionic systems in a controlled manner. The technique, in the leading expansion order, provides results comparable to those coming from random-phase-approximation, but allows to incorporate systematically the effects of quantum fluctuations at higher orders. Here we highlight main features of the approach by focusing on the canonical case of the one-band Hubbard model, described by the Hubbard Hamiltonian

\begin{align}
\hat{\mathcal{H}} =  -|t| \sum \limits_{\left<i, j\right>, \sigma} (\hat{a}_{i \sigma}^\dagger \hat{a}_{j \sigma} + \mathrm{H.c.})  + U \sum \limits_i \hat{n}_{i\uparrow} \hat{n}_{i\downarrow}, \label{pni_eq:Hubbard_model_Hamiltonian}
\end{align}

\noindent
controlled by the hopping integral $t$ and the on-site Coulomb repulsion $U$. In summation over pairs of nearest neighbors ($\langle ij\rangle$), we use the convention that each pair is counted only once (explicit Hermitian conjugate of the hopping term is thus needed). To make the problem analytically tractable, and to stay clear of instabilities to incommensurate spin- and charge-ordered phases \cite{IgoshevPhysRevB2010,IgoshevJPCM2015}, we restrict to the half-filled situation or, equivalently, doping $\delta = 0$. The bare electron energy dispersion takes then the form

\begin{align}
  \label{eq:hubbard_dispersion}
  \epsilon_\mathbf{k} \equiv -|t| \sum_{i(i)j} \mathrm{e}^{-\mathbf{k} ({\mathbf{r}_i - \mathbf{r}_j})}  = -2 |t| \left[\mathrm{cos}(k_x) + \mathrm{cos}(k_y)\right]
\end{align}

  \begin{figure}
    \centering
    \includegraphics[width=0.85\textwidth]{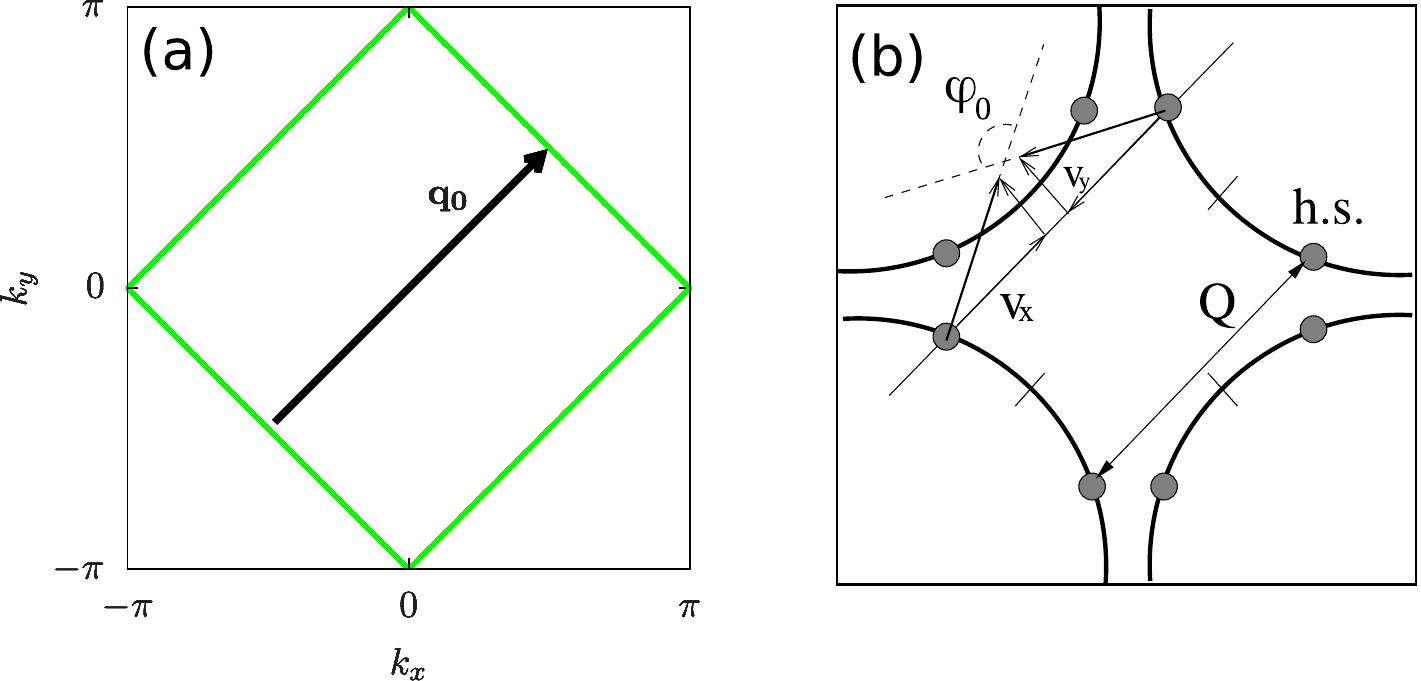}
    \caption{(a) Fermi surface of the Hubbard model with only nearest-neighbor hopping and at half-filling ($\delta = 0$). The nesting wave vector $\mathbf{q}_\mathrm{AF} = (\pi,  \pi)$ is displayed as black arrow. Fermi surface is transformed onto itself after shift by $\mathbf{q}_\mathrm{AF}$. After Ref.~\cite{FidrysiakPhDThesis2016}. (b) Illustration of the realistic Fermi surface for the cuprates. Even though no perfect nesting occurs in this case, the so called ``hot spots'' (h.s.) may be identified, which are separated by antiferromagnetic vector (here marked as $\mathbf{Q}$). After~\cite{AbanovAdvPhys2003}.}
    \label{fig:hubbard_fs}
  \end{figure}

\noindent
and Fermi surface at half-filling forms a square, as illustrated in Fig.~\ref{fig:hubbard_fs}(a). The bare Fermi surface transforms into itself after translation by the antiferromagnetic  wave vector $\mathbf{q}_\mathrm{AF} = (\pi, \pi)$. This property is sometimes referred to as perfect nesting, and it favors the instability to the commensurate antiferromagnetic state. The analysis for non-zero next-nearest-neighbor hopping is more challenging as the perfect nesting property is lost. Nonetheless, the so-called ``hot spots'' that are connected by antiferromagnetic vector may still be defined \cite{AbanovAdvPhys2003} (cf. Fig.~\ref{fig:hubbard_fs}(b)).

The starting point of the expansion is the path-integral representation of the generating functional

\begin{align}
Z[\mathbf{J}] = \int \mathcal{D} [\bar{\eta}, \eta] \exp\Bigg[-\mathcal{S}[\bar{\eta}, \eta] + \sum_i \mathbf{J}_i \cdot 
\left(\hat{\mathbf{S}}_i - \left<\hat{\mathbf{S}}_i\right>\right)\Bigg], \label{pni_eq:Hubbard_gen_functional}
\end{align} 

\noindent
where

\begin{align}
  \label{eq:original_action_large_n}
  \mathcal{S}[\bar{\eta}, \eta] = & \sum_i \int \limits_0^\beta d\tau \bar{\eta}_{i\sigma}(\tau) \partial_\tau \eta_{i\sigma}(\tau)  -|t| \sum_{\langle i, j\rangle} \int\limits_0^\beta d\tau \left(\bar{\eta}_{i\sigma}(\tau) \eta_{j\sigma}(\tau) + \bar{\eta}_{j\sigma}(\tau) \eta_{i\sigma}(\tau)\right) \nonumber \\ &+ U \sum_i \int \limits_0^\beta d\tau  \bar{\eta}_{i\uparrow}(\tau) \bar{\eta}_{i\downarrow}(\tau) \eta_{i\downarrow}(\tau) \eta_{i\uparrow}(\tau)    - \mu \sum_i \int \limits_0^\beta d\tau  \bar{\eta}_{i\sigma}(\tau) \eta_{i\sigma}(\tau)
\end{align}

\noindent
denotes Euclidean action that has been expressed in terms of anticommuting Grassman fields $\eta_{i\sigma}(\tau)$ and  $\bar{\eta}_{i\sigma}(\tau)$. The external currents, $\mathbf{J}_i$, have been introduced as a tool to generate spin- and charge correlation functions as derivatives of $Z[\mathbf{J}]$. The equilibrium expectation values has been subtracted from spin operators coupled to $\mathbf{J}_i$ in order to cancel out the Bragg-peak contribution and single-out dynamic susceptibilities. In particular, one obtains

\begin{align}
  \label{eq:generating_functional_differentiation}
 \langle T_\tau \delta \hat{S}_{i_1}^{\alpha_1}(\tau_1) \delta\hat{S}_{i_2}^{\alpha_2}(\tau_2) \ldots \delta\hat{S}_{i_n}^{\alpha_n}(\tau_n) \rangle = Z^{-1} \frac{\delta^n Z}{\delta J_{i_1}^{\alpha_1}(\tau_1) \ldots \delta J_{i_n}^{\alpha_n}(\tau_n)},
\end{align}

\noindent
where $T_\tau$ is time ordering operator and $\delta \hat{S}_i^\alpha(\tau) \equiv \hat{S}_i^\alpha(\tau) - \langle\hat{S}_i^\alpha\rangle$. 

By employing Hubbard-Stratonovich transformation, one can decouple the interaction part in Eq.~\eqref{eq:original_action_large_n} as follows

\begin{align}
  \label{eq:hs_decoupling}
\exp\Bigg(-U \int d\tau\sum_i  \bar{\eta}_{i\uparrow} \eta_{i\uparrow} \bar{\eta}_{i\downarrow} \eta_{i\downarrow}\Bigg) \propto &\int \mathcal{D} \phi^{(\alpha)} \exp\Bigg[ \int d\tau \sum_{i\mu} \Big(  U \bar{\eta}_i \sigma^\mu \eta_i \cdot \phi^\mu_i -  U (\phi^\mu_i)^2\Big)\Bigg],
\end{align}

\noindent
where $\mu = 0, 1, 2, 3$ and $\sigma^\mu$ are Pauli matrices ($\sigma_0$ denotes identity matrix), and $\phi_i^\mu$ are real fields. The decoupling \eqref{eq:hs_decoupling} is not the only one possible, which is the source of the so-called Fierz (or mean-field) ambiguity. Those aspects and their implications are discussed below. By substituting Eq.~\eqref{eq:hs_decoupling} into Eq.~\eqref{eq:original_action_large_n}, one obtains an equivalent representation of the generating functional

\begin{align}
Z[\mathbf{J}] = \int \mathcal{D} [\bar{\eta}, \eta] \mathcal{D} \phi \exp\left(\SumInt_{ij} d\tau \bar{\eta}_i [(-\partial_\tau + \mu)\delta_{ij} - t_{ij} + U \sigma^\mu \phi_i^\mu \delta_{ij}] \eta_j  - U \SumInt_i d\tau \phi_i^\mu \phi_i^\mu + \SumInt_i d\tau \mathbf{J}_i \cdot \delta \hat{\mathbf{S}}_i \right). \label{pni_eq:gen_functional_hubbardd_stratonovich}
\end{align} 

\noindent
To evaluate the path integral \eqref{pni_eq:gen_functional_hubbardd_stratonovich} in a controlled manner, we employ the $1/\mathcal{N}_f$ technique. As the first step , we create $\mathcal{N}_f$ copies of fermionic species as $\eta_{i\sigma} \rightarrow \eta_{i\sigma}^s$, where now $s = 1, \ldots, \mathcal{N}_f$. At the end of calculation, $\mathcal{N}_f$ should be set to unity in order to retrieve the original model, but $1/\mathcal{N}_f$ serves as a formal small parameter throughout the calculation. To ensure that the theory is non-trivial already in the large-$\mathcal{N}_f$ limit, the interaction needs to be scaled accordingly as $U \rightarrow U/N_f$. Moreover, since $\phi_i^\mu$ are dummy integration variables, it is useful to rescale them according to $\phi_i^\mu \rightarrow \mathcal{N}_f \phi_i^\mu$. By carrying out the above transformations and integrating over the Grassmann fields, one obtains

\begin{align}
Z[\mathbf{J}] = \int \mathcal{D} \phi \exp\Big(&\mathcal{N}_f \mathrm{Tr} \mathrm{log} [(-\partial_\tau + \mu)\delta_{ij} - t_{ij} + U \sigma^\mu \phi_i^\mu \delta_{ij} + S J_i^\alpha \sigma_i^\alpha \delta_{ij}]  - \nonumber \\ & \mathcal{N}_f U \SumInt_i d\tau \phi_i^\mu \phi_i^\mu - \mathcal{N}_f \SumInt_i d\tau \mathbf{J} \cdot \langle \hat{\mathbf{S}}_i\rangle \Big), \label{pni_eq:gen_functional_hubbardd_stratonovich_integrated}
\end{align} 

\noindent
where $S = \frac{1}{2}$ is the spin. Since all terms in the exponent are proportional to the flavor count $\mathcal{N}_f$, the integral \eqref{pni_eq:gen_functional_hubbardd_stratonovich_integrated} can be now expanded in a formal series in $1/\mathcal{N}_f$.  The large-$\mathcal{N}_f$ solution is found by locating the saddle-point of the integral with respect to the auxiliary fields $\phi_i^\alpha$ for external currents set to zero, $\mathbf{J}_i \equiv \mathbf{0}$. We thus split the Hubbard-Stratonovich field as $\phi_i^\mu \rightarrow \bar{\phi}_i^\mu + \delta\phi_i^\mu$ into its saddle-point value, $\bar{\phi}_i^\mu$, and perturbation, $\delta \phi_i^\mu$. Also, we introduce the non-interacting two-time fermionic Green's function $\hat{G}_{ij \sigma\sigma^\prime}(\tau, \tau^\prime) = -\langle T_\tau \hat{a}_{i\sigma}(\tau) \hat{a}^\dagger_{j\sigma}(\tau^\prime) \rangle_0$, where the subscript ``0'' means that the average is taken at the saddle point. By inspecting the quadratic part of Eq.~\eqref{pni_eq:gen_functional_hubbardd_stratonovich}, one arrives for the explicit formula for the Green's function inverse

\begin{align}
  \label{eq:green_function_hubbard}
  \hat{G}^{-1}_{ij}(\tau, \tau^\prime) \equiv -\partial_\tau(\tau-\tau^\prime) \delta_{ij} + \left(\mu \delta_{ij} - t_{ij} + U \sigma^\mu \bar{\phi}_i^\mu \delta_{ij}\right) \delta(\tau-\tau^\prime),
\end{align}

\noindent
where the spin indices have been dropped in favor of $2 \times 2$ matrix notation in terms of Pauli matrices $\sigma^\mu$. The Fourier transform of Eq.~\eqref{eq:green_function_hubbard} is

\begin{align}
  \label{eq:greens_function_fourier_transform}
  \hat{G}^{-1}_{\mathbf{K} \mathbf{K}^\prime} = \sum_{ij} \int d\tau d\tau^\prime  \mathrm{e}^{i\omega_n \tau - i \omega_{n^\prime} \tau^\prime - i \mathbf{k} r_i + i \mathbf{k}^\prime r_j} \hat{G}^{-1}_{ij}(\tau, \tau^\prime) = (i \omega_n + \mu - \epsilon_\mathbf{k}) \bar{\delta}_{\mathbf{K}\mathbf{K}^\prime} + U \sigma^\mu \bar{\phi}_{\mathbf{K} - \mathbf{K}^\prime}^\mu,
\end{align}

\noindent
where $\mathbf{K} \equiv (\mathbf{k}, \omega_n)$, $\mathbf{K}^\prime \equiv (\mathbf{k}^\prime, \omega_{n^\prime})$ are Euclidean four-vectors and

\begin{align}
  \label{eq:fourier_transform_hs_field}
  \bar{\phi}_\mathbf{K}^\alpha \equiv \sum_i \int d\tau \mathrm{e}^{-i \mathbf{k} \mathbf{r}_i + i\omega\tau} \bar{\phi}_i^\alpha(\tau).
\end{align}

\noindent
We have defined $\bar{\delta}_{\mathbf{K}\mathbf{K}^\prime} \equiv \beta N \delta_{\mathbf{k}\mathbf{k}^\prime} \delta_{n n^\prime}$. The saddle-point condition generates the set of self-consistent equations

\begin{align}
  \label{eq:saddle_point_hubbard}
  \bar{\phi}_\mathbf{P}^\mu = \frac{1}{2 \beta N} \sum_{\mathbf{K}}  \mathrm{tr} \hat{G}_{\mathbf{K}, \mathbf{K} - \mathbf{P}} \sigma^\mu
\end{align}

\noindent
that need to be solved numerically. The lowercase trace (tr) acts only on the spin and orbital indices.

To study antiferromagnetic solution, we make an ansatz 

\begin{align}
  \label{eq:af_order_parameter}
  U \bar{\phi}_{\mathbf{K}}^\mu = \delta_{\mu 3} \Delta \bar{\delta}_{\mathbf{K}, \mathbf{q}_\mathrm{AF}} + \delta_{\mu 0} (\delta\mu) \bar{\delta}_{\mathbf{K}, \mathbf{0}},
\end{align}

\noindent
where $\delta \mu$ is the constant energy shift that can be incorporated into the chemical potential  $\mu \rightarrow \bar{\mu} \equiv \mu + \delta\mu$, and $\Delta$ is the order parameter with dimension of energy. With the use of the Kronecker symbol, $\delta_{\mu 3}$, the staggered magnetization has been oriented along the $z$-axis. Importantly, due to doubling of the unit cell in the N\'{e}el state, magnetic ordering generates umklapp terms that are reflected by non-zero off-diagonal matrix elements in the Green's function $\hat{G}_{\mathbf{K}\mathbf{K}^\prime}$. The latter becomes thus a $4 \times 4$ matrix

\begin{align}
  \label{eq:green_function_inverse}
  \hat{G}^{-1} =
  \left(
  \begin{tabular}{cc}
    $(i\omega_n - \epsilon_\mathbf{k} + \bar{\mu}) \sigma^0$ & $\Delta \sigma^3$  \\
    $\Delta \sigma^3$ & $(i\omega_n - \epsilon_{\mathbf{k}+\mathbf{q}_\mathrm{AF}} + \bar{\mu})\sigma^0$
  \end{tabular}
\right)
\end{align}

\noindent
with two states originating from spin-up and spin-down $z$-axis spin projections, and further doubled by emergence of the magnetic Brillouin zone. The Green's function can be evaluated as

\begin{align}
\hat{G}_{\mathbf{K} \alpha, \mathbf{K}' \beta} = g_\mathbf{0}(\mathbf{K}) \bar{\delta}_{\mathbf{K}, \mathbf{K}'} \delta_{\alpha \beta}+ g_{\mathbf{q}_\mathrm{AF}}(\mathbf{K}) \bar{\delta}_{\mathbf{K}, \mathbf{K}' - \mathbf{q}_\mathrm{AF}} \sigma^3_{\alpha \beta},
\end{align}

\noindent
where

\begin{align}
  g_\mathbf{0}(\mathbf{K}) &=\frac{i\omega_n - \epsilon_{\mathbf{k}+\mathbf{q}_\mathrm{AF}} + \bar{\mu}}{(i \omega_n + \bar{\mu} - \epsilon_\mathbf{k})(i \omega_n + \bar{\mu} - \epsilon_{\mathbf{k}+\mathbf{q}_\mathrm{AF}}) - \Delta^2},
\end{align}

\noindent
and

\begin{align}
g_{\mathbf{q}_\mathrm{AF}}(\mathbf{K}) &= \frac{-\Delta}{(i \omega_n + \bar{\mu} - \epsilon_\mathbf{k})(i \omega_n + \bar{\mu} - \epsilon_{\mathbf{k}+\mathbf{q}_\mathrm{AF}}) - \Delta^2}.
\end{align}

\noindent
The two above functions are further simplified if one explicitly makes use of the perfect nesting condition $\epsilon_\mathbf{k} = - \epsilon_{\mathbf{k} + \mathbf{q}_\mathrm{AF}}$, which results in

\begin{align}
g_\mathbf{0}(\mathbf{K}) &=\frac{i\omega_n + \epsilon_{\mathbf{k}} + \bar{\mu}}{(i \omega_n + \bar{\mu})^2 - E_\mathbf{k}^2}, \\
g_{\mathbf{q}_\mathrm{AF}}(\mathbf{K}) &= \frac{-\Delta}{(i \omega_n + \bar{\mu})^2 - E_\mathbf{k}^2}.
\end{align}

\noindent
In the above equations $E_\mathbf{k} \equiv \sqrt{\epsilon_\mathbf{k}^2 + \Delta^2}$, i.e., The quasiparticles acquire non-zero energy gap $\Delta$. Equation~\eqref{eq:saddle_point_hubbard} leads then to system of three mutually consistent equations

\begin{align}
    \Delta =& \frac{U \Delta}{\beta N} \sum_\mathbf{K} \frac{1}{-(i \omega_n + \bar{\mu})^2 + E_\mathbf{k}^2},   \label{eq:saddle_point_equations_hubbard} \\
    \delta \mu =& U \frac{n_e}{2}, \\
    n_e = & \frac{1}{N} \sum_\mathbf{k} n_F(E_\mathbf{k} - \bar{\mu}) + \frac{1}{N}\sum_\mathbf{k} n_F(-E_\mathbf{k} - \bar{\mu}),   \label{eq:saddle_point_ne}
\end{align}

\noindent
where $n_F(E) \equiv 1/(\mathrm{exp}(\beta E) + 1)$ with $\beta \equiv 1/(k_B T)$ is the Fermi function and target total electronic density $n_e$ is evaluated per flavor (i.e. normalized by $\mathcal{N}_f$) so that it is well-defined in the large-$\mathcal{N}_f$ limit. From the particle-hole symmetry of the model and Eq.~\eqref{eq:saddle_point_ne}, it follows that the shifted chemical potential $\bar{\mu} = 0$ at the half-filling ($n_e = 1$). Note that, from the above equations, one obtains $\mu < 0$ as long as $U > 0$. This is \emph{inconsistent} with the Hartree-Fock analysis and may be understood as an artifact of the particular Hubbard-Stratonovich decoupling scheme, employed in the calculation. For the same reasons, charge fluctuations are also described incorrectly (cf., e.g., Ref.~\cite{PhysRevB43John} for discussion of those aspects). As we demonstrate below, such inconsistencies are common within $1/\mathcal{N}_f$ approach. On the other hand, the saddle-point static and dynamic spin properties remain in agreement with random-phase-approximation analysis. Indeed, assuming non-trivial state ($\Delta \neq 0$) and taking $\bar{\mu} = 0$, Eq.~\eqref{eq:saddle_point_equations_hubbard} can be recast in the usual antiferromagnetic gap equation form 

\begin{align}
  \label{eq:af_order_parameter_hubbard}
  1 = \frac{U}{N} \sum_\mathbf{k} \frac{1 - 2 n_F(E_\mathbf{k})}{2 E_\mathbf{k}} \overset{T \rightarrow 0}{\rightarrow} \frac{U}{N} \sum_\mathbf{k} \frac{1}{2 E_\mathbf{k}}.
\end{align}

\noindent
As the final step, we perform the shift $\delta\phi^\alpha_\mathbf{K} \leftrightarrow \delta\phi^\alpha_\mathbf{K} - S /U \cdot J^\alpha_\mathbf{K}$, which eliminates external currents from the trace term. The resultant generating functional is $Z = \int \mathrm{exp}(-\mathcal{S}_\mathrm{eff})$, with effective action

\begin{align}
\mathcal{S}_\mathrm{eff} = & - \mathrm{Tr} \ln (\hat{G}^{-1} + U \hat{F}) + 2 \Delta  \delta\phi^3_{\mathbf{q}_\mathrm{AF}} + 2 \delta \mu  \delta\phi^0_{\mathbf{0}} + \frac{\beta N (\Delta^2 + \delta\mu^2)}{U} + U  \bar {\sum \limits_\mathbf{K}}  \delta\phi^\alpha_\mathbf{K} \delta\phi^\alpha_\mathbf{-K}  + \nonumber \\ 
& + \bar{\sum \limits_\mathbf{K}} (- 2  S  \delta\phi^\alpha_\mathbf{K} J^\alpha_{-\mathbf{K}} + \frac{ S^2}{U}  J_{\mathbf{K}}^\alpha J_{-\mathbf{K}}^\alpha), \label{pni_eq:ActionHubbard}
\end{align}

\noindent
In equation~(\ref{pni_eq:ActionHubbard}) $\hat{F}_{\mathbf{K} \mathbf{K}^\prime} = \delta\phi^\alpha_{\mathbf{K}-\mathbf{K'}} \sigma^\alpha$. The diagonal component of the dynamical spin susceptibility can be now obtained by differentiation of $Z[\mathbf{J}]$ over background currents $\mathbf{J}_\mathbf{P}$, namely

\begin{align}
\chi_\mathbf{0}^{\alpha \beta}(i \omega_n, \mathbf{p}) = \frac{\mathcal{N}}{T Z[0]}  
\frac{\delta^2 Z[J]}{\delta J_{(i\omega_n, \mathbf{p})}^\alpha \delta J_{(-i 
\omega_n, -\mathbf{p})}^\beta}. \label{eq:Hubbard_susceptibility_currents}
\end{align}

\noindent
A direct implication of Eq.~(\ref{pni_eq:ActionHubbard}) is that the dynamical properties of spin and auxiliary fields $\phi^\alpha_\mathbf{K}$ are closely related, but no quite equivalent. Indeed, the structure of the coupling between $\delta\phi^\alpha_\mathbf{K}$ and the background currents $J_{\mathbf{K}}^\alpha$ allows to express the spin and auxiliary field correlation functions as

\begin{align}
\frac{\left< \delta \hat{S}^\alpha_{\mathbf{P}_1} \delta \hat{S}^\beta_{\mathbf{P}_2} \right>}{4 S^2 } =   \left< \delta\phi^\alpha_{\mathbf{P}_1} \delta\phi^\beta_{\mathbf{P}_2} \right> - \frac{1}{2  U} \delta_{\alpha \beta} \bar{\delta}_{\mathbf{P}_1 + \mathbf{P}_2}.\label{pni_eq:spin_aux_relation_Hubbard}
\end{align}

\noindent
The technical part is now to expand Eq.~\eqref{pni_eq:ActionHubbard} to quadratic terms in Hubbard-Stratonovich fields and evaluate resultant Gaussian path integrals. We skip this part of the analysis here and discuss the results below.

\subsubsection{Transverse spin susceptibility}

  \begin{figure}
    \centering
    \includegraphics[width=0.8\textwidth]{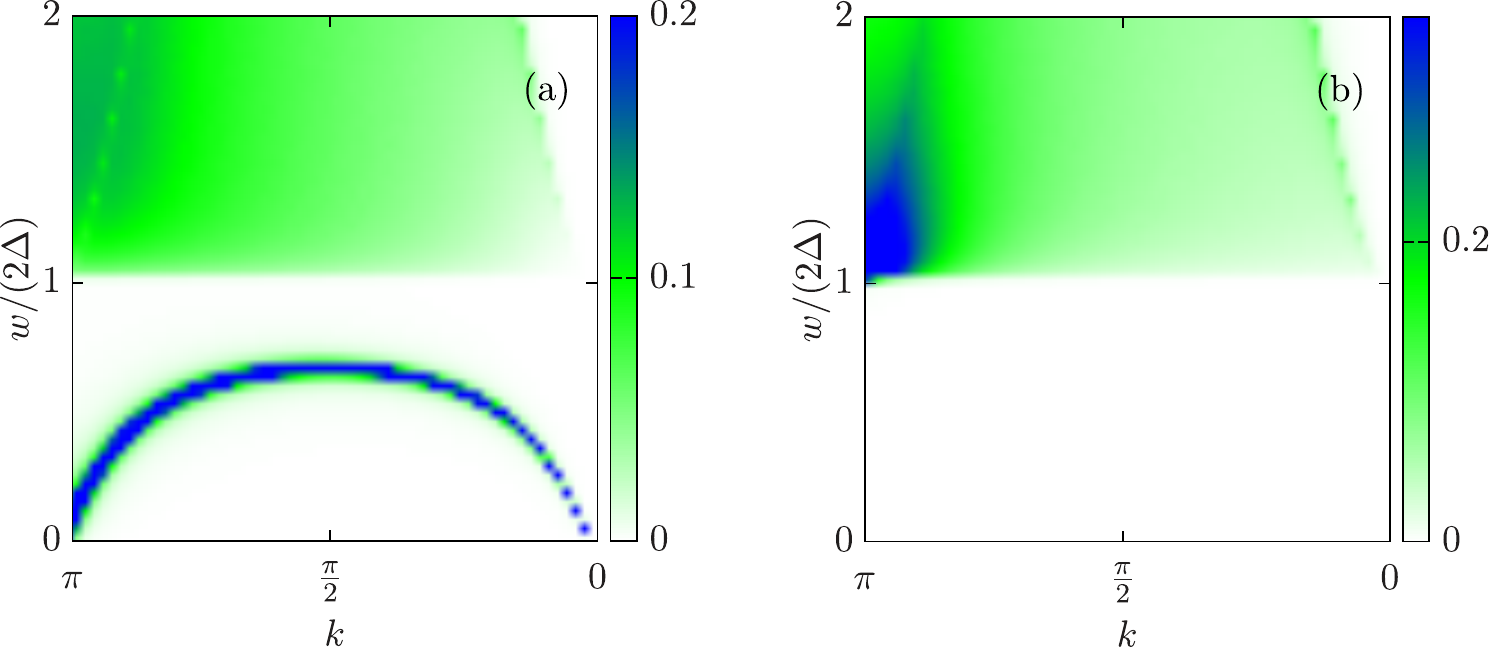}
    \caption{Imaginary part of the (a) transverse $t \cdot 
\mathrm{Im} \chi_\mathbf{0}^{+-}(\omega, \mathbf{k})$ and (b) longitudinal $t 
\cdot \mathrm{Im} \chi_\mathbf{0}^{33}(\omega, \mathbf{k})$ dynamical spin susceptibilities at large-$\mathcal{N}_f$ along the direction $\mathbf{k} = (k, k)$. The model parameters are for $\Delta = 0.5 |t|$ ($|t|/U \approx 
0.44$). Wave vector summations in the large-$\mathcal{N}_{f}$ expressions (\ref{pni_eq:RPA_transverse_susc_Hubbard}) and 
(\ref{pni_eq:chi_zz_Hubbard}) have been performed on a $5000 \times 5000$  $\mathbf{k}$-space mesh.  A small imaginary energy shift $\omega \rightarrow \omega + i \cdot 0.001 |t|$ has been introduced for numerical purposes. Adapted from Ref.~\cite{FidrysiakPhDThesis2016}.}
    \label{fig:hubbard_susc}
  \end{figure}

The leading contribution to the transverse spin susceptibility appears already in the large-$N_f$ limit and is given by\footnote{In this subsection, we define $\hat{S}^{\pm} \equiv (\hat{S}^1 \pm i \hat{S}^2)/\sqrt{2}$.}

\begin{align}
\chi^{+ -}_\mathbf{0}(\mathbf{P}) = &\frac{2 S^2 N_f U^{-1} \cdot [1 + U \chi^0_\mathbf{0}(\mathbf{P} + \mathbf{Q_\mathrm{AF}})]}{[1 + U \chi^0_\mathbf{0}(\mathbf{P})] [1 + U \chi^0_\mathbf{0}(\mathbf{P} + \mathbf{Q_\mathrm{AF}})] - U^2 [\chi^0_\mathbf{Q_\mathrm{AF}}(\mathbf{P})]^2} - \frac{2 S^2 N_f}{U}, \label{pni_eq:RPA_transverse_susc_Hubbard}
\end{align}

\noindent
where (for $T \rightarrow 0$)

\begin{align}
\chi^0_\mathbf{0}(\mathbf{P}) = & \bar{\sum \limits_\mathbf{k}} \frac{E_\mathbf{k} + E_\mathbf{k + p}}{2 E_\mathbf{k} E_\mathbf{k + p}} \cdot \frac{- \Delta^2 -  E_\mathbf{k} E_\mathbf{k + p} +  \epsilon_\mathbf{k} \epsilon_\mathbf{k + p}}{\omega_n^2 + (E_\mathbf{k} + E_\mathbf{k + p})^2} \label{pni_eq:RPA_chi00}
\end{align}

\noindent
and

\begin{align}
\chi^0_{\mathbf{q}_\mathrm{AF}}(\mathbf{P}) = & \bar{ \sum \limits_\mathbf{k}} \frac{i \omega_n \Delta (E_\mathbf{k} + E_\mathbf{k + p})}{2 E_\mathbf{k} E_\mathbf{k + p}} \cdot \frac{1}{\omega_n^2 + (E_\mathbf{k} + E_\mathbf{k + p})^2}. \label{pni_eq:RPA_chi0Q0}
\end{align}

\noindent
Equations~(\ref{pni_eq:RPA_transverse_susc_Hubbard})-(\ref{pni_eq:RPA_chi0Q0}) are equivalent to the \emph{random-phase-approximation} (RPA) expressions which represent the lowest-order dynamics corrections to the Hartree-Fock saddle-point solution. The continuum for $\omega > 2\Delta$ represents electron-hole excitations across the antiferromagnetic gap, with the deep blue part in (b) corresponding to damped longitudinal (Higgs-type) magnetic mode. It can be verified that Eq.~\eqref{pni_eq:RPA_transverse_susc_Hubbard} properly describes Goldstone modes in the broken symmetry phase. Indeed, the denominator of this expression vanishes for both $\mathbf{p} = \mathbf{0}$ and $\mathbf{p} = \mathbf{q}_\mathrm{AF}$ if one makes use of the gap equation \eqref{eq:af_order_parameter_hubbard}. The imaginary part of transverse susceptibility is displayed in Fig.~\ref{fig:hubbard_susc}(a). Two distinctive features of the magnetic response should be noted. First, at low energies, a dispersive feature that corresponds to spin-wave excitation appears. Here, contrary to the spin-wave theory, where spin-waves are elementary excitations, they arise due to resonant scattering of particle-hole pairs.  Second, at high energies (above twice the spin-density-wave gap, $2\Delta$), incoherent \emph{particle-hole continuum} emerges. Within the fermionic model one observes thus richer behavior that in the effective local-spin description.

It is now instructive to look closely at the low-energy structure of the spin-wave modes, as obtained within the Hubbard model (cf. Eq.~\eqref{pni_eq:RPA_transverse_susc_Hubbard}). For energies small compared to the spin-density-wave gap ($|\omega| \ll 2 \Delta$), we obtain an asymptotic formula

\begin{align}
\mathrm{Im} \chi^{+-}_\mathbf{0}(\omega + i\epsilon, \mathbf{p}) \approx \pi \mathcal{N}_{f}\mathcal{R}_{\mathrm{1M}} \cdot 2 d S^2 J_\mathrm{eff} \frac{1}{\mathcal{E}_{\delta\mathbf{p}}} \cdot [\delta(\omega - \mathcal{E}_{\delta\mathbf{p}}) - \delta(\omega + \mathcal{E}_{\delta\mathbf{p}})] \label{pnieq:Hubbard_low_energy_transverse_susceptibility}
\end{align}

\noindent
which takes the same functional form as the spin-wave-theory result for a square-lattice Heisenberg antiferromagnet in two-dimensions ($d=2$), but with renormalized exchange coupling

\begin{align}
J_\mathrm{eff}^2 \equiv \frac{U \xi t^2 \gamma_0}{4 S^2 x d}
\label{eq:effective_exchange_large_n}  
\end{align}

\noindent
The latter is  related  to the low-energy spin-wave dispersion as $\mathcal{E}_{\delta\mathbf{p}} \approx \sqrt{2} J_\mathrm{eff} |\delta \mathbf{p}|$, where $\delta \mathbf{p} \equiv \mathbf{p} - \mathbf{q}_\mathrm{AF}$. Moreover, there appears an overall prefactor

\begin{align}
  \mathcal{R}_{\mathrm{1M}} =  \frac{2 S^2 J_\mathrm{eff}}{U^2 \gamma_0 t^2}
  \label{eq:eq:one-magnon_ren_factor}
\end{align}

\noindent
which is responsible for renormalization of the spin-wave spectral weight. Such spectral weight redistribution is expected as, in the itinerant-electron case, it is split up between coherent and incoherent excitations. To write down those quantities in a compact form, auxiliary parameters $x \equiv \frac{1}{2N} \sum_\mathbf{k} E_\mathbf{k}^{-3}$, $\xi \equiv 4 \left(1/U - \Delta^2 x  \right)$ and

\begin{align}
\gamma_0 \equiv & \frac{1}{N} {\sum \limits_\mathbf{k}} \left[ \frac{-\cos^2(k_x) - \cos(k_x) \cos(k_y)}{2 E_\mathbf{k}^3}  + \frac{- \epsilon_\mathbf{k}^2 + 2 \Delta^2}{2 E_\mathbf{k}^5} \sin^2(k_x)\right]. 
\end{align}

\noindent
have been defined \cite{BrydonPhysRevB2009}. Note that Eq.~\eqref{pnieq:Hubbard_low_energy_transverse_susceptibility} is equivalent to the spin-wave theory result, given by Eq.~\eqref{eq:transverse_susceptibility_swt_low_energy}, except for a multiplicative factor of two which results from different definition of $\hat{S}^{\pm}$ in the present subsection for more convenient handling of relevant path integrals ($\hat{S}_i^{\pm} = (\hat{S}_i^x \pm i \hat{S}_i^y)/\sqrt{2}$ rather than $\hat{S}_i^{\pm} = (\hat{S}_i^x \pm i \hat{S}_i^y)$). The role of $J$ is taken over by effective exchange, $J_\textrm{eff}$, which now acquires substantially more complex dependence on $U$ and $t$ than in the Heisenberg model limit.

\subsubsection{Longitudinal spin susceptibility}

The longitudinal susceptibility in the large-$N_f$ limit reads 

\begin{align}
\chi_\mathbf{0}^{33}(\mathbf{P}) = \frac{2 S^2 N_{f}}{U} \frac{1}{1 + U \chi_{L}^0(\mathbf{P})} - \frac{2 S^2 N_{f}}{U}, \label{pni_eq:chi_zz_Hubbard}
\end{align}

\noindent
with

\begin{align}
\chi^0_{L}(\mathbf{P}) = &  \bar{\sum \limits_\mathbf{k}} \frac{E_\mathbf{k} + E_\mathbf{k + p}}{2 E_\mathbf{k} E_\mathbf{k + p}} \cdot \frac{\Delta^2 -  E_\mathbf{k} E_\mathbf{k + p} +  \epsilon_\mathbf{k} \epsilon_\mathbf{k + p}}{\omega_n^2 + (E_\mathbf{k} + E_\mathbf{k + p})^2}. \label{pni_eq:chi_0L}
\end{align}

  \begin{figure}
    \centering
    \includegraphics[width=0.45\textwidth]{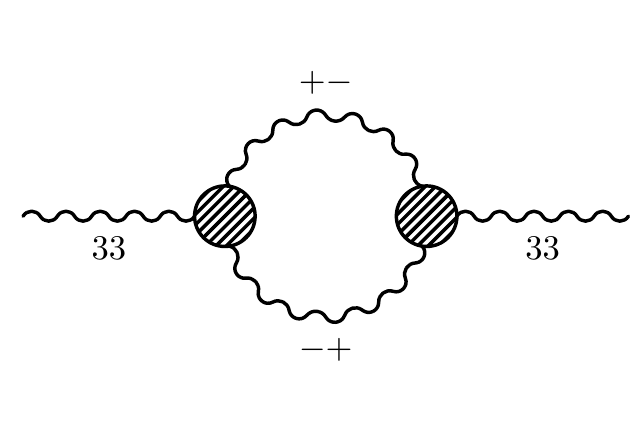}
    \caption{Diagram representing dominant low-energy contribution to the low energy longitudinal magnetic response. This is an example of a subleading term in the $1/\mathcal{N}_f$ expansion scheme. Adapted from Ref.~\cite{FidrysiakPhDThesis2016}.}
    \label{fig:hubb_diagram_2m}
  \end{figure}

\noindent
Equations~(\ref{pni_eq:chi_zz_Hubbard})-(\ref{pni_eq:chi_0L}), once again, take the from equivalent to the RPA result. The imaginary part of the longitudinal response is displayed in Fig.~\ref{fig:hubbard_susc}(b). The coherent peak in now absent, but particle-hole continuum appears above the threshold of $2\Delta$ instead, leading to the conclusion that there is no spectral intensity at low energies. This is not consistent with the conclusions reached within the spin-wave theory analysis for a local-moment antiferromagnet, which yields two-magnon continuum extending down to zero energy, cf. Eq.~\eqref{eq:longitudinal_susceptibility_swt}. In fact, two-magnon continuum appears also in the  $1/N_f$ expansion, yet as a subleading contribution \cite{FidrysiakEPJB2016}. This can be explicitly shown by expanding the $\mathrm{Tr} \mathrm{ln}$ term in Eq.~\eqref{pni_eq:ActionHubbard} up to cubic terms, which generates interaction vertices between the Hubbard-Stratonovich fields. In particular, the vertex $V_{\phi^3 \phi^{+} \phi^{-}} \propto \phi^3 \phi^{+} \phi^{-}$ appears, with $\phi^{\pm} \equiv (\phi^1 \pm i \phi^2)/\sqrt{2}$. Physically, the latter accounts for kinematically-allowed decays of the longitudinal magnon into two transverse magnons, as illustrated diagrammatically in Fig.~\ref{fig:hubb_diagram_2m}. This is also the only decay process active at low energies and temperatures, since all other modes are gapped in the magnetic insulator state. 

\subsubsection{Discussion of the large-$\mathcal{N}_f$ results for Hubbard model at half-filling}

\begin{figure}
    \centering
    \includegraphics[width=0.45\textwidth]{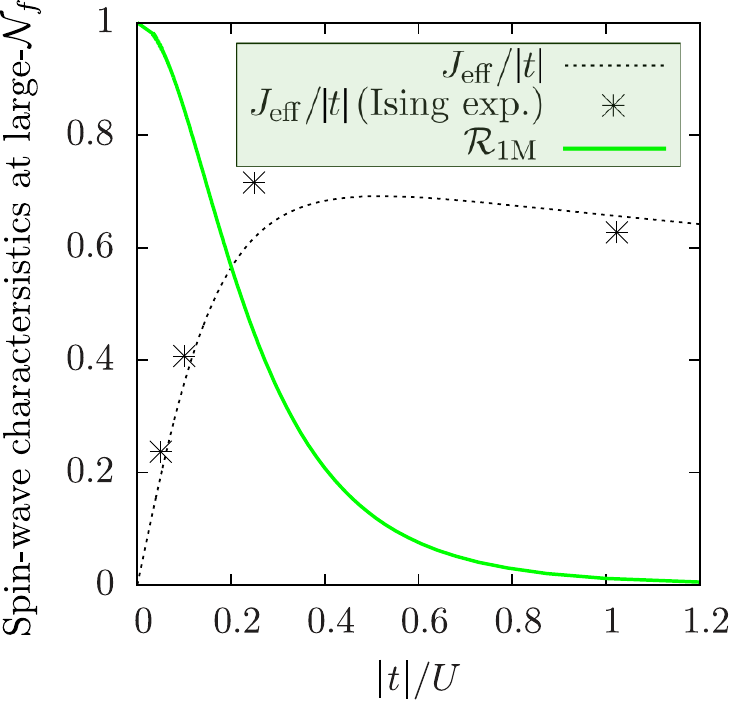}
    \caption{One- and two-magnon intensity renormalization factors $\mathcal{R}_{\mathrm{1M}}$ and $\mathcal{R}_{\mathrm{2M}}$ for the Hubbard model as a function of $|t|/U$, obtained within the large-$\mathcal{N}_{f}$ expansion.  The qualitative change in the $|t|/U$ dependence of the effective exchange $J_\mathrm{eff}$ signals crossover from strong- to weak-coupling regime. The stars denote $J_\mathrm{eff} / |t|$ obtained within the Ising expansion method in Ref.~\cite{PhysRevB52Shi}. In the strong-coupling limit $|t|/U \rightarrow 0$, Heisenberg model results are reproduced since $\mathcal{R}_{\mathrm{1M}} \rightarrow 1$ and $J_\mathrm{eff} \approx 4 t^2 /U$. The figure has been adapted from \cite{FidrysiakPhDThesis2016}.}
    \label{fig:hubbard_ren}
  \end{figure}

In Fig.~\ref{fig:hubbard_ren} we display the summary of the large-$\mathcal{N}_f$ characteristics, describing low-energy magnetic dynamics of the Hubbard model, as a function of the $|t|/U$. The limit $|t|/U \rightarrow 0$ corresponds to the strong-coupling situation, whereas strong- to weak-coupling crossover is expected to occur around $|t|/U \sim 0.125$, where the on-site Coulomb interaction becomes comparable to bare single-particle bandwidth, $W = 8 t$. The effective exchange coupling constant $J_\mathrm{eff}$, displayed as a dotted line, indeed exhibits a crossover behavior, yet for somewhat larger $|t|/U \approx 0.4$. By making use of explicit analytic formula for $J_\mathrm{eff}$ (cf. Eq.~\eqref{eq:effective_exchange_large_n}), one can show that $J_\mathrm{eff} = \frac{4 t^2}{U}$ asymptotically at strong-coupling. This agrees with the result for the kinetic exchange interaction, obtained by applying the canonical perturbation expansion to the Hubbard model at half-filling, see Appendix~\ref{appendix:derivation_of_the_tj_model}. The large-$\mathcal{N}_f$ calculation captures thus implicitly the second-order kinetic exchange processes. At weak-coupling, the exchange interaction (as expressed in the units of $t$) gradually diminishes for increasing $|t|/U$. In both weak- and strong-coupling regime, the value of $J_\mathrm{eff}$ obtained at large-$\mathcal{N}_f$ matches quantitatively the values obtained from Ising expansion \cite{PhysRevB52Shi}, indicated by stars in Fig.~\ref{fig:hubbard_ren}. 

An additional insight into the structure of magnetic excitations may be gained by the analysis of the spectral-weight of the spin-wave excitations as the system evolves from the strong- to weak-coupling regime, which should provide insight into the interplay of itinerant electrons with collective magnetic excitations. Such an analysis goes beyond what can be achieved within the spin-wave theory for a local-moment system, discussed in previous subsections. The principal finite-$U$ effect, besides renormalization of the kinetic exchange integral, is reduction of the low-energy transverse dynamical susceptibility magnitude. The latter may be accounted for by introducing the one-magnon intensity normalization factor $\mathcal{R}_\mathrm{1M}$ (cf. Eq.~\eqref{eq:eq:one-magnon_ren_factor}), displayed in Fig.~\ref{fig:hubbard_ren} as a green solid line. As expected, $\mathcal{R}_\mathrm{1M} \rightarrow 1$ at strong coupling ($|t|/U \rightarrow 0$), which means that the spin-wave-theory result is retrieved in this limit. As one moves away from the strong-coupling regime, the renormalization factor $\mathcal{R}_\mathrm{1M}$ rapidly decreases and the magnetic response becomes dominated by incoherent particle-hole excitations.

To recapitulate, the essential results regarding magnetic collective modes, accessible within the Heisenberg model description, are also reproduced within the $1/\mathcal{N}_f$ technique. The latter approach is more general though, allowing to study intermediate-correlation regime and the structure of incoherent particle-hole excitations. Similar conclusions can be obtained for more complex multi-band models that emerge in the context of iron-based superconductors \cite{PhysRevB81Knolle,PhysRevB80Brydon,FidrysiakEPJB2016}. Nonetheless, the large-$\mathcal{N}_f$ gap equation for the order parameter (cf. Eq.~\eqref{eq:af_order_parameter_hubbard}) is of the Hartree-Fock type and misses the essential physics of strong-local correlations, e.g., there is no metal to insulator transition in the large-$\mathcal{N}_f$ limit. Those effects may be effectively studied using other techniques, such as diagrammatic variational wave function approach, reviewed in Sec.~\ref{sec:de_gwf_expansion}. The effects related to electron localization, can be to some extent incorporated by evaluating higher-order $1/\mathcal{N}_f$-expansion corrections to thermodynamic quantities, yet this a computationally demanding task. In Sec.~\ref{sec:VWF+1/N} we describe a modified version of $1/\mathcal{N}_f$ technique, capable of preserving the key advantages of both $1/\mathcal{N}_f$ expansion and variational-wave-function approach already at the leading large-$\mathcal{N}_f$ order.

\subsubsection{Fierz identities and mean-field ambiguity} \label{pni_sec:Mean_Field_Ambiguity}

The large-$\mathcal{N}_f$ approach suffers from a subtle problem related to multiple equivalent ways of Hubbard-Stratonovich decoupling of the four-fermion interaction term. This ambiguity originates from the so-called \emph{Fierz identities} for fermionic models, discussed extensively in high-energy physics literature (cf., e.g., Refs.~\cite{JaeckelPhysRevD2003,BraunPhysRevD2017} and references therein). Whereas all such decoupling schemes are completely equivalent if the model is solved rigorously, it turns out that the large-$\mathcal{N}_f$ phase diagrams are highly sensitive to unphysical parameters controlling the particular representation of the interaction term. This circumstance severely limits reliability of the theoretical schemes that are based on Hubbard-Stratonovich transformation, particularly in the intermediate- and strong-coupling regimes. Here we concisely review this aspect of large-$\mathcal{N}_f$ approximations, as well as the common workarounds employed to sidestep the problems.

For the Hubbard model, one can write down an exact relation (up to a multiplicative, field independent constant)

\begin{align}
\mathrm{e}^{-\frac{1}{2} U \int d\tau\sum_i  \hat{n}_{i}^2} \propto &\int \mathcal{D} \phi^{(\alpha)} \exp\Bigg[ U \int d\tau \sum_i \Big(  2 c_1 \hat{S}^0_i \phi^0_i + 2 c_2  \hat{\mathbf{S}}_i \cdot \boldsymbol{\phi}_i -  c_1  (\phi^0_i)^2 -  c_2  (\boldsymbol{\phi}_i)^2\Big)\Bigg], \label{pni_eq:mean_field_ambiguity_decoupling}
\end{align}

\noindent
where $\hat{S}^0_i \equiv \frac{1}{2} \hat{n}_i$ and the coefficients $c_1$ and $c_2$. In order to retrieve the original expressions by integrating out the Hubbard-Stratonovich fields, one needs to impose the constraint $3 \cdot c_2 - c_1 = 2$, which still leaves out one arbitrary real parameter. Here  $\hat{n}_i = \sum_\sigma \bar{\eta}_{i\sigma} \eta_{i\sigma}$ and $\hat{S_i}^\alpha = \sum_{\sigma\sigma^\prime}\bar{\eta}_{i\sigma^\prime} \sigma^\alpha_{\sigma\sigma^\prime}\eta_{i\sigma}$ are not regarded as operators, but bilinears expressed in terms of Grassman fields. By decoupling the Hubbard interaction term with the use of equation~(\ref{pni_eq:mean_field_ambiguity_decoupling}) one obtains the effective action of the type

\begin{align}
\mathcal{S}_\mathrm{eff} \propto \int d\tau \Bigg[ & \sum_{ij} \bar{\eta}_i \left( (\partial_\tau  - \mu)\delta_{ij} + t_{ij}\right) \eta_j  + U \sum_i \Big(-2 c_1  \hat{S}^0_i \phi^0_i - 2 c_2  \hat{\mathbf{S}}_i \cdot \boldsymbol{\phi}_i +  c_1  (\phi^0_i)^2 +  c_2  (\boldsymbol{\phi}_i)^2\Big)\Bigg]. \label{pni_eq:mean_field_ambiguity_Seff}
\end{align}

\noindent
In Sec.~\ref{sec:large_n_expansion} we have employed identity \eqref{pni_eq:mean_field_ambiguity_Seff} for the specific choice $c_1 = c_2 = 1$ (cf. Eq.~\eqref{eq:hs_decoupling}). This decoupling manifestly preserves spin-rotational invariance and hence the Goldstone's theorem is fulfilled in the broken symmetry antiferromagnetic state, even for the leading order solution. However, the transformation \eqref{pni_eq:mean_field_ambiguity_Seff} is parameterized by one unphysical real parameter (if $c_1$ is selected as an independent variable, $c_2$ is then determined as $c_2 = (2 + c_1)/3$).  In the general situation, the antiferromagnetic gap equation for the Hubbard model becomes modified with respect to Eq.~\eqref{eq:af_order_parameter_hubbard} as

\begin{align}
1 = c_2 U \frac{T}{N} \sum \limits_{n, \mathbf{k}}  \frac{1}{{\omega_n^2 + \epsilon_\mathbf{k}^2 + \Delta^2}},
\end{align}

\noindent
which, for $T \rightarrow 0$, reduces to $1 = c_2 U \bar{\sum}_\mathbf{k} E_\mathbf{k}^{-1}$. The mean field phase diagram acquires thus an explicit dependence on the  decoupling procedure through the arbitrary coefficient $c_2$ (see, e.g., Ref.~\cite{PhysRevB70Baier} for explicit calculations).

The choice of the Hubbard-Stratonovich decoupling needs to be thus based on additional physical arguments, going beyond the intricacies of the formalism alone. Namely, the resulting model should respect the symmetries of the original Hamiltonian at each desired expansion order. This is necessary to make the decoupling compatible with certain rigorous results, such as Goldstone's theorem. If application of the symmetry argument is not sufficient to specify the decoupling, then the values of remaining free parameters may be selected so that the saddle-point gap equations coincide with those obtained within Hartree-Fock approximation in an appropriate limiting situation. However, the latter choice often leads to further ambiguities, since the agreement cannot be usually reached for all particle-hole scattering channels. Consistency with RPA for selected set of Hubbard-Stratonovich fields comes at a price of sacrificing the agreement in the other channels. In Sec.~\ref{sec:large_n_expansion}, we have selected $c_1 = c_2 = 1$ which leads to agreement with RPA for magnetic degrees of freedom, but not for the charge Hubbard-Stratonovich fields. This choice was also adopted in several previous studies of the Hubbard model (see, e.g., Refs.~\cite{PhysRevB43John, PhysRevB70Baier}).

Another common decoupling is based on the identity

\begin{align}
\hat{n}_{i \uparrow} \hat{n}_{i \downarrow} = \left(\frac{\hat{n}_{i\uparrow} + \hat{n}_{i\downarrow}}{2}\right)^2 - \left(\hat{S}^3\right)^2 =  (\hat{S}^0_i)^2 - (\hat{S}^3_i)^2,
\end{align}

\noindent
which yields

\begin{align}
\mathrm{e}^{-U \int d\tau\sum_i  \hat{n}_{i \uparrow} \hat{n}_{i \downarrow}} \propto & \int \mathcal{D} \phi^{(0)} \mathcal{D} \phi^{(3)} \exp\Bigg( U \int d\tau\sum_i \Big[-\frac{1}{4}  (\phi^{0}_i)^2  -\frac{1}{4} (\phi^{3}_i)^2 - \hat{S}_i^3 \phi^{3}_i - i \hat{S}_i^0 \phi^0_i\Big]\Bigg), \label{pni_eq:mean_field_ambiguity_decoupling_broken_symmetry}
\end{align}

\noindent
so that the corresponding action reads

\begin{align}
\mathcal{S}_\mathrm{eff} =  \int d\tau \Bigg[ &\sum_{ij} \bar{\eta}_i \left( (\partial_\tau  - \mu)\delta_{ij} + t_{ij}\right) \eta_j +   U \sum_i \Big(\frac{1}{4} \cdot  (\phi^{0}_i)^2 + \frac{1}{4} \cdot (\phi^{3}_i)^2 +  \hat{S}_i^3 \phi^{3}_i + i  \hat{S}_i^0 \phi^0_i\Big) \Bigg]. \label{pni_eq:mean_field_ambiguity_Seff_broken_symmetry}
\end{align}

\noindent
Equation~\eqref{pni_eq:mean_field_ambiguity_Seff_broken_symmetry} leads to the saddle point solution equivalent to the Hartree-Fock results both in spin- and charge- channels. This has been achieved at the price of discarding the transverse spin-modes (note the lack of $\phi_i^1$ and $\phi_i^2$ Hubbard-Stratonovich fields). Therefore, the model \eqref{pni_eq:mean_field_ambiguity_Seff_broken_symmetry} manifestly violates the spin rotational symmetry which is restored in an indirect manner by higher-order quantum fluctuations around the saddle point. A different decoupling scheme has been proposed in Ref.~\cite{MaitiPhysRevB1994}.

Over the years, several approaches have been developed to reduce severeness of the Fierz-ambiguity problem. First, specialized symmetrization procedures may be applied to make the decoupling scheme consistent in otherwise conflicting particle-hole channels, e.g., by means of the fluctuating reference frame approach to the Hubbard model \cite{PhysRevLett65Schulz}. It has been also demonstrated that incorporating fluctuation effects via renormalization group approach indeed reduces the dependence of the decoupled model on unphysical decoupling parameters \cite{JaeckelPhysRevD2003}. Specialized multi-channel decouplings \cite{StepanovPhysRevB2019} and hybrid fermion-boson schemes \cite{DenzPhysRevB2020} have also been proposed. Moreover, in specific circumstances, Fierz ambiguity may be turned into a convergence criterion for a computational scheme \cite{AyralPhysRevLett2017}. In the following subsections, we introduce an alternative way to avoid Fierz problem by specialized resummation of the $1/\mathcal{N}_f$ expansion, which interfaces well with the DE-GWF variational, introduced in Sec.~\ref{sec:vwf_solution}. 

\subsection{Fluctuations and strong correlations: Variational wave function approach combined with $1/\mathcal{N}_f$ expansion}
\label{sec:VWF+1/N}

\subsubsection{General characterization}
\label{subsubsec:general_characterization}

\begin{figure}
    \centering
    \includegraphics[width=0.8\textwidth]{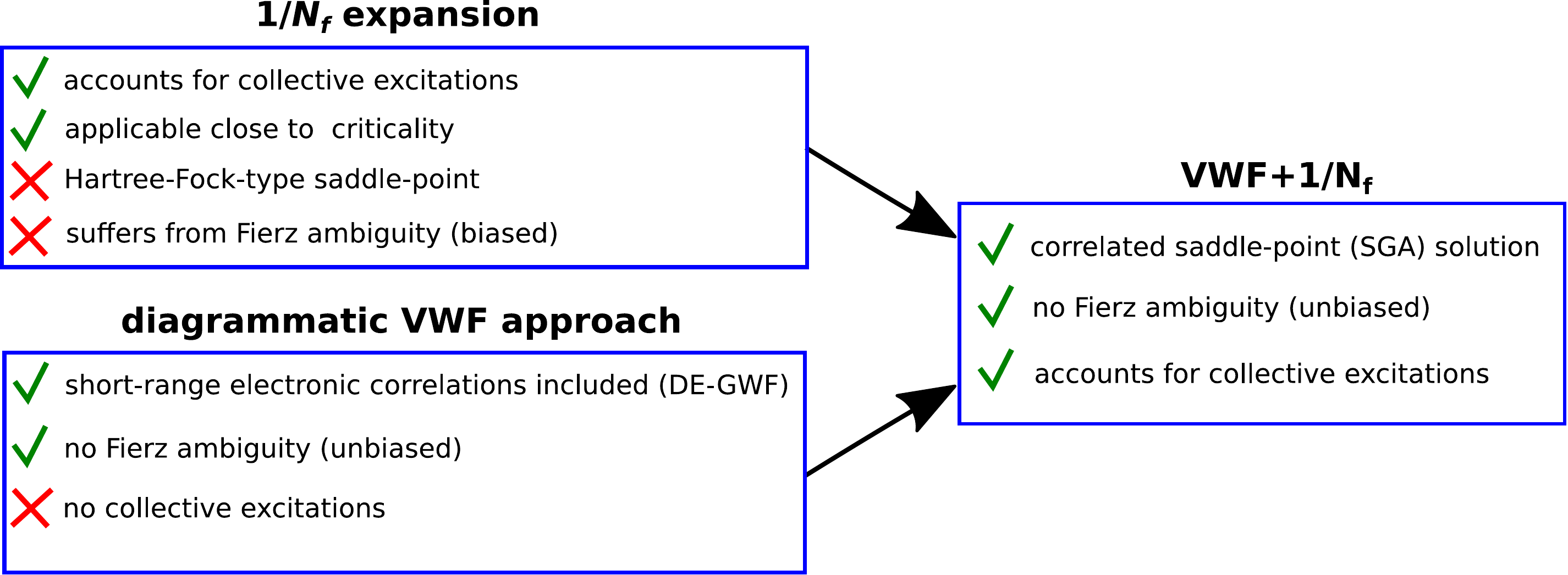}.
    \caption{Schematic comparison of the variational-wave function (VWF) technique, $1/\mathcal{N}_f$ expansion, and the combined VWF+$1/\mathcal{N}_f$ scheme. Green ticks and red crosses mark desirable and undesirable features of the methods, respectively. Note that VWF and $1/\mathcal{N}_f$ approaches account well for complementary system properties, equilibrium and dynamic.}
    \label{fig:sga_fluct_scheme}
  \end{figure}

We now overview a more sophisticated approximation, combining \textbf{V}ariational \textbf{W}ave \textbf{F}unction approach (VWF) in its diagrammatic form (cf. Sec.~\ref{sec:vwf_solution}), and $1/\mathcal{N}_f$ expansion, introduced in Sec.~\ref{sec:large_n_expansion}. The applicability range and limitations of both techniques are relatively well understood at this point, if treated separately. Diagrammatic VWF approach allows to explore the effects of short-range electronic correlations in normal and superconducting states, with a route for systematic improvement by enlarging the variational space (e.g., by promoting the correlator from Gutzwiller- to  Jastrow-form). In practice though, the number of variational parameters is rather limited for computational reasons. A clear advantage of the VWF scheme over the techniques based on applied at the start Hubbard-Stratonovich decoupling is that it is free of the Fierz ambiguity (cf. Sec.~\ref{pni_sec:Mean_Field_Ambiguity}) and thus unbiased with respect to competing particle-hole scattering channels, unless such a bias is introduced manifestly at the correlator level. However, VWF approximation does not describe well the collective modes if one is restricted to the most commonly used family of the correlated Fermi-sea variational states. On the other hand, $1/\mathcal{N}_f$ expansion is particularly well suited for studying collective excitation effects as the latter emerge already in the leading (large-$\mathcal{N}_f$) order, see Sec.~\ref{sec:large_n_expansion}. Also, thermodynamic and quantum fluctuations around the saddle-point solution may be systematically included as higher-order corrections. In particular, for localized-spin systems, approximations of this sort have been effectively utilized to study critical behavior in antiferromagnetic systems \cite{IrkhinPhysRevB1997,KataninPhysUspekhi2007}. The essential disadvantages of the $1/\mathcal{N}_f$ technique are as follows. First, it suffers from the Fierz problem, making the phase diagram highly-sensitive to unphysical parameters. In effect, this approach is \emph{biased} by the selection of decoupling scheme and of limited reliability in quantitative studies. Second, the common variants of the $1/\mathcal{N}_f$ expansion, such as those discussed above, yield Hartree-Fock-type saddle point solutions. This is an essential limitation, since robust collective excitations are universally observed in strongly correlated systems, where the Hartree-Fock analysis in inadequate. A VWF+$1/\mathcal{N}_f$ approach, being a combination of the two above schemes, is intended to unify their desirable features and avoid some of the limitations (e.g., those related to Fierz ambiguity). A summary of VWF and $1/\mathcal{N}_f$ approaches, and the desired characteristics of the combined VWF+$1/\mathcal{N}_f$ scheme is schematically characterized in Fig.~\ref{fig:sga_fluct_scheme}.

\subsubsection{Formulation of the VWF+$1/\mathcal{N}_f$ scheme}

The VWF+$1/\mathcal{N}_f$ approach has been formulated in Refs.~\cite{FidrysiakPhysRevB2020,FidrysiakPhysRevB2021}. Here we limit ourselves to a brief summary of main steps of its derivation. The starting point, as before, is the diagrammatic VWF function scheme introduced in Sec.~\ref{sec:vwf_solution}, based on minimization of the energy functional

\begin{align}
  \label{eq:variational_energy}
  E_\mathrm{var} \equiv \langle \hat{\mathcal{H}} \rangle_\mathrm{var} \equiv \frac{\langle\Psi_\mathrm{var}| \hat{\mathcal{H}} |\Psi_\mathrm{var}\rangle}{\langle\Psi_\mathrm{var}|\Psi_\mathrm{var}\rangle} \equiv \frac{\langle\Psi_0| \hat{P}\hat{\mathcal{H}}\hat{P} |\Psi_0\rangle}{\langle\Psi_0|\hat{P}^2|\Psi_0\rangle}.
\end{align}

\noindent
The latter formula is expressed in terms of the Hamiltonian $\hat{\mathcal{H}}$ and the variational wave function $|\Psi_\mathrm{var}\rangle \equiv \hat{P} |\Psi_0\rangle$, where $\hat{P}$ and $|\Psi_0\rangle$ denote the correlator and uncorrelated state, respectively. To account for collective spin fluctuations, one needs to consider a generalized form of the correlator~\eqref{eq:Gutz_operator}, including also off-diagonal terms. Namely, we take $\hat{P} \equiv \prod_i \hat{P}_i$ with

\begin{align}
  \label{eq:Gutzwiller_correlator}
  \hat{P}_{G, i} \equiv& \lambda_{i0} \lvert 0\rangle_{i}{}_{i}\langle 0\rvert + \lambda_{i\uparrow\uparrow} \lvert\uparrow\rangle_{i}{}_{i}\langle \uparrow\rvert + \lambda_{i\downarrow\downarrow} \lvert\downarrow\rangle_{i}{}_{i}\langle \downarrow\rvert + \lambda_{i\uparrow\downarrow} \lvert\uparrow\rangle_{i}{}_{i}\langle \downarrow\rvert + \lambda_{i\downarrow\uparrow} \lvert\downarrow\rangle_{i}{}_{i}\langle \uparrow\rvert + \lambda_{id} \lvert\uparrow\downarrow\rangle_{i}{}_{i}\langle \uparrow\downarrow\rvert,
\end{align}

\noindent
where there are now six variational $\lambda$-parameters per site rather than four as in Sec.~\ref{sec:vwf_solution}. The two additional complex coefficients, $\lambda_{i\uparrow\downarrow}$ and $\lambda_{i\downarrow\uparrow}$, are needed to account for magnon excitations that are associated with local precession of magnetic moments, even if they are absent in the saddle-point solution. Specifically, neglecting the off-diagonal correlator terms $\propto \lvert\uparrow\rangle_{i}{}_{i}\langle \downarrow\rvert$ and $\lvert\downarrow\rangle_{i}{}_{i}\langle \uparrow\rvert$, would lead to violation of the spin-rotational symmetry by electronic correlations and violation of Goldstone's theorem in magnetically ordered states. We demand that the correlator $\hat{P}$ is Hermitian which may be assured by imposing the condition $\lambda_{i\downarrow\uparrow}^{*} = \lambda_{i\uparrow\downarrow}$. The correlator \eqref{eq:Gutzwiller_correlator} can be generalized to a multi-band situation in a straightforward manner by introducing an additional orbital index and enriching the starting local-wave-function basis.

The correlator parameters are not arbitrary, but should be subjected to the constraints

\begin{align}
  C^1_{i} \equiv & \langle\Psi_0| \hat{P}_{G, i}^2 |\Psi_0\rangle \equiv 1, \label{eq:c1}\\
  C^2_{i} \equiv & \langle\Psi_0| \hat{P}_{G, i} \hat{c}^\dagger_{i\uparrow}\hat{c}_{i\uparrow} \hat{P}_{G, i} |\Psi_0\rangle - \langle\Psi_0| \hat{c}^\dagger_{i\uparrow}\hat{c}_{i\uparrow}|\Psi_0\rangle \equiv 0, \label{eq:c2}\\
  C^3_{i} \equiv & \langle\Psi_0| \hat{P}_{G, i} \hat{c}^\dagger_{i\downarrow}\hat{c}_{i\downarrow} \hat{P}_{G, i} |\Psi_0\rangle - \langle\Psi_0| \hat{c}^\dagger_{i\downarrow}\hat{c}_{i\downarrow}  |\Psi_0\rangle \equiv 0, \label{eq:c3}\\
  C^4_{i} \equiv & \mathrm{Re} \left(\langle\Psi_0| \hat{P}_{G, i} \hat{c}^\dagger_{i\uparrow}\hat{c}_{i\downarrow} \hat{P}_{G, i} |\Psi_0\rangle - \langle\Psi_0| \hat{c}^\dagger_{i\uparrow}\hat{c}_{i\downarrow}  |\Psi_0\rangle \right) \equiv 0, \label{eq:c4}\\
  C^5_{i} \equiv & \mathrm{Im} \left(\langle\Psi_0| \hat{P}_{G, i} \hat{c}^\dagger_{i\uparrow}\hat{c}_{i\downarrow} \hat{P}_{G, i} |\Psi_0\rangle - \langle\Psi_0| \hat{c}^\dagger_{i\uparrow}\hat{c}_{i\downarrow}  |\Psi_0\rangle \right) \equiv 0, \label{eq:c5}
\end{align}

\noindent
that allow for efficient diagrammatic evaluation of the energy functional and effectively reduce to those discussed in Sec.~\ref{sec:vwf_solution} \cite{FidrysiakPhysRevB2021}. In effect, only one correlator degree of freedom remains per orbital.

We now explicitly decompose a general Hamiltonian $\hat{\mathcal{H}} = \hat{\mathcal{H}}_\mathrm{kin} + \hat{\mathcal{H}}_\mathrm{int}$ into the kinetic-energy part $\hat{\mathcal{H}}_\mathrm{kin}$ (quadratic in the fermionic creation/annihilation operators) and the quartic interaction, $\hat{\mathcal{H}}_\mathrm{int}$. The VWF+$1/\mathcal{N}_f$ approach is formulated in terms of three functionals, $E_{0, \mathrm{kin}} \equiv \langle\Psi_0|\hat{\mathcal{H}}_\mathrm{kin}|\Psi_0\rangle$, $E_{0, \mathrm{int}} \equiv \langle\Psi_0|\hat{\mathcal{H}}_\mathrm{int}|\Psi_0\rangle$, and $E_\mathrm{var}$ (cf. Eq.~\eqref{eq:variational_energy}). Physically, $E_{0, \mathrm{kin}}$ and $E_{0, \mathrm{int}}$ represent uncorrelated expectation values of the kinetic- and interaction-energy part of the Hamiltonian, whereas $E_\mathrm{var}$ is (correlated) variational energy. By application of Wick's theorem, all those quantities become functionals of the vector of correlator parameters and two-point instantaneous correlation functions, $P_{i\sigma,j\sigma^\prime} \equiv \langle\Psi_0|\hat{a}_{i\sigma}^\dagger \hat{a}_{j\sigma^\prime}|\Psi_0\rangle$. Direct application of the Wick's theorem to the numerator and denominator of Eq.~\eqref{eq:variational_energy} is ineffective from computational perspective. For this reason, specialized linked cluster expansion in real-space, making use of the constraints~\eqref{eq:c1}-\eqref{eq:c5} have been developed to evaluate those terms \cite{BuenemannEurophysLett2012,FidrysiakPhysRevB2021}. The latter allow for expressing all above functional in terms of a vector of correlator-parameter fields, $\boldsymbol{\lambda} \equiv \{\lambda_{i\alpha}\}$, and $\mathbf{P} \equiv \{P_{i\sigma, j\sigma^\prime}\}$. Specifically, one obtains $E_\mathrm{var} = E_\mathrm{var}(\mathbf{P}, \boldsymbol{\lambda}, \mu)$, $E_{0, \mathrm{kin}} = E_{0, \mathrm{kin}} (\mathbf{P}, \mu)$, and $E_{0, \mathrm{int}} = E_{0, \mathrm{int}} (\mathbf{P})$ (to make notation concise, we have included the chemical potential term in the kinetic energy part).

The action for the VWF+$1/\mathcal{N}_f$ expansion is then taken in the form

\begin{align}
  \label{eq:sga_1n_action}
  \mathcal{S}[\mathbf{P}, \boldsymbol{\xi}, \boldsymbol{\rho}, \bar{\boldsymbol{\eta}}, \boldsymbol{\eta}, \mathbf{J}, \mu] = & \int d\tau \bar{\boldsymbol{\eta}}^s \left( \partial_\tau + (i \boldsymbol{\xi}^\dagger - \mathbf{J}^\dagger)\hat{\boldsymbol{\mathcal{O}}} \right) \boldsymbol{\eta}^s + \mathcal{N}_f \int d\tau \left( E_\mathrm{var}(\mathbf{P}, \boldsymbol{\lambda}, \mu) - i \boldsymbol{\xi}^\dagger\mathbf{P} \right) 
+ \nonumber \\ & \int d\tau E_{0, \mathrm{kin}}(\mathbf{P}, \mu) + \frac{1}{2} \int d\tau E_{0, \mathrm{int}}(\mathbf{P}) - \int d\tau E_\mathrm{var}(\mathbf{P}, \boldsymbol{\lambda}, \mu) - \nonumber \\ &
i \mathcal{N}_f \int d\tau \boldsymbol{\rho}^T\mathbf{C}(\mathbf{P}, \boldsymbol{\lambda}) + \frac{\mathcal{N}_f}{2} \kappa \int d\tau \mathbf{C}(\mathbf{P}, \boldsymbol{\lambda})^T \mathbf{C}(\mathbf{P}, \boldsymbol{\lambda}) - \ln \left|\mathrm{det} \frac{\partial \mathbf{C}}{\partial \boldsymbol{\lambda}}\right|,
\end{align}

\noindent
where $\boldsymbol{\eta}^s$ are Grassman fields with $s=1, \ldots, \mathcal{N}_f$ being fermionic flavor index. The Lagrange multiplier fields, $\boldsymbol{\xi}$, ensure consistency between the dynamics of the Grassman fields and lines, $\mathbf{P}$, whereas $\boldsymbol{\rho}$ has been introduced to impose the constraints \eqref{eq:c1}-\eqref{eq:c5} on the (imaginary-time) dynamical level. The vector of matrices $\boldsymbol{\hat{\mathcal{O}}} = \{\hat{\mathcal{O}}_{i\sigma,j\sigma^\prime}\}$ is defined by the conditions $\bar{\boldsymbol{\eta}} \hat{\mathcal{O}}_{i\sigma,j\sigma^\prime} \boldsymbol{\eta} \equiv \bar{\eta}_{i\sigma} \eta_{j\sigma^\prime}$. The coefficient $\kappa > 0$ is needed to ensure convergence, whereas $\ln \left|\mathrm{det} \frac{\partial \mathbf{C}}{\partial \boldsymbol{\lambda}}\right|$ cancels unphysical terms that may otherwise arise in the $1/\mathcal{N}_f$ expansion, and may be related to the Faddeev-Popov ghost fields \cite{FaddeevPhysLettB1967,FaddeevBook2018}.

The detailed form of the action~\eqref{eq:sga_1n_action} is motivated by several physical and technical requirements. First, one demands that local electronic correlations are included already in the $\mathcal{N}_f \rightarrow \infty$ limit. It can be shown that the model defined by Eq.~\eqref{eq:sga_1n_action}  reproduces exactly the VWF result at large $\mathcal{N}_f$. Second, the solution is required to be free of Fierz ambiguity problem related to selection of the Hubbard-Stratonovich decoupling channel, which is essential for an unbiased analysis of fluctuations in complementary particle-hole channels. This is indeed the case here, as the action is expressed in terms of variational functionals ($E_\mathrm{var}$, $E_{0, \mathrm{kin}}$, and $E_{0, \mathrm{int}}$), constructed as expectation values taken in general variational states, which does not favor any of the scattering channels (i.e., it may be regarded as all-channel decoupling). Finally, we impose the condition that the exact solution is reproduced as all terms in the series expansion are included. The above properties are shown in Ref.~\cite{FidrysiakPhysRevB2021}, where also a constructive derivation the action~\eqref{eq:sga_1n_action} is provided.

Starting from the model, defined by Eq.~\eqref{eq:sga_1n_action}, one can now proceed along the lines of standard $1/\mathcal{N}_f$ expansion, outlined in Sec.~\ref{sec:large_n_expansion}. Namely, the Grassman fields should be integrated out and the resultant action is to be expanded in a power series in the remainign real- and complex fields. Integrating out the Gaussian fluctuations yields then the final expression for the generating functional, $Z[\mathbf{J}]$, which is used to evaluate dynamical susceptibilities. The VWF+$1/\mathcal{N}_f$ method may be considered in several, closely related, variants, depending on the selection truncation scheme for the diagrammatic energy functional, as well as the detailed way of treating the correlator-parameter fluctuations. Specifically, the $\boldsymbol{\lambda}$-field may or may not have their own dynamics in the large-$\mathcal{N}_f$ limit, which does not affect the results if the expansion is carried out exactly, but has an impact on the leading-order results \cite{FidrysiakPhysRevB2021}. 

\subsubsection{$\mathrm{SGA}_x$+$1/\mathcal{N}_f$ variant of $\mathrm{VWF}$+$1/\mathcal{N}_f$}

The so called $\mathrm{SGA}_x$+$1/\mathcal{N}_f$ variant of the VWF+$1/\mathcal{N}_f$ is based on explicit evaluation of the constraints in the action~\eqref{eq:sga_1n_action}, which leads to several simplifications allowing to express the energy functionals in a transparent manner, and thus also to acquire a semi-analytical insight into the structure of the method. Therefore, even though the $\mathrm{SGA}_x$+$1/\mathcal{N}_f$ generally provides less reliable results for the charge dynamics than the other truncation schemes, we briefly overview it here. The starting point of $\mathrm{SGA}_x$+$1/\mathcal{N}_f$ is the correlator~\eqref{eq:Gutzwiller_correlator}, supplemented with constraints~\eqref{eq:c1}-\eqref{eq:c5}. Since $\hat{P}_{G,i}$ depends on six parameters, only one independent real correlator coefficient remains per lattice site. The latter may be selected in a rotationally-invariant form

\begin{align}
  x_i = \frac{\left(\lambda_{i0}\right)^2 - 1}{n_{i\uparrow} n_{i\downarrow} - m_i m_i^{*}},
\end{align}

\noindent
where $n_{i\sigma} \equiv \langle\Psi_0|\hat{c}^{\dagger}_{i\sigma} \hat{c}_{i\sigma}|\Psi_0\rangle$ and $m_{i} \equiv \langle\Psi_0|\hat{c}^{\dagger}_{i\uparrow} \hat{c}_{i\downarrow}|\Psi_0\rangle$. The remaining task is to express all the coefficients $\lambda_{i0}$, $\lambda_{i\sigma\sigma'}$, and $\lambda_{id}$ in terms of $x_i$. The most transparent way of doing that is to perform a rotation in the spin space so that spin-expectation value is aligned with the $z$-axis. In this coordinate system, it is straightforward to solve equations~\eqref{eq:c1}-\eqref{eq:c5} in a closed form. By rotating the resultant expressions back to original coordinates, one obtains

\begin{align}
  \lambda_{i0} =& \sqrt{1 + x_i \left(n_{i\uparrow} n_{i\downarrow} - m_i m_i^{*}\right)}, \\
  \lambda_{id} =& \sqrt{1 + x_i \left[(1 - n_{i\uparrow}) (1 - n_{i\downarrow}) - m_i m_i^{*}\right]}, \\
  \lambda_{i\uparrow\uparrow} = & \frac{1}{2} \left(\lambda_{i\uparrow\uparrow}^{\parallel}+\lambda_{i\downarrow\downarrow}^{\parallel} + \frac{\left(n_{i\uparrow}-n_{i\downarrow}\right)\left(\lambda_{i\uparrow\uparrow}^{\parallel}-\lambda_{i\downarrow\downarrow}^{\parallel}\right)}{\sqrt{\left(n_{i\uparrow}-n_{i\downarrow}\right)^2 + 4 m_i m_i^{*}}}\right), \\
  \lambda_{i\downarrow\downarrow} = & \frac{1}{2} \left(\lambda_{i\uparrow\uparrow}^{\parallel}+\lambda_{i\downarrow\downarrow}^{\parallel} - \frac{\left(n_{i\uparrow}-n_{i\downarrow}\right)\left(\lambda_{i\uparrow\uparrow}^{\parallel}-\lambda_{i\downarrow\downarrow}^{\parallel}\right)}{\sqrt{\left(n_{i\uparrow}-n_{i\downarrow}\right)^2 + 4 m_i m_i^{*}}}\right), \\
  \lambda_{i\uparrow\downarrow} = & \frac{m_i^{*} \left(\lambda_{i\uparrow\uparrow}^{\parallel}-\lambda_{i\downarrow\downarrow}^{\parallel}\right)}{\sqrt{\left(n_{i\uparrow}-n_{i\downarrow}\right)^2 + 4 m_i m_i^{*}}}, \\
  \lambda_{i\uparrow\downarrow} = & \lambda_{i\downarrow\uparrow}^{*},
\end{align}

\noindent
where

\begin{align}
  \lambda_{i\uparrow\uparrow}^{\parallel} = \sqrt{1 - \frac{x_i}{2} \left(n_{i\uparrow}+n_{i\downarrow} - \sqrt{\left(n_{i\uparrow}-n_{i\downarrow}\right)^2 + 4 m_i m_i^{*}}\right) + x_i \left(n_{i\uparrow} n_{i\downarrow} - m_i m_i^{*}\right)}, \\
  \lambda_{i\downarrow\downarrow}^{\parallel} = \sqrt{1 - \frac{x_i}{2} \left(n_{i\uparrow}+n_{i\downarrow} + \sqrt{\left(n_{i\uparrow}-n_{i\downarrow}\right)^2 + 4 m_i m_i^{*}}\right) + x_i \left(n_{i\uparrow} n_{i\downarrow} - m_i m_i^{*}\right)}. 
\end{align}

\noindent
In a multi-band system, all the correlator parameters and line field should acquire an additional orbital index, $\alpha$. In particular, the variational parameter $x_i \rightarrow x_i^\alpha$. Note that this is also the case for antiferromagentic phase, where $\alpha$ plays the role of sublattice index. We have omitted $\alpha$ in above equations for brevity of notation, but reintroduce it below. 

We now focus on a class of generalized $t$-$J$-$U$-$V$ models, given by the Hamiltonian (cf. also Sec.~\ref{sec:theoretical_models})

\begin{align}
  \label{eq:tju_model}
  \hat{\mathcal{H}} = \sum\limits_{ij\alpha\beta\sigma} t_{ij}^{\alpha\beta} \hat{a}^{\alpha\dagger}_{i\sigma} \hat{a}^{\beta}_{j\sigma} + \sum\limits_{i\alpha} U^\alpha \hat{n}_{i\uparrow}^\alpha \hat{n}_{i\downarrow}^\alpha + \frac{1}{2} \sum \limits_{ij\alpha\beta}  J_{ij}^{\alpha\beta} \hat{\mathbf{S}}_i^\alpha \hat{\mathbf{S}}_j^\beta + \frac{1}{2} \sum \limits_{ij\alpha\beta} V_{ij}^{\alpha\beta} \hat{n}_i^\alpha \hat{n}_j^\beta,
\end{align}

\noindent
where $\hat{n}_{i\sigma}^\alpha \equiv \hat{a}^{\alpha\dagger}_{i\sigma} \hat{a}^{\alpha}_{i\sigma}$, $\hat{n}_i^\alpha = \sum_\sigma \hat{n}_{i\sigma}^\alpha$, and $\hat{\mathbf{S}}_i^\alpha = (\hat{S}_i^{\alpha x}, \hat{S}_i^{\alpha y}, \hat{S}_i^{\alpha z})$ is spin operator. The indices $\alpha$ and $\beta$ enumerate the orbitals. In the single-band case $\alpha = 1$ and can be disregarded altogether.

The $\mathrm{SGA}_x$+$1/\mathcal{N}_f$ method utilizes a formal infinite lattice coordination-number limit to evaluate the correlated energy functional, $E_\mathrm{var}$ in a closed form. In that case, for $(i, \alpha) \neq (j, \beta)$ the consecutive terms of the Hamiltonian contribute as follows

\begin{align}
  \langle{\hat{a}^{\alpha\dagger}_{i\sigma}  \hat{a}^{\beta}_{j\sigma}}\rangle_G =& \sum\limits_{\sigma\sigma'} q_{i\sigma j\sigma'}^{\alpha\beta} \langle{\hat{a}^{\alpha\dagger}_{i\sigma}  \hat{a}^{\beta}_{j\sigma'}}\rangle_0, \label{eq:expansion_hopping}\\
  \langle{\hat{n}_{i\uparrow}^\alpha \hat{n}_{i\downarrow}^\alpha}\rangle_G = & \left(1 + x_i^\alpha \left[(1 - n_{i\uparrow}^\alpha) (1 - n_{i\downarrow}^\alpha) - m_i^\alpha m_i^{\alpha*}\right]\right) \langle{\hat{n}_{i\uparrow}^\alpha \hat{n}_{i\downarrow}^\alpha}\rangle_0, \\
  \langle{\hat{\mathbf{S}}_i^\alpha \hat{\mathbf{S}}_j^\beta}\rangle_G = & g_{S0, ij}^{\alpha\beta} +  \sum\limits_{\sigma\sigma'} g_{S, i\sigma j\sigma'}^{\alpha\beta} \langle\hat{n}_{i\sigma}^\alpha \hat{n}_{j\sigma'}^\beta\rangle_0^c + g_{S, imjm^{*}}^{\alpha\beta} \langle\hat{m}_{i}^\alpha \hat{m}_{j}^{\beta\dagger}\rangle_0^c + g_{S, im^{*}jm}^{\alpha\beta} \langle\hat{m}_{i}^{\alpha\dagger} \hat{m}_{j}^{\beta}\rangle_0^c \nonumber \\ & g_{S, imjm}^{\alpha\beta} \langle\hat{m}_{i}^{\alpha} \hat{m}_{j}^{\beta}\rangle_0^c + g_{S, im^{*}jm^{*}}^{\alpha\beta} \langle\hat{m}_{i}^{\alpha\dagger} \hat{m}_{j}^{\beta\dagger}\rangle_0^c + \sum\limits_\sigma g_{S, in_\sigma jm}^{\alpha\beta} \langle\hat{n}_{i\sigma}^{\alpha} \hat{m}_{j}^{\beta}\rangle_0^c + \sum\limits_\sigma g_{S, in_\sigma jm^{*}}^{\alpha\beta} \langle\hat{n}_{i\sigma}^{\alpha} \hat{m}_{j}^{\beta\dagger}\rangle_0^c + \nonumber \\ & \sum\limits_\sigma g_{S, imjn_\sigma}^{\alpha\beta} \langle \hat{m}_{i}^{\alpha} \hat{n}_{j\sigma}^{\beta} \rangle_0^c + \sum\limits_\sigma g_{S, im^{*}jn_\sigma}^{\alpha\beta} \langle \hat{m}_{i}^{\dagger\alpha} \hat{n}_{j\sigma}^{\beta} \rangle_0^c, \label{eq:expansion_SS} \\
  \langle{\hat{n}_i^\alpha \hat{n}_j^\beta}\rangle_G = & g_{N0, ij}^{\alpha\beta} +  \sum\limits_{i\sigma j\sigma'} g_{N, i\sigma j\sigma'}^{\alpha\beta} \langle\hat{n}_{i\sigma}^\alpha \hat{n}_{j\sigma'}^\beta\rangle_0^c + g_{N,  imjm^{*}}^{\alpha\beta} \langle\hat{m}_{i}^\alpha \hat{m}_{j}^{\beta\dagger}\rangle_0^c + g_{N, im^{*}jm}^{\alpha\beta} \langle\hat{m}_{i}^{\alpha\dagger} \hat{m}_{j}^{\beta}\rangle_0^c \nonumber \\ & g_{N, imjm}^{\alpha\beta} \langle\hat{m}_{i}^{\alpha} \hat{m}_{j}^{\beta}\rangle_0^c + g_{N, im^{*}jm^{*}}^{\alpha\beta} \langle\hat{m}_{i}^{\alpha\dagger} \hat{m}_{j}^{\beta\dagger}\rangle_0^c + \sum\limits_\sigma g_{N, in_\sigma jm}^{\alpha\beta} \langle\hat{n}_{i\sigma}^{\alpha} \hat{m}_{j}^{\beta}\rangle_0^c + \sum\limits_\sigma g_{N, in_\sigma jm^{*}}^{\alpha\beta} \langle\hat{n}_{i\sigma}^{\alpha} \hat{m}_{j}^{\beta\dagger}\rangle_0^c + \nonumber \\ & \sum\limits_\sigma g_{N, imjn_\sigma}^{\alpha\beta} \langle \hat{m}_{i}^{\alpha} \hat{n}_{j\sigma}^{\beta} \rangle_0^c + \sum\limits_\sigma g_{N, im^{*}jn_\sigma}^{\alpha\beta} \langle \hat{m}_{i}^{\dagger\alpha} \hat{n}_{j\sigma}^{\beta} \rangle_0^c, \label{eq:expansion_nn} 
\end{align}

\begin{figure}
  \centering
  \includegraphics[width=0.5\textwidth]{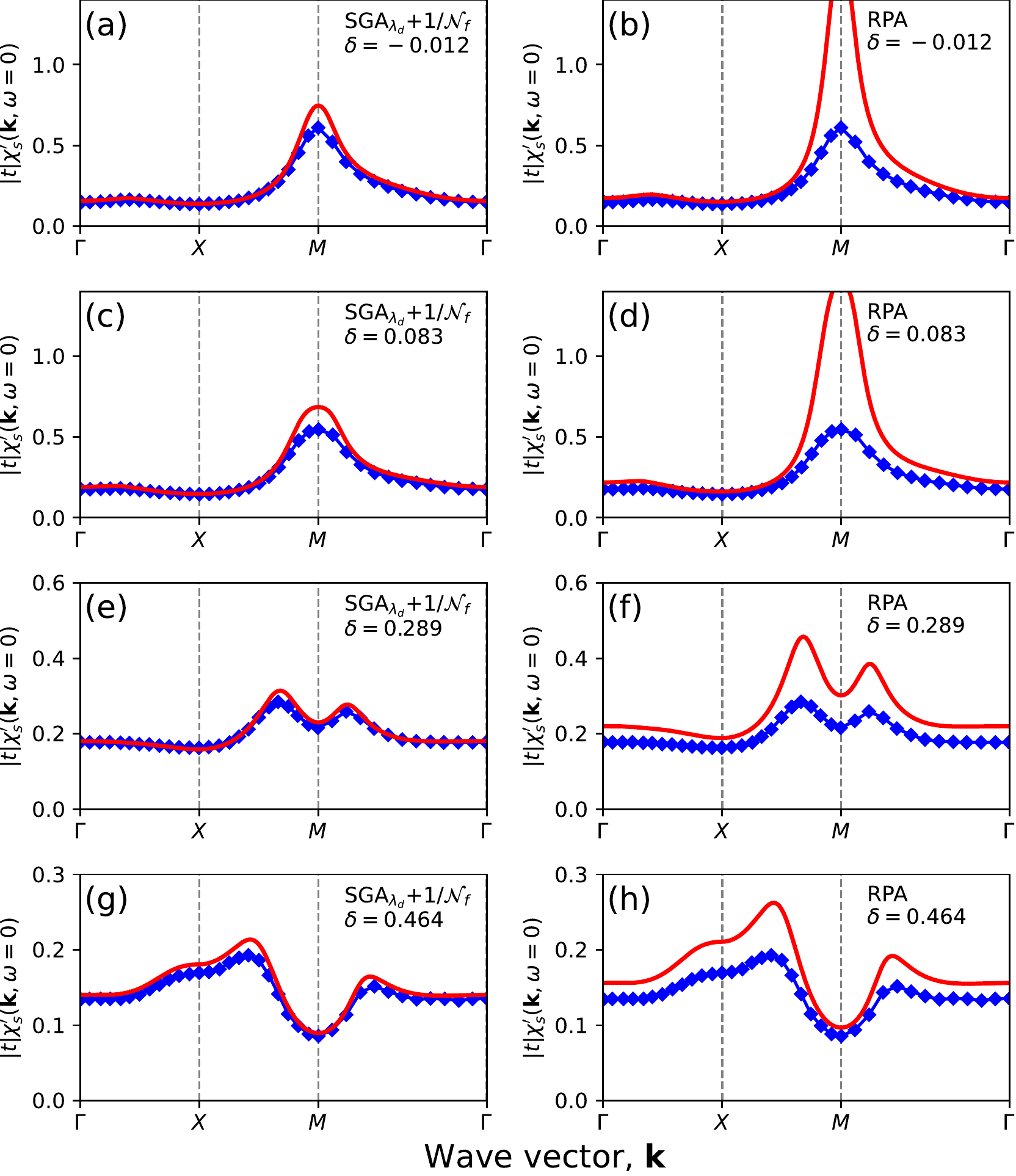}
  \caption{Static spin susceptibility along the $\Gamma$-$X$-$M$-$\Gamma$ contour for the two-dimensional Hubbard model (nearest- and next-nearest-neighbor hopping integrals are set to $t < 0$ and $t^\prime = 0.2 |t|$, $U/|t| = 2$, and temperature is taken as $k_B T = 0.2 |t|$). Red lines in (a), (c), (e), and (g) are $\mathrm{SGA}_{\lambda_d}$+$1/\mathcal{N}_f$ results, whereas panels (b), (d), (f), and (g) represent random-phase approximation (RPA) calculation. Hole doping levels are selected as $\delta=-0.017$ in panels (a)-(b), $\delta=0.083$ in (c)-(d), $\delta=0.289$ in (e)-(f), and $\delta=0.464$ in (g)-(h). Blue solid points are the corresponding  determinant quantum Monte-Carlo data \cite{HillePhysRevResearch2020}. Adapted from Ref.~\cite{FidrysiakPhysRevB2021}.}
  \label{fig:weak_coupling_benchmark}
\end{figure}

\noindent
where we use the notation $\hat{m}_i^\alpha \equiv \hat{a}_{i\uparrow}^{\alpha\dagger} \hat{a}_{i\downarrow}^{\alpha}$ and the superscript ``c'' means that only connected graphs should be retained in Wick's decomposition. The (in general complex) coefficients $q_{i\sigma j\sigma'}^{\alpha\beta}$, $g_{S/N0, ij}^{\alpha\beta}$, $g_{S/N, i\sigma j\sigma'}^{\alpha\beta}$, $g_{S/N, imjm}^{\alpha\beta}$, $g_{S/N, imjm^{*}}^{\alpha\beta}$, $g_{S/N, im^{*}jm}^{\alpha\beta}$, $g_{S/N, im^{*}jm^{*}}^{\alpha\beta}$, $g_{S/N, injm}^{\alpha\beta}$, $g_{S/N, injm^{*}}^{\alpha\beta}$, $g_{S/N, imjn}^{\alpha\beta}$, and $g_{S/N, im^{*}jn}^{\alpha\beta}$ are polynomials of twelve correlator coefficients $\lambda_{i0}^\alpha$, $\lambda_{i\uparrow\uparrow}^\alpha$, $\lambda_{i\uparrow\downarrow}^\alpha$, $\lambda_{i\downarrow\uparrow}^\alpha$, $\lambda_{i\downarrow\downarrow}^\alpha$, $\lambda_{id}^\alpha$, $\lambda_{j0}^\beta$, $\lambda_{j\uparrow\uparrow}^\beta$, $\lambda_{j\uparrow\downarrow}^\beta$, $\lambda_{j\downarrow\uparrow}^\beta$, $\lambda_{j\downarrow\downarrow}^\beta$, and $\lambda_{jd}^\beta$ (which, in turn, may be expressed in terms of local density matrix elements $n_{i\sigma}^\alpha$, $m_{i}^\alpha$, $n_{j\sigma}^\beta$, $m_{j}^\beta$, and parameters $x_i^\alpha$, $x_j^\beta$). Nonetheless, not all of the above factors are independent. Due to Hermitian form of the correlator $\hat{P}_G$, we have $g_{S/N, im^{*}jm^{*}}^{\alpha\beta} = g_{S/N, imjm}^{\alpha\beta*}$, $g_{S/N, im^{*}jm}^{\alpha\beta} = g_{S/N, imjm^{*}}^{\alpha\beta*}$, $g_{S/N, injm^{*}}^{\alpha\beta} = g_{S/N, injm}^{\alpha\beta*}$, and $g_{S/Nm^{*}n}^{\alpha\beta} = g_{S/Nmn}^{\alpha\beta*}$. This reduces the number of factors for both magnetic exchange and Coulomb repulsion.

Within the $\mathrm{SGA}_x$+$1/\mathcal{N}_f$ variant of the VWF+$1/\mathcal{N}_f$ scheme, the correlator parameters $x_i^\alpha$ are treated as time-independent variables, whose values are determined by minimization of the free energy functional. The other, related, variants are based on the condition that $\lambda_{id}^\alpha$ is time-independent ($\mathrm{SGA}_{\lambda_d}$+$1/\mathcal{N}_f$), or that all variational parameters acquire imaginary-time dependence ($\mathrm{SGA}_f$+$1/\mathcal{N}_f$).

\subsubsection{Relation of VWF+$1/\mathcal{N}_f$ to determinant quantum Monte-Carlo}

Here we briefly assess the VWF+$1/\mathcal{N}_f$ approach by comparing the static- and dynamic susceptibilities with available determinant quantum Monte-Carlo (DQMC) data. The latter scheme yields numerically exact static quantities for relatively small lattices, but requires ill conditioned analytic continuation of the simulated data to access dynamic properties. The procedure of analytic continuation introduces poorly controlled errors into the scheme and thus reduces reliability of calculated spectral properties (cf., e.g., Ref.~\cite{TripoltComPhysCommun2019} for detailed discussion of those aspects). Benchmarking the dynamical susceptibilities against DQMC between the two techniques should be thus regarded as a comparison of two approximate data sets.

In Fig.~\ref{fig:weak_coupling_benchmark} we compare $\mathrm{SGA}_{\lambda_d}$+$1/\mathcal{N}_f$ (red lines in left panels) and random phase approximation (red lines in right panels) static spin susceptibilities with the DQMC results of Ref.~\cite{HillePhysRevResearch2020} for the Hubbard model. The nearest- and next-nearest hopping integrals are selected as $t < 0$ and $t^\prime = 0.2 |t|$, and the on-site Coulomb repulsion is set to $U/|t| = 2$. This corresponds to weak-coupling situation. Temperature is taken as $k_B T = 0.2 |t|$ for all methods. Hole doping levels are selected as $\delta=-0.017$ in panels (a)-(b), $\delta=0.083$ in (c)-(d), $\delta=0.289$ in (e)-(f), and $\delta=0.464$ in (g)-(h), so that Fig.~\ref{fig:weak_coupling_benchmark} encompasses all relevant doping regimes, form underdoped to overdoped. The $\mathrm{SGA}_{\lambda_d}$+$1/\mathcal{N}_f$ follows closely the DQMC data, whereas random phase approximation results substantially overestimate static magnetic response. Even for relatively weak-coupling, the correlation effects included in the variational saddle point solution, cannot be neglected.

\begin{figure}
  \centering
  \includegraphics[width=\linewidth]{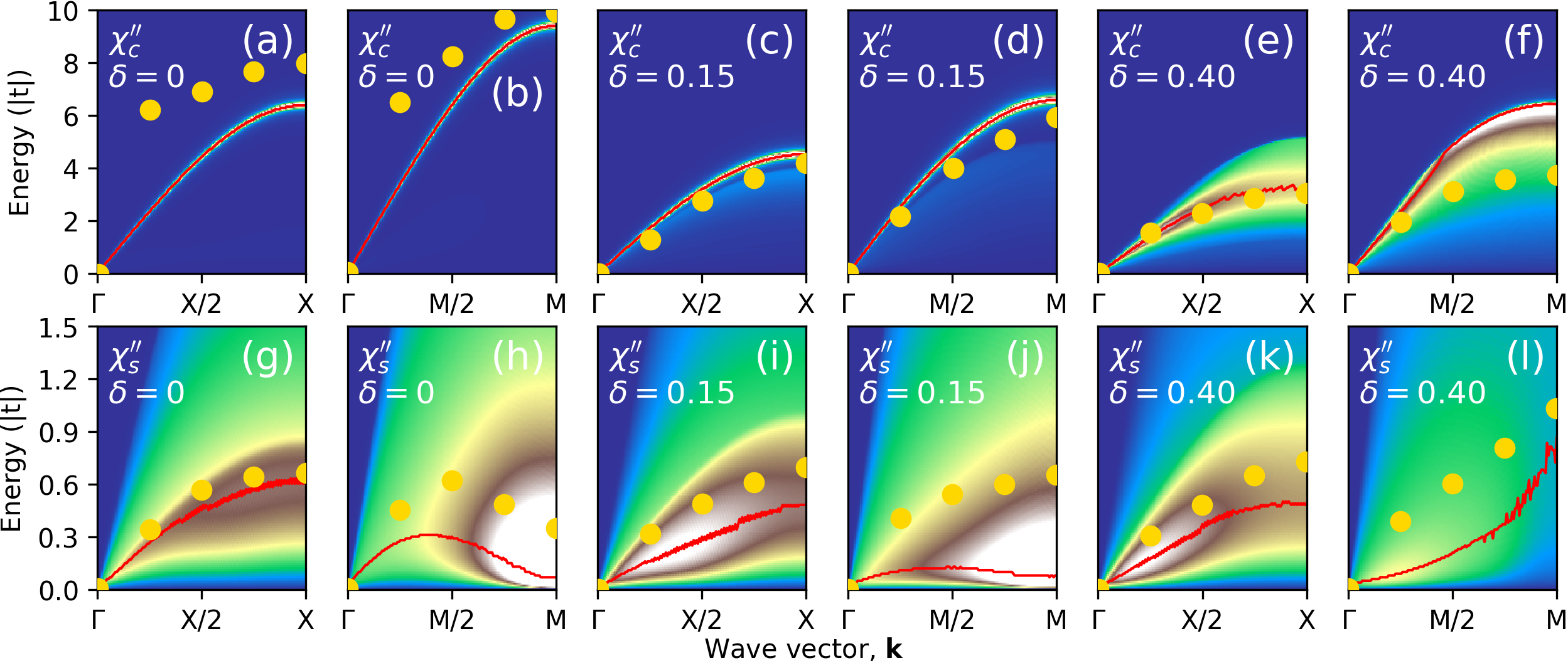}
  \caption{Imaginary parts of the dynamical charge- and spin susceptibilities (top and bottom panels), evaluated using $\mathrm{SGA}_{\lambda_d}$+$1/\mathcal{N}_f$ and DQMC for doping levels $\delta \rightarrow 0$, $\delta = 0.15$, $\delta = 0.40$. The model parameters have been selected as follows: $t < 0$, $t^\prime = 0.3 |t|$, $U = 8|t|$, $k_B T = |t|/3$ (DQMC), $k_BT = 0.35 |t|$ ($\mathrm{SGA}_{\lambda_d}$+$1/\mathcal{N}_f$ at $\delta = 0$), and $k_BT = 0.333 |t|$ ($\mathrm{SGA}_{\lambda_d}$+$1/\mathcal{N}_f$ at $\delta > 0$). Color maps show $\mathrm{SGA}_{\lambda_d}$+$1/\mathcal{N}_f$ dynamical susceptibilities, whereas red lines follow intensity maximal values. Yellow points are the peak-maxima for analytically continued DQMC data \cite{KungPhysRevB2015}. Adopted from Ref.~\cite{FidrysiakPhysRevB2021}.}
  \label{fig:benchmatk-ld}
\end{figure}

In Fig.~\ref{fig:benchmatk-ld} we perform an analogous comparison between $\mathrm{SGA}_{\lambda_d}$+$1/\mathcal{N}_f$ at strong coupling, this time for imaginary parts of dynamical charge- and spin susceptibilities (top and bottom panels, respectively). Color maps represent  $\mathrm{SGA}_{\lambda_d}$+$1/\mathcal{N}_f$ intensities, with red lines following maximum-intensities. The peak-intensity values for DQMC data of Ref.~\cite{KungPhysRevB2015} are displayed as yellow circles. The Hubbard model parameters have been selected as follows: $t < 0$, $t^\prime = 0.3 |t|$, $U = 8|t|$, $k_B T = |t|/3$ (DQMC), $k_BT = 0.35 |t|$ ($\mathrm{SGA}_{\lambda_d}$+$1/\mathcal{N}_f$ at $\delta = 0$), and $k_BT = 0.333 |t|$ ($\mathrm{SGA}_{\lambda_d}$+$1/\mathcal{N}_f$ at $\delta > 0$). The hole-doping levels are displayed inside the panels. The agreement of the maximum intensity profiles is semi-quantitative between two methods in the situation, where charge/spin modes are not excessively broad (either within $\mathrm{SGA}_{\lambda_d}$+$1/\mathcal{N}_f$ or DQMC). This is the case for the spin response along the anti-nodal $\Gamma$-$X$ line, but not in the nodal direction, where larger discrepancies are observed (cf. the detailed discussion in Ref.~\cite{FidrysiakPhysRevB2021}). 

The VWF+$1/\mathcal{N}_f$ approach at large $\mathcal{N}_f$ yields thus reliable static spin susceptibilities up to intermediate interaction strengths, as well as at strong coupling (i.e. for $U$ comparable with bare single-particle bandwidth) if wave-vector-independent intensity renormalization factors are introduced \cite{FidrysiakPhysRevB2021}. Even though both VWF+$1/\mathcal{N}_f$ and analytically continued DQMC data sets should be regarded as approximate, in the regime of coherent spin and charge dynamics, one obtains semi-quantitative agreement between the two techniques.

\subsubsection{Application to the two-dimensional Hubbard model and relation to experiments}
\label{sec:2d_hubbard_fluctations}

  \begin{figure}
    \centering
    \includegraphics[width=\textwidth]{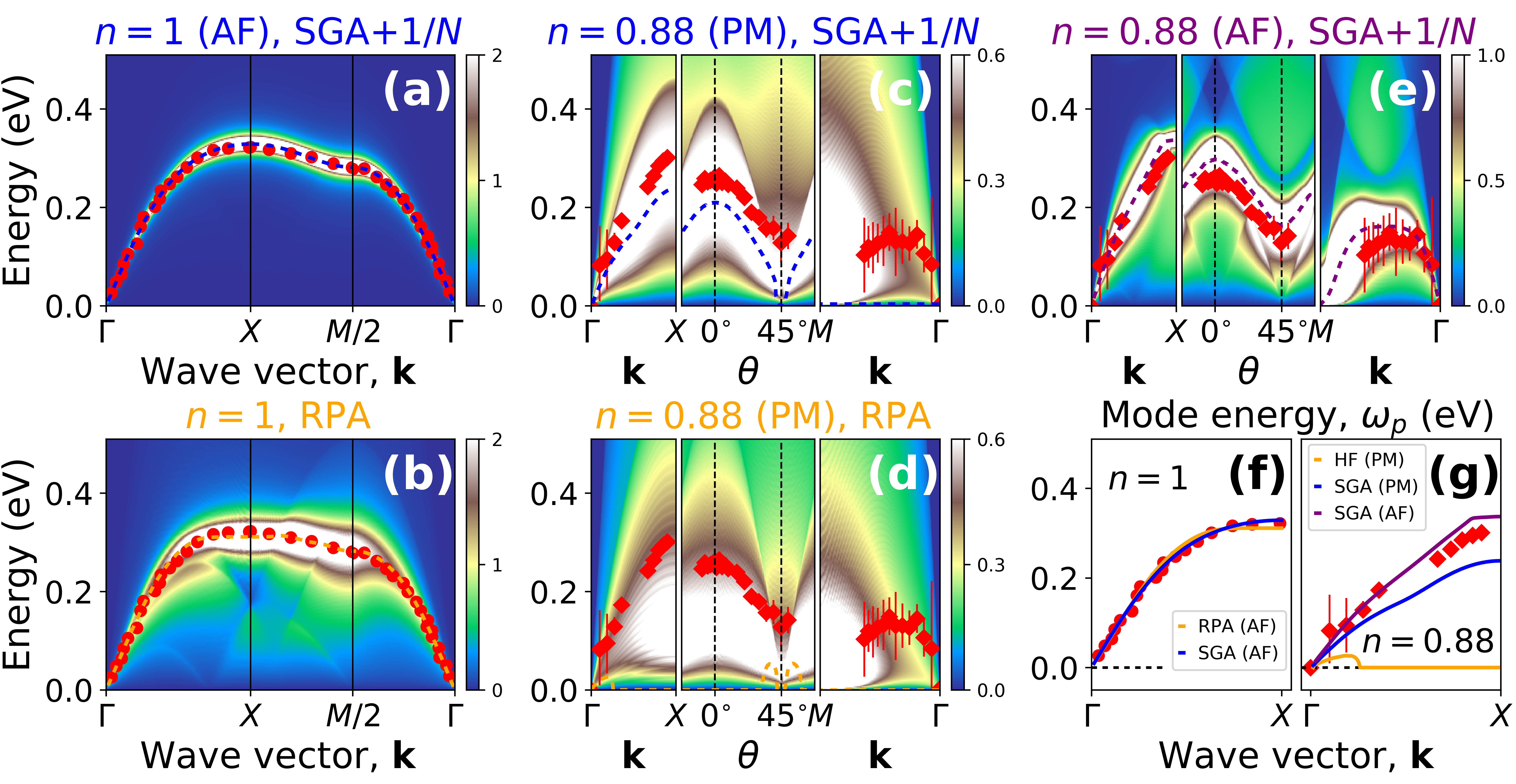}
    \caption{Imaginary parts of transverse dynamical spin susceptibility for obtained within one-orbital Hubbard model with nearest- and next-nearest-neighbor hopping integrals included, and comparison with experiment for $\mathrm{La_{2-\delta}Sr_{\delta}CuO_4}$ (LSCO). The model parameters are $t = -0.34\,\mathrm{eV}$, $U = 7|t|$. Panels (a) and (b) represent magnetic response at half filling ($n = 1$) in the antiferromagnetic state, obtained within $\mathrm{SGA}_x$+$1/\mathcal{N}_f$ and RPA, respectively. The spectra are similar and both of them match neutron scattering data \cite{HeadingsPhysRevB2010} for LSCO (red circles). The dashed lines are the paramagnon energies obtained from theoretical intensities using damped harmonic oscillator model. Panels (c) and (d) result from the same analysis, but for hole-doped system ($n = 0.88$) in the paramagnetic state. Here, the differences are qualitative: the $\mathrm{SGA}_x$+$1/\mathcal{N}_f$ method yields propagating magnetic excitations along $\Gamma$-$M$ line, whereas within RPA one obtains overdamped dynamics (see dashed curves and discussion in the text). The agreement of the $\mathrm{SGA}_x$+$1/\mathcal{N}_f$ with RIXS data \cite{IvashkoPhysRevB2017} (diamonds) is semi-quantitative. Panel (e) shows $\mathrm{SGA}_x$+$1/\mathcal{N}_f$ results in the antiferromagnetic phase at lower temperature. In (f)-(g) we compare the theoretical RPA and $\mathrm{SGA}_x$+$1/\mathcal{N}_f$ paramagnon dispersions with experiment. Adapted from Ref.~\cite{FidrysiakPhysRevB2020}.}
    \label{fig:lsco_sga1n}
  \end{figure}

  We now turn to the discussion of the VWF+$1/\mathcal{N}_f$ results for the Hubbard model and relate them to current spectroscopic experiments.  In Fig.~\ref{fig:lsco_sga1n}, the imaginary part of the dynamical spin susceptibilities, obtained for the two dimensional Hubbard model within $\mathrm{SGA}_x$+$1/\mathcal{N}_f$ (top panels) and random phase approximation (bottom panels), is compared with inelastic neutron scattering and RIXS data for $\mathrm{La_{2-\mathit{\delta}}Sr_\mathit{\delta}CuO_4}$ (red symbols). The model parameters have been selected as $t = -0.34\,\mathrm{eV}$, $t^\prime = 0.25 |t|$, and $U = 7|t|$. Moreover, relatively large temperatures $k_B T = 0.1224\,\mathrm{eV}$ (panels (a) and (c)), $k_B T = 0.1122\,\mathrm{eV}$ (panel (e)), $k_B T = 0.4420\,\mathrm{eV}$ (panel (b)), and $k_B T = 0.5168\,\mathrm{eV}$ (panel (d)) have to be adopted to stay clear of incommensurate broken symmetry states, particularly in the RPA analysis (cf. Ref.~\cite{IgoshevJPCM2015}). Panels (a) and (b) represent magnetic susceptibility at half filling ($n = 1$) in the antiferromagnetic state, obtained within $\mathrm{SGA}_x$+$1/\mathcal{N}_f$ and RPA, respectively. Both spectra are quite similar and match neutron scattering data \cite{HeadingsPhysRevB2010} for LSCO (red circles), with the RPA result being noticeably more incoherent. This may be attributed to a finite-temperature effects that affect RPA to a larger degree.

To make the discussion quantitative, it is still necessary to extract magnetic excitation energies and lifetimes from the intensity maps, displayed in Fig.~\ref{fig:lsco_sga1n}. This can be done by employing the harmonic oscillator model \cite{LamsalPhysRevB2016} for the resonant spin-susceptibility contribution. Specifically, we adopt the following representation for the imaginary part of dynamical spin susceptibility

  \begin{align}
  \label{eq:damped_oscillator}
  \mathrm{Im} \chi(\mathbf{k}, \omega) = \frac{2 F(\mathbf{k}) \gamma(\mathbf{k}) \omega}{\left[\omega^2 - \omega_0^2(\mathbf{k})\right]^2 + 4 \gamma(\mathbf{k})^2 \omega^2} + \chi_\mathrm{inc}(\mathbf{k}, \omega),
\end{align}

\noindent
where $\chi_\mathrm{inc}(\mathbf{k}, \omega)$ accounts for the incoherent contribution, providing background to the oscillator peak. Such a decomposition is not unique and depends of the precise form of $\chi_\mathrm{inc}(\mathbf{k}, \omega)$. As the guiding principle we adopt the requirement that $\chi_\mathrm{inc}(\mathbf{k}, \omega)$ should be smooth and featureless; see, e.g., Ref.~\cite{FidrysiakPhysRevB2020} for a specific ansatz. Also, supplemental physical information  can be utilized to construct the expression for $\chi_\mathrm{inc}(\mathbf{k}, \omega)$ in a more systematic manner, such as fermiology of the underlying electronic subsystem \cite{FidrysiakArXiV2021}. In Eq.~\eqref{eq:damped_oscillator}, $\omega_0(\mathbf{k})$ denotes ``bare'' paramagnon energy and $\gamma(\mathbf{k})$ is wave-vector-dependent damping coefficient. The quantity which is of physical interest is the so-called propagation frequency, defined as as the real part of the quasiparticle pole and equal to $\omega_p(\mathbf{k}) \equiv \sqrt{\omega_0(\mathbf{k})^2 - \gamma\mathbf{(k)}^2}$ if $\omega_0(\mathbf{k}) > \gamma(\mathbf{k})$. If the opposite case of $\omega_0(\mathbf{k}) < \gamma(\mathbf{k})$, the paramagnon is \emph{overdamped} and does not represent a propagating excitation. Therefore, the bare energy, $\omega_0(\mathbf{k})$, has no direct physical significance and is generally non-zero even if magnetic dynamics is overdamped. Also, distinguishing between resonant and overdamped excitations by visual inspection of the imaginary part of dynamical magnetic susceptibility is difficult, since in either case $\chi(\mathbf{k}, \omega)^{\prime\prime}$ exhibits maxima at non-zero energy in both cases. A careful analysis is thus needed to characterize magnetic excitations. The magnon (paramagnon) energies extracted from such analysis are displayed in Fig.~\ref{fig:lsco_sga1n}(a)-(b) by dashed lines. Since in the antiferromagnetic state, spin waves are coherent and well defined, the propagating energies closely follow the calculated intensity pattern.

In Fig.~\ref{fig:lsco_sga1n}(c)-(d), the outcome of a similar analysis, but for hole-doped system ($n = 0.88$) in the paramagnetic state is presented. Contrary to the half-filled case, now one observes qualitative differences between $\mathrm{SGA}_x$+$1/\mathcal{N}_f$ (panel (b)) and RPA (panel (c)). Within RPA, the propagation energy, $\omega_p(\mathbf{k})$,  is exactly zero, except for a narrow $\mathbf{k}$-space regions. This indicates that coherent magnetic excitations are rapidly suppressed by hole doping. Within the $\mathrm{SGA}_x$+$1/\mathcal{N}_f$, on the other hand, one obtains strongly anisotropic paramagnon damping rates. Specifically, along the $\Gamma$-$M$ line, the paramagnons remain coherent ($\omega_p(\mathbf{k}) > 0$), whereas they are overdamped into the anti-nodal, $\Gamma$-$X$, direction. Experimentally, such a selective overdamping of magnetic excitations has been observed in multiple families of copper-oxide superconductors (see, e.g., Refs.~\cite{GuariseNatCommun2014,Robarts_arXiV_2019}). The red diamonds in panels (b)-(c) represent RIXS data of Ref.~\cite{IvashkoPhysRevB2017}, indicating that the paramagnons remain stable in doped samples. The agreement with the $\mathrm{SGA}_x$+$1/\mathcal{N}_f$ result is semi-quantitative along $\Gamma$-$M$ line. In Fig.~\ref{fig:lsco_sga1n}(e), the $\mathrm{SGA}_x$+$1/\mathcal{N}_f$ magnetic response in the antiferromagnetic phase and for lower temperature is shown and compared with RIXS data. The agreement is now fully quantitative along all considered contours. It is not clear if such a remnant magnetic order might be present in the system as the experiment has been performed remarkably compared to the so-called magnetic stripe phase \cite{YamadaPhysRevB1998}. In Fig.~\ref{fig:lsco_sga1n}(f)-(g), a summary of paramagnon propagation energies, obtained using both RPA and $\mathrm{SGA}_x$+$1/\mathcal{N}_f$ methods, is shown and compared with available INS and RIXS data.

In summary, whereas the RPA approach provides reasonable and qualitatively correct picture of magnetic excitations in the antiferromagnetically ordered cuprates, it is not applicable in hole-doped magnetically disordered samples. Strong electronic correlations turn out to be a crucial ingredient that needs to be taken into account in order to properly describe magnetic spectra away from half-filling. This, in turn, points towards an interplay between high-energy processes (local-correlation physics) and those governing the collective dynamics at the scale $\sim J \approx 4 t^2/U \ll U$.

\subsubsection{Collective-mode dynamics in layered cuprates: Three-dimensional plasmon excitations}
\label{subsection:plasmons_3d_model}

\begin{figure}
  \centering
  \includegraphics[width=0.6\linewidth]{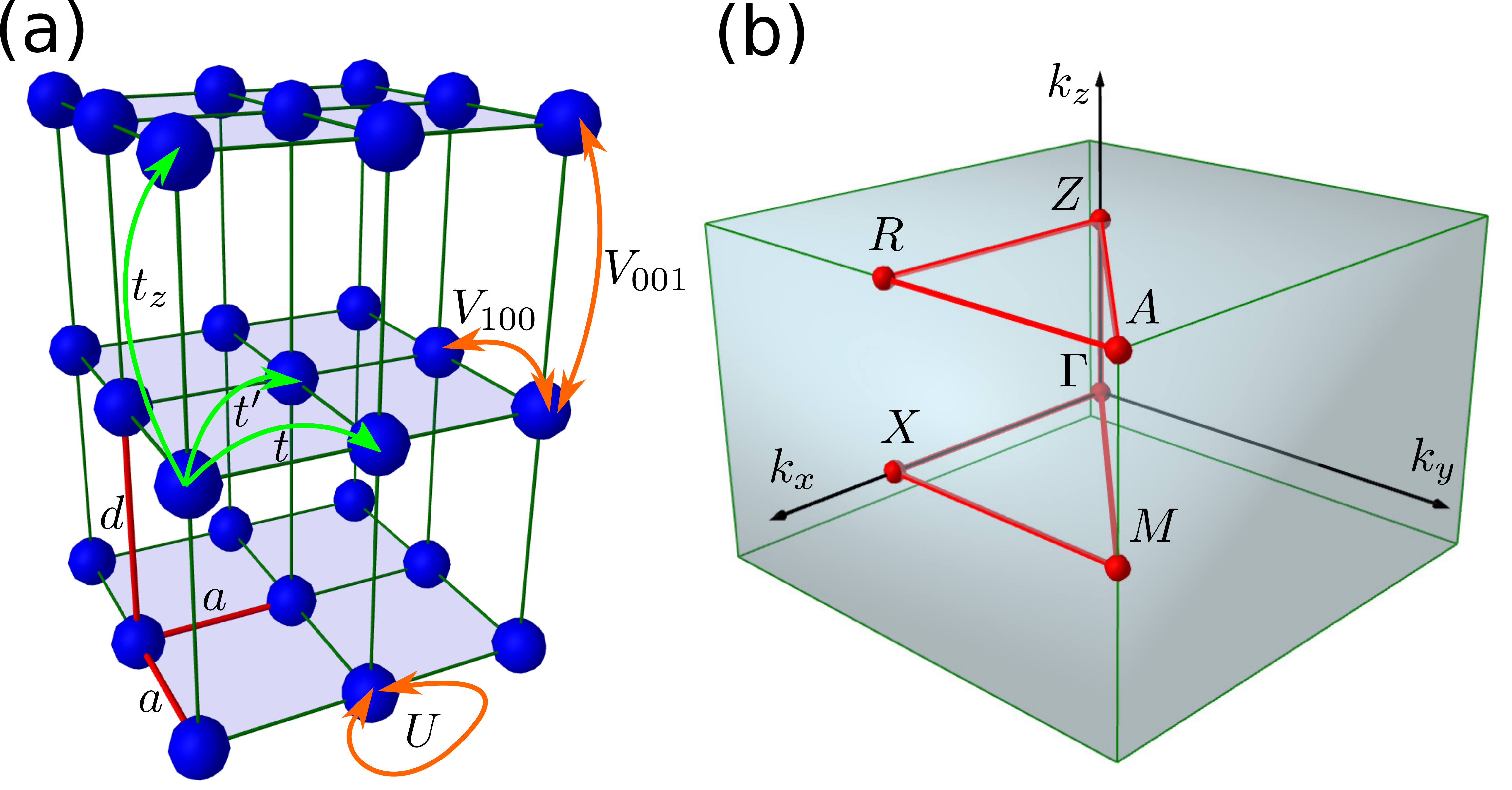}
  \caption{(a) Illustration of extended three-dimensional Hubbard model with in-plane and out-of-plane lattice constants,  $a$ and $d$, respectively. The included hopping and interaction integrals are marked inside. Only two out of infinite number of long-range Coulomb integrals, $V$, are displayed explicitly. (b) First Brillouin zone with marked $\Gamma$-$X$-$M$-$\Gamma$-$Z$-$R$-$A$-$Z$ contour. After Ref.~\cite{FidrysiakArXiV2021}.} \label{fig:lattice_3d_hubbard}
\end{figure}

In addition to magnetic fluctuations, signatures of low energy charge excitations are also observed by RIXS in cooper-oxide superconductors \cite{IshiiPhysRevB2017,HeptingNature2018,IshiiJPhysSocJapan2019,LinNPJQuantMater2020,SinghArXiV2020,NagPhysRevLett2020}. The latter modes may be interpreted as acoustic plasmons and are affected both by local electronic correlations and by long-range Coulomb interactions. The algebraic tail of the electron-electron interaction, combined with large anisotropy between in-plane and out-of-plane directions for layered systems, results in sharp dimensional crossovers in the charge excitation properties \cite{KresinPhysRevB1988,IshiiPhysRevB1993,MarkiewiczPhysRevB2008}. Therefore, the planar model of Sec.~\ref{sec:2d_hubbard_fluctations} is not sufficient to study combined collective spin- and charge- dynamics in layered correlated systems. A better starting point for such an analysis is an extended Hubbard model

\begin{align}
  \label{eq:hamitlonian_hubbard_3d}
  \hat{\mathcal{H}} = \sum_{ij\sigma} t_{ij} \hat{c}_{i\sigma}^\dagger \hat{c}_{j\sigma} + U \sum_i \hat{n}_{i\uparrow}  \hat{n}_{i\downarrow} + \frac{1}{2}\sum_{i \neq j} V_{ij} \hat{n}_i \hat{n}_j,
\end{align}

\noindent
where the lattice summations run over three-dimensions lattice composed of stacked two-dimensional planes, as illustrated in Fig.~\ref{fig:lattice_3d_hubbard}(a). The in-plane and out-of-plane lattice constants are set to $a$, and $d > a$, respectively. To reproduce fermiology of high-temperature copper-oxide superconductors, we retain non-zero nearest- and next-nearest in-plane hopping integrals, $t$ and $t^\prime$, as well as nearest-neighbor out-of-plane hopping, $t_z$. Moreover, the Hamiltonian~\eqref{eq:hamitlonian_hubbard_3d} incorporates long-range Coulomb interactions, $V_{ij}$ (the summation is restricted to $i \neq j$, since the on-site repulsion term has been written separately). In Fig.~\ref{fig:lattice_3d_hubbard}(a), the structure of included hopping and interaction integrals is illustrated, and Fig.~\ref{fig:lattice_3d_hubbard}(b) shows the first Brillouin zone with marked high-symmetry $\Gamma$-$X$-$M$-$\Gamma$-$Z$-$R$-$A$-$Z$ contour.

The remaining task is to determine the structure of the interaction integrals $V_{ij} \equiv V(\mathbf{r}_i - \mathbf{r}_j)$. This in a non-trivial problem due to translational symmetry breaking induced by the underlying lattice structure, which may be solved by considering discretized Laplace equation \cite{BeccaPhysRevB1996}

\begin{align}
  \label{eq:discretized_laplace_equation}
  \epsilon_{||} \sum_{\boldsymbol{\delta} = \mathbf{a}_1, \mathbf{a}_2} \frac{- V(\mathbf{r}_i - \boldsymbol{\delta}) + 2 V(\mathbf{r}_i) - V(\mathbf{r}_i + \boldsymbol{\delta})}{a^2} + \epsilon_{\perp} \frac{- V(\mathbf{r}_i - \mathbf{a}_3) + 2 V(\mathbf{r}_i) - V(\mathbf{r}_i + \mathbf{a}_3)}{d^2} = \frac{e^2}{\epsilon_0} \frac{\delta_{i0}}{a^2 d},
\end{align}

\noindent
where $\epsilon_{||}$ and $\epsilon_{\perp}$ are high-energy in-plane and out-of-plane dielectric constants. The lattice vectors are defined as $\mathbf{a}_1 \equiv (a, 0, 0)$, $\mathbf{a}_2 \equiv (0, a, 0)$, and $\mathbf{a}_3 \equiv (0, 0, d)$.  By Fourier-transforming $V(\mathbf{r}_i)$, one arrives at

\begin{align}
  \label{eq:long_range_coulomb_k_space}
  V_\mathbf{k} = \frac{V_c}{\gamma \left(2 - \cos(k_xa) - \cos(k_ya)\right) + 1 - \cos\left(k_z d\right)}
\end{align}

\noindent
with $V_c = e^2 d / (2 a^2\epsilon_\perp \epsilon_0)$ and $\gamma = \epsilon_{||} d^2 / (\epsilon_\perp a^2)$. Physically, the parameter $\gamma$ reflects anisotropy between in-plane and out-of-plane directions, and it is controlled both by the geometric factor $d/a$, and the ratio of dielectric constants $\epsilon_{||}/\epsilon_{\perp}$. The other parameter, $V_c$, has the units of energy and controlled the overall scale of the long-range Coulomb repulsion. Notably, the plasmon gap at the $\Gamma$-point may be estimated as $E_p \propto \sqrt{ \frac{V_c}{\gamma} \frac{n_c}{m^{*}}}$, where $n_c$ is the conduction electron concentration, and $m^{*}$ denotes effective carrier mass. The plasmon gap is thus determined by the ratio $\frac{V_c}{\gamma}$ rather than directly by $V_c$. Also, both the long-range Coulomb interaction parameters and the effective mass affect the numeric values of plasmon energy. Since $m^{*}$ is strongly renormalized by correlations close to metal-insulator transition, the latter need to be incorporated in order properly describe charge dynamics in high-temperature superconductors. For this reason, we employ VWF+$1/\mathcal{N}_f$ to study spin- and charge modes in the model~\eqref{eq:hamitlonian_hubbard_3d}. 

It should be noted that an alternative form of the long-range Coulomb potential

\begin{align}
  {V}^\prime_\mathbf{k} = \frac{2 \pi e^2}{k_{||}} \frac{\sinh k_{||} d}{ \cosh k_{||} d - \cos k_z d},
  \label{eq:energy_cont}
\end{align}

\noindent
where $k_{||}$ and $k_z$ are in-plane and out-of-plane wave vectors, respectively, is also used in literature to study plasmon excitation in layered systems \cite{CotePhysRevB1993,FetterAnnPhys1973,FetterAnnPhys1974}. Equation~\eqref{eq:energy_cont} is based on the model of stacked continuous layers, in contrast to Eq.~\eqref{eq:long_range_coulomb_k_space} which also takes into account the discrete structure of the planes. Yet, for small $k_{||}$, Eq.~(\ref{eq:energy_cont}) reduces to

\begin{align}
  {V}^\prime_\mathbf{k} \approx \frac{2 \pi e^2 d}{\frac{\left(k_{||} d\right)^2}{2} + 1 - \cos k_z d} = \frac{2 \pi e^2 d}{\frac{d^2}{a^2} \cdot \frac{\left(k_{||} a\right)^2}{2} + 1 - \cos k_z d} \approx \frac{2 \pi e^2 d}{\frac{d^2}{a^2} \cdot (2 - \cos k_xa - \cos k_y a) + 1 - \cos k_z d}.
  \label{eq:energy_cont_simplified}
\end{align}

\noindent
To make relation between Eqs.~\eqref{eq:energy_cont_simplified} and \eqref{eq:long_range_coulomb_k_space}, one needs to convert Eq.~\eqref{eq:energy_cont_simplified} to SI units and relate the continuous/discrete Fourier transforms over $k_{||}$ that have been used in respective derivations. In effect, one arrives at

\begin{align}
  \tilde{V}_\mathbf{k} \rightarrow \frac{1}{4 \pi \epsilon_0} \cdot \frac{1}{a^2} \cdot \tilde{V}_\mathbf{k} \approx \frac{\frac{e^2 d}{2 a^2 \epsilon_0}}{\frac{d^2}{a^2} \cdot (2 - \cos k_xa - \cos k_y a) + 1 - \cos k_z d},
  \label{eq:energy_cont_simplified_SI}
\end{align}

\noindent
which is equivalent to Eq.~\eqref{eq:long_range_coulomb_k_space} if $\epsilon_{||} = \epsilon_\perp = 1$. Both expressions yield thus the same results (up to anisotropic screening effects) for small $k_{||}$, where the discrete layer structure becomes irrelevant, but only Eq.~\eqref{eq:long_range_coulomb_k_space} accounts for the lattice symmetry at short distances. Even though, for large in-plane momenta, the discrete Laplace equation is not expected to quantitatively describe the interaction potential, we select the long-range interaction term in the form of Eq.~\eqref{eq:long_range_coulomb_k_space} rather then that of Eq.~\eqref{eq:energy_cont} due to its better compatibility with lattice calculations.

\begin{figure}
  \centering
  \includegraphics[width=0.6\linewidth]{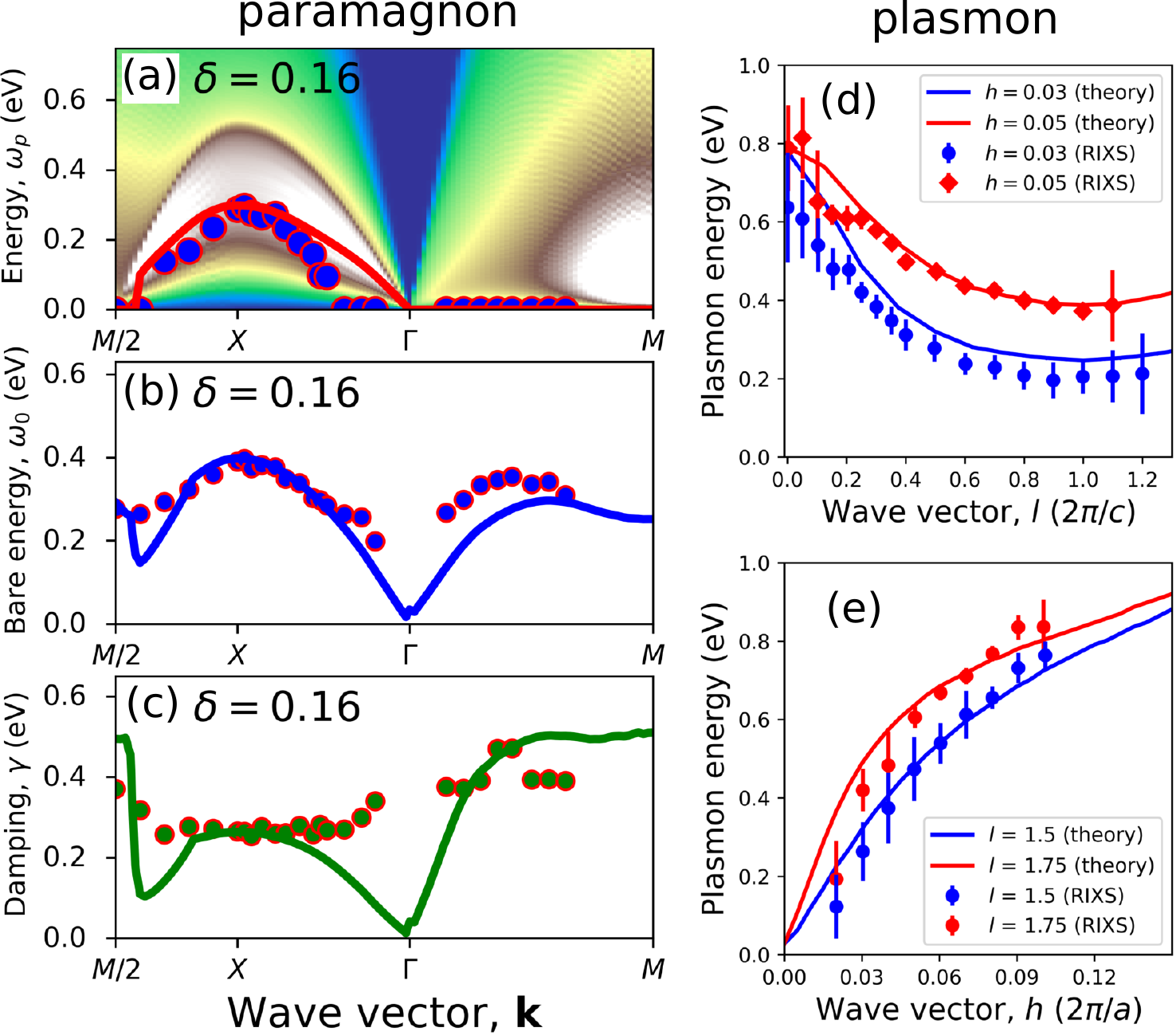}
  \caption{$\mathrm{SGA}_f$+$1/\mathcal{N}_f$ results for the three-dimensional extended Hubbard model~\eqref{eq:hamitlonian_hubbard_3d} at hole doping $\delta=0.16$, and their comparison with RIXS data (points) for $\mathrm{(Bi, Pb)_2 (Sr, La)_2 CuO_{6+\delta}}$ [(a)-(c)] \cite{PengPhysRevB2018} and $\mathrm{Bi_2Sr_{1.6}La_{0.4}CuO_{6+\delta}}$ [(d)-(e)] \cite{NagPhysRevLett2020}. Panels (a) shows the calculated imaginary part of the dynamical spin susceptibility as a color map, and the paramagnon propagation energy by red solid line. In panels (b)-(c), bare paramagnon energy and damping coefficient are displayed, respectively. Panels (d)-(e) show the plasmon energies along selected Brillouin-zone contours (details are given inside the panels). Wave vectors are represented as $\mathbf{k} = (h \frac{2\pi}{a}, 0, l \frac{2 \pi}{c})$, with $c = 2d$. After Ref.~\cite{FidrysiakArXiV2021}.}
  \label{fig:plasmons_hubbard_3d}
\end{figure}

In Fig.~\ref{fig:plasmons_hubbard_3d} we display the result of VWF+$1/\mathcal{N}_f$ calculation in the $\mathrm{SGA}_f$+$1/\mathcal{N}_f$ variant for the model~\eqref{eq:hamitlonian_hubbard_3d} at hole doping $\delta = 0.16$, and compare them with RIXS data of Refs.~\cite{PengPhysRevB2018,NagPhysRevLett2020} for $\mathrm{(Bi, Pb)_2 (Sr, La)_2 CuO_{6+\delta}}$ (points). The model parameters have been taken as: $t = -0.35\,\mathrm{eV}$, $t^\prime=0.25 |t|$, $t_z = -0.01 |t|$, $U = 6|t|$, $V_c = 46 |t|$, and $\gamma = 10$. The scale of $\gamma$ reflects the ratio out-of-plane to in-plane spatial anisotropy for bismuth based cuprates. The temperature is set to $k_B t = 0.4 |t|$ to stay clear of magnetic instabilities. In panel (a), the calculated imaginary part of dynamic spin susceptibility is shown as a color map. Red solid line represents paramagnon propagation energy, as obtained from the harmonic oscillator model~\eqref{eq:damped_oscillator}, whereas points are the corresponding experimental data. The agreement is quantitative along all considered Brillouin-zone direction. Strongly anisotropic damping of spin excitations is apparent, both in theory and experiment. Namely, propagating paramagnons persist only along the nodal ($\Gamma$-$M$) line, whereas they are overdamped in the nodal direction. This finding is consistent with the former analysis, carried out for a planar model (cf. \cite{FidrysiakPhysRevB2020} and Sec.~\ref{sec:2d_hubbard_fluctations}), pointing toward predominantly two-dimensional nature of magnetic fluctuations. The calculated intensity maps provide also other characteristics of the paramagnon peak, including bare energy, $\omega_0(\mathbf{k})$ and damping coefficient, $\gamma(\mathbf{k})$. Those are presented and compared with RIXS values in Fig.~\ref{fig:plasmons_hubbard_3d}(b) and (c), respectively. In panels (c) and (d),  plasmon energies along selected contours in the three-dimensional Brillouin zone are displayed. Wave vectors are represented as $\mathbf{k} = (h \frac{2\pi}{a}, k \frac{2\pi}{a}, l \frac{2\pi}{c})$, where $c \equiv 2d$ to account for two primitive cells in crystallographic cell \cite{KovalevaPhysRevB2004}. Fig.~\ref{fig:plasmons_hubbard_3d}(c) represents thus out-of-plane scans for two selection of the in-plane wave vector, which exhibits a substantial dispersion as a function of $l$, with plasmon minimum energy located at $l = 1$. This value corresponds to antiphase charge oscillation on neighboring copper-oxygen planes and evidences a three-dimensional character of charge excitations. Panel (e) demonstrates sensitivity of the in-plane cuts to the value of $l$ (parameters are detailed inside the panels). As can be in panels in Fig.~\ref{fig:plasmons_hubbard_3d}(a)-(e), extended three-dimensional Hubbard model, treated within VWF+$1/\mathcal{N}_f$ approach, allows to reproduce both spin- and charge dynamics in hole-doped Bi-based cuprates (points).

\subsection{A generalization: Collective excitations in the $t$-$J$-$U$ model}

In Secs.~\ref{sec:theoretical_models}-\ref{sec:selected_equilibrium_properties}, we overviewed the most relevant models of correlated lattice electrons, encompassing Hubbard, $t$-$J$, $t$-$J$-$U$, as well as those based on extended multi-orbital Hamiltonians. In regard to equilibrium quantities and single-particle dynamics, the $t$-$J$-$U$ model has been argued to provide a more consistent overall description of cooper-based high-temperature superconductors then either Hubbard or pure $t$-$J$ model (see Sec.~\ref{sec:selected_equilibrium_properties}). A question remains, whether the $t$-$J$-$U$ model is capable of simultaneous quantitative description of equilibrium properties and collective excitation spectra of those materials. Here we elaborate on the extension of the VWF+$1/\mathcal{N}_f$ formalism to the case of two-dimensional square-lattice $t$-$J$-$U$ model

\begin{align}
  \label{eq:tju-model-flukt}
  \hat{\mathcal{H}} = \sum_{i\neq j, \sigma} t_{ij} \hat{a}_{i\sigma}^\dagger \hat{a}_{j\sigma} + U \sum_i \hat{n}_{i\uparrow} \hat{n}_{i\downarrow} + J \sum_{\langle i, j\rangle} \hat{\mathbf{S}}_i \cdot \hat{\mathbf{S}}_j,
\end{align}

\noindent
where the notation is standard (cf. Sec.~\ref{sec:theoretical_models}). Such an analysis has also a methodological motivation. Namely, the Hamiltonian~\eqref{eq:tju-model-flukt} encompasses both the Hubbard- and $t$-$J$ models as special cases ($J=0$, $U \neq 0$ and $J \neq 0$, $U\rightarrow\infty$, respectively), and thus allows for a systematic comparison between them. In order to interpolate between those two extreme limits, we introduce effective exchange interaction $J_\mathrm{eff} \equiv J + \frac{4t^2}{U}$ \cite{Fidrysiakarxiv2021_2} that combines explicit exchange, entering Eq.~\eqref{eq:tju-model-flukt}, with the kinetic exchange interaction that arises in the canonical perturbation expansion for the Hubbard model (cf. Appendix~\ref{appendix:derivation_of_the_tj_model}).

\begin{figure}
  \centering
  \includegraphics[width=\linewidth]{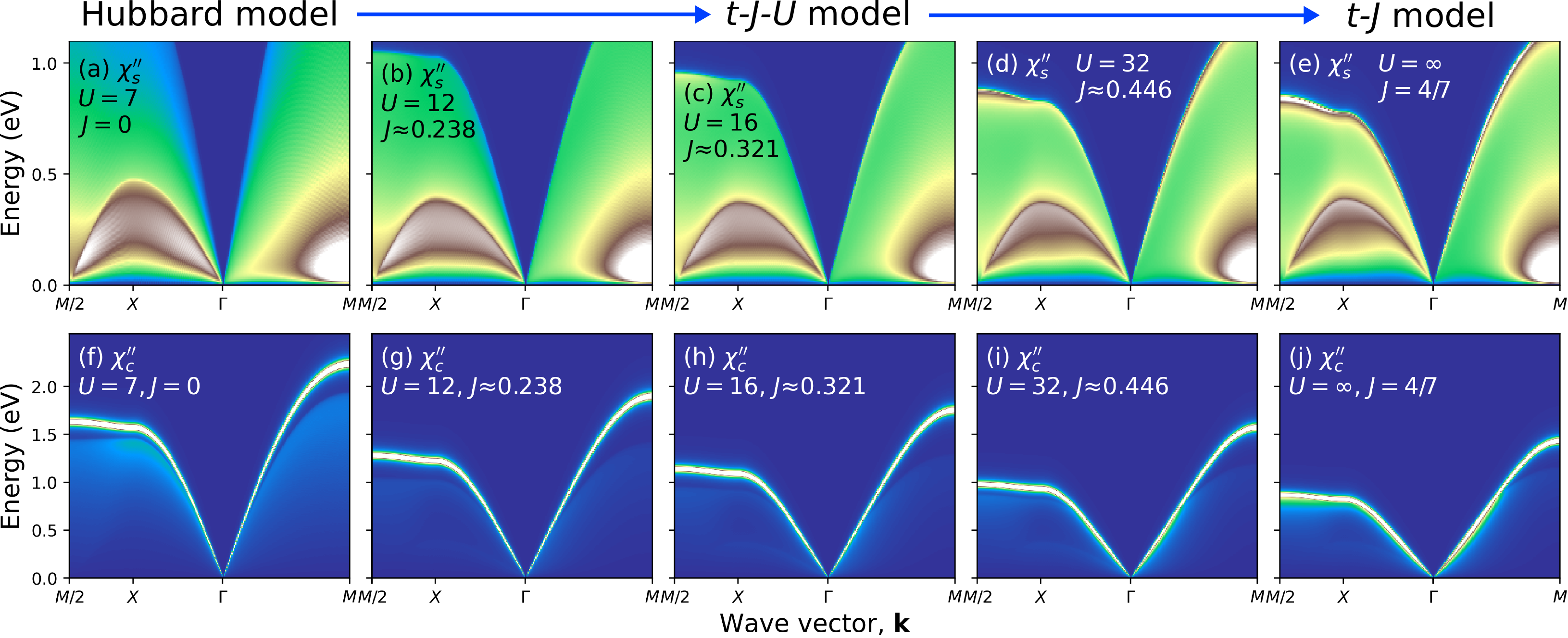}
  \caption{Imaginary parts of the dynamical spin (to panels) and charge (bottom panels) susceptibilities for the $t$-$J$-$U$ model, obtained using the VWF+$1/\mathcal{N}_f$ approach. The model parameters have been set to $t \equiv - 0.35\,\mathrm{eV}$ and $t^\prime \equiv 0.25 |t|$, and $J_\mathrm{eff} \equiv 0.2\,\mathrm{eV}$. The values of $U$ and $J$ are detailed inside the panels. After~\cite{Fidrysiakarxiv2021_2}.}
  \label{fig:collective_excitations_tju_model}
\end{figure}

In the VWF+$1/\mathcal{N}_f$ calculation, we retain only nearest- and next-nearest-neighbor hopping integrals, $t \equiv - 0.35\,\mathrm{eV}$ and $t^\prime \equiv 0.25 |t|$, respectively. By keeping the constant value of the effective exchange, $J_\mathrm{eff} \equiv 0.2\,\mathrm{eV}$, in Fig.~\ref{fig:collective_excitations_tju_model} we show the evolution of the imaginary parts of the dynamical spin- and charge susceptibilities from the Hubbard (left panels) to the $t$-$J$-model limits (right panels). The spectra are presented along high-symmetry $M/2$-$X$-$\Gamma$-$M$ Brillouin-zone contour. In the Hubbard model limit (panels (a) and (f)), we retrieve the results obtained previously for La- and Bi-cuprates \cite{FidrysiakPhysRevB2020}. A well-defined paramagnon emerges along the $\Gamma$-$X$ line, but is absent in the nodal ($\Gamma$-$M$) direction. Remarkably, as one moves towards the $t$-$J$-model limit with fixed $J_\mathrm{eff}$ (panels (a)-(f)), the paramagnon peak energy remains essentially unchanged along the $\Gamma$-$X$ direction, suggesting that the same level of agreement with experiment, obtained within the Hubbard model, could be also achieved within the $t$-$J$-$U$-model scheme. Moreover, this behavior suggests that $J_\mathrm{eff}$ may be an appropriate microscopic parameter for a quantitative interpretation of magnetic spectra within the class of $t$-$J$-$U$ Hamiltonians. In Fig.~\ref{fig:collective_excitations_tju_model}(f), one can also see a sharp charge excitation emerging above the continuum. In contrast to the layered model, discussed in Sec.~\ref{subsection:plasmons_3d_model}, now the charge excitations are gapless and approach zero energy close to the $\Gamma$-point (see Fig.~\ref{fig:plasmons_hubbard_3d} and Ref.~\cite{FidrysiakArXiV2021}). The reason is that the long-range Coulomb repulsion has not been included in the $t$-$J$-$U$ model~\eqref{eq:tju-model-flukt}. As follows from Fig.~\ref{fig:collective_excitations_tju_model}(f)-(j), the charge dynamics is highly sensitive to the magnitude of the Hubbard $U$, as opposed to spin excitations (panels (a)-(f)).

Those preliminary results suggest that the $t$-$J$-$U$ model should be appropriate for a quantitative analysis of the spin- and charge excitations in layered copper oxides, and show that magnetic dynamics remains sensitive to the overall scale of kinetic exchange interactions. The latter observation rationalizes remarkably similar structure of the spin excitations seen experimentally in both magnetic-insulator- and paramagnetic-metal state. Moreover, the same kinetic exchange process provides also the pairing potential within the local-pairing scenario of high-$T_c$ superconductivity (cf. Sec.~\ref{sec:selected_equilibrium_properties}). Obviously, the correlations induced by the Hubbard term are equal, or even higher, importance in the resultant picture of both collective excitations and pairing.

\section{Summary and Outlook} \label{sec:summary}

\subsection{Selective overview of results from other approaches}

Whereas this review is focused predominantly on variational-wave-function perspective, over the years a number of other theoretical schemes have been developed to study ground state collective excitation dynamics and their contribution to thermodynamics in strongly-correlated electronic systems, as mentioned in Sec.~\ref{sec:introduction}. For completeness, we provide a brief and selective overview of those methods. The latter can be generally classified according to how the interparticle interactions are handled. Perturbative methods are based on a small parameter of physical significance (e.g., the ratio of the single-particle hopping integral to  interaction magnitude). On the other hand, non-perturbative approximations are constructed on the basis of less obvious expansion parameters that are not related directly to the interaction magnitude (e.g., spatial dimension, number of field components or range of correlations, that need to be set to a finite value at the end of calculation). Finally, there are truly non-perturbative approaches without small parameters, which are typically implemented as large-scale numerical simulation techniques, such as quantum Monte-Carlo or numerical renormalization group methods. 

The most direct approach to study the effects of quantum fluctuation is weak-coupling expansion, where the ratio of the interaction magnitude to single-particle hopping $U/|t| \ll 1$ is selected as an expansion parameter. Application of the loop expansion directly to the free energy $F = - \frac{1}{\beta} \ln \mathrm{Tr} \exp(-\beta\hat{\mathcal{H}})$, with $\beta=1/(k_BT)$, results in weak-coupling diagrammatic series in the powers of $U/t$, making the procedure by construction suitable for weak-coupling regime. Yet, there is a subtle point that may render the naive version of this technique inapplicable even at an \emph{arbitrarily weak-coupling}. Namely, both in the Hartree-Fock theory of antiferromagnetism and the Bardeen-Cooper-Schrieffer theory of conventional superconductivity, order parameter is an exponentially small function of the inverse of the interaction strength (i.e., of $1/(\rho V)$) and, as effect of that, any power series expansion cannot reliably describe them. For this reason, a class of more refined weak-coupling schemes were developed and subsequently applied to model Hamiltonians \cite{VanDongenPhysRevLett1991,VanDongenPhysRevB1994,vanDongenPhysRevB1996}. The idea is based on carrying out Legendre transform of the free energy with respect to the time-independent auxiliary fields, conjugate to physical observables of interest (e.g., the superconducting gap, staggered magnetization or staggered charge density). The latter are then kept constant throughout the expansion of the resulting functional in the power series in $U/t$. This is effectively equivalent to a non-perturbative resummation of Feynman diagrams, resulting in expansion of the two-particle-point-irreducible (2PPI) type \cite{RentropJPhysA2015,SmetPhysRevD2002}. The main difference between the naive and improved scheme is that in the latter the mean-fields are regarded as independent of interaction magnitude throughout the weak-coupling expansion, but are obtained by minimization of the appropriate thermodynamic potential. Parenthetically, the Landau expansion with respect to the order parameter belongs to this class of approach on the phenomenological level.

We now turn to selected techniques not relying the interaction magnitude as an expansion parameter. One of them is slave-boson theory due to Kotliar and Ruckenstein \cite{KotilarPhysRevLett1986}. It is based on introduction of four families of boson fields, as well as the corresponding creation (annihilation) operators, $\hat{e}_i^\dagger$ ($\hat{e}_i$), $\hat{p}_{i\sigma}^\dagger$ ($\hat{p}_{i\sigma}$), and $\hat{d}_i^\dagger$ ($\hat{d}_i$).  The former represent empty, singly-occupied (either spin-up or spin-down), and doubly occupied sites, respectively. The formulation, based on slave bosons, operates on the enlarged Hilbert space, hence the unphysical states must be removed by local constraints, $\hat{p}_{i\uparrow}^\dagger\hat{p}_{i\uparrow} + \hat{p}_{i\downarrow}^\dagger\hat{p}_{i\downarrow} + \hat{e}_{i}^\dagger\hat{e}_{i} + \hat{d}_{i}^\dagger\hat{d}_{i} \equiv 1$, and $\hat{a}_{i\sigma}^\dagger \hat{a}_{i\sigma} = \hat{p}_{i\sigma}^\dagger\hat{p}_{i\sigma}  + \hat{e}_{i}^\dagger\hat{e}_{i}$. The last condition relates the occupation number operator, expressed in terms of the fermionic original creation (annihilation) operators $\hat{a}_{i\sigma}^\dagger$ ($\hat{a}_{i\sigma}$), to those slave bosons. The method is based on a functional integral representation of the model, with the constraints enforced by means of time-dependent Lagrange multipliers. This technique yields a modified Gutzwiller-type (SGA) saddle-point solution and allows to systematically incorporate fluctuations as higher order corrections in the path integral scheme. Yet, special care is needed in order not to violate symmetries of the original Hamiltonian while working in extended Hilbert space, involving unphysical field configurations \cite{LiPhysRevB1989}. For recent applications of slave-bosons formalism to one-orbital and extended Hubbard models, see Refs.~\cite{GeorgescuPhysRevB2015,RieglerPhysRevB2020}.

Another class of approximations is based on Hubbard operators, defined as $X_i^{pq} \equiv |ip\rangle\langle iq|$, where the indices $p$ and $q$ denote local basis states for site $i$ (i.e. they run over the following values: empty site, spin up/down electron, and doubly occupied configuration). The latter follow a simple multiplication rule $X_i^{p_1 q_1} X_i^{p_2 q_2} = \delta_{q_1 p_2} X_i^{p_1 q_2}$, and obey the algebra

\begin{align}
  \left[X_i^{p_1 q_1}, X_j^{p_2 q_2}\right] = \delta_{ij} \left(\delta_{q_1 p_2} X_i^{p_1 q_2} - \delta_{q_2 p_1} X_i^{p_2 q_1}\right),   \label{eq:hubbard_ops_algebra_commutator} \\
  \left\{X_i^{p_1 q_1}, X_j^{p_2 q_2}\right\} = \delta_{ij} \left(\delta_{q_1 p_2} X_i^{p_1 q_2} + \delta_{q_2 p_1} X_i^{p_2 q_1}\right). \label{eq:hubbard_ops_algebra_anticommutator}
\end{align}

\noindent
By construction, Hubbard operators are particularly well suited for the Hubbard model with infinite on-site repulsion, since then the doubly occupied sites are excluded from the Hilbert space, effectively reducing the number of independent $X_i^{pq}$ fields. The same simplifications occur for the $t$-$J$ models, where the subspace involving doubly-occupied sites is decoupled from the low-energy sector of the Hilbert space by the canonical transformation. The circumstance that the Hubbard operators commute (anticommute) to linear combinations of themselves, allows to prove generalized Wick's theorem \cite{WestwanskiPhysLettA1973}, and devise specialized diagrammatic techniques for models expressed in terms of $X_i^{pq}$ \cite{ZaitsevSovPhysJETP1976,IzyumovJPCM1991,IzyumovPhysRevB1992,OvchinnikovBook2004}. Fluctuations are then incorporated by resummation of loop diagrams, leading to the so-called generalized random-phase-approximation \cite{IzyumovJPCM1992}. This technique is formulated within the operator formalism. Alternative, path-integral-based approximations for Hubbard-operators have been also developed \cite{FoussatsPhysRevB2002,FoussatsPhysRevB2004} and subsequently applied to interpret charge dynamics in high-temperature copper-oxide superconductors \cite{GrecoCommunPhys2019,GrecoPhysRevB2020,NagPhysRevLett2020}.

Another family of techniques is based on generalization of the variational wave function approach to the dynamical situation. A feasible route to carry out this program is to adopt real-time equation of motion approach to the Gutzwiller functional, resulting in the so-called GA+RPA technique \cite{SeiboldPhysRevLett2001,SeiboldPhysRevB2004,SeiboldPhysRevB2005,SchiroPhysRevLett2010}. The latter has been applied to address spin fluctuations in correlated systems of lattice fermions \cite{MarkiewiczPhysRevB2010,BunemannNJP2013}. A generalization of this type of approach, involving time-dependent canonical transformation of the variational states, has been developed subsequently \cite{WysokinskiPhysRevB2017,WysokinskiPhysRevB2017_2}. The path integral VWF+$1/\mathcal{N}_f$ method, reviewed in Sec.~\ref{sec:VWF+1/N}, also falls into this category.

The aforementioned theoretical schemes are intended for a microscopic analysis, i.e., they are directly applicable to microscopic Hamiltonians. Yet, complexity of real materials of interest often renders those approaches unreliable or impractical, particularly if the problem involves multiple competing energy scales or a large number of interrelated degrees of freedom. Under such circumstances, a feasible strategy is then to start with a semi-phenomenological model which for the cuprates may take the form

\begin{align}
  \label{eq:semi-phenomenological_model}
  \mathcal{S} = -\int d\tau \int d\tau^\prime \left\{\sum_{\mathbf{k}, \sigma} \bar{\eta}_{\mathbf{k}\sigma} \hat{G}_{0, \mathbf{k}}^{-1}(\tau-\tau^\prime) \eta_{\mathbf{k}\sigma}  + g^2 \frac{2}{3} \chi_\mathbf{q}(\tau-\tau^\prime) \hat{\mathbf{S}}_\mathbf{q} \hat{\mathbf{S}}_{-\mathbf{q}}\right\},
\end{align}

\noindent
where $\hat{S}_\mathbf{q}$ are spin operators expressed in terms of Grassmann fields, $\eta_{\mathbf{k}\sigma}$ and $\bar{\eta}_{\mathbf{k}\sigma}$, $\hat{G}_{0, \mathbf{k}}(\tau-\tau^\prime)$ is the bare single-particle fermionic Green function, and $g$ plays he role of a coupling constant between quasiparticles and collective excitations. The detailed form of the dynamic spin susceptibility, $\chi_\mathbf{q}(\tau-\tau^\prime)$, is determined by fitting to available neutron scattering data. This makes the model phenomenological, but allows for describing coupled spin- and single-particle excitations with realistically modeled magnetic subsystem. In particular, the model~\eqref{eq:semi-phenomenological_model} has been used to study pseudogap behavior in cuprate superconductors \cite{SchmalianPhysRevLett1998,SchmalianPhysRevB1999}. Analogous approach has been extensively applied also to other materials, such as heavy-fermion superconductors (see, e.g., \cite{TadaJPhysConfSeries2013}). Moreover, the semi-phenomenological approach combining experimental fits with microscopic calculations is a common way of evaluating fluctuation-driven superconducting pairing mechanisms. An example of such analysis is provided in Ref.~\cite{LeTaconNatPhys2011}, where Eliashberg theory with measured magnetic spectra as an input was demonstrated to reproduce experimental superconducting transition temperatures for YBCO up to a factor of two.

Finally, there exist several large-scale numerical simulation techniques, suitable for addressing both static properties and collective excitations. The approach that stand out in this category is determinant quantum Monte-Carlo (DQMC), based on statistical sampling of bosonic field configurations \cite{BookBecca}. It is well suited for finite-temperature (imaginary-time) computations and provides equilibrium expectations values in and controlled manner. Also, since arbitrary field configurations are allowed throughout the simulation, DQMC scheme does not suffer from Fierz problem, outlined in Sec.~\ref{pni_sec:Mean_Field_Ambiguity}, which is a notable advantage. Application of DQMC to fermionic systems poses, however, a challenge due to the sign problem \cite{LohPhysRevB1990} and an analytic continuation procedure needed to access dynamical properties (such as real-time Green's functions or dynamical susceptibilities) \cite{TripoltComPhysCommun2019}. DQMC has been used extensively to study spin fluctuations in the cuprates \cite{KungPhysRevB2015,JiaNatCommun2014,PengPhysRevB2018}. Also, the method can be applied to relatively small lattices, typically up to $10\times 10$ lattice sites. In effect, plasmon dynamics is more challenging from the DQMC perspective as it requires considering large three-dimensional lattices and long-range Coulomb interactions. Nevertheless, it should be emphasized once more again that all the above methods assume that the systems are strongly correlated at the outset.

\subsection{Theoretical models: Their methodology and selection}

All physically relevant microscopic models of the ordering and fluctuations in the cuprates must take into account the strong correlations emerging from the fact that their reference state is that of the Mott or Mott-Hubbard insulator, i.e., the underlying fermions must undergo a localization $\rightarrow$ delocalization (Mott insulator to correlated metal) transition upon their doping or other external influence (e.g., pressure). Parenthetically, whereas the cuprates can be placed on the strong-correlation side of the Mott-Hubbard boundary, the iron pnictides are on the opposite side, involving moderately correlated fermions. Additional factor to the relative-to-bare-bandwidth strength is the concomitant atomic disorder \cite{ByczukPhysRevLett2005}, which should be incorporated into any reliable description of the phase diagram involving SC, CDW, and other states. However, there still exist models which provide attractive overall picture of whole classes of phenomena for those systems without taking into account the atomic disorder, as elaborated in detail above. Here we briefly overview those principal models, sometimes repeating their main features, to specify their fundamental (also historical) characteristics.

\subsubsection{Hubbard model}

The Hubbard model (1963) is qualitatively based on starting from atomic states as a reference macroscopic state of electrons. This picture differs in a fundamental manner from that starting from the electron gas and adding (screened) interaction amongst them in a perturbative way, culminating in the Luttinger theorem (1961). The starting point of the Hubbard model differed also essentially from the Landau Fermi liquid picture \cite{VarmaPhysRep2002}, in which the starting point was also that of Fermi gas which, in turn, under influence of interaction exhibits renormalized properties of electron mass, enhanced paramagnetic susceptibility, as well as, in its normal phase, possesses two branches of collective charge excitations (sounds) and one of magnetic (paramagnons). The statistical entropy in the low-$T$ regime is that of fermion gas of renormalized particles (quasiparticles), i.e., the statistics consists of a modified Fermi-Dirac distribution. No phase transition of the Mott-Hubbard type was discussed within the original Fermi-liquid picture.

In supplement to the Fermi-liquid properties \cite{VollhardRevModPhys1984}, in the subsequent solution of the Hubbard model (1964) \cite{HubbardProcRoySoc1964}, the concept of the Hubbard subbands was introduced, i.e., of the splitting into two halves of the original band in the normal metallic state of ordinary metals, explaining automatically why the systems with odd ($n = 1$) filling are insulators. The concept of Hubbard subbands is extremely important also for non-integer filling ($n \neq 1, 2$) as then the metallic state is stable even in the strong-correlation limit, but the fermionic states are not those of ordinary electrons, as during their motion throughout the system they avoid double occupancies, i.e., the fermionic annihilation and creation operators ($\hat{a}$ and $\hat{a}^\dagger$, respectively) obey non-fermion anticommutation relations (cf. Appendix~\ref{appendix:derivation_of_the_tj_model}). Those two principal characteristics, the Hubbard subbands structure and non-fermion anticommutation relations for projected fermions form the essence of strongly-correlated behavior from a formal standpoint. We introduce next the third principal ingredient, the \emph{kinetic exchange interaction}, summarized below within the $t$-$J$ model. Parenthetically, the limitation of the Coulomb interactions to the Hubbard intraatomic part for strongly- correlated (near localized) electrons may be explained in elementary fashion as follows. As we have discussed throughout the present review, the intersite Coulomb interaction, characterized by parameter $V$, is defined as

\begin{align}
  \label{eq:Vij_discussion}
  V_{ij} = \int d^3\mathbf{r} \int d^3\mathbf{r}^\prime |w_i(\mathbf{r})|^2 \frac{e^2}{|\mathbf{r} - \mathbf{r}^\prime|} |w_j(\mathbf{r}^\prime)|^2.
 \end{align}

 \noindent
 This resembles the well-known classical expression of the Lenard-Wiechert potential from classical electromagnetism, in which $n_i(\mathbf{r}) \equiv n(\mathbf{r} - \mathbf{R}_i) \equiv |w_i(\mathbf{r})|^2$ represents the electron density of charge centered at site $\mathbf{R}_i$ and $n_j(\mathbf{r}^\prime) \equiv |w_j(\mathbf{r}^\prime)|^2$ is the charge density centered at site $j$. Now, if we start with atomic states of electrons, the electron density $n_i(\mathbf{r})$ can be roughly estimated as $\sim e/a_B$ for $|\mathbf{r} - \mathbf{R}_i| \leq a_B$  and vanishing otherwise. In that reference situation if only $|\mathbf{R}_i - \mathbf{R}_j| > a_B$, we can approximate the interaction to the form $V_{ij} = U \delta_{ij}$, reducing it to the Hubbard form. An essential feature is thus the almost atomic character of the Wannier functions with $|\mathbf{R}_i - \mathbf{R}_j|$ substantially larger that their effective Bohr-like size  amounting physically to the condition that the overlap between the neighboring reference atomic states is becoming small. However, in real systems, the intersite interactions are important, as discussed in the main text.

 \subsubsection{$t$-$J$ model}

 As already mentioned above, the third signature of strong correlations is the antiferromagnetic kinetic exchange interaction among electrons in the lower Hubbard subband, reducing to the Heisenberg-type antiferromagnetic exchange interaction for the localized electrons in the Mott insulator. The kinetic exchange in the last case was introduced by Anderson (1959) \cite{AndersonPhysRev1959}. In fact, in this original work the Hubbard model was introduced for the first time, albeit only in passing and regarded as a starting point when deriving the kinetic exchange interaction. This analysis has been extended (1976) \cite{SpalekPreprint_tJ,ChaoJPCM1977} to treat the magnetism of strongly-correlated metallic systems for partially filled ($n \leq 1$) Mott Hubbard subband. The effective Hamiltonian obtained in this manner has been termed the $t$-$J$ model, as it contains the restricted (projected) hopping term $\sim \sum_{ij} t_{ij} \hat{b}_i^\dagger \hat{b}_j$ and the kinetic exchange interaction $\sim \sum_{ij} J_{ij} \left(\hat{\mathbf{S}}_i \hat{\mathbf{S}}_j - \frac{1}{4} \hat{\nu}_i \hat{\nu}_j\right)$ (cf. Appendix~\ref{appendix:derivation_of_the_tj_model} for details). In the analysis of $t$-$J$ model it is assumed that the Hubbard subband structure which appears naturally in the Mott insulator limit (with inter-subband splitting $\sim U-K$) survives when the metallic state is stable for which the Hubbard contribution to the system energy, $U d^2$ is strongly reduced upon increasing the hole concentration (for the optimally doped HTS, i.e., for $\delta \sim 0.2$ it is reduced by about factor of 2, i.e., it is $U/8$, whereas the bare hopping-energy magnitude increase is $\sim \frac{W}{2} \cdot (1 -\delta) \cdot \delta$, i.e. at the optimal doping $\sim 0.1 W$). In effect, for the carriers near the Fermi energy the carriers may be moderately correlated while retaining their strong-correlation nature deeper below the Fermi surface (see Sec.~\ref{subsection:k-DE-GWF} for detailed discussion). 

 The fundamental insight into the $t$-$J$ model, originally invented to describe magnetism, has been provided by Anderson \cite{AndersonScience1987,AndersonProcIntSchollEnricoFermi1988} who made a suggestion that spin-singlet nature of antiferromagnetic kinetic exchange can be understood in categories of the local spin-singlet pairing. This type of pairing shows up as either the floating spin-singlets in the Mott insulator (termed as \emph{resonating valence bond} or spin-liquid state) or else as the itinerant state involving the spin-singlet pairs, which may condense into the unconventional superconductor with real space pairing $\sim J_{ij}$. An avalanche of theoretical papers followed this suggestion and the \emph{renormalized mean-field theory} has been created (see, e.g., Refs.~\cite{RuckensteinPhysRevB1987,ZhangSpuSciTech1988,KotliarPhysRevB1988,ParamekantiPhys2004,AndersonJPhysCondensMatter2004,EdeggerAdvPhys2007,JedrakPhsRevB2011,RanderiaBook2011}) which is based on perturbation expansion of the Hubbard model \cite{ChaoJPCM1977,ChaoPhysRevB1978} and on the Gutzwiller-type approximation, adopted to the situation with the double occupancies excluded from the unprotected $t$-$J$ model.

At this point, critical remarks are in place. First, the $t$-$J$ model in its explicit form can be derived only in the strong-correlation limit $|t| \ll U$ and then the removal of the double occupancies must be carried out on the Hamiltonian level. In such a situation, using formally much more involved atomic (projected) representation may provide better results, as doing the Gutzwiller approximation on the unprojected $t$-$J$ Hamiltonian reintroduces back double occupancies and, in effect, leads to Fermi-liquid type of the resultant state of the system. Second, the plain Gutzwiller approximation (``plain-vanilla'' state \cite{AndersonJPhysCondensMatter2004}) is not statistically consistent in the Bogoliubov sense, i.e., the equilibrium state obtained by solving self-consistent \emph{does not} coincide with that obtained from related variational principle (see, e.g., \cite{JedrakPhsRevB2011}). In this respect, we mention again, that in order to retain the consistency one has either to solve the $t$-$J$ Hamiltonian within the slave-boson approach in the saddle-point approximation or start from the \emph{statistically-consistent Gutzwiller approximation} \cite{SpalekPhysRevB2017,ZegrodnikPhysRevB2018,JedrakArXiV2010}. The latter way is clearer physically as it does not involve any spurious Bose condensation of the slave-boson fields. SGA is regarded as a zero-order approximation to both the systematic DE-GWF analysis (cf. Sec.~\ref{sec:selected_equilibrium_properties}), as well as to that including quantum spin and charge fluctuations (cf. Sec.~\ref{sec:fluctuations}).

 \subsubsection{$t$-$J$-$U$-$(V)$ model as the most general single-band model}

 To avoid some of the questions raised above, we have proposed to start from the $t$-$J$-$U$ model \cite{SpalekPhysRevB2017}, analyzed extensively in Sec.~\ref{sec:selected_equilibrium_properties}. As already said, the model in the limit $U \rightarrow \infty$ reduces to the (projected) $t$-$J$ model and for $J = 0$ the Hubbard model is recovered. Furthermore, by including the intersite Coulomb interactions $\sim V_{ij} \hat{n}_i \hat{n}_j$, some results such as the upper critical hole concentration for disappearance of the $d$-wave SC state or analysis of CDW states, can be put on a quantitative agreement with experiment. The obvious formal advantage of discussing the $t$-$J$-$U$-$(V)$ model is the circumstance that then we do not have to introduce the projected-fermion representation. The other is related to the fact that, at least within the variational approach, the model fits better experiment in quantitative manner for the cuprates. In brief, $t$-$J$-$U$-$(V)$ model represents the most general single-band model of high-temperature superconductivity in the cuprates. The presence of quite few microscopic parameters in it is caused by the circumstance that, in the present situation, there is a number of physical processes for which their energy scales are close to each other. This state of affairs manifests itself with the emergence of a number of phases competing among themselves (CDW with SC, antiferromagnetism with superconductivity, localization with itinerancy, etc.).

One formal aspect of the $t$-$J$-$U$-$(V)$ model and its relation to the (extended) Hubbard model is worth mentioning. Namely, in the former, the projected fermion pairing operators have been introduced first in \cite{SpalekPhysRevB1988_2}, whereas their unprojected correspondants in \cite{BaskaranSolStateCommun1987}. In that manner, the condensed SC solution appears already on the (renormalized) mean-filed theory level, unlike in the Hubbard model. This situation is consistent with the fact that $t$-$J$ model contains already some dynamic pairing correlation (the virtual intersite hopping), whereas in the latter it must be obtained still in a more laborious process of solving the model. However, as $U$ is not much larger than the bare bandwidth ($U \sim 2.5$-$3 W$), the choice of $t$-$J$-$U$ model is, in our view, better appealing to the physical intuition then the exact $t$-$J$ model with all its intricacies.
 
\subsubsection{Multiband models}

Here we have two classes of models, namely the hybridized models of Anderson-lattice type as applied to heavy fermions, and the $d$-$p$ models applicable to the cuprates and iron SC. We characterize briefly the $d$-$p$ model case only. Then, the three-band model involves atomic $d_{x^2-y^2}$ states due to $3d$ ions ($\mathrm{Cu}^{2+}$ in this case), hybridized with the itinerant $p_x$ and $p_y$ states due to oxygen $\mathrm{O}^{2-}$. The model seems to contain an essential physics of the system composed of $\mathrm{CuO_2}$ clusters arranged into a two-dimensional copper-oxygen plane. In some realistic calculations, including that discussed by us in Sec.~\ref{subsec:results_for_the_3_band_model}, no concept of \emph{Zhang-Rice singlet} is involved to describe principal electronic properties. Instead, crucial is the accordance of the three-band model results with those obtained within single-band approach which already provides a quite successful description of principal experimental data (cf. Sec.~\ref{sec:selected_equilibrium_properties}).

At this stage, a further comparison of the three-band model results with inclusion of both the Kondo and antiferromagnetic superexchange with those of the single-band $t$-$J$-$U$-$(V)$ model would be desirable, but not so simple to analyze. This is because such an extended model would contain even two more parameters, the Kondo and superexchange interaction constants. In effect, constructing a consistent quantitative scheme of comparison with experiment is much more intricate.
 
\subsection{Outlook}

The purpose of the present report was to determine and analyze from the variational approach perspective the non-local amendment to mean-field theory down to the point where a (semi)quantitative comparison with experiment is possible. Parenthetically, the original formulation of the BCS theory \cite{BardeenPhysRev1957} had also a self-consistent variational character, and this coincidence is not accidental. This is because any theory of superconducting pairing involves introducing emergent anomalous averages of the type $\langle \hat{a}^\dagger \hat{a}^\dagger\rangle$ which must be obtained from a proper optimization procedure. Here we have to extend this original, essentially Hartree-Fock, procedure in several respects. The first and foremost of them is to include \emph{strong correlations} in both normal and broken-symmetry states. This means that strong real-space correlations determine the $\mathbf{k}$-space properties and nature of pairing. The second is that it must encompass both the (magnetic) Mott-Hubbard insulator and normal (non-superconducting) border cases with no or sufficient number of carriers, respectively. The third is the complex phase diagram involving a number of phases, with the chemical potential being adjusted self-consistently in each of them.

In the process of executing the above program, it turned out that even the mean-field variational approach must be reformulated in the sense that the so-called renormalized mean-field theory (RMFT) must be corrected. The first correction relies on making the whole approach \emph{statistically consistent}, i.e., satisfying the \emph{Bogoliubov paradigm} that the self-consistent fields and correlations calculated either from the corresponding self-consistent equations or obtained variationally, must coincide \cite{JedrakArXiV2010}.  In turn, this leads to the \emph{statistically-consistent-Gutzwiller approximation} (SGA) which requires extra conditions which have lead to introducing subsidiary constraints. Those constraints may be interpreted physically. In effect, SGA method provides results equivalent to those obtained within a proper slave-boson approach (SBA) in the saddle point approximation \cite{KotilarPhysRevLett1986}. However, in distinction to SBA, our scheme contains only physical fields and averages so there are no auxiliary boson fields needed. In effect, no spurious Bose-Einstein condensation appears as only expectation values of fermionic composite fields are present.

The principal message of the report is to overview the systematic extension of predetermined SGA in the form of a diagrammatic expansion of the Gutzwiller wave function (DE-GWF, cf. Sec.~\ref{sec:vwf_solution}), as well as to incorporate the quantum spin and charge excitations next within the lowest order of $1/\mathcal{N}_f$ expansion, starting from the correlated stationary state in the form of SGA (cf. Sec.~\ref{sec:fluctuations}). The full approach, combining DE-GWF with $1/\mathcal{N}_f$ expansion still needs to be formulated, hopefully in the near future. Even a fully developed SGA+$1/\mathcal{N}_f$ theory remains still to be carried out in detail, as here the effect of fluctuations on the physical properties such as, e.g., electrical resistivity in the normal state or the system free energy, have not been addressed as yet. Minimally, the overviewed approach may be looked upon as the first step towards a realistic account of both strong correlations and quantum fluctuations combined into a single microscopic picture. This statement is based on the fact that our theory is compared directly with experiment and, for selected properties, in a fully quantitative manner. We have chosen this type of \emph{modus operandi}, since there are quite few other methods to provide reasonable (but not the same) theoretical results (e.g., cluster DMFT \cite{SchaferPhysRevB2015}, DQMC \cite{KungPhysRevB2015,KungPhysRevB2017}, functional renormalization group \cite{HillePhysRevResearch2020}, etc.), but our subsequent attempt to quantitatively compare the results with experiment for both equilibrium and dynamical excitations spectra within \emph{a single model with fixed set of microscopic parameters}, seems so far unique.

What concerns the model selection: the Hubbard, $t$-$J$, $t$-$J$-$U$, and multiple-band models, all have been discussed. In the case of high-$T_c$ superconducting cuprates, the $t$-$J$-$U$-$(V)$ model seems to be the most general single-band choice to encompass both the $t$-$J$ and the Hubbard-model limits as particular cases, with $U \rightarrow \infty$ and $J = 0$, respectively. The choice of such a general form of single-band model has its own formal merits (cf. Appendix~\ref{appendix:sga_and_slave_bosons}). Here we  would like to stress only its two features. First, formal, that by adding the Hubbard $U$ term to the $t$-$J$ model, we do not necessarily have to use the projected-fermion (or Hubbard-operator) language explicitly and hence we do not have to struggle with the projected-fermion representation at each stage of the calculations, and thus avoid unnecessary algebraic complications. Second, the estimate of the relative value of $U$ with respect to bare bandwidth $W$ is $U/W \sim 2.5$-$3$, which is not yet the limit $U \gg W$ for doping of the order $\delta = 0.1$. So, keeping both $U$ and $J$ terms is advantageous and physically attractive, as exchange term introduces the real-space pairing in the simplest and probably clearest form. Also, with the $t$-$J$-$U$ model as the starting point, we can take a realistic value of $U \approx 8 \div 10\,\mathrm{eV}$ and hence the model in certain ways implements directly the ingenious Anderson idea \cite{AndersonProcIntSchollEnricoFermi1988} of real-space pairing. As to the three-band model, its principal results match qualitatively those of a single band case when the latter is  composed of the antibonding hybridized $p$-$d$ states (cf. Sec.~\ref{subsec:results_for_the_3_band_model}). It should be emphasized that no concept of the Zhang-Rice singlet needs to be involved when relating the three- and single-band models. On the basis of all the principal results for the cuprates, reported here, we can strongly support the statement that the models based on strong correlations reflect the principal physics of the cuprates. Obviously, some of the results such as interpretation of NMR spectra \cite{RybickiNatCommun2016} or the question of evolution from the carrier concentration from $\delta$ in underdoped samples ($\delta$ is the nominal doping) to $1+\delta$ for overdoped samples and its relation to Mott-Hubbard transition \cite{TabisArXiV2021} require further studies. At the end, it is tempting to suggest that the three-band ($d$-$p$) model with the Kondo-type and superexchange interactions would be natural and quite interesting generalization of $t$-$J$-$U$ model on one hand, and the proper explicit treatment of the orbital structure  on the other. Such a model principally differs from that of Emery and Reiter \cite{EmeryPhysRevB1988}, because here the hybridization term survives the introduction of those exchange interactions (cf. Appendix~\ref{appendix:derivation_of_the_tj_model}). In effect, the carriers are not holes in the $2p$ band, but hybridized $p$-$d$ hole states.

    \begin{figure}
    \centering
    \includegraphics[width=\textwidth]{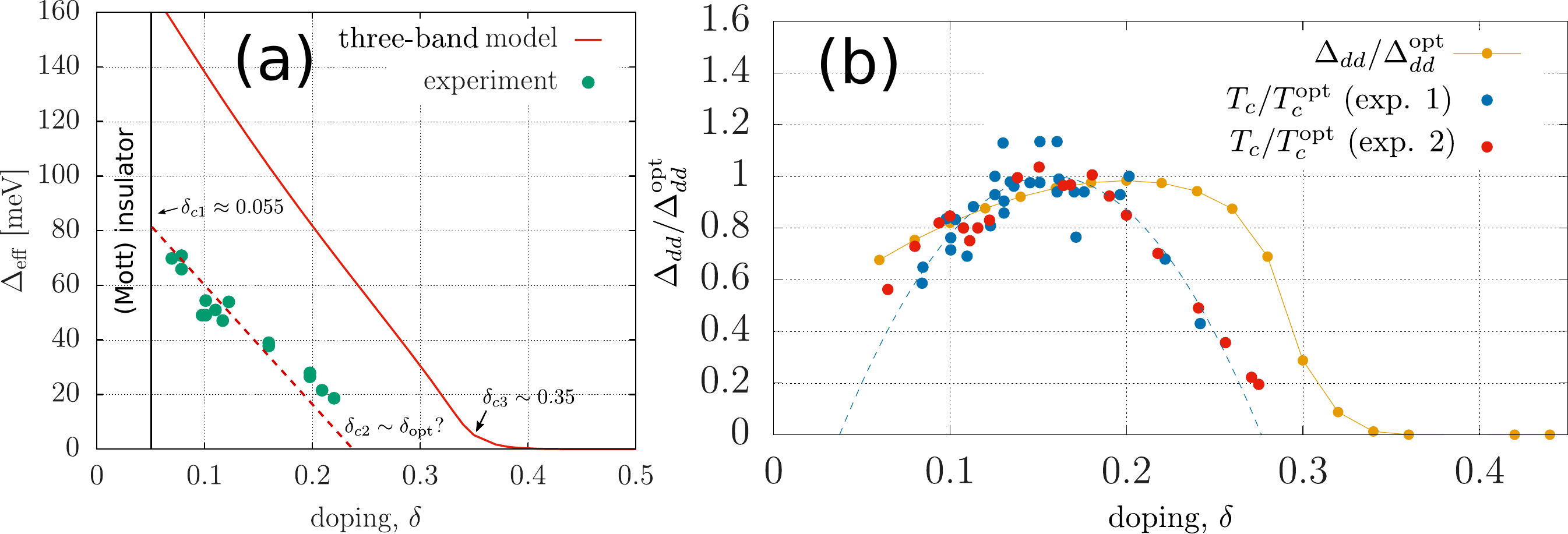}.
    \caption{(a) Plot of the experimentally observed pseudogap \cite{AnzaiNatCommun2013} (points) as compared to theoretical results for the effective single-particle gap obtained within the three-band model \cite{ZegrodnikPhysRevB2019}. The Hamiltonian parameters are: $U_d = 11\,\mathrm{eV}$, $U_p = 4.1\,\mathrm{eV}$, $\epsilon_{dp} = 3.2\,\mathrm{eV}$, $t_{pp} = 0.4\,\mathrm{eV}$. The critical doping levels and the Mott-insulator boundary are also marked.  (b) Correlated $d$-wave gap component \cite{ZegrodnikJPCM2021} with intersite Coulomb interaction $V_{dd} = 0.7\,\mathrm{eV}$. Experimental data sets 1 and 2 are taken from Refs.~\cite{HufnerRepProgPhys2008} and \cite{deMelloPhysRevB2003}, respectively. For a brief discussion of the role of quantum fluctuations in bringing the theoretical results to those obtained from experiment see main text.}
    \label{fig:pseudogap_theory_vs_experiment}
  \end{figure}

Finally, we make a remark of a more tentative character. Namely, to illustrate the importance of the two-step (SGA$\rightarrow$DE-GWF) character of the approach and its relation to equilibrium properties, we have plotted in Fig.~\ref{fig:pseudogap_theory_vs_experiment}(a) the observed exemplary data for the pseudogap \cite{AnzaiNatCommun2013}, together with our theoretical results for the effective gap as obtained in the three-band model. In Fig.~\ref{fig:pseudogap_theory_vs_experiment}(b) the exemplary data for the doping dependence of the correlated gap \cite{HufnerRepProgPhys2008,deMelloPhysRevB2003} are shown. We see that both $\Delta_\mathrm{eff}$ (uncorrelated) and renormalized gap $\Delta_{dd}$ at the optimal doping are too large to match quantitatively  the experiment. We ascribe this discrepancy in both cases to not including the effect of spin and charge (collective) excitations on the SC ordering, since the non-local correlations have not been accounted for in that situation, as detailed analysis provided in Sec.~\ref{sec:selected_equilibrium_properties} shows. Note, however, that the data trends are reproduced still correctly in a qualitative manner.

\section*{Acknowledgments}

This review is based on the papers from our groups at the Jagiellonian University and AGH University of Science and Technology, both located in Krak\'{o}w, Poland. The research originated and was a part of PhD thesis series in the years 2011-2019 at Jagiellonian University (\url{http://th-www.if.uj.edu.pl/ztms/eng/phdTheses.php}).\footnote{See~\url{http://web.archive.org/web/20220203171045/http://th-www.if.uj.edu.pl/ztms/eng/phdTheses.php} for archived version of the website.} We are particularly grateful to the former students of JS: Drs. Olga Howczak, Jakub J\k{e}drak, Jan Kaczmarczyk, Marcin Abram, Marcin M. Wysoki{\'n}ski, Andrzej P. K\k{a}dzielawa, and Ewa K\k{a}dzielawa-Major for their day-to-day questions and cooperation. We thank Professor J\"{o}rg B\"{u}nemann for a help with DE-GWF on early stage of this project. Discussions with experimentalists: Professors Neven Bari\v{s}i\'{c} and Wojciech Tabi\'{s} (AGH) have also been very useful. JS is grateful to Professors Jurgen M. Honig and Krzysztof Byczuk for many fruitful discussions over the years. We thank Professor Krzysztof Wohlfeld for discussions and for turning our attention to some relevant references. Last, but not least, the technical help and everyday support of Dr. Danuta-Goc Jag{\l}o and Maciej Hendzel is appreciated. The work summarized here has been supported over the years by a number of grants: Grant TEAM (2011-16) from Foundation for Polish Science, Grant MAESTRO (2012-18), Grant OPUS (2019-22) No. UMO-2018/29/ST3/02646, Grant OPUS (2022-23) No. UMO-2021/41/B/ST3/04070, and grant Miniatura 5 (2021-22) No.~DEC-2021/05/X/ST3/00666, all from Narodowe Centrum Nauki (NCN) of Poland. We also acknowledge two special grants from SciMat Priority Research Area under the Strategic Programme Excellence Initiative at the Jagiellonian University (2021). The first two authors, JS \& MF, contributed equally to this report.

\appendix

\section{Hubbard and extended Hubbard models}
\label{appendix:hubbard_model}
In our considerations we utilize models in the second-quantization scheme. Below we provide a detailed derivation of the single and hybridized-band models. It helps to understand various versions of parametrized microscopic Hamiltonians, introduced in the main text.

\subsection{Hubbard and extended Hubbard models}

We derive below the expression for the Hamiltonian operator for interacting electrons in single band separated energetically from other states. Our starting point is the expression for the many-particle Hamiltonian in the second quantization representation \cite{BookFetterWalecka}:

\begin{align}
\hat{\mathcal{H}}&= \sum_{\sigma} \int d^3\!\mathbf{r}\, \hat{\Psi}_{\sigma}^{\dagger}(\mathbf{r}) \mathcal{H}_1(\mathbf{r})\, \hat{\Psi}_{\sigma}(\mathbf{r})
+ \frac{1}{2} \sum_{\sigma_{1}\sigma_{2}} \iint d^3\! \mathbf{r}_1\,d^3\!\mathbf{r}_2\,\hat{\Psi}_{\sigma_1}^{\dagger}(\mathbf{r}_{1})\, \hat{\Psi}_{\sigma_2}^{\dagger}(\mathbf{r}_{2})\, V\!\left(\mathbf{r}_{1}-\mathbf{r}_{2}\right)  \hat{\Psi}_{\sigma_2}(\mathbf{r}_2)\hat{\Psi}_{\sigma_1} (\mathbf{r}_1)
\nonumber\\& \equiv \hat{\mathcal{H}}_1 + \hat{\mathcal{H}}_2,
\label{eq:app:sec_quant_hamitlonian}  
\end{align}

\noindent
where $\hat{\Psi}_{\sigma}(\textbf{r})$ is the field operator for particle of spin $\sigma$. In Eq.~\eqref{eq:app:sec_quant_hamitlonian}, $\mathcal{H}_1(\mathbf{r})$ denotes single-particle Hamiltonian and $V\!\left(\mathbf{r}_{1}-\mathbf{r}_{2}\right)$ is the pair-interaction potential energy, both represented in the wave-mechanics language (position representation). In Eq.~\eqref{eq:app:sec_quant_hamitlonian} we have explicitly decomposed $\hat{\mathcal{H}}$ into the single- and many-particle contributions,  $\hat{\mathcal{H}}_1$ and $\hat{\mathcal{H}}_2$, respectively (note the hat in the definition of the second-quantized term $\hat{\mathcal{H}}_1$ that distinguishes it from its first-quantized counterpart,  $\mathcal{H}_1(\mathbf{r})$). The field operator may be expanded in a complete orthonormal basis set $\{\Phi_{i\sigma}(\mathbf{r})\}$ of single-particle wave functions; for the present purpose we choose $\Phi_{i\sigma}(\mathbf{r})=\Phi_i (\mathbf{r})\chi_{\sigma}(\mathbf{r} )$, where $\Phi_i(\mathbf{r})$ presents a member in the set of Wannier functions for a single band and $\chi_{\sigma}(\mathbf{r})$ is the spin function for the electron, labeled by the space coordinate $\mathbf{r}$. In effect,

\begin{equation}
\hat{\Psi}_{\sigma}(\mathbf{r}) = \sum_{i} \Phi_{i\sigma}(\mathbf{r})\,\chi_{\sigma}(\mathbf{r})\, \hat{a}_{i\sigma},
\label{c2}
\end{equation}

\noindent
where $\hat{a}_{i\sigma}$ is the annihilation operator that removes an electron with spin index $\sigma$ from the single-particle state centered at the lattice site $i$. We now substitute Eq.~\eqref{c2} into Eq.~\eqref{eq:app:sec_quant_hamitlonian} and examine first the one-particle part $\hat{\mathcal{H}}_1$ that becomes
\begin{equation}
\hat{\mathcal{H}}_1 = \sum_{ij\sigma} \int d^3\!\mathbf{r}\,\Phi_{i}^{*}(\mathbf{r})\,H_{1}(\mathbf{r})\, \Phi_{j}(\mathbf{r})\,\chi_{\sigma}^{\dagger}(\mathbf{r})\,\chi_{\sigma}(\mathbf{r})\,\hat{a}_{i\sigma}^{\dagger}\,\hat{a}_{j\sigma}.
\label{c3}
\end{equation}

\noindent
By recalling that for a system with a global spin quantization axis $\chi_{\sigma}^{\dagger}(\mathbf{r})\, \chi_{\sigma}(\mathbf{r}) =\chi_{\sigma}^{\dagger} \chi_{\sigma} \equiv 1$, and defining the hopping matrix element as

\begin{equation}
t_{ij} \equiv \int d^3\!\mathbf{r}\,\Phi_{i}^{*}(\mathbf{r})\, \mathcal{H}_1(\mathbf{r})\,\Phi_{j}(\mathbf{r}),
\label{c4}
\end{equation}

\noindent
one arrives at the simple expression

\begin{equation}
\hat{\mathcal{H}}_{1} = \sum_{ij\sigma } t_{ij}\, \hat{a}_{i\sigma}^{\dagger}\, \hat{a}_{j\sigma } \equiv \sum_{i\sigma} t_{ii}\,\hat{n}_{i\sigma} + \sum_{ij}\!^{'} \sum_{\sigma} t_{ij}\, \hat{a}_{i\sigma}^{\dagger}\,\hat{a}_{j\sigma},
\label{c5}
\end{equation}

\noindent
where the primed summation means that the terms $i=j$ are omitted.

The two-particle part is transformed in a similar manner. We first write 

\begin{align}
\hat{\mathcal{H}}_2 =& \,\frac{1}{2} \sum_{ijkl} \sum_{\sigma_{1}\sigma_{2}} \hat{a}_{i\sigma_{1}}^{\dagger}\,\hat{a}_{j\sigma_2}^{\dagger}\, \hat{a}_{l\sigma_2}\,\hat{a}_{k\sigma_1} \chi_{\sigma_1}^{\dagger}\!(1)\,\chi_{\sigma_2}^{\dagger}\!(2)\,\chi_{\sigma_2}\!(2)\,\chi_{\sigma_1}\!(1)  \times \nonumber \\ \times & \iint d^3\!\mathbf{r}_1\,d^3\mathbf{r}_2\,\Phi_i^*(\mathbf{r}_1)\, \Phi_j^*(\mathbf{r}_2)\, V(\mathbf{r}_1-\mathbf{r}_2)\, \Phi_k(\mathbf{r}_1)\, \Phi _l (\mathbf{r}_2 ) = \nonumber \\
=  & \, \frac{1}{2} \sum_{ijkl} \sum_{\sigma_{1}\sigma_{2}}\, V_{ijkl}\, \hat{a}_{i\sigma_1}^{\dagger}\, \hat{a}_{j\sigma_2}^{\dagger}\, \hat{a}_{l\sigma_2}\, \hat{a}_{k\sigma_1}.
  \label{c6} 
\end{align}

\noindent
Note the interchange of  indices in the $\hat{a}$ operators. The product of the corresponding $\chi$'s in~\eqref{c6} yields unity and we have introduced a compact notation for the four-center integral, $V_{ijkl} \equiv \int d^3\textbf{r}_1d^3\textbf{r}_2 \Phi_i^*(\textbf{r}_1)\Phi_j^*(\textbf{r}_2)V(\textbf{r}_1-\textbf{r}_2)\Phi_k(\textbf{r}_1)\Phi_l(\textbf{r}_1)$. We then obtain 
\begin{equation}
\begin{split}
\hat{\mathcal{H}}_2 =& \, \frac{1}{2} \sum_{ijkl} V_{ijkl} \sum_{\sigma_{1}\sigma_{2}} \hat{a}_{i\sigma_1}^{\dagger}\,\hat{a}_{j\sigma_2}^{\dagger} \hat{a}_{l\sigma_2}\, \hat{a}_{k\sigma_1}= \, \frac{1}{2} \sum_{ijkl} V_{ijkl} \sum_{\sigma} \left(\hat{a}_{i\sigma}^{\dagger}\,\hat{a}_{j\sigma}^{\dagger}\,\hat{a}_{l\sigma}\,\hat{a}_{k\sigma} + \hat{a}_{i\sigma}^{\dagger}\,\hat{a}_{j\bar{\sigma}}^{\dagger}\,\hat{a}_{l\bar{\sigma}}\,\hat{a}_{k\sigma}\right)
\end{split}
\label{c7}
\end{equation}

\noindent
with $\bar{\sigma} \equiv - \sigma$.

Equation~\eqref{c7} may be simplified by assuming that the three- and four center integrals may be neglected except when:
 (\emph{i}) $i=j=k=l$; (\emph{ii}) $i=j \neq k = l$, $i=k \neq j = l$, $i=l \neq j = k$; (\emph{iii}) $i=j=k \neq l$, $i=j=l \neq k$, $i=k=l \neq j$, $j=k=l \neq i$.
Each of these cases should be treated separately. Before considering them explicitly note that the term \emph{i} contains interactions between the electrons in the same $\Phi_i(\mathbf{r} )$ state, the so called \emph{intraatomic interactions}. The terms corresponding to (\emph{ii}) and (\emph{iii}) contain two-center integrals.
The fact that we disregard three- and four-center interactions is in accord with the physical assumption that overlap of neighboring orbitals $\Phi_i(\mathbf{r} )$ and $\Phi_j (\mathbf{r} )$ is small. This is the essence of the tight-binding approximation. Hence, the interaction part may be described adequately using a relatively small number of parameters, namely, the largest of the set of coefficients $V_{ijkl}$.

The terms of Eq.~\eqref{c7}, satisfying the condition (\emph{i}), may be compactly written as

\begin{equation}
I^{(i)} = \frac{1}{2} \sum_{i} V_{iiii} \sum_{\sigma} \hat{a}_{i\sigma}^{\dagger}\,\hat{a}_{i\bar{\sigma}}^{\dagger}\, \hat{a}_{i\bar{\sigma}}\, \hat{a}_{i\sigma} \equiv \frac{1}{2} \sum_{i\sigma} V_{iiii}\, \hat{n}_{i\sigma}\, \hat{n}_{i\bar{\sigma}}
= \sum_{i} V_{iiii}\, \hat{n}_{i \uparrow}\, \hat{n}_{i\downarrow} \equiv U\sum_{i} \hat{n}_{i\uparrow}\, \hat{n}_{i\downarrow},
\label{c8}
\end{equation}

\noindent
where $U$ is the so-called intraatomic Coulomb interaction energy, often referred to as the \emph{Hubbard U}. Equation~\eqref{c7} with (\emph{ii}) yields

\begin{equation}
\begin{split}
I^{(ii)} = &\, \frac{1}{2} \sum_{ik}\!^{'} V_{iikk} \sum_{\sigma} \left(\hat{a}_{i\sigma}^{\dagger}\, \hat{a}_{i\sigma}^{\dagger}\, \hat{a}_{k\sigma}\, \hat{a}_{k\sigma} + \hat{a}_{i\sigma}^{\dagger}\, \hat{a}_{i\bar{\sigma}}^{\dagger}\, \hat{a}_{k\bar{\sigma}}\,\hat{a}_{k\sigma}\right) +\\
&+ \frac{1}{2} \sum_{ij}\!^{'} V_{ijji} \sum_{\sigma} \left(\hat{a}_{i\sigma}^{\dagger}\, \hat{a}_{j\sigma}^{\dagger}\, \hat{a}_{i\sigma}\, \hat{a}_{j\sigma}\, + \hat{a}_{i\sigma}^{\dagger}\, \hat{a}_{j \bar{\sigma}}^{\dagger}\, \hat{a}_{i\bar{\sigma}}\, \hat{a}_{j\sigma} \right) +\\
& + \frac{1}{2} \sum_{ij}\!^{'} V_{ijij} \sum_{\sigma} \left( \hat{a}_{i\sigma}^{\dagger}\, \hat{a}_{j\sigma}^{\dagger}\, \hat{a}_{j\sigma}\, \hat{a}_{i\sigma} + \hat{a}_{i\sigma}^{\dagger}\, \hat{a}_{j\bar{\sigma}}^{\dagger}\, \hat{a}_{j\bar{\sigma}}\, \hat{a}_{i\sigma} \right).
\end{split}
\label{c9}
\end{equation}

\noindent
The first sum in Eq.~\eqref{c9} vanishes identically because of the Pauli principle. Through (repeated) anticommutation operations the summands in the third, fifth, and sixth terms can be recast in terms of the particle number operators $\hat{n}_{i\sigma}, \hat{n}_{j\sigma}$, and $\hat{n}_{j\bar{\sigma}}$; in the fourth sum we introduce circular components of the spin operators: $\hat{S}_i^{+} \equiv \hat{a}_{i\uparrow}^{\dagger} \hat{a}_{i\downarrow}$ and $\hat{S}_i^{-} \equiv \hat{a}_{i\downarrow}^{\dagger}\hat{a}_{i\uparrow}$. Finally, we change the dummy index $k$ to $j$. Then

\begin{equation}
I^{(ii)} = \sum_{ij} \!^{'} \left\{ V_{iijj}\,\hat{a}_{i\uparrow}^{\dagger}\, \hat{a}_{i\downarrow}^{\dagger}\, \hat{a}_{j\downarrow}\,\hat{a}_{j\uparrow} + \frac{1}{2} V_{ijji} \sum_{\sigma} \left(-\hat{n}_{i\sigma}\, \hat{n}_{j\sigma} - \hat{S}_i^{\sigma}\, \hat{S}_j^{\bar{\sigma}} \right)
+ \frac{1}{2} V_{ijij} \sum_{\sigma} \left(\hat{n}_{i\sigma}\, \hat{n}_{j\sigma} + \hat{n}_{i\sigma}\,\hat{n}_{j\bar{\sigma}} \right)  \right\}.
\label{c10}
\end{equation}

\noindent
This expression formally represents the Heisenberg-Dirac exchange interaction in the occupation-number representation. We now invoke the identity $\mathbf{\hat{S}}_i\cdot \mathbf{\hat{S}}_j = \frac{1}{2}\left(\hat{S}_i^+ \hat{S}_j^- + \hat{S}_i^- \hat{S}_j^+\right) + \hat{S}_i^z \hat{S}_j^z$, with $\hat{S}_i^z = \frac{1}{2} \left(\hat{n}_{i\uparrow} - \hat{n}_{i\downarrow} \right)$ and $\hat{n}_i = \hat{n}_{i\uparrow} + \hat{n}_{i\downarrow}$. Then $\hat{n}_{i\sigma} = \frac{1}{2} \hat{n}_i + \sigma \hat{S}_i^z$. On now summing over $\sigma$ Eq.~\eqref{c10} is reduced by elementary steps to the form
\begin{equation}
I^{(ii)} = \sum_{ij}\!{'} \left\{ V_{iijj} \hat{a}_{i\uparrow}^{\dagger}\, \hat{a}_{i\downarrow}^{\dagger}\, \hat{a}_{j\downarrow}\, \hat{a}_{j\uparrow} - V_{ijji} \left( {\mathbf{\hat{S}}}_i \cdot \mathbf{\hat{S}}_j + \frac{1}{4} \hat{n}_i\, \hat{n}_j \right) + \frac{1}{2} V_{ijij}\, \hat{n}_i\hat{n}_j \right\}.
\label{c11}
\end{equation}
Under the condition (\emph{iii}),  Eq.~\eqref{c7} reduces to

\begin{equation}
\begin{split}
I^{(iii)} = &\, \frac{1}{2} \sum_{il}\!^{'} V_{iiil} \sum_{\sigma} \hat{a}_{i\sigma}^{\dagger}\, \hat{a}_{i\bar{\sigma}}^{\dagger}\, \hat{a}_{l\bar{\sigma}}\, \hat{a}_{i\sigma} + \frac{1}{2} \sum_{ik}\!^{'} V_{iiki} \sum_{\sigma} \hat{a}_{i\sigma}^{\dagger}\, \hat{a}_{i\sigma}^{\dagger}\, \hat{a}_{i\bar{\sigma}}\, \hat{a}_{k\sigma} +\\
& + \frac{1}{2} \sum_{ij}\!^{'} V_{ijii} \sum_{\sigma} \hat{a}_{i\sigma}^{\dagger}\, \hat{a}_{j\bar{\sigma}}^{\dagger}\, \hat{a}_{i\bar{\sigma}}\, \hat{a}_{i\sigma} + \frac{1}{2} \sum_{ij}\!^{'} V_{ijjj} \sum_{\sigma} \hat{a}_{i\sigma}^{\dagger}\, \hat{a}_{j\bar{\sigma}}^{\dagger}\, \hat{a}_{j\sigma}\, \hat{a}_{j\sigma}=\\
= &\, \frac{1}{2} \sum_{il}\!^{'} V_{iiil} \sum_{\sigma} \hat{n}_{i\sigma}\, \hat{a}_{i\bar{\sigma}}^{\dagger}\, \hat{a}_{l\bar{\sigma}} + \frac{1}{2} \sum_{ik}\!^{'} V_{iiki} \sum_{\sigma} \hat{n}_{i\sigma}\, \hat{a}_{i \bar{\sigma}}^{\dagger}\, \hat{a}_{k \bar{\sigma}}+\\
& +\frac{1}{2} \sum_{ij}\!^{'} V_{ijii} \sum_{\sigma} \hat{n}_{i\sigma}\, \hat{a}_{j\bar{\sigma}}^{\dagger}\, \hat{a}_{i\bar{\sigma}} + \frac{1}{2} \sum_{ij}\!^{'}V_{ijjj} \sum_{\sigma} \hat{n}_{j\sigma}\, \hat{a}_{i\bar{\sigma}}^{\dagger}\, \hat{a}_{j\bar{\sigma}}.
\end{split}
\label{c12}
\end{equation}

\noindent
In the terms involving $V_{iiki}$ and $V_{ijjj}$, the summation over $\sigma$ has been replaced by a summation over $\bar{\sigma} $. As long as the Wannier functions are chosen as real (one may always be done since these functions are translationally invariant and because their phase may be arbitrary chosen), one has $V_{iiil} = V_{iiki} = V_{ijii}= V_{ijjj} \equiv V_{iiij} \equiv V_{ij}$. By making use of those identities and exchanging the dummy indices in \eqref{c12} appropriately, one can transform Eq.~\eqref{c12} into

\begin{equation}
I^{(iii)} = \sum_{ij}\!^{'} V_{iiij} \sum_{\sigma} \hat{n}_{i\sigma} \left(\hat{a}_{i\bar{\sigma}}^{\dagger} \hat{a}_{j\bar{\sigma}} + \hat{a}_{j\bar{\sigma}}^{\dagger} \hat{a}_{i\bar{\sigma}}\right).
\label{c13}
\end{equation}

\noindent
Combining (\ref{c13}), (\ref{c11}), (\ref{c8}), (\ref{c5}) with \eqref{eq:app:sec_quant_hamitlonian} (in (\ref{c13}) the $\sigma (\bar{\sigma} )$ subscripts have been replaced by $\bar{\sigma} (\sigma )$, one finally obtains

\begin{align}
\hat{\mathcal{H}} = &\, t_0 \sum_{i\sigma} \hat{n}_{i\sigma} + \sum_{ij\sigma}\!^{'} t_{ij}\, \hat{a}_{i\sigma}^{\dagger}\, \hat{a}_{j\sigma} + U \sum_{i} \hat{n}_{i\uparrow}\, \hat{n}_{i\downarrow}+\nonumber\\
+ &\frac{1}{2} \sum_{ij}\!^{'} \left(K_{ij} - \frac{1}{2} J_{ij}\right) \hat{n}_i \hat{n}_j - \sum_{ij}\!^{'} J_{ij}\, \mathbf{S}_i \cdot \mathbf{S}_j + \sum_{ij}\!^{'} J_{ij}\,\hat{a}_{i\uparrow}^{\dagger}\, \hat{a}_{i\downarrow}^{\dagger}\, \hat{a}_{j\downarrow}\, \hat{a}_{j\uparrow}+\nonumber\\
+ &\frac{1}{2} \sum_{ij\sigma}\!^{'} V_{ij} \left(\hat{n}_{i\bar{\sigma}} + \hat{n}_{j\bar{\sigma}} \right) \left( \hat{a}_{i\sigma}^{\dagger}\, \hat{a}_{j\sigma} + \hat{a}_{j\sigma}^{\dagger}\, \hat{a}_{i\sigma} \right).
\label{c14}
\end{align}

\noindent
This is the central result of this Appendix. As usual, primed summation means that we omit the terms with $i=j$. We also have introduced the following notation:

\begin{eqnarray}
t_{ii}&  \equiv & \int d^3\!\mathbf{r}\, \Phi_i^* (\mathbf{r}) \left[ -\frac{\hbar^2}{2m} \nabla_{\mathbf{r}}^2 + V(\mathbf{r}) \right] \Phi_i (\mathbf{r} ) \equiv t_0,
\label{c15} \\
t_{ij} & \equiv & \int d^3\!\mathbf{r}\, \Phi_i^* (\mathbf{r} ) \left[ -\frac{\hbar^2}{2m} \nabla_{\mathbf{r}}^2 + V(\mathbf{r} ) \right] \Phi_j (\mathbf{r}),
\label{c16} \\
V_{iiii} & \equiv & \iint d^3\!{\mathbf{r}_1}\, d^3\!\mathbf{r}_2\:\left|\Phi_i (\mathbf{r}_1 )\right|^2 V(\mathbf{r}_1 - \mathbf{r}_2 ) \left| \Phi_i (\mathbf{r}_2)\right|^2 \equiv U,
\label{c17}\\
V_{ijij} & \equiv & \iint d^3\!{\mathbf{r}_1}\,d^3\!{\mathbf{r}_2}\, \left|\Phi_i (\mathbf{r}_1 ) \right|^2 V(\mathbf{r}_1 - \mathbf{r}_2 ) \left|\Phi_j (\mathbf{r}_2) \right|^2 \equiv K_{ij},
\label{c18}\\
V_{ijji} & \equiv & \iint d^3\!{\mathbf{r}_1}\:d^3\!{\mathbf{r}_2}\: \Phi_i^* (\mathbf{r}_1 ) \Phi_j^* (\mathbf{r}_2 ) V(\mathbf{r}_1 - \mathbf{r}_2) \Phi_i (\mathbf{r}_2) \Phi_i (\mathbf{r}_1 ) \equiv J_{ij},
\label{c19}\\
V_{iijj} & \equiv & \iint d^3\!{\mathbf{r}_1}\:d^3\!{\mathbf{r}_2}\: \Phi_i^* (\mathbf{r}_1 ) \Phi_i^* (\mathbf{r}_2 ) V(\mathbf{r}_1 - \mathbf{r}_2) \Phi_j (\mathbf{r}_1)\Phi_j (\mathbf{r}_2 ) \equiv J_{ij},
\label{c20}\\
V_{iiij} & \equiv & \iint d^3\!{\mathbf{r}_1}\: d^3\!{\mathbf{r}_2}\: \Phi_i^* (\mathbf{r}_1 ) \Phi_i^* (\mathbf{r}_2 ) V(\mathbf{r}_1 - \mathbf{r}_2) \Phi_i (\mathbf{r}_1) \Phi_j (\mathbf{r}_2 ) \equiv V_{ij},
\label{c21}
\end{eqnarray}

\noindent
where $V(\mathbf{r})$  is the potential energy of electrons in the lattice.

One should observe that $t_{ii}=t_0$ is the translationally invariant energy of an electron in a one-particle state centered on site $i$; the first term in (\ref{c14}) then specifies the one-electron energy for the set of $\hat{n}_e = \sum_{i\sigma} \hat{n}_{i\sigma}$ electrons. This assumption is customarily made in one-band model of high-$T_c$ superconductors which are, strictly speaking, weakly disordered systems at non-zero doping. In view of (\ref{c16}), the second term in (\ref{c14}) represents the sum of one-electron energies when each electron, independently of all others, has access to all $N$ sites; this includes but is not limited to the kinetic energy of each electron. The third term in (\ref{c14}) is the Hubbard (1963) expression \cite{HubbardProcRoySoc1963} for the Coulomb energy whenever two electrons with opposite spins occupy the same single particle state centered on a typical site $i$. Eq.~(\ref{c18}) represents the Coulomb energy for two interacting electrons placed on different sites. The corresponding exchange integral is specified by Eq.~(\ref{c19}) and by the equivalent integral (\ref{c20}). The fourth summation in Eq.~(\ref{c14}) represents thus the total pairwise Coulomb and exchange interaction energy for two electrons in two single particle states located on two different sites.
The next term specifies the Heisenberg exchange interaction. As is well established, this quantity induces further splitting of electron energy levels, based on the resultant total spin configuration. However, one should notice that in the present formulation the operators $\{ \mathbf{\hat{S}}_i \} $ do not necessarily specify atomic spins.

The second-quantization representation of $\hat{\mathbf{S}}_i$ is compatible with a variable number of electrons on a given site $i$. Hence, both the configurations with ${\hat{\mathbf{S}}}_i=0$ and ${\hat{\mathbf{S}}}_i \neq 0$ contribute to the dynamics of the system. The operator $\mathbf{\hat{S}}_i$ has the properties of the (Pauli) spin operators if a given site $i$ is singly occupied. The next to last sum in (\ref{c14}) represents an energy contribution resulting from the hopping of an electron pair with opposite spins moving from site $j$ to site $i$; this pairwise contribution has the same coupling constant as the exchange energy.
The final term represents an effect where a single electron hops under conditions where a second electron of opposite spin is present either at the initial or at the destination site. One should realize that {\it all} interaction terms in (\ref{c14}) derive from the electrostatic interaction $V(\mathbf{r} _1 - \mathbf{r} _2 )$ between the electrons, as can be seen by inspection of the definitions (\ref{c17}-\ref{c21}). The division into Coulomb and exchange parts arises because the interactions represented by the parameters (\ref{c17}) and (\ref{c18}) have the classical form of density-density (direct) interactions, with the charge density $\rho _i (\mathbf{r} ) \equiv \rho (\mathbf{r} - \mathbf{R}_i ) \equiv \vert \Phi_i (\mathbf{r}) \vert^2$, whereas the exchange term $\sim J_{ij}$ has {\it no} classical analog, hence a different name is attached to this particular integral. It emerges because of antisymmetry of the total wave function with respect to transposition of particle spatial coordinates and spins.

One should note the generality of the derivation leading to Eq.~\eqref{c14}. It was assumed only that single particle and pairwise interactions are sufficient to describe the physics of the system. Also, that a model of nondegenerate atomic and band states, not overlapping with any other states, is applicable, and the effects associated with all three- and four-center integrals may be ignored. Depending on the relative magnitudes of the parameters $t_{ij}, U, K_{ij}, V_{ij}$, and $J_{ij}$, Eq.~\eqref{c14} specializes to a Hamiltonian appropriate for the nearly free electron case, the Hubbard Hamiltonian, the Heisenberg Hamiltonian, or a Hamiltonian appropriate to the description of chemical bonds. The last two terms in Eq.~\eqref{c14} are small if $U$ is large as, in such circumstances, the expectation values of $\hat{n}_{i\uparrow }\hat{n}_{i\downarrow }$ are negligible. Finally, it is interesting to observe that the summation involving $t_{ij}$ is the kinetic counterpart to the atomic term involving $t_0$.

The expression (\ref{c14}) is called the \emph{parametrized narrow band Hamiltonian}; overlap of wave functions between more distant sites is usually neglected.
Also, while the six distinct parameters, listed in Eqs.~\eqref{c15}-\eqref{c21}, can be calculated in principle, such numerical calculations of the Wannier functions may be challenging for collections of strongly interacting electrons. It is therefore customary to treat these quantities as adjustable parameters, to be evaluated by fitting the theoretical predictions to experimental measurements. From the theoretical viewpoint, it is interesting to consider possible situations corresponding to different values of the parameters. This provides information about nontrivial solutions of a given class of models. In particular, it is important to ascertain under what circumstances the Hamiltonian~\eqref{c14} reduces to the Heisenberg model with localized spins or how the concept of exchange interactions can be incorporated into band theory, in addition to much stronger Coulomb interaction $U$.

\subsection{Model of hybrized fermions}

To derive the Hamiltonian of the type~\eqref{c14} for hybridized fermions, we introduce two field operators, $\hat{\Psi}^{(c)}_{\sigma}(\mathbf{r})$ and $\hat{\Psi}^{(f)}_{\sigma}(\mathbf{r})$. The first one corresponds to the conduction-band states $\{\Phi^{(c)}_{i\sigma}\}$ and the second one to atomic like-states $\{\Phi^{(f)}_{m\sigma}\}$, and may be represented explicitly as follows \cite{AnisimovBook2010}

\begin{equation}
\hat{\Psi}^{(c)}_{\sigma}(\mathbf{r})=\sum_{m}\Phi^{(c)}_{m\sigma}(\mathbf{r})\,\hat{c}_{m\sigma}=\sum_{m}\Phi^{(c)}_{m}(\mathbf{r})\,\chi_{\sigma}(\mathbf{r})\,\hat{c}_{m\sigma},
\label{cc11}
\end{equation}

\noindent
and

\begin{equation}
\hat{\Psi}^{(f)}_{\sigma}(\mathbf{r})=\sum_{i}\Phi^{(f)}_{i\sigma}(\mathbf{r})\,\hat{f}_{i\sigma}=\sum_{i}\Phi^{(f)}_{i}(\mathbf{r})\,\chi_{\sigma}(\mathbf{r})\,\hat{f}_{i\sigma},
\label{cc12}
\end{equation}

\noindent
where $\hat{c}_{m\sigma}$ and $\hat{f}_{i\sigma}$ are annihilation operators of the orthogonalized Wannier states, $\Phi^{(c)}_{m\sigma}(\mathbf{r})$ and $\Phi^{(f)}_{i\sigma}(\mathbf{r})$, respectively. The many-particle Hamiltonian in the second quantization representation can be constructed in a straightforward manner by noting that the quantity $\hat{\Psi}_{\sigma}(\mathbf{r})\equiv \hat{\Psi}^{(f)}_{\sigma}(\mathbf{r})+\hat{\Psi}^{(c)}_{\sigma}(\mathbf{r})$ plays formally the role of the field operator for this two-orbital system\footnote{Equivalently, we could have also introduced a two component orbital representation, $\{\Phi^{(f)}_{i\sigma}, \Phi^{(c)}_{m\sigma}\}$, with the orbital index $l=f,c$ plays the role of an additional (orbital) quantum number. The states $\{\Phi^{(f)}_{i\sigma}\}$  and $\{\Phi^{(c)}_{m\sigma}\}$ are not independent anymore as hybridization between them takes place and thus they must be also taken in the orthogonalized form of $c$+$f$ orbitals.} (note that the basis contains two Wannier functions per site). Hence, on the basis of general form of the Hamiltonian in second quantization representation, we have

\begin{align}
 \hat{\mathcal{H}} =  & \sum_{\sigma}\int d^{3}\!\textbf{r}\,\left\{ \hat{\Psi}^{(c)\dagger}_{\sigma}(\mathbf{r})+\hat{\Psi}^{(f)\dagger}_{\sigma}(\mathbf{r})\right\}\mathcal{H}_1(\mathbf{r}) \left\{ \hat{\Psi}^{(c)}_{\sigma}(\mathbf{r})+\hat{\Psi}^{(f)}_{\sigma}(\mathbf{r})\right\} \nonumber + \\
 + & \,\frac{1}{2} \sum_{\sigma\sigma'}\iint d^{3}\!\textbf{r}\,d^{3}\!\textbf{r}'\left\{ \hat{\Psi}^{(c)\dagger}_{\sigma}(\mathbf{r})+\hat{\Psi}^{(f)\dagger}_{\sigma}(\mathbf{r})\right\} \left\{\hat{\Psi}^{(c)\dagger}_{\sigma'}(\mathbf{r})\,\hat{\Psi}^{(f)\dagger}_{\sigma'}(\mathbf{r})\right\}V\!\left(\mathbf{r}-\mathbf{r}'\right) \times\nonumber\\\times & \left\{\hat{\Psi}^{(c)}_{\sigma'}(\mathbf{r}')+\hat{\Psi}^{(f)}_{\sigma'}(\mathbf{r})\right\} \left\{\hat{\Psi}^{(c)}_{\sigma}(\mathbf{r})+\hat{\Psi}^{(f)}_{\sigma}(\mathbf{r})\right\},
 \label{cc13}
\end{align}

\noindent
where the $\mathcal{H}_1(\mathbf{r})$ and $V\!\left(\mathbf{r}-\mathbf{r}'\right)$ have the same meaning as in Eq.~\eqref{eq:app:sec_quant_hamitlonian}. Note that in Eq.~\eqref{cc13} we have all $f$-$f$, $f$-$c$, and $c$-$c$ interactions. The assumption usually made at this point is that the subsystem $c$ is regarded as composed of noninteracting electrons. In other words, $c$ electrons are treated as a high-density metallic gas. On the other hand, the subsystem $f$ is regarded as composed of narrow-band or quasiatomic electrons, for which the mutual interactions are important and they may be of either quasiatomic  of itinerant character.

By substituting Eqs.~(\ref{cc11}) and (\ref{cc12}) into Eq.~(\ref{cc13}) we obtain a parameterized Hamiltonian in the Wannier or atomic representation. Here we consider only intraatomic interaction part for $f$ electrons and the dominant $f$-$c$ interaction terms. Using exactly the same procedure for interorbital interactions as for intersite terms in the narrow-band case, we obtain in the real-space representation

\begin{eqnarray}
\mathcal{H} =\sum_{mn\sigma}t_{mn}\,\hat{c}^{\dagger}_{m\sigma}\,\hat{c}_{n\sigma}+ \sum_{ij\sigma}T_{ij}\,\hat{f}^{\dagger}_{i\sigma}\hat{f}_{j\sigma} + \sum_{im\sigma}\left(V_{im}\,\hat{f}^{\dagger}_{i\sigma}\,\hat{c}_{m\sigma} + V^{*}_{im}\,\hat{c}^{\dagger}_{m\sigma}\,\hat{f}_{i\sigma}\right) \nonumber \\
 + U\sum_{i}\hat{N}_{i\uparrow}\,\hat{N}_{i\downarrow}+K_{fc}\sum_{\langle im\rangle} \hat{N}_{i}\,\hat{n}_{m}-J_{c} \sum_{\langle im\rangle} \left( \mathbf{\hat{S}}_{i}\cdot\mathbf{\hat{s}}_{m} - \frac{1}{2}\hat{N}_{i}\,\hat{n}_{m}  \right),
\label{cc14}
\end{eqnarray}

\noindent
where $\hat{N}_{i\sigma}$, $\hat{\mathbf{S}}_i$ and $\hat{n}_{i\sigma}$, $\hat{\mathbf{s}}_i$ are occupation number and spin operators for $f$ and $c$ electrons, respectively. In Eq.~\eqref{cc14}, $\langle im\rangle$ labels pairs of $f$- and $c$-electrons located on the same or on nearest neighboring sites. Here, again, we have neglected 
interaction between the $c$-carriers; this approximation may by unjustified, e.g., in a two-band situation, where narrow-band
electrons interact with localized electrons (particularly when both types of states derive from  $3d$ or $4f$ orbitals). The Hamiltonian in the
limit $T_{ij}(i\neq j) = K_{fc}=J_c=0$ is referred to as \emph{periodic Anderson} (or \emph{Anderson-lattice}) model \cite{CzychollPhysRep1986}. To avoid complications, we 
assume at this point that the states $\Phi^{(c)}_{m\sigma}(\textbf{r})$ and $\Phi^{(f)}_{i}(\textbf{r})$ represent two orthogonalized (hybridized)
states already, i.e., we assume that $\braket{\Phi^{(c)}_{n}|\Phi^{(f)}_{i}} = 0$. Otherwise, one would have to take the anticommutation relations in the
form $\left\{\hat{f}_{i\sigma}, \hat{c}^{\dagger}_{m\sigma'}\right\}=\delta_{\sigma\sigma'}\braket{\Phi^{(c)}_m|\Psi^{(f)}_i}$.

The microscopic parameters of Hamiltonian are defined as
\begin{subequations}
\begin{align}
& t_{mn}=\int d^{3}\!\textbf{r}\,\Phi^{(c)*}_{m}(\mathbf{r})\,\mathcal{H}_1(\mathbf{r})\,\Phi^{(c)}_{n}(\mathbf{r}),  \\
& T_{ij}=\int d^{3}\!\textbf{r}\,\Phi^{(f)*}_{i}(\mathbf{r})\,\mathcal{H}_1(\mathbf{r})\,\Phi^{(f)}_{j}(\mathbf{r}), \\
& V_{im}=\int d^{3}\!\textbf{r}\,\Phi^{(f)*}_{i}(\mathbf{r})\,\mathcal{H}_1(\mathbf{r})\,\Phi^{(c)}_{m}(\mathbf{r}), \\
& K_{fc}=\iint d^{3}\!\textbf{r}\,d^{3}\!\textbf{r}'\, \left| \Phi^{(f)}_{i}(\mathbf{r})\right|^{2} V\!\left(\mathbf{r}_{1}-\mathbf{r}_{2}\right) \left|\Phi^{c}_{m}(\mathbf{r}')\right|^{2}, \\
& J_{c}=\iint d^{3}\!\textbf{r}\,d^{3}\!\textbf{r}'\, \Phi^{(f)*}_{i}(\mathbf{r})\,\Phi^{(c)*}_{m}(\mathbf{r}') V\!\left(\mathbf{r}-\mathbf{r}'\right)  \Phi^{(f)}_{i}(\mathbf{r}')\,\Phi^{(c)}_{m}(\mathbf{r}).
\end{align}
\end{subequations}

In terms of creation- and annihilation operators, the particle number operators are expressed as

\begin{align}
&\hat{N}_{i\sigma} \equiv \hat{f}^{\dagger}_{i\sigma}\,\hat{f}_{i\sigma}\;;\;\;\hat{N}_{i}\equiv \sum_{\sigma}\hat{N}_{i\sigma},\\
&\hat{n}_{i\sigma} \equiv \hat{c}^{\dagger}_{m\sigma}\,\hat{c}_{m\sigma}\;;\;\;\hat{n}_{m}\equiv \sum_{\sigma}\hat{n}_{m\sigma},
\end{align}

\noindent
whereas the spin operators are

\begin{align}
&\mathbf{\hat{s}}_{m} \equiv \left(\hat{s}^{+}_{m},\hat{s}^{-}_{m},\hat{s}^{z}_{m}\right)\equiv \left(\hat{c}^{\dagger}_{m\uparrow}\,\hat{c}_{m\downarrow}, \hat{c}^{\dagger}_{m\downarrow}\,\hat{c}_{m\uparrow}, \frac{1}{2}\left(\hat{n}_{m\uparrow}-\hat{n}_{m\downarrow}\right)\right),\\
&\mathbf{\hat{S}}_{i} \equiv \left( \hat{S}^{+}_{i},\hat{S}^{-}_{i},\hat{S}^{z}_{i} \right)\equiv \left( \hat{f}^{\dagger}_{i\uparrow}\,\hat{f}_{i\downarrow}, \hat{f}^{\dagger}_{i\downarrow}\hat{f}_{i\uparrow},\frac{1}{2}\left(\hat{N}_{i\uparrow}-\hat{N}_{i\downarrow}\right) \right).
\end{align}

\noindent
The quantity $t_{mn}$ for $m\neq n$ represents hopping integral for the conduction electrons. The whole term containing $t_{mn}$ describes band kinetic energy for the $c$-electrons. Note that the diagonal part

\begin{equation}
\sum_{m\sigma} t_{mm}\, \hat{c}^{\dagger}_{m\sigma}\,\hat{c}_{m\sigma} = t_{0}\sum_{m\sigma} \hat{n}_{m\sigma}
\label{cc17}
\end{equation}

\noindent
of the first term in Eq.~(\ref{cc14}) is \emph{not} a constant since the mixing of the states of the subsystems takes place. In the other words, only the total number of particles $N_e$
\begin{equation}
    \sum_{i\sigma} \hat{N}_{i\sigma} + \sum_{m\sigma} \hat{n}_{m\sigma} = N_e
\end{equation}

\noindent
is conserved. Therefore, Eq.~(\ref{cc17}) may be rewritten in the form

\begin{equation}
t_{0} \sum_{m\sigma} \hat{n}_{m\sigma} = t_{0}\,N_{e} - t_{0}\sum_{i\sigma} \hat{N}_{i\sigma}.
\label{cc19}
\end{equation}

In Eq.~\eqref{cc14}, The quantity $T_{ii}$ is denoted as $\varepsilon_{a}$ and describes the atomic energy of atomic or $\{i\}$ electrons with respect to the reference position $t_{0}$, since the second term in (\ref{cc19}) may be incorporated into the $i=j$ single-particle part. $T_{ij}$ has the same meaning as $t_{mn}$, but for the $f$ states. The third term describes the hybridization (mixing) between the two types of states since $V_{im}$ represents matrix element for the $c\rightarrow f$ transfer of electron between the states $\Phi^{(c)}_{m}$ and $\Phi^{(f)}_{i}$. In other words, this term expresses the attraction of $f$ electron to the nuclear charge of the atom, on which $c$ states reside. The fourth term expresses the intraatomic interaction between $f$ electrons. The last two terms of Eq.~\eqref{cc14} describe interorbital Coulomb and the exchange interactions, respectively.

\subsection{Multiorbital model of correlated electrons}

The starting point for consideration of a general multiorbital model is the orthogonalized basis $\{\Phi^{(l)}_{i\sigma}(\textbf{r})\}$ of Wannier functions and the corresponding definition of the field operator through the sum of (\ref{c2}) type, in which the discrete orbital index $l=1,2,3,\ldots,n$ plays the role analogous to the spin index. The whole analysis is practically the same as that provided above for the single-band and hybridized models, so here we write down only the final result. The parametrized Hamiltonian is composed of single particle part $\mathcal{\hat{H}}_1$ and the interaction part $\mathcal{\hat{H}}_2$, with $\mathcal{\hat{H}} = \mathcal{\hat{H}}_1 + \mathcal{\hat{H}}_2$. One arrives at 

\begin{align}
    \mathcal{\hat{H}}_1 \equiv \sum_{il\sigma} \varepsilon_{il} \hat{n}_{il\sigma} + \sum_{i\neq j, l\sigma} t^{ll'}_{ij} \hat{a}^{\dagger}_{il\sigma}\hat{a}_{il'\sigma},
    \label{32}
\end{align}

\noindent
and

\begin{align}
   & \mathcal{\hat{H}}_2 \equiv \sum_{il}\hat{U}_l\hat{n}_{il\uparrow}\hat{n}_{il\downarrow} + \frac{1}{2}\sum_{i,l\neq l'}\hat{U}'_{il}\hat{n}_{il}\hat{n}_{il'} 
    - \sum_{i,l\neq l'} J^{H}_{ll'}(\hat{\mathbf{S}}_{il} \cdot \hat{\mathbf{S}}_{il'} + \frac{1}{4}\hat{n}_{il}\hat{n}_{il'}) +  \\
    &+ \sum_{i, l\neq l'} J_{ll'}\hat{c}^{\dagger}_{il\uparrow}\hat{c}^{\dagger}_{il\downarrow}\hat{c}_{il\uparrow}\hat{c}_{il\downarrow} + \sum_{i\sigma, l\neq l'}
    V_{ll'}\hat{n}_{il\bar{\sigma}} (\hat{c}^{\dagger}_{il\sigma}\hat{c}_{il'\sigma} + \mathrm{H.c.}).
    \label{A33}
\end{align}

\noindent
The first term in Eq.~(\ref{32}) describes the atomic energy of particle placed on orthogonalized orbital \emph{l}, $t^{ll}_{ij}$ and $t^{ll'}_{ij}$ ($l \neq l'$) represent the intraband and interband (interorbital) hopping, respectively. In the case with $d$-fold degenerate band the atomic part is the same for all electrons, i.e.,  $\varepsilon_l = \varepsilon_{at}$ which can be taken as reference (zero) energy in a transitionally invariant system. The part, given by Eq.~(\ref{A33}), contains all pair-particle intraatomic interactions. The third term represents the interorbital ferromagnetic exchange interaction (the so-called \emph{Hund's rule} interaction). Particularly interesting is the modeling situation when the microscopic parameters are not explicitly dependent on the orbital index \emph{l}. This is the so-called model of with equivalent \emph{d} orbital and has some interesting global properties (cf. Klejnberg and Spa{\l}ek (1998) \cite{KlejnbergPhysRevB1998}).

\subsection{Concluding remarks}

The above model Hamiltonians contain microscopic parameters which are composed of integrals defined in terms of properly orthogonalized Wannier functions. Their determination is a separate task. In particular, they should be renormalized in the correlated state, i.e., if the interaction part is the dominant part in $\hat{\mathcal{H}}$. They can be determined explicitly in some model situations \cite{SpalekPhysRevB2000,SpalekJPCM2007}. Usually, we assume that they are adjustable and estimated either from experiment or from specific model calculations. Here their value is taken on the basis on the estimate based on atomic calculations and readjusted subsequently to match selected experimental results (see Sec.~\ref{sec:selected_equilibrium_properties}).

\section{Derivation and discussion of $t$-$J$ and $t$-$J$-$U$ models; real-space pairing representation}
\label{appendix:derivation_of_the_tj_model}

In this Appendix we discuss the principal effective Hamiltonians, originating from the Hubbard and related models. Particular emphasis is placed on reviewing the approach applicable in the strong-correlation limit. To define more concretely the strong-correlation limit, we derive first the form of exchange interaction among strongly correlated fermions (\emph{the kinetic exchange interaction}). The latter allows for a clear definition of simple mean-field approach in that limit, which is known under the name of \emph{renormalized mean-field theory} (RMFT). We also define an invariant (operator) form of the so-called \emph{real-space paring} and demonstrate its non-BCS character that results from the projected-fermion nature of the creation and annihilation operators in the limit of strong correlations.

\subsection{An elementary example: Mott-Hubbard insulators}

Here we derive the explicit form of the exchange interactions among localized electrons using the perturbation theory which is valid for $U \gg W$. We also define the {\it virtual\/} and {\it real\/} hopping processes as representation of interparticle correlations in the second-quantization language. Namely, for the spin lattice depicted in Fig.~\ref{fig:metal-isulator-transition}(b) we estimate the magnitude of antiferromagnetic spin-spin interactions.

We start again from standard form of the Hubbard Hamiltonian

\begin{align}
\mathcal{\hat{H}} = \sum_{ij\sigma}\!^{'}  t_{ij}\, \hat{a}_{i\sigma}^{\dagger}\, \hat{a}_{j\sigma} + U \sum_{i} \hat{n}_{i\uparrow}\,\hat{n}_{i\downarrow}
= \sum_{\mathbf{k}\sigma}\epsilon_{\mathbf{k}}\,\hat{n}_{\mathbf{k}\sigma}+U\sum_{i}\hat{n}_{i\uparrow}\,\hat{n}_{i\downarrow}
\label{8.1}
 \end{align}

\begin{figure}
\begin{center}
\includegraphics[width=0.5\textwidth]{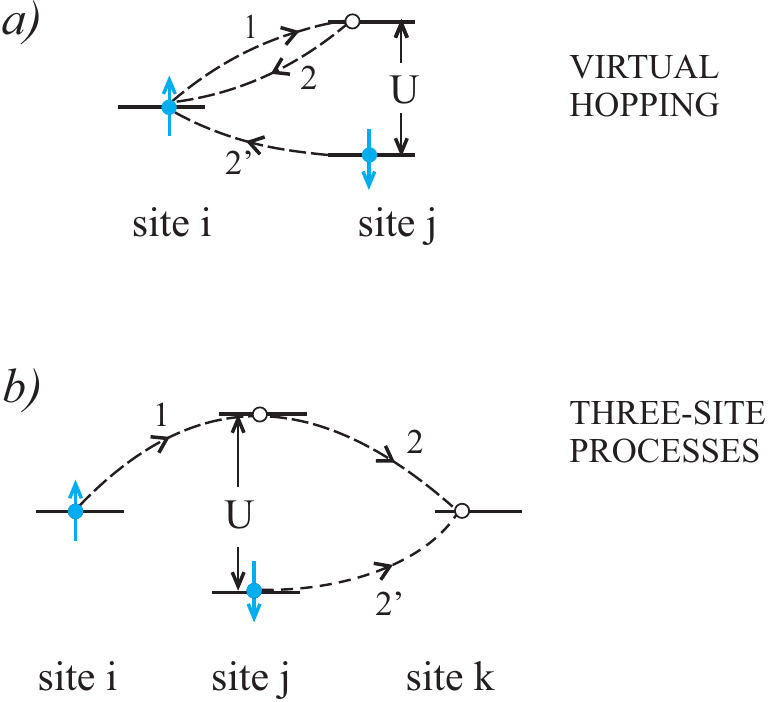}
\end{center}
\caption{
Illustration of virtual hopping between the singly occupied neigboring sites with opposite spins, leading either to the processes providing the two-site kinetic exchange (a) or inducing the three-site real hoping (b). The arrows marked as $2$ and $2^\prime$ represent alternative hopping processes. The processes (b) are possible only in non-half-filled situation ($n < 1$).}
\label{Virtual_hopping}
\end{figure}

\noindent
 In the limit $U \gg W$ and for $n=1$, only virtual hoppings are allowed (i.e., there is no net electron transfer between the sites: particles hop forth and back). This is the canonical principle behind Mott-insulator (quasi-atomic) state. Both real and virtual electron transfer are depicted in Fig.~\ref{Virtual_hopping}, but even the hopping processes depicted in Fig.~\ref{Virtual_hopping}(b) are possible only if the spins are antiparallel in the intermediate (double occupied Wannier) state. We will attend to this subtle point later. Here we examine the virtual processes in the second order. If the hopping takes place between two specific sites, $i$ and $j$, we can write down the effective Hamiltonian in the first nontrivial order and in the invariant (operator) language (not obvious at this stage) as\footnote{The operator form of (\ref{8.2}) expresses actually the matrix element in the second order: $t_{ij}\hat{a}_{i\sigma'}^{\dagger},\hat{a}_{j\sigma'}|m\rangle\langle m|\hat{a}_{j\sigma}^{\dagger},\hat{a}_{i\sigma}$, where $|m\rangle$ represents intermediate state. The part $|m\rangle\langle m|$ is not necessary, since the algebra of $a$-operators accounts for the dynamics correctly.}

\begin{align}
\mathcal{\hat{H}}_{ij \sigma}^{(2)} =
\sum_{\sigma'}\; \frac{ t_{ij}\: \hat{a}_{j\sigma'}^{\dagger}\: \hat{a}_{i\sigma'}\: t_{ij}\: \hat{a}_{i \sigma}^{\dagger}\: \hat{a}_{j \sigma }}{E_0 - E_m},
\label{8.2}
\end{align}

\noindent
where $E_{0}$ and $E_{m}$ are the system energies in the initial and intermediate states, respectively. The sum over $\mbox{$\sigma$} '$ takes into account the fact that the electron hopping back from the site $i$ back to $j$ can have either spin $\mbox{$\sigma$} ' = \mbox{$\sigma$}$, or $\mbox{$\sigma$} ' =  \mbox{$\bar{\sigma}$} $. Since process (\ref{8.2}) involves only one hop then in the limit $|t_{ij}| \ll U$, the energy difference to the first nontrivial order is $E_0 - E_m \simeq -U$. Therefore, considering separately the processes with $\mbox{$\sigma$} ' = \pm \mbox{$\sigma$}$ one obtains an effective two-site Hamiltonian after summing over $\mbox{$\sigma$} $ to account for the two possible processes. Explicitly,

\begin{align}
\mathcal{\hat{H}}_{ij}^{(2)} = - \frac{t_{ij}^2 }{U} \sum_{\mbox{$\sigma$}}
\left(\hat{a}_{i\sigma}^{\dagger}\: \hat{a}_{j\sigma}\:\hat{a}_{j\sigma}^{\dagger}\: \hat{a}_{i\sigma} +
\hat{a}_{i\bar{\sigma}}^{\dagger}\: \hat{a}_{j\bar{\sigma}}\: \hat{a}_{j\sigma}^{\dagger}\:\hat{a}_{i\sigma} \right).
\label{8.3}
\end{align}

\noindent
Using the relations: $\hat{n}_{i\sigma} = \hat{a}_{i\sigma}^{\dagger}\:\hat{a}_{i\sigma}$, $\;\hat{a}_{j\sigma}\: \hat{a}_{j\sigma}^{\dagger} = 1 -\hat{n}_{j\sigma}$, and $\hat{S}_i^{\sigma} = \hat{a}_{i\sigma}^{\dagger}\: \hat{a}_{i\bar{\sigma}}$, we can reduce Eq. (\ref{8.3}) to the form

\begin{align}
\mathcal{\hat{H}}_{ij}^{(2)} = -\frac{t_{ij}^2} {U} \sum_{\sigma}
\left[\hat{n}_{i\sigma} \left( 1 - \hat{n}_{j\sigma}\right) - \hat{S}_i^{\bar{\sigma}} \hat{S}_j^{\sigma} \right].
\label{8.4}
\end{align}

\noindent
Finally, we introduce the relation $\hat{n}_{i\sigma} = \frac{n_i}{2} + \sigma S_i^z$ to find
\begin{align}
\mathcal{\hat{H}}_{ij}^{(2)} = - \frac{t_{ij}^2}{U} \hat{n}_i + \frac{2t_{ij}^2}{U}
\left[ \frac{1}{4}\hat{n}_i\hat{n}_j + \hat{S}_i^z\hat{S}_j^z + \frac{1}{2} \left( \hat{S}_i^+\hat{S}_j^- + \hat{S}_i^-\hat{S}_j^+ \right) \right],
\label{8.5}
\end{align}

\noindent
where $\hat{n}_i = \hat{n}_{i\uparrow} + \hat{n}_{i\downarrow}$ is the total number of particles on site $i$. For the Mott insulator, we can set explicitly $\hat{n}_i=1$. Then, taking also into account the identity $\mathbf{\hat{S}}_{i}\cdot \mathbf{\hat{S}}_{j}\equiv \hat{S}_{i}^{z} \hat{S}_{j}^{z} + (1/2)\, (\hat{S}_{i}^{+} \hat{S}_{j}^{-} + \hat{S}_{i}^{-} \hat{S}_{j}^{+})$, we can cast Eq.~(\ref{8.5}) into the rotationally invariant form in the spin space, i.e.,

\begin{align}
\mathcal{\hat{H}}_{ij}^{(2)} = \frac{2t_{ij}^2}{U} \left( \mathbf{\hat{S}}_i \cdot \mathbf{\hat{S}}_j - \frac{1}{4} \right) .
\label{8.6}
\end{align}

\noindent
In the situation where many lattice sites are involved, we have thus the total additive perturbative contribution in the form

\begin{align}
\mathcal{\hat{H}}^{(2)} = \sum_{ij}\!^{'}  \; \frac{2t_{ij}^2}{U} \left( \hat{\mathbf{S}}_i \cdot {\hat{\mathbf{S}}}_j -\frac{1}{4} \right).
\label{8.7}
\end{align}

\noindent
Note that each pair of lattice sites is counted twice, since in (\ref{8.2}) we could have started enumerations either from site $i$, or from site $j$ (the two sites are equivalent and the particles are \emph{indistinguishable}). The Hamiltonian~(\ref{8.7}) describes the intersite antiferromagnetic exchange interactions, the so-called {\it kinetic exchange interactions} for the case $\hat{n}_i\equiv 1$ \cite{AndersonPhysRev1959}. Thus, the corresponding total kinetic exchange integral is equal to $J_{ij}\equiv 4t_{ij}^{2}/U$. Hence, in any subsequent application of Eq.~\eqref{8.7} we have to assume explicitly that each lattice site is singly occupied (and must then project out local double occupancies explicitly if we work in the fermionic language representing the pure spin lattice).

The inclusion of higher order processes requires a more powerful tool -- the canonical perturbation expansion (Chao \emph{et. al.}, (1978) \cite{ChaoPhysRevB1978}). Note that if $W=1\div 3\,\mathrm{eV}$ and $U=8\div 12\,\mathrm{eV}$, the ground state energy per site in the second-order is changed by the amount

\begin{align}
\frac{4t^2}{U} z = \frac{W^2}{Uz} \sim 10^{-3}\div 10^{-2}\,\mathrm{eV} \sim 10\text{-}10^{2}\,\mathrm{K}.
\end{align}

\noindent
Therefore, the higher-order processes are usually regarded as irrelevant. Clearly, the energy is lowered in the antiferromagnetic state, since then for a singlet pair $|s\rangle = \frac{1}{\sqrt{2}}\left(\hat{a}_{i\uparrow}^{\dagger}\hat{a}_{j\downarrow}^{\dagger}- \hat{a}_{i\downarrow}^{\dagger} \hat{a}_{j\uparrow}^{\dagger}\right)\ket{0}$ we have that $\langle s | \mathbf{\hat{S}}_i \cdot \mathbf{\hat{S}}_j - \frac{1}{4} |s\rangle = -1$ (we assume that the neighboring spins are not frustrated). The reader can check that for any of the three triplet states $\{|t\rangle\}$, $\langle t|\mathcal{\hat{H}}_{ij}^{(2)}|t\rangle=0$. This means, that the operator $(\mathbf{\hat{S}}_i \cdot \mathbf{\hat{S}}_j - \frac{1}{4})$ is a pair projection operator onto the spin-singlet state of an arbitrary two-electron $|ij\rangle$ state for $i\neq j$. If the condition $n_{i}=1$ is enforced on every site, the three-site processes shown in Fig.~\ref{Virtual_hopping} do not take place.

\subsection{Kinetic exchange and $t$-$J$ Hamiltonian in the leading (second) order: Original derivation}

\begin{figure}[pos=h]
\begin{center}
\includegraphics[width=0.8\textwidth]{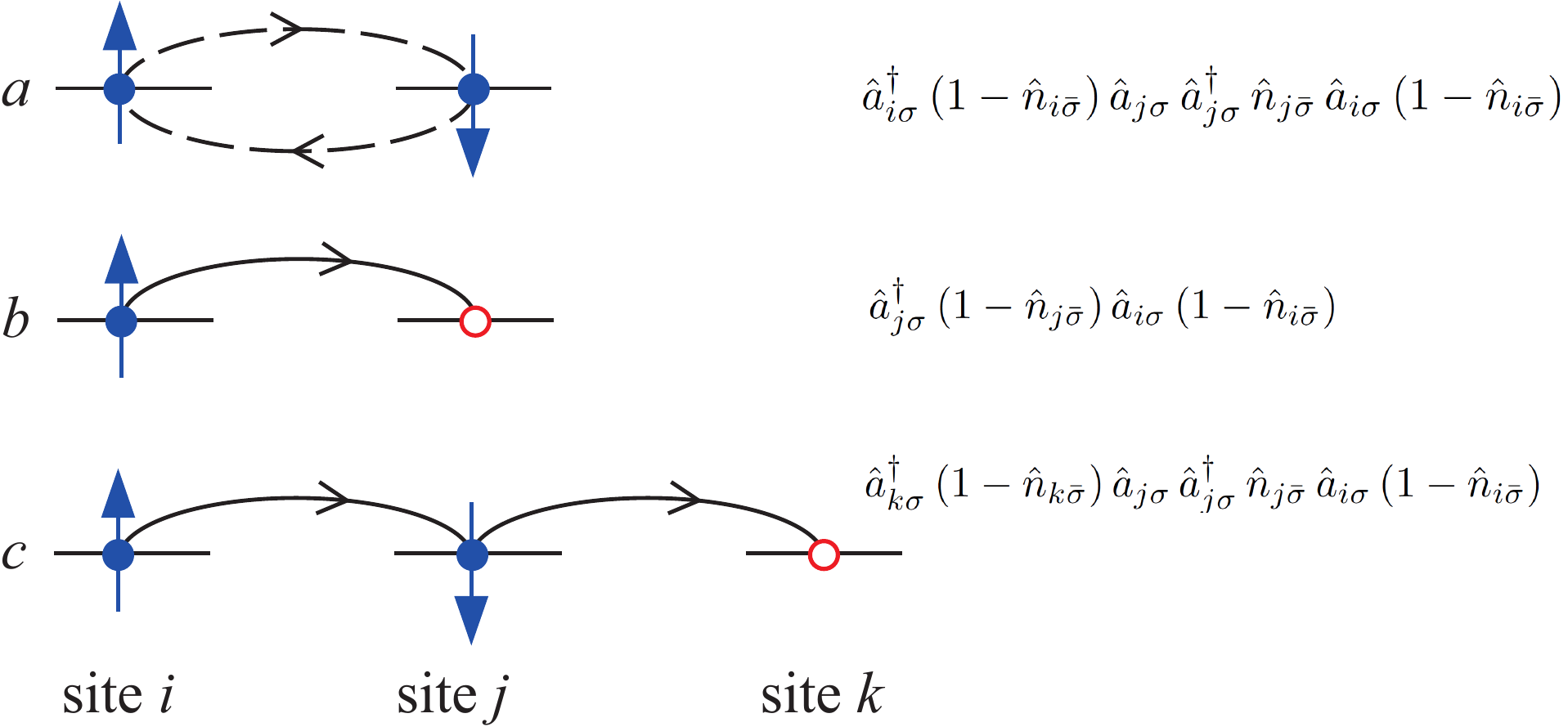}
\end{center}
\caption{
Real hopping process (panels (b) and (c)) as opposed to the virtual processes (a) in the case when holes in the Mott insulator are present. On the right, the formal expressions corresponding to those processes in terms of the respective parts of the projected hopping are shown (cf. Eq.~\eqref{8.8}).}
\label{Real_hopping}
\end{figure}

In this subsection, based on original papers~\cite{SpalekPhysicaB1977,ChaoJPCM1977}, we derive the effective Hamiltonian for $n < 1$ so that both the virtual processes considered in the previous subsection and the real hopping processes are allowed. The possible processes between neighboring sites are depicted schematically in Fig.~\ref{Real_hopping}. In the present situation, the process (a) corresponds to virtual hopping, whereas the processes (b) and (c) represent hopping of electron from site $i$ to empty sites $j$ or $k$ (this may happen because $\langle n_{i}\rangle<1$). Therefore, from our present treatment, we should obtain the effective Hamiltonian~\eqref{8.7} in the limit $n \rightarrow 1$.

We start formally by rewriting the hopping term in the following manner

\begin{align}
\hat{a}_{i\sigma}^{\dagger}\hat{a}_{j\sigma}  &\equiv \;
\hat{a}_{i\sigma}^{\dagger} \left(1 - \hat{n}_{i\bar{\sigma}} + \hat{n}_{i\bar{\sigma}}\right) \hat{a}_{j\sigma} \left(1 - \hat{n}_{j\bar{\sigma}} + \hat{n}_{j\bar{\sigma}}\right)
 \nonumber \\ &\equiv \hat{a}_{i\sigma}^{\dagger}\left(1 -\hat{n}_{i\bar{\sigma}}\right)\hat{a}_{j\sigma}\left(1-\hat{n}_{j\bar{\sigma}}\right) 
 + \hat{a}_{i\sigma}^{\dagger}\left(1-\hat{n}_{i\bar{\sigma}}\right)\hat{a}_{j\sigma}\hat{n}_{j\bar{\sigma}}
+ \hat{a}_{i\sigma}^{\dagger}\:\hat{n}_{i\bar{\sigma}}\:\hat{a}_{j\sigma}\left(1-\hat{n}_{j\bar{\sigma}}\right)
+ \hat{a}_{i\sigma}^{\dagger}\:\hat{n}_{i\bar{\sigma}}\:\hat{a}_{j\sigma}\:\hat{n}_{j\bar{\sigma}} .
\label{8.8}
\end{align}

\noindent
The first term in the second line of Eq.~\eqref{8.8} describes electron transfer from a singly occupied sites $j$ to an empty site $i$, the last deals with a transfer from a doubly occupied site $j$ to an already singly occupied site. The intermediate two terms represent respectively electron transfer from a doubly occupied site to an empty site and from a singly occupied site to an already singly occupied site. By such a decomposition, we distinguish the processes involving the high-energy transfers from those which do not. In this manner, we can decompose the whole dynamics into (\emph{i}) those in the subspace of physical (low) energies (i.e., with no energy scale $U$ involved), (\emph{ii}) those in the high-energy subspace (realized only as virtual configurations), and (\emph{iii}) those corresponding to the transitions between them. This decomposition in real space represents the starting point for the discussion of the strong-correlation physics and represents a systematic elaboration of the earlier ideas of Harris and Lange (1967) for infinite $U$ \cite{HarrisPhysRev1967}.

\begin{figure}[pos=t]
\begin{center}
\includegraphics[width=0.6\textwidth]{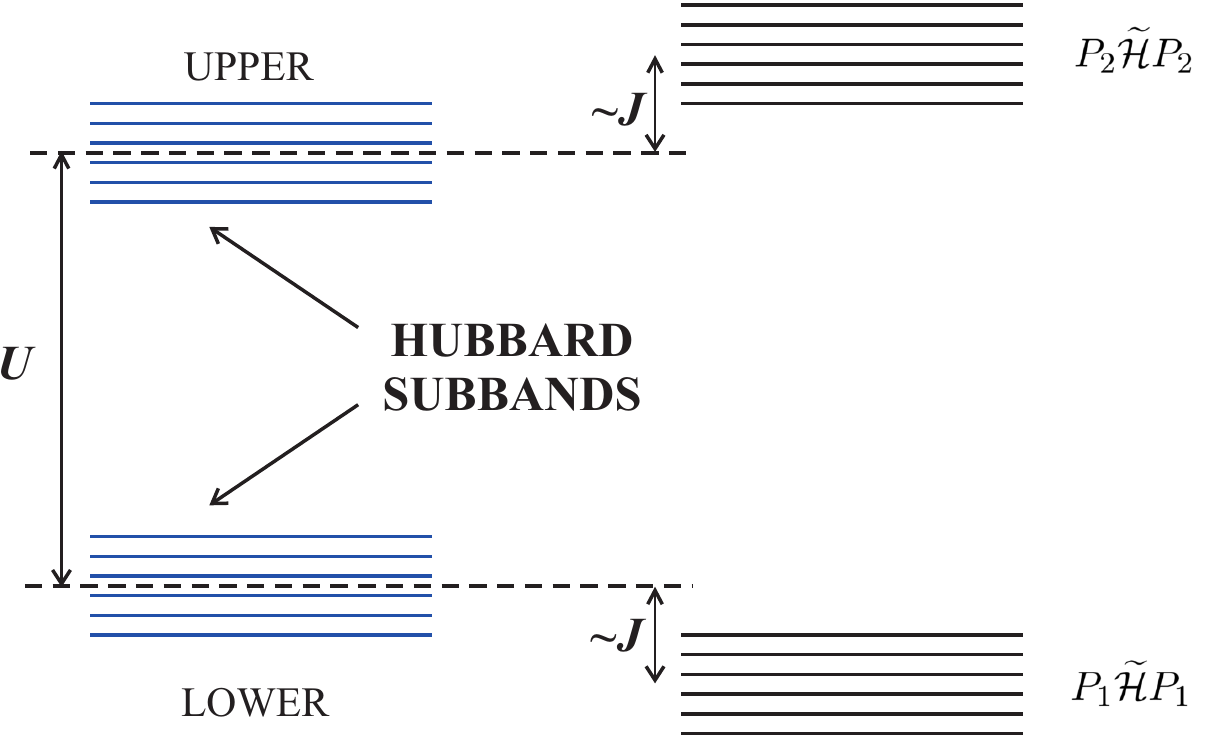}
\end{center}
\caption{
The Hubbard subbands (manifolds) before (left) and after taking the virtual hopping processes in the second order with $J\equiv 4t_{ij}^{2}/U$ (the lower subband is filled in the antiferromagnetic state). The subbands are represented formally by the projected effective Hamiltonian (see main text) with the kinetic exchange $\sim J$ included.}
\label{Subbands}
\end{figure}

To place the above reasoning on more formal grounds, we introduce two operators, $\hat{P}_1$ and $\hat{P}_2$, such that $\hat{P}_1 + \hat{P}_2 = \mathbb{1}$. Those operators are constructed as projectors onto orthogonal Fock subspaces so that $\hat{P}_{\mu}\hat{P}_{\nu} = \delta _{\mu \nu}$ with $\mu , \nu = 1,2$. In effect, $\hat{P}_1$ and $\hat{P}_2$ will represent subspaces without and with double occupancies, respectively. Now, in analogy to Eq.~\eqref{8.8}, we write the Hubbard Hamiltonian in the form

\begin{align}
\mathcal{\hat{H}} \equiv (\hat{P}_1 + \hat{P}_2)\mathcal{\hat{H}}(\hat{P}_1+\hat{P}_2) = \hat{P}_1\mathcal{\hat{H}}\hat{P}_1 + \hat{P}_2\mathcal{\hat{H}}\hat{P}_2 + \hat{P}_1\mathcal{\hat{H}}\hat{P}_2 + \hat{P}_2\mathcal{\hat{H}}\hat{P}_1,
\label{8.9}
\end{align}

\noindent
where the various projected parts are defined as follows:

\begin{align}
\hat{P}_{1}\mathcal{\hat{H}}\hat{P}_1 \equiv \hat{P}_1 \sum_{ij\sigma}\!^{'} t_{ij}\:\hat{a}_{i\sigma}^{\dagger}\left(1-\hat{n}_{i\bar{\sigma}}\right)\hat{a}_{j\sigma}\left(1-\hat{n}_{j\bar{\sigma}}\right)\hat{P}_1
\label{8.10}
\end{align}

\noindent
describes the dynamics of electrons in the lowest energy manifold (the lower Hubbard subband). The second term in Eq.~\eqref{8.10},

\begin{align}
\hat{P}_2\mathcal{\hat{H}}\hat{P}_2 \equiv \hat{P}_2 \left[ \sum_{ij\sigma}\!^{'} t_{ij}\:\hat{a}_{i\sigma}^{\dagger}\:\hat{n}_{i\bar{\sigma}}\:\hat{a}_{j\sigma}\:
\hat{n}_{j\bar{\sigma}} + U\sum_{i}\:\hat{n}_{i\uparrow}\:\hat{n}_{i\downarrow}\right]\hat{P}_2,
\label{8.11}
\end{align}

\noindent
deals with electronic properties in the higher Hubbard subbands, i.e., with one, two, and larger number of double occupancies, the first of them placed at the energy $\sim U$ above the lowest. Finally, the last two terms,

\begin{align}
\hat{P}_1\mathcal{\hat{H}}\hat{P}_2 \equiv \hat{P}_1 \left[\sum_{ij\sigma}\!^{'} t_{ij}\:\hat{a}_{i\sigma}^{\dagger}(1-\hat{n}_{i\bar{\sigma}})\:\hat{a}_{j\sigma}\:\hat{n}_{j\bar{\sigma}}\right]\hat{P}_2
\label{8.12}
\end{align}

\noindent
and

 \begin{align}
 \hat{P}_2\mathcal{\hat{H}}\hat{P}_1 \equiv
\left(\hat{P}_2\mathcal{\hat{H}}\hat{P}_1\right)^{\dagger} = \hat{P}_2 \left[ \sum_{ij\sigma}\!^{'} t_{ij}\:\hat{a}_{i\sigma}^{\dagger}\:\hat{n}_{i\bar{\sigma}}\:\hat{a}_{j\sigma}\left(1-\hat{n}_{j\bar{\sigma}}\right) \right]\hat{P}_1,
\label{8.13}
 \end{align}

\noindent
represent transitions between those low- end high-energy subspaces. In the above expressions we use real space (Wannier) representation of the electronic states. By direct inspection, one can check that $\mathcal{\hat{H}}=(\hat{P}_{1}+\hat{P}_{2})\mathcal{\hat{H}}(\hat{P}_{1}+\hat{P}_{2})$. Physically, the subbands should be well separated energetically in the limit $U \gg W$. The explicit expression for the projectors, $\hat{P}_l$, can be constructed by using the relation \cite{SpalekPhysicaB1977}

\begin{align}
\prod_{i}\left[\left(1-\hat{n}_{i\uparrow}\hat{n}_{i\downarrow} \right) + \hat{n}_{i\uparrow}\hat{n}_{i\downarrow}x\right] \equiv \sum_{l=0}^{N_e/2} \hat{P}_{l+1}x^l,
\label{8.14}
\end{align}

\noindent
where we have slightly altered the notation for the projection operators:\; now $\hat{P}_l$ projects onto the Fock subspace containing \emph{exactly} \emph{l} doubly occupied sites. In this language, $\hat{P}_1 \equiv \hat{P}_{l=0}$ and $\hat{P}_2 \equiv 1 - \hat{P}_1$. In the regime of strong correlations, when only the second-order (two-site) processes are considered and the intermediate state contains one doubly occupied site, we can take that $\hat{P}_2 = \hat{P}_{l=1}$. However, we should be taking the full form of the operator $\hat{P}_2 \equiv \hat{P}_{l=1} + \sum_{l>1}\hat{P}_l$ at the end, when we calculate the expectation value of the effective Hamiltonian 
(the system energy in the correlated state). The division of the states into the two categories, as represented by the projections $\hat{P}_{l=0}\mathcal{\hat{H}}\hat{P}_{l=0}$ and $\hat{P}_{l=1}\mathcal{\hat{H}}\hat{P}_{l=1}$, reflects the original concept of Hubbard subbands (Hubbard (1964) \cite{HubbardProcRoySoc1964}), depicted schematically in Fig.~\ref{Subbands}.

We now generalize slightly Eq.~\eqref{8.9} by introducing Hamiltonian operator in the form \cite{ChaoJPCM1977} 

\begin{align}
\mathcal{\hat{H}}(\varepsilon ) \equiv \mathcal{\hat{H}}_0 + \varepsilon\mathcal{\hat{H}}_1,
\label{8.15}
\end{align}

\noindent
where

\begin{align}
\mathcal{\hat{H}}_0 \equiv \hat{P}_1\mathcal{\hat{H}}\hat{P}_1 + \hat{P}_2\mathcal{\hat{H}}\hat{P}_2 ,
\label{8.16}
\end{align}

\noindent
and
\begin{align} \mathcal{\hat{H}}_1 \equiv \hat{P}_1\mathcal{\hat{H}}\hat{P}_2 + \hat{P}_2\mathcal{\hat{H}}\hat{P}_1 .
\label{8.17}
\end{align}

\noindent
The extra variable $\varepsilon$ will serve as a formal perturbation-expansion parameter, allowing us to group terms of the same magnitude. This factor differentiates between the parts of the hopping terms, contained in $\mathcal{\hat{H}}_0$, from those contained in $\mathcal{\hat{H}}_1$; only the latter part is regarded as a perturbation. Such a procedure is admissible if these two parts, each $\sim t_{ij}$, lead to the dynamical processes of different magnitude in the effective Hamiltonian. Ultimately, we set $\varepsilon= 1$ at the end of calculations.

We next search for a Hermitian operator $\hat{S}=\hat{S}^{\dagger}$, such that the canonically transformed (effective) Hamiltonian

\begin{align}
\widetilde{\hat{\mathcal{H}}} (\varepsilon ) \equiv \exp (-i\varepsilon \hat{S}) \mathcal{\hat{H}}(\varepsilon) \exp (i\varepsilon \hat{S})
\label{8.18}
\end{align}

\noindent
contains no term linear in $\varepsilon$. Io first order, such a procedure removes the hopping processes between the two well separated subbands, as they correspond to transfers from singly occupied to doubly-occupied (high-energy) states. Introduction of the variable $\varepsilon$ in the exponential functions in (see Eq.~\eqref{8.18}) means that we perform the transformation only on the part of $\mathcal{\hat{H}}(\varepsilon)$ involving the terms $\sim t$. Namely, we carry out the expansion

\begin{align}  \exp ( \pm i\varepsilon \hat{S}) = 1 \pm i\varepsilon S - \frac{1}{2} {\varepsilon}^2 \hat{S}^2 + \ldots,
\label{8.19}
\end{align}

\noindent
i.e., perform the canonical perturbation expansion and substitute it into Eq.~(\ref{8.18}) to obtain the expansion 

\begin{align}
\widetilde{\hat{\mathcal{H}}} (\varepsilon ) = \mathcal{\hat{H}}_0 + \varepsilon\left( \mathcal{\hat{H}}_1 + i\varepsilon[\mathcal{\hat{H}}_0, \hat{S}]\right) +
\frac{1}{2} {\varepsilon}^2\left( 2i[\mathcal{\hat{H}}_1, \hat{S}] - [ [\mathcal{\hat{H}}_0,\hat{S}], \hat{S}] \right) + \ldots. 
\label{8.20}
\end{align}

\noindent
The transformation $\hat{S}$ is determined through the requirement that the term linear in $\varepsilon$ should be removed from $\widetilde{\mathcal{\hat{H}}}$, i.e., we should have

\begin{align}
 \mathcal{\hat{H}}_1 + i[\mathcal{\hat{H}}_0,\hat{S}] = 0,
\label{8.21}
\end{align}

\noindent
from which it follows that

\begin{align}
\widetilde{\hat{\mathcal{H}}} (\varepsilon ) = \mathcal{\hat{H}}_0 + \frac{1}{2} i {\varepsilon}^2[\mathcal{\hat{H}}_1, \hat{S}] + ...  \equiv \mathcal{\hat{H}}_0 + \sum_{n=2}^{\infty} \frac{(n-1)(-i)^{n-1}\varepsilon^n}{n!} [[\hat{S},\mathcal{\hat{H}}_1]]_{n-1},
\label{8.22}
\end{align}

\noindent
where we have defined $[[A,B]]_n \equiv [A,[A[..., [A,B]...]]$. The principal difficulty in solving Eq.~\eqref{8.21} is connected with evaluation of the commutator on the right-hand side, under the circumstance that $\mathcal{\hat{H}}_0$ is not in diagonal form (it contains the projected hopping part \eqref{8.10}). To address this problem, we multiply Eq.~(\ref{8.21}) by $\hat{P}_{\mu}$ from the left and by $\hat{P}_{\nu}$ from the right, and replace $\hat{S}$ with the projected expression $(\hat{P}_1+\hat{P}_2)\hat{S}(\hat{P}_1+\hat{P}_2)$. One then arrives at

\begin{align}
\left(\hat{P}_{\mu}\mathcal{\hat{H}}\hat{P}_{\mu}\right)(\hat{P}_{\mu}\hat{S}\hat{P}_{\nu}) -
(\hat{P}_{\mu}\hat{S}\hat{P}_{\nu})(\hat{P}_{\nu}\mathcal{\hat{H}}\hat{P}_{\nu}) = i(\hat{P}_{\mu}\mathcal{\hat{H}}\hat{P}_{\nu})(1 -
\delta_{\mu \nu}).
\label{8.23}
\end{align}

\noindent
Note that the projected part $\hat{P}_\nu\hat{\mathcal{H}}\hat{P}_{\nu}$ is, in general, not diagonal, and this feature contributes to the complexity of solving this operator equation. Yet, for $\mu = \nu$ Eq.~\eqref{8.23} reduces to

\begin{align}
(\hat{P}_{\mu}\mathcal{\hat{H}}\hat{P}_{\mu})(\hat{P}_{\mu}\hat{S}\hat{P}_{\mu}) -
(\hat{P}_{\mu}\hat{S}\hat{P}_{\mu})(\hat{P}_{\mu}\mathcal{\hat{H}}\hat{P}_{\mu}) = 0 .
\label{8.24}
\end{align}

\noindent
Since $(\hat{P}_{\mu}\hat{S}\hat{P}_{\mu})$ commutes with $(\hat{P}_{\mu}\mathcal{\hat{H}}\hat{P}_{\mu}) = \mathcal{\hat{H}}_0$, we may set $\hat{P}_{\mu}\hat{S}\hat{P}_{\mu} \equiv 0$.

The nontrivial part of the solution of (\ref{8.23}) is related to the part $\mathcal{\hat{H}}_1$. For $\mu \neq \nu$, we may rewrite Eq.~\eqref{8.23} in the form

\begin{align}
\hat{P}_1\hat{S}\hat{P}_2 = -i\hat{P}_1\mathcal{\hat{H}}\hat{P}_2 + (\hat{P}_1\mathcal{\hat{H}}\hat{P}_1)(\hat{P}_1\hat{S}\hat{P}_2)(\hat{P}_2\mathcal{\hat{H}}\hat{P}_2)^{-1} .
\label{8.25}
\end{align}

\noindent
This solution is obtained iteratively, starting from the zeroth-order term $\hat{P}_1\hat{S}^{(0)}\hat{P}_2 =0$ to find $\hat{P}_1\hat{S}^{(1)}\hat{P}_2$, etc. We are interested in the {\it first nontrivial} order, i.e., in the creation of a single double occupancy via the elementary process $j \rightarrow i$ contained in $\hat{P}_1\mathcal{\hat{H}}\hat{P}_2$. By rewriting Eq. (\ref{8.25}) as

\begin{align}
\hat{P}_1\hat{S}^{(1)}\hat{P}_2 = \left( 1 + \frac{\hat{P}_1\mathcal{\hat{H}}\hat{P}_1}{\hat{P}_2\mathcal{\hat{H}}\hat{P}_2} \right)
(-i\hat{P}_1\mathcal{\hat{H}}\hat{P}_2) (\hat{P}_2\mathcal{\hat{H}}\hat{P}_2)^{-1} .
\label{8.26}
\end{align}

\noindent
To the first order in $t_{ij}$

\begin{align}
(\hat{P}_1\hat{S}^{(1)}\hat{P}_2)^{\dagger} = \frac{-i\hat{P}_1\mathcal{\hat{H}}\hat{P}_2}{1 - \hat{P}_1\mathcal{\hat{H}}\hat{P}_1(\hat{P}_2\mathcal{\hat{H}}\hat{P}_2)^{-1}}
(\hat{P}_2\mathcal{\hat{H}}\hat{P}_2)^{-1}
\simeq \frac{-i\hat{P}_1\mathcal{\hat{H}}\hat{P}_2}{\langle \hat{P}_2\mathcal{\hat{H}}\hat{P}_2\rangle - \langle \hat{P}_1\mathcal{\hat{H}}\hat{P}_1\rangle } = \frac{-i\hat{P}_1\mathcal{\hat{H}}\hat{P}_2}{U} ,
\label{8.27}
\end{align}

\noindent
where we have replaced the average distance between the Hubbard subbands by its value in the atomic limit, $U$ (in general,  contributions renormalizing intrasubband dynamics in higher orders appear, see, e.g., \cite{ChaoPhysRevB1978}). Similarly,

\begin{align}
\hat{P}_2\hat{S}^{(1)}\hat{P}_1 = \frac{i\hat{P}_2\mathcal{\hat{H}}\hat{P}_1}{U} .
\end{align}

\noindent
By inserting these equations into Eq.~\eqref{8.22}, we obtain the effective Hamiltonian in the form

\begin{align}
\widetilde{\mathcal{\hat{H}}} = \hat{P}_1\mathcal{\hat{H}}\hat{P}_1 + \hat{P}_2\mathcal{\hat{H}}\hat{P}_2 - \frac{1}{U} \left\{ (\hat{P}_1\mathcal{\hat{H}}\hat{P}_2)(\hat{P}_2\mathcal{\hat{H}}\hat{P}_1) - (\hat{P}_2\mathcal{\hat{H}}\hat{P}_1)(\hat{P}_1\mathcal{\hat{H}}\hat{P}_2) \right\}
\equiv \hat{P}_1 \widetilde{\mathcal{\hat{H}}}\hat{P}_1 + \hat{P}_2 \widetilde{\mathcal{\hat{H}}} \hat{P}_2 .
\label{8.28}
\end{align}

The effective Hamiltonian contains the single-particle hopping part $\hat{P}_{\mu}\mathcal{\hat{H}}\hat{P}_{\mu}$ within the two Hubbard subbands, as well as the virtual hopping processes between the two subbands $(\mbox{the}\,\hat{P}_{\mu} \mathcal{\hat{H}}\hat{P}_{\nu} \mathcal{\hat{H}}\hat{P}_{\mu}$ part), separated by the energy $U$. Thus the right-hand-side of Eq.~\eqref{8.28} represents effective Hamiltonian for the Hubbard subbands, with no cross-hopping between them. For $n \leq 1$ only the part $\hat{P}_1\mathcal{\hat{H}}\hat{P}_1 - (1/U)\hat{P}_1\mathcal{\hat{H}}\hat{P}_2\mathcal{\hat{H}}\hat{P}_1$ matters since, in the strong correlation limit, no double occupancies in the ground state occur (i.e., $\hat{P}_{2}\tilde{\mathcal{\hat{H}}}\hat{P}_{2}$ part can be neglected). Explicitly,

\begin{align}
\hat{P}_1\mathcal{\hat{H}}\hat{P}_2\mathcal{\hat{H}}\hat{P}_1 = \hat{P}_1 \sum_{ij\sigma}\!^{'}t_{ij}\: \hat{a}_{i\sigma}^{\dagger} \left(1-\hat{n}_{i\bar{\sigma}}\right) \hat{a}_{j\sigma}\:\hat{n}_{j\bar{\sigma}} \hat{P}_2
\sum_{kl\sigma'}\!^{'} t_{lk}\:\hat{a}_{l\sigma'}^{\dagger}\:\hat{n}_{l\bar{\sigma}'}\:\hat{a}_{k\sigma'}\left(1 -\hat{n}_{k\bar{\sigma}'}\right)\hat{P}_1.
\label{8.29}
\end{align}

\noindent
In the absence of double occupancies in the ground state, the only possible processes are those for which $l=j$. By decomposing the summation over $\mbox{$\sigma$} '$ into that over $\mbox{$\sigma$} ' =
\mbox{$\sigma$}$ and $\mbox{$\sigma$} ' = \mbox{$\bar{\sigma}$}$ separately, we arrive at the expression

\begin{align}
\hat{P}_1\mathcal{\hat{H}}\hat{P}_2\mathcal{\hat{H}}\hat{P}_1 = \hat{P}_1 \sum_{ijk\sigma}\!^{'} t_{ij}t_{jl} \left\{
\hat{a}_{i\sigma}^{\dagger}\left(1-\hat{n}_{i\bar{\sigma}}\right)\hat{a}_{j\sigma}\:\hat{n}_{j\bar{\sigma}}\:\hat{a}_{j\sigma}^{\dagger}\:\hat{n}_{j\bar{\sigma}}\:\hat{a}_{k\sigma}\left(1-\hat{n}_{k\sigma}\right) \right. \nonumber\\
\left. +\, \hat{a}_{i\sigma}^{\dagger}\left(1 - \hat{n}_{i\bar{\sigma}}\right) \hat{a}_{j\sigma}\: \hat{n}_{j\bar{\sigma}}\: \hat{a}_{j\bar{\sigma}}^{\dagger}\: \hat{n}_{j\sigma}\: \hat{a}_{k\bar{\sigma}} \left(1-\hat{n}_{k\sigma}\right)
\right\} \hat{P}_1.
\label{8.30}
\end{align}

\noindent
In the first line of Eq.~\eqref{8.30}, we may transpose $\hat{a}_{j\sigma}^{\dagger}$ and $\hat{n}_{j\bar{\sigma}}$, set $\hat{n}_{j\bar{\sigma}}^2 = \hat{n}_{j\bar{\sigma}}$, and commute $\hat{a}_{j\sigma}\: \hat{a}_{j\sigma}^{\dagger} = 1 - \hat{n}_{j\sigma}$. In the second line we note that $\hat{n}_{j\bar{\sigma}}\:\hat{a}_{j\bar{\sigma}}^{\dagger} = \hat{a}_{j\bar{\sigma}}^{\dagger}$, and that $\hat{a}_{j\bar{\sigma}}\:\hat{a}_{j\sigma}^{\dagger} = - \hat{S}_j^{\sigma}$. Therefore,
\begin{align}
\hat{P}_1\mathcal{\hat{H}}\hat{P}_2\mathcal{\hat{H}}\hat{P}_1 = \hat{P}_1 \sum_{ijk\sigma}\!^{'} t_{ij}\:t_{jk} \left\{
\hat{a}_{i\sigma}^{\dagger} \left(1-\hat{n}_{i\bar{\sigma}}\right)\hat{n}_{j\bar{\sigma}}\left(1-\hat{n}_{j\sigma}\right)\hat{a}_{k\sigma}\left(1 - \hat{n}_{k\bar{\sigma}}\right) \right. \nonumber +\\
\left. +\, \hat{a}_{i\bar{\sigma}}^{\dagger}\left(1-\hat{n}_{i\sigma}\right)\left(-\hat{S}_j^{-\sigma}\right)\hat{a}_{k\bar{\sigma}}\left(1-\hat{n}_{k\sigma}\right) \right\}\hat{P}_1.
\label{8.31}
\end{align}

\noindent
Let us consider separately the terms with $k=i$ (two-site terms) and with $k \neq i$ (three-site terms). In the former case, we can transpose $\hat{a}_{k\sigma}\left(1-\hat{n}_{k\sigma}\right)$ with the operators preceding them to obtain the sequence $\hat{a}_{i\sigma}^{\dagger}\:\hat{a}_{i\sigma}\left(1-\hat{n}_{i\bar{\sigma}}\right) = \hat{n}_{i\sigma}\left(1-\hat{n}_{i\bar{\sigma}}\right)$. In line two, we obtain $\hat{S}_i^{\sigma}$ in this manner. The $k=i$ contribution thus reduces to

\begin{align}
\hat{P}_1\sum_{ij\sigma}\!^{'}t_{ij}^2 \left\{ \hat{n}_{i\sigma} \left(1-\hat{n}_{i\bar{\sigma}}\right)\hat{n}_{j\sigma}\left(1-\hat{n}_{j\bar{\sigma}}\right) - \hat{S}_i^{\sigma} \hat{S}_j^{\bar{\sigma}} \right\}\hat{P}_1.
\label{8.32}
\end{align}

\noindent
Next, by substituting $\hat{n}_{i\sigma} = \frac{\hat{n}_i}{2} + \sigma \hat{S}_i^z$ and taking $\hat{n}_{i\sigma}\left(1 - \hat{n}_{i\bar{\sigma}}\right) = \frac{\hat{\nu}_i}{2} + \sigma \hat{S}_i^z$,
where $\hat{\nu}_i \equiv \sum_{\sigma} \hat{n}_{i\sigma}\left(1 - \hat{n}_{i\bar{\sigma}}\right)$ is the operator describing to the number of single occupancies, we obtain

\begin{align}
\hat{P}_1 \sum_{ij}\!^{'} 2t_{ij}^2 \left\{ \frac{1}{4} \hat{\nu}_i\hat{\nu}_j - \hat{S}_i^z\hat{S}_j^z - \frac{1}{2}\sum_{\sigma} \hat{S}_i^{\sigma}\hat{S}_j^{\bar{\sigma}} \right\} \hat{P}_1 =
\hat{P}_1 \sum_{ij}\!^{'} 2t_{ij}^2 \left( \frac{1}{4} \hat{\nu}_i \hat{\nu}_j - \mathbf{\hat{S}}_i \cdot \mathbf{\hat{S}}_j \right) \hat{P}_1.
\label{8.33}
\end{align}

\noindent
Hence, the two-site (kinetic-exchange) interaction differs from that in the insulating case only by the presence of the factor $\hat{\nu}_i \hat{\nu}_j$ in the first term. Obviously, for the Mott-Hubbard insulator $\hat{\nu}_i = \hat{\nu}_j = 1$, i.e., the projected particle-number operators are just occupation integer numbers at each site. Also, the spin must have the fermion representation.

We can discuss the three-site terms in the same manner. We simply summarize the result here. The total effective Hamiltonian projected onto the lower subspace takes the form \cite{SpalekPhysRevB1988}

\begin{align}
\hat{P}_1\widetilde{\hat{\mathcal{H}}} \hat{P}_1 = \hat{P}_1 \left\{ \sum_{ij\sigma}\!^{'} t_{ij\sigma}\: \hat{a}_{i\sigma}^{\dagger} \left(1 - \hat{n}_{i\bar{\sigma}}\right)\hat{a}_{j\sigma}\left(1-\hat{n}_{j\bar{\sigma}}\right)
+ \sum_{ij}\!^{'} \frac{2t_{ij}^2}{U} \left( {\mathbf{\hat{S}}}_i \cdot \mathbf{\hat{S}}_j - \frac{1}{4} \nu_i \nu_j \right) \right. \nonumber-\\
 \left.
 - \sum_{ijk\sigma}\!^{''} \frac{t_{ij}\:t_{jk}}{U} \left[ \hat{a}_{i\sigma}^{\dagger} \left(1-\hat{n}_{i\bar{\sigma}}\right) \hat{n}_{j\bar{\sigma}}\left(1-\hat{n}_{j\sigma}\right)\hat{a}_{k\sigma}\left(1-\hat{n}_{k\bar{\sigma}}\right) + \hat{a}_{i\sigma}^{\dagger} \left(1-\hat{n}_{i\bar{\sigma}}\right)\hat{S}_j^{-\sigma} \hat{a}_{k\bar{\sigma}}\left(1-\hat{n}_{k\sigma}\right) \right] \right\}\hat{P}_1 .
\label{8.34}
\end{align}

\noindent
In the above equation, the first term on the right-hand-side describes the hopping in the lowest subspace, i.e., between single occupied and empty sites. The second accounts for the antiferromagnetic kinetic exchange interaction. The third is the three-site hopping, without and with spin flip of the intermediate electron (located at site $j$).

The above Hamiltonian motivates definition of a consistent algebra of the projected operators, namely

\begin{align}
\left\{
\begin{array}{ll}
\hat{b}_{i\sigma} \equiv \hat{a}_{i\sigma} \left(1-\hat{n}_{i\bar{\sigma}}\right), \; & \; \hat{b}_{i\sigma}^{\dagger} \equiv \hat{a}_{i\sigma}^{\dagger} \left(1- \hat{n}_{i\bar{\sigma}}\right), \\
\hat{\nu}_{i\sigma} \equiv \hat{b}_{i\sigma}^{\dagger}\:\hat{b}_{i\sigma} = \hat{n}_{i\sigma}\left(1-\hat{n}_{i\bar{\sigma}}\right), \; & \; \hat{\nu}_i \equiv \sum_{\sigma}
\hat{b}_{i\sigma}^{\dagger}\:\hat{b}_{i\sigma},
\end{array}
\right.
\label{8.35}
\end{align}

\noindent
Now one can write Eq.~\eqref{8.34} in a more compact form

\begin{align}
\widetilde{\hat{\mathcal{H}}}= \hat{P}_1\left\{\sum_{ij\sigma}\!^{'} t_{ij}\: \hat{b}_{i\sigma}^{\dagger}\:\hat{b}_{j\sigma} + \sum_{ij}\!^{'} \frac{2t_{ij}^2}{U} \left( \mathbf{\hat{S}}_i \cdot \mathbf{\hat{S}}_j - \frac{1}{4}
\hat{\nu}_i \hat{\nu}_j \right) - \sum_{ijk \sigma}\!^{''} \frac{t_{ij}\:t_{jk}}{U} \left( \hat{b}_{i\sigma}^{\dagger}\: \hat{\nu}_{j\bar{\sigma}}\: \hat{b}_{k\sigma} + \hat{b}_{i\sigma}^{\dagger}\: \hat{S}_i^{\bar{\sigma}}\: \hat{b}_{k\bar{\sigma}} \right)\right\}\hat{P}_1 .
\label{8.36}
\end{align}

\noindent
Note that the presence of the projection operators $\hat{P}_1$ outside of the curly brackets $\{\}$ is justified, even though the expression inside takes already the projected form. This is because the projector $\hat{P}_1$ contains also the corresponding eliminations of the double occupancies from the sites distinct from any current pair $\braket{i,j}$ or triple $\braket{i,j,k}$, appearing in summations. This projection plays a nontrivial role when taking into account higher order correlations (beyond renormalized mean-field theory), as is the case in the DE-GWF method (see Sec.~\ref{sec:vwf_solution}).  

Equation~\eqref{8.36} represents the full form of the $t$-$J$ Hamiltonian in the fermion representation (i.e., the effective Hamiltonian $\widetilde{\hat{\mathcal{H}}}$ projected onto the lowest Fock subspace $\hat{P}_1$). It contains the kinetic-exchange antiferromagnetic interaction, entangled with a nontrivial hopping part and containing restricted two- and three- site processes, as those terms do not commute among themselves. Note that the single primed summation means that $i\neq j$, whereas doubly primed correspondent comprises the three-site terms with $i\neq j \neq k \neq i$. Note also that, within the algebra~\eqref{8.35}, we can define the spin operators as well, namely

\begin{align}
\left\{
\begin{array}{l}
\hat{S}_i^{\sigma} \equiv \hat{a}_{i\sigma}^{\dagger}\:\hat{a}_{i\bar{\sigma}} \equiv \hat{b}_{i\sigma}^{\dagger}\:\hat{b}_{i\bar{\sigma}}, \\
\hat{S}_i^z \equiv \frac{1}{2} \sum_{\sigma} \sigma \hat{n}_{i\sigma} \equiv \frac{1}{2} \sum_{\sigma} \sigma \hat{\nu}_{i\sigma},
\end{array}
\right.
\label{8.37}
\end{align}

\noindent
as can be checked out by direct calculation. This universality in defining the spin operator through either the original or the projected fermions should be noted.  However, there is one principal difference between the algebra of $b$-operators as compared to that for $a$-operators. Namely, they obey nonstandard (non-fermion) anticommutation relations

\begin{align}
\left\{ \hat{b}_{i\sigma}, \hat{b}_{i\sigma'}^{\dagger} \right\} = \left(1-\hat{n}_{i\bar{\sigma}}\right)\delta_{\sigma\sigma'} + \left(1- \delta_{\sigma\sigma'}\right)
\hat{S}_i^{\bar{\sigma}},
\label{8.38}
\end{align}

\noindent
i.e., they do not anticommute to a number. Additionally, in the $U \rightarrow \infty$ limit $(U \gg W)$ the first term on right-hand-side of Eq.~\eqref{8.38} can be replaced by $\left(1- \hat{\nu}_{i\bar{\sigma}}\right)$ to make the representation self-contained. Also, the remaining anticommutation relations for the projected operators are standard, namely $\{\hat{b}_{i\sigma}, \hat{b}_{j\sigma}\} = \{\hat{b}_{i\sigma}^{\dagger}, \hat{b}_{j\sigma}^{\dagger}\}  =  0$, for $i\neq j$.

The principal message of the effective Hamiltonian~\eqref{8.36} is that it contains the dynamics of holes (empty sites) on a lattice of spins coupled by exchange interactions. The correlation between the dynamics of holes and that of the interacting spins is substantial, since the projected hopping term and the exchange part do not commute with each other. 

In a similar manner one can derive the Hamiltonian for the upper Hubbard subband which becomes populated when $n>1$. As the method of approach is completely identical, as above we only provide the final result for the effective Hamiltonian

\begin{align}
& \hat{P}_{2}\widetilde{\hat{\mathcal{H}}}\hat{P}_{2}=\hat{P}_{2} \left\{ \sum_{ij\sigma}\!^{'} t_{ij}\,\hat{c}_{i\sigma}^{\dagger}\, \hat{c}_{j\sigma}+\frac{U}{2}\sum_{i\sigma}
\hat{c}_{i\sigma}^{\dagger}\, \hat{c}_{i\sigma}+\sum_{ij\sigma}\!^{'} \frac{2t_{ij}^{2}}{U} \left(\hat{c}_{i\uparrow}^{\dagger}\,
\hat{c}_{i\downarrow}^{\dagger}\, \hat{c}_{j\downarrow}\,
\hat{c}_{j\uparrow}+\hat{c}_{i\downarrow}\:\hat{c}_{i\uparrow}\: \hat{c}_{j\uparrow}^{\dagger}\, \hat{c}_{j\downarrow}^{\dagger}\right) \hat{P}_{2} \right\} + \\
& +\mbox{(three-site terms)}. \\
\label{8.39}
\end{align}

\noindent
In this expression the new projected operators are

\begin{align}
\left\{
\begin{array}{ll}
  \hat{c}_{i\sigma}\,\equiv\, \hat{a}_{i\sigma}\: \hat{n}_{i\overline{\sigma}}, \\
  \hat{c}_{i\sigma}^{\dagger}\equiv \hat{a}_{i\sigma}^{\dagger}\: \hat{n}_{i\overline{\sigma}} \\
\hat{c}_{i\sigma}^{\dagger}\: \hat{c}_{i\sigma}= \hat{c}_{i\overline{\sigma}}^{\dagger}\: \hat{c}_{i\overline{\sigma}}=
\hat{n}_{i\uparrow}\: \hat{n}_{i\downarrow}.
\end{array}
\right.
\label{8.40}
\end{align}

\noindent
The first term of Eq.~\eqref{8.39} describes the hopping process among the doubly occupied sites, the second expresses the increase of atomic energy due to creation of an extra doubly occupied 
site, and the last reflects the increased system energy by the virtual hopping processes. The change of the narrow-band system energetics due to the virtual hopping 
processes is shown schematically in Fig.~\ref{Subbands}. 

\subsection{Remark: Holes in the Mott insulator once more}

The Hamiltonian~\eqref{8.36} reduces to the spin Hamiltonian~\eqref{8.7} in the limit of Mott insulator. Namely, in that limit the electrons are localized on atoms which means that the number of them is constant on \emph{every} lattice site, i.e., the particle-number operator, takes only one of its values $\hat{n}_{i}=1$. In that limit, the projected hopping term vanishes because $\hat{n}_{i\uparrow}+ \hat{n}_{i\downarrow}=1$. In the same manner, the three-site hopping term in Eq.~\eqref{8.36} is zero, and we are left only with the full Heisenberg term. Similarly, the whole Hamiltonian $\hat{P}_{2}\widetilde{\mathcal{\hat{H}}}\hat{P}_{2}$ vanishes. The Mott insulator corresponds thus to a completely filled lower Hubbard subband, expressing antiferromagnetically-coupled localized spins.

The question is what happens when $n_{i}<1$? Can the holes in the Mott insulating state be treated as ordinary band holes in an almost filled band? In an ordinary band, the sum of the number of carriers, $N_{e}$, and that of empty states, $N_{h}$, is constant, namely $N_{e}+N_{h}=2N$, where $N$ is the number of atomic sites contributing to those band states. Thus, creation of a hole (particle with the spin $\sigma$) means that in that state there is still particle with the opposite spin (for each occupied state below the Fermi energy). In the case of doped Mott insulator, the number of states in the lower Hubbard subband $N_{e}+N_{h}$ may not be a {\it constant of motion} (only in the Mott insulating limit we have that $N_{e}+N_{h}=N$). Therefore, we have to understand the hole states in the present situation in a different sense. Namely, we can perform the electron-hole transformation utilizing that the electron removal from given atomic site is equivalent to creation of a hole as an empty site. It is important to realize that the discussion of high-$T_c$ superconductivity uses the term ``holes'' in that sense as they represent empty sites in the Mott insulators, \emph{not} the holes in almost filled band. We analyze the paired states in real space analyzed with the use of Eq.~\eqref{8.36}, but the results are presented as a function of the number of holes $\delta\equiv 1-\langle \hat{n}_{i}\rangle$.

\subsection{Kinetic exchange in the extended Hubbard model}

The Hubbard Hamiltonian represents only the simplest model of correlated electrons in a single narrow band. One can write down the full narrow-band Hamiltonian, including all two-site interactions (cf. Appendix~\ref{appendix:hubbard_model}). Namely, we now have the starting Hamiltonian for the analysis of strongly correlated electrons in the form

\begin{align}
\mathcal{\hat{H}} =&\sum_{ij\sigma}\!^{'}  t_{ij}
\hat{a}_{i\sigma}^{\dagger} \, \hat{a}_{j\sigma}+U\sum_{i}\,\hat{n}_{i\uparrow}\, \hat{n}_{i\downarrow}+\frac{1}{2}
\sum_{ij\sigma \sigma'}\!\!^{'} \left( K_{ij}-\frac{1}{2} J_{ij} \right) \hat{n}_{i\sigma} \hat{n}_{j\sigma^{\prime}} \nonumber\\
&- \sum_{ij}\!^{'} J_{ij} \mathbf{\hat{S}}_{i}\cdot \mathbf{\hat{S}}_{j} +\sum_{ij}\!^{'}J_{ij}^{H} \hat{a}_{i\uparrow}^{\dagger} \hat{a}_{i\downarrow}^{\dagger} 
\hat{a}_{\downarrow} \hat{a}_{j\uparrow} + \sum_{ij\sigma}\!^{'} V_{ij} \hat{n}_{i\bar{\sigma}} \left(\hat{a}_{i\sigma}^{\dagger}
\hat{a}_{j\sigma} + \hat{a}_{j\sigma}^{\dagger}  \hat{a}_{i\sigma}\right).
\label{8.41}
\end{align}

\noindent
Apart from the first two terms representing the Hubbard model, the third term expresses the direct intersite Coulomb interaction, the fourth the direct intersite (Heisenberg-Dirac) exchange interaction, the fifth the pair-hopping processes, and the last one the so-called correlated hopping processes (i.e., the hopping under the presence of an extra electron with opposite spin). The intersite Coulomb interaction may be of particular importance for low-dimensional systems, for which screening of the long-range Coulomb part is not very effective. Also, the direct exchange integral, which usually leads to ferromagnetic interaction $(J_{ij}>0)$, may be compared to the kinetic exchange, which is antiferromagnetic, as we have seen above.

To derive the effective Hamiltonian from Eq.~\eqref{8.41} one should carry out the procedure analogous to that discussed in this Appendix for the Hubbard model (cf.~\cite{SpalekPSS1981} for details). In effect, the effective Hamiltonian $\hat{P}_{1}\widetilde{\mathcal{\hat{H}}}\hat{P}_{1}$ takes the form

\begin{align}
\hat{P}_{1}\widetilde{\mathcal{\hat{H}}}\hat{P}_{1}= \sum_{ij\sigma}\!^{'} t_{ij}\, \hat{b}_{i\sigma}^{\dagger}\, \hat{b}_{j\sigma}+\frac{1}{2} \sum_{ij}\!^{'}
\tilde{V}_{ij} \hat{\nu}_{i}\, \hat{\nu}_{j}+\sum_{ij}\!^{'}\, \tilde{J}_{ij}\: \hat{\mathbf{S}}_{i} \cdot \hat{\mathbf{S}}_{j}+\; \mbox{(3-site terms)},
\label{8.42}
\end{align}

\noindent
where $\tilde{J}_{ij}\equiv\,2t_{ij}^{2}/(U-K_{ij}+J_{ij}^{H}/2)-J_{ij}^{H}$ is now the effective exchange integral and $\tilde{V}_{ij} \equiv K_{ij} - 
\frac{3}{2}\tilde{J}_{ij}$ the effective intersite Coulomb interaction. Essentially, the physics of the correlated electrons for $n\leq 1$ does not change, apart from the renormalization of the Hubbard gap. This is also because the kinetic exchange is usually much stronger than the direct ferromagnetic exchange, $\sim J_{ij}$. However, the presence of the intersite Coulomb interactions may lead to the formation of the charge-density wave state, particularly near the quarter filling of the band and when $zV_{1}\gtrsim U$, where $V_{1}$ is the magnitude of the Coulomb interaction between $z$ nearest neighboring pairs. One can see that the presence of the term $\sim V_{ij}$ enhances the value of $\widetilde{V}_{ij}$ with respect to $J_{ij}$, since usually $V_{ij} \gg J_{ij}^{H}$.

\subsection{Methodological remark:  Meaning of the canonical perturbation expansion}

As the canonical perturbation expansion \eqref{8.20}-\eqref{8.22} is unitary, the physics contained in the effective Hamiltonian~\eqref{8.36} should coincide with that of the original Hubbard model~\eqref{8.1} in the limit $U/W\gg 1$ and for $n\leq 1$. Second, one may ask why we call the expansion \eqref{8.20}-\eqref{8.22} the (canonical) perturbation expansion, since both $\hat{P}_{1}\mathcal{\hat{H}}\hat{P}_{1}$ and $\hat{P}_{1}\mathcal{\hat{H}}\hat{P}_{2}$ contain the terms of the same order $\sim t_{ij}$?

The answer to this question is as follows. One can represent the projected Hamiltonian in the following symbolic matrix form

\begin{align}
\mathcal{\hat{H}}=
\left(
\begin{array}{ccc}
\hat{P}_{1}\mathcal{\hat{H}}\hat{P}_{1}\: &\vdots & \:\hat{P}_{1}\mathcal{\hat{H}}\hat{P}_{2} \\
\cdots\cdots & & \cdots\cdots\\
\hat{P}_{2}\mathcal{\hat{H}}\hat{P}_{1}\: & \vdots&\:\hat{P}_{2}\mathcal{\hat{H}}\hat{P}_{2}
\end{array}
\right),
\label {8.44}
\end{align}

\noindent
where $\hat{P}_{1}\mathcal{\hat{H}}\hat{P}_{1}$ and $\hat{P}_{2}\mathcal{\hat{H}}\hat{P}_{2}$ represent two Fock subspaces intermixed by the terms $\hat{P}_{1}\mathcal{\hat{H}}\hat{P}_{2}$ and $\hat{P}_{2}\mathcal{\hat{H}}\hat{P}_{1}$. Now, the important thing is that a representative distance between the eigenvalues of $\hat{P}_{1}\mathcal{\hat{H}}\hat{P}_{1}$ and $\hat{P}_{2}\mathcal{\hat{H}}\hat{P}_{2}$ is of the order of $U\gg |t_{ij}|$. Therefore, that part $\hat{P}_{1}\mathcal{\hat{H}}\hat{P}_{2}$ is a small admixture $\sim t_{ij}/U$ to the dynamics of the electrons in the Hubbard subbands. The canonical perturbation expansion is thus understood as a way of calculating the successive contributions to the subband dynamics in the powers of $t_{ij}/U$. First, these contributions are calculated in an operator (invariant) form and second, they turn out crucial as they are the only contributions remaining in the limit of the Mott insulator. However, by executing this whole procedure, there is also a price to pay. Namely, the fermion operators acquire non-standard anticommutation relations which makes the whole analysis much more cumbersome from the formal point of view. Physically, this complexity signals the possibility of emergence of non-Landau quasiparticles, non-Fermi (non-Landau) liquid behavior, and other exotic phenomena. To make the problem easier to tackle, we introduce in the main text the $t$-$J$-$U$ form of the model as a starting point; this is the subject of Appendix~\ref{appendix:sga_and_slave_bosons}.

\subsection{Canonical transformation beyond the leading order}

Here we discuss a generalization of the canonical transformation scheme that allows for derivation of higher-order corrections to the effective Hamiltonian, including also the corrections due to intraband hopping terms, not explicitly included in previous subsections of this Appendix. As the starting point, we take decomposed Hamiltonian in the form $\hat{\mathcal{H}} = \hat{\mathcal{H}}_0 + \hat{\mathcal{H}}_1$, where $\hat{\mathcal{H}}_0$ does not mix the states with different number of doubly occupied sites. This condition may be formally stated as $\left[\hat{\mathcal{H}}_0, \sum_i \hat{n}_{i\uparrow} \hat{n}_{i\downarrow}\right] = 0$. The other contribution, $\hat{\mathcal{H}}_1$, accommodates all the remaining terms. For the Hubbard-type models, $\hat{\mathcal{H}}_0$ can be further separated into the on-site energy $\hat{U} \equiv U \sum_i \hat{n}_{i\uparrow} \hat{n}_{i\downarrow}$ that is diagonal in the occupation number representation and the projected intersite hopping, $\hat{T}_0$, fulfilling $\left[\hat{T}_0, \hat{U}\right] = 0$. Moreover, $\hat{\mathcal{H}}_1 = \hat{\mathcal{H}}^{+} + \hat{\mathcal{H}}^{-}$, where  $\hat{\mathcal{H}}^{+}$ and $\hat{\mathcal{H}}^{-}$ increase and decrease the number of doubly occupied sites, respectively. By hermiticity of the Hamiltonian, those two operators are related as $\hat{\mathcal{H}}^{-} = (\hat{\mathcal{H}}^{+})^\dagger$. To illustrate the convention, for the single-orbital Hubbard model one can take

\begin{align}
  \hat{T}_0 &= -t \sum_{\langle i, j\rangle} \hat{n}_{i\bar{\sigma}}\hat{a}_{i\sigma}^\dagger\hat{a}_{j\sigma} \hat{n}_{j\bar{\sigma}} -t \sum_{\langle i, j\rangle} \left(1-\hat{n}_{i\bar{\sigma}}\right)\hat{a}_{i\sigma}^\dagger\hat{a}_{j\sigma} \left(1 - \hat{n}_{j\bar{\sigma}}\right), \\
   \hat{\mathcal{H}}^{+} & = -t \sum_{\langle i, j\rangle} \hat{n}_{i\bar{\sigma}}\hat{a}_{i\sigma}^\dagger\hat{a}_{j\sigma} \left(1-\hat{n}_{j\bar{\sigma}}\right), \\
 \hat{\mathcal{H}}^{-} & = -t \sum_{\langle i, j\rangle} \left(1-\hat{n}_{i\bar{\sigma}}\right)\hat{a}_{i\sigma}^\dagger\hat{a}_{j\sigma} \hat{n}_{j\bar{\sigma}},
\end{align}

\noindent
so that the total kinetic energy operator reads $-t\sum_{\langle i, j\rangle} \hat{a}_{i\sigma}^\dagger \hat{a}_{j\sigma} = \hat{T}_0 + \hat{\mathcal{H}}^{+} + \hat{\mathcal{H}}^{-}$.

The aim of the canonical transformation is to eliminate the high energy processes, contained in $\hat{\mathcal{H}}_1$ by means of a properly constructed unitary transformation $\exp(-i\hat{S})$, where $\hat{S} = \hat{S}^\dagger$. The rotated Hamiltonian $\widetilde{\hat{\mathcal{H}}} = \exp(-i\hat{S}) \hat{\mathcal{H}} \exp(i\hat{S})$ will then provide an effective low-energy description of the system. Here we show carry out this program to given order $m$ so that the mixing terms up to the order $O(t \cdot (t/U)^{m - 1})$ are eliminated.

We start from the formula for the canonically transformed Hamiltonian

\begin{align}
\widetilde{\hat{\mathcal{H}}} = \hat{\mathcal{H}}_0 + (\hat{\mathcal{H}}_1 - i [\hat{S}, \hat{\mathcal{H}}_0]) + \sum \limits_{n = 2}^{\infty} \frac{(-i)^{n - 1}}{n!} \left( n \cdot [\hat{S}, \hat{\mathcal{H}}_1]_{n - 1} - i [\hat{S}, \hat{\mathcal{H}}_0]_{n} \right).
\label{eq:general_formula_effective_hamiltonian_original}
\end{align}

\noindent
If one chooses $\hat{S}$ so that

\begin{align}
\hat{\mathcal{H}}_1 - i [\hat{S}, \hat{\mathcal{H}}_0] = 0
\label{eq:condition_for_the_S_operator}
\end{align} 

\noindent
then $\hat{\mathcal{H}}_1$ is removed and $\widetilde{\hat{\mathcal{H}}}$ takes the form

\begin{align}
\widetilde{\hat{\mathcal{H}}} = \hat{\mathcal{H}}_0 + \sum \limits_{n = 2}^{\infty} \frac{(-i)^{n - 1} \cdot (n - 1)}{n!} [\hat{S}, \hat{\mathcal{H}}_1]_{n - 1}.
\label{eq:general_formula_effective_hamiltonian}
\end{align}

\noindent
so we have essentially rederived formula~\eqref{8.22}, discussed earlier, yet without the epsilon terms that will not be used here. Equation~\eqref{eq:condition_for_the_S_operator} may be solved iteratively, by expanding $\hat{S} = \sum \limits_{\alpha = 1}^\infty \hat{S}^{(\alpha)}$, where $\hat{S}^{(\alpha)} = O\left(\hat{\mathcal{H}}_1 \cdot (\hat{T}_0)^{\alpha - 1} \cdot U^{-\alpha}\right)$. By substituting $\hat{\mathcal{H}}_0 = \hat{T}_0 + \hat{U}$ and collecting the terms according to their order, one obtains

\begin{align}
\underbrace{\hat{\mathcal{H}}_1 - i[\hat{S}^{(1)}, \hat{U}]}_{O(\hat{\mathcal{H}}_1)} - i \sum \limits_{\alpha = 1}^{\infty} \underbrace{ \left( [\hat{S}^{(\alpha + 1)}, \hat{U}] + [\hat{S}^{(\alpha)}, \hat{T}_0] \right) }_{O[\hat{\mathcal{H}}_1 \cdot (\hat{T}_0 / U)^{\alpha}]} = 0.
\label{eq:condition_for_S_operator_expanded}
\end{align} 

\noindent
Equation~(\ref{eq:condition_for_S_operator_expanded}) may be now solved using the identity $[\hat{\mathcal{H}}^{\pm}, \hat{U}] = \mp U \hat{\mathcal{H}}^\pm$. The latter physically means that $\hat{\mathcal{H}}^{\pm}$ increases (decreases) the number of doubly occupied sites by one, resulting in energy difference $U$. The terms  $O(\hat{\mathcal{H}}_1)$  are eliminated if one takes

\begin{align}
\hat{S}^{(1)} = \frac{i}{U} \left( \hat{\mathcal{H}}^{+} - \hat{\mathcal{H}}^{-}  \right).
\label{eq:recursive_S_1}
\end{align}

\noindent
To write the formula for higher-order contributions to $\hat{S}$ in a compact manner, we carry out a decomposition $\hat{S}^{(\alpha)} = \hat{S}^{(\alpha) +} + \hat{S}^{(\alpha) -}$ and define $\hat{S}^{(1) \pm} \equiv \pm \frac{i}{U} \cdot \hat{\mathcal{H}}^{\pm}$ (cf. Eq.~\eqref{eq:recursive_S_1}). One can now show that taking

\begin{align}
\hat{S}^{(\alpha + 1) \pm} = \pm \frac{1}{U} [\hat{S}^{(\alpha) \pm}, \hat{T}_0] \:\:\: \text{for $\alpha \geq 1$}
\label{eq:recursive_formula_S}
\end{align}

\noindent
makes the left-hand-side of Eq.~\eqref{eq:condition_for_S_operator_expanded} vanish to all orders. Indeed, from Eq.~(\ref{eq:recursive_formula_S}) it follows that $\hat{S}^{(\alpha + 1)} = \hat{S}^{(\alpha + 1)+} + \hat{S}^{(\alpha + 1)-} =  1/U \cdot [\hat{S}^{(\alpha) +}, \hat{T}_0] - 1/U \cdot [\hat{S}^{(\alpha) -}, \hat{T}_0]$. The commutator $[\hat{S}^{(\alpha + 1)}, \hat{U}]$, which appears under the summation symbol in Eq.~(\ref{eq:condition_for_S_operator_expanded}), can now be reorganized by recalling that $[\hat{T}_0, \hat{U}] = 0$. One then arrives at

\begin{align}
[\hat{S}^{(\alpha + 1)}, \hat{U}] & =  \left[\frac{1}{U} \cdot [\hat{S}^{(\alpha) +}, \hat{T}_0] - \frac{1}{U} \cdot [\hat{S}^{(\alpha) -}, \hat{T}_0], \hat{U}\right]   = \frac{1}{U} \cdot \left[ [\hat{S}^{(\alpha) +}, \hat{U}], \hat{T}_0 \right] - \frac{1}{U} \cdot \left[ [\hat{S}^{(\alpha) -}, \hat{U}], \hat{T}_0 \right].
\label{eq:commutator_S_U}
\end{align}

\noindent
The commutators $[\hat{S}^{(\alpha) \pm}, \hat{U}]$ can be readily evaluated. From Eq.~\eqref{eq:recursive_formula_S} it follows that for any $\alpha \geq 1$, $\hat{S}^{(\alpha) \pm}$ contains exactly one $\hat{\mathcal{H}}^\pm$ (entering through the leading-order term $\hat{S}^{(1) \pm}$) and $\alpha$-dependent number of operators $\hat{T}_0$ (which commute with $\hat{U}$ by construction). Since $[\hat{\mathcal{H}}^{\pm}, \hat{U}] = \mp U \hat{\mathcal{H}}^{\pm}$, one obtains $[\hat{S}^{(\alpha) \pm}, \hat{U}] = \mp U \hat{S}^{(\alpha) \pm}$. Eq.~(\ref{eq:commutator_S_U})~now reads

\begin{align}
[\hat{S}^{(\alpha + 1)}, \hat{U}]  =  - [\hat{S}^{(\alpha) +} + \hat{S}^{(\alpha) -}, \hat{T}_0] = - [\hat{S}^{(\alpha)}, \hat{T}_0],
\label{eq:commutator_S_U_2}
\end{align}

\noindent
hence $\left( [\hat{S}^{(\alpha + 1)}, \hat{U}] + [\hat{S}^{(\alpha)}, \hat{T}_0] \right) = 0$ for each $\alpha \geq 1$ and the sum on the left-hand-side of Eq.~(\ref{eq:condition_for_S_operator_expanded}) vanishes. In brief, $\hat{S} = \sum_{\alpha = 1}^\infty \hat{S}^{(\alpha)}$, with $\hat{S}^{(\alpha)}$ defined by Eqs.~(\ref{eq:recursive_S_1}) and (\ref{eq:recursive_formula_S}), satisfies Eq.~(\ref{eq:condition_for_the_S_operator}).

Equations~(\ref{eq:recursive_S_1}) and (\ref{eq:recursive_formula_S}) can be combined into an explicit formula 

\begin{align}
\hat{S}^{(\alpha)} = \frac{i}{U^\alpha} \left( [\hat{\mathcal{H}}^{+}, \hat{T}_0]_{\alpha - 1}^R + (-1)^{\alpha} [\hat{\mathcal{H}}^{-}, \hat{T}_0]_{\alpha - 1}^R \right),
\label{eq:non_recursive_formula_S}
\end{align}

\noindent
where $[\hat{A}, \hat{B}]^R_n \equiv [ \ldots [ [\hat{A}, \underbrace{ \hat{B}], \hat{B}], \ldots, \hat{B}}_\text{$n$ operators $\hat{B}$} ]$.

We now demonstrate derive the effective Hamiltonian $\widetilde{\hat{\mathcal{H}}}$ to the order $O(t^4/U^3)$ so that all the terms up to the order $O(t^4/U^3)$ between subspaces with different number of doubly occupied sites are eliminated. To accomplish this goal, two subsequent canonical transformations are actually needed, i.e.~$\widetilde{\hat{\mathcal{H}}} \equiv \exp(-i \hat{S}_2) \exp(-i \hat{S}_1) \hat{\mathcal{H}} \exp(i \hat{S}_1) \exp(i \hat{S}_2)$. We hence introduce the notation

\begin{align}
{\hat{\mathcal{H}}}^{(1)} \equiv \exp(-i \hat{S}_1) \hat{\mathcal{H}} \exp(i \hat{S}_1),
\end{align}

\noindent
and

\begin{align}
\hat{\mathcal{H}}^{(2)} \equiv \widetilde{\hat{\mathcal{H}}} \equiv \exp(-i \hat{S}_2) {\hat{\mathcal{H}}}^{(1)} \exp(i \hat{S}_2).
\end{align}

By the fist canonical transformation we want to fulfill the original condition \label{eq:condition_for_S_operator} to the order $O(t^3/U^2)$. To achieve that, it is sufficient to keep the first three terms in the expansion of the $\hat{S}_1$ operator so that $\hat{S}_1 \equiv \hat{S}^{(1)}_1 + \hat{S}^{(2)}_1 + \hat{S}^{(3)}_1$. Indeed, the only term in the expansion of Eq.~\eqref{eq:condition_for_S_operator_expanded} that remains uncompensated is $[\hat{S}^{(3)}_1, \hat{T}_0]$, as there is no $[\hat{S}^{(4)}_1, \hat{U}]$ to cancel it. Since $\hat{S}^{(3)}_1 = O(\hat{\mathcal{H}}_1 \cdot (\hat{T}_0)^{2} \cdot U^{-3}) = O(t^3/U^3)$, we get $\hat{\mathcal{H}}_1 - [\hat{S}_1, \hat{\mathcal{H}}_0] = O(t^4/U^3)$. By substituting $ [\hat{S}_1, \hat{\mathcal{H}}_0] = -i \hat{\mathcal{H}}_1 + O(t^4/U^3)$ into the right-hand-side Eq.~(\ref{eq:general_formula_effective_hamiltonian_original}) and recalling that $\hat{S}_1 = O(t/U)$ (as $\hat{S}_1^{(1)} = O(t/U)$), we get

\begin{align}
\hat{\mathcal{H}}^{(1)} = &\hat{\mathcal{H}}_0 + (\hat{\mathcal{H}}_1 - i [\hat{S}_1, \hat{\mathcal{H}}_0]) + \sum \limits_{n = 2}^{\infty} \frac{(-i)^{n - 1}}{n!} \left( n \cdot [\hat{S}_1, \hat{\mathcal{H}}_1]_{n - 1} - i [\hat{S}_1, \hat{\mathcal{H}}_0]_{n} \right) = \nonumber \\ = & \hat{\mathcal{H}}_0 + \underbrace{(\hat{\mathcal{H}}_1 - i [\hat{S}_1, \hat{\mathcal{H}}_0])}_{O(t^4/U^3)} + \sum \limits_{n = 2}^{\infty} \frac{(-i)^{n - 1} (n - 1)}{n!} \cdot [\hat{S}_1, \hat{\mathcal{H}}_1]_{n - 1}  + O(t^5/U^4).
\label{eq:general_formula_effective_hamiltonian_after_truncation}
\end{align}

\noindent
It now becomes apparent why Eq.~(\ref{eq:condition_for_the_S_operator}) has to be fulfilled to the order $O(t^3/U^2)$. Otherwise, the corrections to the sum in Eq.~\eqref{eq:general_formula_effective_hamiltonian_after_truncation} would be of lower order than $O(t^5/U^4)$ and it would be necessary to systematically include them in further discussion.

The effective Hamiltonian $\hat{\mathcal{H}}^{(1)}$ still contains terms mixing the subspaces with different numbers of doubly occupied sites. The lowest order term comes from the $O(t^2/U)$ contribution to (\ref{eq:general_formula_effective_hamiltonian_after_truncation}), i.e.,

\begin{align}
-\frac{i}{2} \cdot [\hat{S}_1, \hat{\mathcal{H}}_1] = & -\frac{i}{2} \cdot \left[\frac{i}{U}\cdot (\hat{\mathcal{H}}^{+} - \hat{\mathcal{H}}^{-}) + O(t^2/U^2), \hat{\mathcal{H}}^{+} + \hat{\mathcal{H}}^{-}\right] = \nonumber \\ = &\frac{1}{U} \cdot \left( \hat{\mathcal{H}}^{+} \hat{\mathcal{H}}^{-} - \hat{\mathcal{H}}^{-} \hat{\mathcal{H}}^{+}\right) + O(t^3/U^2).
\label{eq:lack_of_second_order_contribution_to_mixing_terms}
\end{align}

\noindent
From Eq.~(\ref{eq:lack_of_second_order_contribution_to_mixing_terms}) one can see that, even though $-\frac{i}{2} \cdot  [S_1, \mathcal{H}_1] = O(t^2/U)$, its lowest order contribution to the mixing terms is $O(t^3/U^2)$. The  Hamiltonian $\hat{\mathcal{H}}^{(1)}$ can hence be decomposed as

\begin{align}
\hat{\mathcal{H}}^{(1)} = \hat{\mathcal{H}}^{(1)}_0 + \hat{\mathcal{H}}^{(1)}_1 = \hat{U} + \underbrace{\hat{T}_0^{(1)}}_{O(t)} + \underbrace{\hat{\mathcal{H}}^{(1)}_1}_{O(t^3/U^2)},
\end{align}

\noindent 
where $\hat{\mathcal{H}}^{(1)}_1$ and $\hat{\mathcal{H}}^{(1)}_0$ are subspace mixing and non-mixing terms, respectively. As before we write $\hat{\mathcal{H}}^{(1)}_0 = {\hat{U}} + \hat{T}_0^{(1)}$.

We now carry out the second canonical transformation to eliminate $\hat{\mathcal{H}}^{(1)}_1$ to the order $O(t^4/U^3)$. To do that, it is sufficient to retain only the first two terms in the expansion of the $\hat{S}_2$ operator, i.e.,  $\hat{S}_2 \equiv \hat{S}_2^{(1)} + \hat{S}_2^{(2)}$. Since $\hat{\mathcal{H}}^{(1)} = O(t^3/U^2)$ and $\hat{S}^{(\alpha)}_2 = O(\hat{\mathcal{H}}^{(1)}_1 \cdot (\hat{T}^{(1)}_0)^{\alpha - 1} \cdot U^{-\alpha})$, we find that $\hat{S}_2^{(1)} = O(t^3/U^3)$ and $\hat{S}_2^{(2)} = O(t^4/U^4)$. By application of  Eqs.~(\ref{eq:condition_for_the_S_operator}) and (\ref{eq:condition_for_S_operator_expanded}), and evoking the same argument as for the first canonical transformation, one can show that $\hat{\mathcal{H}}^{(1)}_1 - i [\hat{S}_2, \hat{\mathcal{H}}^{(1)}_0] = O(t^5/U^4)$. By referring to general formula of Eq.~(\ref{eq:general_formula_effective_hamiltonian_original}) and recalling that $\hat{S}_2 = O(t^3/U^3)$, the final effective Hamiltonian $\hat{\mathcal{H}}^{(2)} = \widetilde{\hat{\mathcal{H}}}$ takes the form

\begin{align}
\hat{\mathcal{H}}^{(2)} = &\hat{\mathcal{H}}_0^{(1)} + (\hat{\mathcal{H}}_1^{(1)} - i [\hat{S}_2, \hat{\mathcal{H}}_0^{(1)}]) + \sum \limits_{n = 2}^{\infty} \frac{(-i)^{n - 1}}{n!} \left( n \cdot [\hat{S}_2, \hat{\mathcal{H}}_1^{(1)}]_{n - 1} - i [\hat{S}_2, \hat{\mathcal{H}}_0^{(1)}]_{n} \right) = \nonumber \\ = & \hat{\mathcal{H}}_0^{(1)} + \underbrace{(\hat{\mathcal{H}}_1^{(1)} - i [\hat{S}_2, \hat{\mathcal{H}}_0^{(1)}])}_{O(t^5/U^4)} + \underbrace{\sum \limits_{n = 2}^{\infty} \frac{(-i)^{n - 1} (n-1)}{n!}  [\hat{S}_2, \hat{\mathcal{H}}_1^{(1)}]_{n - 1}}_{O(t^6/U^5)} + O(t^8/U^7) = \nonumber \\ =&  \hat{\mathcal{H}}^{(1)}_0 + O(t^5/U^4)
\end{align}

\noindent
The final effective Hamiltonian at the order $O(t^4/U^3)$ is then $\widetilde{\hat{\mathcal{H}}} = \hat{\mathcal{H}}^{(2)}$. Moreover, $\hat{\mathcal{H}}^{(2)}$ is equal to the non-mixing part of $\hat{\mathcal{H}}^{(1)}$. The latter observation significantly simplifies the discussion as the explicit form of the operator $\hat{S}_2$ is not needed.

The effective Hamiltonian to the order $O(t^4/U^3)$ thus reads

\begin{align}
\widetilde{\hat{\mathcal{H}}} = \Bigg( \hat{\mathcal{H}}_0 + (\hat{\mathcal{H}}_1 - i[\hat{S}_1, \hat{\mathcal{H}}_0]) -\frac{i}{2} [\hat{\mathcal{S}}_1, \hat{\mathcal{H}}_1] - \frac{1}{3} \Big[\hat{S}_1, [\hat{S}_1, \hat{\mathcal{H}}_1]\Big] + \frac{i}{8} \Bigg[\hat{S}_1, \Big[\hat{S}_1, [\hat{S}_1, \hat{\mathcal{H}}_1] \Big] \Bigg] \Bigg)_\text{non-mixing},
\label{eq:effective_hamiltonian_final}
\end{align}

\noindent
where

\begin{align}
\hat{S}_1 = \frac{i}{U} (\hat{\mathcal{H}}^{+} - \hat{\mathcal{H}}^{-}) + \frac{i}{U^2} [\hat{\mathcal{H}}^{+} + \hat{\mathcal{H}}^{-}, \hat{T}_0] + \frac{i}{U^3}  \Big[ [\hat{\mathcal{H}}^{+} - \hat{\mathcal{H}}^{-}, \hat{T}_0], \hat{T}_0 \Big].
\label{eq:S1}
\end{align}

\noindent
The subscript  ``non-mixing'' means that only the terms that do change the number of doubly occupied sites need to be retained. We now note that the second term on the right-hand-side of Eq.~\eqref{eq:effective_hamiltonian_final},

\begin{align}
\hat{\mathcal{H}}_1 - i[\hat{S}_1, \hat{\mathcal{H}}_0] = \frac{1}{U^3}  \Bigg[ \Big[ [\hat{\mathcal{H}}^{+} - \hat{\mathcal{H}}^{-}, \hat{T}_0], \hat{T}_0 \Big], \hat{T}_0 \Bigg],
\end{align}

\noindent
necessarily mixes the subspaces and hence can be dropped.

Equations~(\ref{eq:effective_hamiltonian_final}) and (\ref{eq:S1}) can be combined and expanded. The part of the effective Hamiltonian, restricted to the low-energy sector, reads then

\begin{align}
\widetilde{\hat{\mathcal{H}}} =& \hat{\mathcal{H}}_0 - \frac{1}{U} \hat{\mathcal{H}}^{-} \hat{\mathcal{H}}^{+} - \frac{1}{2 U^2} \hat{T}_0 \hat{\mathcal{H}}^{-} \hat{\mathcal{H}}^{+} + \frac{1}{U^2} \hat{\mathcal{H}}^{-} \hat{T}_0  \hat{\mathcal{H}}^{+} - \frac{1}{2 U^2}  \hat{\mathcal{H}}^{-} \hat{\mathcal{H}}^{+} \hat{T}_0 \nonumber \\ & - \frac{1}{2 U^3} (\hat{T}_0)^2 \hat{\mathcal{H}}^{-} \hat{\mathcal{H}}^{+} - \frac{1}{U^3} \hat{\mathcal{H}}^{-} (\hat{T}_0)^2 \hat{\mathcal{H}}^{+}  - \frac{1}{2 U^3} \hat{\mathcal{H}}^{-} \hat{\mathcal{H}}^{+} (\hat{T}_0)^2 \nonumber \\ & + \frac{1}{U^3} \hat{T}_0 \hat{\mathcal{H}}^{-} \hat{T}_0 \hat{\mathcal{H}}^{+} + \frac{1}{U^3}  \hat{\mathcal{H}}^{-} \hat{T}_0 \hat{\mathcal{H}}^{+} \hat{T}_0 \nonumber \\ & + \frac{1}{U^3}  \hat{\mathcal{H}}^{-} \hat{\mathcal{H}}^{+} \hat{\mathcal{H}}^{-} \hat{\mathcal{H}}^{+} - \frac{1}{2 U^3}  \hat{\mathcal{H}}^{-} \hat{\mathcal{H}}^{-} \hat{\mathcal{H}}^{+} \hat{\mathcal{H}}^{+} + O(t^5/U^4).
\label{eq:effective_hamiltonian_final_expanded}
\end{align}

\noindent
Note that two consecutive canonical transformations were necessary to derive the effective Hamiltonian to the $\frac{t^4}{U^3}$ order, but an explicit form of $\hat{S}_2$ was never used. However, in higher-order calculations, computation of $\hat{S}_n$ with $n > 1$ would be needed. Here we point out that, in this case, the recursive formulas of Eqs.~(\ref{eq:recursive_S_1}) and (\ref{eq:recursive_formula_S}) need to be modified. The reason is that the mixing terms in the effective Hamiltonian might, in general, change the energy of the system by some integer multiplicity of $U$, not necessarily by $\pm U$.

\subsection{$t$-$J$ model with explicitly real-space pairing}

The $t$-$J$ and its extension are used extensively to discuss the pairing among strongly correlated electrons. Here we introduce its explicit pairing representation (cf.~Spa{\l}ek (1988) \cite{SpalekPhysRevB1988}). We regarded the operator $\left[\mathbf{\hat{S}}_{i}\cdot\mathbf{\hat{S}}_{j}-\frac{1}{4}\hat{\nu}_{i}\hat{\nu}_{j}\right]$ as the full exchange operator. This operator  when acting on pair $\langle ij\rangle$ spin singlet state produces eigenvalue $-1$, whereas when acting on any of the three spin-triplet states yields zero. In other words, it favors the singlet  two-particle wave function on the lattice with the spin $1/2$ fermions. It is therefore natural to rewrite the spin-singlet creation and annihilation operators in the following manner

\begin{align}
\left\{
\begin{array}{l}
\hat{B}_{ij}^{\dagger}\equiv \frac{1}{\sqrt{2}}\left(\hat{b}_{i\uparrow}^{\dagger}\:\hat{b}_{j\downarrow}^{\dagger}- \hat{b}_{i\downarrow}^{\dagger}\:\hat{b}_{j\uparrow}^{\dagger}\right), \\
\hat{B}_{ij}\equiv \left(\hat{B}_{ij}^{\dagger}\right)^{\dagger}\equiv\frac{1}{\sqrt{2}}\Big(\hat{b}_{i\downarrow}\:\hat{b}_{j\uparrow}- \hat{b}_{i\uparrow}\:\hat{b}_{j\downarrow}\Big).
\end{array}
\right.
\label{8.44}
\end{align}

\noindent
Note that here we have projected fermion operators $\hat{b}_{i\sigma}$ and $\hat{b}^{\dagger}_{i\sigma}$, so the $\hat{B}_{ij}$ and $\hat{B}^\dagger_{ij}$ are nonzero only when $i\neq j$ (this is because $\hat{b}_{i\uparrow}^{\dagger}\:\hat{b}_{j\downarrow}^{\dagger} \equiv \hat{a}_{i\uparrow}^{\dagger}\left(1-\hat{n}_{i\downarrow}\right) \hat{a}_{i\downarrow}^{\dagger}\left(1-\hat{n}_{i\uparrow}\right)\equiv 0$). The B-operators represent thus spin-singlet bonds without the ionic admixture. One should also note that for the pair singlet state $|\Psi_{ij}^{(s)} \rangle\equiv \hat{B}_{ij}^{\dagger}|0 \rangle$, we have

\begin{align}
\langle\Psi^{(s)}_{ij}|\mathbf{\hat{S}}_\mathrm{tot}^{2}|\Psi^{(s)}_{ij}\rangle = \langle\Psi^{(s)}_{ij}|{\hat{S}}_\mathrm{tot}^{z}|\Psi^{(s)}_{ij}\rangle = 0,
\label{8.45}
\end{align}

\noindent
as can be checked readily by taking $\mathbf{\hat{S}}_\mathrm{tot}\equiv \mathbf{\hat{S}}_{i}+\mathbf{\hat{S}}_{j}$ and the projected fermion representation \eqref{8.37} of the spin operators $\mathbf{\hat{S}}_{i}$ and $\mathbf{\hat{S}}_{j}$.

The principal advantage of using the representation \eqref{8.44} explicitly is in reexpressing the full $t$-$J$ Hamiltonian \eqref{8.34} in the following form \cite{SpalekPhysRevB1988}

\begin{align}
\mathcal{\hat{H}}_{t\text{-}J} \equiv \hat{P}_{1}\widetilde{\hat{\mathcal{H}}}\hat{P}_{1}=\hat{P}_{1} \left\{ \sum_{ij}t_{ij}\:\hat{b}_{i\sigma}^{\dagger}\:\hat{b}_{j\sigma} - \sum_{ijk}\frac{2t_{ij}\:t_{kj}}{U} \hat{B}_{ij}^{\dagger}\: \hat{B}_{kj} \right\} \hat{P}_{1}.
\label{8.46}
\end{align}

\noindent
We see that, in this form, the single-particle and the two-particle parts, with those two aspects intertwined as the corresponding terms in the effective Hamiltonian do not commute with each other. We have also omitted the prime summation as the relevant nonzero terms are singled out by the corresponding nonzero hopping integral values $t_{ij}$ and $t_{kj}$. Note also that the second term has a minus sign so the system energy is decreased by creating spin-singlet bonds, with their number operator expressed by $\hat{B}_{ij}^{\dagger}\:\hat{B}_{ij}$. Furthermore, as the part $\hat{b}_{i\sigma}^{\dagger}\:\hat{b}_{j\sigma}$ represents direct hopping between the single occupied and the empty sites, $\hat{B}_{ij}^{\dagger}\:\hat{B}_{kj}$ for $k\neq i$ describes a transmutation of spin-singlet pair from the pair located at sites $(kj)$ into that on $(ij)$. To put it lightly, the first term represents hopping of individual particles, whereas the latter, motion of the pair bound by the energy $\left(-4t_{ij}^{2}/U\right)$. It is also important to note that this local-pair coupling leads indeed to the pair-binding for sufficiently large $J \equiv 4 t^2/U$, even for the case of electron pair in an empty narrow band \cite{ByczukPhysRevB1992}. 

The other principal question is whether the pairs are able to condense into same sort of paired state, described by the condensation amplitude $\langle \hat{B}_{ij}^{\dagger}\rangle$ on the bonds, and how to distinguish it from the quantum spin-liquid state? This question is, in part, addressed further in Secs.~\ref{sec:vwf_solution} and \ref{sec:selected_equilibrium_properties}. Nevertheless, one has two stress two features of this seemingly obvious step. First, the representation of the ``pairing'' part in Eq.~\eqref{8.46} and the spin part in Eq.~\eqref{8.36} are equivalent in mathematical sense. This is clearly seen by noting that

\begin{align}
\hat{B}_{ij}^{\dagger}\:\hat{B}_{ij}\equiv -\left( \mathbf{\hat{S}}_{i}\cdot \mathbf{\hat{S}}_{j}-\frac{1}{4}\hat{\nu}_{i}\:\hat{\nu}_{j}\right).
\label{8.47}
\end{align}

\noindent
Therefore, the spin degrees of freedom and the local spin-singlet pairing should be treated on the same footing. From this we draw the second conclusion, namely that the antiferromagnetism and the unconventional superconductivity (i.e., that with $\langle \hat{B}_{ij}^{\dagger}\rangle\neq 0$) appear as either coexisting or competing states in a natural manner. In this two-faced character of the $t$-$J$ model lies its principal attractiveness in applying it to the pairing in strongly correlated systems. Additionally, the unconventional superconducting states emerge from the Mott insulator upon hole doping (i.e., as a result of removing spins from that system). The combined restrictions on the single-particle hopping, relatively strong pairing effects in real space, and the Mott insulating state as the reference state, constitute the essence of strongly correlated state and an unconventional paired state of this correlated quantum matter.  

\subsection{Supplement: Hubbard atomic representation}

The projected-fermion operator picture with $\hat{b}^{\dagger}_{i\sigma} \equiv \hat{a}^{\dagger}_{i\sigma}(1-\hat{n}_{i\bar{\sigma}}) $, and $\hat{c}^{\dagger}_{i\sigma} \equiv \hat{a}_{i\sigma}\hat{n}_{i\bar{\sigma}}$ may be alternatively expressed within the framework of general Hubbard atomic representation \cite{HubbardProcRoySocA1965}. Namely, with the help of atomic states $\{\ket{p,i}\}$ located on site \emph{i}, we can define the transition operators

\begin{equation}
  \mathrm{X}^{pq}_i \equiv \ket{p,i}\bra{q,i},
  \label{eq:x_operators}
  \end{equation}

  \noindent
  where $\{\ket{p,i}\}$ represents the orthogonalized basis on $n$ atomic states $\ket{p=1,2,\ldots,n}$. The operators~\eqref{eq:x_operators} satisfy the commutation relations relations \cite{OvchinnikovBook2004}

      \begin{equation}
        \left[\mathrm{X}^{p_1q_1}_i,\mathrm{X}^{p_1q_2}_j\right]_{\pm} = \delta_{ij}\left(\delta_{q_1p_2}\mathrm{X}^{p_1q_1}_i \pm \delta_{p_1q_2}, \mathrm{X}^{p_1q_2}_j   \right)
    \end{equation}

\noindent
and the completeness condition (sum rule)

    \begin{equation}
        \sum_p\mathrm{X}^{pp}_i = 1.
    \end{equation}

\noindent
The $X$-operators are either fermion-like and boson-like, depending on whether the relative occupancy of $p$ and $q$ states, $|n_{ip} -n_{iq}|$, is odd or even, respectively. 

We next represent those four atomic states $\ket{i0}$, $\ket{i\sigma} \equiv\hat{a}^{\dagger}_{i\sigma}(1-\hat{n}_{i\bar{\sigma}}) \ket{0}$, and $\ket{i2} \equiv \sigma 
\hat{a}^{\dagger}_{i\sigma}\hat{a}^{\dagger}_{i\bar{\sigma}}\ket{0}$ (empty, singly occupied by a particle with spin $\sigma$, and doubly occupied) by the $X$-operators and subsequently also the $t$-$J$ Hamiltonian. We obtain

\begin{eqnarray}
\begin{cases}
 \mathrm{X}^{00}_i \equiv (1-\hat{n}_{i\uparrow})(1-\hat{n}_{i\downarrow}),\\
 \mathrm{X}^{\sigma 0}_i \equiv \hat{a}_{i\sigma}^{\dagger}(1-\hat{n}_{i\bar{\sigma}}),\\
 \mathrm{X}^{0\sigma}_i \equiv \hat{a}_{i\sigma}(1-\hat{n}_{i\bar{\sigma}}),\\
 \mathrm{X}^{\sigma\sigma}_i \equiv \hat{n}_{i\sigma}(1-\hat{n}_{i\bar{\sigma}}) =\mathrm{X}^{\sigma0}_i\mathrm{X}^{0\sigma}_i,\\
 \mathrm{X}^{22}_i  \equiv \hat{n}_{i\uparrow}\hat{n}_{i\downarrow},\\
 \mathrm{X}^{\sigma\bar{\sigma}}_i \equiv \hat{a}^{\dagger}_{i\sigma}\hat{a}_{i\bar{\sigma}} = \hat{S}^{\sigma}_{i} = (\mathrm{X}^{\bar{\sigma}\sigma}_i)^{\dagger},\\
 \mathrm{X}^{2\sigma}_i \equiv 2\sigma \hat{a}^{\dagger}_{i\bar{\sigma}} \hat{a}_{i\sigma},\\
 \mathrm{X}^{20}_i \equiv 2\sigma \hat{a}^{\dagger}_{i\bar{\sigma}}\hat{a}^{\dagger}_{i\sigma}.
 \end{cases}
\end{eqnarray}

\noindent
so the projected fermion representation, used in main text, and the Hubbard atomic representation are formally equivalent. In effect, the $t$-$J$ model takes the form

\begin{equation}
    \widetilde{{\hat{H}}} = \hat{P}_1 \left\{ \sum_{ij} t_{ij }\mathrm{X}^{\sigma 0}_i\mathrm{X}^{0\sigma}_j + \frac{1}{2} \sum_{ij}\!^{'} \sum_{ij}J_{ij} \left( \mathrm{X}^{\sigma\bar{\sigma}}_i \mathrm{X}^{\bar{\sigma}\sigma}_j + \mathrm{X}^{\sigma\sigma}_i\mathrm{X}^{\bar{\sigma}\bar{\sigma}}_j\right) \right\} \hat{P}_1, 
  \end{equation}

  \noindent
with the total number of particles

\begin{equation}
    \hat{N}_e = \sum_{i\sigma} \mathrm{X}^{\sigma\sigma}_i. 
\end{equation}

\noindent
One can see that within this formalism both the single-particle aspect (as described by the operators $\mathrm{X}^{\sigma 0}_i$ and $\mathrm{X}^{0\sigma}_i$) and two-particle aspects (expressed $\mathrm{X}^{\sigma\bar{\sigma}}_i$ and $\mathrm{X}^{\sigma\sigma}_i$ operators) appear on the same footing, albeit with involved (anti)commutation relations as in the original projected fermion representation.

\section{Meaning of the $t$-$J$-$U$-$(V)$ model}
\label{appendix:sga_and_slave_bosons}

We would like to estimate the physical relevance of the $t$-$J$-$U$ model. In the canonically transformed extended Hubbard model \cite{SpalekPhysRevB1988,SpalekActaPhysPolon2007,SpalekPSS1981} the antiferromagnetic exchange interaction is of the form $J_{ij}=2t_{ij}^2/(U-V_{ij})$ and therefore no Hubbard extra term should appear, if we are in the strong-correlation limit $W \ll U$ (not only $|t_{ij}| \ll U$). Namely, the contribution to the $N$-particle wave function coming from double occupancies is of the order of $t/|U|$ \cite{HarrisPhysRev1967}. Before discussing the application of that limit in real calculations, let us estimate its value by taking the standard microscopic-parameter: $U=8$-$10\,\mathrm{eV}$, $t=-0.35\div$-$0.4\,\mathrm{eV}$, and $t'=|t|/4$. When neglecting $V$ in above formula for $J_{ij}$ we obtain the value of $J\simeq 150\;$K at most. If the bare parameter of the intersite Coulomb interaction, $V$, is taken as $U/3$ (maximum), then the value of $J$ increases by $50\%$, which is still much lower than the typical value of measured $J\sim 0.13\;$eV$\approx 1.
5\cdot 10^3\,\mathrm{K}$ in the insulating phase \cite{HarrisPhysRev1967,KastnerRevModPhys1998}. On the other hand, the typical bare bandwidth of the planar states is $W=8|t|\simeq 2.8\,\mathrm{eV} \sim 3\, \mathrm{eV}$. Therefore, the $U/W$ ratio is in the interval $2.9$-$3.6\sim 3$, which is not in the asymptotic limit of being $\gg 1$. Hence, one may expect that the double occupancy probability is not exactly vanishing, particularly for $\delta > 0$ as then the admixture of double occupancy to the single-particle state is of the order \cite{HarrisPhysRev1967} of $|t|/U \sim 0.04$. In our calculations $d\lesssim 10^{-2}$ for $\delta = 0.1$ and $U/W = 2.5$. Such a small value does not influence at all the spin magnitude in the Mott insulating state, since then $\langle \mathbf{S}^{2}\rangle=\frac{3}{4}(1-2d^2)\sim \frac{1}{2}(\frac{1}{2}+1)+o(10^{-2})$ \cite{SpalekJSolStateChem1990} and the zero-point spin fluctuations are much more important. 

After mentioning the relevance of the Hubbard term, the basic question still remains as to what is the dominant contribution to $J$. As said earlier, this is due to the superexchange \cite{ZaanenJSolStateChem1990,EskesPhysRevB1993,JeffersonPhysRevB1992,FeinerPhysRevB1996,AvellaEurPhysJB2013} via $p$ orbitals with inclusion of the fact that HTS are charge transfer insulators with the corresponding gap $\Delta=\epsilon_p-\epsilon_d\simeq 3\;$eV and the $p-d$ hybridization magnitude $t_{pd}\simeq 1.3\,\mathrm{eV}$, as well as the $p$-$d$ Coulomb interaction $U_{pd}\simeq 1\,\mathrm{eV}$. In effect, the nearest neighbors superexchange can be estimated as \cite{ZaanenJSolStateChem1990,EskesPhysRevB1993,JeffersonPhysRevB1992,FeinerPhysRevB1996}

\begin{equation}
\begin{split}
 J&=\frac{2\;t_{pd}^4}{(\Delta_{pd}+U_{pd})^2}\bigg(\frac{1}{U_{d}}+\frac{1}{\Delta_{pd}+U_{pp}/2}\bigg) \simeq 0.13\;\textrm{eV}=1560\;\textrm{K},
 \end{split}
\end{equation}

\noindent
a value close to that determined experimentally \cite{KastnerRevModPhys1998}. This reasoning provides a direct support for the predominant value of $J$ as not coming from the large-$U$ expansion of the Hubbard model \cite{ChaoJPCM1977,SpalekPhysRevB1988,SpalekActaPhysPolon2007}. 

One should note that the mechanism introduces also the Kondo-type coupling between the $p$ holes and $d$ electrons with the corresponding Kondo exchange integral

\begin{equation}
 J^K=2\;t_{pd}^2\bigg(\frac{1}{\Delta_{pd}}+\frac{1}{{\Delta_{pd}+U}}\bigg)\simeq 2.5\;\textrm{eV}\gg J.
 \label{eq:Kondo}
\end{equation}

\noindent
This coupling may cause a bound configuration of the hole and $d$-electron of Cu$^{2+}$ ion composing the Zhang-Rice singlet \cite{ZhangPhysRevB1990,ZhangPhysRevB1988}. Also, the hopping amplitude for $d$ electron between the nearest neighboring sites $\langle{i,j}\rangle$ can be estimated as

\begin{equation}
 t\sim\frac{t^2_{pd}}{(\Delta_{pd}+U_{pp}-U_{pd})^2}t_{pp}.
\end{equation}

\noindent
Taking $t_{pp}=0.1\,\mathrm{eV}$ and $U_{pp}=3\,\mathrm{eV}$, we obtain $t=-0.34\,\mathrm{eV}$, also a quite reasonable value. In effect, we have $J/|t|=0.38$ which is a reasonable ratio in view of simplicity of our estimates. In such a reduction procedure to the one-band model the effective Hubbard interaction is $U\simeq U_{dd}-U_{pp}\simeq 7.5\,\mathrm{eV}$, if we take the value $U_{dd}=10.5\,\mathrm{eV}$ for the original $d$ atomic states. In the fitting to experiment we have obtained a slightly larger value of $U=8\,\mathrm{eV}$ and $|t|=0.35\,\mathrm{eV}$. This brief discussion summarizes the meaning of the starting Hamiltonian.

The general three-band model would include a direct single-particle hopping between the oxygen sites $\sim t_{pp}$. Under these circumstances, the Kondo-type coupling (\ref{eq:Kondo}) between the oxygen and copper sites must be also taken into account explicitly and we end up in the Emery-Reiter type of model \cite{EmeryPhysRevB1988,ValkovJETPLett2016} in the limit of localized $3d$ electrons. In this respect, we have selected here the most general one-band model of correlated $d$ electrons with the Zhang-Rice singlet idea disregarded. Note also that the present estimate differs from the mapping of many-band model onto the single-band case, with the artificially small value of $U \sim 3$-$3\,\mathrm{eV}$, in order to obtain a realistic value of $J$.

\section{Effective Hamiltonian for correlated and hybridized electrons: Heavy-fermion limit}
\label{appendix:kondo-exchange}

\subsection{Motivation}

\begin{figure}
\begin{center}
\includegraphics[width=0.99\textwidth]{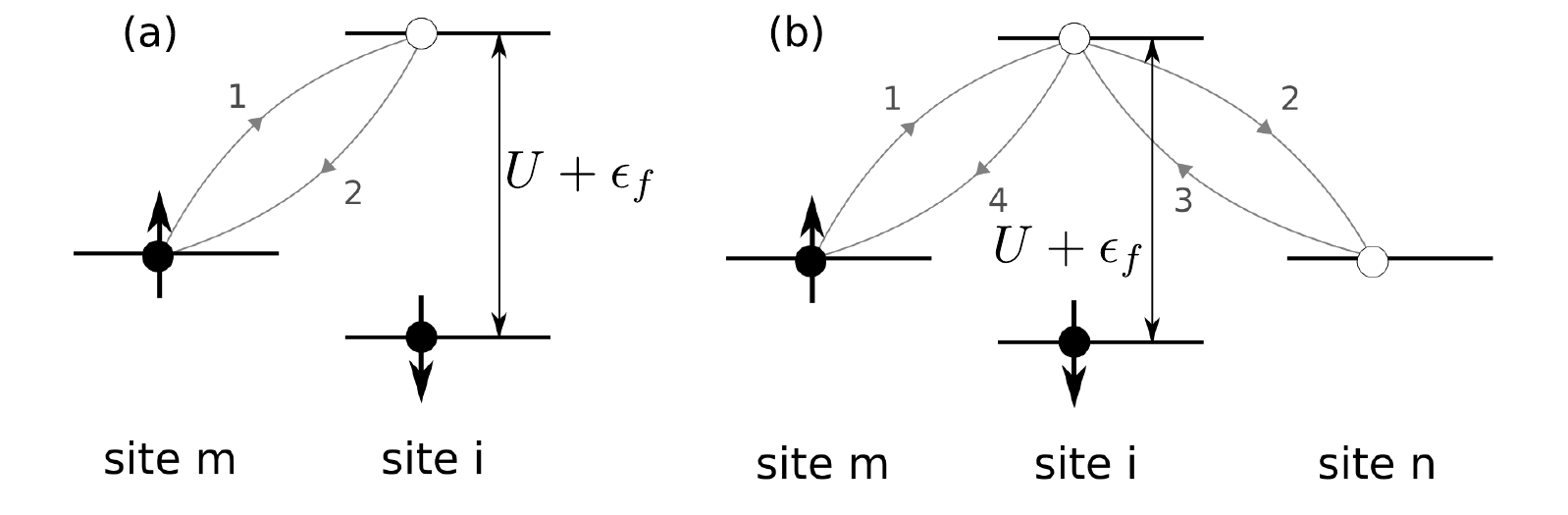}
\end{center}
\caption{Virtual processes in the second (a) and fourth (b) orders of the canonical perturbation expansion for the periodic Anderson model. After~\cite{KadzielawaMajorActaPhysPolon2014}.}
\label{fig:cpe_processes_EKM}
\end{figure}

In Appendices~\ref{appendix:derivation_of_the_tj_model} and \ref{appendix:sga_and_slave_bosons} we have dealt with effective single-band Hamiltonians for the cuprates. In those treatments, the $2p$-$3d$ hybridization of the relevant $d_{x^2-y^2}$ and $2p_x$/$2p_y$ has been reduced to their influence on the $d$ states. In the case of heavy-fermion systems, this hybridization is taken into account explicitly, since the $f$ electrons (in, e.g., $\mathrm{Ce}^{3+}$) ions are located close to the Fermi level due to the $c$ carriers. In the situation with strong correlations, the $f$-$f$ intraatomic interaction is by far the largest parameter in the system. The two above conditions lead to the situation that the Kondo-lattice model and the general Anderson-lattice model should not be equivalent, as in the appearing then $f$-electron itinerancy requires that only the total number of electrons $n_f + n_c$ is conserved, but not $n_f$ and $n_c$ separately, which is the case in the Kondo-lattice model. In other words, the lattice generalization of the Schrieffer-Wolff of the Anderson-lattice Hamiltonian into its Kondo-lattice form is inapplicable. Below we summarize our earlier results \cite{SpalekJPhysFrance1989,HowczakPSS2013,KadzielawaMajorActaPhysPolon2014} which provide the answer how a modified Schieffer-Wolff transformation can be carried out and what is the effective Hamiltonian in this case, containing both effective Kondo and superexchange interaction up to the fourth order, carried out with respect to hybridization to $U$ ratio, $V/U$ (cf. Fig.~\ref{fig:cpe_processes_EKM}).

\subsection{The canonical perturbation expansion in heavy-fermion limit}

As already mentioned in Sec.~\ref{sec:theoretical_models}, we start with decomposition of the Hamiltonian~\eqref{eq:anderson_model} with the chemical potential included explicitly, i.e.,

\begin{align}
  \label{eq:anderson_model_appendix}
  \hat{\mathcal{H}} = \sum_{ij\sigma} \left(t_{ij} - \mu \delta_{ij}\right) \hat{c}_{i\sigma}^\dagger \hat{c}_{j\sigma} + \left(\epsilon_f - \mu\right) \sum_{i\sigma} \hat{f}_{i\sigma}^\dagger \hat{f}_{i\sigma} + U \sum_i \hat{n}_{i\uparrow}^f \hat{n}_{i\downarrow}^f + \sum_{ij\sigma} V_{ij} \left(\hat{f}_{i\sigma}^\dagger \hat{c}_{j\sigma} + \hat{c}_{i\sigma}^\dagger \hat{f}_{j\sigma}\right).
\end{align}

\noindent
Next, we decompose the last term into low- and high-energy processes (cf.~Fig.~\ref{efU}) according to

\begin{align}
  \label{eq:high_low_en_decomposition}
  \hat{f}^\dagger_{i\sigma} \hat{c}_{m\sigma} \equiv \left(1 - \hat{n}_{i\sigma}^f + \hat{n}_{i\sigma}^f \right) \hat{f}^\dagger_{i\sigma} \hat{c}_{m\sigma} = \left(1 - \hat{n}_{i\sigma}^f\right) \hat{f}^\dagger_{i\sigma} \hat{c}_{m\sigma} + \hat{n}_{i\sigma}^f  \hat{f}^\dagger_{i\sigma} \hat{c}_{m\sigma},
\end{align}

\noindent
and similarly for the Hermitian conjugate term. Next, we define the projection operators which single out the subspaces with zero, one, two, etc. double $f$-electron state occupancies. They are

\begin{align}
	\hat{{P}}_0 &= \prod_{i} \left( \left( 1 - \hat{n}^f_{i \uparrow} \right) \left( 1 - \hat{n}^f_{i \downarrow} \right) + \sum_{\sigma}^{} \hat{n}^f_{i \sigma} \left( 1 - \hat{n}^f_{i \bar{\sigma}} \right)	\right) = \prod_i \left( 1 - \hat{n}^f_{i \uparrow} \hat{n}^f_{i \downarrow}  \right),               \label{eq:P0_Anderson_lattice}
	\\
	\hat{{P}}_1 &= \sum_j \left( \hat{n}^f_{j \uparrow} \hat{n}^f_{j \downarrow} \times \prod_{i \neq j} \left( 1 - \hat{n}^f_{i \uparrow} \hat{n}^f_{i \downarrow}  \right) \right),           \label{eq:P1_Anderson_lattice}
	\\
  \hat{{P}}_2 &= \sum_{\substack{j,k \\ j \neq k}} \left( \hat{n}^f_{j \uparrow} \hat{n}^f_{j \downarrow} \hat{n}^f_{k \uparrow} \hat{n}^f_{k \downarrow} \times \prod_{\substack{i \neq j \\ i \neq k}} \left( 1 - \hat{n}^f_{i \uparrow} \hat{n}^f_{i \downarrow}  \right) \right).                  \label{eq:P2_Anderson_lattice}
\end{align}

\begin{figure}
\begin{center}
\includegraphics[width=0.4\textwidth]{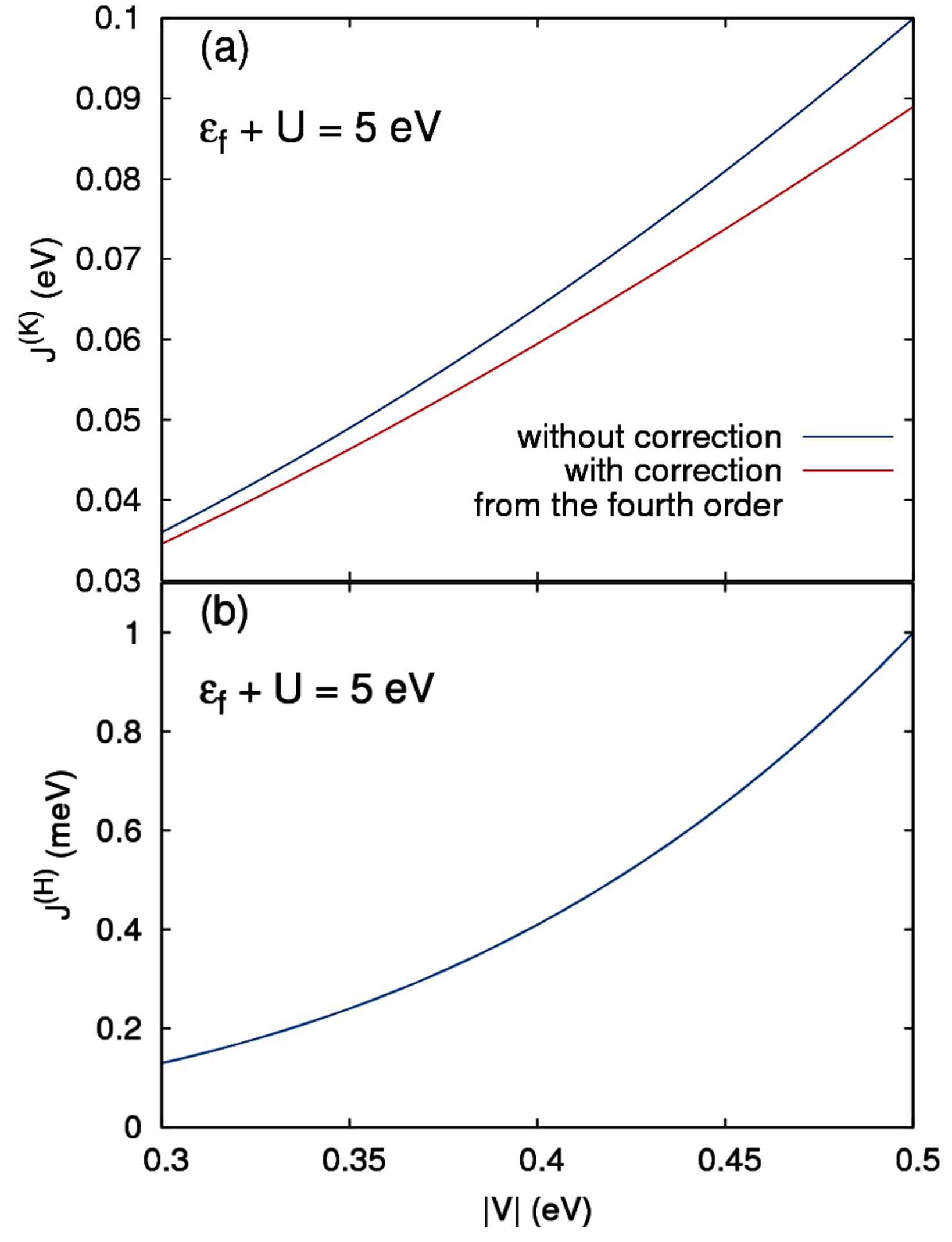}
\end{center}
\caption{(a) Values of the Kondo exchange $J^{(K)}$, with and without correction from the fourth expansion order included (blue and red lines, respectively). (b) The superexchange integral $J^{H}$ (b). The data are plotted vs. the hybridization magnitude, $|V|$, and for $U + \epsilon_f = 5\,\mathrm{eV}$. Note the eV scale for $J^{K}$ and meV for $J^{H}$. After~\cite{KadzielawaMajorActaPhysPolon2014}.}
\label{fig:cpe_values_of_integrals_EKM}
\end{figure}

\noindent
Now, the further procedure is to remove by the canonical perturbation expansion only the high-energy process in Eq.~\eqref{eq:high_low_en_decomposition} and replace them by virtual hopping processes which lead to the effective Hamiltonian with exchange interactions of the Kondo ($c$-$f$) and superexchange ($f$-$f$) type. Such a procedure has been proposed some time ago \cite{SpalekJPhysFrance1989} and is analogical to that employed in Appendix~\ref{appendix:derivation_of_the_tj_model} to derive the $t$-$J$ model. It should be distinguished from the Schrieffer-Wolff-type of transformation \cite{SchriefferPhysRev1966}, where whole hybridization term is transformed out and hence reflects the physical situation in which the $f$-electron spins are localized and coupled to $c$-carriers only via Kondo interaction. By contrast, within the present formulation the low-energy hybridization survives and leads to the heavy $f$-electron quasiparticles in the limit when $n_f = 1-\delta$, and $\delta < 0.1$. Physically, the effective Hamiltonian, obtained within such a procedure, can be calculated systematically up to the fourth order in $V/U + \epsilon_f$ and takes the form of Eq.~\eqref{eq:anderson_hamiltonian_fourth_order_canonical_transformation} in the main text. It contains the Kondo, superexchange, as well as the terms of the Dzialoshinskii-Moriya-type of purely electronic character. The corresponding exchange integrals have been evaluated explicitly \cite{KadzielawaMajorActaPhysPolon2014} and drawn in Fig.~\ref{fig:cpe_values_of_integrals_EKM}. One sees that the fourth-order contribution reduces moderately the antiferromagnetic character of the Kondo-type interaction.

\begin{figure}
\begin{center}
\includegraphics[width=0.4\textwidth]{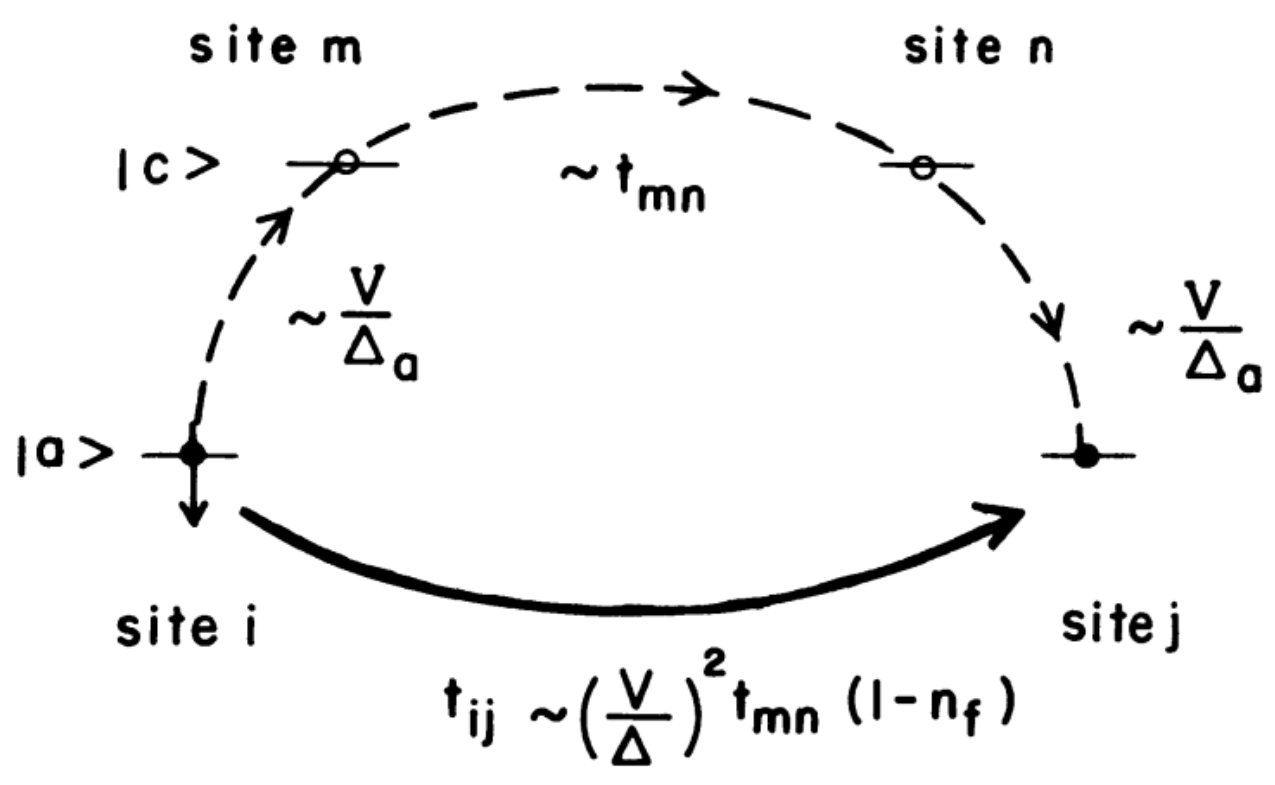}
\end{center}
\caption{Schematic representation of the $f$-$f$ hopping via the conduction ($c$) states. This hopping is induced by residual hybridization. Here $\Delta_a = \epsilon_f - \mu$. After~\cite{SpalekJPhysFrance1989}.}
\label{fig:cpe_residual_hopping}
\end{figure}

One feature of the results for the effective Hamiltonian should be stressed. The effective Hamiltonian with the residual hybridization term

\begin{align}
  \label{eq:residual_hybrydization_term}
  \sum_{i,m} \left( V_{im} (1- \hat{n}_{i\bar{\sigma}}) \hat{f}^\dagger_{i\sigma} \hat{c}_{m\sigma} + \mathrm{H.c.}\right)
\end{align}

\noindent
leads implicitly to the very narrow band behavior of $f$-electrons. The corresponding process in fourth order is schematically sketched in Fig.~\ref{fig:cpe_residual_hopping} with the $f$-$f$ hopping $t^f_{ij}$ marked. It is the band narrowing factor $\propto (1- n_f)$ in conjunction with the narrow width that leads to extremely high density of states at the system Fermi energy and, as a consequence, very heavy masses of quasiparticles with enhancement factor $m^{*}/m_e \sum 10^2$.

The real space pairing, containing both the hybrid $f$-$c$ pairing \cite{SpalekPhysRevB1988} and $f$-$f$ pairings. leads to two-gap behavior discussed in Sec.~\ref{subsec:basic_properties_heavy_fermions}.

\section{Details of DE-GWF expansion: Real-space \textit{versus} $\mathbf{k}$-space versions}
\label{appendix:de-gwf}

In this Appendix we provide supplemental technical information on the diagrammatic expansion of the Gutzwiller wave function in real-space (DE-GWF), as well as in $\mathbf{k}$-space ($\mathbf{k}$-DE-GWF). The structure of the diagrammatic contributions to the variational energy functional $E_G$ was characterized in Sec.~\ref{sec:vwf_solution}. Specifically, $E_G$ was expressed in terms of several diagram classes, including $T_{ij}^{11}$, $T_{ij}^{13}$, $T_{ij}^{33}$, $S_{ij}^{11}$, $S_{ij}^{13}$, $S_{ij}^{33}$, $I_i^{2}$, $I_i^{4}$. The latter are defined explicitly as infinite sums

\begin{align}
  T^{11}_{ij} &= \sum \limits_{k=0}^{\infty} \frac{x^k}{k!} \sideset{}{'}\sum_{l_1, \ldots, l_k}  \langle\Psi_0| \hat{a}^\dagger_{i\uparrow} \hat{a}_{j\uparrow} \hat{d}^\mathrm{HF}_{l_1} \ldots \hat{d}^\mathrm{HF}_{l_k}|\Psi_0\rangle_c, \label{eq:t11} \\
  T^{13}_{ij} &= \sum \limits_{k=0}^{\infty} \frac{x^k}{k!} \sideset{}{'}\sum_{l_1, \ldots, l_k} \langle\Psi_0| \hat{a}^\dagger_{i\uparrow} \hat{a}_{j\uparrow} \hat{n}^\mathrm{HF}_{j\downarrow} \hat{d}^\mathrm{HF}_{l_1} \ldots \hat{d}^\mathrm{HF}_{l_k}|\Psi_0\rangle_c, \\
  T^{33}_{ij} &= \sum \limits_{k=0}^{\infty} \frac{x^k}{k!} \sideset{}{'}\sum_{l_1, \ldots, l_k} \langle\Psi_0| \hat{a}^\dagger_{i\uparrow}\hat{n}^\mathrm{HF}_{i\downarrow} \hat{a}_{j\uparrow} \hat{n}^\mathrm{HF}_{j\downarrow} \hat{d}^\mathrm{HF}_{l_1} \ldots \hat{d}^\mathrm{HF}_{l_k}|\Psi_0\rangle_c, \label{eq:T33_definition}\\
  S^{11}_{ij} &= \sum \limits_{k=0}^{\infty} \frac{x^k}{k!} \sideset{}{'}\sum_{l_1, \ldots, l_k} \langle\Psi_0| \hat{a}_{i\uparrow} \hat{a}_{j\downarrow} \hat{d}^\mathrm{HF}_{l_1} \ldots \hat{d}^\mathrm{HF}_{l_k}|\Psi_0\rangle_c, \\
  S^{13}_{ij} &= \sum \limits_{k=0}^{\infty} \frac{x^k}{k!} \sideset{}{'}\sum_{l_1, \ldots, l_k} \langle\Psi_0| \hat{a}_{i\uparrow} \hat{a}_{j\downarrow} \hat{n}^\mathrm{HF}_{j\uparrow} \hat{d}^\mathrm{HF}_{l_1} \ldots \hat{d}^\mathrm{HF}_{l_k}|\Psi_0\rangle_c, \\
  S^{33}_{ij} &= \sum \limits_{k=0}^{\infty} \frac{x^k}{k!} \sideset{}{'}\sum_{l_1, \ldots, l_k} \langle\Psi_0| \hat{a}_{i\uparrow}\hat{n}^\mathrm{HF}_{i\downarrow} \hat{a}_{j\downarrow} \hat{n}^\mathrm{HF}_{j\uparrow} \hat{d}^\mathrm{HF}_{l_1} \ldots \hat{d}^\mathrm{HF}_{l_k}|\Psi_0\rangle_c, \label{eq:T33_definition}\\
  S^{22}_{ij} &= \sum \limits_{k=0}^{\infty} \frac{x^k}{k!}  \sideset{}{'}\sum_{l_1, \ldots, l_k} \langle\Psi_0| \hat{S}^{+}_i \hat{S}^{-}_j \hat{d}^\mathrm{HF}_{l_1} \ldots \hat{d}^\mathrm{HF}_{l_k}|\Psi_0\rangle_c, \\
  I_i^{2} &= \sum \limits_{k=0}^{\infty} \frac{x^k}{k!} \sideset{}{'}\sum_{l_1, \ldots, l_k} \langle\Psi_0| \hat{n}_{i\sigma}^\mathrm{HF} \hat{d}^\mathrm{HF}_{l_1} \ldots \hat{d}^\mathrm{HF}_{l_k}|\Psi_0\rangle_c, \\
  I_i^{4} &= \sum \limits_{k=0}^{\infty} \frac{x^k}{k!}  \sideset{}{'}\sum_{l_1, \ldots, l_k}  \langle\Psi_0| \hat{d}_i^\mathrm{HF} \hat{d}^\mathrm{HF}_{l_1} \ldots \hat{d}^\mathrm{HF}_{l_k}|\Psi_0\rangle_c, \label{eq:i4}
\end{align}

\noindent
where primed summation means that $l_m$, $i$, and $j$ must be all different, and the subscript $c$ indicates that only connected diagrams should be retained in Wick's decomposition of the expectation values. The  particle-number and double occupancy operators with the subscript ``HF'' (or Hartree-Fock) in those expressions are modified in such a way that the local tadpole diagrams cancel out (cf. Sec.~\ref{sec:vwf_solution}). In brief, to evaluate expressions \eqref{eq:t11}-\eqref{eq:i4}, one needs to apply Wick's theorem and discard all disconnected graphs and those involving tadpoles.

The above procedure reduces then to a combinatorial problem of finding inequivalent diagrams and their multiplicities. Each of such graphs consists of internal vertices (corresponding to the operators $\hat{d}_i$, and the external ones, related to the operators acting on the sites $i$ and $j$). Those vertices are connected by lines corresponding to the two-site expectation values, such as $\langle{\hat{a}_{i\sigma}^\dagger\hat{a}_{j\sigma^\prime}}\rangle$, $\langle{\hat{a}_{i\sigma}\hat{a}_{j\sigma^\prime}}\rangle$, and $\langle{\hat{a}_{i\sigma}^\dagger\hat{a}_{j\sigma^\prime}^\dagger}\rangle$. Depending on the symmetry of the problem, some classes of lines may be absent, e.g., in the paramagnetic state with no spin-flip terms in the Hamiltonian, only one family of lines prevails ($\langle{\hat{a}_{i\sigma}^\dagger\hat{a}_{j\sigma}}\rangle$).

  \begin{table}
    \caption{Number of diagrams contributing to selected diagrammatic sums out of \eqref{eq:t11}-\eqref{eq:i4} ($N_\mathrm{diag}$), as well as the corresponding to them $\mathbf{k}$-space  integral dimensions ($D$) in the paramagnetic (PM) and $d$-wave superconducting ($d$-SC) states up to fourth expansion order ($k \leq 4$). The number  $N_\mathrm{total}$ is the total diagram number at given order. After Ref.~\cite{FidrysiakJPhysCondensMatter2018}.}
    \label{table:diagram_numbers}
    \centering
  \begin{tabular}{ c  c  c c c c c  c c c c c}
    \hline\hline
   \multicolumn{2}{c}{} & \multicolumn{5}{c  }{PM phase} & \multicolumn{5}{c}{$d$-SC phase} \\
   \multicolumn{2}{c}{} & \multicolumn{5}{c  }{Expansion order, $k$} & \multicolumn{5}{c}{Expansion order, $k$} \\
   \multicolumn{2}{c}{} & $0$ & $1$ & $2$ & $3$ & $4$ & $0$ & $1$ & $2$ & $3$ & $4$ \\
    \hline \hline 
    $T^{11}_{ij}$ &  $N_\mathrm{diag}$ & 1 & 0 & 1 & 2 & 7 & 1 & 0 & 8 & 40 & 644 \\
                        & $D$ & 2 & N/A  & 6  & 8  & 10  & 2  & N/A  & 6  & 8  & 10  \\
    $T^{13}_{ij}$ & $N_\mathrm{diag}$ & 0 & 1 & 2 & 7 & 27 & 0 & 4 & 20 & 322 & 3262 \\
                            & $D$ & N/A & 6 & 8 & 10 & 12 & N/A & 6 & 8 & 10 & 12 \\
    $T^{33}_{ij}$ & $N_\mathrm{diag}$ & 1 & 1 & 5 & 20 & 91 & 2 & 4 & 133 & 1238 & 27187 \\
                            & $D$ & 6 & 8 & 10 & 12 & 14 & 6 & 8 & 10 & 12 & 14 \\
    $S^{22}_{ij}$ & $N_\mathrm{diag}$ & 1 & 1 & 3 & 8 & 34 & 2 & 4 & 36 & 306 & 4388 \\
                            & $D$ & 2 & 4 & 6 & 8 & 12 & 2 & 4 & 6 & 8 & 12 \\
    $I^2_i$ &  $N_\mathrm{diag}$ & 0 & 0 & 1 & 2 & 7 & 0 & 0 & 6 & 28 & 393 \\
                            & $D$ & N/A & N/A & 6 & 8 & 10 & N/A & N/A & 6 & 8 & 10 \\
    $I^4_i$ &  $N_\mathrm{diag}$ & 0 & 1 & 1 & 3 & 12 & 0 & 3 & 6 & 103 & 888 \\
                            & $D$ & N/A & 6 & 8 & 10 & 12 & N/A & 6 & 8 & 10 & 12 \\
                        & $N_\mathrm{total}$ & 3 & 4 & 13 & 42  & 178  & 5  & 15  & 209  & 2037 & 36762  \\
    \hline\hline
  \end{tabular}
\end{table}

In Table~\ref{table:diagram_numbers}, we summarize the numbers ($N_\mathrm{diag}$) of graphs contributing to respective families up to fourth expansion order ($k \leq 4$), both in the paramagnetic (PM) and $d$-wave superconducting state ($d$-SC). In the former case, there is only one family of lines ($\langle{\hat{a}_{i\sigma}^\dagger\hat{a}_{j\sigma}}\rangle$), whereas in the latter both paramagnetic and anomalous lines prevail ($\langle{\hat{a}_{i\sigma}^\dagger\hat{a}_{j\sigma}}\rangle$ and $\langle{\hat{a}_{i\sigma}^\dagger\hat{a}_{j\bar{\sigma}}^\dagger}\rangle$, respectively). Note that other anomalous line are related to $\langle{\hat{a}_{i\sigma}^\dagger\hat{a}_{j\bar{\sigma}}^\dagger}\rangle$ by symmetry). This results in substantially larger number of graphs in the $d$-SC phase (total number of diagrams at given expansion order is listed as $N_\mathrm{tot}$). In addition to the numbers, we also provide dimensions of $\mathbf{k}$-space integrals arising in calculations using $\mathbf{k}$-DE-GWF variant of the method, with N/A indicating no diagrams at given order \cite{FidrysiakJPhysCondensMatter2018}. By inspection of Table~\ref{table:diagram_numbers}, it follows that high-order computations become computationally prohibitive with increasing order, $k$. In practice, $k=5$ is attainable within the real-space variant of DE-GWF, and $k=3$ within $\mathbf{k}$-DE-GWF.

Finally, in Figs.~\ref{fig:graphs1}-\ref{fig:graphs6} we show explicitly the structure of low-order diagrams from families~\eqref{eq:t11}-\eqref{eq:i4} in the $d$-wave superconducting state. The green circles represent external vertices, $i$ and $j$, whereas pink circles are the internal ones. Solid black and dashed blue lines represent paramagnetic and anomalous (superconducting) lines, respectively. Numbers given next to diagrams are respective combinatorial factors, including the sign due to Fermi statistics. The value of $k$ inside the plot in the expansion order, and the empty set symbol means that there are no graphs in given category. Note that the number of diagrams is consistent with the data presented in Table~\ref{table:diagram_numbers}.

\begin{minipage}{1.0\linewidth}
    \centering
    \includegraphics[width=0.95\textwidth]{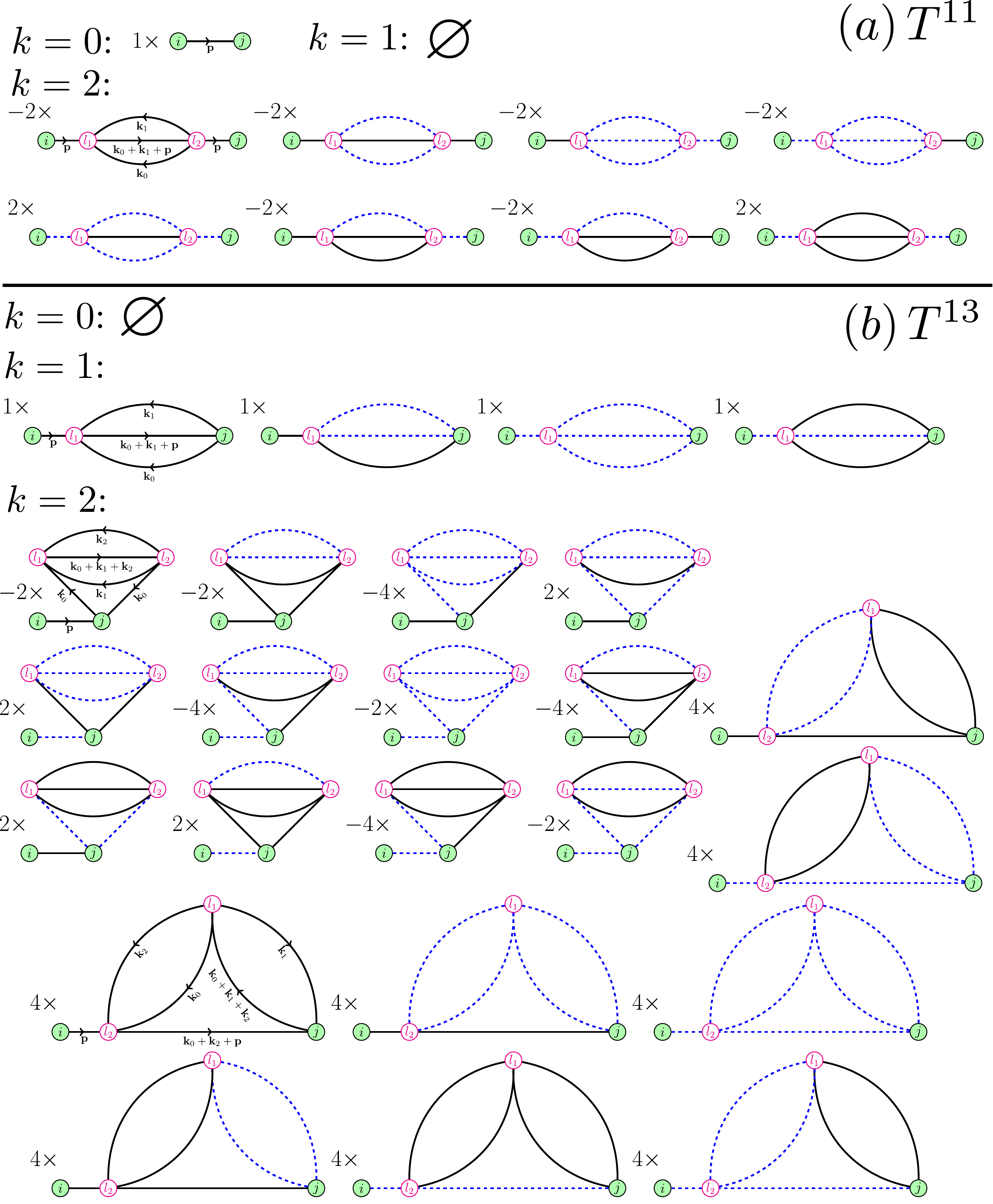}.
    \captionof{figure}{Diagrams contributing to the expectation values $T^{11}$ (a) and $T^{13}$ (b) up to the second expansion order ($k = 0, 1, 2$). Black solid  and blue dashed lines represent normal and anomalous (superconducting) lines, respectively. External vertices, $i$ and $j$, are marked as shaded circles, whereas empty circles correspond to internal vertices, $l_i$. The summation over indices $l_i$ is carried out in real-space within the DE-GWF method. The external ($\mathbf{p}$) and internal wave vectors ($\mathbf{k}_i$), along with the corresponding quasi-momentum flow direction (used in $\mathbf{k}$-DE-GWF method), are marked on representative graphs of distinct topologies. Numbers next to the graphs are combinatorial factors. Empty set symbol $\varnothing$ indicates that no graphs contribute at given order.}
    \label{fig:graphs1}
  \end{minipage}
  \newpage

\begin{minipage}{1.0\linewidth}
    \centering
    \includegraphics[width=\textwidth]{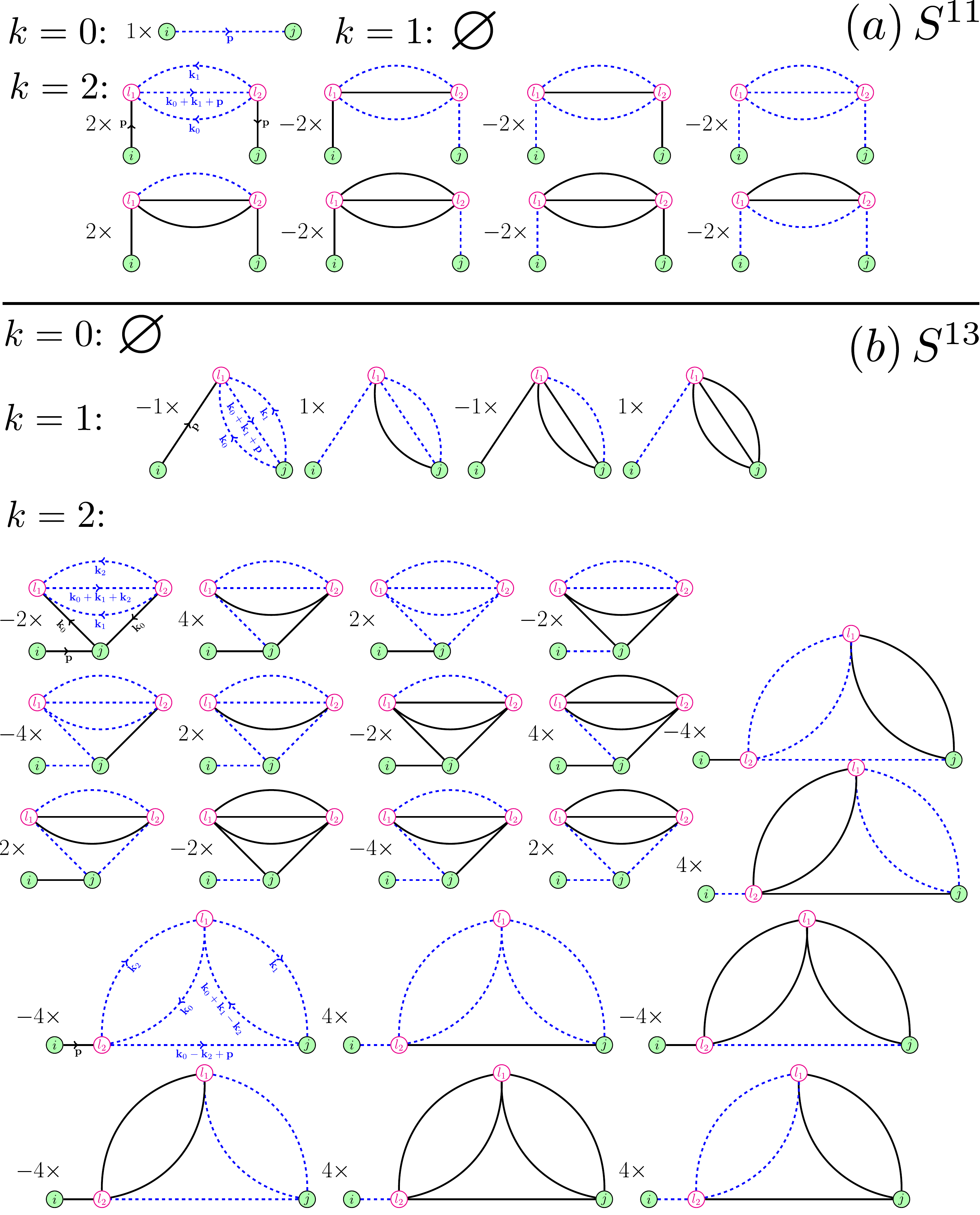}.
    \captionof{figure}{Diagrams contributing to the expectation values $S^{11}$ (a) and $S^{13}$ (b) up to the second expansion order ($k = 0, 1, 2$). The meaning of symbols is the same as in Fig.~\ref{fig:graphs1}.}
    \label{fig:graphs2}
  \end{minipage}
  \newpage
  
\begin{minipage}{1.0\linewidth}
    \centering
    \includegraphics[width=\textwidth]{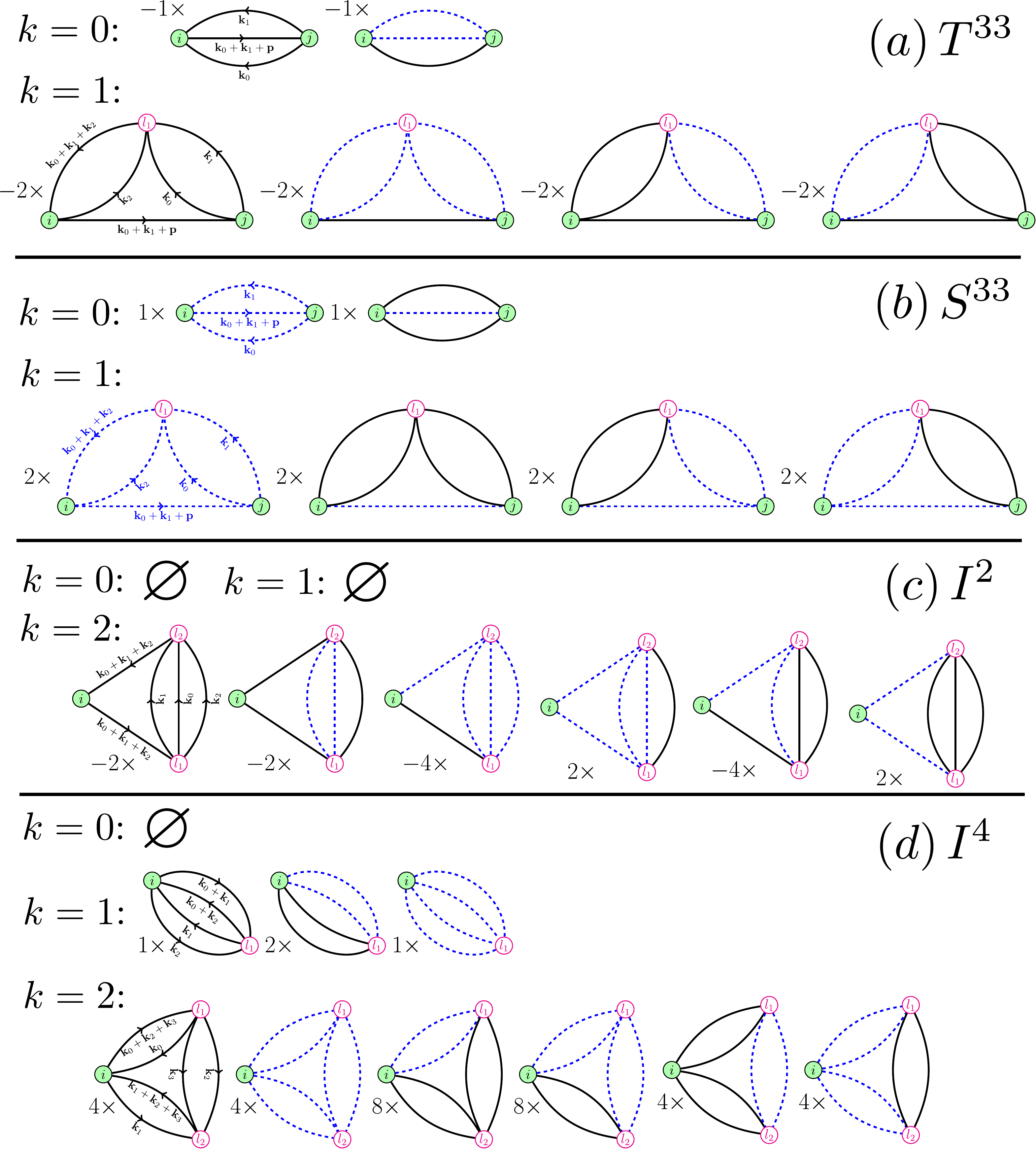}.
    \captionof{figure}{Diagrams contributing to the expectation values $T^{33}$ (a) and $S^{33}$ (b) up to the first expansion order ($k = 0, 1$), as well as $I_2$ (c) and $I^4$ (d) up to the second expansion order ($k = 0, 1, 2$). The meaning of symbols is the same as in Fig.~\ref{fig:graphs1}. The number of graphs contributing to $T^{33}$ and $S^{33}$ at second order ($k = 2$) is substantial, hence they are displayed as separate graphs (Figs.~\ref{fig:graphs5} and \ref{fig:graphs6}).}
    \label{fig:graphs3}
  \end{minipage}
  \newpage
  
  \begin{minipage}{1.0\linewidth}
    \centering
    \includegraphics[width=\textwidth]{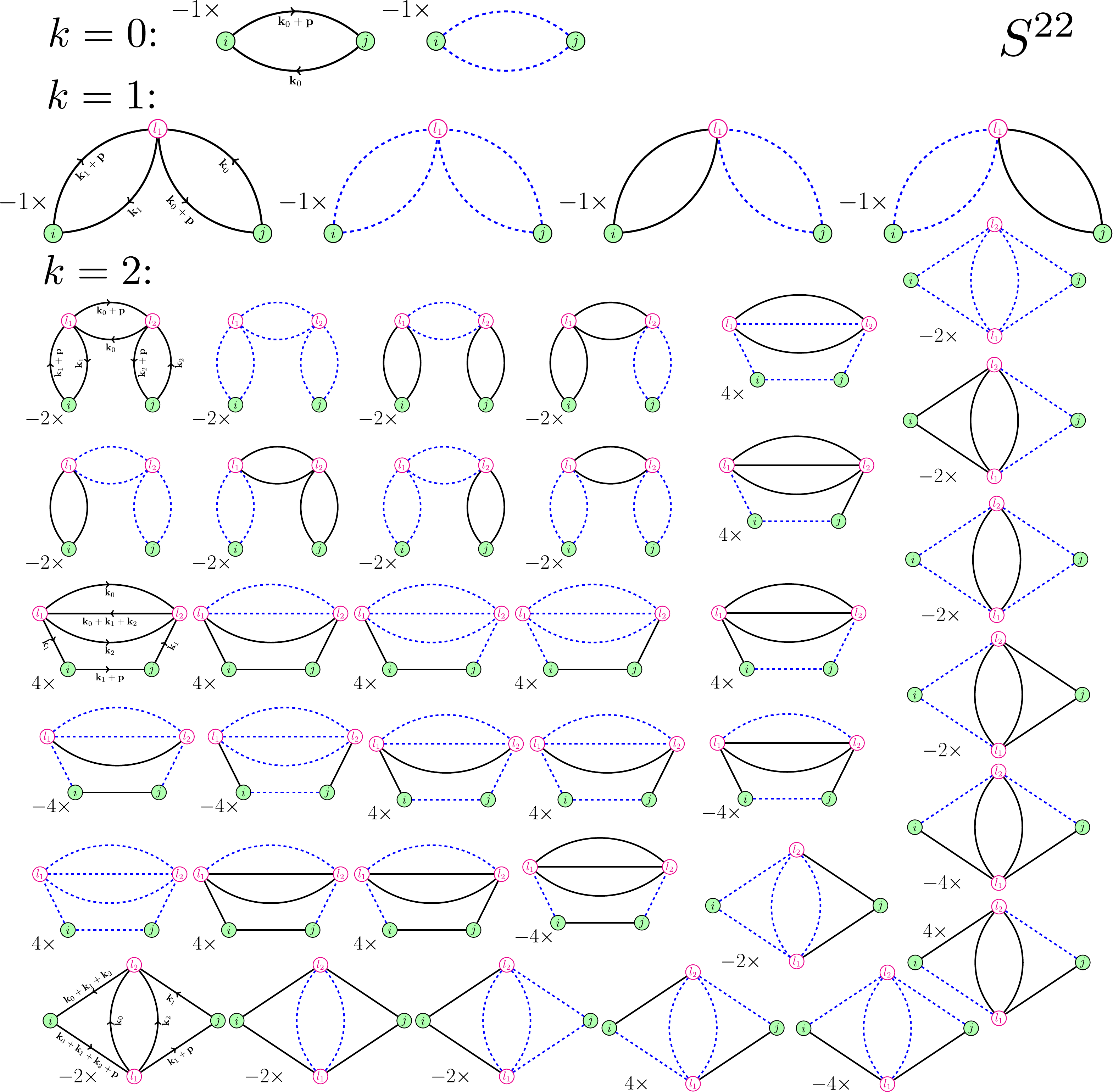}.
    \captionof{figure}{Diagrams contributing to the expectation value $S^{22}$ up to the second expansion order ($k = 0, 1, 2$). The meaning of symbols is the same as in Fig.~\ref{fig:graphs1}.}
    \label{fig:graphs4}
  \end{minipage}
    \newpage

  \begin{minipage}{1.0\linewidth}
    \centering
    \includegraphics[width=\textwidth]{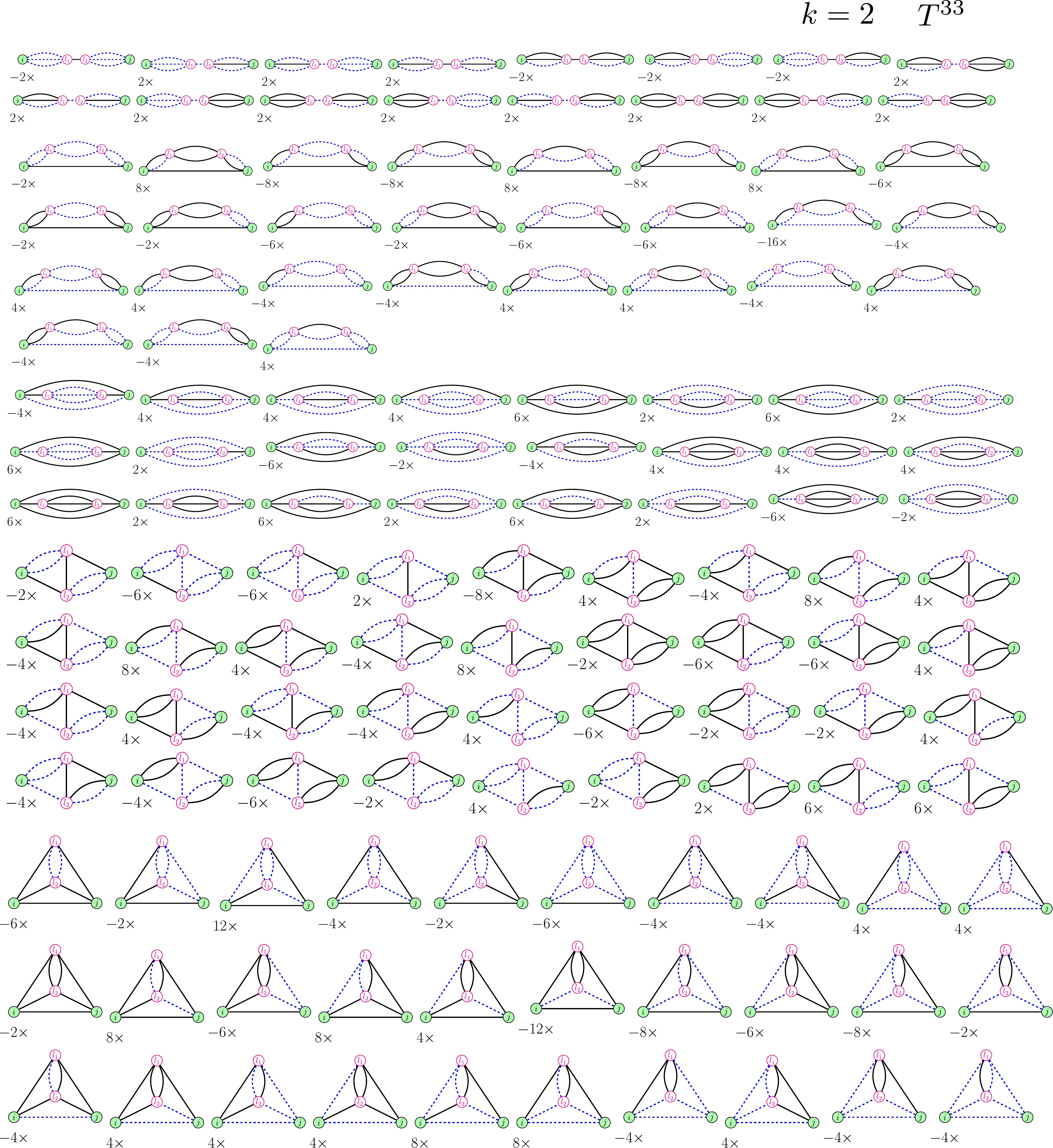}.
    \captionof{figure}{Diagrams contributing to the expectation value $T^{33}$ at the second expansion order ($k = 2$). The meaning of symbols is the same as in Fig.~\ref{fig:graphs1}.}
    \label{fig:graphs5}
  \end{minipage}
    \newpage

  \begin{minipage}{1.0\linewidth}
    \centering
    \includegraphics[width=0.98\textwidth]{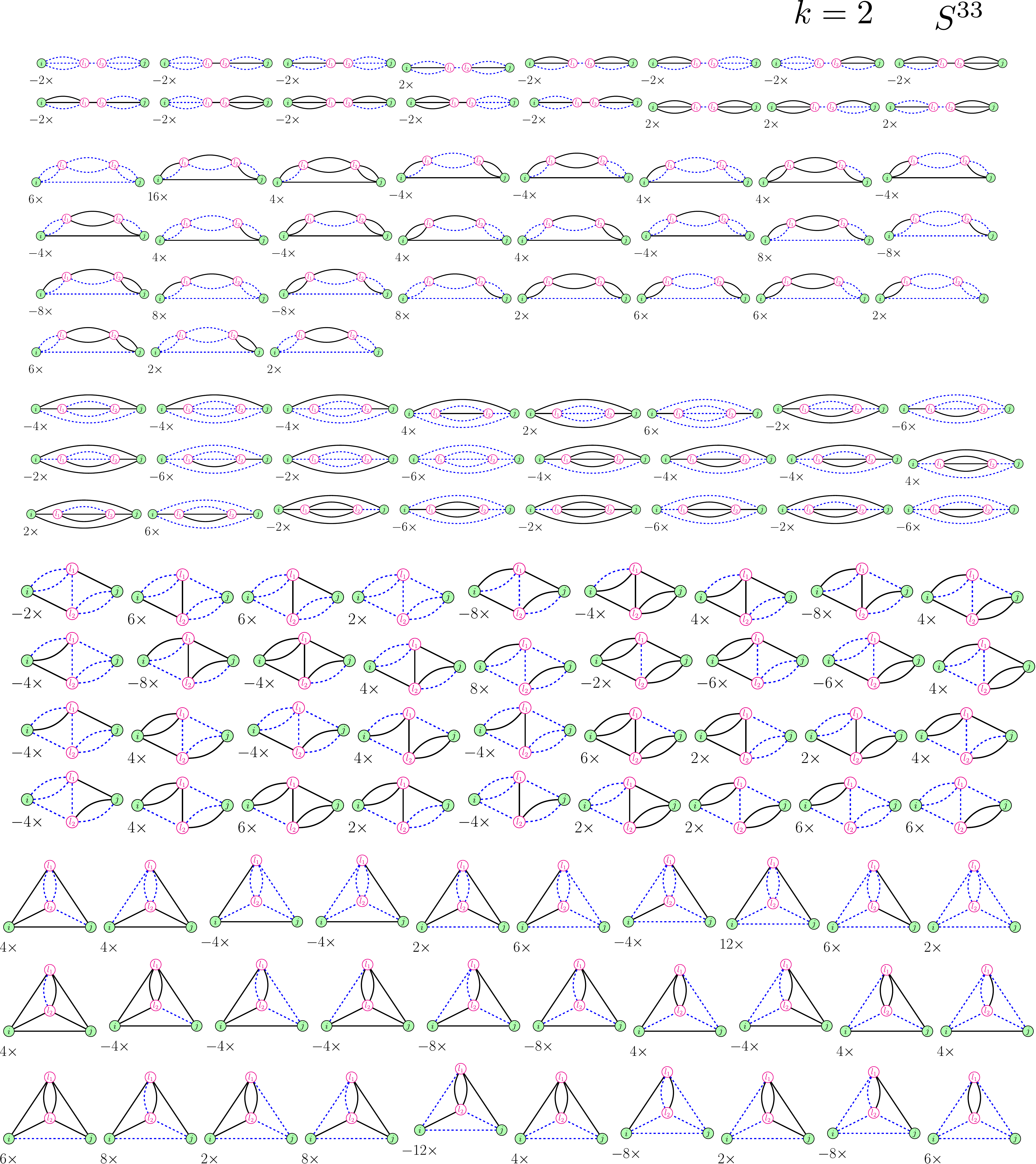}.
    \captionof{figure}{Diagrams contributing to the expectation value $S^{33}$ at the second expansion order ($k = 2$). The meaning of symbols is the same as in Fig.~\ref{fig:graphs1}.}
    \label{fig:graphs6}
  \end{minipage}

\newpage

\end{document}